\documentclass[oneside,12pt]{extbook}
\usepackage{amssymb,epsfig}
\usepackage{amsfonts}
\usepackage{amsmath}
\usepackage[T2A]{fontenc}
\usepackage[cp1251]{inputenc}
\usepackage{thesis}
\usepackage{float}
\usepackage{afterpage}

\newcommand{\dd}{\mbox{\rm d}}

\newcommand{\beqn}{\begin{eqnarray}}
\newcommand{\eeqn}{\end{eqnarray}}

\def\mys#1{\Sigma_{ } { }_{\bf #1}}
\def\bc{\begin{center}}
\def\ec{\end{center}}
\def\vp{\vspace*{0.2cm}}
\def\vm{\vspace*{-0.2cm}}

\newfont{\cyrfnt}{wncyr10 scaled 1150}

\newfont{\cyrsm}{wncyr10 scaled 1050}

\newfont{\cyrbf}{wncyb10 scaled 1350}

\newfont{\cyrnf}{wncyr10 scaled 1350}

\newfont{\cyrB}{wncyb10 scaled 1740}

\newfont{\cyrH}{wncyb10 scaled 2500}

\newfont{\cyrI}{wncyi10 scaled 1440}

\begin{document}

\thispagestyle{empty}

\vspace{6ex}
\begin{center}

{\large  \bf  National Academy of Sciences of Ukraine  \\
Bogolyubov Institute for  Theoretical Physics } \\

\end{center}

\vspace{4ex}
\hspace*{10cm}
{\small  Has  the rights of a manuscript}

\begin{center}

\vspace{4ex}
{\Large \bf    Bugaev Kyrill Alekseevich} \\

\end{center}

\vspace{4ex}
\hspace*{8cm}
{\bf UDC:} \ \ 532.51; 533.77; 539.125/126; 544.586.6

\begin{center}

\vspace{8ex}
{\huge\bf     Equation of State and Phase Transitions} \\

{\huge\bf     in the Nuclear and Hadronic Systems}\\

\vspace{6ex}
{\small  Speciality  01.04.02 - theoretical physics}

\vspace{8ex}
{ DISSERTATION}\\
{ to receive  a scientific degree of }\\
{the Doctor of Science in  physics and mathematics}\\

\vspace{30ex}
\noindent
{Kiev} - 2009
\
\end{center}


\newpage

\centerline{\bf Abstract}

\begin{minipage}[t]{16.2cm} 
An investigation of strongly interacting matter equation of state remains one of the major tasks of modern high energy nuclear physics for almost a quarter of century. The present work is my doctor of science thesis which contains my contribution (42 works) to this field made between 1993 and 2008. \\

Inhere I mainly discuss the common physical and mathematical features
of several exactly solvable statistical models which describe the nuclear liquid-gas phase transition and the deconfinement phase transition. Luckily, in some  cases it was possible to rigorously extend the solutions found in thermodynamic limit to finite volumes and to formulate the finite volume analogs of phases directly from the grand canonical partition. It turns out that finite volume (surface) of a system  generates also the temporal constraints, i.e. the finite formation/decay time of possible states in this finite system. \\

Among other results I would like to mention the calculation of upper and lower bounds for the surface entropy of physical clusters within the Hills and Dales model; evaluation of the second virial coefficient which accounts  
for the Lorentz contraction of the hard core repulsing potential between hadrons; inclusion of large width of heavy quark-gluon bags into statistical description. \\

I believe that the suggested mathematical solution of the freeze-out problem in relativistic hydrodynamic model and in hydro-cascade model has not only an academic interest, but also has some practical value. In addition I hope that the experience gained in working out some partly successful signals of deconfinement transition can be useful for other researchers to go further in this direction. 
\end{minipage}

\tableofcontents

\newpage

\chapter*{Abbreviations}

\addcontentsline{toc}{chapter}{Abbreviations}

\begin{enumerate}

\item
PT \hspace*{2.25cm} phase transition
\item
A+A \hspace*{1.85cm} nuclear heavy-ion (collision; reaction)
\item
QGP \hspace*{1.8cm} quark-gluon plasma
\item
EOS \hspace*{2.0cm} equation of state 
\item
HS \hspace*{2.25cm} hypersurface
\item
FO \hspace*{2.25cm} freeze-out 
\item
SMM \hspace*{1.8cm} Statistical Multifragmentation Model
\item
FDM \hspace*{1.8cm} Fisher Droplet  Model
\item
GBM \hspace*{1.8cm} Gas of Bags  Model
\item
QGBST \hspace*{1.3cm} Quark-Gluon Bags with Surface Tension 
\item
GCE \hspace*{1.95cm} grand canonical ensemble
\item
HDM \hspace*{1.8cm} Hills and Dales Model
\item
VdW \hspace*{1.8cm} Van der Waals 
\item
GCSP \hspace*{1.75cm} grand canonical surface partition
\item
CCSP \hspace*{1.75cm} canonically constrained surface partition
\item
SGCSP \hspace*{1.5cm} semi-grand canonical surface partition
\item
CSMM \hspace*{1.65cm}  constrained Statistical Multifragmentation Model
\item
GSMM \hspace*{1.65cm}  generalized Statistical Multifragmentation Model
\item
SBM \hspace*{2.1cm} Statistical Bootstrap  Model
\item
HTM \hspace*{1.95cm} Hagedorn Thermostat   Model
\item
BD \hspace*{2.35cm} Bass-Dumitru 
\item
TLS \hspace*{2.15cm} Teaney-Lauret-Shuryak 
\item
t.l. \hspace*{2.4cm} time-like
\item
s.l. \hspace*{2.4cm} space-like
\item
RFG \hspace*{1.95cm} reference frame of gas
\item
RFF \hspace*{1.95cm} rest frame of fluid 
\item
AT \hspace*{2.25cm} apparent temperature

\end{enumerate}

\def\smh{\hspace*{0.3cm}}

\def\medh{\hspace*{0.6cm}}


%
%
\def\sj{\mbox{$\hspace{4.15pt}\check{}$\hspace{-4.15pt}i}}
\def\cj{I$\hspace{-6.0pt}^{^{\breve{}}}$}
\def\ts{t{\hspace*{0.1pt}}s}
%
%
%
\def\I2{{\large \rm \"{I}}}
%
\def\i2{\mbox{\footnotesize\rm \"l\hspace*{-2.45pt}l}}

\def\bi2{\mbox{\footnotesize\rm \bf \"l\hspace*{-2.75pt}l}}

%




\chapter*{Introduction}

\addcontentsline{toc}{chapter}{Introduction}


In this dissertation, the author investigates the equation of state of strongly interacting matter.
The main attention is devoted to theoretical research  of two major phase transitions (PT),
which  are accessible for the experimental  studies in
the nuclear heavy-ion (A+A) collisions. They are
the liquid-gas PT in nuclear media observed in A+A reactions at 
intermediate energies known as  nuclear multifragmentation,
and  the  deconfinement  PT from usual hadrons to 
quark-gluon plasma (QGP).

The major difficulties of experimental and theoretical studies of these PTs are 
determined  by the spatial and temporal finiteness  of strongly  interacting systems created 
during the A+A collision process.  Thus, the characteristic number of  particles 
at the beginning of a collision is about a few hundred nucleons which collide in 
practically vanishing volume due to the Lorentz contraction of original nuclei.
After a complicated  evolution during 10 to 20 fm/c, which includes thermalization of highly excited  hadronic/nuclear matter, its expansion accompanied by  a  possible PT and cooling  down  to free streaming particles, the system typically stops its nontrivial evolution in the volume of  about a few thousand fm$^3$   which is filled up with 
a few thousand secondary hadrons. 
Therefore, theoretical investigations of these PTs  in the course of A+A  collision inevitably combine three relatively independent elements:

I) Development of the statistical description of the PT  which accounts for 
some specific  features of  the interaction between the corresponding constituents and
normalization of the model parameters on the available data;

II) Modeling the PT dynamics in the collision process and transformation of 
strongly interacting constituents into the free moving particles;

III) Analysis of the experimental signatures  of PT and their comparison with the data.

These elements form the phenomenological basis  to study the strongly interacting matter properties and  its phase diagram, i.e. equation of state (EOS),  in the nuclear laboratories.  On one hand, they allow one to make a bridge between the results obtained within the modern theory of strong interaction, the  quantum chromodynamics  and its lattice formulation, and, on the other hand, they provide the experimentalists with explanations of the observed phenomena and  give recommendations for future measurements.

The author's motivation to study the EOS of nuclear and hadronic matter is based  on the following three reasons. 
First, 
despite the recent progress, the lattice quantum chromodynamics  provides us with a very limited information for non-zero baryonic densities. Also the present and planned  experimental measurements 
cannot cover the full range 
of thermodynamic  parameters  of the strongly interacting matter  EOS.  Therefore, for practical purposes, including even understanding of the experimental results,   it is necessary 
to develop the phenomenological models of EOS which can be compared with the existing data and 
extended to the terra incognita regions on the phase diagram.  

Second, since the system under investigation is not a macroscopically large one, the theoretical studies of  phase transformations are extremely difficult. Therefore,  it is necessary to develop the 
approaches that would  allow one to rigorously describe the phases and PTs of strongly interacting matter in finite systems.  Such approaches are  important from the  academic point of view and their results can be, in principle,  used in other branches of physics to study the PTs in systems with long range interaction which may not have a thermodynamic limit. 
In this dissertation, the author presents several exactly solvable models for finite volume EOS along with the new methods and  ideas which allow to rigorously define the finite volume analogs of phases for some classes of models.

Third, since the PTs occur dynamically in the course of a collision, their modeling requires a reliable theoretical  apparatus to describe the expansion, cooling down, phase transformation(s), and emission of secondary particles  at the same time.  This is a real challenge for
theoreticians.  One of the most promising approaches of this type is relativistic hydrodynamics,   
which is best suited to study the strongly interacting matter  EOS.  However,  despite an essential 
progress achieved in hydrodynamic simulations,  as shown below, the  internal problems of the approach require 
an important modification and inclusion of the elements of relativistic  kinetic theory to correctly describe  the particle emission of the perfect fluid. Also, the discussion covers the necessary modifications of relativistic kinetic equations for two domains separated by an arbitrary hypersurface (HS) and their possible consequences for the hydrocascade description of the collision process with several observables. 

Overall,  the present  dissertation is devoted to the development of statistical mechanics for the  static and dynamic processes occurring during A+A collisions at relativistic energies. 

{\bf Relevance of the theme.}  Investigation of the properties of strongly interacting matter
under extreme conditions  is one of the most important subjects of  modern physics. The experimental programs operating   at CERN SPS and BNL RHIC colliders have opened a principal possibility to test  the modern theory  of strong interaction at high temperatures and high energy densities against a vast amount of data  as well as to find a new state of matter, the QGP, and study its properties in the laboratory. Other experimental  programs for studying the strongly interacting matter EOS may open soon.
Thus, the low energy  BNL RHIC,  Darmstadt's  FAIR and, perhaps, Dubna's Nuclotron 
experimental programs will 
search  for the (tri)critical endpoint of the deconfinement PT diagram,
whereas the present CERN SPS program  will study the size and energy dependence 
of existing signatures of the deconfinement PT. On the other hand, the CERN LHC 
experiments will probe the quantum chromodynamics at  high energy  densities. 
In all these cases,  the QGP,  created during the collision, will hadronize either via 
cross-over or via PT and, hence,  the  description and interpretation of  the measurements 
will inevitably require the knowledge of realistic EOS and  adequate dynamic  models for the  collision process. 

Besides the tantalizing technical difficulties,
there exist some unresolved  scientific problems which make these investigations quite hard. 
Thus, the
main theoretical  difficulties  are due to highly nonlinear structure and complexity  of the quantum chromodynamics. Since these difficulties are not yet resolved, the theoretical description of the collision 
process, in principle,  cannot be achieved without the phenomenological EOS models. 
Furthermore, the only known way to dynamically model PTs in the collision process  is through relativistic hydrodynamics because, unfortunately,   the microscopic kinetic theory of PTs even for infinite systems has not been developed so far. The same is valid for  the hydrokinetic or hydrocascade  model, where the hydrodynamic equations are used just to describe the PT.  The hydrodynamic description, however,  has to be interrupted at the HS, where it becomes irrelevant. 
This procedure, the  freeze-out (FO) of momentum distribution functions of particles, 
cannot be 
done arbitrarily, but requires a mathematically correct boundary conditions for relativistic 
hydrodynamics. The correct formulation of these boundary conditions, that does not lead to the recoil paradox, had been a long standing  (over 40 years) problems which found its solution in the author's works.

Another principal problem of the theoretical modeling of relativistic A+A collisions is a necessity 
to study the PT in a finite system. Originally the theory of critical phenomena was 
formulated for infinite systems only. However, the A+A collision experiments are dealing with finite
and, sometimes, small systems and, hence, they require the development of  theoretical approaches which can rigorously define the PT in a finite system. On the other hand, there exist
systems with the long range interaction which do not have thermodynamic limit at all (namely, charged systems with Coulomb interaction). The nuclear 
liquid-gas PT is a typical example of such a system, which requires the correct and rigorous extension of the critical phenomena theory to finite systems. Such approaches are under  extensive discussion in the nuclear multifragmentation community during last 15 years.  
Fortunately, recently in this field  there were  obtained several rigorous results, based
on the exact analytical solutions for specific statistical model EOS in finite systems. 

Therefore, the relevance and importance of the theme of this dissertation is based on 
three facts. First, the description and understanding of the measurements require the development 
of phenomenological models of the strongly interacting matter EOS in a wide range of parameters.
Second, to study the strongly interacting matter PTs in A+A collisions, it is necessary to rigorously define  the finite volume analogs of phases. Finally,  to investigate the many-body reactions with phase transformation(s),  it is  vitally necessary to have an appropriate  theoretical apparatus 
to describe the collective dynamics of a PT in the course of a collision. 

{\bf Relation to other research programs.}   The present work was done at the Department 
of  High Energy Densities of the Bogolyubov Institute for Theoretical Physics of the 
National Academy of Sciences of Ukraine. The major part of this dissertation is a constituent element
of the investigations of the strongly interacting matter properties  under extreme conditions, i.e. at high temperatures and/or high densities,  which are  conducted   at the Department 
of High Energy Densities within the state program.  During last 15 years 
the Department of  High Energy Densities was working on the following state programs: 

1994-1998:   Studies of strongly interacting matter in particle and nuclear collisions. The number of state registration is  0194U025942,   code is 1.4.7.5.

1999-2001: Research of  strongly interacting matter at high energy densities and high baryonic densities.   The number of state registration is  0100U000215,  code is 1.4.7.7.

2002-2004: Research of  strongly interacting matter at high energy densities and high baryonic densties.   The number of state registration is 0101U006428,  code  is 1.4.7.13.

2005-2007:   Studies of strongly interacting matter in  particle and nuclear collisions at high energies. The number of state registration 
0105U000431,  
code is 1.3.1. 

{\bf The aim and tasks of the research.} The major aim of the present dissertation is the development of statistical methods to rigorously study the EOS of finite  strongly interacting matter, its  PTs,  and  their dynamics in the relativistic A+A collisions.  The spatial and temporal finiteness of the system, created during the collision, are unavoidable elements of the 
experimental and theoretical analysis and have to be incorporated into the statistical description
of the deconfinement and nuclear liquid-gas PTs.  

To reach this goal, it was necessary to resolve the following tasks:
\begin{enumerate}

\item To study the EOS and  phase diagram  of  strongly interacting matter on the basis of  exactly solvable  statistical models  which have  PT(s) in thermodynamic limit. They include: 
a simplified version of the Statistical  Multifragmentation Model (SMM), the Fisher Droplet Model (FDM), the Gas of Bags Model (GBM),  the Mott-Hagedorn resonance gas model, and the Quark Gluon Bags with Surface Tension (QGBST) model. 

\item To develop the methods of solving various statistical models with PT in finite volumes and  define the finite volume analogs of phases from the first principles of statistical mechanics.

\item To account for  important  features of strong interaction (the mass spectrum, surface tension,  the short range repulsion with multi hard core radii, the Lorentz contraction of 
hadronic hard core radii e.t.c.)  into statistical description,   and to derive the model EOS. 

\item To include the correct boundary conditions into hydrodynamic equations of finite perfect fluid, and to account for the  emission of  free streaming particles from the surface and/or volume of the fluid.

\item To  generalize the relativistic Boltzmann equation for  finite domains of different 
kinetics which are  separated  by the arbitrary HS, and to derive the hydrokinetic  
equations from the first principles of statistical mechanics. 

\item To apply the statistical and hydrodynamic models to the description of particle yields and hadronic transverse momentum spectra, and  to extract from  the data  an information about the hadronization stage of QGP in central A+A collisions. 

\end{enumerate}

{\it The object  of the research.}  The object of the research of this  dissertation is highly excited 
strongly interacting matter that is created  during the relativistic A+A collision.
At intermediate colliding  energies,  this can be
highly compressed nuclear matter which breaks up into some amount of fragments depending on
its phase  state. At  high  and very high energies (per nucleon) of collision, this can be hot and dense hadronic matter,  QGP, or their mixture.

{\it The subject of the research.} The subject of the research of the present dissertation  is  the statistical mechanics of  PTs, which occur in finite nuclear/hadronic matter during the A+A collisions at intermediate and high energies. The main attention is paid  to finding the exact analytical solutions of a variety of statistical models and to analyzing the PTs in these models  for infinite and finite volume. However, an essential 
part of the dissertation is also devoted to the development of hydrodynamic and hydrokinetic 
approaches  to study the  space-time  dynamics of  PTs during the expansion of highly excited, 
strongly interacting matter created in the A+A collision.  Also, a comparison with the experimental data has been made whenever possible, in order  to fix the model parameters, on one hand, 
and to extract information about the system evolution, on the other hand. 
Therefore, the author concentrates on the following problems: 
\begin{enumerate}

\item  The phase structure of  nuclear matter.  It includes an exact analytical solution of 
a simplified SMM and the analysis of the dependence of  the phase diagram and (tri)critical point   properties  on the model parameters.  Also,  the  mechanism of PT in statistical models, that are similar to the SMM, is discussed in detail. 
A special attention is paid to the calculations of critical indices 
of the model and their dependence on the Fisher index $\tau$. 

\item  Exactly solvable statistical models for finite nuclear systems. 
Here the author focuses on a simplified SMM and three ensembles of the Hills and Dales Model (HDM),  proposed to study 
the  surface deformations of physical clusters. 
The main attention is devoted to the rigorous definitions of the finite volume analogs of phases of 
nuclear matter within the finite volume SMM. The latter  is solved analytically by a newly developed method, the Laplace-Fourier transform. Such a solution allows one, in principle,  to study the statistical mechanics of nuclear systems with Coulomb interaction without taking the thermodynamic limit.
The HDM is  formulated to study the degeneracy factor of large, but finite, physical clusters of fixed volume. It allows one to find out the supremum and infimum for the  surface entropy  of such systems  like the  usual nuclei, 2- and 3-dimensional Ising model clusters, and so on. 

\item The  virial expansion for the Lorentz contracted rigid  spheres and  the EOS  for 
hadronic binary mixtures. The hard core repulsion for hadron gas is necessary, but at high pressures existing at the  deconfinement PT  and/or above the  deconfinement  cross-over regions   the relativistic effects  for the hard core interaction of  light hadrons become important 
and have to be accounted for.  For this purpose, the author proposes and  analyzes the relativistic analog of the VdW   EOS  for the Lorentz contracted hard spheres. 

\item  The exactly solvable  phenomenological EOS  for the deconfinement PT. 
This is the problem of a correct statistical description of the exponential mass spectrum of 
hadrons, suggested by Hagedorn. 
The author critically analyzes  the Hagedorn model in the microcanonical ensemble.
This analysis allows to explain why  the statistical ensembles are not equivalent 
for the exponential mass spectrum,  and to  study the thermostatic properties of the Hagedorn-like systems. Then the two extensions of the Hagedorn model, the Mott-Hagedorn resonance gas model and the QGBST model, are analyzed. 
In the Mott-Hagedorn resonance gas model, the singular properties of the thermodynamic functions 
above the Hagedorn temperature are removed by the assumption that, due to the Mott transition, the discrete hadronic  mass spectrum becomes  continuous  with the width of hadrons being exponentially  dependent on their mass. In another generalization of the Hagedorn model, the QGBST model, which is also solved analytically,  the usual problems are avoided 
because of the hard  core repulsion between the  quark-gluon
bags. It turns out that the null line of the surface tension coefficient  of  quark-gluon bags 
plays a crucial role in the deconfinement  cross-over  existence at high temperatures and 
low  baryonic  densities. 

\item   The FO problem of relativistic hydrodynamics and hydrokinetics. 
This is a fundamental problem of  the  correct formulation of  the boundary conditions for 
the system of nonlinear  partial differential equations of relativistic hydrodynamics at the 
FO HS where they become irrelevant.  It was unresolved for almost 45 years. 
One of the  difficulties  was related to the fact that, while 
for the space-like parts of the  FO  HS the Cooper-Frye prescription works well, for the  time-like parts of the  FO  HS it generates the unphysical negative particle numbers. However, it turns out that, to avoid the recoil problem and negative numbers of particles due to the emission of free streaming particles from the  time-like parts of the  FO  HS, it is necessary to  modify the original  hydrodynamic equations. Therefore, a self-consistent formulation of relativistic hydrodynamics is considered.  
Then this problem is studied  at  the kinetic theory level, where   the finiteness of the domains with two, generally different, kinetics 
requires modification of the relativistic Boltzmann equation. 

\item  The signatures of the deconfinement transition. An essential progress was achieved recently  in solving this problem. The author discusses  the idea of early statistical hadronization of charmonia ($J/\psi$, $\psi^\prime$ mesons), charmed particles and multistrange hadrons ($\phi$ meson, $\Omega$ hyperon e.t.c.) along with  the  experimental consequences. Also, the transverse momentum spectra of K-mesons 
are analyzed, and it is argued that the plateau  of  the inverse slopes of K$^\pm$  mesons as the function of the colliding energy evidences for the onset of the deconfinement transition.

\end{enumerate}

{\it The methods of research.}  To solve the above formulated problems, the author uses  the standard methods of statistical mechanics, like the Laplace transform, cluster and virial expansions,  the maximum term method of Lee-Yang, 
Van der Waals approximation e.t.c., as well as  the  newly proposed ones. The new methods are discussed in this dissertation in detail.  The main essence of the latter methods is related to accounting for  the  spatial and temporal finiteness of  highly excited  strongly interacting systems created in the  A+A collisions.  

The simplified SMM is solved analytically by the standard Laplace transform method. However, to analyze  the divergent sums, which define the critical indices of this model, the author has proposed a rigorous method based on the  Newton-Leibnitz  definition of the integral. The application of this method to the well known FDM shows that the previously stated range of the $\tau$ index, $2< \tau < 3$, must, in fact,
be changed to $2< \tau < 2 \frac{2}{3}$.
With the help of this method, a principal correction is made to Fisher's  derivation of the Fisher-Rushbrooke and Fisher-Griffiths inequalities for the critical exponents.  This finding   indicates  fundamental problems in defining some critical indices of statistical systems, in particlar, the $\alpha$ index. 

The Complement method to analyze the chemical equilibrium in finite systems is 
also proposed. It  is based  on the evaluation of  the free 
energy change occurring when a cluster moves from one 
phase to another. For a finite liquid drop in equilibrium 
with its vapor, this is done by virtually transfering a 
cluster from the liquid drop to the vapor and evaluating 
the energy and entropy changes associated with both the 
vapor cluster and the residual liquid drop (complement). 
The method is applied to the description of the 
clusters in 2- and 3-dimensional Ising models. It allows one to account for the 
finite size effects associated with the the largest (but mesoscopic) drop representing 
the liquid in equilibrium with the vapor and derive the Gibbs-Thomson correction
for  pressure and  cluster density. 
This method can be generalized to incorporate energy 
terms common in the nuclear case: symmetry, Coulomb  
and angular momentum energies. 

The author has developed the Laplace-Fourier method as a generalization of the usual Laplace technique. 
It is based on the integral representation of the Dirac $\delta$-function which allows one to 
rewrite any complicated dependence as an exponential of a corresponding parameter. 
The resulting exponential can be dealt with  the usual Laplace transformation. 
Such an approach allows to exactly  solve a simplified SMM for finite volumes. 
It is shown that, for finite volumes, the grand canonical  ensemble (GCE) partition  can be  identically 
represented via the simple poles  singularities of the isobaric partition. 
The behavior of these simple poles in the complex free energy density plane  as function of the system's volume allows  one to distinguish the  model with PT from the model without it. 
The analysis of the real part of the effective chemical potential allows one to unambiguously define the finite volume analog of the  phase diagram and the corresponding  phases.
These results are also generalized to define the finite volume analogs of phases for the deconfinement PT
in the GBM.  

Also, using this method, the author has exactly solved three ensembles of HDM for the  clusters of finite size. 
This is the statistical model of surface deformations of physical clusters with the constraint of 
the  fixed cluster volume.   This method, for the first time,  allows one to estimate the upper and lower bounds on the degeneracy of different physical clusters consisting of the  finite size constituents. 
In order to elucidate  the influence  of the volume conservation condition on  the surface entropy,
the author has introduced and studied  two new isochoric ensembles, semigrand canonical and specially constrained canonical ensembles, of the surface partition. 

The usual cluster and  virial expansion methods are generalized to the momentum dependent 
interparticle potentials. This generalization is necessary to account for the Lorentz contraction of the hard core 
repulsion of light hadrons at the deconfinement PT region and/or  above the deconfinement cross-over. 
It is shown that, in contrast with the previous unsuccessful  attempts, the obtained relativistic analog of the Van der Waals (VdW) EOS  obeys causality.

To solve the FO problem in relativistic hydrodynamics, the author has extended the equations of motion of the perfect fluid with the help of  generalized functions to the system consisting of two subsystems, the fluid and the gas of free particles emitted by the fluid, which are separated by a FO HS.  Then the  extended equations of motion generate the equations of motion
of each subsystem and the  boundary conditions between them which have to be solved consistently with the equations of the perfect fluid. Such an approach allows to avoid the problems and paradoxes faced by the previous  researches  of this problem,  and to build up the paradox-free, self-consistent relativistic hydrodynamics. 

This method was refined further and applied to derive the kinetic and hydrokinetic equations for two domains of generally different  kinetics separated by an arbitrary HS. Using this method, the author has accounted for the exchange of particles via the boundary HS and has derived the source terms in the relativistic Boltzmann equations for  both  domains.  Such an approach has allowed for the first time to   self-consistently  derive
the boundary conditions for the hydrocascade model and to  discover a discontinuity of the most general  form, the three flux discontinuity. The latter generates  all previously   known kinds of hydrodynamic discontinuities as limiting cases.

{\bf Scientific novelty   of the  obtained results.} Applications of all these methods  to  the statistical mechanics of strongly interacting matter  and  to  solving specific problems, discussed above, have led to multiple new and original  results.  The most important and new  of these results   are the following:
\begin{enumerate}

\item 
An  exact analytical solution of 
a simplified SMM  is found in the thermodynamic limit. A rigorous mathematical
proof of the 1-st order nuclear liquid-gas PT has been
given and the possibility of the existence of a 2-nd order PT
has been found. The critical indices $\alpha^\prime, \beta,\gamma^\prime, \delta$
as functions of the Fisher parameter $\tau$
are calculated for this  analytical solution of  the SMM.
For the first time,  it is found that these indices 
differ from those 
of   the FDM.
It is  also shown that, in contrast to general expectations,
 the scaling relations for the SMM critical indices
differ from the corresponding relations of  the FDM.
In contrast to the  FDM,  
the SMM predicts
a narrow range of values for index $\tau$, $1.799< \tau < 1.846$,
which is consistent with both the ISiS Collaboration data  and EOS Collaboration data.
As shown in the dissertation, such a range of  $\tau$ index  is of a principal importance because 
it  gives a direct evidence
that nuclear matter has a tricritical rather than a critical point.

\item 
For the first time the cluster and virial expansions are  derived  
for the momentum dependent inter-particle potentials. 
Using these results, 
the two models of the relativistic  EOS  of the VdW  gas are proposed and analyzed.
In both of these EOS the Lorentz contraction of the sphere's volume
is taken into account.
The first EOS, which does not obey causality in the limit of high densities,
is generalized to the two-component excluded volume  model  and applied to
describe the hadron yield ratios for BNL AGS and CERN SPS data. 
The first EOS is refined further to  obey the causality
at  high densities, i.e., the  sound velocity of this model is 
subluminar.
As shown  for the high values of  chemical potential,  the pressure in the 
second EOS  has  an 
interesting kinetic interpretation. 
The second  EOS indicates  that for high densities the most probable configuration
corresponds to the smallest value of the relativistic excluded volume. 
In other words, for high densities  the configurations with the collinear velocities of the neighboring 
hard core particles are the most probable ones. This, perhaps, may shed light 
on the coalescence process of any relativistic hard core constituents.

\item
A novel powerful mathematical method,  the Laplace-Fourier transform,
is proposed.  It has allowed to find an  analytical
solution of a simplified version of the SMM  with the
restriction that the largest fragment size cannot exceed the finite volume of the system.
A complete analysis of the singularities of the isobaric partition function is done for 
the finite volumes. The finite
size effects for large fragments and the role of metastable (unstable) states are discussed. 
These results allowed the author, for the first time, to rigorously  define
the finite volume analogs of   nuclear liquid phase, gaseous phase, and the mixed phase.
Also these findings  explicitly demonstrate that the famous T. Hill approach to  define the PT  in finite systems is  based on a wrong assumption.

\item
To elucidate the origin  of surface entropy  of large 
physical clusters the HDM 
for their  surface deformations is proposed.  It is solved analytically for 
large  finite  clusters by  the Laplace-Fourier transformation method.  
The requirement of volume conservation of the deformed cluster is automatically accounted for. 
In the limit of  small amplitude  of  deformations,
the HDM    reproduces the leading term of the famous Fisher
result for the surface entropy within  a few percent. In other words, 
in this limit the HDM generates a linear temperature dependence of surface 
tension coefficient of large clusters. 
Also, the model  gives the degeneracy prefactor of large fragments 
which was unknown to Fisher.
The surface entropy coefficient of two other  surface partitions  with the different 
volume conservation constraints are  exaclty solved and  compared with the 
surface entropy coefficients of 2- and 3-dimensional Ising models.

\item
The effects of the finite size of a liquid drop undergoing a PT are described in terms of the complement, the largest (while still mesoscopic) drop representing liquid in equilibrium with vapor. 
Vapor cluster concentrations, pressure and density from fixed mean density lattice gas (Ising) model calculations are explained in terms of the complement. The complement approach generalizes the FDM and SMM by accounting for  generic features of chemical equilibrium in finite systems. 
Accounting for this finite size effect is a key element in constructing the infinite nuclear matter phase diagram from experimental data.

\item
A microcanonical treatment of Hagedorn systems, i.e. finite mass hadronic resonances with an exponential mass spectrum controlled by the Hagedorn temperature $T_H$, is performed. For the first time it is shown   that, in the absence of any restrictions, a Hagedorn system is a perfect thermostat,  i.e. it imparts its temperature $T_H$ to any other system in thermal contact with it.  It is found that, if the particles of different masses are generated by 
a Hagedorn system,
then they are in chemical equilibrium with each other. Thus, a Hagedorn system is a perfect particle reservoir. 
The thermodynamic effects of the lower mass cut-off in the Hagedorn mass spectrum is analyzed.
By a direct calculation of  the microcanonical partition, it is found   that in the presence of a single Hagedorn resonance the temperature of any number of $N_B$ Boltzmann particles  only slightly differs from $T_H$ up to the kinematically allowed limit $N_B^{kin}$. 
For $N_B > N_B^{kin}$ however, the low mass cut-off leads to a decrease of the temperature as $N_B$ grows. The properties of Hagedorn thermostats naturally explain a single value of hadronization temperature observed in elementary particle collisions at high energies which is a consequence of a finite volume PT.

\item
The statistical bootstrap model is critically revised in order to
 include a medium-dependent resonance width into it.
It is  shown that a thermodynamic model with a vanishing width below the
Hagedorn temperature $T_H$ and a Hagedorn spectrum-like width above
$T_H$ not only eliminates the divergence of the thermodynamic
functions above $T_H$, but also gives a satisfactory description
of  lattice quantum chromodynamics  data on the energy density
above the chiral/deconfinement transition. 
The proposed  model allows one to explain the absence of heavy resonance
contributions in the fit of the experimentally measured particle
ratios at SPS and RHIC energies.
This approach is applied to the description of the NA50
experiment and the analysis suggests  that the anomalous suppression of J/$\psi$
production can  be explained by the increase of the effective number of
degrees of freedom at the Hagedorn temperature.

\item
The temperature and  chemical potential  dependent  surface tension of bags  is introduced into  
the gas of quark-gluon bags model. This resolves a long standing problem  of  a unified description of
the first and second order PT with the cross-over. 
Such an approach  is  necessary to  model the complicated  properties of
the QGP and hadronic matter from the first principles of statistical mechanics. 
As shown in this dissertation, the proposed model has an exact analytical solution and allows one to rigorously study 
the vicinity of the critical endpoint  of the deconfinement  PT. The existence of  higher order PT at the critical endpoint is  discussed.
In addition, it is  found that, at the curve  of  a zero surface tension coefficient, 
there must exist a surface induced  PT of the 2$^{nd}$ or higher order, which separates 
the pure QGP  from the cross-over states, that are the mixed states of hadrons and QGP bags. 
Thus, the suggested  model predicts that for $\tau \le 2$ the  critical endpoint of quantum chromodynamics  is the tricritical endpoint.
Using the other results of the dissertation, one can extend this analytical solution to finite volumes. 

\item
The FO  problem of relativistic hydrodynamics is solved in the zero width approximation.
For the solution, the author has derived the correct generalization of  the Cooper-Frye formula, the cut-off distribution, 
for the invariant momentum  spectra of secondary particles emitted
from the time-like HSs. Then, for the first time, the conservation laws are formulated not  just for a perfect fluid alone, but for a perfect  fluid  and the gas of emitted particles.
The conservation laws at  the  boundary  of the fluid and gas of free particles are 
analyzed, and the  new kind of a discontinuity  in relativistic hydrodynamics, the FO shock, is discovered.  
It is shown that the FO condition in relativistic hydrodynamics has to be formulated
not for a fluid, but exclusively  for a gas of free particles. For a wide class of EOS  it is proven that 
the FO shock resolves the recoil problem. Thus, 
a self-consistent, paradox-free  relativistic hydrodynamics 
with particle emission  is formulated for the first time.

\item
The  relativistic kinetic equations for  two
domains, separated by the HS with both the space- and time-like
parts, are derived for the first time. 
The particle exchange between these domains, occuring at the time-like boundaries,
generates  the source terms and modifies the collision term of the kinetic equations.
By integrating the derived set of transport  equations, 
the correct boundary conditions between the hydro and cascade domains are obtained.  
Remarkably,  the conservation laws at  the
boundary between these domains conserve   both  the incoming
and outgoing components of energy, momentum and baryonic charge separately. Thus,
 the  relativistic kinetic theory generates twice the number of  conservation laws
compared to traditional hydrodynamics.
Further analysis shows that these boundary conditions between domains,
the  three flux discontinuity,
can be  satisfied only by a special superposition of two  cut-off  distribution functions for
the ``out'' domain.
All these results are applied to the case of the PT between
the QGP and hadronic matter. The possible  consequences for an improved
hydrocascade description of the relativistic nuclear collisions are  discussed.
The unique properties   of the  three flux discontinuity  and  their effect on the
space-time evolution of the transverse expansion are also analyzed.
The possible modifications of both transversal radii  of   pion correlations,
generated by a correct hydrocascade approach,
are discussed.

\item
A hypothesis of early hadronization of charmonia and multistrange 
hadrons  is formulated for the first time.
In this dissertation,  it 
is shown that this hypothesis naturally explains  the data on the apparent 
temperature (inverse slope) of the 
transverse mass spectra of $J/\psi$ and $\psi^{\prime}$
mesons at SPS energies. 
The transverse mass spectra of $\Omega, J/\psi$ and $\psi^\prime$  
(Pb+Pb  at 158 AGeV)  
and $\phi$, $\Omega$ (Au+Au at $\sqrt{s_{NN}} = 130$~GeV)  particles 
are perfectly reproduced within the hypothesis
of  simultaneous kinetic and chemical FO of
those  particles from the hadronizing QGP  and predictions
for the $J/\psi$ and $\psi^{\prime}$ spectra at RHIC are made. 
Also
it is shown  that the RHIC data for (anti)$\Lambda$ and (anti)proton 
transverse momentum spectra are in contradiction with  the predictions of a popular 
Single Freeze-out Model. 

\item
The analysis of the experimental values of  the inverse slopes of transverse momentum spectra
of kaons 
produced in central Pb+Pb (Au+Au) interactions shows an anomalous dependence on the
collision energy. 
The inverse slopes of the spectra  increase with energy 
in the low (AGS) and high (RHIC) energy domains, whereas they are constant in the
intermediate (SPS) energy range. It is shown  that this anomaly is caused 
by a modification of the EOS  in the transition region between confined and
deconfined matter, and is similar to the usual caloric curves. 
Therefore,  such a behavior of the transverse caloric curves  can be considered as  a new signature
of the onset of deconfinement  located in the low SPS energy domain.
Nowadays this signature  is known to the relativistic heavy ion  community as a Step.

\end{enumerate}

{\bf Practical value of the obtained results.}  The results obtained in the present dissertation  give 
a principally new understanding of the strongly interacting matter EOS and its  PTs.  
Thus, the new scaling relations 
between critical exponents, found for a simplified SMM,  predict  a very narrow range of the $\tau$ index
which, in contrast to the FDM, is perfectly consistent with the experimental findings of the ISiS and EOS collaborations. 
Such a finding is of principal importance because it gives a direct evidence that nuclear matter has a tricritical point rather than the critical one. 

A hypothesis of statistical production of charmonia at hadronization  of  QGP has not only helped to explain the  inverse momentum spectra of early hadronized particles,  measured  by NA50  and WA97  experiments  performed  at CERN SPS,   and  by   STAR Collaboration at BNL  RHIC,  
but  also to  predict  the  inverse slopes of  $J/\psi$ and $\Psi^\prime$ mesons which are to be measured at BNL RHIC at the center of mass energy 130 GeV$\cdot$A. 

Despite  almost  three decades of searches for the QGP signals, the transverse caloric curves (the inverse slopes of transverse momentum spectra ) of K$^\pm$ mesons as the  function of the  collision energy  is one of the only three experimentally established signals of the deconfinement transition. 

Besides the practical importance of these results for understanding the  experiments, the results obtained in this dissertation have a principal  theoretical  importance. Thus,  the theoretical description of PTs in finite system is, on one hand,  demanded  by  the  experiments  on  A+A collisions, in which the finite or  even small number of particles is produced.  On the other hand, it is necessary  to study the critical phenomena  in the systems with  the Coulomb-like interaction  which, in principle,  do not have the thermodynamic limit. 
The analysis of several exactly solvable models of  realistic EOS of  strongly interacting matter, for instance,  shows  that in finite systems the GCE partition is defined by 
a set of states having (in general) the complex values of free energy, and that these states
are not in a true chemical equilibrium for the same value of the chemical potential  due to the interaction of  the constituents. Such a feature cannot be obtained within a famous FDM  due to lack of the hard core repulsion between the constituents. 

Also the obtained results indicate that,  in contrast to Hill's expectations, the mixed phase at finite volumes is not just a composition of two  pure phases. It is rigorously shown that a finite volume analog of the mixed phase is a superposition of three and more collective states, each of them is characterized by its own value of (complex) free energy, and, consequently, the difference between the free energies of these states is not a surface-like, as Hill  argued in his many works and several books on statistical mechanics of small systems, but volume-like. It is necessary to stress that such results cannot be, in principle, obtained either numerically or by the  mean-field models, which 
are not truly statistical.  Since there are only a few statistical models with PT which are exactly   solved
for finite systems, the results of the present dissertation are indispensable for building up the 
theory of critical phenomena in finite systems. 

The solution of such a long standing problem as the FO of  relativistic hydrodynamics is also of a great theoretical significance because it completes the relativistic hydrodynamics for the fluid emitting the 
gas of free particles and converts the relativistic hydrodynamics into a powerful and reliable tool of research of  
the strongly interacting matter EOS and/or the collective  phenomena associated with the matter under extreme conditions (nuclear A+A collisions, the Big Bang, the supernova explosions e.t.c.). 
The results, obtained in solving  the FO problem, have   initiated the ten-year long activity of the Bergen and San Paulo hydro groups on further refinement of this approach.  Also these results were used 
by U. Heinz and collaborators to study the FO effects in  pion interferometry and by E. Shuryak and collaborators for their  hydrocascade model. 

Furthermore, this activity had increased the author's personal  interest in  the hydrokinetic (hydrocascade)  approach which has led to the rigorous formulation of such a model based on the firm theoretical ground.
As discussed in the present dissertation, the hydrokinetic approach is based on the generalized relativistic Boltzmann equations which are rigorously  derived  for the finite domains separated by an arbitrary   HS.  The derived equations include the boundary conditions at this HS, and it is shown that the relativistic kinetics generates, in general, twice the number of conservation laws compared to the usual relativistic hydrodynamics. The analysis of these boundary conditions shows that they could  be important for understanding the source of turbulence which up to now remains a true challenge for theoreticians. 
Also, the developed hydrokinetic approach can be used to model  the dynamics of  PTs in the 
inhomogeneous  media which is rather hard  to describe by  the usual hydrodynamic equations.


{\bf The personal contribution of the candidate.\,}  The personal contribution of the candidate in 
obtaining  the results 1-12, listed above, is decisive for the development of major ideas, main approaches 
and  strategic lines of the research.  In particular: 
\begin{enumerate}

\item   In analyzing the thermodynamic limit  of  a simplified SMM  (a series of works in collaboration with M. I. Gorenstein, W. Greiner and  I. N. Mishustin),  the candidate suggested to apply the Laplace transform technique to solve this model, made all the analytical and numerical calculations, and  
gave a correct physical interpretation of the obtained results.

\item  During the studies of the GCE partition of the relativistic  gas of  Lorentz contracted  rigid spheres (in collaboration with  M. I. Gorenstein, H. St\"ocker and  W. Greiner), the candidate suggested 
to generalize the cluster and virial expansions for the momentum dependent potentials,  derived the analytical formulas, and performed their numerical analysis.  Later on  a  candidate alone has reformulated the acausal EOS, obtained with his collaborators,  and made it causal. 

\item  To study  the problem of surface entropy (one paper in collaboration with J. B. Elliott and L. Phair and another in collaboration with J. B. Elliott) the candidate formulated the HDM and solved it analytically by  the Laplace-Fourier methods which he proposed earlier.  

\item  An analytic   analysis of   magnetization of the 2- and 3-dimensional Ising models with the help of  a Complement method (in collaboration with L.G. Moretto, J. B. Elliott, 
R. Ghetti, J. Helgesson  and  L. Phair) was done by the candidate. Also he  found the derivation of the Gibbs-Thomson correction of the system's  free energy  due to the presence 
of a finite liquid drop. 

\item  An analytic  treatment of the microcanonical systems with a Hagedorn mass spectrum (a series of works in collaboration with L.G. Moretto, J. B. Elliott, and L. Phair) 
was  proposed and performed  by the candidate. Using  these results,  the candidate also showed that 
such an analysis justifies the Statistical Hadronization Model of Becattini and explains  the constant value of hadronization temperature, observed in elementary particle collisions at high energies.

\item  A modification of the EOS of the Mott-Hagedorn model above $T_H$ (a series of works in collaboration with
D. B. Blaschke)  was proposed by the candidate. He also fitted the data of lattice quantum 
chromodynamics, and,   using this model, he suggested an  explanation of the absence of
heavy resonances  (heavier  than 5 GeV) in the relativistic A+A  collisions.  The latter occurs due to a very short life time of heavy resonances in the media with the temperature higher than that  one of the  Mott-Hagedorn transition. 

\item  The  solution of the FO problem was proposed by the candidate alone and is published
in his  paper without coauthors. In the subsequent papers on this subject,  written in collaboration with M. I. Gorenstein and W. Greiner,   the candidate developed his original ideas further.
All the  analytical and numerical results  of these subsequent publications were obtained  by  a candidate himself. 

\item  The hypothesis of early kinetic FO  of charmonia at   the QGP hadronization   was 
suggested by the candidate. He derived the necessary expressions and  fitted  the experimental data  
to describe the transverse momentum spectra of early hadronized paticles (a series of works in collaboration with M. I. Gorenstein and M. Gazdzicki).  He also predicted the results of the future  measurements  of  the  transverse momentum spectra 
of $J/\psi$ and $\psi^{\prime}$ mesons at the center of mass energy  $\sqrt{s_{NN}} = 130$~GeV.

\item The  argumentation based on the mixed phase  softest point     to  justify  the plateau in the 
dependence  of the 
 inverse slopes of the soft kaons  transverse momentum spectra on the colliding energy per nucleon 
as a signal of the deconfinement  was suggested by a candidate. 

\end{enumerate}

{\bf Approbation  of   the dissertation results.\,}  All the results of this dissertation were presented 
at  more than 20  international conferences and schools  on phenomenology of A+A collisions and close themes, at the  seminars of the Bogolyubov Institute for Theoretical Physics of the National  Academy of Sciences of Ukraine and many leading scientific centers in this field of research. The latter include:   Brookhaven National Laboratory (New York, USA), Lawrence Berkeley  National Laboratory (California, USA),  Columbia  University (New York, USA), Institute for Nuclear Physics (Seattle, Washington, USA), GSI Darmstadt, 
University of Regensburg (Germany), University of Bielefeld (Germany), University of Frankfurt (Germany), University of  Hannover (Germany), University of  Giessen (Germany), ECT Trento (Italy), 
University of  Salamanca (Spain), etc. The most important  presentations were made at the following conferences and schools: 

\begin{itemize}

\item  International conference ``Structure of the Nucleus at the Dawn of the Century'',
Bologna, Italy, May 29 - June 3, 2000, 

\item  Int. conference ``Nucleus Nucleus Collisions 2000'',
Strasbourg, France, July 3-7, 2000, 

\item  International conference ``Quark Matter in Astro- and Particle Physics'',
Rostock, Germany, 
November 27--29, 2000,

\item  International conference ``Dynamical Aspects of the QCD Phase Transition", 
Trento, Italy,  March 12 - 15, 2001,

\item  ``International Conference on Nucleus and Particle Collisisons (INPC) 2001'',
Berkeley, California, USA, July 30 -- August 3, 2001,

\item  Conference ``Strange Quarks in Matter  2001'',
Frankfurt, Germany, September 24--29, 2001,

\item  50-th Gordon Research Conference ``Nuclear Chemistry'',
New London, New Hampshire, USA, June 16-22, 2002, 

\item  International Program INT-03  ``Three Years of RHIC Physics",
Seattle, Washington, USA, April 15 - October 15, 2003,

\item  International workshop  ``Aspects of Nonperturbative QCD - Hadrons and Thermodynamics",
Rostock, Germany, July 14-16, 2003,

\item  International  workshop  ``Modern Methods in Relativistic Nuclear Physics'', 
Dubna, Russia, August 19-20, 2003, 

\item  International Workshop on Progress in Particle and Nuclear Physics: 
 ``Heavy Ion Reactions from Nuclear to Quark Matter", 
Erice, Italy, September 16-23, 2003,

\item  International conference ``Tracing the Onset of Deconfinement in Nucleus-Nucleus Collisions'',
Trento, Italy,   April 23-30, 2004,

\item  Gordon Research Conference  ``Nuclear Chemistry'',
New London, New Hampshire, USA, June 13-18, 2004, 

\item  International conference ``World Consensus Initiative III'', Texas A \& M University, College Station, Texas, USA, February 11-17, 2005, 

\item  ``School of Collective Dynamics in High-Energy Collisions'',
Berkeley, USA, May 19-27, 2005,  

\item  International school-seminar ``New Physics and Quantum Chromodynamics  
at External Conditions", Dniepropetrovsk, Ukraine, May 3-6, 2007,

\item  International  Workshop  ``Relativistic Nuclear Physics: from Nuclotron to LHC  Energies'', Kiev, Ukraine, June 18-22, 2007.

\end{itemize}

Using  the  results obtained in the present dissertation, the candidate gave a series of lectures  for the advanced students 
at the Center for  Physics of Ultra-relativistic  Nuclear Collisions at the Faculty of Nuclear Sciences and Physical Engineering
of the Czech Technical University in Prague, Czech Republic, April 15-26, 2007.

{\bf Publications.} The results of this dissertation are published  in 42 works, including the articles in the leading scientific journals, conference proceedings and preprints. In particular, 30 papers are published in  scientific journals, Refs. \cite{Bugaev:93,Bugaev:00,Bugaev:01,Reuter:01,Complement,Bugaev:04a,BugaevReuter,Bugaev:07b,Bugaev:04b,BugaevElliott,HThermostat:1,Bugaev:RVDW1,Zeeb:02,Bugaev:RVDW2,Bugaev:96,Blaschke:03,Bugaev:01d,Bugaev:02,Bugaev:02a,Bugaev:02b,Bugaev:03,Blaschke:04,Blaschke:05,Bugaev:07, Bugaev:08new,Bugaev:99, Bugaev:09,Bugaev:02HC,
Bugaev:04HC,Step}, 
6 works are published as conference proceedings, 
Refs.  \cite{Bugaev:00b,Bugaev:01b,Bugaev:01c,Reuter:01b, Bugaev:07new,Bugaev:99b}, and  
6 works are preprints, Refs.  
\cite{Reuter:01b,Elliott:05wci,Bugaev:05csmm,
HThermostat:2,Moretto:05,Bugaev:99a}.  12  publications  are prepared by K. A. Bugaev alone. 

{\bf The structure and volume of the dissertation.\,} The dissertation consists of an introduction, 
six  chapters, conclusions, two appendices  and  bibliography which contains 358 references.  The dissertation 
has 55 figures and 8 tables. The volume is 341 pages of the printed text.


\chapter{Nuclear Matter in Thermodynamic Limit}



It was noted long ago that statistical methods could be applied to nuclear processes, if 
the energies involved are large when compared to the lowest excitation energies of nuclei \cite{Weisskopf:37}.
Assuming this,  Weisskopf was able to formulate expressions for the probability of neutron (or charged particle)
emission from excited nuclei.

Following Bohr,  Weisskopf    divided process initiated by nuclear collisions into two stages: the first stage was the formation of a compound  and the second one was the disintegration of the compound nucleus. 
Both stages could be treated independently. The energy of compound nucleus is similar to the heat energy in a solid or liquid and the emission of particles from the compound nucleus is analogous to an evaporation process.
Exploiting this idea,  Weisskopf   derived a general statistical formula for evaporation of particles from an excited nucleus accounting for finiteness of the nucleus and for the fact that  the evaporation of a particle takes away significant energy from the compound nucleus.  In that regard, Weisskopf
worked out the formulas to describe the evaporation neutrons  from a hot nucleus, i.e. he was describing a 
1$^{st}$ order PT with a neutron leaving (or evaporating from) the condensed phase (the hot nucleus) and 
entering the dilute phase (a very low density neutron vapor).

Already this example of neutron evaporation shows that for some processes the Coulomb interaction can be 
neglected at leading order because of its weakness compared to strong interaction. 
Such a view was refined further in a concept of infinite  nuclear matter. This is a helpful  mathematical construct which corresponds to the  nucleons  without electrical charge. In nuclear physics  such a  concept  is very similar to the concept of idea gas: like in many applications the dilute real gases behave similarly to the ideal gas,
in many respects  the  hot compound  nucleus  reminds a finite piece of nuclear matter. 

Historically this was the first concept allowing one to study the nuclear liquid-gas PT in nuclear reactions. 
Like in  Weisskopf's treatment the compound nucleus represent a nuclear matter, whereas the emitted 
fragments of different masses (charges) correspond to a nuclear vapor. Then, studying the fragment distributions
at different  excitation energies one can learn the information about  the  emission stage of the reaction and 
probe the nuclear matter EOS. 
This was exactly the case: for excitation energy below 4-5 MeV per nucleon the fission probability $W_A$ of compound nucleus emitting  a A-nucleon fragment  shows  an exponential behavior of as the function of  inverse square root of excitation energy $E^* = \varepsilon/\rho - M$, i.e.  $W_A \sim \exp( - B_A / \sqrt{E^*} )$, where $B_A$ is a fission barrier for a fragment $A$, $\varepsilon$ is the energy density of a nucleus  and $\rho$ is its particle density.  Such a behavior of the compound nucleus  temperature $T \sim \sqrt{E^*} $  is typical for a Fermi 
liquid with the EOS  (caloric curve) $E^* = a\, T^2$  \cite{CalorCurve:Exp1,CalorCurve:Exp2,CalorCurve:Exp3,CalorCurve:Exp4}.
If, however, the excitation energy per nucleon is increased by a few MeVs only, the nucleus breaks down into 
several fragments and its  EOS,  $E^* =  \frac{3}{2}\, T $ 
\cite{CalorCurve:Exp1,CalorCurve:Exp2,CalorCurve:Exp3,CalorCurve:Exp4}, corresponds to the nonrelativistic gas of  massive  Boltzmann particles.  Also the mass distribution of fragments is similar to that one of the molecule clusters in the vapor of real  liquids.  

The change of the caloric curve behavior from a Fermi liquid regime to the nonrelativistic  massive Boltzmann gas
can be naturally explained \cite{NuclearPT:1, NuclearPT:2, StockerGreiner} by a PT from the nuclear liquid  phase representative  to the vapor of nuclear fragments. 
The multi particle  nature of this phenomenon is reflected in its name - nuclear multifragmentation. 

The early  theoretical arguments on the nuclear liquid-gas PT \cite{NuclearPT:1, NuclearPT:2,  StockerGreiner} initiated the theoretical and experimental  efforts to study the nuclear matter EOS  which are going on over the two decades.  The theoretical approaches can be divided into two different categories: 
analytical/semi-analytical models  \cite{Analyt:1,Analyt:2,Analyt:3,Analyt:4,Analyt:5,Analyt:6,Analyt:7,Analyt:8,Analyt:9}  and computational models: both on a lattice \cite{ComputLat:0, ComputLat:1, ComputLat:2, ComputLat:3, ComputLat:4, ComputLat:5, ComputLat:6, ComputLat:7, ComputLat:8, ComputLat:9,ComputLat:10,ComputLat:11,ComputLat:12,ComputLat:13}  and off it  \cite{Comput:1,Comput:2,Comput:3, Comput:4, Comput:5, Comput:6, Comput:7, Comput:8, Comput:9, Comput:10, Comput:11}.

The analytical/semi-analytical theories employed various methods (e.g. particles interacting through a Skyrme force, 
finite temperature Hartree-Fock theory and various nuclear extensions of  the  FDM) to determine the critical point 
of bulk (i.e. infinite, uncharged and symmetric) nuclear matter and the liquid-vapor phase boundary. This lead to 
estimates of the critical temperature in the range of 12.6 MeV to 28.9 MeV depending on the theoretical techniques 
employed. Once estimates were made for bulk nuclear matter, the effects of a finite number of nucleons and a fluid 
made up of two components (one which carries an electric charge) were studied. Those effects generally lead to a 
lower critical temperature with estimates between 8.1 MeV and 20.5 MeV. 
Computational models on the lattice attempted to study the process of nuclear cluster formation from ``the bottom 
up"  by modeling in a simple way the short range interaction of the nucleons. This was done both geometrically with 
percolation models \cite{ComputLat:0, ComputLat:1, ComputLat:2, ComputLat:3, ComputLat:4, ComputLat:5, ComputLat:6, ComputLat:7} and thermally with lattice gas (Ising) models  \cite{ComputLat:8, ComputLat:9, ComputLat:10,ComputLat:11,ComputLat:12,ComputLat:13}. 

Despite a great variety of the theoretical models they can be divided into two groups: the mean-field models and 
the statistical models. The vast majority of  these models are mean-fieled ones, whereas there are only two successful  statistical models, and both of them  are the cluster models. The latter are the extensions of the FDM \cite{Fisher:67, Moretto:97} and SMM \cite{Bondorf:95}.  These cluster models are used to study the nuclear liquid-gas PT both in finite and infinite nuclear matter. In this chapter I consider the infinite nuclear matter. It is investigated by  a
mean-fieled model, which is  a phenomenological generalization \cite{Bugaev:93}  of  a famous Walecka model \cite{Analyt:1}, and by an exact solution \cite{Bugaev:00, Bugaev:00b, Bugaev:01, Bugaev:01b, Bugaev:01c} of a  simplified SMM  \cite{Gupta:98, Gupta:99} for symmetric (i.e. with equal number of protons and neutrons) nuclear matter.  

The  suggested mean-fieled model   \cite{Bugaev:93}  has three parameters  only which are fixed to reproduce three major  properties of the nuclear matter (for details see below). However, surprisingly  this model simultaneously reproduces 
the value of the nucleon effective mass  and the incompressibility factor  at the normal nuclear density.  
The suggested   model  belongs to the class of self-consistent EOS of  nuclear matter. 
Since its partition does not  obey  the first Van Hove   axiom \cite{Pathria},  
it is necessary 
to  assume  some additional conditions on the EOS which will provide the validity of thermodynamic identities. 

In contrast to the mean-field models an exact analytical solution of statistical models \cite{Bugaev:00, Bugaev:00b, Bugaev:01, Bugaev:01b, Bugaev:01c} 
are much  more valuable because they 
allow one not only to study the nuclear matter  phase diagram, but also to study the properties of the (tri)critical point of the PT diagram, which, as we know from 
the theory of critical phenomena, depend only on  the most general characteristics of 
short range interaction and the dimensionality of the system. 
Therefore, investigation of a statistical  model with somewhat simplified interaction can provide us with the results 
valid for the (tri)critical point of  more realistic models and/or  some substances. 
The found critical exponents of the SMM  \cite{Reuter:01, Reuter:01b} allow  one to verify the scaling relations for  the SMM critical indices and compare them with the experimental findings. 
Such a comparison enables us  to determine the universality class of the SMM and that one of nuclear matter. 

This chapter is based on the results obtained in   
\cite{Bugaev:93, Bugaev:00, Bugaev:00b, Bugaev:01, Bugaev:01b, Bugaev:01c, Reuter:01, Reuter:01b,Elliott:05wci}.

\section{A Self-consistent Mean-Field EOS for Nuclear Matter}

The determination of the nuclear matter EOS is one of  the foremost goals in todays heavy-ion physics.
Up to now our knowledge of the the nuclear EOS is restricted to one point in the plane of the 
independent thermodynamical variables temperature  $T$ and net baryon density $\rho$. 
This point is the so-called ground state of nuclear matter: at $T = 0$ nuclear matter 
saturates (i.e. the pressure $ p = p_{\rm o} =0$) at a density  
$\rho_{\rm o} \approx 0.16$~fm$^{-3}$. 
From  nuclear physics data one derives the following value for the energy per particle 
$W(\rho)   \equiv   (\varepsilon/\rho)_{T=0} - M $  
($\varepsilon$ is the energy density, $M$ is the nucleon mass) of infinite nuclear matter:
\begin{align} \label{BindingE}
W(\rho=\rho_{\rm o})   \equiv  W_{\rm o} \approx - 16 ~{\rm MeV}.
\end{align}
It is just  bulk part in a Bethe-Weitsz\"acker formula for  nuclear binding energy.   

An  analysis of the incompressibility   \cite{Incompressibility} 
\begin{align} \label{BindingEII}
K_{\rm o} ~\equiv~9\, \frac{\partial ~p}{\partial~ \rho} \biggl|_{T=0, ~\rho=\rho_{\rm o}}~ \equiv  ~
9\, \rho_{\rm o}^2 \, \frac{\partial^2 ~W}{\partial~ \rho^2} \biggl|_{T=0, ~\rho=\rho_{\rm o}} 
\end{align}
shows that  a large value of  $K_{\rm o}$ (=300~MeV, with 
considerable error) may be more compatible with the data than the previously 
reported low one,  $K_{\rm o} = 180 -240$ MeV  
\cite{OldIncompressibility:1,OldIncompressibility:2}. 
The recent analysis shows that the situation is still not very clear.
Thus, the  simulations of the sideward anisotropy observed in the A+A collisions at low and intermediate energies require  $K_{\rm o} \approx 210$~MeV,  whereas the the elliptic flow
anisotropy observed in the same experiment  requires  $K_{\rm o} \approx 300$~MeV \cite{IncompressibilityNew:1, IncompressibilityNew:2}.  The present  situation is hoped to be clarified soon at the GSI accelerator  FAIR. 

However,  to fix the coupling constants of the phenomenological model  it is possible to use 
the estimations of the effective 
nucleon mass $M^*_{\rm o}$ at  $T = 0$ and $\rho=\rho_{\rm o}$  which can be found in the literature 
\cite{BogutaEOS:1, ManyBodyEOS:1} are  $M^*_{\rm o}= (0.7 \pm 0.15) M$. 
Therefore, keeping in mind that $K_{\rm o}$ data are 
not too conclusive, we will pay more attention to the  $M^*_{\rm o}$ value.

Any reasonable model for the nuclear matter EOS must be thermodynamically 
self-consistent and reproduce the above quantities $\rho_{\rm o},~W_{\rm o}, ~ K_{\rm o}$  and 
$M^*_{\rm o}$. These properties have clear meaning: at normal nuclear density the binding 
energy (\ref{BindingE}) is defined, but it must be a minimum (then the system is mechanically stable because of vanishing pressure $ p = p_{\rm o} =0$) and the width of this minimum is defined by the incompressibility  factor (\ref{BindingEII}). 
Once the model is worked out in the vicinity of normal nuclear matter,  its behavior in other regions of the n-T plane can be then probed via heavy-ion collisions.
The aim of this section is to formulate a phenomenological generalization of  the 
mean-field theory approach and obtain a class of nuclear matter equations of state. 
This  EOS is  compared with 
the non-relativistic many-body theory of \cite{ManyBodyEOS:1} and various relativistic equations of 
state.

Following early theoretical suggestions \cite{Schide:78}, experiments which measure the 
$\pi$-meson multiplicity in heavy-ion collisions [7] have been performed in order to 
extract the nuclear EOS directly from data. For this purpose the following 
decomposition for the energy per baryon 
\begin{align} \label{EtoRho}
\frac{\varepsilon}{\rho} ~=~ M ~+~ W_{th}~+~ W_c\,,
\end{align}
was introduced with some phenomenological anz\"atzen for the thermal energy $W_{th}$, 
and the compression energy $W_{c}$. The whole construction was, however, not 
physically self-consistent: for the calculation of $W_{th}$ the momentum distribution of an 
ideal nucleon gas was used, but when introducing, in addition, the `compression 
energy' one took into account the interaction between nucleons. The interaction, 
however, modifies the ideal-gas momentum distribution, and one faces the problem 
of adjusting the relativistic Fermi distribution of the nucleons to the functional form of the `compression energy'.

To find a solution of this problem we remember that in the relativistic mean-field 
theory of Walecka \cite{Analyt:1}  (see also \cite{Analyt:2}) 
the interaction is described by scalar $\phi$ and 
vector $U^\mu$ mesonic fields with baryon-meson interaction terms in the Lagrangian: 
$g_s\, \bar{\psi}\, \psi\, \phi$  and $g_v\, \bar{\psi}\, \gamma^\mu\, \psi\, U_\mu$. For nuclear matter in thermodynamical equilibrium these 
meson fields are considered to be constant classical quantities. The scalar field 
describes the attraction between nucleons and changes the nucleon mass 
$M \rightarrow M* = M - g_s \, \phi$. The nucleon repulsion is described by a zero component 
(the spatial components must vanish due to the translational invariance) of a vector field which 
adds $U(\rho) = C_v^2 \rho $  ($C_v$ = const) to the nucleon energy ($-U(\rho)$) for the antinucleon). 
Following \cite{Bugaev:89} it is possible to  formulate a generalized nuclear matter EOS which 
includes the mean-field theory and pure phenomenological models as special cases. 
Restricting ourself at the moment to nucleonic degrees of freedom we suggest the 
following general form for the nuclear EOS in GCE:
\begin{align} \label{pEOS}
p (T, \mu) ~=~ 
\frac{\gamma_N}{3} \int ~ \frac{d^3 k}{(2 \pi)^3} ~ \frac{k^2}{\sqrt{k^2~+~{M^*}^2}}~
(f_+ + f_-) ~+~ \rho\, U(\rho)~ - ~ \int\limits_0^\rho~ d\, n ~U(n)~+~ P(M^*)  \,,
\end{align}
where $f_+$ and $f_{-}$ are the distribution functions of nucleons and antinucleons 
\begin{align} \label{fEOS}
f_{\pm}~ \equiv ~ \left[  \exp \left(  \frac{\sqrt{k^2~+~{M^*}^2} ~\mp~ \mu~ \pm ~ U(\rho) }{T} \right)  ~+~1  \right]^{-1}\, , 
\end{align}
$\mu$ is the baryonic chemical potential and $\gamma_N$ is the number of spin-isospin nucleon 
states, which equals four for symmetric nuclear matter. The dependence of the 
effective nucleon mass $M^*$ on  $T$ and $\mu$ is defined by extremizing the thermodynamical potential (maximum of the pressure): 
\begin{align} \label{dpEOS}
\left( \frac{\delta~ p (T, \mu) }{\delta~ ~~M^*} \right)_{T,~ \mu} ~\equiv ~  \frac{d~ P(M^*)}{d~~M^*}
~-~ {\gamma_N} \int ~ \frac{d^3 k}{(2 \pi)^3} ~ \frac{M^*}{\sqrt{k^2~+~{M^*}^2}}~
(f_+ + f_-) ~=~ 0    \,.
\end{align}
The baryonic number density and energy density can be found from (\ref{pEOS})-(\ref{dpEOS}) using the 
well-known thermodynamical relations:
\begin{align} \label{nEOS}
\rho(T, \mu) ~ &\equiv~ \left( \frac{\partial~ p }{\partial~\mu} \right)_{T} ~= ~  
 {\gamma_N} \int ~ \frac{d^3 k}{(2 \pi)^3} ~ 
(f_+ - f_-)    \,, \\
\varepsilon(T, \mu) ~ &\equiv~ T \left( \frac{\partial~ p }{\partial~T} \right)_{\mu} ~+~ \mu\, \left( \frac{\partial~ p }{\partial~\mu} \right)_{T}  \nonumber \\
&~=~{\gamma_N} \int ~ \frac{d^3 k}{(2 \pi)^3} ~ \sqrt{k^2~+~{M^*}^2}~(f_+ - f_-) ~+~  
\int\limits_0^\rho~ d\, n ~U (n)~-~ P(M^*)\,. 
\label{eEOS}
\end{align}

It follows from (\ref{fEOS}) that the nucleon (antinucleon) momentum distribution has 
the form of the ideal Fermi distribution in `external fields': the scalar field changes 
the nucleon (antinucleon) mass $M$  to the effective mass $M^*$ and the vector field adds 
the energy  $U(\rho)$ (-$U(\rho)$ for the antinucleon). It is important, however, that 
additional terms in  (\ref{pEOS}) and  (\ref{eEOS}) appear and represent thermodynamically 
self-consistent `field' contributions to the pressure and the energy density. The form of 
these additional contributions to the pressure (\ref{pEOS})  is adjusted to the Fermi 
distributions (\ref{fEOS}) through the general thermodynamical relation (\ref{dpEOS})! 
Formulas (\ref{pEOS})-(\ref{eEOS}) define, therefore, a special class of thermodynamically 
self-consistent equations of state for nuclear matter. It is a phenomenological 
extension of the mean-field theory. Models of this class are fixed by specifying the 
two functions $U(\rho)$ and $P(M^*)$. General physical restrictions on these functions 
have the form: 
\begin{align} 
U(-\rho) ~=~  -  ~U(\rho), \quad \quad &U(\rho)_{\rho \rightarrow \infty} ~\sim~ \rho^a
\quad {\rm for} \quad 0~ \le~ a~ \le ~1 \,, \nonumber \\
&U(\rho)_{\rho \rightarrow 0}~ ~\sim~ \rho^b \,, 
\quad {\rm for} \quad 0 ~\le~ b   \nonumber \\
 P(M^*) ~=~ &\sum\limits_{k \ge 2} a_k\, (M ~-~ M^*)^k \quad  {\rm with} \quad a_2 < 0 \,.
\label{UpEOS}
\end{align}
For instance, the upper bound $a \le 1$  for  the power $a$ provides the causality condition at high baryonic densities, i.e.  that speed of sound does not exceed that of light.

Particular choices of $U(\rho)$ and $P(M^*)$  satisfying  (\ref{UpEOS}) reproduce a great variety of nuclear EOS models known from the literature. The models of \cite{EOS1, EOS2} correspond to 
$M^* = M$ ($P(M^*)\equiv 0$) in (\ref{pEOS})   and special forms of $U(\rho)$. 
For the mean-field theory 
models \cite{Analyt:1, Analyt:2, BogutaEOS:1,  BogutaEOS:2, Waldhauser:88}  
$U(\rho) = C_V^2\, \rho$. 
The choice $a_2 = -1/2\, C_s^2$,~ $a_k = 0 ~(k > 3)$ corresponds to 
the linear mean-field theory (in the following also referred to as `Walecka model')  \cite{Analyt:1}, while considering $a_2 = -1/2\, C_s^2$, ~$a_3 = 0$,~ $ a_4 \neq 0$, ~$ a_k = 0~ (k \ge  5)$ the model reproduces the  non-linear mean-field theory  \cite{BogutaEOS:1, BogutaEOS:2, Waldhauser:88}.

At $T = 0$ we find the general relation for the models  (\ref{pEOS})-(\ref{eEOS}): 
\begin{align} 
U(\rho)  ~+~ \sqrt{ \left[ \frac{3}{2}  \pi^2 \, \rho  \right]^{ \frac{2}{3} } ~+~ {M^*}^2 } = M~+~ W(\rho)~+~\rho\, \frac{d~W}{d~n}  
\end{align} 
which corresponds to the Hugenholtz-Van Hove theorem \cite{EOS1} for an interacting 
Fermi gas at zero temperature. As $\left( \frac{d~W}{d~n} \right)_{ \rho= \rho_{\rm o} } = 0$, at saturation density we obtain the original Weisskopf relation \cite{EOS2}  between Fermi energy and the energy per 
particle 
\begin{align} \label{WeiskopfRel:1}
U(\rho_{\rm o})  ~+~ \sqrt{ \left[ \frac{3}{2}  \pi^2 \, \rho_{\rm o}   \right]^{ \frac{2}{3} } ~+~ {M^*_{\rm o}}^2 } = M~+~ W_{\rm o}~ \approx~  922~{\rm MeV}.
\end{align} 
Equation (\ref{WeiskopfRel:1}) gives us the relation between $ U(\rho_{\rm o}) $ and $M^*_{\rm o}$, therefore only one of  these quantities (e.g. $M^*_{\rm o}$) is free. 

I consider now one new example for the nuclear matter EOS from the 
generalized mean-field theory class (\ref{pEOS})-(\ref{eEOS}). One can  choose 
\begin{align} \label{OrigMy:1}
P(M^*) ~=~ - ~\frac{1}{2} \, C_s^2\, (M - M^*)^2 
\quad \quad \quad  U(\rho)~ =~ C_v^2\, \rho~ -~ C_d^2\,  \rho^{ \frac{1}{3} }. 
\end{align} 
Thus, the Walecka model is extended by an attractive term in the potential $U(\rho)$. 
Such a modification of  $U(\rho)$ is to some extent similar to the approach of the models 
\cite{Rischke:88, Zimanyi:88}. However, unlike to these models we now account for the fact that $ M^* \neq M $
in the nuclear medium. The introduction to the third parameter $C_d$ allows us to 
choose $M_{\rm o}^*$  freely in addition to the required values of $\rho_{\rm o}$ and  $W(\rho_{\rm o})$. 
It is necessary to  stress that, 
if one requires that our `coupling constant' has the dimension of an integer power 
of the fundamental units, only the power $\frac{1}{3}$ of  $\rho$ in the new attractive term of   $U(\rho)$
has the property to satisfy the constraints (\ref{UpEOS}). 
The additional term in the potential  $U(\rho)$ could be derived (in mean-field 
approximation) from a Lagrangian containing an additional nucleon-nucleon 
self-interaction term of the form 
\begin{align} \label{OrigMy:2}
\frac{3}{4} \, C_d^2\, \left( \bar{\psi}\, \gamma^\mu\,  \psi \, \bar{\psi}\, \gamma_\mu\,  \psi   \right)^ \frac{2}{3}
\,. 
\end{align} 
Of course, there is no immediate physical motivation for such a term on 
field-theoretical grounds. However, such a motivation is certainly not strictly 
required for a phenomenological  EOS. As justification, it is sufficient 
that the equation of state has physically resonable properties. This can be shown in 
the following. 

Although the incompressibility $K_{\rm o}$ cannot be chosen independently from $M^*_{\rm o}$, one 
finds  that reasonable values of $ M^*_{\rm o}$ lead to values of  $K_{\rm o}$  which lie in the 
experimentally found range (see  Table~\ref{EOSTable}). The first line in 
the Table~\ref{EOSTable}  corresponds to the 
Walecka model for which we have a too small value of M, and a too large value of 
$K_{\rm o}$.

\begin{table}[h!] \label{EOSTable}
\caption{Different sets  of `coupling constants' for fixed  $\rho_{\rm o}$ =0.15891 fm$^{-3}$ \cite{ManyBodyEOS:1} and $W_{\rm o}$= -16~MeV. }
\begin{center}
\begin{tabular}{||c|c|c|c|c||}\hline
$M^*/M$  &  $C_v^2$ (GeV$^{-2}$) &  $C_s^2$ (GeV$^{-2}$) &  $C_d^2$ &  $K_{\rm o}$ (MeV) \\
\hline
0.543 &  285.90 &  377.56  & 0        &  553 \\
0.600 &  257.40 &  326.40  & 0.124 & 380 \\
0.635 &  238.08 &  296.05  & 0.183 & 300 \\
0.688 &  206.79 & 251.14   & 0.254 & 210 \\
0.720 & 186.94  & 244.52   &~ 0.288~ & 170 \\
\hline
\end{tabular}

\end{center}

\end{table}

The surprisingly good correlation between $M^*_{\rm o}$ and  $K_{\rm o}$ implies that the above 
model accounts for four ground-state properties of nuclear matter with only three 
independent parameters. The energy density in present  model  \cite{Bugaev:93} has the form 
\begin{align} \label{OrigMy:3}
\hspace*{-0.3cm}\varepsilon(T, \mu) 
= {\gamma_N} \int  \frac{d^3 k}{(2 \pi)^3} ~ \sqrt{k^2 +{M^*}^2}~(f_+ - f_-) ~+~  \frac{1}{2}\,C_v^2\, \rho^2 ~ -~  \frac{3}{4}\, C_d^2\,  \rho^{ \frac{4}{3} }
 ~+~ \frac{1}{2} \, C_s^2\, (M - M^*)^2 \,. 
\end{align}

\begin{figure}[ht]
%

\centerline{
\hspace*{0.0cm}\epsfig{figure=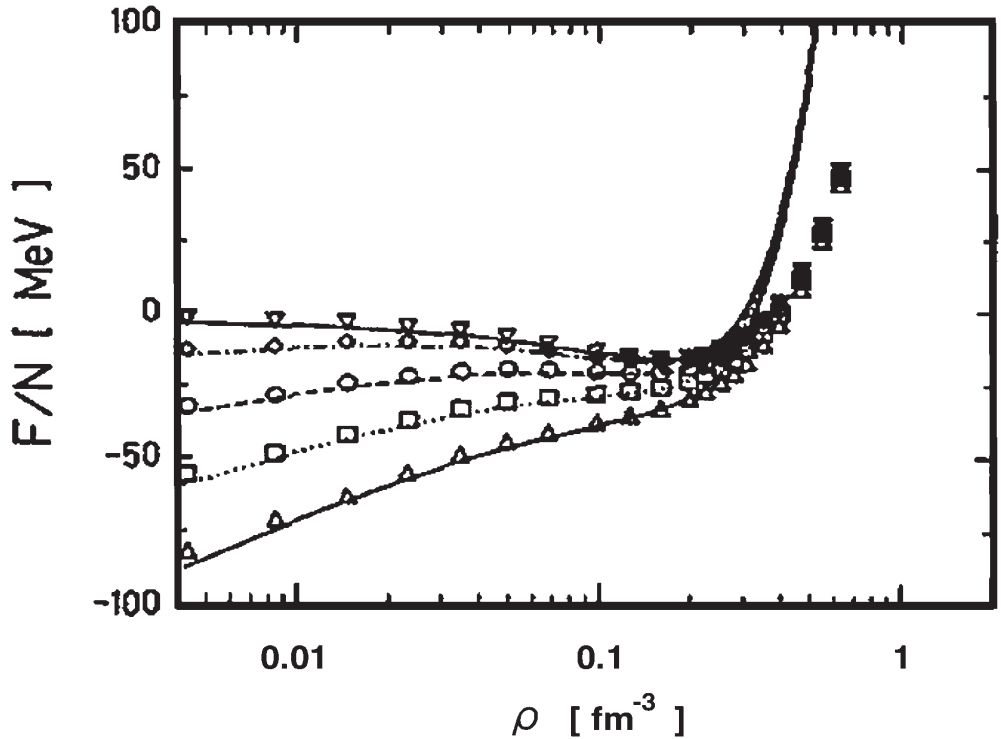,height=10.5cm,width=14.5cm} 
}

\vspace*{0.3cm}

\caption{\label{fig:oneA} 
The free energy per baryon as a function of $\rho$ for constant $T$ is shown  for the 
new EOS with  $M^*_{\rm o} = 0.635 M,~ K_{\rm o} = 300$ MeV (lines) in comparison to the calculations of 
[5] (symbols). The upper full line (and symbol $\nabla$) corresponds to $ T = 0$, the lower full 
line (and $\Delta$) to $T = 20$ MeV. Dashed-dotted (and $\Diamond$): T = 5 MeV; dashed (and {\Large o}): 
T = 10 MeV; dotted (and $\square$): $T = 15 MeV$.
}
\end{figure}

\begin{figure}[ht]

\centerline{
\hspace*{-0.0cm}\epsfig{figure=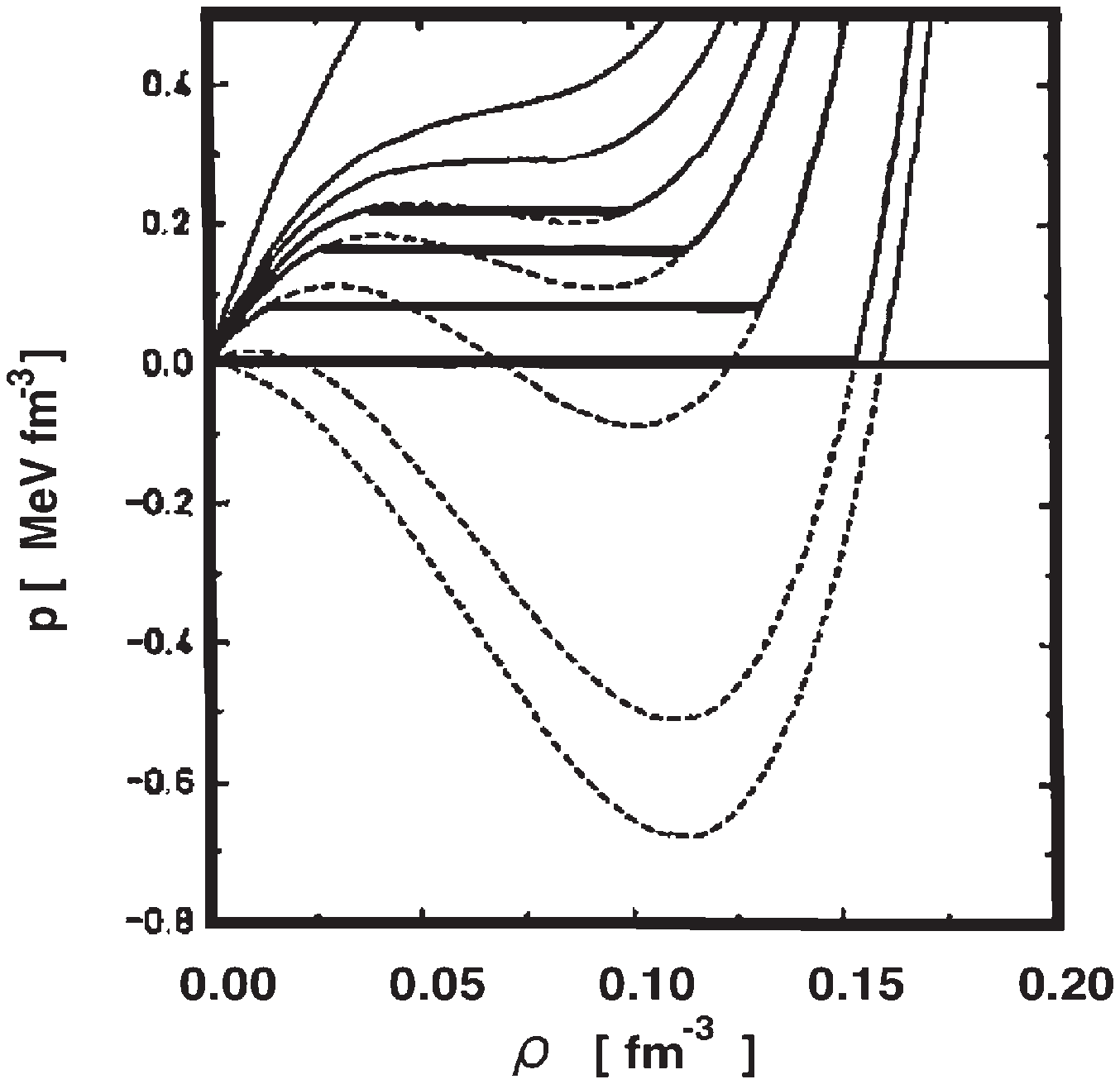,height=11.5cm,width=14.5cm} 
}

\vspace*{0.3cm}

\caption{\label{fig:oneB}
$\rho-p$ diagram showing the liquid-vapour phase transition in the self-consistent EOS. 
The curves are isotherms for $T = 0,~5,~10,~12,~13,~14, ~15, ~20$ MeV (from below to above). Dashed 
lines mark  the unstable part of the isotherms in 
the phase coexistence region, which are replaced by the full horizontal lies according to 
Maxwell's construction. 
}
\end{figure}

For further analysis we fix  $K_{\rm o} =$ 300 MeV (cf  \cite{Incompressibility}). 
Note that for this value of  $K_{\rm o} $  the effective mass 
$M^*_{\rm o} $ is in very good agreement with that obtained in the non-relativistic many-body 
calculations of Friedman and Pandharipande  \cite{ManyBodyEOS:1}. In Fig. \ref{fig:oneA} I compare the free 
energy per baryon in their calculations \cite{ManyBodyEOS:1} with that of  the present model. 
One  finds  rather 
good agreement at small $\rho$ and systematic deviations from the non-relativistic results 
at large   $\rho$. This happens at  $\rho \ge  3 \, \rho_{\rm o}$ where the nucleon Fermi momentum is large, 
$k_F \ge M$  and, therefore,  relativistic effects become important. 
At low temperatures and densities our EOS exhibits a liquid-vapour nuclear phase 
transition, as shown in  Fig.~\ref{fig:oneB}. The critical temperature beyond which there is no 
two-phase equilibrium is $\sim 14$ MeV.  This value  is somewhat by a few MeVs lower than the expected value,  however, one should remember that the present model is rather simple and does not take  into 
account any clustering effects (e.g. deuterons, tritons  and a-particles) which are necessary. 
Nevertheless, since the present model correctly accounts for the fact that the domination of the internucleon attraction  at  around  normal nuclear density changes to the domination of  short range repulsion at high baryonic densities. As a consequence the model predicts a PT existence.


The non-linear mean-field theory \cite{BogutaEOS:1, BogutaEOS:2} with four parameters $C_v^2$, $C_s^2$,
$\alpha_3$, $\alpha_4$
allows 
one to choose $M^*_{\rm o}$  and  $K_{\rm o} $ at will. However, the value of  $\alpha_4$  is positive for 
experimentally reasonable sets of  $M^*_{\rm o}$  and  $K_{\rm o} $ (see  \cite{Waldhauser:88}). 
This means that the energy 
density of the system is not bounded from below (because of the term $-P(M^*)$) 
with respect to variations of M*: such a theory is unstable, since its energetic 
minimum is $- \infty$ which demands an infinite value of the scalar field, i.e. $\phi \rightarrow \pm \infty$.

Thus,  the above  results  generalize the mean-field theory approach to the 
nuclear matter EOS.  They  give us the rules  (\ref{pEOS})-(\ref{eEOS}) to construct a class of thermo- 
dynamically self-consistent phenomenological models. As an example,  
a simple version of the nuclear EOS from this class is 
suggested and investigated. It allows for a reasonable value of the nucleon effective mass   $M^*_{\rm o}$ and simultaneously 
allows from an incompressibility  $K_{\rm o}$ in the range of the experimental values. The 
non-relativistic many-body calculations of  \cite{ManyBodyEOS:1}  coincide with the suggested  EOS up to a few 
percent in the low-density and low-temperature region. Its agreement with known 
nuclear matter properties is better than for either the phenomenological models with 
$M^* = M$, or the Walecka model as well as stable versions ($a_4 \le 0$) of the non-linear 
mean-field theory.

\section{Statistical Multifragmentation Model  in Thermodynamic Limit}

Here I present a more specialized and profound  model of the nuclear EOS. 
During more than two decades it was playing a guiding role in the research of 
nuclear liquid-gas PT.  This success is based on two facts: first,  since it is a cluster model,
the short range  attraction between fragments is accounted by the reactions between all fragments, and, second, the hard-core repulsion is an essential improvement compared to the FDM 
\cite{Fisher:67}  which,
as will be shown below, leads to very different scaling relations  in the SMM compared to   the FDM.
The infinite number of the degrees of freedom results in a qualitatively different PT mechanism 
compared to the mean-field models, whereas the different scaling relations  of  the SMM 
signals about another universality class  compared to   the FDM.
However,  despite  the great success  the proof of a PT existence in SMM and the SMM critical indices 
were missing  until recently. 

The system states in the SMM are specified by the multiplicity
sets  $\{n_k\}$
($n_k=0,1,2,...$) of $k$-nucleon fragments.
The partition function of a single fragment with $k$ nucleons is
\cite{Bondorf:95}:
$
V \phi_k (T) = V\left(m T k/2\pi\right)^{3/2}~z_k~
$,
where $k=1,2,...,A$ ($A$ is the total number of nucleons
in the system), $V$ and $T$ are, respectively, the  volume
and the temperature of the system,
$m$ is the nucleon mass.
The first two factors  on the right hand side (r.h.s.) 
of 
the single fragment partition 
originate from the non-relativistic thermal motion
and the last factor,
 $z_k$, represents the intrinsic partition function of the
$k$-nucleon fragment. Therefore, the function $\phi_k (T)$ is a phase space
density of the k-nucleon fragment. 
For \mbox{$k=1$} (nucleon) we take $z_1=4$
(4 internal spin-isospin states)
and for fragments with $k>1$ we use the expression motivated by the
liquid drop model  (see details in \mbox{Ref. \cite{Bondorf:95}):}
$
z_k=\exp(-f_k/T),
$ with fragment free energy
\begin{equation}\label{one}
f_k = - W(T)~k 
+ \sigma (T)~ k^{2/3}+ (\tau + 3/2) T\ln k~,
\end{equation}
with $W(T) = W_{\rm o} + T^2/\epsilon_{\rm o}$.
Here $W_{\rm o}=16$~MeV is the bulk binding energy per nucleon.
$T^2/\epsilon_{\rm o}$ is the contribution of
the excited states taken in the Fermi-gas
approximation ($\epsilon_{\rm o}=16$~MeV). $\sigma (T)$ is the
temperature dependent surface tension parameterized
in the following relation:
$
\sigma (T) = \sigma (T)|_{SMM}  \equiv \sigma_{\rm o}
[(T_c^2~-~T^2)/(T_c^2~+~T^2)]^{5/4},
$
with $\sigma_{\rm o}=18$~MeV and $T_c=18$~MeV ($\sigma=0$
at $T \ge T_c$). The last contribution in Eq.~(\ref{one}) involves the famous Fisher's term with
dimensionless parameter
$\tau$.  
As will be shown below, at the critical (tricritical) point the fragment mass distribution 
will lose it exponential form and will become a power law $k^{-\tau}$.

{ It is necessary to stress that  the  SMM parametrization  of  the surface tension coefficient is 
not   a unique one. For instance, the FDM successfully  employs  another one
$\sigma (T)|_{\rm FDM}  = \sigma_{\rm o} [ 1~ - ~T/T_c].$ 
As we shall see later   the temperature dependence of the surface tension 
coefficient  in the vicinity of the critical point will define the critical indices of the model, 
but   the following mathematical  analysis of the SMM  is general and  is valid for an arbitrary
$\sigma (T)$ function. 
}

The canonical partition function (CPF) of nuclear
fragments in the SMM
has the following form:
\begin{equation} \label{two}
\hspace*{-0.2cm}Z^{id}_A(V,T)=\sum_{\{n_k\}} \biggl[
\prod_{k=1}^{A}\frac{\left[V~\phi_k(T) \right]^{n_k}}{n_k!} \biggr] 
{\textstyle \delta(A-\sum_k kn_k)}\,.
\end{equation}
In Eq. (\ref{two}) the nuclear fragments are treated as point-like objects.
However, these fragments have non-zero proper volumes and
they should not overlap
in the coordinate space.
In the excluded volume (Van der
Waals) approximation
this is achieved
by substituting
the total volume $V$
in Eq. (\ref{two}) by the free (available) volume
$V_f\equiv V-b\sum_k kn_k$, where
$b=1/\rho_{{\rm o}}$
($\rho_{{\rm o}}=0.16$~fm$^{-3}$ is the normal nuclear density).
Therefore, the corrected CPF becomes:
$
Z_A(V,T)=Z^{id}_A(V-bA,T)
$.
The SMM defined by Eq. (\ref{two})
was studied numerically in Refs. \cite{Gupta:98,Gupta:99}.
This is a simplified version of the SMM, since  the symmetry and
Coulomb contributions are neglected.
However, its investigation
appears to be of  principal importance
for studies of the nuclear  liquid-gas phase transition.

The calculation of $Z_A(V,T)$  
is difficult due to  the constraint $\sum_k kn_k =A$.
This difficulty can be partly avoided by 
evaluating
the GCE partition)
\begin{equation}\label{three}
{\cal Z}(V,T,\mu)~\equiv~\sum_{A=0}^{\infty}
\exp\left({\textstyle \frac{\mu A}{T} }\right)
Z_A(V,T)~\Theta (V-bA) ~,
\end{equation}
where $\mu$ denotes a chemical potential.
The calculation of ${\cal Z}$  is still rather
difficult. The summation over $\{n_k\}$ sets
in $Z_A$ cannot be performed analytically because of
additional $A$-dependence
in the free volume $V_f$ and the restriction
$V_f>0 $.
The presence of the theta-function in the GCE partition (\ref{three}) guarantees
that only configurations with positive value of the free volume 
are counted. However,
similarly to the delta function restriction in Eq.~(\ref{two}),
it makes again
the calculation of ${\cal Z}(V,T,\mu)$ (\ref{three}) to be rather
difficult. This problem  was  resolved   \cite{Bugaev:00,Bugaev:01} 
by performing the Laplace
transformation of ${\cal Z}(V,T,\mu)$. This introduces the so-called
isobaric
partition function (IP)  \cite{Goren:81}:
\begin{eqnarray} \label{four}
\hat{\cal Z}(s,T,\mu)~\equiv ~\int_0^{\infty}dV~{\textstyle e^{-sV}}
~{\cal Z}(V,T,\mu)
&=&
\hspace*{-0.0cm}\int_0^{\infty}\hspace*{-0.2cm}dV^{\prime}~{\textstyle e^{-sV^{\prime} } }
\sum_{\{n_k\}}\hspace*{-0.1cm}\prod_{k}~\frac{1}{n_k!}~\left\{V^{\prime}~\phi_k(T)~
{\textstyle e^{\frac{(\mu  - sbT)k}{T} }}\right\}^{n_k} \nonumber \\
&=&\hspace*{-0.0cm}\int_0^{\infty}\hspace*{-0.2cm}dV^{\prime}
~{\textstyle e^{-sV^{\prime} } }
\exp\left\{V^{\prime}\sum_{k=1}^{\infty}\phi_k ~
{\textstyle e^{\frac{(\mu  - sbT)k}{T} }}\right\}~.
\end{eqnarray}

\vspace*{-0.1cm}

\noindent
After changing the integration variable $V \rightarrow V^{\prime}$,
the constraint of $\Theta$-function has disappeared.
Then all $n_k$ were summed independently leading to the exponential function.
Now the integration over $V^{\prime}$ in Eq.~(\ref{four})
can be  done resulting in
\begin{equation}\label{five} 
\hat{\cal Z}(s,T,\mu)~=~\frac{1}{s~-~{\cal F}(s,T,\mu)}~,
\end{equation}

\vspace*{-0.3cm}

\noindent
where

\vspace*{-0.5cm}

\begin{eqnarray} 
\hspace*{-0.9cm}
{\cal F}(s,T,\mu) & = &  \sum_{k=1}^{\infty}\phi_k ~ 
\exp\left[\frac{(\mu - sbT)k}{T}\right]  \nonumber \\
& = &
 \left( \frac{mT }{2\pi}\right)^{\frac{3}{2} } 
\left[z_1 \exp\left(\frac{\mu-sbT}{T}\right) + 
\sum_{k=2}^{\infty}
k^{ -\tau} \exp\left(
\frac{(\tilde\mu - sbT)k -
\sigma k^{2/3}}{T}\right)\right]. 
\label{Osix}
\end{eqnarray}
%
Here we have introduced the shifted chemical potential
$\tilde{\mu}~\equiv~\mu ~+~W (T) $.
From the definition of pressure in the grand canonical ensemble
it follows that, in the thermodynamic limit,
the GCE partition of the system  behaves as 
\begin{equation}\label{gcpfunc}
p(T,\mu)~\equiv~ T~\lim_{V\rightarrow \infty}\frac{\ln~{\cal Z}(V,T,\mu)}
{V}
\quad \Rightarrow \quad 
{\cal Z}(V,T,\mu)\biggl|_{V \rightarrow \infty } \sim 
\exp\left[\frac{p(T,\mu)V}{T} \right]~. \biggr.
\end{equation}  
An exponentially over $V$ increasing part of ${\cal Z}(V,T,\mu)$
on the right-hand side of Eq.~(\ref{gcpfunc}) generates
the rightmost singularity $s^*$ of the function
$\hat{\cal Z}(s,T,\mu)$, because for $s<p(T,\mu)/T$ the
$V$-integral for $\hat{\cal Z}(s,T,\mu)$ (\ref{four}) diverges at its upper
limit. Therefore, 
in the thermodynamic limit, $V\rightarrow \infty$ the system pressure
is defined by this rightmost  singularity, $s^*(T,\mu)$, of  
IP $\hat{\cal Z}(s,T,\mu)$ (\ref{four}): 
\begin{equation}\label{ptmuii}
p(T,\mu)~=~T~s^*(T,\mu)~.
\end{equation}
Note that this simple connection of the rightmost \mbox{$s$-singularity}
of $\hat{\cal Z}$, Eq.~(\ref{four}), to the asymptotic,
 $V\rightarrow\infty$,
behavior of ${\cal Z}$, Eq.~(\ref{gcpfunc}), is a general mathematical
property
of the Laplace transform. Due to this property the study of the 
system behavior in the thermodynamic limit
$V\rightarrow \infty$ can be reduced to the investigation of
the singularities of $\hat{\cal Z}$.


\section{Singularities of  Isobaric Partition  and the Mechanism of   Phase Transitions}

The IP, Eq.~(\ref{four}),  has two types of singularities:
1) the simple pole singularity
defined by the equation
\begin{equation}\label{pole}
s_g(T,\mu)~=~ {\cal F}(s_g,T,\mu)~,
\end{equation}
2)  the singularity  of the function ${\cal F}(s,T,\mu)$ 
 it-self at the point $s_l$ where the coefficient 
in linear over $k$ terms in the exponent is equal to zero,
\begin{equation}\label{sl}
s_l(T,\mu)~=~\frac{\tilde{\mu}}{Tb}~.
\end{equation}

The simple pole singularity corresponds to the gaseous phase 
where pressure is determined by the 
equation
\begin{eqnarray}\label{pgas}
p_g(T,\mu) &=& \left( \frac{mT }{2\pi}\right)^{3/2} T
\left[z_1 \exp\left(\frac{\mu-bp_g}{T}\right)
+ \sum_{k=2}^{\infty}
k^{ -\tau} \exp\left(
\frac{(\tilde{\mu} - bp_g)k - \sigma k^{2/3} }{T}\right)\right]~.
\end{eqnarray}
The singularity $s_l(T,\mu)$ of the function ${\cal F}(s,T,\mu)$
(\ref{Osix}) defines the liquid pressure
\begin{equation}\label{pl}
p_l(T,\mu)~\equiv~ T~s_l(T,\mu)~=~
\frac{\tilde{\mu}}{b}~.
\end{equation}

In the considered model the liquid phase is represented by an
infinite fragment, i.e. it corresponds to the macroscopic population
of the single mode $k = \infty$. Here one can see the analogy
with the Bose condensation where the  macroscopic population
of a single mode occurs in the momentum space.

\begin{figure}

\mbox{
\hspace*{0.0cm}\epsfig{figure=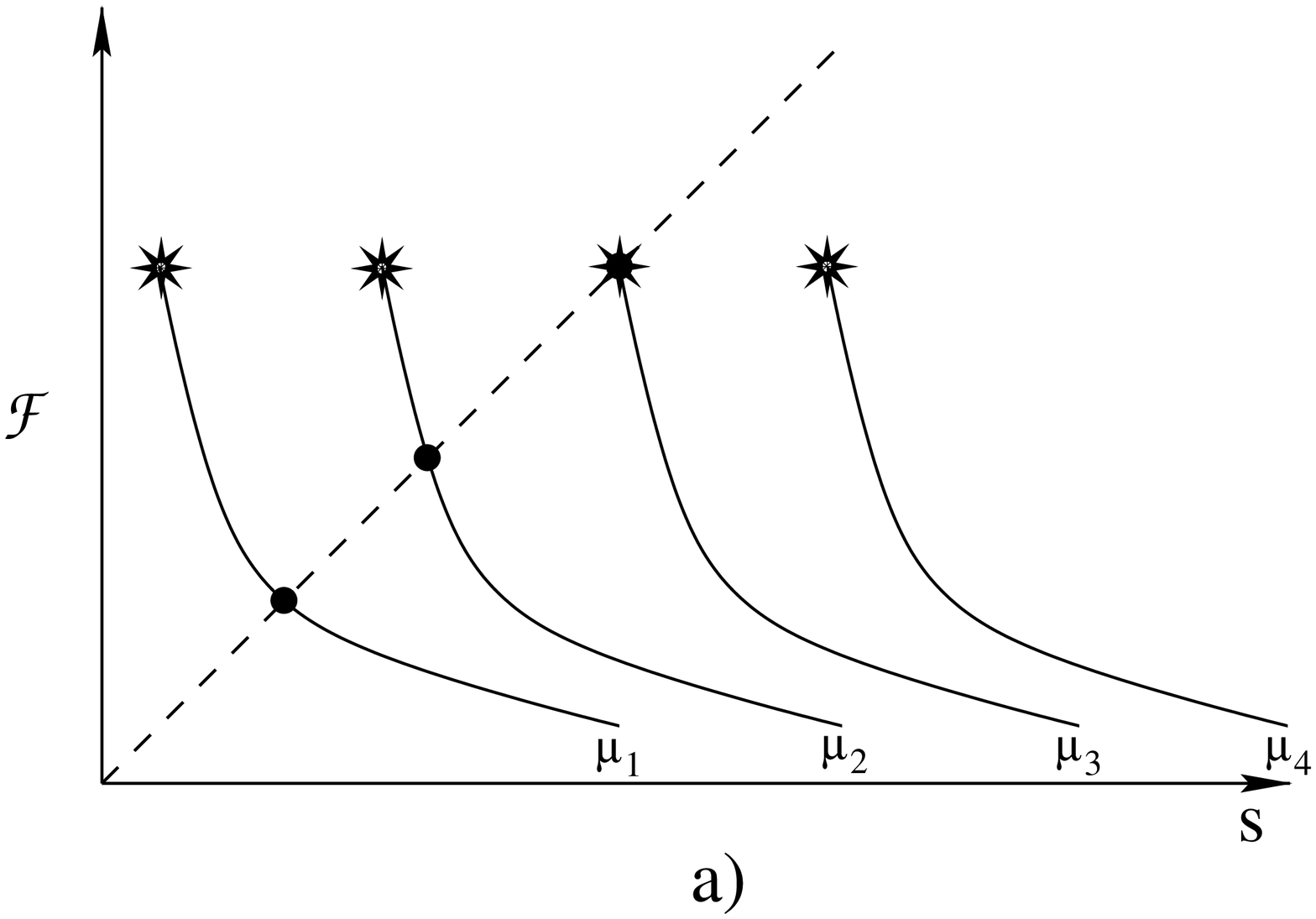,height=5.5cm,width=7.5cm} 
\hspace*{0.5cm} \epsfig{figure=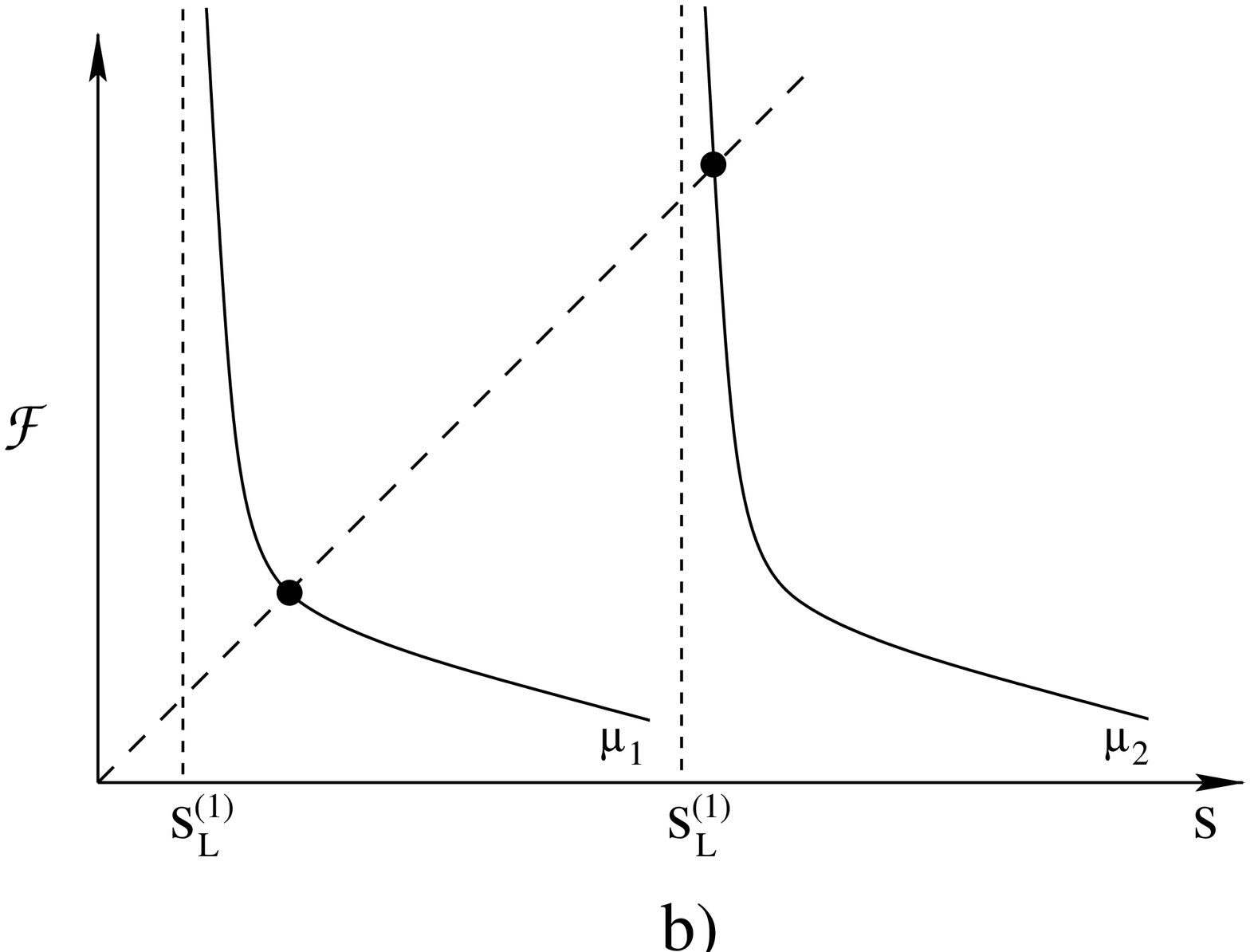,height=5.5cm,width=7.cm}
}

\vspace*{0.3cm}

\caption{\label{fig:one}
Schematic view of singular points of the Isobaric Partition, Eq. (\ref{five}),
at $T < T_c$ (a) and $T > T_c$ (b).
Full lines show ${\cal F}(s,T,\mu)$ as a function of $s$ at fixed $T$ and $\mu$,
$\mu_1 < \mu_2 < \mu_3 < \mu_4$.
Dots and asterisks indicate the simple poles ($s_g$) and the singularity of
function ${\cal F}$ it-self ($s_l$), respectively.
At $\mu_3 = \mu^*(T)$ the two singular points coincide signaling a phase transition.
}
\end{figure}

In the $(T,\mu)$-regions where $\tilde{\mu} < bp_g(T,\mu)$ the gas phase
dominates ($p_g > p_l$), while  the liquid phase
corresponds to $\tilde{\mu} > b p_g(T,\mu)$. The liquid-gas phase transition
occurs when  two singularities coincide,
i.e. $s_g(T,\mu)=s_l(T,\mu)$.
A schematic  view of singular points is shown 
in Fig.~\ref{fig:one}a for $T <T_c$, i.e. when $\sigma > 0$.
The two-phase coexistence region is therefore defined by the
equation
\begin{equation}\label{ptr}
p_l(T,\mu)~=~p_g(T,\mu)~,~~~~{\rm i.e.,}~~ \tilde{\mu}~=~b~p_g(T,\mu)~.
\end{equation} 
One can easily see that ${\cal F}(s,T,\mu)$ is monotonously decreasing
function of $s$. 
The necessary condition for the phase
transition is that this function
remains finite in its singular
point \mbox{$s_l=\tilde{\mu}/Tb$:}
\begin{equation}\label{Fss}
{\cal F}(s_l,T,\mu)~<~\infty~.
\end{equation}
The convergence of ${\cal F}$ is determined
by $\tau$ and $\sigma$.
At $\tau=0$ the condition (\ref{Fss}) requires $\sigma(T) >0$.
Otherwise, ${\cal F}(s_l,T,\mu)=\infty$ and the simple pole
singularity $s_g(T,\mu)$ (\ref{pole}) is always the
rightmost $s$-singularity
of $\hat{\cal Z}$ (\ref{four}) (see Fig.~\ref{fig:one}b). 
At $T>T_c$, where $\sigma(T)|_{SMM}=0$, the considered system
can exist only in the one-phase state. It will be shown below
that for $\tau>1 $  the condition (\ref{Fss})
can be satisfied even at $\sigma(T)=0$.

{
It is possible to generalize the SMM with its specific parameterization for 
the liquid phase pressure to more general expressions \cite{Bugaev:05csmm}.
Indeed, if  in the SMM equations above one  substitutes $W(T)  \longrightarrow  b\,  p_{lg} (\mu, T)- \mu $ and $\tilde{\mu} \longrightarrow  b\,  p_{lg}(\mu, T) $, where some general  expression
is used for the  liquid phase pressure $p_{lg}(\mu, T)$, then it can be checked that all the  SMM remain valid.
Therefore,  the resulting exactly solvable  model can be called the generalized SMM, or GSMM hereafter. 
As was discussed at the beginning of this chapter, to be realistic the generalized SMM should reproduce the properties of normal  nuclear matter. The usual SMM  was formulated in such a way that these properties are reproduced automatically.  This choice is, however, not a unique one. 
Moreover, it turns out that by a proper choice of the  liquid phase pressure the GSMM can account for the compressibility of the nuclear matter.  In particular,  for 
$p_{lg} (\mu, T)$ it is possible to use either the self-consistent mean-field  EOS  discussed 
at the beginning of this chapter  or the hadronic matter  EOS which will be considered in the chapter 3.

}


\vspace*{0.3cm}

At $T<T_c$ the system undergoes the 1-st order phase transition
across the line $\mu^*=\mu^*(T)$ defined by Eq.(\ref{ptr}).
Its explicit form is given
by the expression:
\begin{eqnarray}\label{muc}
\mu^*(T)~ & = & - W (T) ~
+~\left(\frac{mT}{2\pi}\right)^{3/2} T b
\left[z_1\exp\left(-~\frac{W(T)}{T}\right) 
+  \sum_{k=2}^{\infty}
k^{ -\tau} \exp\left(-~ \frac{\sigma~ k^{2/3}}{T}\right)
\right]\,.~
\end{eqnarray}
The points on the line
$\mu^*(T)$ correspond to the mixed phase
states. First we  consider the case  $\tau=- 1.5$ because it is  the standard SMM choice.

The
baryonic density 
is found as $(\partial p/\partial \mu)_T$ and
is given by the following  formulae in the liquid and gas phases
\begin{eqnarray}
& & \rho_l~  \equiv  ~
\left(\frac{\partial  p_l}{\partial \mu}\right)_{T}~
= ~ \frac{1}{b}~, \quad  \quad 
\rho_g  \equiv  ~
\left(\frac{\partial  p_g}{\partial \mu}\right)_{T}~=~
 \frac{ \rho_{id} }{ 1 + b\, \rho_{id} } ~,\label{rhog}
\end{eqnarray}
respectively. Here the function $ \rho_{id}$ is defined as
\begin{eqnarray}\label{rhoid}
\rho_{id}(T,\mu) & = & \left( \frac{mT }{2\pi}\right)^{3/2} 
\left[z_1 \exp\left(\frac{\mu-bp_g}{T}\right) 
 + \sum_{k=2}^{\infty} 
k^{1 -\tau} \exp\left(
\frac{( \tilde\mu - bp_g)k - \sigma k^{2/3}}{T} \right)\right]~.
\end{eqnarray}


\noindent
Due to the condition
(\ref{ptr})
this expression 
is simplified 
in the mixed phase: 
\begin{eqnarray}\label{rhoidmix}
\hspace*{-0.6cm}\rho_{id}^{mix}(T)~ &\equiv&~\rho_{id}(T,\mu^*(T))~
=  \left( \frac{ mT }{2 \pi}\right)^{3/2}
 \left[z_1 \exp\left(-~\frac{W (T) }{T}\right)
~+~\sum_{k=2}^{\infty}
k^{1 - \tau}
\exp\left(-~ \frac{\sigma ~k^{2/3}}{T}\right)\right]~.
\end{eqnarray}
This formula clearly shows that the bulk (free) energy acts in favor of the composite
fragments, but the surface term favors single nucleons.


\begin{figure}[ht] 
\mbox{
\hspace*{-0.7cm}\epsfig{figure=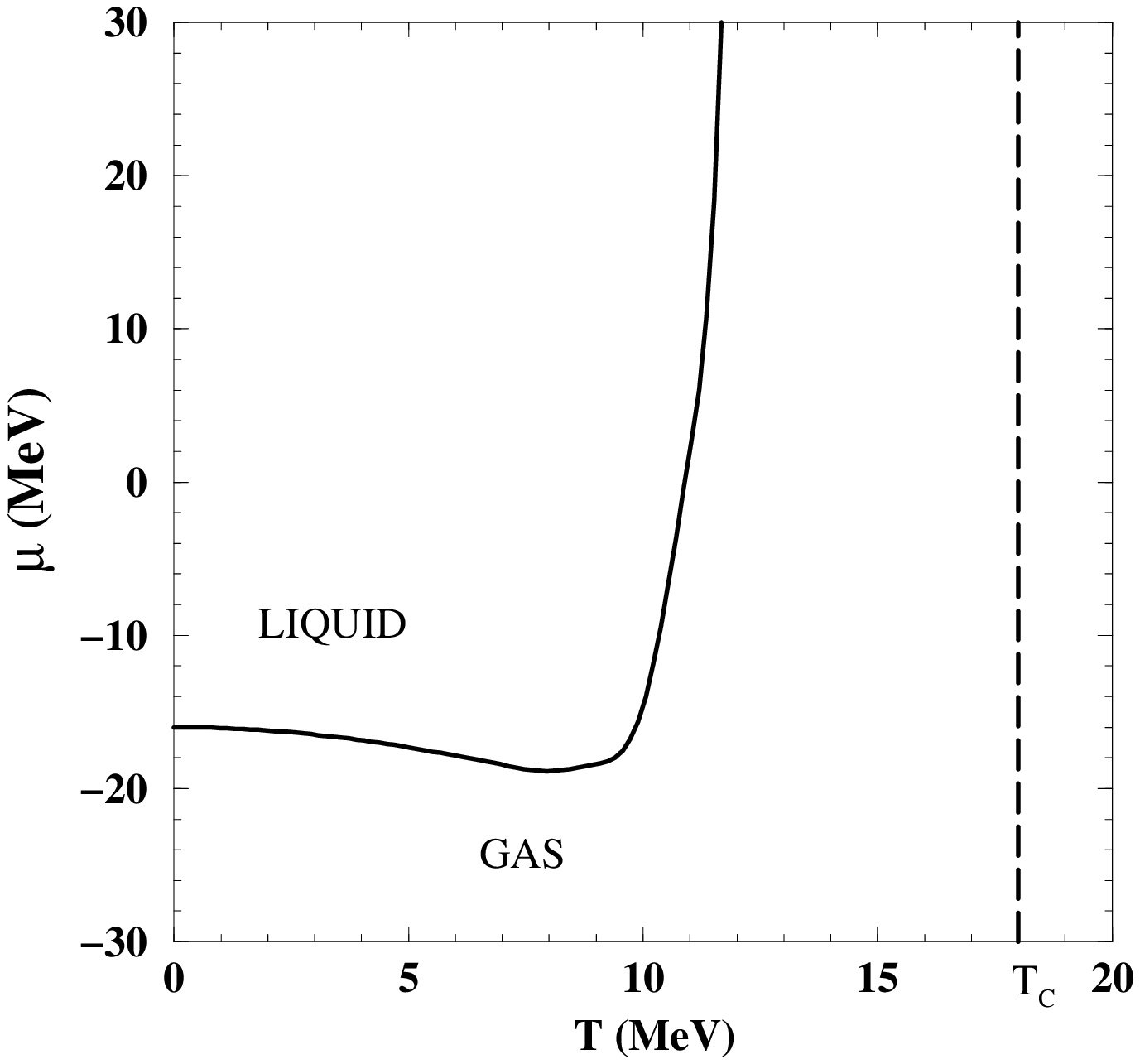,width=8.4cm}
\hspace*{-3.0cm} \epsfig{figure=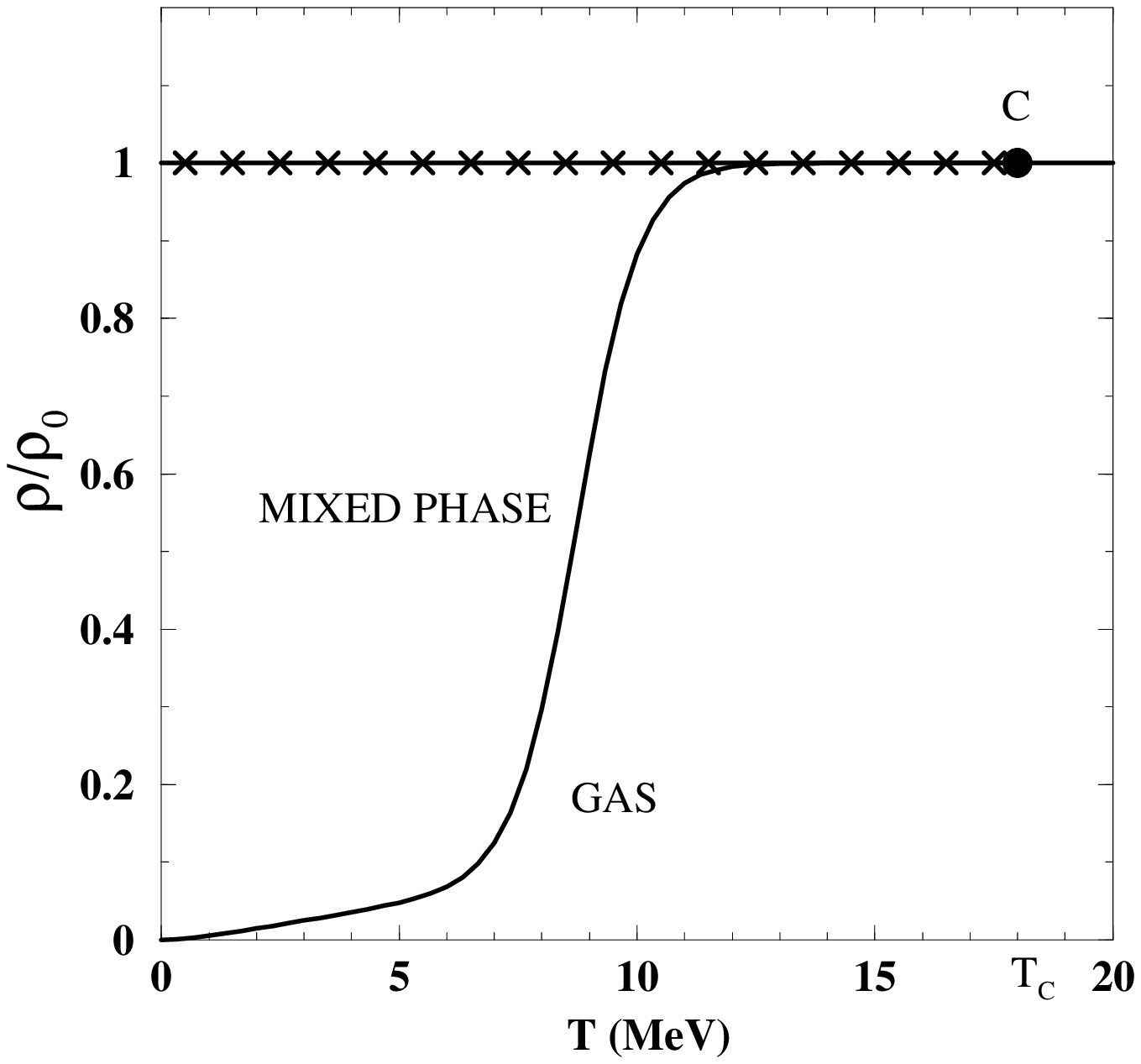,width=8.4cm}
\hspace*{-3.0cm} \epsfig{figure=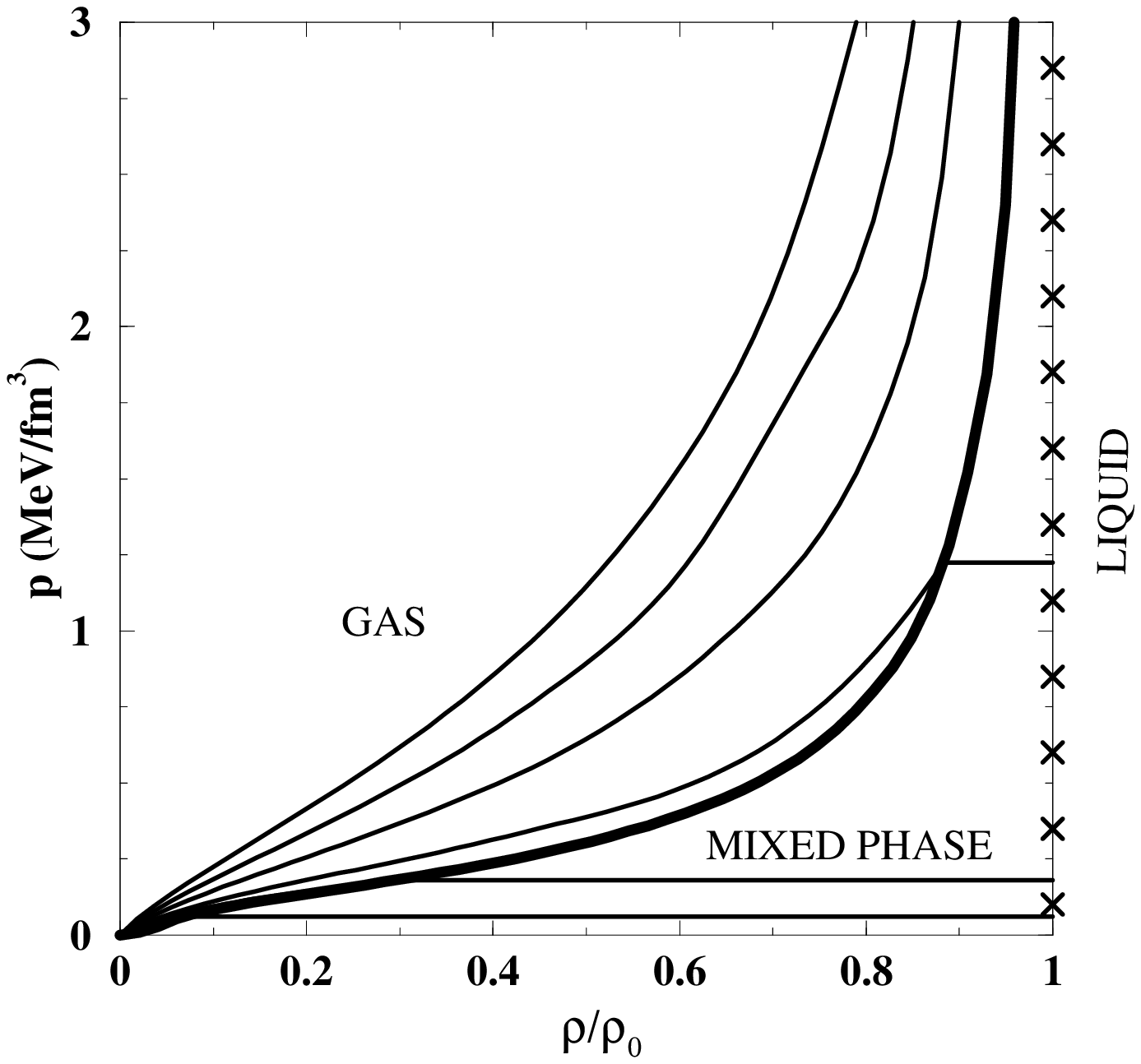,width=8.4cm}
}

\vspace*{0.3cm}

\caption{\label{fig:two}
Phase diagram in $T-\mu$ (left panel),  $T-\rho$ (middle  panel) and $\rho/\rho_{\rm o}-p$ 
(right panel)
planes for $\tau = -1.5$.
The mixed phase is represented by the line $\mu^*(T)$ in the left panel and by 
the extended region in the middle panel.
Liquid phase (shown by crosses) exists at density $\rho = \rho_{\rm o}$.
Point $C$ is the critical point.
In the right pane the isotherms (thin solid lines) are shown for $T = 4, 8, 10, 14, 18\,\,{\rm and}\,\, 22$ MeV from bottom to top.
The boundary of the mixed and gaseous phases is shown by the thick solid line. 
Liquid phase is indicated by crosses.
The critical point $T = T_c = 18$ MeV, $\rho = \rho_{\rm o}$ corresponds to infinite pressure.
}
\end{figure}

Since at $\sigma >0$ the sum in Eq.~(\ref{rhoidmix})
converges at any $\tau$, $\rho_{id}$ is finite and according to
Eq.~(\ref{rhog}) $\rho_g<1/b$. Therefore,
the baryonic density has a discontinuity $\Delta\rho =\rho_l-\rho_g >0$
across the line $\mu^*(T)$ (\ref{muc}) for  any $\tau$. 
The discontinuities
take place also for
the energy and entropy densities. 
The phase diagram of the system in the $(T,\rho)$-plane 
is shown in the middle  panel of  Fig.~\ref{fig:two}
The line
$\mu^*(T)$ (\ref{muc}) (see the left panel of  Fig.~\ref{fig:two})  corresponding to the mixed phase
states 
is transformed into
the finite region in the $(T,\rho)$-plane. As usual,  in this mixed phase
region  of the
{ phase diagram the baryonic density $\rho$ and the energy density
are   superpositions of the corresponding densities of  liquid and gas:}
\begin{equation}\label{mixed}
\rho~=~\lambda~\rho_l~+~(1-\lambda)~\rho_g~, \quad \quad  
\varepsilon ~=~\lambda~\varepsilon_l~+~(1-\lambda)~\varepsilon_g~.
\end{equation}
Here $\lambda$ ($0<\lambda <1$) is a fraction of the system volume
occupied by the liquid  inside the mixed phase, 
{ and the  partial 
energy densities 
for $(i=l,g)$ can be found from the thermodynamic identity \cite{Bugaev:00}:}
\begin{equation}\label{eps}
\varepsilon_i~\equiv~T\frac{\partial p_i}{\partial T}~+~
\mu\frac{\partial p_i}{\partial \mu}~-~p_i~.
\end{equation}

One finds
\begin{eqnarray}
\hspace*{-1.5cm}&&\varepsilon_l~=~
\frac{T^2/\epsilon_{\rm o}~-~W_{\rm o}}{b}~,\label{epsl}\\
%
%
\hspace*{-1.5cm}&&\varepsilon_g~=~\frac{  1}{1~+~b\rho_{id} } 
\,\,\left\{ \,\,
\frac{3}{2} \,p_g~
+~ \left(    {T^2}/{\epsilon_{\rm o}} -W_{\rm o} \right)\,\rho_{id} 
 \right. \label{epsg} 
 \nonumber
 \\
\hspace*{-1.5cm}&&+~\left. {\left(\frac{mT}{2\pi}\right)^{3/2}
\left(\sigma -T\frac{d\sigma}{dT} \right)
\left[ z_1 \, \exp\left(\frac{\mu -bp_g}{T} \right) +\sum_{k=2}^{\infty}
k^{\frac{2}{3}-\tau}\exp\left(\frac{(\nu -bp_g)k-\sigma k^{2/3}}{T}
\,\,\,\right)
\right] }\,\, \right\}\,.
%
\end{eqnarray}
The pressure on the phase transition line $\mu^* (T)$ (\ref{muc}) is
a monotonically 
increasing function of $T$
\begin{equation}\label{pcr}
 p^*(T) \equiv p_g(T, \mu^* (T) ) =  
\left( \frac{  m T}{2 \pi} \right)^{3/2} T 
\left[z_1\exp\left(-\frac{W}{T}\right)+
\sum_{k=2}^{\infty}  k^{ -\tau}
\exp\left(- ~\frac{\sigma~k^{2/3}}{T}\right)\right]~.
\end{equation}
The right panel of Fig.~\ref{fig:two} 
shows the pressure isotherms as functions of the reduced density
$\rho/\rho_{\rm o}$  for $\tau=-1.5$.
Inside the mixed phase
the obtained pressure isotherms 
are horizontal straight lines  in accordance with the Gibbs
criterion.
These straight lines go up to infinity when $T\rightarrow T_c-0$.
This formally corresponds to the critical point, $T=T_c$,
$\rho=\rho_c=1/b$ and $p_c=\infty$, 
in the considered case of $\tau=-1.5$. 
For $T>T_c$ the pressure isotherms never enter into the mixed phase region.
Note that, if $\sigma(T)$ would never vanish, the mixed
phase would extend up to infinite temperatures.

\begin{figure}[ht]

\mbox{
\hspace*{0.0cm}\epsfig{figure=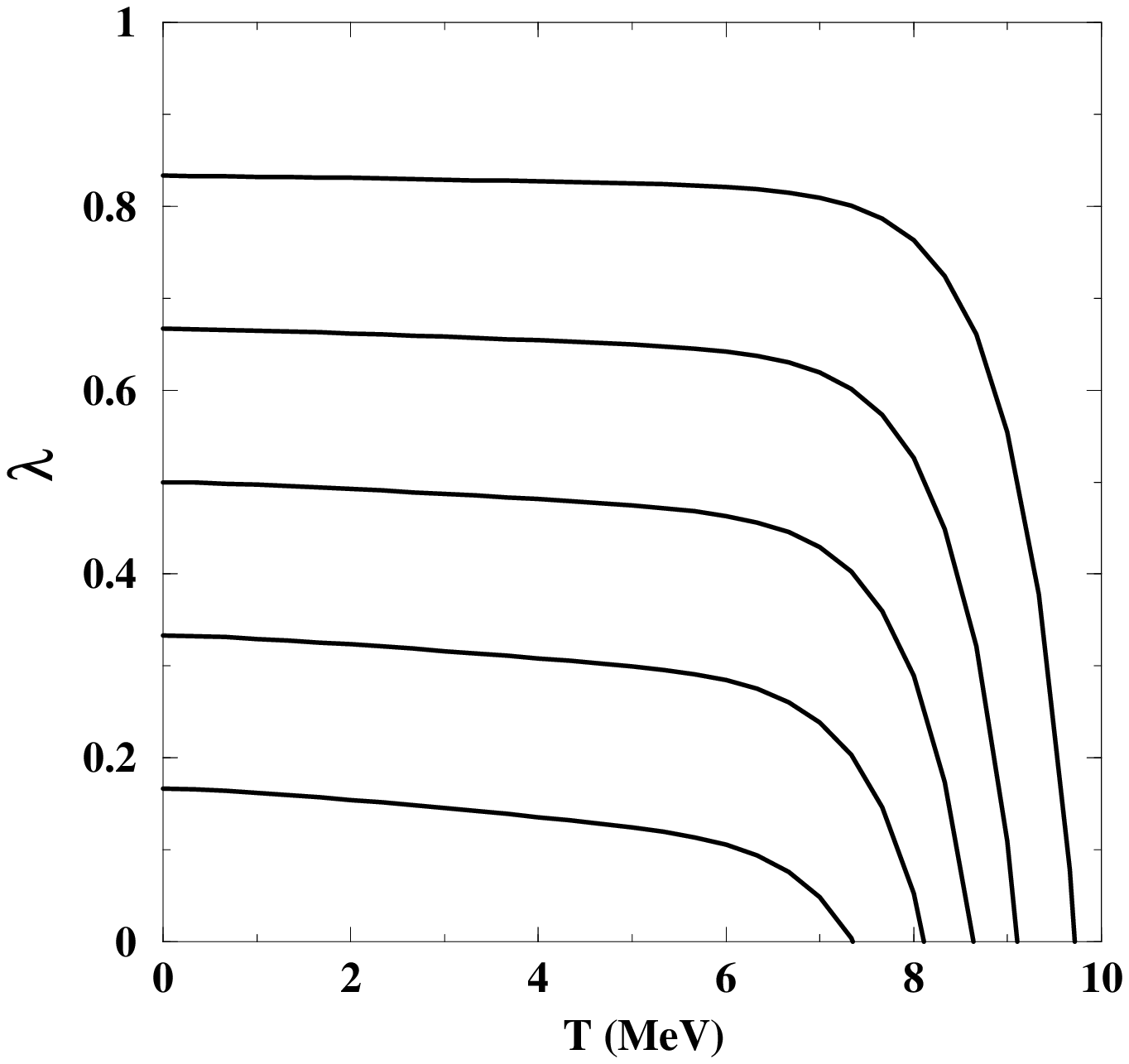,width=8.4cm}
\hspace*{-3.0cm} \epsfig{figure=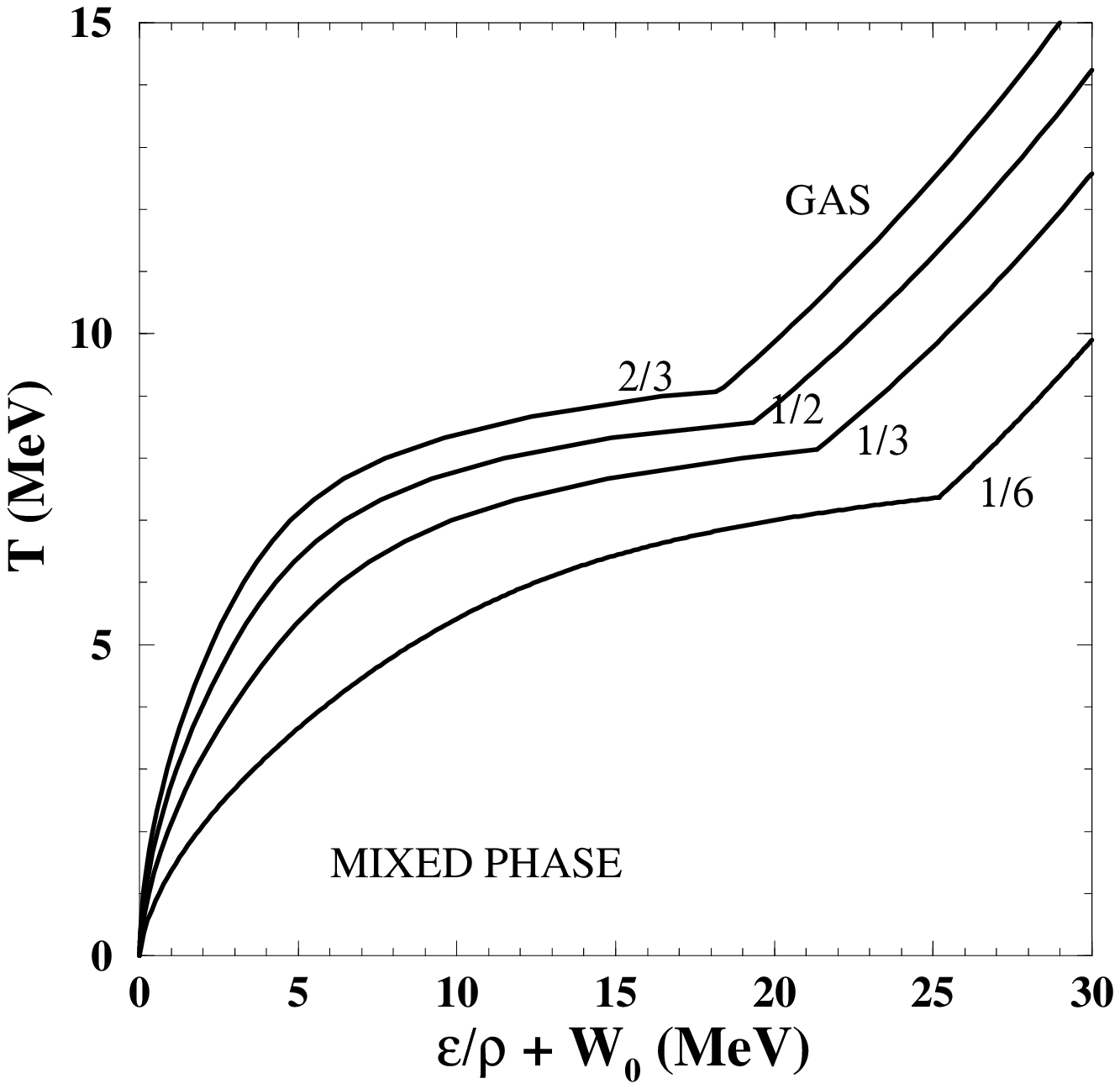,width=8.4cm}
\hspace*{-3.0cm} \epsfig{figure=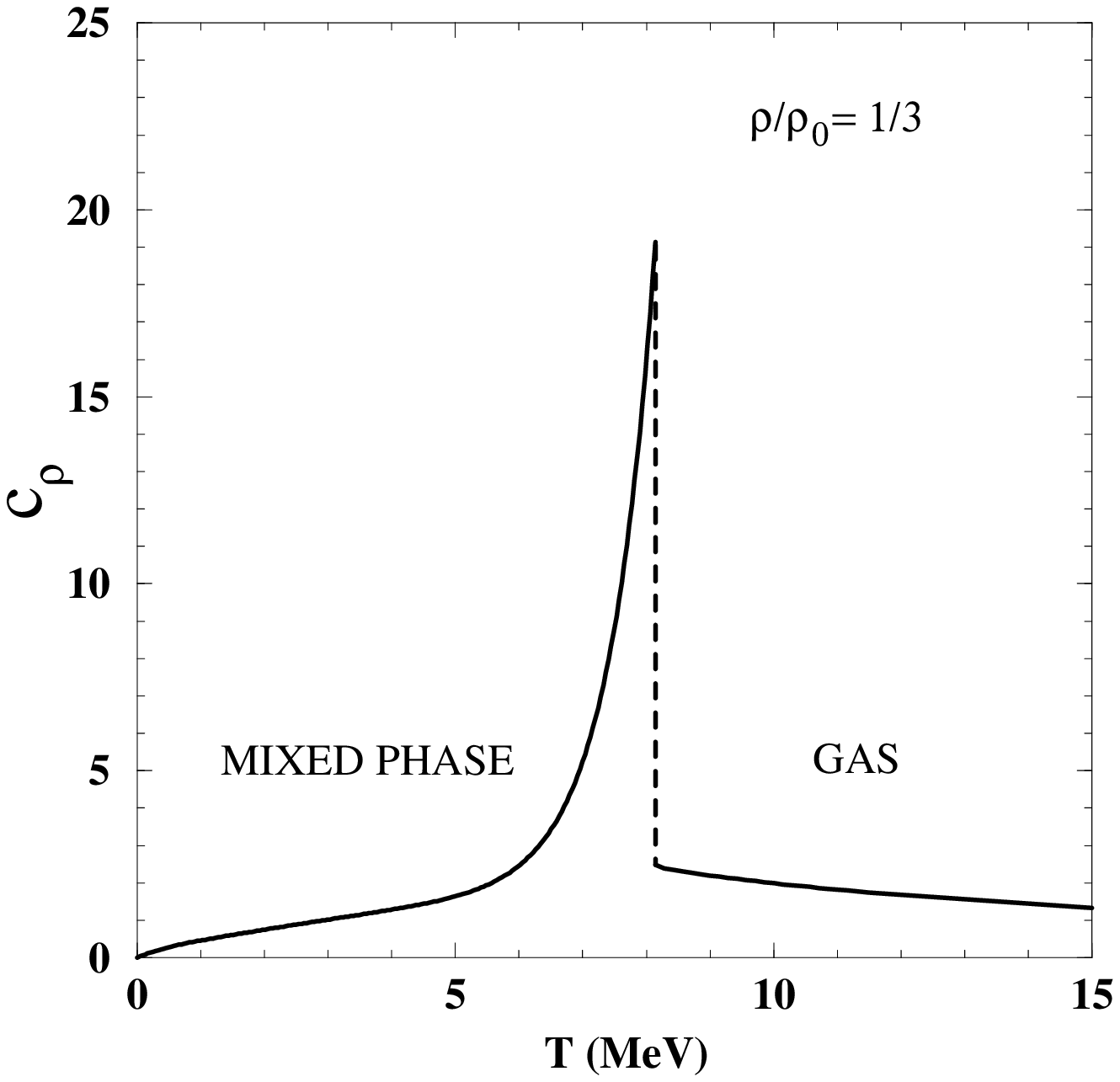,width=8.4cm}
}

\vspace*{0.3cm}

\caption{\label{fig:three}
{\bf Left panel:} Volume fraction $\lambda(T)$ of the liquid
inside the mixed phase is
shown as a function of temperature
for fixed nucleon densities ${\rho}/{\rho_{\rm o} } = 1/6, 1/3, 1/2, 2/3, 5/6$
(from bottom to top) and  $\tau = -1.5$.  \hfill \newline
{\bf Middle panel:} 
Temperature as a function of energy  per nucleon
(caloric curve)
is shown for fixed nucleon densities ${\rho}/{\rho_{\rm o} } = 1/6, 1/3,
1/2, 2/3$ and $\tau = -1.5$. {  Note that the shape  of the model caloric curves is very similar
to the experimental finding \cite{CalorCurve:Exp4}, although  our estimates for the excitation 
energy is somewhat larger due to oversimplified interaction. For  a quantitative comparison between 
the simplified SMM the full SMM interaction should be accounted for. 
}
 \newline
{\bf Right panel:} 
Specific heat per nucleon as a function of temperature
at fixed nucleon density ${\rho}/{\rho_{\rm o} } = 1/3$. The dashed line
shows the finite discontinuity of $c_{\rho}(T)$
at the boundary of the mixed and gaseous phases for $\tau = -1.5$. 
}
\end{figure}

Inside the mixed phase at constant density $\rho$ the
parameter $\lambda$ has a specific temperature dependence
shown in the left  panel of  Fig.~ \ref{fig:three}:
from an approximately
constant value $\rho/\rho_{\rm{o}}$ at small $T$ the function 
$\lambda(T)$ drops to zero in a narrow
vicinity of the boundary separating the  mixed phase and 
the pure gaseous phase.
This corresponds to a fast change of the configurations from
the state which is  dominated by one infinite liquid fragment to 
the gaseous multifragment configurations. It happens inside the
mixed phase  without
discontinuities in the thermodynamical functions.



An abrupt decrease of $\lambda(T)$ near this boundary
causes a strong
increase of the energy density as a function of temperature.
This is evident from the middle panel of  Fig.~\ref{fig:three}  which shows the caloric curves at different
baryonic densities. One can clearly see a 
leveling of temperature at energies per nucleon between 10 
and 20 MeV.
As a consequence this  
leads to a sharp peak 
in the specific heat per nucleon at constant density,
$c_{\rho}(T)\equiv (\partial \varepsilon/\partial T)_{\rho}/\rho~$,
presented in Fig.~\ref{fig:three}.
A finite discontinuity of $c_{\rho}(T)$ arises
at the boundary between the mixed phase and  the gaseous phase.
This finite discontinuity
is caused by the fact that
$\lambda(T)=0$, but
$(\partial\lambda/\partial T)_{\rho} \neq 0$
at this boundary  (see Fig.~\ref{fig:three}).

It should be emphasized that the energy density is continuous
at the boundary of the mixed phase and the gaseous phase, hence
the sharpness of the
peak in $c_{\rho}$ is entirely due to the strong temperature
dependence
of $\lambda(T)$ near this boundary. 
Moreover, at any $\rho < \rho_{\rm o}$
the maximum value of $c_{\rho}$ remains finite
and the peak width in $c_{\rho}(T)$ is nonzero in the thermodynamic
limit considered in our study. 
This is in contradiction with the expectation of Refs. \cite{Gupta:98,Gupta:99}
that an infinite peak of zero width will appear in $c_{\rho}(T)$ in this
limit.
Also a comment about the so-called ``boiling point''
is appropriate here.
This is a discontinuity in the energy density $\varepsilon(T)$ 
or, equivalently, a plateau in the
temperature as a function of the excitation energy. 
Our analysis shows that this type of behavior indeed happens 
at constant pressure, but not at constant density! This is similar to
the usual picture of a liquid-gas phase transition.
In Refs. \cite{Gupta:98,Gupta:99} a rapid  
increase of the energy density as a function of temperature
at fixed $\rho$ near the boundary of the mixed and gaseous phases
(see the middle  panel of  Fig.~\ref{fig:three})
was misinterpreted as a manifestation of the ``boiling point''.

\begin{figure}[ht] 
\centerline{\mbox{
\hspace*{0.0cm} \epsfig{figure=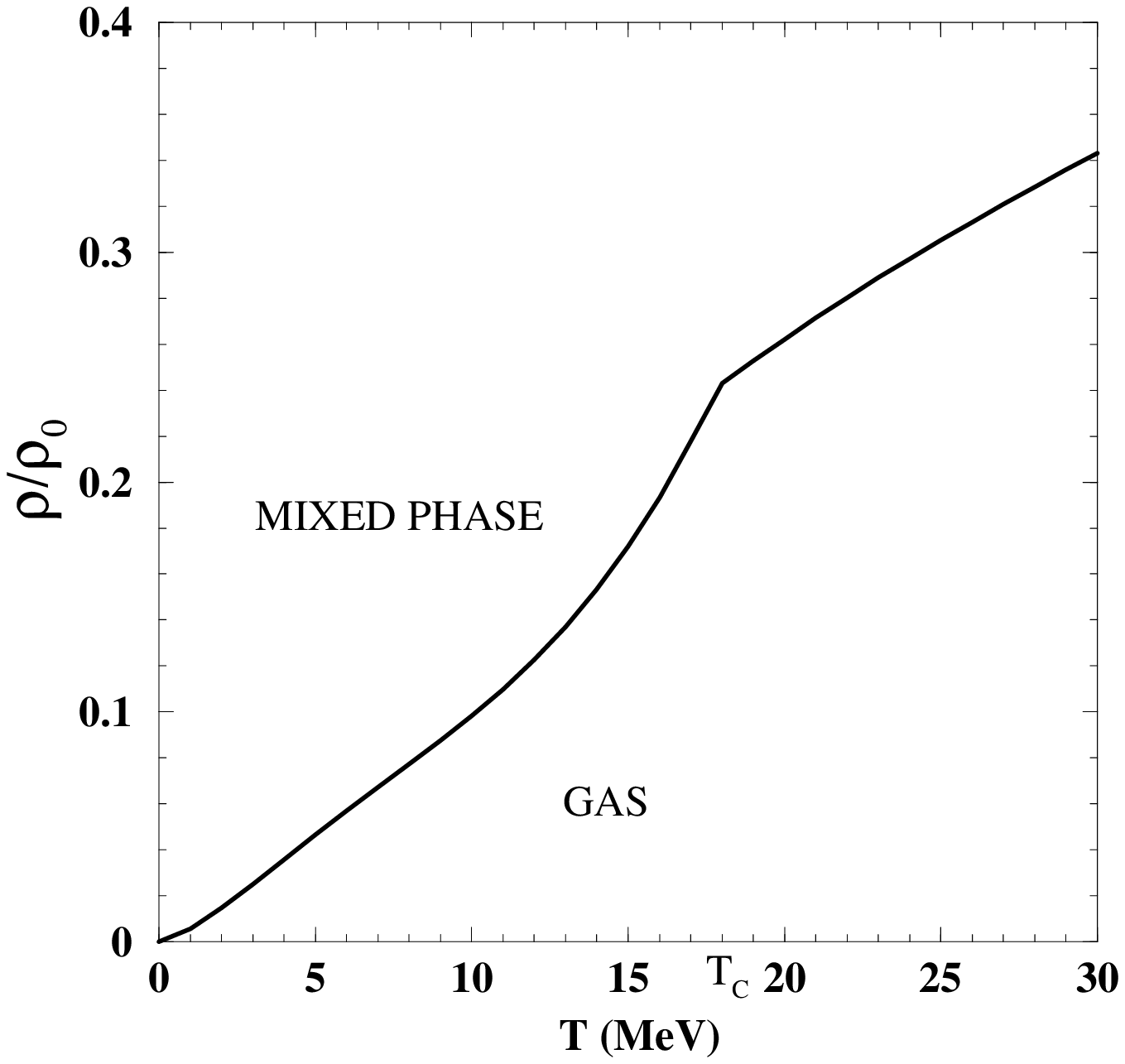,width=8.4cm}
\hspace*{-3.0cm} \epsfig{figure=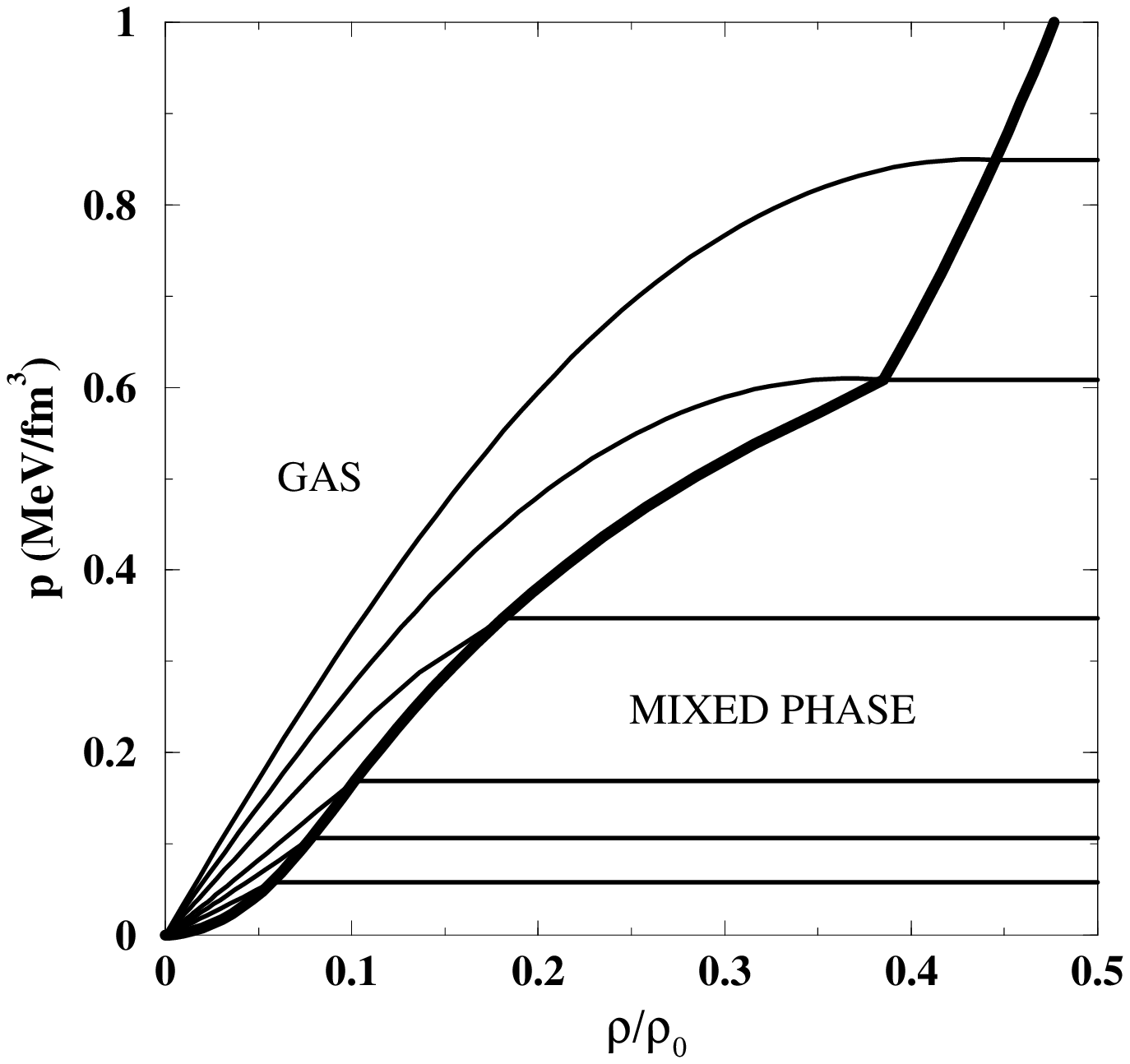,width=8.4cm}
}}


\caption{\label{fig:four}
Phase diagram in  $T-\rho$ (left  panel) and $\rho/\rho_{\rm o}-p$ 
(right panel)
planes for $\tau = 2.1$.
The isotherms in the right panel are  the same as in Fig.~\ref{fig:two}. 
There is no critical point in this case.
For $\tau > 2$   the PT exists for all temperatures $T \ge 0$.
}
\end{figure}

New possibilities appear
at non-zero values of the parameter $\tau$.  
At $0<\tau \le 1$ the qualitative picture remains the same
as discussed above, although there are some
quantitative changes. 
For $\tau > 1$  the condition (\ref{Fss}) is also satisfied 
at $T>T_c$ where $\sigma(T)|_{SMM} =0$.
Therefore, the liquid-gas phase transition
extends now to all temperatures. Its properties
are, however, different for $\tau >  2 $ and for $\tau \le  2$.
If $\tau > 2$ the 
gas density is always lower than $1/b$ as $\rho_{id}$ is finite (see Fig.~\ref{fig:four}).
Therefore, the liquid-gas transition at $T>T_c$ remains
the 1-st order phase transition with discontinuities
of baryonic density, entropy and energy densities.
The pressure isotherms as functions of the reduced density
$\rho/\rho_{\rm o}$ are shown for this case in Fig.~\ref{fig:four}
%


\begin{figure}[ht] 
\centerline{\mbox{
\hspace*{0.0cm} \epsfig{figure=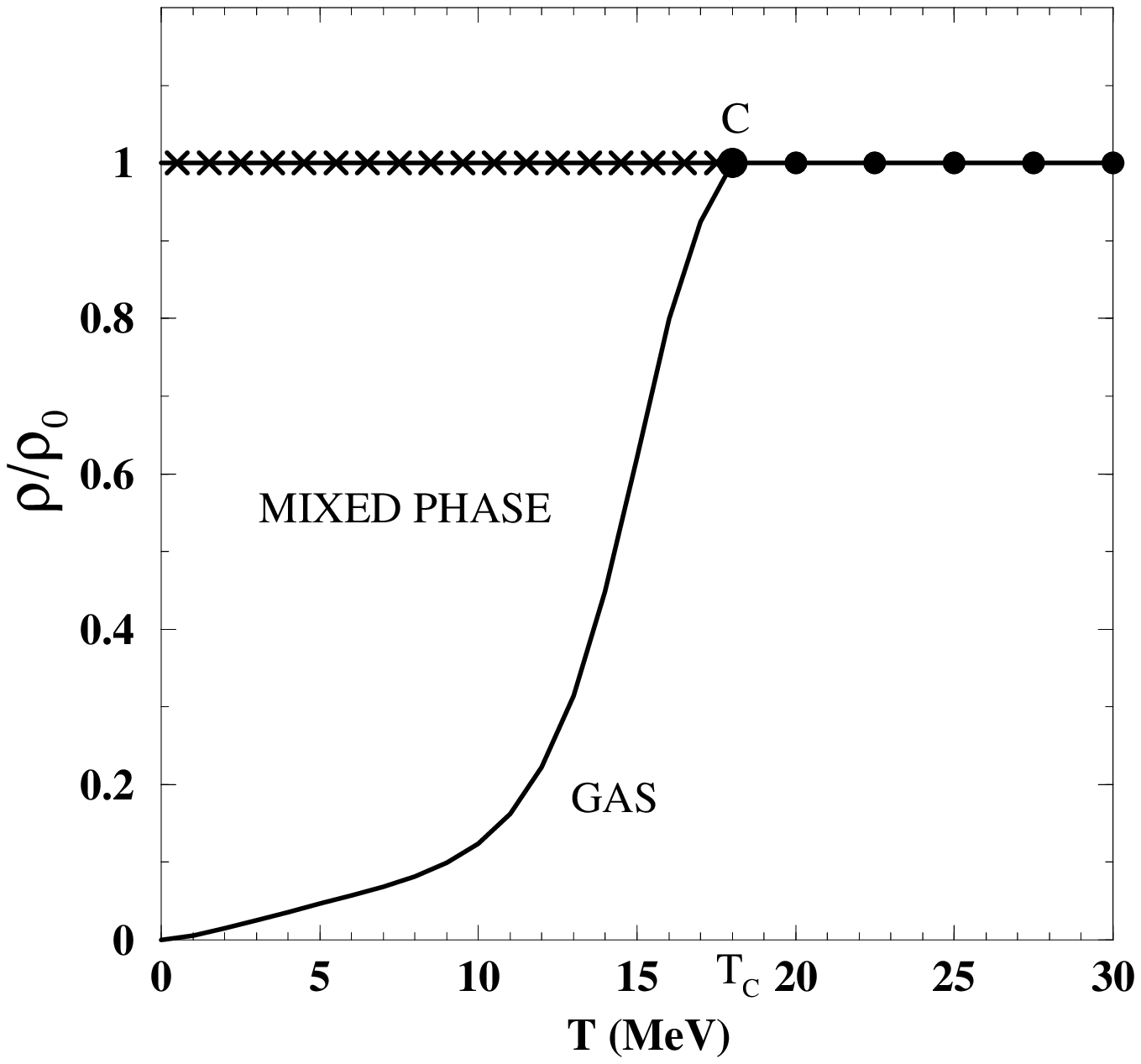,width=8.4cm}
\hspace*{-3.0cm} \epsfig{figure=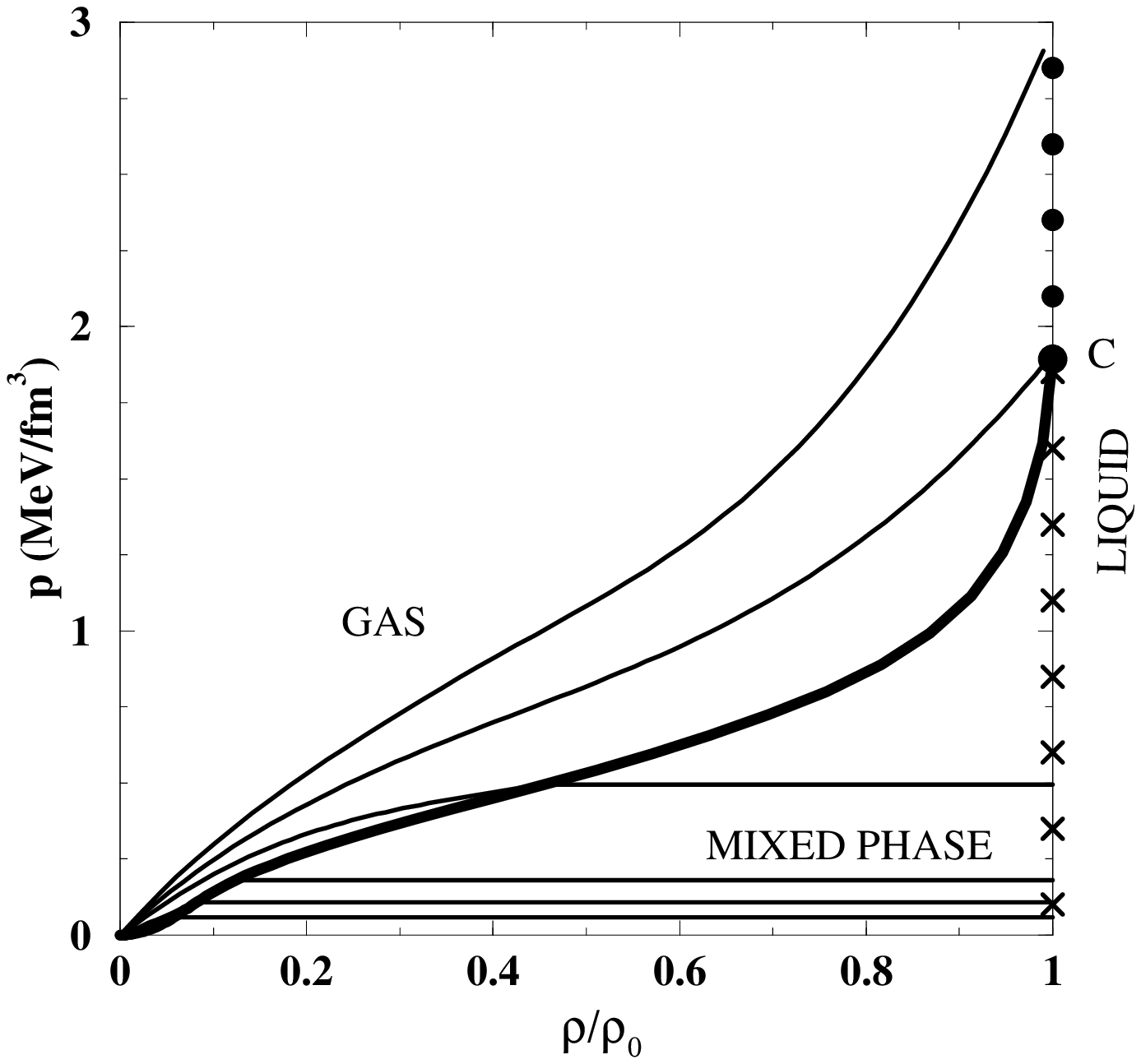,width=8.4cm}
}}


\caption{ \label{fig:five}
Phase diagram in  $T-\rho$ (left  panel) and $\rho/\rho_{\rm o}-p$ 
(right panel)
planes for $\tau = 1.1$.
The isotherms in the right panel are  the same as in Fig.~\ref{fig:two}. 
The mixed phase is represented by
the extended region.
Liquid phase (shown by crosses) exists at density $\rho = \rho_{\rm o}$.
The 2-nd order PT line is shown by circles.
Point $C$ in the left panel  is the tricritical point.  
}
\end{figure}

At $1 < \tau \le 2$ the baryonic density of the gas
in the mixed phase, see Eqs.~(\ref{rhog}), (\ref{rhoidmix}),
becomes equal to that of the liquid at $T>T_c$,  since 
$\rho_{id}\rightarrow \infty$ and 
$\rho_g^{mix}=1/b\equiv \rho_{\rm o}$ (see also Fig.~{fig:five}). 
It is easy to prove that the entropy and energy densities
for the liquid and gas phases are also equal to each other.
There are discontinuities only in the derivatives of these densities
over $T$ and $\mu$, i.e., $p(T,\mu)$  has discontinuities
of its second derivatives.
Therefore,  the liquid-gas transition at $T>T_c$ for $1 < \tau \le 2$
becomes the 2-nd order phase transition. 
According to standard definition, the point $T=T_c$,
$\rho = 1/b$ separating the first and second order
transitions is the tricritical point.
One can see that this point is now at a finite pressure.
Fig.~\ref{fig:five} shows the pressure isotherms as functions of the reduced density
$\rho/\rho_{\rm o}$.


It is interesting to note that at $\tau >0$ the mixed
phase boundary in $T-\rho$ plane shown in Figs.~\ref{fig:four} and \ref{fig:five}   is not so steep function of $T$ as
in the case $\tau= -1.5$ presented in Fig.~\ref{fig:three} Therefore, the peak in the specific
heat discussed above becomes less pronounced.

The density of fragments with $k$ nucleons 
can be obtained by differentiating
the gas pressure (\ref{pgas}) with respect to the $k$-fragment
chemical potential $\mu_k=k\mu$. This leads to
the fragment mass distribution $P(k)$ in the gas phase
\begin{equation}\label{pkgas}
P_g(k)~=~a_{\rm o}~k^{ -\tau}~\exp(- a_1 k - a_2 k^{2/3})~,
\end{equation}
where $a_1\equiv (bp_g -\nu)/T \ge 0$, $a_2\equiv \sigma/T$
and $a_{\rm o}$ is the normalization constant. 
Since the coefficients $a_{\rm o}, a_1, a_2$ depend on
$T$ and $\mu$ the distribution $P(k)$ (\ref{pkgas})
has different shapes in different points of the phase diagram.
In the mixed phase the condition (\ref{ptr}) leads to
$a_1=0$ and Eq.~(\ref{pkgas}) is transformed into
\begin{equation}\label{pkgasmix}
P_g^{mix}(k)~=~a_{\rm o}~k^{ -\tau}~\exp( - a_2
k^{2/3})~.
\end{equation}
 
The liquid inside the mixed phase is one infinite fragment
which occupies a fraction $\lambda$ of the total system volume.
Therefore, in a large system with $A$ nucleons
in volume $V$ ($A/V=\rho$) the mixed phase
consists of one big fragment with $\lambda V \rho_{\rm o}$
nucleons (liquid) and $(1-\lambda)V\rho_g $ nucleons distributed
in different $k$-fragments according to Eq.~(\ref{pkgasmix}) (gas). 
At low $T$ most nucleons are inside one big liquid-fragment 
with only few small gas-fragments distributed according to
Eq.~(\ref{pkgasmix}) with large $a_2$.
At increasing temperature the fraction of the gas fragments increases
and their mass distribution becomes broader since $a_2(T)$
in Eq.~(\ref{pkgasmix}) decreases.  
Outside the mixed phase region the liquid disappears
and the fragment mass distribution acquires an exponential falloff,
Eq.(\ref{pkgas}). Therefore, the fragment mass distribution is widest
at the boundary of the mixed phase.  
At even higher temperatures, $T>T_c$, the coefficient $a_2$
vanishes. 

Details of the fragment mass distribution depend on the parameter 
$\tau$. At $\tau <  1$ one observes a sudden transformation of the large
liquid fragment into light and intermediate mass fragments
in the narrow vicinity of the mixed phase boundary.
This sudden change of the fragment composition has the same origin as a
narrow peak in the specific heat, i.e. a sharp drop of $\lambda(T)$
near the mixed phase boundary (see Fig.~\ref{fig:five}).
For larger $\tau$ all these changes are getting smoother.

An interesting possibility opens when $1 <\tau \le 2$.
As shown in Fig.~\ref{fig:four} the mixed phase in this case ends at the tricritical
point $T=T_c$, $\rho=\rho_{\rm o}$.
In this point both the coefficients $a_1$ and $a_2$ vanish and the mass
distribution becomes
a pure power law
\begin{equation}\label{power}
P_g(k)~=~a_{\rm o}~k^{  -\tau}~.
\end{equation}
At $\tau > 2$
the mixed phase exists at all $T$ (see Fig\ref{fig:four}). Thus 
the mass distribution 
of gaseous fragments inside the mixed phase
fulfills a  power-law (\ref{power}) at all $T>T_c$.



\section{The Critical Indices  and Scaling Relations of the SMM}


Scaling has been called a pillar of modern critical phenomena \cite{Stanley:99}. The scaling hypothesis used in the study of 
critical phenomena was independently developed by several scientists, including Widom, Domb, Hunter, Kadanoff, 
Fisher, Patashinskii and Pokrovskii (see reference \cite{Fisher:69} for an authoritative review). 
Much of scaling is contained in 
the renormalization group work of Wilson \cite{Wilson:74}. 
The most important prediction of
scaling hypothesis, which have been verified experimentally for a 
variety of physical systems,  is a set of relations called scaling laws. These scaling laws relate the 
critical exponents $\alpha^\prime, \, \beta$ and $\gamma^\prime$ which describe, for instance, the behavior of the specific heat, density differences of the phases (order parameter)  and isothermal compressibility  
for fluid systems; specific heat, magnetization  
and isothermal susceptibility   for magnetic systems or the singular 
part of the zeroth, first and second moment of the cluster distribution percolating systems near a critical point 
($\varepsilon = \frac{T_c -T}{T_c}$
for physical systems and  $\varepsilon = \frac{p_c -p}{p_c}$ for percolating systems). In all the systems mentioned here, 
and more, these exponents are related via the scaling laws. 
The FDM, which is often used to describe the nuclear multifragmentation data, belongs to the universality class of 3-dimensional Ising model. Since the SMM is also used for the same purpose it would be interesting, first,
to determine its critical exponents and, second, to compare them with the available experimental data.

The exact  results described above allow one to find 
the critical exponents $\alpha^\prime, \, \beta$ and $\gamma^\prime$ of the simplified SMM. 
These exponents  describe the temperature dependence of the system near the critical point on the coexistence curve $\mu^* = \mu^*(T)$  (\ref{ptr}), where the effective chemical potential   
$\nu \equiv  \mu^*(T) + W(T) -  b p(T,  \mu^*(T) )  = 0 $ vanishes
\begin{eqnarray}\label{alpha}
c_\rho &~ \sim ~& 
\left\{
\begin{tabular}{ll}
$\mid \varepsilon \mid^{-\alpha}$\,, \hspace*{0.975cm} ${\rm for} ~ \varepsilon  < 0 $~, \\
$\varepsilon^{-\alpha^\prime}$\, ,\hspace*{1.25cm} ${\rm for} ~ \varepsilon  \ge 0$~,
\end{tabular}
\right.
\\ \label{beta}
\Delta\rho &~ \sim ~ & 
\varepsilon^{\beta}\,,   \hspace*{2.15cm}	 {\rm for} ~ \varepsilon  \ge 0 ~,
\\ \label{gamma}
\kappa_T &~ \sim ~&  
\varepsilon^{-\gamma^\prime}\,,\hspace*{1.925cm} 	 {\rm for} ~  \varepsilon  \ge 0 ~,
\end{eqnarray}
where $\Delta\rho \equiv \rho_l - \rho_g$ defines the order parameter, $c_\rho \equiv \frac{T}{\rho} \left(\frac{\partial s}{\partial T} \right)_\rho$ denotes the specific heat at fixed particle density and $\kappa_T \equiv \frac{1}{\rho} \left(\frac{\partial \rho}{\partial p} \right)_T$ is the  isothermal compressibility.
The shape of the critical isotherm for $\rho \leq \rho_c$ is given by the critical index $\delta$
(the tilde indicates $\varepsilon = 0$ hereafter)
\begin{equation}\label{delta}
p_c - \tilde{p}  ~\sim ~
(\rho_c  - \tilde{\rho})^{\delta}		\hspace*{1.1cm} {\rm for} ~  \varepsilon = 0~. 
\end{equation}

The calculation of $\alpha$ and $\alpha^\prime$ requires the specific heat $c_{\rho}$. With the formula \cite{Ya:64}
\begin{equation}\label{crho}
\frac{c_{\rho}(T, \mu)}{T}~= ~ \frac{1}{\rho}\left(\frac{\partial^2 p}{\partial T^2} \right)_{\rho} - \left(\frac{\partial^2 \mu}{\partial T^2} \right)_{\rho}
\end{equation}
one obtains the specific heat on the PT curve by replacing the partial derivatives by the total ones \cite{Fi:70}. The latter can be done for every state inside or on the boundary of the mixed phase region.
For the chemical potential $\mu^*(T) = bp^*(T) - W(T)$ one gets
%
$
\frac{c_{\rho}^*(T)}{T}~ = ~ 
\left( \frac{1}{\rho}- b \right) \frac{ {\rm d}^2 p^*(T)}{ {\rm d} T^2} + \frac{{\rm d}^2 W(T)}{{\rm d} T^2}.
$
%
Here the asterisk indicates the condensation line ($\nu = 0$) hereafter.  
Fixing $\rho = \rho_c = \rho_l = 1/b$ one finds $c_{\rho_l}^*(T) = T\frac{ {\rm d^2}W(T)} {{\rm d}T^2}$ and, hence, obtains 
$ \alpha ~= ~\alpha^\prime ~= ~ 0.$
To calculate $\beta$, $\gamma^\prime$  and $\delta$ the behavior  
of the series 
\begin{equation}\label{I}
\mys{q}( \varepsilon, \nu)  \equiv 
\sum_{k=2}^{\infty}~k^{q-\tau}~e^{\textstyle \frac{\nu}{T_c} k - A \varepsilon^\zeta k^\sigma}
\end{equation}
should be analyzed 
for  small positive values of $\varepsilon$  and $- \nu$ \mbox{$(A \equiv a_{\rm o}/T_c)$}.
In the limit $\varepsilon \rightarrow 0$ the function $\mys{q}( \varepsilon, 0)$ remains finite, if $\tau > q+1$, and diverges otherwise. For $\tau = q+1$ this divergence is logarithmic. 
{ The case $\tau < q + 1$ is analyzed in some details, since even in Fisher's papers it was 
performed incorrectly.}

With the substitution ${\mbox z_{k}\equiv k\left[ A \varepsilon^\zeta \right]^{1/\sigma} }$
{ one can prove \cite{Reuter:01}   that in the limit  $\varepsilon \rightarrow 0$ 
the series on the r.\,h.\,s. of (\ref{I}) converges to an integral}

\vspace*{-0.5cm}

\begin{equation}\label{series}
\hspace*{1.4cm}
\mys{q}( \varepsilon, 0)~= ~\left[ A  \varepsilon^\zeta \right]^{\textstyle \frac{\tau - q }{\sigma}}~ 
\sum_{k=2}^{\infty}~
z_k^{q-\tau}~e^{\textstyle -z_k^\sigma}~ \rightarrow ~
\left[ A  \varepsilon^\zeta \right]^{\textstyle \frac{\tau-q-1}{\sigma}}
\hspace*{-3 mm}\int \limits_{2 \left[ A  \varepsilon^\zeta \right]^{\frac{ 1}{\sigma}}}^{\infty}
\hspace*{-3 mm} {\rm d}z~
z^{q-\tau}~e^{\textstyle -z^\sigma} ~.
\end{equation}

\vspace*{-0.1cm}

\noindent
The assumption $q-\tau > -1$ is { sufficient} to guarantee  the convergence of the integral at its lower limit.
{ Using this representation, one finds the following general results}  \cite{Reuter:01}
\begin{equation}\label{Prop1}
\mys{q}( \varepsilon, 0) \sim \left\{ 
\begin{array}{ll}
\varepsilon^{\textstyle\, \frac{\zeta}{ \sigma }(\tau - q- 1) }\,,
&
{\rm if} ~ \tau < q + 1 \,,  \\
\ln\mid\varepsilon\mid\,,  
&
{\rm if} ~ \tau = q + 1 \,,
\\
\varepsilon^{\,0}\,, 
&
{\rm if} ~  \tau > q + 1 \, .
\end{array}
\right. \quad {\rm and} \quad 
\mys{q}( 0, \tilde{\nu})~\sim ~\left\{ 
\begin{array}{ll}
\tilde{\nu}^{\textstyle\, \tau - q- 1}\,,
&
{\rm if} ~ \tau < q +1\,,  
\\
\ln\mid\tilde{\nu}\mid\,,  
&
{\rm if} ~ \tau = q + 1 \,,
\\
\tilde{\nu}^{\,0}\,,
&
{\rm if} ~  \tau > q + 1 \,,
\end{array}
\right.
\end{equation}
which allowed us to find out the critical indices of the SMM (see  Table~\ref{tab:a}).


\begin{table}[htdp] 

\caption{Critical exponents of the SMM and FDM as functions of Fisher index $\tau$
for  the general parametrization of the  surface free  energy 
$ \sigma(T) k^{\frac{2}{3} }  \rightarrow  \varepsilon^\zeta k^\sigma $  with $\varepsilon = (T_c - T)/T_c$.
}

\begin{tabular}{|c|c|c|c|c|c|}
\hline
  & { $\alpha^\prime$ } 
  & { $\alpha^\prime_s$ }
  & \hskip6mm {$\beta$  } \hskip7mm
  & \hskip5mm {$\gamma^\prime$ }  \hskip5mm
  & \hskip8mm { $\delta $ } \hskip8mm  \\
\hline
SMM for  $\tau < 1 + \sigma$ & $0$ &  $2 -  \frac{\zeta}{\sigma}$     &  $ \frac{\zeta}{\sigma}  (2 - \tau) $    
& $ \frac{2 \zeta}{\sigma}  \left(\tau - \frac{3}{2} \right) $   &  $\frac{\tau - 1}{2 - \tau}$   \\ 
 & & & & &  \\
SMM for  $\tau \ge 1 + \sigma$ & $0$ &  $2 -  \frac{\zeta}{\sigma}(\sigma + 2 - \tau) $     &  $ \frac{\zeta}{\sigma}  (2 - \tau) $    
& $ \frac{2 \zeta}{\sigma}  \left(\tau - \frac{3}{2} \right) $   &  $\frac{\tau - 1}{2 - \tau}$   \\ 
 & & & & &  \\
FDM  &  $2 -  \frac{\zeta}{\sigma}{ (\tau - 1)}$ & N/A & $ \frac{\zeta}{\sigma}  (\tau -2 ) $  & $ \frac{\zeta}{\sigma}  (3 - \tau) $ &     $\frac{1}{\tau - 2 }$   \\
\hline
\end{tabular}
\\
 \label{tab:a}
\end{table}

\vskip8mm

%

In the special case $\zeta = 2\sigma$
 the well-known exponent inequalities proven for real gases by
\begin{eqnarray}
\label{F1}
{\rm Fisher  \cite{Fi:64}:} & \hspace{.5cm}	\alpha^\prime + 2\beta + \gamma^\prime ~& \ge~2 ~,\\
\label{G}
{\rm Griffiths \cite{Gri:65}:}& 		\alpha^\prime + \beta(1 + \delta) & \ge~ 2 ~,\\
\label{L}
{\rm Liberman \cite{Li:66}:}&			\gamma^\prime + \beta(1-\delta) & \ge ~0 ~,
\end{eqnarray}
are fulfilled exactly for any $\tau$. (The corresponding exponent inequalities for magnetic systems are often called Rushbrooke's, Griffiths' and Widom's inequalities, respectively.) For $\zeta > 2\sigma $,  Fisher's and Griffiths' exponent inequalities are fulfilled as inequalities and for $\zeta < 2\sigma $ they are not fulfilled. The contradiction to Fisher's and Griffiths' exponent inequalities in this last case is not surprising. This is due to the fact that in the present version of the SMM the critical isochore $\rho = \rho_c = \rho_l$ lies on the boundary of the mixed phase to the liquid. Therefore, in  expression (2.13) in Ref. \cite{Fi:64} for the specific heat only the liquid phase contributes and, 
{ therefore, Fisher's 
 proof of Ref. \cite{Fi:64} 
 } following (2.13) cannot be applied for the SMM. Thus, the exponent inequalities (\ref{F1}) and (\ref{G}) have to be modified for the SMM. Using  results of  Table~\ref{tab:a},  
one finds the following scaling relations
\begin{eqnarray}
\alpha^\prime + 2\beta + \gamma^\prime  =   \frac{\zeta}{\sigma} 
\hspace{0.5cm} {\rm and} \hspace{0.5cm}
\alpha^\prime + \beta(1 + \delta)  =  \frac{\zeta}{\sigma} ~.
\end{eqnarray}
Liberman's exponent inequality (\ref{L}) is fulfilled exactly for any choice of $\zeta$ and $\sigma$.

Since the coexistence curve of the SMM is not symmetric with respect to $\rho = \rho_c$, it is interesting with regard to the specific heat to consider the difference $\Delta c_\rho(T) \equiv c^*_{\rho_g}(T) - c^*_{\rho_l}(T)$, following the suggestion of Ref. \cite{Fi:70}. 
Using  Eq. (\ref{crho}) for gas and liquid  and noting that $1/\rho_g^* - b = 1/\rho_{id}^*$,  one obtains a specially defined index  $\alpha^\prime_s$
from the most divergent term for $\zeta>1$
\begin{eqnarray}\label{alphas}
\Delta c_\rho(T) ~=~ \frac{T}{\rho_{id}^*(T)} \frac{ {\rm d}^2 p^*(T)}{ {\rm d} T^2}~ \quad \Rightarrow
\quad  \alpha^\prime_s ~= ~
\left\{ 
\begin{array}{ll}
\vspace{0.1cm}2 - \frac{\zeta}{\sigma}\,,
& 
{\rm if} ~ \tau < \sigma + 1 \,,  \\
2 - \frac{\zeta}{\sigma}( \sigma + 2 - \tau)\,,	
&
{\rm if} ~  \tau \ge \sigma + 1 \,.
\end{array}
\right.
\end{eqnarray}
Then it is $\alpha_s^\prime > 0$ for $\zeta/\sigma <2$.
Thus, approaching the critical point along any isochore within the mixed phase region except for $\rho = \rho_c = 1/b$ the specific heat diverges for $\zeta/\sigma <2$ as defined by $\alpha^\prime_s$ and remains finite for the isochore $\rho = \rho_c = 1/b$. This demonstrates the exceptional character of the critical isochore in this model. 

In the special case that $\zeta = 1$ one finds $\alpha^\prime_s = 2 - 1/\sigma$ for $\tau \leq 1 + 2 \sigma$ and $\alpha^\prime_s = -\beta$ for $\tau > 1 + 2 \sigma$.
Therefore, using $\alpha_s^\prime$ instead of $\alpha^\prime$, the exponent inequalities (\ref{F1}) and (\ref{G}) are fulfilled
exactly if $\zeta >1$ and  $\tau \leq \sigma + 1$ or if $\zeta =1$ and  $\tau \leq 2\sigma + 1$. In all other cases (\ref{F1}) and (\ref{G}) are fulfilled as inequalities.
Moreover, it can be shown that the SMM belongs to the universality class of real gases for $\zeta >1$ and $\tau \ge \sigma + 1$.

The comparison of 
the above derived formulae for the critical exponents of the SMM
for $\zeta = 1$
with  
those obtained within the FDM 
(Eqs. 51-56 in \cite{Fisher:67}) shows that these models
belong to  different  universality classes (except for the singular case $\tau = 2$). 

Furthermore, one has to note that for $\zeta = 1\,,~\sigma \leq 1/2$ and $1 + \sigma < \tau \leq 1 + 2 \sigma$ the critical exponents of the SMM coincide with those of the exactly solved one-dimensional FDM with non-zero droplet-volumes \cite{Fi:70}. 

For the usual parameterization of the SMM \cite{Bondorf:95} 
one obtains with $\zeta = 5/4$ and $\sigma = 2/3$ the exponents
\begin{eqnarray}
\hspace*{-0.3cm}
\alpha^\prime_s &=& \left\{ 
\begin{array}{ll}
\vspace{0.2cm} \frac{1}{8}\,, 
&
{\rm if} ~ \tau < \frac{5}{3}   \\
\frac{15}{8}\tau - 3\,, 
&
{\rm if} ~  \tau \ge \frac{5}{3}  
\end{array}
\right.,\hspace*{0.2cm}
\beta~=~
\frac{15}{8}\left(2 - \tau \right), \hspace*{0.2cm}
\gamma^\prime ~ = ~
\frac{15}{4}\left( \tau - \frac{3}{2} \right),\hspace*{0.2cm}
\delta ~ = ~ \frac{\tau - 1}{2 - \tau}~. 
\end{eqnarray}
{Thus, Fisher suggestion to use $\alpha^\prime_s$ instead of  $\alpha^\prime$
allows one to  {\it ``save''  the exponential inequalities}, however,  it is not a final  solution of the problem.
 }

The critical indices of the nuclear liquid-gas PT were determined 
from the multifragmentation of gold nuclei \cite{Eos:94} and
found to be close to those ones of real gases.  
The method used to extract the critical exponents 
$\beta$ and $\gamma^\prime$  in Ref. \cite{Eos:94} was, however,
found to have large uncertainties of about 25 per cents \cite{Bau:94}.
Nevertheless, those results allow us  
to estimate the value of $\tau$
from the experimental values of the critical exponents  of real
gases taken with large error bars.
Using the above results we 
generalized \cite{Reuter:01}  the exponent relations of Ref. \cite{Fi:70}
\begin{eqnarray}
\label{ntau1}
\tau ~= ~ 2 - \frac{\beta}{ \gamma^\prime + 2 \beta} ~~~ {\rm and} \hspace{0.5cm} 
\tau ~= ~ 2 - \frac{1}{1 + \delta}
\end{eqnarray}
for arbitrary $\sigma$ and $\zeta$. 
Then, one obtains with \cite{Hu} $\beta = 0.32-0.39$\,, $\gamma^\prime = 1.3-1.4$ and $\delta = 4-5$ the estimate $\tau =  1.799 - 1.846$. 
This demonstrates also that the value of $\tau$ is rather 
insensitive to the special choice of $\beta\,,~\gamma^\prime$ and $\delta$, which leads to  $\alpha^\prime_s \cong 0.373 - 0.461$ for the SMM.
Theoretical values for $\beta\,,~\gamma^\prime$ and $\delta$ for Ising-like systems within the renormalized $\phi^4$ theory \cite{Zi:98} lead to the narrow range $\tau = 1.828 \pm 0.001$\,.
The values of  $\beta\,,~\gamma^\prime$ and $\delta$ indices for nuclear matter
and percolation of  two- and three-dimensional clusters are reviewed in \cite{Elliott:05wci}.

This finding  is  not only of a  principal theoretical  importance, since it allows one to find out the universality
class of the nuclear liquid-gas phase transition, if  $\tau$  index  can be determined from 
experimental mass distribution of fragments, but also  it   
enhanced  a great activity  in extracting the value of  $\tau$  exponent from 
the data \cite{Karnaukhov:tau,Ogul:tau} and by  a full SMM. The latter numerical  analysis  agrees with our estimate 
for a reasonable values of $T_c$.

There was a decent  try to study 
the critical indices of  the SMM  numerically 
\cite{El:00}. The version V2 of Ref. \cite{El:00} corresponds  
precisely to 
our model with $\tau = 0$, $\zeta = 5/4$ and $\sigma = 2/3$,
but their results contradict our analysis.
Their results for version V3 of Ref. \cite{El:00} are in contradiction with
our proof presented in Ref. \cite{Bugaev:00}. There it was shown that for non-vanishing surface energy (as in version V3)
the critical point does not exist at all.
The latter was found in \cite{El:00}
for the finite system
and the critical indices were analyzed.
Such a strange result, on the one hand,  
 indicates 
that the numerical  methods used in Ref. \cite{El:00} 
are not  self-consistent, and, on the other hand, it shows an indispensable value of the analytical
calculations, which can be used as a test problem  for numerical algorithms.

It is widely believed that the effective value of $\tau$ defined 
as $\tau_{\rm eff} \equiv  - \partial \ln n_k(\varepsilon) / \partial \ln k$ 
 attains
its minimum at the critical point (see references in \cite{EOS:00}).
This 
can be easily  shown for the SMM. Indeed, taking the SMM fragment distribution  
$n_k(\varepsilon) = g(T) k^{-\tau} \exp[{\textstyle \frac{\nu}{T} k 
- \frac{a(\varepsilon)}{T} k^\sigma }] \sim k^{- \textstyle \tau_{\rm eff}}  $ 
one finds
\begin{equation}\label{teff}
\tau_{\rm eff}~ = ~ \tau - \frac{\nu}{T} k + \frac{\sigma a(\varepsilon)}{T} k^\sigma  
\quad \Rightarrow \quad \tau = \min( \tau_{\rm eff}) ~,
\end{equation}
where the last step follows from the fact that  
the inequalities 
$a(\varepsilon) \ge 0$\,, $\nu \leq 0$ become equalities  
at the critical point $\nu = a(0) = 0$. 
Therefore,
the SMM predicts
that the minimal value of $\tau_{\rm eff}$ corresponds to the critical point,  where, as we see now,  
the mass distribution of fragments  is power-like.

In the E900 $\pi^-+$\,Au multifragmentation experiment \cite{ISIS} 
the ISiS collaboration measured
the dependence of
$\tau_{\rm eff}$ upon the excitation energy and 
found the minimum value ${\rm min}(\tau_{\rm eff} ) \cong 1.9$ 
(Fig.\,5 of Ref. \cite{ISIS} ). 
Also the EOS collaboration \cite{EOS:00} performed an analysis of the minimum of $\tau_{\rm eff}$ on Au\,+\,C multifragmentation data. The fitted $\tau_{\rm eff}$, plotted in Fig.\,11.b of Ref. \cite{EOS:00} versus the fragment multiplicity, exhibits a minimum in the range 
${\rm min}(\tau_{\rm eff}) \cong 1.8 - 1.9$\,. 
Both results contradict the original FDM \cite{Fisher:67}, but agree well
with the above estimate of $\tau$ for real gases and for Ising-like systems in general.
The main consequence of such a  comparison is the fact the nuclear matter EOS has a tricritical point rather than the 
critical one. 


\section{Major Conclusions}

In this chapter I presented  a mean-field  model based on a concept of nuclear matter.  In nuclear physics this concept plays a role which is  comparable to the ideal gas concept in statistical mechanics. 
From the discussed model \cite{Bugaev:93} which is a phenomenological generalization of the Walecka model \cite{Analyt:1}  it is seen that in the absence of the Coulomb interaction between nucleons  
even a 
model inspired by a simple field theory predicts the existence of the nuclear liquid-gas PT, although its critical temperature is somewhat lower than expected.  A suggested model  \cite{Bugaev:93}  correctly 
reproduces the four known  properties of nuclear matter with three parameters only  and even the low density results of numerical simulations based on a realistic internuclear interaction  \cite{ManyBodyEOS:1}. Also  this model  simultaneously reproduces the value of  effective mass of the nucleon in the normal nuclear matter and its compressibility  of  about $K_{\rm o} \approx 300$ MeV, which is observed in the experiments on elliptic flow  anisotropy  \cite{IncompressibilityNew:1, IncompressibilityNew:2}.  Of course, the question why 
the sideward anisotropy observed in the A+A collisions shows $K_{\rm o} \approx 210$ MeV remains puzzling, but there is hope that  in the coming years it will be clarified at the accelerator FAIR.

Then I presented an exact analytical solution \cite{Bugaev:00, Bugaev:00b} of a simplified SMM found with my collaborators.
The  SMM \cite{Bondorf:95} is  a   more profound statistical model and for many years it was 
a guide for both theoreticians and experimentalists in nuclear physics at intermediate energies. 
These results not only allowed me to explain the pitfalls of several numerical studies (``boiling point" 
\cite{Gupta:98,Gupta:99},  numerical extraction of  the SMM critical indices \cite{El:00}), but also 
allowed me, for the first time,  to find out the critical  exponents of the SMM \cite{Reuter:01}, to derive the scaling relations between them and show explicitly that SMM is in the other universality class ($\tau \le  2$) than the famous FDM \cite{Fisher:67} ($ \tau>2$). The predicted  range of the Fisher  exponent $\tau =  1.799 - 1.846$ contradicts  the FDM value $\tau \approx 2.16$, but is in a good agreement,
$\tau_{EXP} \cong 1.8 - 1.9 $, with  the results of the ISiS collaboration \cite{ISIS} and the EOS collaboration \cite{EOS:00}.  Thus,  for the first time,  my results  showed  that the nuclear matter has 
the tricritical, rather than the critical, endpoint. 

In addition, the  calculation of the SMM critical exponents indicated that, despite the common beliefs,
the fundamental scaling inequalities \cite{Fi:64, Gri:65} found  in mid sixties, are not that well  established as one might  expect. Fisher's  suggestion  \cite{Fi:70} to introduce a special index $\alpha_s$ instead of  $\alpha$ ``saves'' the   scaling inequalities, but it looks like a palliative.
Hopefully, new  experimental and theoretical studies of these inequalities may clarify this problem.

The SMM itself  should be improved and  developed further. One generalization of the SMM, the 
GSMM \cite{Bugaev:05csmm},  
which enables to include a more sophisticated EOS for the nuclear liquid phase, is discussed above.
It gives  a possibility to avoid the problem with causality which exists  in the present SMM at high baryonic densities.  For this one can use, for instance, the mean-field EOS of nuclear matter \cite{Bugaev:93} or 
similar models to describe the nuclear  liquid phase. Another possibility to resolve the causality problem
of such  a  typical nuclear EOS  like the SMM, is to include  the Lorentz contraction of the hard core repulsion of nuclear fragments. The latter is studied in the chapter 3. 
Another generalization of the SMM, the CSMM, is discussed in the next chapter concerning the 
possibility to rigorously  study the PTs in finite systems.


\chapter{Exactly Solvable Statistical Models for Finite Nuclear Systems}



This chapter  is devoted to the investigation of phase equilibrium in finite systems and accounting for finite size effects. 
Finite size effects are essential in the study of nuclei and other mesoscopic systems for opposite, but complementary reasons. In modern cluster physics, the problem of finite size arises when attempts are made to relate known properties of the infinite system to cluster properties brought to light by experiment \cite{schmidt-01}. For nuclear physics, the problem is the opposite: finite size effects dominate  the physics at all excitations and one of the most important  challenges is to generalize specific properties of a \emph{drop} (nucleus) to a description of uncharged, symmetric infinite nuclear matter. This goal has been achieved already for cold nuclei by the liquid drop model. Finite size effects are also encountered in nuclear physics in efforts to generate a liquid-vapor phase diagram from heat capacity measurements \cite{dagostino-00} and fragment distributions \cite{elliott-02}.

The second  of the most important  challenges of nuclear physics  is to formulate a rigorous  
theory of critical phenomena in finite systems. 
The point is that  the experimental studies of PTs in nuclear systems 
require the development of theoretical approaches which  would allow us to study
the  critical phenomena without going into the thermodynamic limit  $V \rightarrow \infty$ ($V$ is 
the volume of the system) because  at finite nuclear densities  such a limit does not exist due the long range Coulomb repulsion.

The general situation in the theory of  critical phenomena for finite (small) 
systems  is not very optimistic  at the moment because  
theoretical  progress in this field has been slow.
It is well known that
the mathematical theory of phase transitions was worked out by 
T. D. Lee and C. N. Yang \cite{lee-52.1, LeeYang}. 
Unfortunately, there is no direct generic  relation between the physical observables and zeros of the grand canonical partition  in a complex fugacity  plane.
Therefore, we know  very well what are the  gaseous phase and liquid at infinite volumes: 
mixture of multiparticle clusters of all sizes and, say, an ocean, respectively.  
This  is known both for pure phases and for their mixture, but, 
despite some limited  success \cite{Chomaz:03},  
this general approach is not useful for  the specific problems of critical phenomena in  finite systems.

A tremendous complexity  of   critical  phenomena in finite systems prevented 
 their  systematic and rigorous  theoretical  study.
For instance, even the best   formulation of  the statistical mechanics and 
thermodynamics of finite systems  by Hill \cite{Hill} is not rigorous while discussing PTs.
As a result,
the absence of  a  well established definition of the liquid and mixed phase for finite volumes
delays the progress of several related fields, including  the theoretical and  experimental 
searches for 
the reliable signals  of  several  PTs which are expected to exist in  strongly interacting matter. 

Since the  analysis of finite volume systems is very difficult, the problem was
attacked  by numerical codes.
During more than two decades there were many successes achieved in this way 
\cite{Bondorf:95, Gross:97, Moretto:97,Randrup:04}, but  the problem is  that 
 the numerical simulations  of this level do not provide us with any proof.  
 On the other hand the developed  analytical  support  was no match for  the demands. 
As a result  the absence of a firm theoretical ground led to formulation of such  highly speculative 
``signals'' of  the  nuclear liquid-vapor PT  as negative heat capacity \cite{Negheat:1, Negheat:2}, bimodality \cite{Bimodality:1}, 
which later on were disproved, in Refs \cite{Negheat:3} and \cite{Bimodality:2}, respectively.

Thus,  there is  a paradoxic   situation in the nuclear multifragmentation community: there are many experimental data and 
facts, but there is no a single theoretical  approach which is able to describe them. 
In fact, a similar state of the art  is  with   the  searches for QGP 
\cite{QM:04}:  there is  
a lack of a firm and rigorous theoretical approach
to  describe the PTs in finite systems.

From these facts I concluded that to improved the  present situation 
it is necessary to develop new theoretical methods which could  break down the model EOS with PT in finite systems and use  their exact  solutions  to understand the PT mechanism
in finite system.
Exactly solvable models with PTs  always played  a special role in  statistical
mechanics  because  they provide us with 
the information obtained directly  from the first principles of statistical mechanics being 
unspoiled by mean-field or other simplifying approximations without which the analytical 
analysis is usually impossible.  On the other hand an exact analytical solution gives the physical
picture  of  PT, which cannot be obtained by numerical evaluation.  Therefore, one can expect 
that an extension of the exact analytical solutions to finite systems may provide us with  the 
ultimate and reliable experimental signals of the nuclear liquid-vapor PT  which are established
on a firm theoretical ground of statistical mechanics. This, however, is a very difficult  general task of the critical phenomena theory in finite systems.

Fortunately,  we do not need to solve this very general task, but to find  its solution for  a specific 
problem of nuclear liquid-gas PT, which is less complicated and more realistic. In this case
the straightforward way is to start from a few statistical models, like the FDM and/or SMM, which are successful  in describing the main part  of the experimental data.  A systematic study of the various
modifications of the FDM for finite volumes was performed by  
Moretto and collaborators \cite{Elliott:05wci}  and 
it led to a discovery of thermal reducibility  of the fragment  charge spectra  \cite{Moretto:97}, 
to a determination  of  a quantitative liquid-vapor phase diagram containing the coexistence line
up to critical temperature for small systems \cite{Elliott:02,Elliott:03},  to the generalization  of the FDM for  finite systems and to a  formulation of the complement concept 
\cite{Precomplement,Complement} 
which allows one  to account for  finite size effects of finite  liquid drop on the properties 
of its vapor.  This finding is very important for quantum chromodynamics because its universality class is expected to be the same as for 3-dimensional Ising model \cite{Misha, fodorkatz, karsch} studied 
in  \cite{Complement}.

A comparable  systematic  analysis for the  SMM was, however,  not  possible  until recently, when 
its finite volume  analytical solution was found in \cite{Bugaev:04a}.  
This exact solution was found with the help of new powerful method, which I named the Laplace-Fourier transform  \cite{Bugaev:04a}.  With this method I was able to analyze the Constrained SMM
 \cite{Bugaev:04a,  BugaevReuter, Bugaev:05csmm} and find out the rigorous definitions of the finite volume analogs of phases for the nuclear liquid-gas PT. However, the very same method 
I applied to the analytical solution of the GBM  in finite volume \cite{Bugaev:07b}, which will be discussed  in the chapter 4  in connection with QGP searches, and to breaking down three statistical ensembles, including a novel one,   
for the HDM  \cite{Bugaev:04b, BugaevElliott}.  The latter explains the origin of surface entropy and is 
very necessary to understand the linear temperature dependence of surface tension coefficient  for a variety of cluster models 
(2- and 3-dimensional clusters of Ising model, nuclear clusters, the molecule clusters in real liquids 
and so on).  
From the above list of exact results it is clear that the Laplace-Fourier transform method   
\cite{Bugaev:04a} can be used to form a common theoretical language for a few physical communities which study critical phenomena in finite systems,  statistical mechanics of  surfaces   and   physics of clusters. 

Therefore, this chapter is devoted to such  exact methods  as the Complement \cite{Complement}  and  Laplace-Fourier transform  \cite{Bugaev:04a} methods and their applications to finite systems 
being in a phase equilibrium.  
The chapter is based on the following works:   \cite{Complement, Elliott:05wci, Bugaev:04a, BugaevReuter, Bugaev:05csmm, Bugaev:04b, BugaevElliott, Bugaev:07b}.



\section{The Complement Method for Finite Drop}

Here I present a general approach to deal with finite size effects in phase transitions and illustrate it for liquid-vapor phase coexistence. A dilute, nearly ideal vapor phase is in equilibrium with a denser liquid phase; finiteness is realized when liquid phase is a finite drop. A finite drop's vapor pressure is typically calculated by including the surface energy in the molar vaporization enthalpy \cite{moretto-02}. This   concept of the complement is introduced to extend and quantify finite size effects down to drops as small as atomic nuclei. It generalizes Fisher's model \cite{Fisher:69, moretto-03}, deriving an expression for cluster concentrations of a vapor in equilibrium with a finite drop and recovers from it the Gibbs-Thomson formulae \cite{krishnamachari-96}.  I demonstrate this approach with the lattice gas (Ising) model.

The complement method consists of evaluating the free energy change occurring when a cluster moves from one phase to another. For a finite liquid drop in equilibrium with its vapor, this is done by virtually transfering a cluster from the liquid drop to the vapor and evaluating the energy and entropy changes associated with \emph{both} the vapor cluster \emph{and} the residual liquid drop (complement). This method can be generalized to incorporate energy terms common in the nuclear case: symmetry, Coulomb (with caution \cite{moretto-03}) and angular momentum energies.

In the framework of physical cluster theories of non-ideal vapors (which assume the monomer-monomer interaction is exhausted by the formation of clusters), clusters behave ideally and are independent of each other. The complement method is based upon this independence. Physical cluster theories state that the concentrations of vapor clusters of $A$ constituents $n_A(T)$ depend on the free energy of cluster formation $\Delta G_A(T) = \Delta E_A(T) - T \Delta S_A(T)$. The epigon of physical cluster theories is Fisher's model \cite{Fisher:69} which writes, at coexistence, $\Delta E = c_0 A^{\sigma}$ and $\Delta S_A(T) = \frac{c_0}{T_c}A^{\sigma} - \tau \ln A $. Thus
\begin{eqnarray}
	n_A(T)  =  \exp \left[-\frac{\Delta G_A(T)}{T} \right] = q_0A^{-\tau} \exp \left( - \frac{c_0 A^{\sigma} \varepsilon }{T} \right)
\label{eq:fisher-1}
\end{eqnarray} 
where $q_0$ is a normalization, $\tau$ is Fisher's topological exponent, $c_0$ is the surface energy coefficient, $\sigma$ is the surface to volume exponent, and $\varepsilon=(T_c-T)/T_c$. The leading term in $\Delta S_A(T)$, proportional to $A^{\sigma}$, permits the vanishing of the cluster free energy at a $T=T_c$ independent of size. Equation~(\ref{eq:fisher-1}) (and the extension below) is valid only at phase coexistence for $T \le T_c$. The direct physical interpretation of the parameters in $\Delta G_A(T)$ and their application to the nuclear case is the reason for this choice here, despite its limitations \cite{mader-03}.

 Eq.~(\ref{eq:fisher-1}), valid for infinite liquid-vapor equilibrium, is generalized to the case of a vapor in equilibrium with a finite liquid drop. For each vapor cluster one can perform the gedanken experiment of extracting it from the liquid, determining entropy and energy changes of the drop and cluster, and then putting it back into the liquid (the equilibrium condition), as if, according to physical cluster theories, no other clusters existed. Fisher's expressions for $\Delta E_A(T)$ and $\Delta S_A(T)$ can be written for a drop of size $A_{\rm d}$ in equilibrium with its vapor as $\Delta E_A(T) = c_0 \left[ A^{\sigma} + (A_{\rm d}-A)^{\sigma}-A_{\rm d}^{\sigma} \right]$ and $\Delta S_A(T) =  \frac{c_0}{T_c} \left[ A^{\sigma} + (A_{\rm d}-A)^{\sigma}-A_{\rm d}^{\sigma} \right]- \tau \ln \left[ {A(A_{\rm d}-A)}/{A_{\rm d}} \right]$ giving
\begin{eqnarray}
	n_A(T) & = & q_0\left[ \frac{A \left(A_{\rm d}-A\right)}{A_{\rm d}}\right]^{-\tau}
	 \exp \left\{- \frac{c_0 \varepsilon }{T}\left[ A^{\sigma} + (A_{\rm d}-A)^{\sigma}-A_{\rm d}^{\sigma} \right] 
	 \right\}.
\label{eq:complement-2}
\end{eqnarray}
The free energy cost of complement ($A_{\rm d}-A$) formation is determined just as the free energy cost of cluster ($A$) formation is determined. The resulting expression Eq.~(\ref{eq:complement-2}) reduces to Eq.~(\ref{eq:fisher-1}) \emph{when $A_{\rm d}$ tends to infinity} and contains exactly the same parameters. One can rewrite Eq.~(\ref{eq:complement-2}) as
	\begin{equation}
	n_A(T) = n_{A}^{\infty}(T) \exp \left( \frac{A\Delta \mu_{\text{fs}}}{T} \right) 
	\label{fisher2}
	\end{equation} 
with $n_{A}^{\infty}(T)$ given by Eq.~(\ref{eq:fisher-1}). The finite size of the drop acts as an {\it effective} chemical potential, $\Delta \mu_{\rm fs} = - \left\{  c_0 \varepsilon \left[ \left(A_{\rm d}-A \right)^{\sigma}-A_{\rm d}^{\sigma} \right] - T \tau \ln\left[ \left( A_{\rm d}-A\right) / A_{\rm d} \right] \right\} / A$, increasing the vapor pressure \cite{krishnamachari-96}.

In order to quantitatively demonstrate this method, I apply it to the canonical lattice gas (Ising) model \cite{lee-52.1,LeeYang} with a fixed number of up spins, i.e. a fixed mean occupation density $\rho_{\rm fixed}$ lattice gas (equivalently, a fixed magnetization $M_{\rm fixed}$ Ising model) \cite{Precomplement}. Up spins represent particles of the fluid forming monomers, dimers, drops etc. Down spins are empty space; the lattice is the container enclosing the fluid.  The large lattices with periodic boundary conditions are chosen to minimize finite lattice effects, irrelevant to our study.  For $d=2$ ($3$) we used a square (simple cubic) lattice of side $L=80$ ($25$) which leads to a shift in $T_c$ of $\lesssim 0.5$\% \cite{ferdinand-69,landau-76.1} ($\lesssim 0.5$\% \cite{landau-76.2,ferrenberg-91}). The $M_{\rm fixed}$ calculations were performed according to ref. \cite{heringa-98}. For every $(T,{\rho}_{\rm fixed})$ over $10^5$ thermalized realizations were generated to produce the cluster concentrations.  The  calculations of $M_{\rm free}$ are  performed  for the same lattices as a benchmark in order to differentiate effects of a finite lattice from those of a finite drop.

\begin{figure}
\centerline{\includegraphics[width=8.4cm]{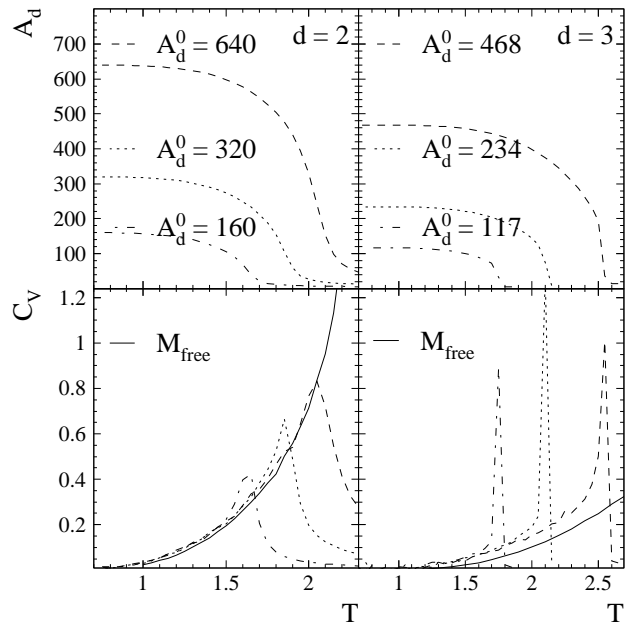}}
\caption{On the left (right), top to bottom: the $d=2$ ($3$) liquid drop size $A_{\rm d}$, specific heat $C_V$. See text for details.}
\label{systematics}
\end{figure}

For the $M_{\rm fixed}$ calculations at $T=0$, the up spins aggregate into a single liquid drop in a vacuum: the ground state. At higher temperatures, the vacuum is filled with a vapor made of up spin clusters. Clusters in the vapor were identified via the Coniglio-Klein algorithm \cite{coniglio-80} to insure that their behavior is physical (i.e. cluster concentrations return Ising critical exponents and not percolation exponents). The largest cluster represents the liquid drop and is identified geometrically (all like spin nearest neighbors bonded) in order to capture the skin thickness associated with liquid drops \cite{Precomplement}. Our choices of ground state liquid drops $A_{\rm d}^{0}$ (shown in Fig.\ref{systematics}) insure that the ground state is approximately square (cubical) for $d=2$ ($3$). Due to periodic boundary conditions the ground state shape changes with $M_{\rm fixed}$ \cite{binder-81}.

Figure~\ref{systematics} shows that as $T$ increases, the drop's size $A_{\rm d}$ decreases from $A_{\rm d}^{0}$; the evaporating drop fills the container with vapor. At a temperature $T_X$, corresponding to the end of two phase coexistence, $A_{\rm d}$ falls quickly. The value of $T_X$ varies with $A_{\rm d}^{0}$. For $T \lesssim T_X$ the specific heat $C_V$ (measuring spin-spin interaction energy fluctuations only) agrees approximately with the $M_{\rm free}$ results (solid lines in Fig.~\ref{systematics}) for all $A_{\rm d}^{0}$ until fluctuations in $A_{\rm d}$ at $\sim T_X$ produce a  $C_V$ greater than that of the $M_{\rm free}$ results. As $T$ increases further, $C_V$ decreases.

To evaluate the efficacy of the complement, I examine the scaled cluster concentrations for our calculations:  $n_A(T)/q_0 A^{-\tau}$ vs. $c_0 A^{\sigma} \varepsilon / T$. For $M_{\rm free}$ calculations it has been shown that this scaling collapses concentrations of clusters over a wide range in $A$ and $T$ \cite{mader-03}. Finite size liquid drop effects will be manifested in the cluster concentrations of the $M_{\rm fixed}$ calculations which should scale better according to Eq.~(\ref{eq:complement-2}) than to Eq.~(\ref{eq:fisher-1}).

\begin{table}[htdp] 
\caption{Fit results for $M_{\rm free}$ calculations}
\begin{center}
\begin{tabular}{||c|c|c|c|c||}
\hline
          -                             &     Onsager & this analysis       & theoretical values                      & this analysis        \\ \hline
          -                             &    $d=2$ & $d=2$        & $d=3$                      & $d=3$ \\
          -                             &     $L \rightarrow \infty$ & $L=80$        & $L \rightarrow \infty$                      & $L=25$         \\ \hline
        ${\chi}^{2}_{\nu}$ &          -               & $4.7$                         & -                                          & $1972.2$                       \\
        $T_c$                     & $2.26915$      & $2.283 \pm 0.004$ & $4.51152\pm0.00004$  & $4.533 \pm 0.002$  \\
        $c_0$                     &        $\ge8$         & $8.6 \pm 0.2$           & $\ge12$                               & $12.63 \pm 0.04$      \\
        $\sigma$                &    $8/15$         & $0.56\pm0.01$    & $0.63946\pm0.0008$    & $0.725 \pm 0.003$ \\
        $\tau$                     &    $31/15$       & $2.071\pm0.002$    & $2.209\pm0.006$           & $2.255 \pm 0.001$ \\
\hline
\end{tabular}
\end{center}
\label{FreeTable}
\end{table}



\begin{table}[htdp] 
\caption{${\chi}^{2}_{\nu}$ results for $M_{\rm fixed}$ calculations.}
\begin{center}
\begin{tabular}{||c|c|c|c||}
\hline
        $A_{\rm d}^{0}$ ($d=2$)                                    & $640$   &$320$   & $160$  \\
        ${\chi}^{2}_{\nu}$ w/o complement & $10.3$ & $10.6$ & $18.2$  \\
        ${\chi}^{2}_{\nu}$ w complement   & $1.7$    & $1.9$   & $4.8$\\
\hline
         $A_{\rm d}^{0}$ ($d=3$) &$468$    & $234$       &$117$ \\
        ${\chi}^{2}_{\nu}$ w/o complement &$14~825.9$ & $7~938.6$ & $3~516.0$\\
         ${\chi}^{2}_{\nu}$ w complement &$1~553.4$     & $838.6$    & $258.1$ \\
\hline
\end{tabular}
\end{center}
\label{FixedTable}
\end{table}

Only clusters of $A\ge9$ are included in the $M_{\rm free}$ fits for the $d=2$ calculations. This is because only large clusters obey Fisher's ansatz for the cluster surface energy: $E_A = c_0 A^{\sigma}$. Small clusters are dominated by geometrical shell effects \cite{Elliott:05wci}. For the $d=3$ $M_{\rm fixed}$ calculations, large clusters are very rare, so clusters of $A\ge2$ are included in our analysis. Thus, the magnitudes of the $\chi^{2}_{\nu}$ values from the $d=3$ calculations are due to the clusters analyzed not following closely Fisher's ansatz.

In the thermodynamic limit, the highest temperature admitted by Eq.~(\ref{eq:fisher-1}) or (\ref{eq:complement-2}) is the temperature at which the system leaves coexistence.  For the $M_{\rm free}$ calculations this occurs at $T=T_c$, while for the $M_{\rm fixed}$ calculations this occurs at $T \approx T_X$. However, due to the small size of our drops, fluctuations grow large before $T_X$ and it is better to consider only $T \lesssim 0.85T_X$ ($0.75T_X$) for $d=2$ ($3$).

\begin{figure}
\centerline{\includegraphics[width=10.7cm]{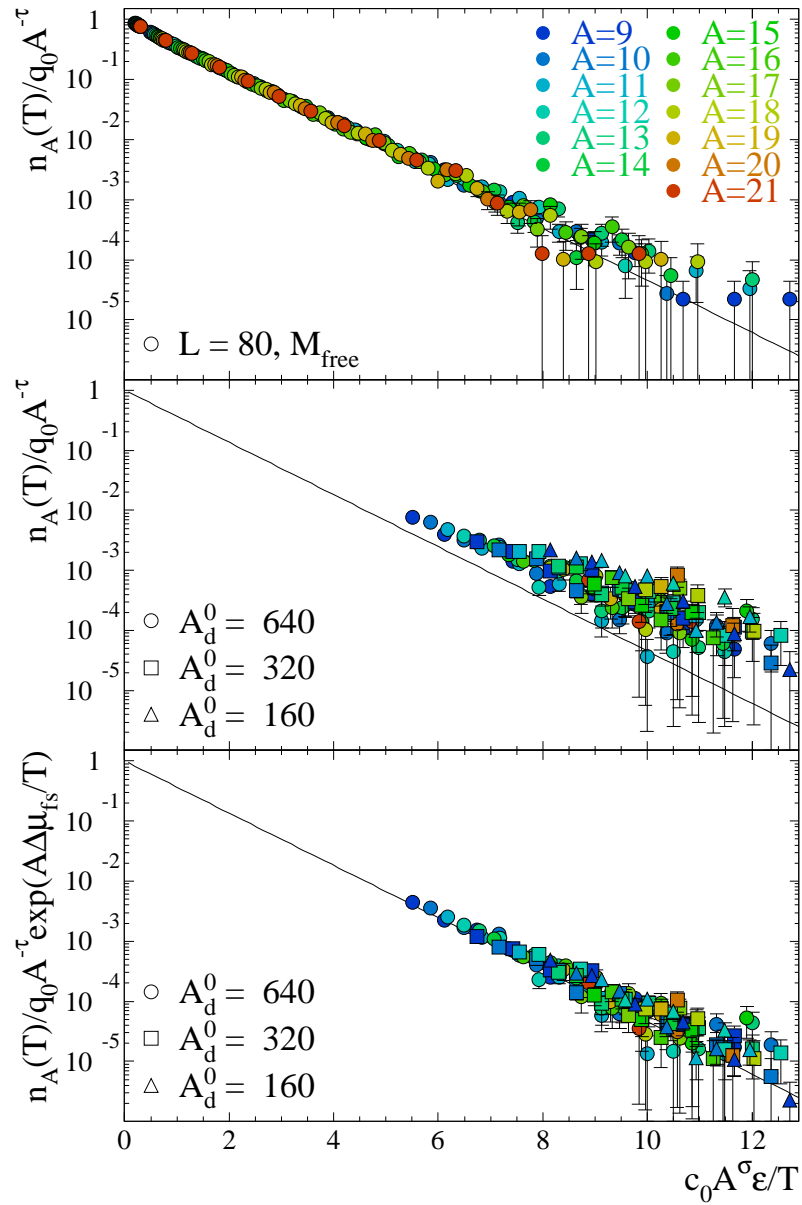}}
\caption{The cluster concentrations of the $d=2$, $L=80$ periodic boundary condition square lattice for: $M_{\rm free}$ calculations (top); $M_{\rm fixed}$ calculations with no complement (middle); $M_{\rm fixed}$ calculations with the complement (bottom).}
\label{cluster-yields-figure}
\end{figure}

\begin{figure}
\centerline{
\includegraphics[width=10.7cm]{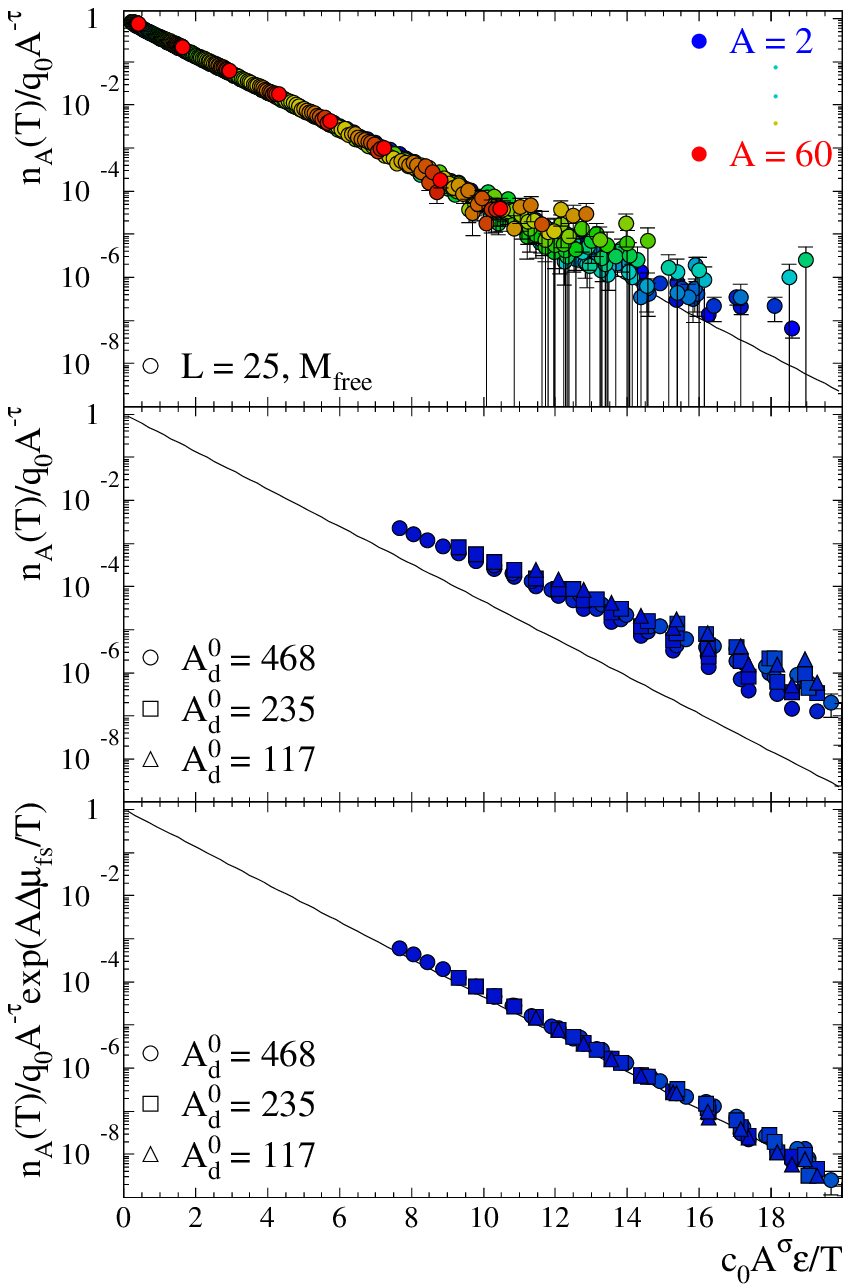}
}
\caption{Same as Fig.~\ref{cluster-yields-figure} but for the $d=3$, $L=25$  periodic boundary condition simple cubic lattice.}
\label{fig:$d=3$}
\end{figure}

To make a comparison between the scaling achieved with the $M_{\rm free}$ clusters and with the $M_{\rm fixed}$ clusters, the fit of  the $M_{\rm free}$ clusters with Eq.~(\ref{eq:fisher-1}) with the free parameters $T_c$, $c_0$, $\sigma$ and $\tau$; $q_0=\zeta(\tau-1)/2$ was performed. Results are given in Table~\ref{FreeTable} and shown in top panels of Figs.~\ref{cluster-yields-figure} and \ref{fig:$d=3$}. The values of the $T_c$ and $\tau$ returned by this procedure are within $1$\% of their established values. The value of $c_0$ shows that all clusters are not perfect squares (cubes) in $d=2$ ($3$) for which $c_0=8$ ($12$) \cite{mader-03,stauffer-99}.  The values of $\sigma$ are within $5\%$ ($15\%$) of their established values for $d=2$ ($3$). This level of inaccuracy arises from shell effects \cite{Elliott:05wci}.

Next it is necessary to calculate $\chi^{2}_{\nu}$ for the $M_{\rm fixed}$ clusters using Eq.~(\ref{eq:fisher-1}) and Eq.~(\ref{eq:complement-2}) with parameters fixed to the Table~\ref{FreeTable} values for the infinite system. This procedure frees one as much as possible from the drawbacks of Fisher's model so  one can concentrate on the effect of the complement. Results are given in Table~\ref{FixedTable} and shown in the middle panels (without complement: concentrations scaled as $n_A(T)/q_0 A^{-\tau}$) and the bottom panels (with complement: concentrations scaled as $n_A(T)/q_0 A^{-\tau}\exp\left[A\Delta \mu_{\rm fs}/T\right]$) of Figs.~\ref{cluster-yields-figure} and \ref{fig:$d=3$}. The $M_{\rm fixed}$ $\chi^{2}_{\nu}$ values for the calculation with the complement are an order of magnitude smaller than the results without the complement and the data collapse is better.

\begin{figure}
\centerline{
\includegraphics[width=9.7cm]{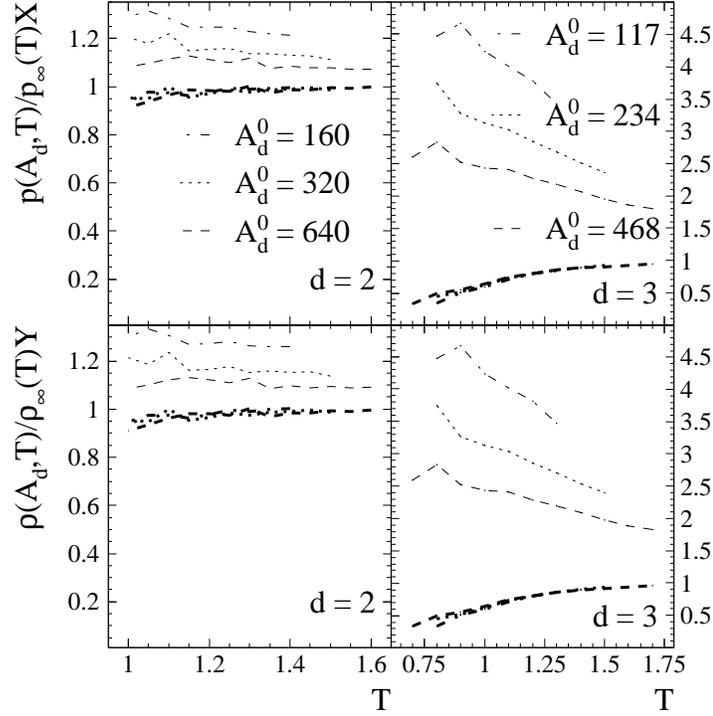}
}
\caption{ Left (right) the normalized pressure and density of a vapor in coexistence with a drop $A_{\rm d}$ for $d=2$ ($3$). Thin [thick] lines show no complement results, $X=Y=1$ [complement results,  $X$ and $Y$  from Eqs.~(\ref{pressure}) and (\ref{density})].}
\label{fig:pressure}
\end{figure}

The middle and bottom panels of Figs.~\ref{cluster-yields-figure} and \ref{fig:$d=3$} show the major point of this method: generalizing Fisher's model with the complement accounts for the finiteness of the liquid. The middle panels of Figs.~\ref{cluster-yields-figure} and \ref{fig:$d=3$} show that the $M_{\rm fixed}$ calculations do not scale as their $M_{\rm free}$ counterparts. The bottom panels of Figs.~\ref{cluster-yields-figure} and \ref{fig:$d=3$} show that when the complement effect is taken into account, the $M_{\rm fixed}$ calculations scale as their $M_{\rm free}$ counterparts.

Now in the integrated quantities, pressure $p(A_{\rm d},T) = T \sum_A n_A(T)$ and density ${\rho}(A_{\rm d},T) = \sum_A n_A(T)A$ . For $A_{\rm d} \gg A$, the expansion of  $\Delta \mu_{\rm fs}$ gives
	\begin{equation}
	\Delta \mu_{\rm fs} = 1 + A\left( \frac{\tau}{A_{\rm d}} + \frac{\sigma c_0 \varepsilon}{TA_{\rm d}^{1-\sigma}} \right) + \cdots
	\label{expansion}
	\end{equation}
This leads to
	\begin{eqnarray}
	p(A_{\rm d},T) & \approx&  p_{\infty}(T) \exp \left[ \left( \frac{\tau}{A_{\rm d}} + \frac{\sigma c_0 \varepsilon}{TA_{\rm d}^{1-\sigma}} \right) \frac{\sum_{A} n_{A}^{\infty}(T) A}{\sum_{A} n_{A}^{\infty}(T)} \right]  \equiv p_{\infty}(T) X
	\label{pressure}
	\end{eqnarray}
and
	\begin{eqnarray}
	{\rho}(A_{\rm d},T) & \approx & {\rho}_{\infty}(T) \exp \left[ \left( \frac{\tau}{A_{\rm d}} + \frac{\sigma c_0 \varepsilon}{TA_{\rm d}^{1-\sigma}} \right) \frac{\sum_A n_{A}^{\infty}(T) A^2}{\sum_A n_{A}^{\infty}(T)A} \right]  \equiv  {\rho}_{\infty}(T) Y.
	\label{density}
	\end{eqnarray}
For a vapor of monomers as $A_{\rm d} \gg \tau$ equations~(\ref{pressure}) and (\ref{density}) yield the Gibbs-Thomson formulae \cite{krishnamachari-96}.

Figure~\ref{fig:pressure} shows the behavior of $p(A_{\rm d},T)$ and ${\rho}(A_{\rm d},T)$ for the $M_{\rm fixed}$ calculations compared to the bulk results. To free ourselves from finite lattice size effects $p_{\infty}(T)$, ${\rho}_{\infty}(T)$ and $n_{A}^{\infty}(T)$ are determined from the $M_{\rm free}$ calculations. As expected $p(A_{\rm d},T)>p_{\infty}(T)$ and ${\rho}(A_{\rm d},T)>{\rho}_{\infty}(T)$ (thin lines in Fig.~\ref{fig:pressure}); i.e. the ratio in question is $>1$. Accounting for the complement via equations~(\ref{pressure}) and (\ref{density}) (using values in Table~\ref{FreeTable}) collapses results from all the calculations to a single line recovering the bulk behavior (thick lines in Fig.~\ref{fig:pressure}); i.e. the ratio in question is $\sim 1$. Deviations at low $T$ are due to the increasing effects of monomers which have $c_0=8$ ($12$) in $d=2$ ($3$).

Thus,  a general approach in terms of the complement has been developed which allows one to account for finite liquid drop size effects and to extrapolate from finite to infinite systems.  It was demonstrated the applicability of the suggested  method using lattice gas model (Ising) calculations. This method can be generalized to include other energy factors present in the nuclear case (e.g. symmetry, Coulomb and angular momentum energies) which is another important step towards determining the liquid-vapor phase boundary of infinite, symmetric nuclear matter from experimental nuclear data. 
Also this general approach  can be applied to a variety of cluster models, like the SMM, GSMM, CSMM and GBM, to  deal  with finite systems.


\section{Constrained SMM in Finite Volumes}

{
Despite the  great
success, the application of the exact  solution \cite{Bugaev:00,Bugaev:01,Reuter:01}
to the description of experimental data is  limited because this solution  
corresponds to an infinite system and due to that it cannot  account for a more
complicated interaction between nuclear fragments. 
Therefore, it was necessary to extend  the exact  solution \cite{Bugaev:00,Bugaev:01,Reuter:01}
to finite volumes.  It is clear that for the finite volume extension it is necessary 
to account for the finite size  and geometrical shape of 
the largest fragments,  when  they  are comparable with the system volume.  
For this one has to abandon the arbitrary size of largest fragment and
consider the constrained SMM (CSMM) in which
the largest fragment  size
is explicitly related to the volume $V$ of the system. 
The latter is assumed to be a cubical one. 
}
Thus, the CSMM  assumes  a more strict constraint
$\sum\limits_k^{K(V)} k~n_k =A$ , where the size
of the largest fragment  $K(V) = \alpha V/b$ cannot exceed the total volume of the system 
(the parameter $\alpha \le 1$  is introduced for convenience).
 The case of the fixed size of the largest fragment, i.e.  $K(V) = Const$,   analyzed  numerically 
  in Ref. \cite{CSMM} 
 is also included in our treatment. 
A similar restriction should be also applied to the upper limit of the product in 
all partitions $Z_A^{id} (V,T)$, $Z_A(V,T)$
and ${\cal Z}(V,T,\mu)$ introduced above 
(how to deal with the real values of $K(V)$, see later). 
Then the  model with this constraint, the CSMM,  cannot be solved by the Laplace 
transform method, because the volume integrals cannot be evaluated due to a complicated 
functional $V$-dependence.  
However, the CSMM can be solved analytically with the help of  the following identity  \cite{Bugaev:04a}
%
\begin{equation}\label{nfour}
G (V) = 
%
\int\limits_{-\infty}^{+\infty} d \xi~ \int\limits_{-\infty}^{+\infty}
  \frac{d \eta}{{2 \pi}} ~ 
{\textstyle e^{ i \eta (V - \xi) } } ~ G(\xi)\,, 
\end{equation}
which is based on the Fourier representation of the Dirac $\delta$-function. 
The representation (\ref{nfour}) allows us to decouple the additional volume
dependence and reduce it to the exponential one,
which can be dealt by the usual Laplace transformation
in the  following sequence of steps
\begin{eqnarray}
\hat{\cal Z}(\lambda,T,\mu) &\equiv  & \int_0^{\infty}dV~{\textstyle e^{-\lambda V}}
~{\cal Z}(V,T,\mu) = 
\hspace*{-0.0cm}\int_0^{\infty}\hspace*{-0.2cm}dV^{\prime}
\int\limits_{-\infty}^{+\infty} d \xi~ \int\limits_{-\infty}^{+\infty}
\frac{d \eta}{{2 \pi}} ~ { \textstyle e^{ i \eta (V^\prime - \xi) - \lambda V^{\prime} } } 
\nonumber \\
&\times & \sum_{\{n_k\}}\hspace*{-0.1cm} \left[\prod_{k=1}^{K( \xi)}~\frac{1}{n_k!}~\left\{V^{\prime}~
{\textstyle \phi_k (T) \,  
e^{\frac{ (\mu  - (\lambda - i\eta) bT )k}{T} }}\right\}^{n_k} \right] \Theta(V^\prime) 
\nonumber \\
& = & 
\hspace*{-0.0cm}\int_0^{\infty}\hspace*{-0.2cm}dV^{\prime}
\int\limits_{-\infty}^{+\infty} d \xi~ \int\limits_{-\infty}^{+\infty}
  \frac{d \eta}{{2 \pi}} ~ { \textstyle e^{ i \eta (V^\prime - \xi) - \lambda V^{\prime} 
+ V^\prime {\cal F}(\xi, \lambda - i \eta) } }\,.
 \label{nfive}
\end{eqnarray}
After changing the integration variable $V \rightarrow V^{\prime} = V - b \sum\limits_k^{K( \xi)} k~n_k $,
the constraint of $\Theta$-function has disappeared.
Then all $n_k$ were summed independently leading to the exponential function.
Now the integration over $V^{\prime}$ in Eq.~(\ref{nfive})
can be straightforwardly done resulting in
%
\begin{equation}\label{six}
\hspace*{-0.4cm}\hat{\cal Z}(\lambda,T,\mu) = \int\limits_{-\infty}^{+\infty} \hspace*{-0.1cm} d \xi
\int\limits_{-\infty}^{+\infty} \hspace*{-0.1cm}
\frac{d \eta}{{2 \pi}} ~ 
\frac{  \textstyle e^{ - i \eta \xi }  }{{\textstyle \lambda - i\eta ~-~{\cal F}(\xi,\lambda - i\eta)}}~,
\end{equation}
where the function ${\cal F}(\xi,\tilde\lambda)$ is defined as follows 
%
\begin{eqnarray}\label{seven}
&&\hspace*{-0.4cm}{\cal F}(\xi,\tilde\lambda) = \sum\limits_{k=1}^{K(\xi) } \phi_k (T) 
~e^{\frac{(\mu  - \tilde\lambda bT)k}{T} }
= 
\hspace*{-0.0cm}\left(\frac{m T }{2 \pi} \right)^{\frac{3}{2} } \hspace*{-0.1cm} \biggl[ z_1
~{\textstyle e^{ \frac{\mu- \tilde\lambda bT}{T} } } + \hspace*{-0.1cm} \sum_{k=2}^{K(\xi) }
k^{-\tau} e^{ \frac{(\mu + W - \tilde\lambda bT)k - \sigma k^{2/3}}{T} }  \biggr]\,.\,
\end{eqnarray}

As usual, in order to find the GCE partition by  the inverse Laplace transformation,
it is necessary to study the structure of singularities of the isobaric partition (\ref{seven}).


\subsection{Isobaric Partition Singularities at Finite Volumes.}
The isobaric partition (\ref{seven}) of the CSMM is, of course, more complicated
than its SMM analog \cite{Bugaev:00,Bugaev:01}
because for finite volumes the structure of singularities in the CSMM 
is much richer than in the SMM, and they match in the limit $V \rightarrow \infty$ only.
To see this let us first make the inverse Laplace transform:
%
\begin{eqnarray}\label{eight}
\hspace*{-0.5cm}{\cal Z}(V,T,\mu)~ = 
\int\limits_{\chi - i\infty}^{\chi + i\infty}
\frac{ d \lambda}{2 \pi i} ~ \hat{\cal Z}(\lambda,  T, \mu)~ e^{\textstyle   \lambda \, V } & = &
\hspace*{-0.0cm}\int\limits_{-\infty}^{+\infty} \hspace*{-0.1cm} d \xi
\int\limits_{-\infty}^{+\infty} \hspace*{-0.1cm}  \frac{d \eta}{{2 \pi}}  
\hspace*{-0.1cm} \int\limits_{\chi - i\infty}^{\chi + i\infty}
\hspace*{-0.1cm} \frac{ d \lambda}{2 \pi i}~ 
\frac{\textstyle e^{ \lambda \, V - i \eta \xi } }{{\textstyle \lambda - i\eta ~-~{\cal F}(\xi,\lambda - i\eta)}}~= 
\nonumber \\
&&\hspace*{-0.0cm}\int\limits_{-\infty}^{+\infty} \hspace*{-0.1cm} d \xi
\int\limits_{-\infty}^{+\infty} \hspace*{-0.1cm}  \frac{d \eta}{{2 \pi}}
\,{\textstyle e^{  i \eta (V - \xi)  } } \hspace*{-0.1cm} \sum_{\{\lambda _n\}}
e^{\textstyle  \lambda _n\, V } 
{\textstyle 
\left[1 - \frac{\partial {\cal F}(\xi,\lambda _n)}{\partial \lambda _n} \right]^{-1} } \,,
\end{eqnarray}
where the contour  $\lambda$-integral is reduced to the sum over the residues of all singular points
$ \lambda = \lambda _n + i \eta$ with $n = 1, 2,..$, since this  contour in the complex $\lambda$-plane  obeys the
inequality $\chi > \max(Re \{  \lambda _n \})$.  
Now both remaining integrations in (\ref{eight}) can be done, and the GCE partition becomes 
%
\begin{equation}\label{nine}
{\cal Z}(V,T,\mu)~ = \sum_{\{\lambda _n\}}
e^{\textstyle  \lambda _n\, V }
{\textstyle \left[1 - \frac{\partial {\cal F}(V,\lambda _n)}{\partial \lambda _n} \right]^{-1} } \,,
\end{equation}
i.e. the double integral in (\ref{eight}) simply  reduces to the substitution   $\xi \rightarrow V$ in
the sum over singularities. 
This is a remarkable result which 
was  formulated in Ref. \cite{Bugaev:04a}  as the following 
\underline{\it theorem:}
{\it if the Laplace-Fourier image of the excluded volume GCE partition exists, then
for any additional $V$-dependence of ${\cal F}(V,\lambda _n)$ or $\phi_k(T)$
the GCE partition can be identically represented by Eq. (\ref{nine}).} 
Now it  clear that the above theorem can be applied to the finite volume formulations of a more general model, the GSMM \cite{Bugaev:05csmm}. However, it the following I will use the CSMM to keep presentation simple. 


The simple poles in (\ref{eight}) are defined by the  equation 
\begin{equation}\label{ten}
\lambda _n~ = ~{\cal F}(V,\lambda _n)\,.
\end{equation}
In contrast to the usual SMM \cite{Bugaev:00,Bugaev:01} the singularities  $ \lambda _n $ 
are (i) 
 are volume dependent functions, if $K(V)$ is not constant,
and (ii) they can have a non-zero imaginary part, but 
in this case there  exist  pairs of complex conjugate roots of (\ref{ten}) because the GCE partition is real.

Introducing the real $R_n$ and imaginary $I_n$ parts of  $\lambda _n = R_n + i I_n$,
one can rewrite  Eq. (\ref{ten})
as a system of coupled transcendental equations 
%
\begin{eqnarray}\label{eleven}
&&\hspace*{-0.2cm} R_n = ~ \sum\limits_{k=1}^{K(V) } \tilde\phi_k (T)
~{\textstyle e^{\frac{Re( \nu_n)\,k}{T} } } \cos(I_n b k)\,,
\\
\label{twelve}
&&\hspace*{-0.2cm} I_n = - \sum\limits_{k=1}^{K(V) } \tilde\phi_k (T)
%
%
~{\textstyle e^{\frac{Re( \nu_n)\,k}{T} } } \sin(I_n b k)\,,
\end{eqnarray}
where I have introduced the 
set of  the effective chemical potentials  $\nu_n  \equiv  \nu(\lambda_n ) $ with $ \nu(\lambda) = \mu + W (T)  - \lambda b\,T$, and 
the reduced distribution functions $\tilde\phi_1 (T) = \left(\frac{m T }{2 \pi} \right)^{\frac{3}{2} }  z_1 \exp(-W(T)/T)$ and 
$\tilde\phi_{k > 1} (T) = \left(\frac{m T }{2 \pi} \right)^{\frac{3}{2} }  k^{-\tau}\, \exp(-\sigma (T)~ k^{2/3}/T)$ for convenience.


Consider the real root $(R_0 > 0, I_0 = 0)$, first. 
For $I_n = I_0 = 0$ the real root $R_0$ exists for any $T$ and $\mu$.
Comparing $R_0$ with the expression for vapor pressure of the analytical SMM solution 
\cite{Bugaev:00,Bugaev:01}
shows   that $T R_0$ is  a constrained grand canonical pressure of the gas. 
 As usual,  for  finite volumes the total mechanical pressure \cite{Hill,Bugaev:04a} differs from   $T R_0$.
Equation (\ref{twelve}) shows that for $I_{n>0} \neq 0$ the inequality $\cos(I_n b k) \le 1$ never 
become the equality for all $k$-values  simultaneously. Then from Eq. (\ref{eleven})  
one obtains ($n>0$)
\begin{equation}\label{thirteen}
R_n < \sum\limits_{k=1}^{K(V) } \tilde\phi_k (T)
~{\textstyle e^{\frac{Re(\nu_n)\, k}{T} } } \quad \Rightarrow \quad R_n < R_0\,, 
\end{equation}
where the second inequality (\ref{thirteen}) immediately follows from the first one.
{ In other words, the gas singularity is always the rightmost one.
This fact }
plays a decisive role in the thermodynamic limit
$V \rightarrow \infty$.

The interpretation of the complex roots $\lambda _{n>0}$  is less straightforward.
According to Eq. (\ref{nine}),   the  GCE partition is a superposition of  the
states of different  free energies $- \lambda _n V T$.  
(Strictly speaking,  $- \lambda _n V T$  has  a meaning of  the change of free energy, but
I  will use the  traditional  term for it.)
For $n>0$ the free energies  are complex. 
Therefore,  
 $-\lambda _{n>0} T$ is   the density of free energy.  The real part  of the free energy density,
  $- R_n T$, defines the significance
 of the state's  contribution to the partition:  due to (\ref{thirteen}) 
 the  largest contribution  always comes from the gaseous state and
 has the smallest  real part  of free energy density. As usual,  the states which do not have
 the smallest value of the (real part of)  free energy, i. e.  $- R_{n>0} T$, are thermodynamically metastable. 
 For  infinite   volume 
 they should not contribute unless they are infinitesimally close to  $- R_{0} T$, 
 but for finite volumes their contribution to the GCE partition may be important.

As one sees from (\ref{eleven}) and (\ref{twelve}), the states of  different  free energies have  
different values of the effective chemical potential $\nu_n$, which is not the case for
infinite volume \cite{Bugaev:00,Bugaev:01},
where there  exists a single value for the effective chemical potential. 
Thus,  for finite $V$
the  states which contribute to the GCE partition (\ref{nine}) are not in a true chemical equilibrium.

{The meaning of the imaginary part of the free energy density  becomes  clear from 
(\ref{eleven}) and (\ref{twelve}) \cite{Bugaev:05csmm}: as one can see from (\ref{eleven})  
the imaginary part $I_{n>0}$
effectively changes the number of degrees of freedom of  each $k$-nucleon fragment ($k \le K(V)$)
contribution to  the free energy  density  $- R_{n>0} T$.  It is clear, that the change 
of the effective number of degrees of freedom can occur virtually only and, if 
$\lambda _{n>0}$ state is accompanied  by 
some kind of  equilibration process. 
Both of these statements become clear,
 if  one recalls  that  
the statistical operator in statistical mechanics and the  quantum mechanical evolution operator 
are related by the Wick rotation \cite{Feynmann}. In other words, the inverse temperature can be
considered as an  imaginary time.  
Therefore, depending on the sign,  the quantity  $ I_n b T \equiv \tau_n^{-1}$  that  appears 
 in the trigonometric  functions      
of  the  equations (\ref{eleven}) and (\ref{twelve}) in front of the imaginary time $1/T$ 
can be regarded  as the inverse decay/formation time $\tau_n$ of the metastable state which corresponds to the  pole $\lambda _{n>0}$ (for more details see next sections).
}
 
This interpretation of  $\tau_n$ naturally explains the thermodynamic  metastability 
of all states except the gaseous one:
the metastable states can exist  in the system only virtually 
because of their finite decay/formation  time,
whereas the gaseous state is stable because it has an infinite decay/formation time.

%
%



\subsection{No Phase Transition Case.}
It is instructive to treat the effective chemical potential $\nu (\lambda)$ as an independent variable
instead of $\mu$. In contrast to the infinite $V$, where  the upper   limit  $\nu \le 0$ defines the liquid phase singularity of the
isobaric partition and  gives the pressure of a liquid phase
$p_l(T,\mu) = T R_0 |_{V \rightarrow \infty}  = (\mu + W(T))/b$ \cite{Bugaev:00,Bugaev:01}, 
for finite  volumes and finite $K(V)$ the effective  chemical potential can
be complex (with either sign for its real part)  and its value defines the number and position of the imaginary roots 
$\{\lambda _{n>0} \}$ 
in the complex plane.
Positive  and negative values of the effective chemical potential  for finite systems  were  considered 
\cite{Elliott:01}
within the Fisher droplet model, but, to our knowledge,  its complex values have never  been  discussed.
From the definition of the effective chemical potential $\nu(\lambda)$ it is evident that  its complex
values for finite systems  exist  only  because of the excluded volume interaction, which is 
not taken into account in the  Fisher droplet model \cite{Fisher:67}.
However, a recent study  of  clusters of 
the $d = 2$ Ising model within the framework of  FDM (see the corresponding section 
in Ref.  \cite{Elliott:05wci})
shows that the excluded volume correction improves 
essentially the description of the thermodynamic functions. 
Therefore, the next step is to consider the complex values of the
effective chemical potential and free energy for the excluded volume correction of the Ising
clusters on finite lattices. 


\begin{figure}[ht]

\mbox{
\hspace*{0.0cm}\epsfig{figure=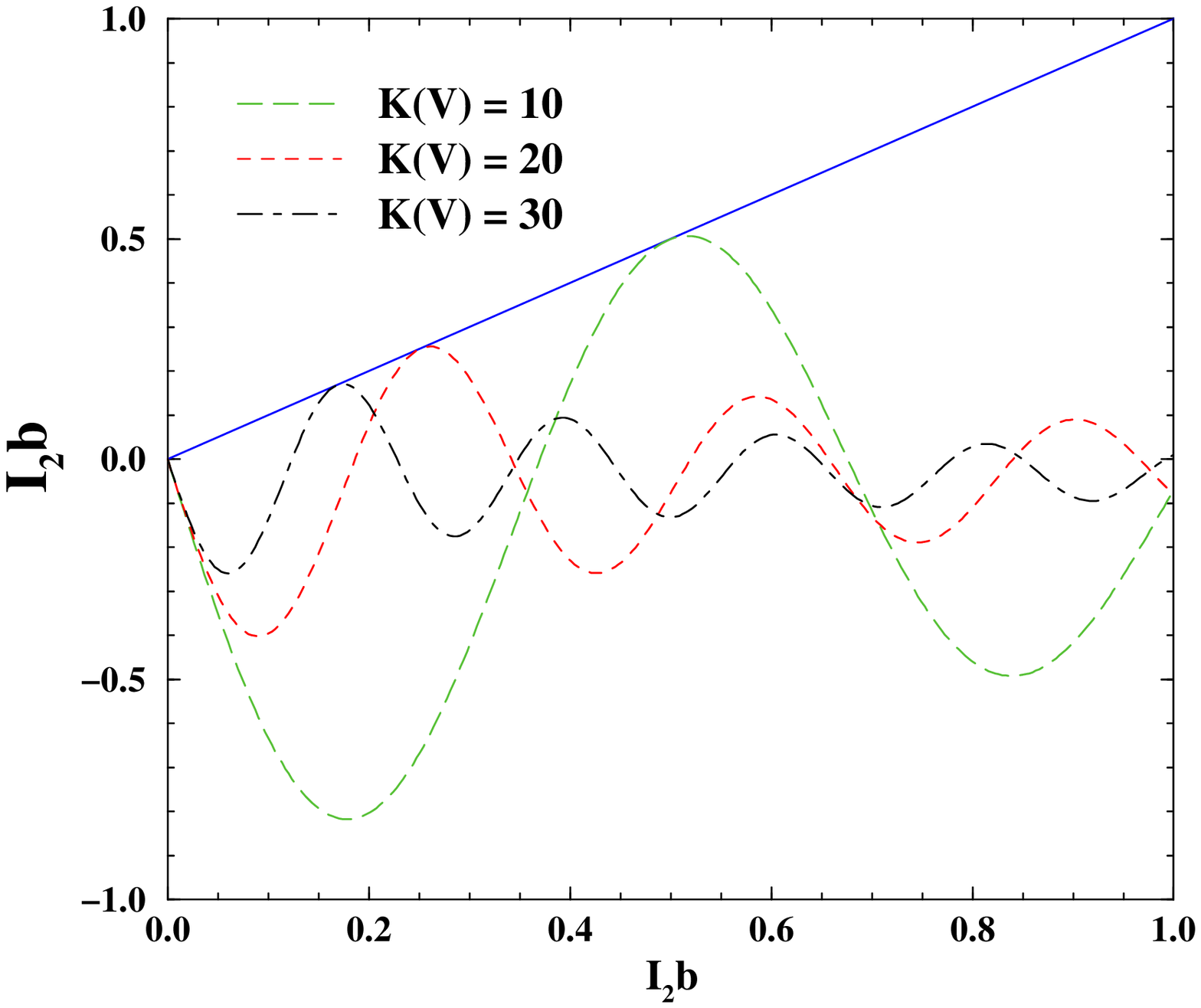,width=7.6cm}
\hspace*{0.5cm}\epsfig{figure=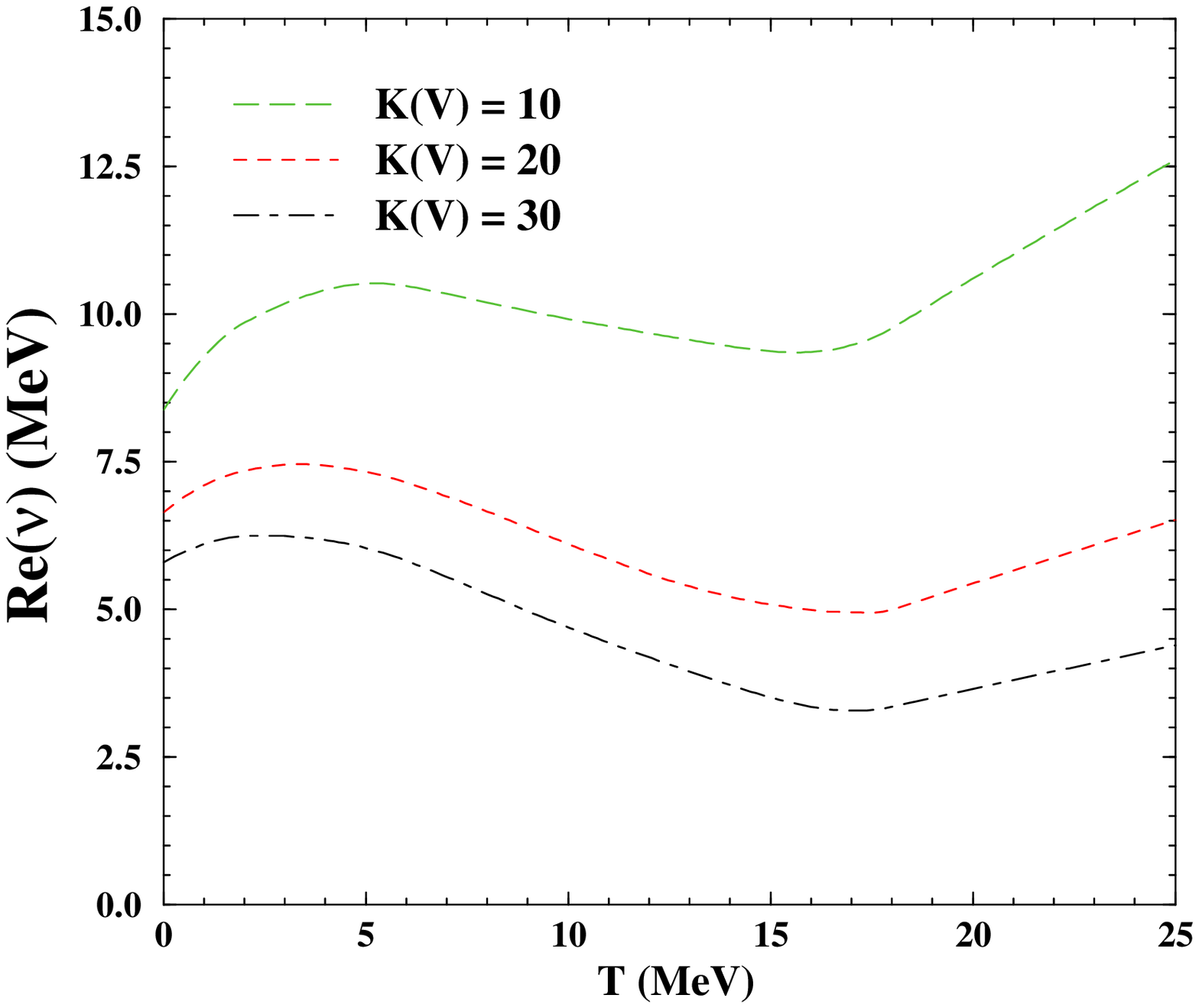,width=7.6cm}
}

\vspace*{0.3cm}

\caption{\label{figPDTnu}
{\bf Left panel:} A graphical solution of Eq. (\ref{twelve}) for $T = 10$ MeV and $\tau = 1.825$.
The l.h.s. (straight line) and  r.h.s. of Eq. (\ref{twelve}) (all dashed curves) are shown
as the function of
dimensionless parameter $I_1\,b$ for the three values of the largest fragment size $K(V)$.
The intersection point at $(0;\,0)$ corresponds to a real root of Eq. (\ref{ten}).
Each tangent point with the straight line generates  two complex  roots of (\ref{ten}). \newline
{\bf Right panel:}  Each curve separates  the $T-Re(\nu_n)$ region of one real root of Eq. (\ref{ten})
(below the curve), three complex roots (at the curve) and five and more roots (above the curve)
for three values of $K(V)$ and the same parameters as in the left panel.
}
\end{figure}


As it is seen from the left  panel of  Fig.~\ref{figPDTnu}, the r.h.s. of Eq. (\ref{twelve})  
is the amplitude and frequency modulated sine-like  
function of dimensionless parameter $I_n\,b$. 
Therefore, depending on $T$ and $Re(\nu)$ values, there may exist 
no 
complex roots $\{\lambda _{n>0}\}$, a finite number of them, or an infinite number of them. 
The left panel of  Fig.~\ref{figPDTnu}  shows  a special case which corresponds to  exactly three 
roots of Eq. (\ref{ten}) for each value of $K(V)$: the real root ($I_0 = 0$) and two complex conjugate
roots ($\pm I_1$). 
Since 
the r.h.s. of (\ref{twelve}) is monotonously increasing function
of  $Re(\nu)$, when the former is positive,  
it is possible to map the $T-Re(\nu)$ plane into
regions of a fixed number of roots of Eq. (\ref{ten}). 
Each curve in the right  panel of \mbox{Fig.~\ref{figPDTnu}} divides the $T-Re(\nu)$ plane
into three parts: for $Re(\nu)$-values below the curve there  is only one real root (gaseous phase), 
for points on  the curve there exist      
three roots, and above the curve there are four or more roots of Eq. (\ref{ten}).

For constant values of  $K(V) \equiv K$  the number of terms in the r.h.s. of (\ref{twelve}) does not depend on
the volume and, consequently, in thermodynamic limit $V \rightarrow \infty$   only the 
rightmost  simple pole in the complex $\lambda$-plane survives out of a finite number of simple poles.
According to the  inequality (\ref{thirteen}), the real root $\lambda_0$ is  the rightmost singularity of isobaric partition (\ref{six}).
However,  there is a  possibility  that the real parts  of other  roots $\lambda_{n>0} $ become infinitesimally 
close to $R_0$, when there is an infinite number of terms which contribute to the GCE partition (\ref{nine}).

Let us show  now that even for an infinite number of simple poles in (\ref{nine})
only the real root $\lambda_0$  survives in the limit $V \rightarrow \infty$.
For this purpose consider  the limit
$Re(\nu_n)  \gg T $.
In this limit   the distance between  the imaginary parts of the nearest roots 
remains finite even for infinite volume.  Indeed,  for $Re(\nu_n)  \gg T   $
the leading contribution to the r.h.s. of (\ref{twelve}) corresponds to the harmonic with $k = K$,
and, consequently,  an exponentially large amplitude of this term
can be only  compensated by  a vanishing value of  $\sin\left( I_n \, b K  \right)$,  
i.e.   $I_n \, b K  =  \pi n + \delta_n$
{  
with $|\delta_n| \ll \pi$ (hereafter I will analyze only 
the branch $I_n > 0$), 
and, therefore, the corresponding decay/formation time $\tau_n  \approx K [ \pi n T ]^{-1}$ 
is volume independent.

Keeping the leading term on the r.h.s. of  (\ref{twelve}) and solving for $\delta_n$, one finds
\begin{eqnarray}\label{Mfourteen}
\hspace*{-0.2cm}
I_n & \approx & (-1)^{n+1}  \tilde\phi_K (T)  ~{\textstyle e^{\frac{Re(\nu_n )\,K}{T} } } ~\delta_n \,, 
\quad {\rm with} \quad 
%
\delta_n \approx    \frac{ (-1)^{n+1}  \pi n }{ K b ~ \tilde\phi_K (T)  }~{
\textstyle e^{- \frac{Re(\nu_n )\,K}{T} } } \,,    \\
\label{Msixteen}
R_n & \approx & (-1)^{n}  \tilde\phi_K (T)  ~{\textstyle e^{\frac{Re(\nu_n )\,K}{T} } }  \,, 
\end{eqnarray}
where in the last step I used Eq. (\ref{eleven}) and condition $|\delta_n| \ll \pi$.
Since for $V \rightarrow \infty$ all  negative values of $R_n$ cannot contribute to the 
GCE partition (\ref{nine}), it is sufficient to analyze even values of $n$ which, according to 
(\ref{Msixteen}),  generate  $R_n > 0$.

Since the inequality  (\ref{thirteen}) can not be broken,   a single possibility,
when $\lambda_{n>0}$ pole can contribute to the partition (\ref{nine}),  corresponds to 
the case 
$ R_n \rightarrow R_0 - 0^+$ 
for  some finite $n$.   
Assuming this, one finds  $Re (\nu (\lambda_n)) \rightarrow Re(\nu (\lambda_0))$
for the same value of $\mu$. 
}
Substituting these results into equation (\ref{eleven}), one gets
\begin{equation}\label{Mseventeen}
\hspace*{-0.2cm}R_n \approx  \sum\limits_{k=1}^{K } \tilde\phi_k (T)
~{\textstyle e^{\frac{Re(\nu (\lambda_0) )\,k}{T} } } \cos\left[ \frac{ \pi n k}{K} \right]  \ll R_0\,.
\end{equation}
The inequality (\ref{Mseventeen})  follows  from the equation for $R_0$ and the fact that, even for
equal leading terms   in the sums above (with $k =  K$ and even  $n$),  the difference between $R_0$  and $R_n$ is  large due to  the next to leading term $k = K - 1$, which is  proportional to 
$e^{\frac{Re(\nu (\lambda_0) )\,(K-1)}{T} } \gg 1$.  
Thus, one arrives  at  a  contradiction with our assumption $R_0 - R_n \rightarrow 0^+$, 
and, consequently,  it cannot be true.  Therefore,
for large volumes  the  real root $\lambda_0$  always gives
the main contribution to  the GCE partition (\ref{nine}), and  this is the only root that survives 
in the limit $V \rightarrow \infty$.
Thus,  I showed that  the model with the fixed  size  of the largest fragment has no phase transition because there is a single singularity of the isobaric partition (\ref{six}), which 
exists in thermodynamic limit.

\subsection{Finite Volume Analogs of Phases.}
If $K(V)$ monotonically grows with the volume,  the situation is different. 
In this case for  positive value of $Re(\nu)  \gg T$ 
the leading exponent in the r.h.s. of (\ref{twelve})  
also corresponds to a largest fragment, i.e. to $k  = K(V)$. 
Therefore,  we can apply 
the same arguments  which were used above  for  the case $K(V) = K =  const$
and derive similarly  equations  (\ref{Mfourteen})--(\ref{Msixteen}) for  $I_n$ and $R_n$. 
From $I_n  \approx \frac{\pi n}{ b\, K( V) }$ it follows that,
when $V$ increases, the number of simple poles in (\ref{eight}) also increases  and
the  imaginary part of 
the closest to the real $\lambda$-axis  poles becomes very small,
 i.e $I_n   \rightarrow 0$ for  $n \ll K(V)$,
 and, consequently, the associated   decay/formation time 
$\tau_n  \approx K(V)  [ \pi n T ]^{-1}$ grows with the volume of the system.
Due to $ I_n  \rightarrow 0$, 
the inequality (\ref{Mseventeen}) cannot be  
established for the poles with $n \ll K(V)$. 
Therefore, in  contrast to the previous case, for large $K(V)$ the  simple poles
with $n \ll K(V)$ will be infinitesimally close to the real axis of the complex $\lambda$-plane.

From Eq.  (\ref{Msixteen}) it follows that 
%
\begin{equation}\label{Meighteen}
R_n ~ \approx ~  \frac{p_l(T,\mu) }{T} -  \frac{ 1}{ K(V) b} 
 \ln \left|  \frac{ R_n}{  \tilde\phi_K (T)     }  \right|  \rightarrow  \frac{p_l(T,\mu) }{T} 
\end{equation}
for  $ | \mu | \gg T $ and $K(V) \rightarrow \infty$.   
Thus, 
it is  proved that
for infinite volume the  infinite number of simple poles moves toward 
the real $\lambda$-axis to the vicinity of liquid phase singularity $\lambda_l = p_l(T,\mu)/T $ 
of the isobaric partition
\cite{Bugaev:00, Bugaev:01} and
generates  an essential singularity of function ${\cal F}(V, p_l/T)$ in (\ref{seven}) 
{\it irrespective to the  sign of 
the chemical potential $\mu$.}
In addition,  as  I showed above, the states with $Re( \nu ) \gg T$  become  stable because they acquire   infinitely large 
decay/formation time $\tau_n$ in the limit $V \rightarrow \infty$.  Therefore, these states should be identified 
as a liquid phase for finite  volumes as well.

Now it is clear 
that each curve in the right  panel of Fig.~\ref{figPDTnu}  is   the  finite volume analog of the phase boundary $T-\mu$ for a given value of $K(V)$:  below the phase boundary there exists a gaseous phase, but at and above each curve there
are  states which can be identified with a finite volume analog of the mixed phase, and,
finally, at $ Re(\nu) \gg T$ there exists a liquid phase.
When  there is no phase transition, i.e. $K(V) = K = const$,  the structure of simple poles is
similar, but, first,  the line which separates the gaseous states from the metastable states does not
change with the volume, and, second, as shown above, the metastable states will never become
stable. 
Therefore,
a systematic study of the 
volume dependence  of free energy (or pressure for very large  $V$)  along with the formation and  decay times may  be  of a crucial importance for  experimental studies of 
the nuclear liquid gas phase transition.

The above results demonstrate that, in contrast  to Hill's expectations \cite{Hill}, the finite volume analog
of the mixed phase does not  
{ consist  just of  two pure phases. }
The mixed phase for finite volumes
consists of  a stable  gaseous phase and  the set of  metastable states which differ by the free energy. 
Moreover, the difference between the free energies of these states is  not surface-like, as Hill 
assumed in his treatment \cite{Hill}, but  volume-like.  Furthermore, 
according to Eqs. (\ref{eleven}) and (\ref{twelve}),  each of these states 
consists of the same fragments, but with different weights. 
As  seen above for the case $ Re(\nu) \gg T$,  
some fragments 
that  belong to 
the states, in which  the largest fragment is  dominant,
may, in principle, have negative weights (effective number of degrees of freedom) in 
the expression  for  $R_{n>0}$  (\ref{eleven}).
This can be understood easily because  higher concentrations of large fragments can be achieved 
 at the  expense of the  smaller fragments and  is reflected in  the corresponding change 
of the real part of  the free energy $- R_{n>0} V T$. 
Therefore, the  actual  structure of the mixed phase at finite volumes is  more complicated
than  was expected in earlier works.  

{
The Hills'  ideas were developed further in Ref.  \cite{Chomaz:03}, where the authors claimed to 
establish the one to one correspondence between the bimodal structure of the partition of measurable
quantity  $B  $ known on average and  the properties of the Lee-Yang zeros of this 
partition in the complex  $  g $-plane. The starting point of  Ref.  \cite{Chomaz:03} is to 
postulate the partition $Z_{ g} $ and  the probability $P_{ g} (  B  )$ of  the following form
\begin{equation}\label{Mnineteen}
Z_{ g}~  \equiv~ \int d { B}~ W ( { B} ) ~ e^{ - { B} \cdot { g}  }  \quad
 \Rightarrow  \quad 
 P_{ g} (  B )~ \equiv ~ \frac{  W ( { B} ) ~ e^{ - { B} \cdot { g}   }  }{  Z_{ g} }\,,
\end{equation}
where $ W ( { B} )$ is the partition  sum of the ensemble of fixed values of the observable  $\{  B \}$ , and
$  g $ is the corresponding Lagrange multiplier. Then the authors of  Ref.  \cite{Chomaz:03}
assume 
the existence of two maxima of the probability  $P_{ g} ( B )$ ($\equiv$ bimodality)  and 
discuss their relation to the Lee-Yang zeros of  $Z_{ g} $  in the complex $ g $-plane.

The CSMM  gives us a unique opportunity  to verify the Chomaz and Gulminelli idea on the bimodality behavior of  $P_{ g} (  B ) $ using the first principle results. Let us use the equation (\ref{nfive})  identifying  the intensive variable $ g $ with  $  \lambda $ and extensive one  $ B  $ with
the available volume $V^\prime \rightarrow \tilde V$. The evaluation of the r.h.s. of  (\ref{nfive}) 
is very difficult in general, but for a special case, when the eigen volume $b$ is small this can be done
analytically. Thus, approximating ${\cal F}(\xi, \lambda - i \eta) \approx {\cal F}(\xi, \lambda) -
i\eta \, \partial {\cal F}(\xi, \lambda)/ \partial \lambda $, one obtains the CSMM analog of  the probability 
(\ref{Mnineteen})
\begin{eqnarray}
 P_{\lambda} ( \tilde V ) ~ \hat{\cal Z}(\lambda,T,\mu) & \equiv &  \int\limits_{-\infty}^{+\infty} d \xi~ 
 \int\limits_{-\infty}^{+\infty}
  \frac{d \eta}{{2 \pi}} ~ { \textstyle e^{ i \eta ( \tilde V - \xi) - \lambda \tilde V
+ \tilde V {\cal F}(\xi, \lambda - i \eta) } }\nonumber \\
\label{Mtwenty}
& \approx &  
\int\limits_{-\infty}^{+\infty} d \xi~  { \textstyle e^{ \tilde V  [   {\cal F}(\xi, \lambda )   -\lambda ]  } 
\delta\left[ \tilde V - \xi -    \frac{\partial {\cal F}(\xi, \lambda ) }{\partial \lambda}   \right]  }
\,,
\end{eqnarray}

\noindent
where I made the $\eta$ integration after  applying the approximation for 
 ${\cal F}(\xi, \lambda - i \eta)$. Further evaluation of (\ref{Mtwenty})  requires to know all 
 possible solutions of the average  volume of the system 
 $\xi^*_\alpha ( \tilde V ) = \tilde V -  \partial {\cal F}(\xi^*_\alpha, \lambda)/ \partial \lambda $
($\alpha = \{1,2,\dots\}$). 
Since the number of these solutions is either one or two \cite{Bugaev:07b},  
 the probability   (\ref{Mtwenty})  can be written as
\begin{equation}\label{Mtwone}
 P_{\lambda} ( \tilde V ) ~ \hat{\cal Z}(\lambda,T,\mu) \approx 
 \sum\limits_{ \alpha }  \frac{ 1 }{ \left|  1 + \frac{\partial^2  {\cal F}(\xi^*_\alpha, \lambda)  }{\partial \lambda ~\partial \xi^*_\alpha}  \right|  }    { \textstyle e^{ \tilde V  [   {\cal F}(\xi^*_\alpha, \lambda )   -\lambda ]  } }  \quad 
 %
%
 %
 \,. 
\end{equation}
%
In contrast to the expectations of  Ref. \cite{Chomaz:03}, the  probability (\ref{Mtwone})
cannot be measured experimentally,  irrespective  to  the sign of  the derivative 
 $ \frac{ \partial \ln P_{\lambda} ( \tilde V )  }{ \partial   \tilde V }$.     Indeed, 
 above it  was rigorously proven  that for
 any  real $\xi$ the IP 
 $\hat{\cal Z}(\lambda,T,\mu)$  is  defined on the real $\lambda$-axis only  for 
 $  {\cal F}(\xi, \lambda )   -\lambda <  0 $, i.e.  on the right hand side  of the gaseous singularity  
 $ \lambda_0$:  $ \lambda > \lambda_0$. However, as one can see from the  equation (\ref{eight}), the ``experimental'' $\lambda_n$
 values belong to the other region of the complex $\lambda$-plane: $ Re( \lambda_{n > 0}) < \lambda_0$.

Thus, it turns out that  the suggestion of   Ref.  \cite{Chomaz:03} to analyze the  probability 
(\ref{Mnineteen}) does not make any sense because, as I explicitly  showed   for the CSMM, it cannot be measured. 
It seems that   the starting point of  the Ref.  \cite{Chomaz:03}  approach,
i.e.  the  assumption that the left  equation  (\ref{Mnineteen}) gives the most general form of
the partition of  finite system, is problematic. 
Indeed, comparing   (\ref{Meighteen})  with  the analytical result
(\ref{Mtwone}), one can see that for finite systems, in contrast to the major  assumption of 
Ref.  \cite{Chomaz:03}, 
the probability  $ W $ of  the  CSMM depends not only on the extensive variable $\tilde V$, but also 
on the intensive variable $\lambda$, which makes unmeasurable  the whole construct of Ref.  \cite{Chomaz:03}. 
Consequently, the conclusions of 
Ref.  \cite{Chomaz:03} on the relation between the bimodality and the phase transition existence  are not general because they have a limited range of validity. In addition,  the suggested  construct  
\cite{Chomaz:03} cannot be verified experimentally. 

}



\vspace*{-0.3cm}

\section{Gas of Bags in Finite Volumes}

\vspace*{-0.2cm}

Now I will apply the formalism of the preceding sections to the 
analysis of the Gas of Bags Model (GBM) \cite{Goren:81,Goren:05}  in finite volumes. 
In the low and high temperature domains 
the GBM  reduces to two well known and
successful  models:
the hadron gas model  \cite{Hgas, Hgas:2}  and the bag model of QGP \cite{BagModel}.
Both of these models are surprisingly successful in describing the bulk properties 
of hadron production in high energy nuclear collisions, and,
therefore, one may hope  that  their generalization, the GBM, may reflect basic features of the nature  
in the  phase transition region.

The van der Waals gas consisting of $n$ hadronic  species,
which are called bags in what follows,  has the following  GCE partition \cite{Goren:81}
\begin{equation}
 Z  (V,T)  =  
 \sum_{\{N_k\}} \biggl[
\prod_{k=1}^{n}\frac{\left[ \left( V -v_1N_1-...-v_nN_n\right)  ~\phi_k(T) \right]^{N_k}}{N_k!} \biggr] 
%
~ \theta\left(V-v_1N_1-...-v_nN_n\right)~, 
\label{qqq}
\end{equation}
\vspace*{-0.2cm}

\noindent
where $\phi_k(T) \equiv g_k ~ \phi(T,m_k)  \equiv  \frac{g_k}{2\pi^2}~\int_0^{\infty}p^2dp~
\exp\left[-~  (p^2~+~m_k^2)^{1/2} / T \right]
~=~  g_k \frac{m_k^2T}{2\pi^2}~K_2\left( \frac{m_k}{T} \right)$  is the particle  density
of  bags of mass $m_k$ and eigen volume $v_k$  and degeneracy $g_k$. 
This expression differs  slightly  form the GCE partition of the simplified SMM  (\ref{three}), where 
$\mu = 0$ and  the  eigen volume of $k$-nucleon fragment $k b$ is changed to 
the  eigen volume of the bag $v_k$.
Therefore,  as  for  a simplified SMM the Laplace transformation  (\ref{four})   with respect to volume  of Eq.~(\ref{qqq}) gives 
%
\begin{equation}\label{Zsn}
 \hat{Z}  (s,T)~=~\left[~s~-~\sum_{j=1}^n \exp\left(-v_j s\right)~g_j\phi(T,m_j)\right]^{-1}~.
 \end{equation}
In preceding sections I showed that 
as long as the number of types of bags, $n$, is finite, the only possible singularities  
of $\hat{Z}(s,T)$ (\ref{Zsn}) are simple   poles. 
However, in the case of an infinite number of types  of bags  an essential  singularity of
$\hat{Z}(s,T)$ may appear.  
This  property is used  the GBM: 
the sum  over different bag states in (\ref{qqq})
can be  replaced  by the integral,
$\sum_{j=1}^{\infty}g_j ...=\int_0^{\infty}dm\, dv ...\rho(m,v)$, 
if   the bag mass-volume spectrum, $\rho(m,v)$,  which defines  
the number of bag states in the mass-volume  region  $[m,v;m+dm,v+dv]$,
is given.  Then, the Laplace transform of $Z(V,T)$ reads \cite{Goren:81}
%
\begin{equation}\label{Zsbag}
 \hat{Z}_{GB} (s,T) \equiv \int\limits_0^{\infty}dV ~ e^{-sV}~Z(V,T)
= \frac{1}{[~s~-~F(T,s)]}\,,
\end{equation}
where $F(s,T)$ is defined as 
\begin{equation}\label{fTs}
 F(s,T)=  \int\limits_0^{\infty} dm\,dv ~\rho(m,v)~e^{-vs}~\phi(T,m)\,.
\end{equation}
Like in the simplified SMM, 
the  pressure  of infinite system is again given by the rightmost singularity: 
$p(T)=Ts^*(T)~=~T\cdot max\{s_H(T),s_Q(T)\}$.  
Similarly to the simplified SMM considered above, the rightmost  singularity $s^*(T)$ of $\hat{Z}(s,T)$ (\ref{Zsbag})
can be either the simple  pole singularity $s_H(T) ~=~F\left(s_H(T), T\right) $ of the isobaric partition
(\ref{Zsbag})  or 
the $s_Q(T)$ singularity of the function $F(s,T)$ (\ref{fTs}) it-self. Note, that all singularities of the 
IP are defined by the equation
\cite{Goren:81,Bugaev:00}:
\begin{align}\label{s*vdw}
s^*(T)~=~ F(s^*,T)~.
 \end{align}

The major mathematical difference between the simplified SMM and the GBM is that the latter 
employs the two parameters  mass-volume spectrum.  Thus, the mass-volume spectrum of the GBM
consists of the  discrete mass-volume spectrum of light hadrons and the continuum contribution
of heavy resonances \cite{Goren:82}
%
\begin{eqnarray}
 \rho(m,v) & =&   \sum_{j=1}^{J_m}~ g_j~ \delta(m-m_j)~\delta(v-v_j) \nonumber \\
\label{rhomv}
&+&  \Theta(v -V_0) \Theta(m -M_0 -Bv) C~v^{\gamma}(m-Bv)^{\delta} ~\exp\left[\frac{4}{3}~\sigma_Q^{  \frac{1}{4} }~
 v^{ \frac{1}{4} }~(m-Bv)^{\frac{3}{4} }\right]~,
\end{eqnarray}
respectively.  Here $m_j < M_0$, $ v_j < V_0$,  $M_0 \approx 2 $ GeV, $V_0 \approx 1$ fm$^3$,
$C, \gamma, \delta$ and $B$ (the so-called bag constant, $B \approx 400$ MeV/fm$^3$) are 
the model parameters and 
\begin{equation}\label{sigmaQ}
 \sigma_Q~=~\frac{\pi^2}{30}\left(g_g~+~\frac{7}{8}g_{q\bar{q}}\right)~
 =~
\frac{\pi^2}{30}\left(2\cdot 8~+~\frac{7}{8}\cdot 2\cdot 2\cdot 3
\cdot 3\right)~=~\frac{\pi^2}{30}~\frac{95}{2}
 \end{equation}
\noindent
is the Stefan-Boltzmann constant counting gluons (spin, color) and (anti-)quarks
(spin, color and $u$, $d$, $s$-flavor) degrees of freedom.

Recently the grand canonical ensemble has been heavily criticized 
\cite{HThermostat:1, HThermostat:2}, when it is used for 
the exponential mass spectrum.  This critique, however, cannot be applied to the mass-volume 
spectrum (\ref{rhomv}) because it  grows less fast than the Hagedorn mass spectrum discussed 
in \cite{HThermostat:1, HThermostat:2}  and  because in the GBM  there is 
an additional suppression of  large and heavy bags due to the van der  Waals  repulsion. 
Therefore, the spectrum (\ref{rhomv})  can be  safely used in  the grand canonical ensemble.

It can be shown \cite{Goren:05}  that  the spectrum (\ref{rhomv})  generates  the
$s_Q (T) = \frac{\sigma_Q}{3} T^3 - \frac{B}{T}$ singularity, which reproduces the bag model  pressure 
$p (T) = T s_Q (T)$ \cite{BagModel} for high temperature phase, and $s_H (T)$ singularity,  which  gives   the pressure of the hadron gas model \cite{Hgas, Hgas:2} for low temperature phase. The transition between 
them can be of the first order or second order, depending on the model parameters.

However, for finite systems the volume of  bags  and their  masses should be finite. 
The simplest finite volume  modification of the GBM  is to introduce the volume dependent 
size of the largest bag $n = n (V)$  in  the partition  (\ref{qqq}).  As we discussed earlier such a modification cannot be handled by the traditional Laplace transform technique used in 
\cite{Goren:82,Goren:05}, but this modification can be easily
accounted for by the Laplace-Fourier method \cite{Bugaev:04a}.  Repeating all the steps 
of the CSMM analysis, one will obtain the  equations  (\ref{seven})-(\ref{ten}), in which the function 
${\cal F}(\xi,\tilde\lambda)$ should be replaced  by its GBM analog 
$f(\lambda,V_B) \equiv   F_H( \lambda) +  F_Q(\lambda, V_B)$ defined via 
%
\begin{equation}\label{Fgbm}
F_H( \lambda) \equiv \sum_{j=1}^{J_m}~ g_j \, \phi(T,m_j) ~ e^{-v_j s }\,, \quad  {\rm and} \quad
F_Q(\lambda, V_B) \equiv  V_0  \int\limits_{1}^{V_B/V_0}  \,dk  ~ a (T,V_0 k )~ e^{ V_0 (s_Q(T) - \lambda) k} \,.
\end{equation}
\noindent
In evaluating (\ref{Fgbm}) I used the mass-volume  spectrum (\ref{rhomv}) with  the maximal volume of the bag $V_B$ and changed integration to a dimensionless variable  $k = v / V_0$. Here the function 
$ a (T,v) = u (T) v^{2 + \gamma + \delta}$ is defined by  
$u(T) = C \pi^{-1} \sigma_Q^{\delta + 1/2}~ T^{4 + 4 \delta} ( \sigma_Q T^4 + B )^{3/4}$. 

The above representation  (\ref{Fgbm}) generates  equations for  the real and  imaginary parts of
$\lambda_n \equiv R_n + i I_n$, which are very similar to the corresponding expressions of the CSMM 
(\ref{eleven})  and (\ref{twelve}).  Comparing (\ref{Fgbm}) with (\ref{ten}), 
one sees that their main difference is that the sum over $k$ in (\ref{ten}) is replaced by the integral
over  $k$ in (\ref{Fgbm}).  Therefore, the equations (\ref{eleven})  and (\ref{twelve}) remain valid for
$R_n$ and $I_n$ of the GBM, respectively, if one replaces the $k$ sum by the integral for  
$K(V) = V_B/V_0$, $b = V_0$, $\nu (\lambda) =  V_0 ( s_Q (T) - \lambda)$ and 
$\tilde\phi_{k>1} (T) =  V_0  ~ a (T,V_0 k)  $. Thus, the results and  conclusions of
our analysis of the $R_n$ and $I_n$ properties of the CSMM should be  valid for the GBM as well.
In particular, for  large  values of $K(V) = V_B/ V_0$  and $R_n < s_Q(T) $ one can immediately 
find out  $I_n \approx \pi n  / V_B $ and the GBM formation/decay time  
$\tau_n = V_B  [ \pi n T V_0]^{-1}$. These equations show that the metastable $\lambda_{n>0}$ 
states can become stable in thermodynamic limit, if and only if  $V_B \sim V$.

The finite volume modification of the GBM equation of state should be used for the quantities which
have $V \lambda_0  \sim 1$.  This may  be important for  the early stage of   the relativistic nuclear collisons when the volume of the system  is small, or for the  systems that have small pressures. 
The latter can be the case for  the  pressure of strange or charm hadrons.

\section{ Hills and Dales Model and  Source of Surface Entropy}

During last forty years the FDM  \cite{Fisher:67}
has been successfully  used to analyze the condensation  of 
a gaseous phase (droplets or clusters  of all sizes)    
into a liquid.  
The systems analyzed with the FDM are many and varied, but
up to now the source of the surface entropy is not absolutely clear.
In his original work  Fisher postulated that 
the surface free-energy $F_A$ of a  cluster  of $A$-constituents  consists 
of surface  ($A^{2/3}$)  and  logarithmic ($\ln A$) parts, i.e. 
$F_A =  \sigma (T)~ A^{2/3} + \tau  T\ln A$.  Its surface part  
$ \sigma (T)~ A^{2/3} \equiv \sigma_{\rm o} [ 1~ - ~T/T_c] ~ A^{2/3}$  consists 
of the  surface energy, i.e. $ \sigma_{\rm o}  ~ A^{2/3}$,  and  
surface entropy $ -  \sigma_{\rm o} / T_c~ A^{2/3}$. 
From the study of the combinatorics of lattice gas clusters in two dimensions,
Fisher  postulated    the  specific 
temperature dependence of the surface tension $\sigma (T)|_{\rm FDM} $
which gives 
naturally an estimate   for the  critical temperature  $T_c$. Surprisingly  
Fisher's estimate works  for  the 3-d Ising model  \cite{mader-03},  
nucleation of real fluids \cite{Dillmann,Kiang}, percolation  clusters \cite{Percolation}
and  nuclear multifragmentation \cite{Moretto:97}.

\subsection{Grand Canonical Surface Partition.}
To understand why the surface entropy has such a form I formulated 
a statistical model of surface deformations of the cluster of $A$-constituents, the Hills and Dales Model (HDM)   \cite{Bugaev:04b}. 
For simplicity I consider
 cylindrical deformations of positive height $h_k>0$ (hills) 
and negative height $-h_k$ (dales), with  $k$-constituents at the base. 
It  is assumed that  
for the deformation of the  base of $k$-constituents 
the top (bottom) of the hill (dale) has the same shape as 
the surface of the original  cluster of $A$-constituents. 
I also  assume that:
(i) the statistical weight of deformations $\exp\left( - \sigma_{\rm o} |\Delta S_k|/s_1 /T  \right) $ 
is given  by the Boltzmann factor due to the  change of the surface $|\Delta S_k|$ in units of 
the surface per  constituent $s_1$;
(ii) all hills of heights $h_k \le H_k$ ($H_k$ is the maximal height of 
a hill with a base of $k$-constituents)
have the same probability $d h_k/ H_k$ besides the statistical one; 
(iii) assumptions (i) and (ii) are valid for the dales. 
Then the  HDM grand canonical surface partition (GCSP)
%
\begin{equation} \label{Oone}
Z_{gc}(S_A)= \hspace*{-0.10cm} \sum\limits_{\{n_k^\pm = 0 \}}^\infty \hspace*{-0.10cm} \left[ \prod_{k=1}^{ K_{max} }
\frac{ \left[ z_k^+ {\cal G}_{gc} \right]}{n^+_k!}^{n^+_k} \frac{ \left[ z_k^- {\cal G}_{gc} \right]}{n^-_k!}^{n^-_k}\right]
\Theta(s_1 {\cal G}_{gc})\, 
\end{equation}
corresponds to the conserved (on average) volume of the cluster because the probabilities 
of hill $z_k^+$  and dale   $z_k^-$
of the same $k$-constituent  base  are identical   \cite{Bugaev:04b}
%
\begin{equation}\label{Otwo}
\hspace*{-0.35cm}
z_k^{\pm} \equiv \hspace*{-0.15cm} \int\limits_0^{\pm H_k} \hspace*{-0.15cm} \frac{ d h_k}{ \pm H_k}\,
{\textstyle e^{ - \frac{\sigma_{\rm o} P_k |h_k| }{T s_1} } }
= \frac{T s_1  }{\sigma_{\rm o} P_k H_k }
\left[1 - {\textstyle e^{ - \frac{\sigma_{\rm o} P_k H_k}{ T s_1} } } \right] .
\end{equation}
Here $P_k$ is the perimeter of the cylinder base. 

The geometrical partition (degeneracy factor) 
of the HDM 
or the number  of ways to place
the center of a  given  deformation on the surface of the $A$-constituent cluster which  is occupied
by the set of $\{n_l^\pm  = 0, 1, 2,...\}$  deformations of the $l$-constituent base I assume to
be  given in the van der Waals approximation  \cite{Bugaev:04b}:
\begin{equation}\label{Othree}
{\cal G}_{gc} = 
{\textstyle \left[ S_A - \sum\limits_{k = 1}^{K_{max} } k\, (n_k^+ ~ + ~ n_k^-) \, s_1 \right] s_1^{-1} } \,, 
\end{equation}
where $s_1 k$ is the area  occupied by the deformation of $k$-constituent base ($k = 1, 2,...$), 
$ S_A$
is the  full surface of the cluster,
and $K_{max} (S_A) $ is the $A$-dependent size of the maximal allowed base on the cluster.

The $\Theta(s_1 {\cal G}_{gc})$-function in (\ref{Oone}) ensures that only configurations
with positive value of the free surface of cluster are taken into account, but makes 
the  analytical evaluation of the  GCSP (\ref{Oone}) very difficult.  However,  it is possible {\it  to solve
this GCSP  exactly}  for any surface dependence of $K_{max} (S_A) $  using the identity (\ref{nfour}) of  the Laplace-Fourier transform  technique \cite{Bugaev:04a} discussed earlier:
%
\begin{equation}\label{Ofour}
 Z_{gc} (S_A)~ = \sum_{\{ \lambda_n\}}
e^{\textstyle  \lambda_n\, S_A }
{\textstyle
\left[1 - \frac{\partial {\cal F}_{gc}(S_A,\lambda_n)}{\partial \lambda_n} \right]^{-1} } \,.
\end{equation}
The  poles  $\lambda_n$  of the isochoric partition  are defined by  
%
\begin{equation}\label{Ofive}
\lambda _n~ = ~{\cal F}_{gc} (S_A,\lambda _n) \equiv  \sum\limits_{k=1}^{ K_{max}(S_A) } 
\left[  \frac{ z_k^+}{s_1} + \frac{ z_k^-}{s_1} \right]
~e^{ - k\,s_1 \lambda_n }
\,,
\end{equation}
\noindent 
which can be cast
as a system of two coupled transcendental equations
\begin{eqnarray}\label{Peleven}
&&\hspace*{-0.2cm} R_n = ~  \sum\limits_{k=1}^{K_{max}(S_A) }  \left[ z_k^+ + z_k^- \right]
~{\textstyle e^{- k\,R_n} } \cos(I_n  k)\,,
\\
\label{Ptwelve}
&&\hspace*{-0.2cm} I_n = -  \sum\limits_{k=1}^{K_{max}(S_A) } \left[ z_k^+ + z_k^- \right]
~{\textstyle e^{-k\,R_n} } \sin(I_n  k)\,, 
\end{eqnarray}
for dimensionless variables $R_n = s_1 Re(\tilde\lambda_n)$ and $I_n = s_1 Im(\tilde\lambda_n)$. 

To this point  Eqs. (\ref{Peleven}) and (\ref{Ptwelve}) are  general and can be used for
particular models which specify the height of hills and depth of dales. 
But there exists an absolute supremum for the real root $(R_0; I_0 = 0)$ of these equations.  
It is sufficient to  consider the  limit $K_{max}(S_A) \rightarrow \infty$,
because for $I_n = I_0 = 0$ the right hand side (r.h.s.) of (\ref{eleven}) is a monotonously
increasing function of $K_{max}(S_A)$.
Since $z_k^+ = z_k^- $ are the monotonously decreasing functions of $H_k$, the maximal value of 
the r.h.s. of (\ref{eleven})  corresponds to the limit of infinitesimally small amplitudes 
of deformations, $H_k \rightarrow 0$.
Then for $I_n = I_0 = 0$  Eq. (\ref{Ptwelve})  becomes an identity and  
Eq. (\ref{Peleven})  becomes 
%
\begin{equation}\label{Pthirteen}
R_0 \rightarrow ~  2 \sum\limits_{k=1}^{\infty } e^{- \frac{ \sigma_{\rm o} P_k H_k}{2 T s_1} }  
~{\textstyle e^{ - k\,R_0} }  = 2 \left[ e^{ R_0} - 1 \right]^{-1} \,,
\end{equation}
and we have $R_0 = s_1 \tilde\lambda_0 \approx 1.06009$. 
Since for $I_n \neq 0$ defined by (\ref{Ptwelve}) the inequality $\cos(I_n k) \le 1$
cannot become the equality for all values of $k$ simultaneously, then
it follows that the real root of (\ref{Peleven}) obeys the inequality $R_0 > R_{n > 0}$.    
The last result means that in the limit of infinite cluster, $S_A \rightarrow \infty$, 
the GCSP is represented by the farthest right singularity among all simple poles $\{\tilde\lambda_n\}$ 
\begin{equation}\label{Pfourteen}
Z(S_A)\biggl|_{S_A \rightarrow \infty} ~ \approx
\frac{e^{\textstyle  \frac{R_0\, S_A}{s_1} } }{1 + \frac{R_0 ( R_0 + 2)}{2} } 
\approx 0.3814~ e^{\textstyle  \frac{R_0\, S_A}{s_1} }
%
%
%
%
\end{equation}
There are two remarkable facts regarding (\ref{Pfourteen}): first, this result is model independent because
in the limit of vanishing amplitude of deformations all model specific parameters vanish; 
second, in evaluating (\ref{Pfourteen}) we did not specify the shape of the cluster under consideration,
but only implicitly required that the cluster surface together with deformations is a regular surface
without  self-intersections.  
Therefore, for vanishing amplitude of deformations the latter means that Eq. (\ref{Pfourteen})
should be valid for any self-non-intersecting surfaces.  

For  spherical clusters the r.h.s. of (\ref{Pfourteen}) becomes familiar, 
$ 0.3814~ e^{\textstyle  1.06009\, A^{2/3}  } $, which, combined with the Boltzmann factor of
the surface energy $e^{\textstyle - \sigma_{\rm o} A^{2/3}/T  } $, generates
the { following temperature dependent surface tension} of the large cluster 
\begin{equation}\label{fifteen}
\sigma (T) = \sigma_{\rm o} \left[ 1 - 1.06009 \frac{T}{\sigma_{\rm o} } \right] 
\end{equation}
which means that the actual critical temperature of the three dimensional Fisher model should be
$T_c = \sigma_{\rm o}/ 1.06009$, i.e. 6.009 \% 
smaller in $\sigma_{\rm o}$ units 
than Fisher originally supposed.
This equation for the critical temperature 
 remains valid for the   temperature dependent $\sigma_{\rm o}$ as well.
 Our result, given in Eq.  (\ref{fifteen}), agrees with Fisher estimate of   $ \sigma(T)$.
 Agreement between our result and $\sigma(T)|_{SMM}$ occurs, if 
 $\sigma_{\rm o} = \sigma(T)|_{SMM} + 1.06009 ~T$.

Also  equation  (\ref{Pfourteen}) allows one, for the first time, 
to find the exact value of the degeneracy prefactor,
$ 0.3814$, which was unknown in the FDM and its extensions.

Now  it is appropriate to ask  which of two surface tension parameterizations, the FDM or SMM, is correct?
On the one hand, the FDM linear $T$-dependence works well and can be derived in the limit of vanishing amplitudes of deformations within HDM,
but on the other hand,  the SMM prescription is based on the hyperscaling  relation 
\cite{HypScaling:1, Fisher:69} for the surface tension of macroscopic fluids 
$\sigma(T)|_{SMM} \rightarrow \Gamma_{\rm o}  \left[ \frac{T_c - T}{T_c} \right]^{(d-1)\, \nu}$, which is nonlinear in $T$.  
Here $\nu$   is the critical exponent that also describes the divergence of the correlation length near the critical point and 
is related to other exponents through the hyperscaling relation \cite{HypScaling:1, Fisher:69}
\begin{equation}\label{Newsixteen}
d\, \nu ~ = ~  \gamma~+~2\, \beta \,. 
\end{equation}
A possible solution of this problem is  that  the hyperscaling relation describes a single macroscopic drop, whereas the FDM and HDM are dealing with the ensemble of clusters or drops of a given volume (in fact, of a given mass) and they account for the mean surface or the surface of mean cluster \cite{Fisher:69}.  The latter is measured  in dimensionless units (in HDM it is written explicitly as the ratio of two surfaces  $S_A/s_1$), whereas the hyperscaling is formulated for  the surface $S_A$ of the  $A$-constituent  drop. 
Therefore, in the relation between the surface tension coefficient used in hyperscaling  and the surface of a mean Fisher cluster 
the temperature dependent density of liquids is involved via $s_1$. 
The analysis performed for  the real fluids 
\cite{Elliott:05wci} shows that the temperature dependent density of liquids, indeed,  `compensates'  
the extra power in temperature and makes the surface tension coefficient linear in $T$.

Before proceeding further  let me consider the left equality in Eq. (\ref{thirteen}) which is valid 
for small deformation heights. It can be shown that for  $S_A \gg s_1$
the deformation energy
\vspace*{-0.4cm}

\begin{equation}\label{twtwo}
\frac{ \sigma_{\rm o} P_k H_k}{2  s_1}  \rightarrow - \frac{3}{2}~T~ k~ \frac{\tau}{\zeta}~ 
%
%
 \frac{s_1}{S_A} ~ \ln \left( \frac{s_1}{S_A} \right)
\end{equation}
\vspace*{-0.3cm}

\noindent
of a $k$-constituent base, 
indeed, generates the Fisher power law $A^{-\tau}$ for the GCSP 
(\ref{Oone}) of { an  $A$-constituent} cluster. 
Now one can see that
besides the coefficient $3 T \tau /(2 \zeta)$ 
[where $\zeta^{-1} = {\textstyle 1 + \frac{2}{R_0 (2 + R_0)} \approx (0.61861)^{-1} }$],  
 the term 
$- k \frac{s_1}{S_A} \ln \left( \frac{s_1}{S_A} \right) $
on the right hand side of (\ref{twtwo})
is the entropy  which gives an {\it a priori} uncertainty 
to measure the position of $k$ constituents each of area $s_1$ 
on the surface of the cluster. 
A comparison of (\ref{twtwo}) with any $k R_n  > 0$ in the left equality 
(\ref{thirteen})
 shows that in the limit $S_A \gg s_1$ the ansatz
(\ref{twtwo}) corresponds to a negligible correction compared to the exponentials $e^{\frac{R_n S_A}{s_1} }$.
Therefore, the Fisher power law is  { too  delicate} 
for the present formulation of the surface partition model. 


\subsection{Special Ensembles for Surface Partition.}
Similarly one can introduce the surface partitions for the other ensembles \cite{BugaevElliott}. 
The canonically constrained surface partition (CCSP) is built up to obey the volume conservation more strictly than it is done in the GCSP.  This  is ensemble of pairs of deformations: the number of the hills $n_k^+$ of
the $k$-constituent base is always identical to the number of corresponding dales, i.e.
$n_k^- \equiv n_k^+ \equiv n_k$. Then the canonical geometrical partition can be cast as 
\begin{equation}\label{Pfour}
{\cal G}_{c} =
{\textstyle \left[ S_{\rm A} - 2 \sum\limits_{k = 1}^{K_{\rm max} } k\, n_k \, s_1 \right] (2 s_1)^{-1} } \,,
\end{equation}
where the factor two in the denominator of the right hand side (r.h.s.) of (\ref{Pfour})
accounts for the fact that it is necessary to place simultaneously the centers of
two $k$-constituent base deformations (hill and dale) out of $2 n_k$ on the surface of cluster.
Using the geometrical partition (\ref{Pfour}), one can
 obtain  the partition function of  canonical ensemble by formally replacing
$ {\cal G}_{\rm gc}  \rightarrow {\cal G}_{c}$ and inserting the Kronecker symbol  $\delta_{n_k^+ \,, n_k^-}$ for
each $k$-multiplier  in (\ref{Oone}).
I, however,  consider  each pair of hills and dales of the same base as a single degree of freedom.
Therefore, the number of ways to place each pair
out of $n_k$ distinguishable pairs  is still given by the canonical geometrical partition
${\cal G}_{c}$. Multiplying it with the probability of a pair of deformations
$z_k^+\, z_k^-$ and  repeating this  for $n_k$ pairs, one obtains
the CCSP,
as follows
\begin{equation} \label{Pfive}
Z_{\rm cc}(S_{\rm A})= \hspace*{-0.10cm} \sum\limits_{\{n_k = 0 \}}^\infty \hspace*{-0.10cm} \left[ \prod_{k=1}^{ K_{\rm max} }
\frac{ \left[ z_k^+ z_k^-  {\cal G}_{c} \right]}{n_k!}^{n_k} \right]
\Theta(2 s_1 {\cal G}_{c})\,.
\end{equation}

Applying the Laplace-Fourier transform technique to this partition, one can find that 
the CCSP has the same form as the GCSP (\ref{Ofour}), but the function 
$ {\cal F}_{\rm gc}$  in  (\ref{Ofive}) must be replaced with $ {\cal F}_{\rm cc}$:
\begin{equation}\label{Pnine}
\lambda_n~ =~ {\cal F}_{\rm cc}(S_A,\lambda_n) = \sum\limits_{k=1}^{ K_{\rm max}(S_A) }
 \frac{ z_k^+\, z_k^-}{2\,s_1}
~e^{ - 2\, k\,s_1\lambda_n  }\,.
\end{equation}

For the limit of the vanishing amplitudes of deformations
it is possible to introduce one more ensemble for the surface deformations \cite{BugaevElliott}
which hereafter will be called as
the {\it semi-grand canonical surface partition} (SGCSP).
This ensemble occupies an intermediate position between the constrained  canonical and grand canonical
formulations. It corresponds to the case, when the hills and dales
of the same base are considered to be indistinguishable. For that  I would like to explore the
fact that according to (\ref{Otwo})  the statistical probabilities of hills and dales
of the same base are equal. Then for the infinitesimally small amplitudes of deformations
the volume conservation constraint is fulfilled trivially.
Below  this ensemble
will be used  for  the deformations of  vanishing amplitude  only, but it may be
used also
for finite amplitudes of deformations,
if the volume is not conserved.
Then  the SGCSP and its geometrical factor  read as
\begin{equation} \label{PPtwelve}
Z_{\rm sg}(S_{\rm A})= \hspace*{-0.10cm} \sum\limits_{\{n_k = 0 \}}^\infty
\hspace*{-0.10cm} \left[ \prod_{k=1}^{ K_{\rm max} }
\frac{ \left[ z_k^+   {\cal G}_{\rm sg} \right]}{n_k!}^{n_k} \right]
\Theta( s_1 {\cal G}_{\rm sg})\,, \quad {\cal G}_{\rm sg} =
{\textstyle \left[ S_{\rm A} -  \sum\limits_{k = 1}^{K_{\rm max} } k\, n_k \, s_1 \right]  s_1^{-1} } \,.
\end{equation}
Again, like in the case of the CCSP,  only  the equation for the simple poles of the isochoric 
partition should be modified
\begin{equation}\label{PPthirteen}
\lambda_n~ =~{\cal F}_{\rm sg}(S_A,\lambda_n) = \sum\limits_{k=1}^{ K_{\rm max}(S_A) }
\frac{ z_k^+}{s_1} ~e^{ - k\,s_1 \lambda_n  }\,.
\end{equation}

Eqs.~(\ref{Pnine}) and  (\ref{PPthirteen}) can be written for real and imaginary parts as follows
\begin{eqnarray}\label{nsixteen}
&&\hspace*{-0.2cm} R_n^\alpha = ~  \sum\limits_{k=1}^{K_{\rm max}(S_{\rm A}) } \phi^\alpha_k
~{\textstyle e^{- k\,R_n^\alpha} } \cos(I_n^\alpha  k)\,,
\\
\label{nseventeen}
&&\hspace*{-0.2cm} I_n^\alpha = -  \sum\limits_{k=1}^{K_{\rm max}(S_{\rm A}) } \phi^\alpha_k
~{\textstyle e^{-k\,R_n^\alpha} } \sin(I_n^\alpha  k)\,,
\end{eqnarray}
for dimensionless variables defined as $R_n^\alpha = s_1 Re(\tilde\lambda_n)$ and
$I_n^\alpha = s_1 Im(\tilde\lambda_n)$ for the GCSP and SGCSP,
and as $ R_n^{cc} = 2 s_1 Re(\tilde\lambda_n)$ and
$I_n^{cc} = 2 s_1 Im(\tilde\lambda_n)$ for the CCSP.
Here the function $\phi^\alpha_k$ is  given by the expression
\begin{eqnarray}\label{neighteen}
\phi^\alpha_k =
\left\{
\begin{tabular}{ll}
\vspace{0.1cm} $   z_k^+ + z_k^-  $\,, &  for $\alpha = gc$\,, \\
$  z_k^+ z_k^-  $\,, & for $\alpha = cc$\,, \\
$ z_k^+  $\,, & for $\alpha = sg$\,.
\end{tabular}
\right.
\end{eqnarray}

Since the equation for the simple poles of  
all three surface partitions  are reduced to the same system (\ref{nsixteen}), (\ref{nseventeen}),
it is clear all the qualitative properties of the solutions
discussed for the GCSP remain valid for the CCSP and SGCSP as well  \cite{BugaevElliott}. 
There is only a quantitative difference for the rightmost singularities $R_0^\alpha$.
Therefore, in the limit of the vanishing amplitudes of deformations for 
an infinite base of the largest deformation $K_{\rm max}(S_A\rightarrow \infty) \rightarrow \infty $ 
each of these surface partitions will reach  an {\it upper limit}  
defined by  the  corresponding value of the surface 
entropy coefficient  $\omega^{\alpha}_{\rm U}$, defined by the corresponding $R_0^\alpha$:
\begin{eqnarray}\label{Ptwenty}
\max\{ Z_\alpha (S_{\rm A}) \} ~ \rightarrow ~ g_\alpha\, e^{\textstyle  \omega^\alpha_{\rm U} \frac{S_{\rm A}}{s_1} } 
\quad {\rm with} \quad
\omega^{\alpha}_{\rm U} =
\left\{
\begin{tabular}{ll}
\vspace{0.1cm}  $ \omega^{\rm gc}_{\rm U} = \max\{R_0^{\rm gc}\} \approx  1.060090 $\,, &  \hspace*{-0.2cm} $\alpha = gc$\,, \\
$ \omega^{\rm cc}_{\rm U} =  \max\{R_0^{\rm cc}/2\} \approx   0.403233 $\,, &   \hspace*{-0.2cm} $\alpha = cc$\,,\\
$\omega^{\rm sg}_{\rm U} = \max\{R_0^{\rm sg}\} \approx  0.806466 $\,, &  \hspace*{-0.2cm} $\alpha = sg$\,,
\end{tabular}
\right.  \hspace*{-0.4cm}
\end{eqnarray}
where the  degeneracy factor $g_\alpha$ is defined as follows:
$g_{\rm gc} \approx 0.38139$ and $g_{\rm cc} = g_{\rm sg} \approx 0.407025$.
The rightmost singularities of these partitions are defined by the equation
\begin{equation}\label{nnineteen}
R_0^\alpha =  B^\alpha \sum\limits_{k=1}^{\infty }
~{\textstyle e^{ - k\,R_0^\alpha} }  = B^\alpha \left[ e^{ R_0^\alpha } - 1 \right]^{-1} \,,
\end{equation}
where $B^{\rm gc} = 2$ and $ B^{\rm cc} = B^{\rm sg} = 1$.



For  large, but finite clusters it is necessary to take into account not only the farthest
right singularity $\tilde\lambda_0 $ in the corresponding surface partition,
but all other roots
of Eqs. (\ref{nsixteen}) and (\ref{nseventeen})
which have
positive real part
$R_{n>0}^\alpha > 0$. In this case for each $R_{n>0}^\alpha$ there are two roots $\pm I_n^\alpha$
of (\ref{nseventeen}) because
each surface  partition function  is real by definition.
The roots of Eqs. (\ref{nsixteen}) and (\ref{nseventeen}) with largest real part are insensitive
to the large values of $K_{\rm max} (S_{\rm A})$, therefore, it is sufficient to  keep
$K_{\rm max} (S_{\rm A}) \rightarrow \infty$.
Then for limit of vanishing amplitude of deformations Eqs. (\ref{nsixteen}) and
(\ref{nseventeen}) can be, respectively, rewritten as
\begin{eqnarray}\label{Aone}
\frac{B^\alpha R_n^\alpha}{(R_n^\alpha)^2 + (I_n^\alpha)^2}~ &=&
~{\textstyle e^{R_n^\alpha} } \cos(I_n^\alpha ) - 1\,,   \\
\label{Atwo}
\frac{B^\alpha I_n^\alpha}{(R_n^\alpha)^2 + (I_n^\alpha)^2} ~ &=&~ -
~{\textstyle e^{R_n^\alpha} } \sin(I_n^\alpha )\,.   %
\end{eqnarray}
After some algebra the system of (\ref{Aone}) and (\ref{Atwo})
can be  reduced to a single equation for $R_n^\alpha$
\begin{equation}\label{Athree}
\cos\left( {\textstyle \left[
\frac{ B^\alpha (B^\alpha + 2 R_n^\alpha)}{e^{2 R_n^\alpha} - 1} - (R_n^\alpha)^2 \right]^{1/2} } \right) =
\cosh R_n^\alpha - \frac{B^\alpha}{B^\alpha + 2~ R_n^\alpha} \sinh R_n^\alpha\,, 
\end{equation}
and  the  quadrature
$I_n^\alpha = \sqrt{\frac{B^\alpha (B^\alpha + 2 R_n^\alpha)}{e^{2 R_n^\alpha}- 1} - (R_n^\alpha)^2 }$.
The analysis shows that besides the
opposite signs there are two branches of solutions, $I^{\alpha\,+}_{n}$ and $I^{\alpha\,-}_{n}$,
for the same $n  \ge 1$ value.
Expanding both sides of (\ref{Athree}) for $R_n^\alpha \ll 1$ and keeping the leading terms,  for $ n \ge 1$
one obtains 
\begin{eqnarray}
\label{ntwtwo}
|I^{\alpha \, \pm}_{n}| ~& \approx &~  2 \pi n  \pm \frac{B^\alpha}{2 \pi n}\,, \\
\label{ntwthree}
R_{n}^\alpha ~& \approx  &~ \frac{(B^\alpha)^2}{8 \pi^2 n^2}\,.
\end{eqnarray}


\begin{figure}[ht]

\centerline{
\hspace*{0.0cm}\epsfig{figure=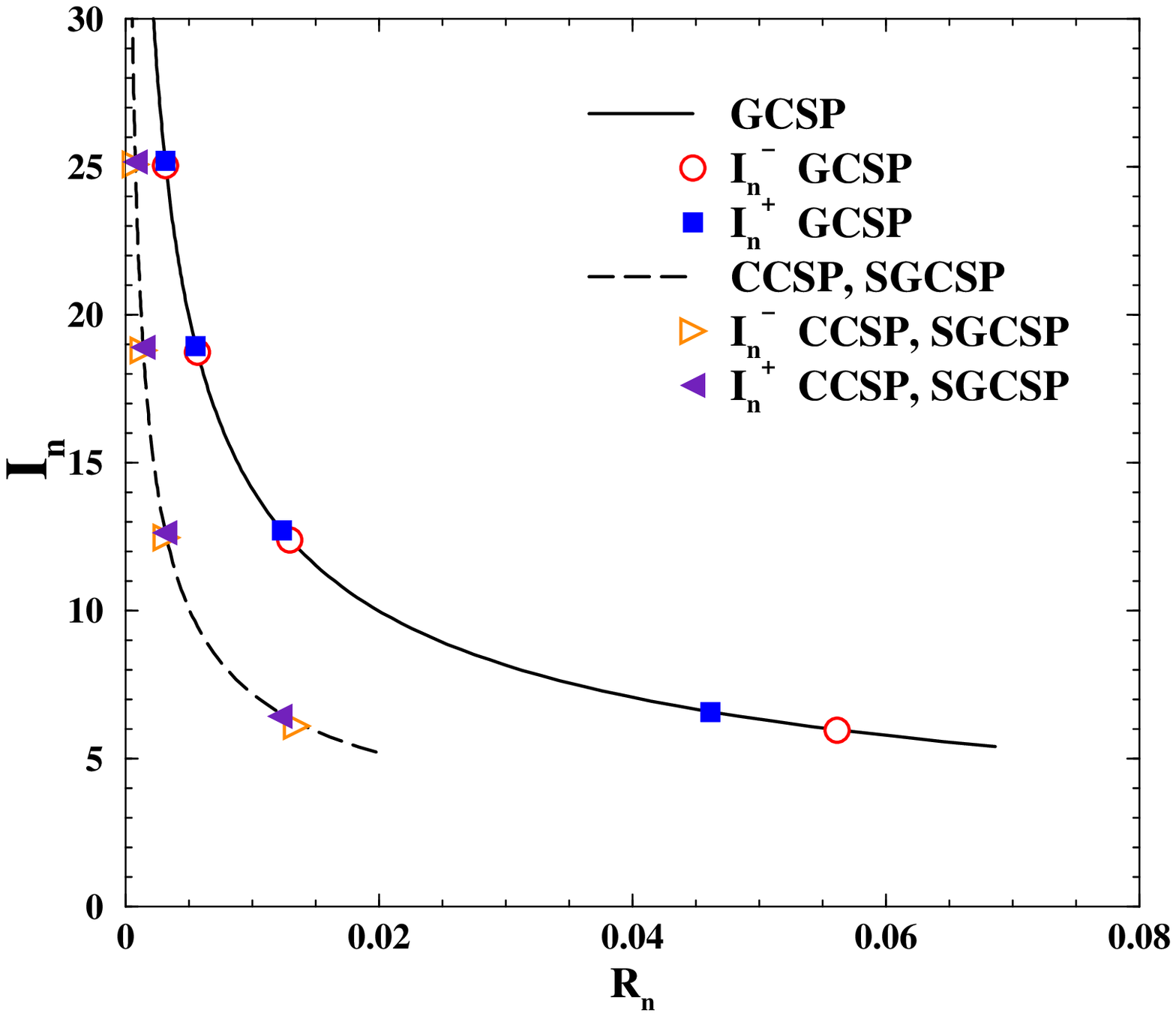,width=7.6cm}
\hspace*{0.5cm}\epsfig{figure=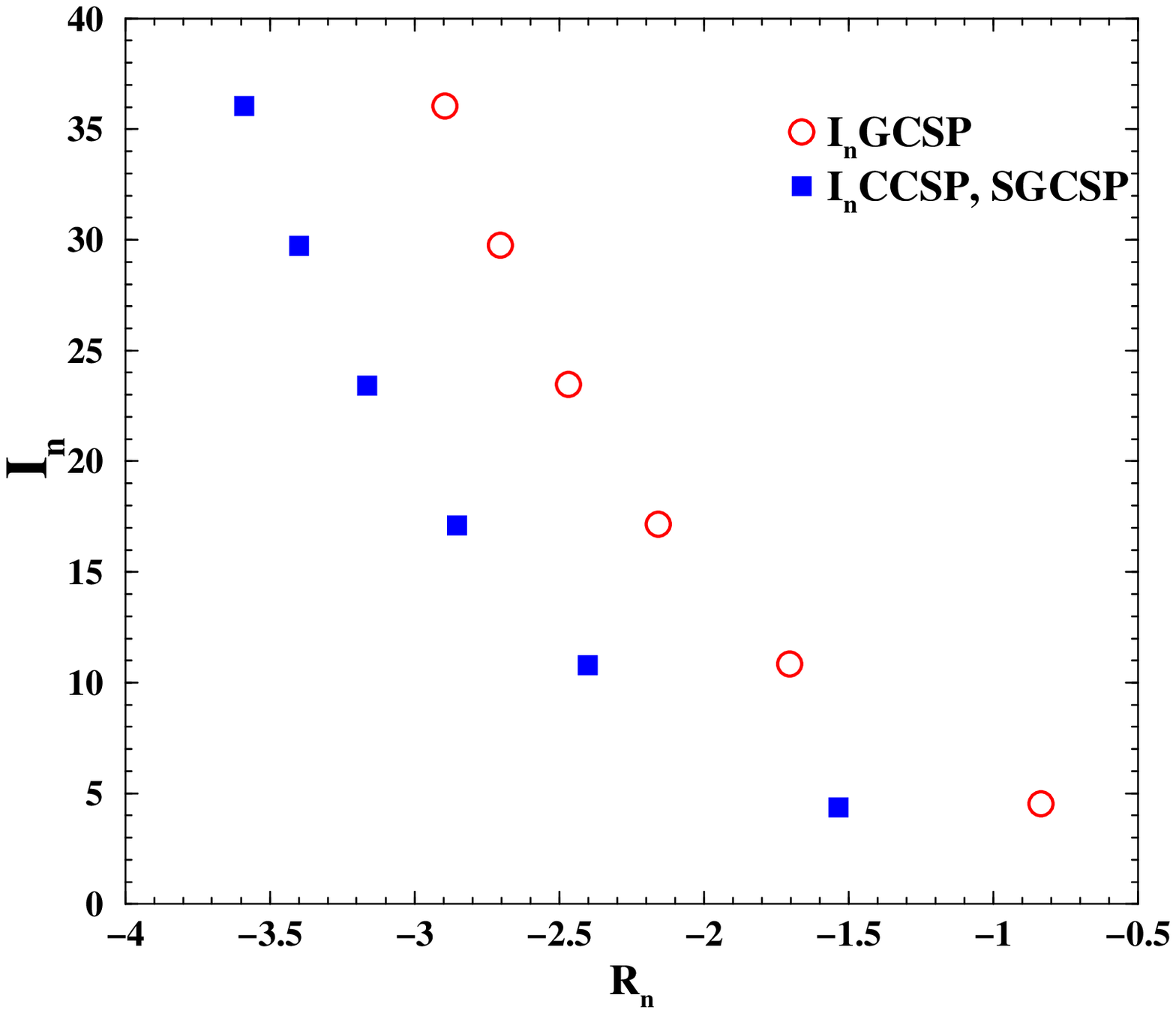,width=7.6cm}
}

\caption{
{\bf Left panel:}
The first quadrant of the complex plane 
$  (R_n^\alpha + i I_n^\alpha)  \equiv  s_1 \lambda_n^\alpha L^\alpha $ shows the 
location of simple poles of the isochoric partitions for $n = 1, 2, 3, 4$. 
The symbols   represent the two branches $I_n^-$ and $I_n^+$ of the roots
for the upper estimate of three surface partitions.
The curves are defined by the approximation suggested in \cite{BugaevElliott}.
\newline
{\bf Right panel:} The  second quadrant of the complex plane $s_1\tilde\lambda_n^\alpha \equiv R_n^\alpha + i I_n^\alpha$ shows the complex roots
of the system of  Eqs. (\ref{ntwfive}) and (\ref{ntwsix}) with the largest real parts.
The circles and squares  represent  the roots
for the lower estimate of the GCSP and CCSP(SGCSP), respectively.
}
\label{fig2.6}
\end{figure}

\noindent
The exact solutions $(R_n^\alpha ; I^{\alpha\, \pm}_n)$ for $ n \ge 1$
which have the largest real part  are shown in the left panel of  Fig.~\ref{fig2.6}
together with the curve  parametrized by functions $I_x^{\alpha\,+}$ and $R_x^\alpha$
taken  from Eqs. (\ref{ntwtwo}) and  (\ref{ntwthree}),
respectively.
From Eq. (\ref{ntwthree}) and the left panel of  Fig.~\ref{fig2.6} it is clear that for the GCSP
the  largest real part $R_1^{\rm gc} \approx 0.0582$ is
about 18 times smaller than $R_0^{\rm gc}$,
whereas for the CCSP and SGCSP the  real part $R_1^{\rm cc} = R_1^{\rm sg}$
of the first most right complex root of Eqs. (\ref{nsixteen}) and (\ref{nseventeen})
is about 63.6 times smaller than $R_0^{\rm cc} = R_0^{\rm sg}$.
Therefore,  for a cluster of a few constituents   the correction
to the leading term  in (\ref{Ptwenty}) is exponentially small for all considered partitions.
Using the approximations (\ref{ntwtwo}) and  (\ref{ntwthree}),
for $n > 2$ one can  estimate
the upper limit of
the $(R_n^\alpha ; I^{\alpha\, \pm}_n)$ root contribution into left hand side Eq. (\ref{Ptwenty})
\begin{equation}\label{ntwfour}
\hspace*{-0.0cm}\biggl| {\textstyle  e^{\tilde\lambda_n\, S_{\rm A}}
\left[1 - \frac{\partial {\cal F}_\alpha (S_{\rm A},\tilde\lambda_n)}{\partial \tilde\lambda_n}
\right]^{-1} }\biggr|  \le
e^{\textstyle  \frac{(B^\alpha)^2\,S_{\rm A}}{ 8 \pi^2 n^2 s_1} }\, /\, (2 \pi^2 n^2) \,.
\end{equation}
This result shows that
for all three considered partition
the total contribution of all complex poles into the corresponding surface partition
is negligibly small compared to the leading term (\ref{Ptwenty}) even 
for a cluster of a few constituents.


\subsection{The Lower Bounds for the Surface Entropy Coefficients.}
To  complete our analysis of the limit of vanishing  deformations
we would like to find the lower estimate for the GCSP, CCSP and SGCSP  for large clusters.
This estimate corresponds to the absence of all other deformations except for
those of smallest base. In other words, one has to substitute $K_{\rm max} (S_{\rm A}) = 1$
in all corresponding expressions.
Then equations (\ref{nsixteen}) and (\ref{nseventeen}), respectively
become
\begin{equation}\label{ntwfive}
R_n^\alpha = ~   \phi^\alpha_1
~{\textstyle e^{- R_n^\alpha} } \cos(I_n^\alpha )\,,
\end{equation}
\begin{equation}
\label{ntwsix}
I_n^\alpha = -  \phi^\alpha_1
~{\textstyle e^{- R_n^\alpha} } \sin(I_n^\alpha )\,.
\end{equation}
Similar to the previous consideration,
the leading term of the lower estimate for the surface partitions  is given by the
real root $(R_0^\alpha ; I^{\alpha}_0 = 0)$ of the system  (\ref{ntwfive}) and (\ref{ntwsix})
\begin{eqnarray}\label{ntwseven}
R_0^\alpha =
\left\{
\begin{tabular}{rl}
\vspace{0.1cm} $ \omega^{\rm gc}_{\rm L} = \min\{\omega^{\rm gc}\} \approx  0.852606 $\,, &\hspace*{-0.2cm}$\alpha = gc$, \\
$2 \omega^{\rm cc}_{\rm L} = 2 \min\{\omega^{\rm cc}\} \approx   0.567143 $\,, &\hspace*{-0.2cm}$\alpha = cc$, \\
$\omega^{\rm sg}_{\rm L} = \min\{\omega^{\rm sg}\} \approx  0.567143 $\,, &\hspace*{-0.2cm}$\alpha = sg$.
\end{tabular}
\right.
\end{eqnarray}

Again, as  in case of upper estimates one can show that the real root
$(R_0^\alpha ; I^{\alpha}_0 = 0)$  approximates well the lower estimate for the  partition function
for a system of a few constituents.
In fact, each of three surface partitions  has only a single root with positive real part which coincides with $(R_0^\alpha ; I^{\alpha}_0 = 0)$.  In the right panel of Fig.~\ref{fig2.6} a few  complex roots  of Eqs.  (\ref{ntwfive}) and (\ref{ntwsix}) with the largest real parts are
shown. Since all these roots have negative real part, they generate an exponentially small contribution
to the  lower estimate of surface partition for a system of a few constituents.

\begin{table}[h!]
\caption{\label{HDMTable} The  maximal and minimal values of the $\omega$-coefficient for
three  HDM  partitions.  
}
\begin{center}
\begin{tabular}{|c|c|c|}\hline
%
\smh Partition \smh &  \smh $\max\{\omega^\alpha\}$ \smh  & \smh  $\min\{\omega^\alpha\}$ \smh  \\
%
 %
\hline
\hline
GCSP   & 1.060090   & 0.852606 \\
\hline
SGCSP  & 0.806466 & 0.567143 \\
\hline
CCSP  & 0.403233  & 0.283572  \\
\hline
\end{tabular}
%

\end{center}

\end{table}

The $\omega$-coefficients  for  upper and lower estimates of  all three surface  partitions are summarized  in the Table~\ref{HDMTable}.
A comparison with the corresponding coefficient for liquids should be made with care because
various contributions to the surface tension, i.e., eigen surface tension
of the liquid  drop, the  geometrical degeneracy factor (surface partition), and the part  
induced by interaction between clusters, are not exactly known.  Therefore, even
the  linear  temperature dependence of the surface tension $\sigma(T) = \sigma_{\rm o} (T_{c} -  T)/T_{c}$ due to Fisher \cite{Fisher:67}
applied to a nuclear  liquid ($ \sigma_{\rm o} \approx 18$ MeV;  $T_{c} \approx  18$ MeV \cite{Bondorf:95}) may be used to estimate the $\omega$-coefficient, if  both
the eigen surface tension and the interaction induced one are non-increasing functions of 
temperature. Under these assumptions one can get  the following inequality for 
nuclear liquid
\begin{equation}\label{ntweight}
 \omega_{nucl}~ \le ~ 1  ~ <~  \omega^{gc}_U~ = ~1.060090\,, 
\end{equation}
i.e. the upper estimate for the GCSP, indeed, provides  the upper limit for surface partition
of nuclear matter.

A similar analysis for real liquids is  difficult  because of  complicated 
temperature dependence of the surface tension.
Therefore, we would like to compare the $\omega$-coefficients from Table~\ref{HDMTable} with 
the $\omega$-coefficients for  the large spin clusters of various 2- and 3-dimensional
Ising models, which are listed in the Tables~\ref{Ising2Table} and \ref{Ising3Table}, respectively \cite{Fisher:69}. 
Such a comparison can be made because  the surface entropy  of large spin clusters 
on the Ising lattices are similar to the considered surface  partitions  \cite{Fisher:69}.

The $\omega$-coefficient for the d-dimensional  Ising model is defined as 
the energy   $ 2 J$ required to flip a given spin interacting with its  $q$-neighbors  to opposite direction 
per  $(d-1)$-dimensional surface  divided by the value of critical temperature
\begin{equation}\label{ntwnine}
\omega_{Lat}  = \frac{q \,\,J}{T_{c} \,d}\,.
\end{equation}
Here $q$ is the coordination number for the lattice, and $J$  denotes the coupling constant of the model. 
A comparison of the Tables~\ref{HDMTable} - \ref{Ising3Table} shows that  all lattice $\omega_{Lat}$-coefficients, indeed,
lie between the  upper estimates for the constrained canonical and grand canonical surface partitions
\begin{equation}\label{nthirty}
0.403233 ~= ~\omega^{cc}_U  ~< ~ \omega_{Lat} ~ <~  \omega^{gc}_U~ = ~1.060090\,, 
\end{equation}
i.e. $\omega^{cc}_U$ and $\omega^{gc}_U$ are the infimum and supremum for 2- and 3-dimensional
Ising models, respectively.



\begin{table}[h!] 
\caption{\label{Ising2Table} The values of the $\omega_{Lat}$-coefficient for various
2-dimensional Ising models.
For more details see the text.
}
\begin{center}
\begin{tabular}{|c|c|}\hline
%
\smh Lattice type \smh &  \smh $ \omega_{Lat} = \frac{\sigma}{T_{c}} $ \smh   \\
%
 %
\hline
\hline
\smh Honeycomb  \smh &\smh  0.987718\smh \\
\hline
Kagome  & 0.933132  \\
\hline
Square  & 0.881374   \\
\hline
Triangular  & 0.823960  \\
\hline
Diamond  & 0.739640   \\
\hline
\end{tabular}
\\
%

\end{center}

\end{table}

\vspace*{0.3cm}

\begin{table}[h!] 
\caption{\label{Ising3Table}
The values of the $\omega_{Lat}$-coefficient for various
3-dimensional Ising models.
}
\begin{center}
\begin{tabular}{|c|c|}\hline
%
\smh Lattice type \smh &  \smh $ \omega_{Lat} = \frac{\sigma}{T_{c}} $ \smh   \\
%
 %
\hline
\hline
\smh Simple cubic  \smh &\smh  0.44342\smh \\
\hline
\smh Body-centered cubic \smh & 0.41989  \\
\hline
Face-centered cubic  & 0.40840   \\
\hline
\end{tabular}
\\
%

\end{center}

\end{table}

The HDM partitions do not have an explicit dependence on the dimension of the surface,
but a comparison of
the HDM and Ising model $\omega$-coefficients shows that the HDM ensembles 
seem to  posses   some sort of  internal 
dimension: the GCSP is close to honeycomb, kagome or square lattices, whereas the SGCSP
is similar to triangular and diamond lattices, and  the $\max\{\omega\}$ of the  CCSP is closer to the 3-dimensional
Ising models.  In some cases the agreement with the lattice data  is remarkable - $\omega^{gc}_L$
coincides with the arithmetical  average of 
the $  \omega$-coefficients for  square and triangular lattices up
to a fifth digit, but in most cases the values agree within a few per cent. 
The latter is not surprising because the HDM estimates  the surface entropy of 
a single cluster, whereas on the lattice the spin  clusters  do interact with each other and this, of course,  changes the surface tension and, consequently, affects the value of  critical temperature.
It is remarkable that so oversimplified estimates of the surface partitions for
a single large cluster 
reasonably  approximate the  $  \omega$-coefficients for 2- and 3-dimensional Ising models.
 
It would be interesting to check whether the lower estimate of the CCSP $\omega_L^{cc} \approx 0.283572$
is an infimum for the Ising lattices of  higher dimensions $d > 3$. 
If this is the case, then we can give an upper limit for the critical temperature of those lattices
using Eq. (\ref{ntwnine})
\begin{equation}\label{nthone}
\frac{T_{c}}{J}~ \le~ \frac{q}{\omega_L^{cc}\, d}~ \approx~  3.5264~ \frac{q}{d} \,. 
\end{equation}
On the other hand,
the lower estimate
for the critical temperature of  Ising lattices, $ \frac{T_{c}}{J}~ \ge~ \frac{q}{\omega_U^{gc}\, d}$,
is provided by 
the supremum of the $\omega$-coefficients of surface partitions.


\section{Conclusions}

An invention of a new powerful mathematical method \cite{Bugaev:04a}, 
the Laplace-Fourier transform,  is, perhaps,  a major theoretical breakthrough in the statistical mechanics
of finite systems of the  last decade  because it  
allowed us   to solve exactly  not only  the simplified SMM  for finite volumes 
\cite{Bugaev:04a}, but  also   a variety of statistical surface partitions  for  finite clusters  \cite{Bugaev:04b,BugaevElliott} 
 and to find out   their surface entropy   and to  shed light
 on a source of the Fisher exponent $\tau$.
It was shown \cite{Bugaev:04a}
that for finite volumes the analysis of the GCE partition  of the simplified SMM
is reduced to the analysis of the simple poles of the corresponding isobaric partition,  obtained 
as a Laplace-Fourier transform of the GCE partition.   Such a  representation of the GCE partition  allows one
not only to show from the first principles  that for finite systems there exist the complex 
values of the effective chemical potential, but 
to define  the finite volume analogs of phases straightforwardly.  Moreover,  this method 
allows one to include into consideration all complicated features of the interaction (including the Coulomb one) which have been
neglected in the simplified SMM because it  was originally  formulated for infinite nuclear matter. 
This can be done, for instance,  by dividing the full system into the set of  subsystems in which the gradient of a Coulomb field inside of each subsystem can be neglected, and considering each 
subsystem in a framework of the CSMM or finite volume formulation of the GSMM with electrostatic potential generated by the external subsystems.

Consequently,   the Laplace-Fourier transform  method opens  a principally new  possibility 
to study the nuclear liquid-gas phase transition directly from the partition of finite system 
without  taking its thermodynamic limit.  Now this method is also applied  \cite{Bugaev:07b} to the finite
volume formulation  of the  GBM  \cite{Goren:81} which is used to  describe the 
PT between the hadronic matter and  QGP. 
Thus,  the Laplace-Fourier transform method not only  gives an analytical  solution for a variety of 
statistical models  with PTs in finite volumes, but 
provides us with  a common framework   for several critical phenomena  in  strongly interacting matter. Therefore, it turns out that further applications and developments of this method are 
very promising and important  not only for the field of   nuclear multifragmentation, but 
for several fields  studying PTs in finite systems  because this method  may provide them with the firm theoretical foundations and a  common theoretical language.

At the moment it is unclear how to  experimentally  verify the existence  of the  complex values of the effective chemical potential and  measure the   decay/formation time  $\tau_n  \approx K(V)  [ \pi n T ]^{-1}$  predicted by the CSMM.  From the expression for  $\tau_n$ one concludes that the 
decay/formation time  decreases with the decrease of the size of  largest fragment in the system $ K(V)$ and with in growth of temperature.  Therefore, by varying these parameters, one can make 
the decay/formation time  much smaller than the equilibration time $\tau_{eq} \gg \tau_n$  for  all modes with $n \ge 1$. In this case only the  stable state with $\lambda_0$ will be described by the hydrodynamics, whereas all the metastable states will not be describe by the hydrodynamics. 
Thus, searching for a failure of hydrodynamic description in the A+A collisions may indicate the metastable collective modes in finite systems. In principle,  this can occur  for  both the nuclear multifragmetation and for the QGP-hadron gas PT.  Perhaps, the systematic study of the A+A reactions with the nucleus of many sizes and in a wide range of energies which will be performed at the  FAIR accelerator can locate the metastable collective modes qualitatively predicted by the CSMM.

Further on, with the help of the Laplace-Fourier transform method it was also possible to formulate and solve analytically three surface partitions within the HDM. 
It is remarkable that such a simple model of surface partition discussed above  gives the upper and lower bounds  of  the  $\omega$-coefficients  for all   2- and 3-dimensional Ising models.  Also the HDM allowed me to resolve the puzzle of  the success of Fisher parameterization of the linear temperature dependence of surface tension coefficient and exactly derive   an analytical  result which is just 6 \% larger than Fisher assumed. 

Also the canonically constrained surface partition, perhaps, is able to explain another long standing puzzle:
why the curvature part of surface tension of clusters is not seen. In principle, this could happen for large 
clusters at (tri)critical  point, where all other parts of free energy vanish, but this is not seen  in the FDM, 
SMM and  many systems described by the FDM.  Moreover, a careful analysis of  of 2- and 3-dimensional Ising clusters performed by a Complement method \cite{Complement} allows one 
not only to  extract  the critical temperature, surface tension coefficient and even the  value of Fisher index $\tau$ of the  infinite system,  but also such a delicate effects as 
the Gibbs-Thomson correction  \cite{krishnamachari-96}  to the free energy of a liquid 
drop. The Gibbs-Thomson corrections  are much weaker than the curvature term, but the latter is not seen. 

Therefore, let us imagine, that for each large cluster of regular shape  the hills and dales of the same base have the same probability and appear in pairs, like in the CCSP,  and that their shape is not a cylinder, but  some smooth shape, say a gaussian like.  Then, while summing up  the free energy associated with a curvature of each surface deformation, one finds that the total curvature  free energy of large cluster is exactly zero because 
the contribution of any  hill is exactly compensated by the curvature  free energy of a similar dale, which has an opposite sign, whereas the eigen  curvature   of  the undeformed  surface of  large  cluster  vanishes  due 
to   large radius of a cluster.


While  further intriguing facts can 
be found in the original work \cite{BugaevElliott}, here I mention only    that the surface tension plays  a very important role in many cluster models like the FDM,   SMM and CSMM, but its influence on the properties of the phase diagram of QGP was realized very recently.  This results I will present  in the chapter 4.  Another interesting problem of surface partition is to consider the fractal deformations within the HDM in order to elucidate the source  of   Fisher power law. 
It is believed that the fractals  may give us  a possible solution of the  Fisher's power law problem.


\chapter{Virial Expansion for the Lorentz Contracted Rigid  Spheres and  Relativistic   VdW EOS  for Hadronic Binary Mixtures.}


\hfill

\def\rmo{r_{\rm o}}
\def\rmot{r^3_{\rm o}}
\def\1#1{{\bf #1}}
\def\lp{\left(}
\def\rp{\right)}

Now I would like to turn from the nuclear matter EOS to that one of hadrons. 
In the chapter 1 the realistic mean-fieled EOS of nuclear matter \cite{Bugaev:96} was discussed in details.
The various extensions of the Walecka model \cite{Analyt:1} are very popular  
\cite{Analyt:2, BogutaEOS:1},
but their application is very limited both at very high and low energy densities. In the first case they are limited by  the deconfinement transition  to QGP, which will be discussed in the chapter 4, whereas in the second case there is a problem of their conversion to the gas of free hadrons. 
Also  this problem is related to the experimental setup of  A+A collisions, in which the 
observed  hadrons are free particles with their vacuum values of masses, charges and dispersion relation.  Therefore, the effective EOS can be applied to description of finite stage of the collision, if and only if they can be transformed into free streaming particles. This is impossible to do for the  Walecka model \cite{Analyt:1} and any its extensions because even a relatively  weak interaction between the nucleons at normal nuclear density is provided by the significant densities of the  mean values of scalar and vector fields. 
Since at the level of EOS  there is no well defined  procedure to convert these mean fields into physical particles, it is unclear what to do with them, at the end of the collision process.  Therefore, it was necessary to develop an alternative statistical  description of the hadronic matter EOS. 

Such  statistical models of hadron gas  \cite{Hgas, Hgas:2} are rather successful in description of
many experimentally observed hardonic  ratios at all available energies above 1 GeV per nucleon. 
On the one hand they are similar to the cluster models discussed in the chapters 1 and 2, where all hadrons and hadronic resonances with masses below 1.8-2 GeV are accounted  for as stable particles. 
On the other hand these statistical models  take into account for  the discrete part of the Hagedorn model  \cite{hagedorn-65, hagedorn-68} with the hard core repulsion. 

One might be very surprised by the fact that the mixture of hadrons and their resonances with the hard core repulsion is able to reproduce a vast amount of experimental data with only 3 or 4 fitting parameters. The explanation was found long ago \cite{Raju:92}, when the direct numerical analysis showed that from the statistical point of view the mixture of stable hadrons, interacting with each other in  all known repulsive and attractive channels,  accounted up to the quantum second  virial coefficient
do behave like the mixture of ideal gases of hadrons and their resonances with the shifted chemical potential due to hard core repulsion. 

This description \cite{Hgas, Hgas:2}, however,  generated the problem of causality \cite{Raju:92}.
In  statistical mechanics of relativistic particles 
such problem is known for awhile. In fact, since more than a century after the discovery of special relativity the statistical mechanics of extended relativistic objects does not exists!  
The problem is that the special relativity does not allow the existence of rigid bodies. If such particles 
would exists, then at their dense packing  the density perturbation would propagate superluminously,
i.e. with the speed higher that the speed of light, but the latter is forbidden by relativity. Thus, the causality problem appears at high densities  in any  hadronic or nuclear matter EOS, if hard core repulsion is used. Although in  the most of  applications for  the  nuclear matter EOS such a problem can be ignored because the studied  systems are nonrelativistic, this problem cannot be ignored for dense and hot hadronic matter.

There were  two unsuccessful tries to resolve the causality problem for the EOS with the relativistic hard core repulsion \cite{Kap:83, Z:95}, but in both of these works  the second virial coeffiecient was mixed up with the eigen volume. Moreover,  whereas in Ref.  \cite{Z:95} the Lorentz transformation was used to contract the eigen volume of hadron,  in  Ref. \cite{Kap:83} the very  same arguments were used to extend the eigen volume of particles. 
There was a discussion of possible temperature dependence of the parameters of the  Van der Waals  EOS \cite{Bugrii:78}, but the weak point of that discussion is the  absence of  a connection between the parameters  of EOS and the interparticle  potentials that could generate  them. 
Therefore, it was necessary to return to the  foundations of statistical mechanics and derive the cluster and virial expansions for the momentum dependent interparticle potentials \cite{Bugaev:RVDW1}. Then it was necessary  to formulate the  Van der Waals  EOS for relativistic particles \cite{Bugaev:RVDW1}.  The first EOS of this type has the temperature dependent second
virial coefficient and can be safely used to describe the experimental data for the  lightest hadrons (pions and kaons) at  temperatures below 200 MeV and  small baryonic densities \cite{Zeeb:02}.

However,  such an improved EOS  \cite{Bugaev:RVDW1} cannot be used for higher temperatures and baryonic densities because at this region it breaks the causality. Therefore, it was necessary to develop an alternative relativistic  Van der Waals extrapolation \cite{Bugaev:RVDW2} to high pressures which can be used 
to describe the hadrons and their resonances and even the effective particles like in the Walecka model \cite{Analyt:1} far  above the cross-over  where, as the lattice quantum chromodynamics  calculations show  \cite{Karsch:03},  they  coexist with quarks and gluons up to  large temperatures like $ 3 T_c$. 
Very recently it was understood that in the deconfined phase 
 there may exist bound states \cite{Shur:04} and resonances \cite{Rapp:05}.

The  relativistic effects on the hard core repulsion may be important for 
a variety of the effective models of hadrons  and hadronic matter
such as the  modified  
Walecka model \cite{Dirk:91}, various extensions of the 
Nambu--Jona-Lasinio model \cite{Na:61}, the quark-meson coupling model  \cite{Gu:X}, 
the chiral SU(3) model \cite{chirm} etc
In fact,  the relativistic hard core repulsion should be  important  for
any effective model in which the strongly interacting particles 
have reduced  masses compared to their vacuum values because with lighter masses 
the large portion of  particles  becomes  relativistic. 
Nevertheless, the relativistic hard core repulsion was, so far,  not incorporated into these models 
due to the absence of  the required formalism.

In addition, the studies  of the Lorentz contracted rigid spheres required  the investigation of the the Van der Waals approximation for  the  different  hard core radii  because the Lorentz contraction  affects  stronger the lighter particles.  If the excluded volumes of particles are different (binary mixture),  then the Van der Waals extrapolation is not unique at high densities and 
one has to carefully  study  its  properties.
In  \cite{Zeeb:02, Zeeb:02b} such an analysis was performed for the Lorentz-Berthelot  binary mixture  \cite{VdW:1889,Muen} and 
for the extrapolation suggested recently  \cite{Gor:99}. It was shown \cite{Zeeb:02, Zeeb:02b} that for  the temperatures below 200 MeV and  small baryonic densities both  extrapolations give the identical results for the experimentally observed  particle ratios. 
Later on  I  showed that  the alternative relativistic  Van der Waals extrapolation 
\cite{Bugaev:RVDW2} resolves the non-uniqueness of  Van der Waals extrapolation for the systems with the multicomponent  hard core repulsion, if  one requires the causal behavior at high pressures.

This chapter is based on the works \cite{Bugaev:RVDW1,  Zeeb:02, Bugaev:RVDW2}. 


\section{The Van der Waals EOS for the Lorentz Contracted Hard Spheres}

The Van der Waals (VdW) excluded volume model has been used
to describe hadron yields in relativistic nucleus--nucleus
collisions (see e.g. \cite{StockerGreiner,  Hgas, Hgas:2, Yen:97, Gor:98, Schide:78, CarstenHorst,rafelski,cleymans,braun-munzinger,becattini:97} and references therein). 
This model treats the hadrons as hard-core spheres and, therefore,
takes into account the hadron
repulsion at short distances. In a relativistic situation one should,
however,
include the Lorentz contraction of the hard core-hadrons.
This problem was discussed in the literature (see e.g. 
Ref.~\cite{Kap:83,Z:95}).
In this  section  the cluster and virial expansions are generalized to
 velocity dependent inter-particle potentials. This extension is
used to construct the VdW model for Lorentz contracted
rigid spheres which may be used to simulate
hadrons.
 
The canonical partition function for the gas of $N$ classical (Boltzmann)
particles takes the form 
\begin{equation}\label{cpf}
Z_N(V,T)~=~\frac{1}{N!}~\int\prod_{i=1}^{N}\left[\frac{g~d{\bf r}_i
d{\bf k}_i}{(2\pi)^3}~\exp\left(- \frac{\omega_i}{T}\right)\right]~
\exp\left(- \frac{U}{T}\right)~,
\end{equation}
where $V$ and $T$ are the system volume and temperature,
$g$ is the number of internal degrees of freedom 
(degeneracy factor) of the particles,
$\omega_i=(m^2+{\bf k}_i^2)^{1/2}$ is the dispersion
relation of free particles
with mass $m$. The particle interactions described by the function
$U$ in Eq.~(\ref{cpf}) are given by
the sum
over pair potentials:
\begin{equation}\label{U}
U~=~\sum_{1\le i <j \le N}~u_{ij}~.
\end{equation}
In contrast to the usual statistical mechanic treatment of
the pair potentials, the $u_{ij}$ are assumed to be both
coordinate and momentum dependent 
$u_{ij}\equiv u({\bf r}_{i},{\bf k}_i;~{\bf r}_j,{\bf k}_j)$.
This generalization is necessary, if Lorentz 
contraction effects of hard spheres are to be taken into account. 
Introducing the Mayer functions 
\begin{equation}\label{fij}
f_{ij}~=~
\left[\exp \left(- \frac{u_{ij}}{T}\right) ~-~1\right]~,
\end{equation}
Eq.~(\ref{cpf}) can be presented as
\begin{equation}\label{cpf1}
Z_N(V,T)~=~\frac{1}{N!}~\int d{\bf x}_1 ... d{\bf x}_N~
\exp\left(-\frac{\omega_1+...+\omega_N}{T}\right)~
\prod_{1\le i < j \le N} (1~+~f_{ij})~,
\end{equation}
with the short notation
$
d{\bf x}_i\equiv g d{\bf r}_i d{\bf k}_i/(2\pi)^3.
$
Similarly to the standard procedure one can introduce
the cluster integrals \cite{May:77} 
\begin{eqnarray}
b_1~&=&~\frac{1}{V}~\int d{\bf x}_1 ~ \exp\left(-\frac{\omega_1}{T}
\right)~=~\frac{g~T^3}{2\pi^2}~K_2\left(\frac{m}{T}\right)~
\equiv ~\phi(T)~,
\label{b1}\\
b_2~&=&~\frac{1}{2!V}~\int d{\bf x}_1 d{\bf x}_2~
\exp\left(-\frac{\omega_1 +\omega_2}{T}\right)~f_{12}~,
\label{b2}\\
b_3~&=&~\frac{1}{3!V}~\int d{\bf x}_1d{\bf x}_2d{\bf x}_3~
\exp\left(-\frac{\omega_1+\omega_2+\omega_3}{T}\right)~
(f_{12}f_{13}~ \label{b3}\\
&+&~f_{12}f_{23}~+~f_{13}f_{23}~+~
f_{12}f_{23}f_{13} 
)~,\nonumber \\
... & & \nonumber
\end{eqnarray}
and present the canonical partition function 
in the familiar form
\begin{equation}\label{cpf2}
Z_N(V,T)~=~\sum_{\{m_l\}} ^{~~~~\prime}~
\prod_{l=1}^N ~\frac{(Vb_l)^{m_l}}{m_l!}~,
\end{equation}
where the summation  in Eq.~(\ref{cpf2}) is taken
over all sets of non-negative integer numbers $\{m_l\}$ 
satisfying the condition
\begin{equation}\label{cond}
\sum_{l=1}^N lm_l~=~N~.
\end{equation}
Note, however, that the cluster integrals defined above are
different from those used in standard statistical mechanics \cite{May:77}
as here nontrivial momentum integrations are included.  
Condition (\ref{cond}) makes the calculation of $Z_N$ (\ref{cpf2})
rather complicated. This problem can be avoided in the grand
canonical ensemble:
the grand canonical partition function can be calculated
explicitly ($z\equiv \exp(\mu/T)$):
\begin{equation}\label{gcpf}
{\cal Z}(V,T,\mu)~\equiv ~\sum_{N=0}^{\infty}
\exp\left(\frac{\mu N}{T}\right)~
Z_N(V,T)~=~\exp\left(~V~\sum_{l=1}^{\infty}b_l z^l~\right)~.
\end{equation}
In the thermodynamical limit the pressure $p$ and particle number density
$n$
are calculated in the grand canonical ensemble
in terms of the asymptotic values of the cluster integrals:
\begin{eqnarray}
p~&=&~T~\lim_{V \rightarrow \infty}~\frac{\ln {\cal
Z}}{V}~=~T~\sum_{l=1}^{\infty}b_l z^l~,
\label{pgce}\\
n~&=&~\lim_{V \rightarrow \infty}~\frac{1}{V}~
\frac{\partial \ln {\cal Z}}{\partial z}~=~
\sum_{l=1}^{\infty}l b_l z^l~. \label{ngce}
\end{eqnarray}
The virial expansion represents the pressure in terms of a series
of particle number density and takes the
form
\begin{equation}\label{virial}
p~=~T~\sum_{l=1}^{\infty}a_ln^l~.
\end{equation}
Substituting $p$ (\ref{pgce}) and $n$ (\ref{ngce})
into Eq.~(\ref{virial}) and equating the coefficients
of each power of $z$, one finds the virial coefficients $a_l$ in terms of
the cluster integrals 
\begin{equation}\label{al}
a_1~=~1~,~~~a_2~=~- \frac{b_2}{b_1^2}~,~~~a_3~=~\frac{4b_2^2}{b_1^4}~-
~\frac{2b_3}{b_1^3}~,~~~...
\end{equation}

\vspace{0.2cm}
Let me  recall, first, the derivation of the standard 
VdW excluded volume model.
Then it is extended by adding the Lorentz contraction of the moving
particles. Keeping the first two terms of the virial expansion
(\ref{virial}) the following result is obtained:
\begin{equation}\label{second}
p(T,n)~=~Tn~(1~+~a_2n)~.
\end{equation}
It is valid for small particle densities (i.e. $n \ll 1/a_2$).
The usual (momentum independent) hard core potential 
for spherical particles with radius $\rmo$ is
$u_{ij}=u(|{\bf r}_i - {\bf r}_j|)$. Here the function $u(r)$
equals to 0 for $r>2 \rmo $ and $\infty$ for $r<2 \rmo$.
The second cluster integral (\ref{b2}) can easily be calculated
in this case:
\begin{equation}\label{b2h}
b_2~=~-~\phi^2(T)~\frac{16\pi}{3}\rmot~.
\end{equation}
Therefore $a_2=4\,v_{\rm o}$, where $v_{\rm o}=4\pi \rmot/3$
is the particle hard core volume.
The VdW excluded volume model is obtained as 
the extrapolation of Eq.~(\ref{second}) to large particle
densities in the form
\begin{equation}\label{vdw}
p(T,n)~=~\frac{Tn}{1~-~a_2n}~.
\end{equation}
For practical use the pressure is given
as a function of $T$ and $\mu$ independent
variables, i.e. in the grand canonical ensemble.
This is done by substituting $n=(\partial p/\partial \mu)_T$
into Eq.~(\ref{vdw}), which then turns into a partial differential
equation for the function $p(T,\mu)$.  For the VdW
model
(\ref{vdw}) the solution of this partial differential equation
can be presented in the form of a transcendental equation
\begin{equation}\label{vdwgc}
p(T,\mu)~=~T\phi(T)e^{\mu/T}\exp\left(-\frac{a_2p}{T}\right)~\equiv
~p_{id}(T,\mu-a_2p)~.
\end{equation}
Eq.~(\ref{vdwgc}) was first obtained in Ref. \cite{Dirk:91}
using the Laplas transform technique.
With $p(T,\mu)$ (the solution of Eq.~(\ref{vdwgc}))
the particle number density, entropy density and energy density
are calculated as ($\nu =\mu-a_2\,p(T,\mu),\,\, a_2=4\,v_{\rm o}$):
\begin{eqnarray}
n(T,\mu)~&\equiv &~\left(\frac{\partial p(T,\mu)}{\partial
\mu}\right)_{T}~=
~\frac{n_{id}(T,\nu)}{1~+~a_2n_{id}(T,\nu)}~,\label{ntmu}\\
s(T,\mu)~&\equiv &~\left(\frac{\partial p(T,\mu)}{\partial
T}\right)_{\mu}~=
~\frac{s_{id}(T,\nu)}{1~+~a_2n_{id}(T,\nu)}~,\label{stmu}\\
\epsilon(T,\mu)~&\equiv &~Ts~-~p~+~\mu n~=~
\frac{\epsilon_{id}(T,\nu)}{1~+~a_2n_{id}(T,\nu)}~.\label{etmu}
\end{eqnarray}
Here the superscripts $id$ in the thermodynamical functions
(\ref{vdwgc}--\ref{etmu}) indicate those of the ideal gas.

The excluded volume effect
accounts for the blocked volume 
of two spheres when they touch each other.
If hard-sphere particles move with relativistic velocities
it is necessary to include their Lorentz contraction
in the rest frame of the fluid.
The model suggested in Ref. \cite{Z:95} 
is not satisfactory: the parameter 
$a_2=4\,v_{\rm o}$ of the VdW excluded volume model
is confused there
with the proper volume of an individual particle --
the contraction effect is introduced for 
the proper volume of each particle. 
In order to get the correct result it is necessary to 
account for the excluded volume of
two Lorentz contracted spheres.

Let ${\bf r}_i$ and ${\bf r}_j$ 
be the coordinates of the $i$-th and $j$-th particle, respectively, 
and ${\bf k}_i$ and ${\bf k}_j$ be their momenta,
${\bf {\hat r}}_{ij}$ denotes 
the unit vector
$ {\bf {\hat r}}_{ij} = {\bf r}_{ij}/|{\bf r}_{ij}|$
($ {\bf r}_{ij}= |{\bf r}_i -  {\bf r}_j|$).
Then for a given  
set of vectors $\left( {\1 {\hat r}}_{ij} , \1 k_i, \1 k_j \right)$
for the Lorentz contracted rigid spheres of radius $\rmo$
there exists the minimum distance between their centers 
$r_{ij} ({\bf {\hat r}}_{ij}; {\bf k}_i, {\bf k}_j) = {\rm min}|\1 r_{ij}|$.
The dependence of the  potentials 
$u_{ij}$ on the coordinates ${\bf r}_i,{\bf r}_j$
and momenta ${\bf k}_i, {\bf k}_j$ 
can be given in terms of the minimal distance
as follows
\begin{equation}
u({\bf r}_{i},{\bf k}_i; {\bf r}_j,{\bf k}_j)  =
\left\{ \begin{array}{rr}
0\,,  &\hspace*{0.3cm}|\1 r_i - \1 r_j| >   r_{ij} \lp
{\1 {\hat r}}_{ij}; \1 k_i, \1 k_j
 \rp  \,, \\
 & \\
\infty\,,  &\hspace*{0.3cm}|\1 r_i - \1 r_j| \le  r_{ij} \lp
{\1 {\hat r}}_{ij}; \1 k_i, \1 k_j
 \rp  \,.
\end{array} \right. 
\end{equation}

\begin{figure}[ht]
\centerline{\epsfig{figure=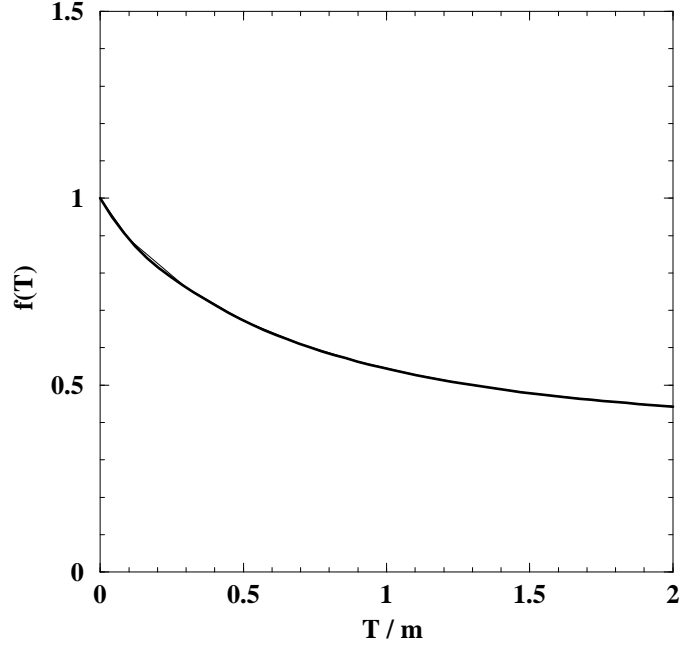,width=7.7cm}
}

\caption{\label{figfT} $f(T)$ 
as the function of the temperature-to-mass ratio. 
For heavy particles (e.g., nucleons $m \gg T$) the volume reduction is just a few per cents,
whereas for pions ($m \approx T$) it is about 50\%. 
}
\end{figure}

The general approach to the cluster- and virial expansions described above
is valid for this momentum dependent potential, and
it leads to Eqs.~(\ref{vdw},\ref{vdwgc}) with
\begin{eqnarray} \label{a2rel}
a_2(T)~&=&~\frac{1}{2\phi^2(T)}~\int \frac{d{\bf k}_1d{\bf k}_2~}
{(2\pi)^6}~
\exp\left(-\frac{\omega_1~+~\omega_2}{T}\right) \int d{\bf r}_{12}~\Theta\left(r_{12}
({\bf {\hat r}}_{12}; {\bf k}_1, {\bf k}_2)~-
~|{\bf r}_{12}|\right)\,. \quad
\end{eqnarray}
The new feature is the temperature dependence of the excluded
volume $a_2$ (\ref{a2rel}) which is due to the Lorentz contraction
of the rigid spheres. The pressure and
particle number density are still given by Eqs.~(\ref{vdwgc},\ref{ntmu}), 
but with temperature dependent $a_2(T)$ (\ref{a2rel}). However,
Eqs.~(\ref{stmu},
\ref{etmu}) are now modified, e.g.
\begin{equation}\label{etmurel}
\epsilon(T,\mu)~=~\frac{\epsilon_{id}(T,\nu)
~-~p^2~da_2(T)/d T}
{1~+~a_2n_{id}(T,\nu)}~.
\end{equation}
In contrast to Eq.~(\ref{etmu}) the energy density (\ref{etmurel})
contains the extra term which appears also in the entropy density.
The excluded volume $a_2(T)$ (\ref{a2rel}) is always smaller than
$4\,v_{\rm o}$. It has been proven rigorously that $a_2(T)$
is a monotonously decreasing function of $T$
and, therefore, the additional term in Eq.~(\ref{etmurel}) is always
positive. 
Let us introduce the notation
\begin{equation}\label{f}
a_2(T)~=~4\,v_{\rm o}~f(T)~.
\end{equation}
The function $f(T)$ depends on the $T/m$ ratio. It can be calculated
numerically and its behavior
is shown in Fig.~\ref{figfT}.
The simple analytical formula
\begin{equation}\label{f1}
f(T)~=~c~+~(1-c)~\frac{\rho_s(T)}{\phi(T)}
\end{equation}
with
$$ 
c~=~\left(1~+~\frac{74}{9\pi}\right)^{-1}~,~~~~
\rho_s~=~\frac{g}{(2\pi)^3}\int d{\bf k}~\frac{m}{\omega}~
exp\left(- \frac{\omega}{T}\right)~,
$$
is found to be valid with an accuracy of
a few percents for all temperatures.
The asymptotic behavior of $f(T)$ is the following:
 $1-{\rm O}(T/m)$ at
$T \ll m$
and $c+{\rm O}(m/T)$ at $T \gg m$.

If one assumes 
that all types
of hadrons 
have at rest the same hard core radius then 
the Lorentz contraction effect leads to different
VdW excluded volumes for moving particles with different masses:
for light particles (e.g. pions) the excluded volume
(at given $T$)  is smaller than
that for heavy ones. 
Fig.~\ref{figfT} shows that at $T\cong 150$~MeV the value
of $a_2$ in the nucleon gas ($m\cong 939$~MeV) decreases by 10\% in
comparison to its nonrelativistic
value $4\,v_{\rm o}$, whereas for pions ($m\cong 140$ MeV) 
$a_2$ shrinks at the same $T$ by almost a factor 2.
This is simply because light particles are more relativistic
than heavy ones at given temperature typical for high
energy nuclear collisions,  in which $T=120\div 170$~MeV.

\vspace*{2.5cm}


\begin{figure}[ht]

\centerline{
\hspace*{0.0cm}\epsfig{figure=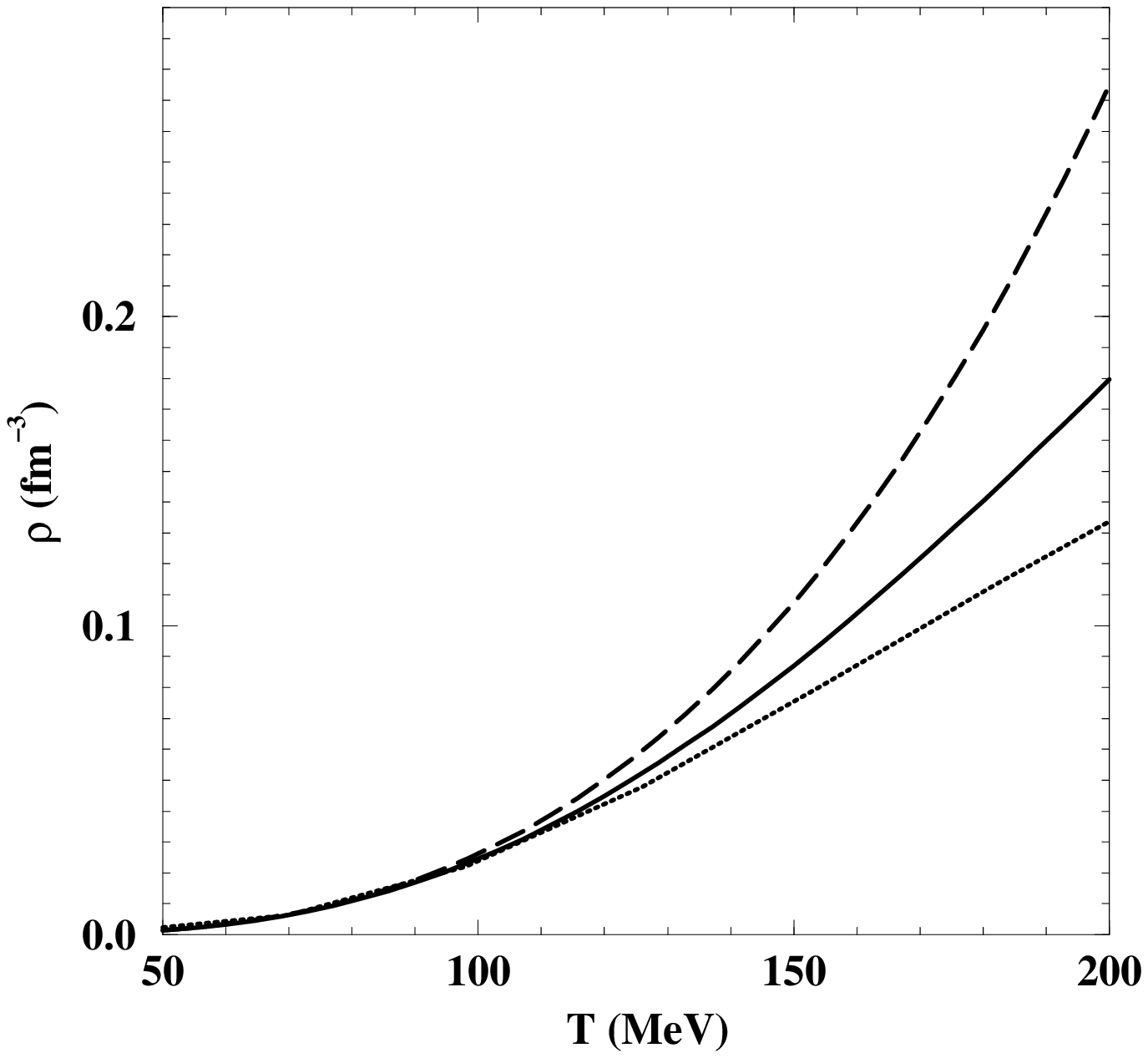,width=6.3cm}
\hspace*{1.5cm}\epsfig{figure=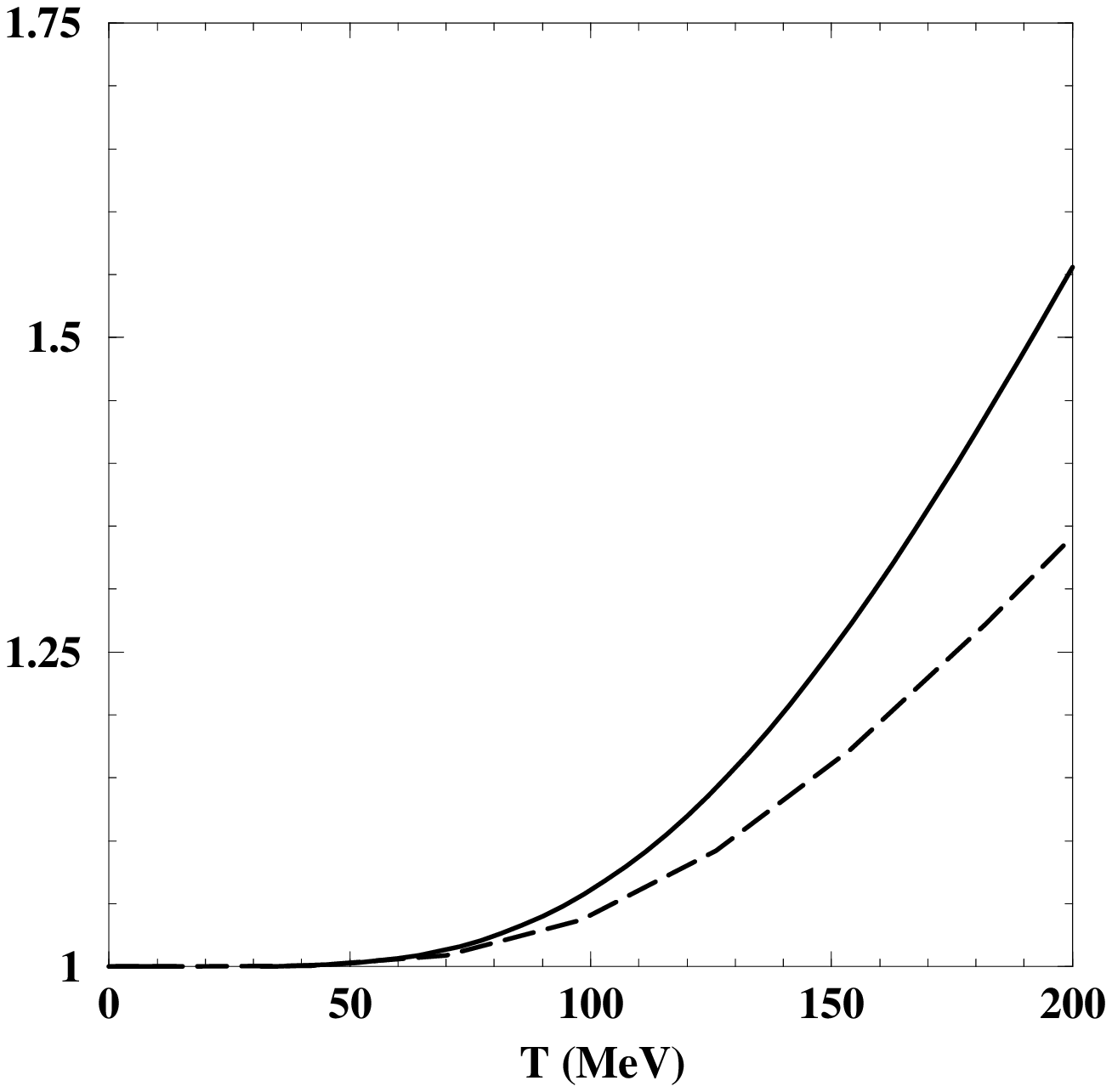,width=6.3cm}
}

\vspace*{0.3cm}

\caption{ \label{RelEOS}
{\bf Left panel:} The  particle number  density
for three models of the pion gas ($\mu =0, g=3$): the
solid line corresponds to the VdW model of the Lorentz 
contracted spheres ($r_{\rm o}=0.5$~fm),
the dashed one  corresponds to the ideal gas of
point-like particles,
and the dotted one corresponds to
the VdW model without Lorentz contraction for the spheres of a constant
radius 0.5~fm.
\newline
{\bf Right panel:}  The dashed line shows the ratio of the particle number densities
of the pion gas
($g=3, \mu=0, r_{\rm o}=0.5$~fm): the VdW model with Lorentz
contraction divided by  the VdW model without Lorentz contraction.
The solid line shows a similar ratio for the energy densities.
}
\end{figure}

As an example, Fig.~\ref{RelEOS} shows  
the EOS 
of the pion gas ($\mu=0, g=3$) with $r_{\rm o}=0.5$~fm.
The particle number density, shown in the left panel of Fig.~\ref{RelEOS},  is calculated according to 
Eq.~(\ref{ntmu})
for three different models: the ideal pion gas ($a_2=0$),
the VdW model with constant excluded volume ($a_2=4v_{\rm o}$)
and the VdW model with
Lorentz contraction ($a_2(T)$ is given by Eq.~(\ref{a2rel})).
It can be seen from the left panel of Fig.~\ref{RelEOS}
that at low $T$ the pion density is small
and excluded volume corrections are unimportant. Therefore, 
all three models are similar. The situation
changes with increasing $T$: the suppression due to
the excluded volume effects are large and different
for $a_2=4v_{\rm o}$ and $a_2(T)$ (\ref{a2rel}).
The ratios of particle number densities and energy densities
of the pion gas
for two versions of the 
VdW model ($a_2=4v_{\rm o}$ and $a_2(T)$ (\ref{a2rel}))
are shown in the right panel of  Fig.~\ref{RelEOS} as functions of the temperature. From 
this panel  one can observe the deviations between these two
models. These deviations increase with temperature. They are
larger for the energy density due to the additional positive term
in Eq.~(\ref{etmurel}). 


\vspace*{2.5cm}

Thus, the traditional cluster and virial expansions
can be consistently generalized to momentum dependent pair potentials.
Hard-core potentials with Lorentz contraction effects
lead to a VdW model
with a temperature dependent excluded volume $a_2(T)$ (\ref{a2rel}).
For light particles the effect of Lorentz contraction
is, evidently, stronger than for heavy ones. Note that smaller
values of the pion hard core radius $r_{\pi}$ were introduced in
Refs.~\cite{Rit:97, Yen:97} within the standard VdW excluded volume model
to fit hadron yield data better. 
The smaller value of the pion excluded volume appears
as a consequence of stronger Lorentz contraction
for light particles.


\renewcommand{\l}{\left} 
\renewcommand{\r}{\right}

\newcommand{\N}{{\cal N}}

\renewcommand{\dd}{{\rm d}}
\newcommand{\pd}{\partial}
%

\section{Hadronic Binary Mixtures in Canonical Ensemble}

As I showed in preceding section the Lorentz contraction affects 
lighter particles stronger.  Therefore, to 
study the experimental particle yields it is necessary to consider 
the VdW description for the particles with different hard core radii.
Since the VdW approximation is an extension of the low density 
expansion  to high densities, the procedure is not unique as long as
there is  no constraints on the high density behavior. 

The qualitative picture for the  gas with the  multicomponent  hard core repulsion  is as follows:
as particle species with smaller hard core radii are closer to the
ideal case, their particle densities are suppressed less.
Consequently, their yield ratios to particle species with larger
hard core radii are enhanced.
This fact has been used in recent efforts \cite{Yen:97} to explain the
experimentally observed pion abundance for AGS and SPS data \cite{PrQM:96}
by introducing a smaller hard core radius for pions $R_\pi$ than for
all other hadrons $R_{\rm o}$\,.
However, the resulting values are quite large, $R_{\rm o}=0.8$ fm and
$R_\pi=0.62$ fm\,.
At the same time in Ref.~\cite{Hgas:2} a reasonable fit of SPS data has been obtained, but 
for a distinctly smaller pair of hard core radii.

The excluded volume models used in \cite{Yen:97,Hgas:2}\,, 
however, are not correct in the case of different hard core radii.
As will be shown below, these models correspond to a system where
the components are separated from each other by a mobile wall and hence
cannot mix.

A more realistic approach requires a two-component partition function
including a term for the repulsion between particles of different hard core
radii.
In the case of two components, however, the { VdW\/} approximation is not
uniquely defined.
The simplest possibility yields the { Lorentz-Berthelot\/} mixture,
which was originally postulated by { Van der Waals\/} for binary mixtures,
see Refs.~\cite{VdW:1889,Muen}\,.
Another { VdW\/} approximation was recently proposed in Ref.~\cite{Gor:99}\,.
These two formulations contain a suppression of particle densities
similar to the one-component VdW gas, which is
{\em reduced to different extend\/} for each formulation.
In this section  I will study and apply both of these formulations.

There is yet another cause for a reduced excluded volume suppression.
Particles are considered to be rigid spheres in the usual  { VdW\/} model.
As was shown above, at high energies as achieved in nuclear collisions, however, relativistic
effects cannot be neglected \cite{Bugaev:RVDW1}\,.
Within the logic of the { VdW\/} model it is  absolutely necessary to take into
account the { Lorentz\/} contraction of the spheres.
Below I  will use an approach developed in Ref.~\cite{Bugaev:RVDW2} providing
approximative formulae for relativistic  excluded volumes:
naturally, they decrease with rising temperature, and the effect is
stronger for lighter particles.
At high temperatures, consequently, it is not possible to use a
{\em one-component\/} { VdW\/} description (i.\,e.~a {\em common\/}
excluded volume for {\em all\/} particle species) for a system of
species with various masses.
Since different masses cause different reductions of the excluded volumes at
a given temperature, a {\em multi-component\/} { VdW\/} description is required.

To illustrate the influence of different excluded volumes in what follows  I will
restrict myself  to the simplest 'multi-component' case,
the two-component case.
The crucial extension from the one- to the two-component case is to include
the repulsion between particles of two {\em different\/} hard core radii.
As it will be illustrated, a generalization to the multi-component case
is straightforward and will yield no essential differences \cite{Zeeb:02b}\,.
%

In this   section a derivation of the one-component canonical
partition function with (constant) excluded volumes is presented.
The generalization to the two-component case is made and two possible
{ VdW\/} approximations are analysed:
the { Lorentz-Berthelot\/} mixture \cite{Muen}
and the recently proposed approximation of Ref.~\cite{Gor:99}\,.
%



First it is necessary to   derive the canonical partition function (CPF) for the
one-component { VdW\/} gas by estimating the excluded volumes of
particle clusters.
Then this procedure will be generalised to the two-component case.

Here I use the   { Boltzmann\/} statistics to avoid the complications due to quantum statistics. 
The deviations from quantum statistics are negligible as long as the density
over temperature ratio is small.
This is the case for the hadron gas at temperatures and densities typical
for heavy ion collisions, see e.\,g.~Ref.~\cite{Yen:97}\,. 

Note that here  the term { VdW\/} is used  for the
{ van der Waals\/} excluded volume model, not for the general
{ van der Waals\/} model which includes attraction.

\subsection{The Van der Waals Excluded Volume Model.}
Let me  consider $N$ identical particles with temperature $T$ kept in
a sufficiently large volume $V$\,, so that finite volume effects can be
neglected.
The partition function of this system 
reads
\begin{eqnarray}
\label{eq:Z-CE}
  Z(T,V,N) &=& \frac{\phi^N}{N!}
               \int_{V^N} \dd^3 x_{1}
                              ~\cdots~ \dd^3 x_N
                              \,\exp \l[{\textstyle -\frac{U_N}{T} }\r] ~.
\end{eqnarray}
Here, $\phi \equiv \phi(T;m,g)$ denotes the momentum integral of the
one particle partition
\begin{equation}
  \phi(T;m,g) = \frac{g}{2 \pi^2} \int\limits_0^{\infty} \dd k \, k^2
                ~\exp\l[{\textstyle - \frac{E(k)}{T} }\r] ~,
\end{equation}
where $E(k) \equiv \sqrt{k^2 + m^2}$ is the relativistic energy and
$g=(2S+1)(2I+1)$ counts the spin and isospin degeneracy.
For a hard core potential $U_N$ of $N$ spherical particles with radii $R$
the potential term in Eq.~(\ref{eq:Z-CE}) reads
\begin{equation}
  \label{eq:pot1c}
  \exp\l[{\textstyle -\frac{U_N}{T} }\r]
    = {\textstyle \prod\limits_{i<j\le N} } \theta(\l|\vec{x}_{ij}\r| - 2R)~,
\end{equation} 
where $\vec{x}_{ij}$ denotes the relative position vector connecting the
centers of the $i$-th and $j$-th particle.
Hence one can write
\begin{eqnarray}
  \label{eq:Intgrls}
&  \int_{V^N} \dd^3 x_1 ~\cdots~ \dd^3 x_N
           \,\exp\l[{\textstyle -\frac{U_N}{T} }\r]
              ~  = ~
    \int_{V^N} \dd^3 x_1 ~\cdots~ \dd^3 x_N
    {\textstyle \prod\limits_{1 \le i<j \le N} }
         \!\!\!\! \theta (\l| \vec x_{ij} \r| - 2R)  \nonumber \\
  & =
    \int_V \dd^3 x_1 ~
    \int_V \dd^3 x_2 ~ \theta (\l| \vec x_{12} \r| - 2R) 
  \cdots~ \int_V \dd^3 x_N
    {\textstyle \prod\limits_{1 \le i \le N-1} }
         \!\! \theta (\l| \vec x_{i, N} \r| - 2R)  \nonumber \\
  & \equiv
    \int_V \dd^3 x_1 \int\limits^{\{ \vec x_1 \}}{\dd^3 x_2}
    ~\cdots \int\limits^{\{\vec x_1 \dots \vec x_{N-1}\}} \!\dd^3 x_N ~.
\end{eqnarray}
Here, $\int\limits^{\{\vec x_1 \dots \vec x_{j}\}} \!\dd^3 x_{j+1}$
denotes the available volume for $\vec x_{j+1}$, which is the center of the
particle with number $j+1$\,, if the $j$ other particles are configurated
as $\{\vec x_1 \dots \vec x_{j}\}$\,.
I will show now that this volume is estimated by
$\int\limits^{\{\vec x_1 \dots \vec x_j\}} \!\dd^3 x_{j+1} \ge (V - 2b\,j)$\,,
where $2b \equiv \frac{4 \pi}{3} (2R)^3$ is the excluded volume of an isolated
particle seen by a second one.
Then, $2b\,j$ estimates the total volume which is excluded by all particle
clusters occuring in the configuration $\{\vec x_1 \dots \vec x_j\}$\,.

It is sufficient to prove that the excluded volume of a cluster of $k$
particles is less than the excluded volume of $k$ isolated particles.
A group of $k$ particles forms a $k$-cluster, if for any of these particles
there is a neighboring particle of this group at a distance less than $4R$\,.
The {\em exact\/} excluded volume of a $k$-cluster, $v_{(k)}$\,, obviously
depends on the configuration of the $k$ particles.
If one considers two isolated particles, i.\,e.~two $1$-clusters, and reduces
their distance below $4R$\,, their excluded volumes will overlap.
They form now a $2$-cluster with the excluded volume
$v_{(2)}=4b-1 v_{\rm ov}$\,,
where $v_{\rm ov}$ denotes the overlap volume.

Evidently, one can construct any $k$-cluster by attaching additional
particles and calculate its excluded volume by subtracting each occuring 
overlap volume from $2b\,k$\,.
It follows that $v_{(k)} < 2b\,k$ is valid for any $k$-cluster, and this
inequality leads to the above estimate.
Obviously, its accuracy improves with the diluteness of the gas.

Using these considerations one can approximate the r.\,h.\,s.~of
Eq.~(\ref{eq:Intgrls})\,:
starting with $j+1=N$ one gradually replaces all integrals
$\int\limits^{\{ \vec x_1 \dots \vec x_{j} \} } {\!\dd^3 x_{j+1} }$
 by $(V - 2b\,j)$\,.
One has to proceed from the right to the left, because only the respective
rightmost of these integrals can be estimated in the described way.
Hence one finds
\begin{eqnarray}
  \label{eq:Zprox}
  Z(T,V,N)
    & \ge & \frac{\phi^N}{N!}\,
            ~ {\textstyle \prod\limits_{j=0}^{N-1} (V - 2b\,j) }~.
\end{eqnarray}

In this treatment the { VdW\/} approximation consists of two assumptions
concerning Eq.~(\ref{eq:Zprox})\,.
Firstly, the product can be approximated by
\begin{eqnarray}
  {\textstyle \prod\limits_{j=0}^{N-1} \l( 1 - \frac{2b}{V}\,j \r)
    ~  \cong~ \exp\l[ {\textstyle - \sum_{j=0}^{N-1} \frac{2b}{V}\,j} \r]}  
      \exp \l[ {\textstyle - \frac{b}{V}\,(N-1)N }\r]
     ~ \cong~  \l( {\textstyle 1 - \frac{b}{V}\,N }\r)^N ~,
   \end{eqnarray}
where $\exp\,[-x] \cong (1-x)$ is used for dilute systems,
i.\,e.\ for low densities $2bN/V \ll 1$\,.
The second assumption is to take the equality
instead of the inequality in Eq.~(\ref{eq:Zprox})\,.
Then the CPF takes the { VdW\/} form,
\begin{equation}
  \label{eq:Z_VdW}
  Z_{\rm VdW}(T,V,N) = \frac{\phi^N}{N!} \,\l( V - bN \r)^N ~.
\end{equation}
As usual, the { VdW\/} CPF is obtained as an approximation for dilute
systems, but when used for high densities it should be considered as
an {\em extrapolation\/}.

Finally, one obtains the well-known { VdW\/} pressure formula from the
thermodynamical identity $p(T,V,N) \equiv T\,\pd \ln[Z(T,V,N)]/\pd V$\,,
\begin{equation}
  \label{eq:p_VdW}
  p_{\rm\,VdW}(T,V,N) = \frac{T \, N}{V - b N} ~,
\end{equation}
using  the logarithm of the { Stirling\/} formula. 

Now let me brief\/ly investigate a system of volume $V$
containing two components with {\em different\/} hard core radii $R_1$ and
$R_2$ which are separated by a wall and occupy the volume fractions $x V$
and $(1-x) V$\,, respectively.
According to Eq.~(\ref{eq:p_VdW}) their pressures read
\begin{eqnarray}
  \label{eq:p-sp_x}
  p_{\rm\,VdW}(T,xV,N_1)   &=& \frac{T \, N_1}{x V - N_1 b_{11}} ~,  \\
  \label{eq:p-sp_1-x}
  p_{\rm\,VdW}(T,(1-x)V,N_2) &=& \frac{T \, N_2}{(1-x) V - N_2 b_{22}} ~,
\end{eqnarray}
where the particle numbers $N_1, N_2$ and the excluded volumes
$b_{11}=\frac{16 \pi}{3}\,R_1^{\,3}\,,\ b_{22}=\frac{16 \pi}{3}\,R_2^{\,3}$
correspond to the components 1 and 2\,, respectively.

If the separating wall is mobile, the pressures (\ref{eq:p-sp_x})
and (\ref{eq:p-sp_1-x}) must be equal.
In this case the fraction $x$ can be eliminated and
one obtains the common pressure of the whole system
\begin{eqnarray}  \label{eq:p-sp_CE}
   p_{\rm\,VdW}(T,xV,N_1) ~= ~ p_{\rm\,VdW}(T,(1-x)V,N_2) 
  ~= ~p^{\rm\,sp}(T,V,N_1,N_2)
      \equiv \frac{T\,(N_1+N_2)}{V - N_1 b_{11} - N_2 b_{22}} ~.
\end{eqnarray}
Since the components are separated in this model system
it will be referred to as the {\em separated\/} model \cite{Zeeb:02b}\,.

The pressure formula (\ref{eq:p-sp_CE}) corresponds to the
{ Boltzmann\/} approximation of the commonly used two-component
{ VdW\/} models of Refs.\ \cite{Yen:97,Hgas:2}.  
It is evident that $p^{\rm\,sp}$ (\ref{eq:p-sp_CE}) does not
describe the general two-component situation without a separating wall.
Therefore, it is necessary to find a more realistic model, i.\,e.~an
approximation from a {\em two-component\/} partition function.
This will be done in the following.

\subsection{Generalization to the Two-component Case.}
Recall the simple estimate (\ref{eq:Intgrls}--\ref{eq:Z_VdW})\,,
which gives a physically transparent derivation of the one-component CPF
in the { VdW\/} approximation.
Let us use it now for a {\em two-component\/} gas of spherical particles
with radii $R_1$ and $R_2$\,, respectively.
It is important to mention that each component may consist of several
particle species as long as these species have one common hard core radius,
i.\,e.~the number of necessary { VdW\/} components is determined by
the number of different excluded volume terms $b_{qq}$\,.
In the case of two radii the potential term (\ref{eq:pot1c}) becomes 
\begin{eqnarray}
  \label{eq:pot2c}
  \exp \l[ {\textstyle - \frac{U_{N_1+N_2}}{T} } \r]
  = {\textstyle \prod\limits_{i < j \le N_1} }
           \!\!\theta (\l| \vec x_{ij} \r| - 2 R_1) \times    {\textstyle \prod\limits_{k < \ell \le N_2} }
             \!\!\theta (\l| \vec x_{k \ell} \r| - 2 R_2)
  \times {\textstyle \prod\limits_{\scriptstyle m \le N_1 \hfill \atop
                     \scriptstyle ~\, n \le N_2  \hfill} }
             \!\!\theta (\l| \vec x_{mn} \r| - (R_1+R_2)) ~. 
   \end{eqnarray}
The integration is carried out in the way described above:
e.\,g.~firstly over the coordinates of the particles of the second component,
then over those of the first component.
For the estimation of the excluded volume of a $k$-cluster now
{\em two different\/} particle sizes have to be considered.
One obtains
\begin{eqnarray}
  Z(T,V,N_1,N_2)   \quad & \ge &
    \frac{\phi_1^{\,N_1}}{N_1!} \,\frac{\phi_2^{\,N_2}}{N_2!}
    \,\l\{{\textstyle \prod\limits_{i=0}^{N_1-1} }\!\l( V - 2b_{11}\,i \r) \r\}
 \l\{{\textstyle \prod\limits_{j=0}^{N_2-1} }
    \! \l( V - 2b_{12}\,N_1 - 2b_{22}\,j \r) \r\}  \nonumber \\
  \label{eq:twoc_CPF}
  &\quad \cong&
    \frac{\phi_1^{\,N_1}}{N_1!}
    \,\frac{\phi_2^{\,N_2}}{N_2!} ~ V^{N_1+N_2} 
    ~ \exp\l[{\textstyle  -\frac{N_1^{\,2} b_{11} + 2 N_1 N_2 b_{12}
                                         + N_2^{\,2} b_{22}}{V} }\r] ~,
\end{eqnarray}
where it is $\phi_q \equiv \phi(T;m_q,g_q)$\,, and
$2b_{pq} \equiv \frac{4 \pi}{3}\,(R_p+R_q)^3$ denotes the excluded volume
of a particle of the component $p$ seen by a particle of the component $q$
($p,~q=1, 2$ hereafter)\,.
Approximating the above exponent by $\exp[-x] \cong (1-x)$ 
yields the {\em simplest\/} possibility of a { VdW\/} approximation
for the {\em two-component\/} CPF\,,
\begin{eqnarray}
  \label{eq:twoc_CPF_nl}
  Z^{\rm\,nl}_{\rm VdW}(T,V,N_1,N_2) 
    &\quad \equiv&
    \frac{\phi_1^{\,N_1}}{N_1!}\,\frac{\phi_2^{\,N_2}}{N_2!} 
     \l(V - \frac{N_1^{\,2}b_{11} +2N_1 N_2 b_{12} +N_2^{\,2}b_{22}}
                         {N_1+N_2}\r)^{\!N_1+N_2}   \\
  \label{eq:twoc_CPF_nl-D}
  &=&
    \frac{\phi_1^{\,N_1}}{N_1!} \,\frac{\phi_2^{\,N_2}}{N_2!}
  \l(V - N_1 b_{11} - N_2 b_{22}
                 + \frac{N_1\,N_2}{N_1+N_2}\,D\r)^{\!N_1+N_2} ~,
\end{eqnarray}
where the non-negative coefficient $D$ is given by
\begin{equation}
  \label{eq:Def_D}
  D \equiv b_{11} + b_{22} - 2\,b_{12} ~.
\end{equation}
This approximation will be called the {\em non-linear\/} approximation
as the volume correction in (\ref{eq:twoc_CPF_nl-D}) contains non-linear
terms in $N_1, N_2$\,.
The corresponding pressure follows from the thermodynamical identity,
\begin{eqnarray}
  \label{eq:p-nl}
p^{\rm\,nl}(T,V,N_1,N_2)
           ~=~ p^{\rm\,nl}_1 + p^{\rm\,nl}_2  
  & \equiv&
    \frac{T\,(N_1+N_2)}{V - N_1 b_{11} - N_2 b_{22}
                        + \frac{N_1\,N_2}{N_1+N_2}\,D}\, ~. 
 \end{eqnarray}
This canonical formula corresponds to the { Lorentz-Berthelot\/} mixture
(without attraction terms) known from the theory of fluids \cite{Muen}\,.
It was postulated by { van der Waals\/} \cite{VdW:1889} and studied
as well by { Lorentz\/} and
{ Berthelot\/} \cite{Muen}\,.

The crucial step from the one- to the two-component gas is to include $b_{pq}$
terms ($p \neq q$) additionally to the $b_{qq}\equiv b|_{R=R_q}$ terms.
For the multi-component gas no further essential extension is necessary.
Consequently, the generalization of the above procedure to the
multi-component case, i.\,e.~an arbitrary number of different
hard core radii, is straightforward \cite{Zeeb:02b}\,.

In Ref.~\cite{Gor:99} a {\em more involved\/} approximation has been
suggested for the two-component { VdW\/} gas.
This follows from splitting the exponent in the CPF (\ref{eq:twoc_CPF}) by
introducing {\em two generalised\/} excluded volume terms $\tilde{b}_{12}$ and
$\tilde{b}_{21}$ (instead of a {\em single\/} and symmetric term $2\,b_{12}$)
for the mixed case,
\begin{eqnarray}
  Z(T,V,N_1,N_2) 
  & \cong&
    \frac{\phi_1^{\,N_1}}{N_1!} \,\frac{\phi_2^{\,N_2}}{N_2!}
    ~V^{N_1+N_2}
   \exp\l[{\textstyle -\frac{N_1^{\,2} b_{11}
                      + N_1 N_2 \l( \tilde{b}_{12} + \tilde{b}_{21} \r)
                      + N_2^{\,2} b_{22}}{V} }\r] ~,
\end{eqnarray}
which leads to an alternative two-component { VdW\/} CPF\,,
\begin{eqnarray}
\label{eq:twoc_CPF_lin}
{Z_{\rm VdW}^{\rm\,lin}(T,V,N_1,N_2)} 
  & \equiv&
    \frac{\phi_1^{\,N_1}}{N_1!}
    ~ \l( V - N_1 b_{11} - N_2 \tilde{b}_{21} \r)^{N_1} \frac{\phi_2^{\,N_2}}{N_2!}
    ~ \l( V - N_2 b_{22} - N_1 \tilde{b}_{12} \r)^{N_2} ~. 
\end{eqnarray}
Since the particle numbers $N_1, N_2$ appear solely linearly in the volume
corrections, these formulae will be referred to as the {\em linear\/}
approximation.
In this approximation one obtains \cite{Gor:99} for the pressure
\begin{eqnarray}
  \label{eq:p-lin}
  {p^{\rm\,lin}(T,V,N_1,N_2)
           ~=~ p_1^{\rm\,lin} + p_2^{\rm\,lin}}  
  &\equiv&
    \frac{T \, N_1}{V - N_1 b_{11} - N_2 \tilde{b}_{21}}
    + \frac{T \, N_2}{V - N_2 b_{22} - N_1 \tilde{b}_{12}} \,~. 
\end{eqnarray}

The choice of the generalised excluded volume terms $\tilde{b}_{pq}$ is
not unique in the sense that all choices which satisfy the basic
constraint $\tilde{b}_{12} + \tilde{b}_{21} = 2 b_{12}$ are consistent
with the second order virial expansion \cite{Gor:99}\,.
Therefore, additional conditions are necessary to fix these generalised
excluded volumes.
In Ref.~\cite{Gor:99} they were chosen as
\begin{eqnarray}
  \label{eq:bTi-s}
  \tilde{b}_{12} \equiv b_{11} \,\frac{2\,b_{12}}{b_{11}+b_{22}} ~,
  &\quad&
  \tilde{b}_{21} \equiv b_{22} \,\frac{2\,b_{12}}{b_{11}+b_{22}} ~.
\end{eqnarray}
For this choice, the linear approximation reproduces a traditional
{ VdW\/} gas behavior, i.\,e.~{\em one-component-like\/}, in the two
limits $R_2=R_1$ and $R_2=0$ as readily checked.
The factor $2 b_{12}/(b_{11}+b_{22})=1-D/(b_{11}+b_{22})$ is always smaller
than unity for $R_1 \neq R_2$\,, consequently, the $\tilde b_{pq}$ terms
are smaller than the corresponding terms $b_{pp}$\,.
Note that there are many possible choices for $\tilde b_{12}$
and $\tilde b_{21}$\,, e.\,g.~additionally dependent on the particle numbers
$N_1$ and $N_2$\,, whereas the non-linear approximation (\ref{eq:twoc_CPF_nl})
contains no such additional parameters.

The formulae of the linear approximation are generally valid for {\em any\/}
choice of $\tilde{b}_{12}$ and $\tilde{b}_{21}$ satisfying the constraint
$\tilde{b}_{12}+\tilde{b}_{21}=2 b_{12}$\,.
In the following, however, I will restrict this  study to the special choice
given in the Eqs.~(\ref{eq:bTi-s})\,.
The canonical (and grand canonical) formulae for the multi-component case
are given in Ref.~\cite{Gor:99}\,.

\subsection{Comparison of  Two-component { VdW\/} Approximations.}
As the { VdW\/} approximation is a low density approximation it is evident
that the linear and non-linear formulae are equivalent for such densities.
Deviations, however, occur at high densities, where any { VdW\/}
approximation generally becomes inadequate.

The differences between both approximations result from the fact that the
linear pressure (\ref{eq:p-lin}) has two poles, $v^{\rm\,lin}_1=V$ and
$v^{\rm\,lin}_2=V$\,, whereas the non-linear pressure (\ref{eq:p-nl}) has
solely one pole, $v^{\rm\,nl}=V$\,.
For constant volume $V$ these poles define limiting densities,
e.\,g.~$\hat{n}_1=\max(N_1/V)$ as functions of $n_2=N_2/V$\,,
\begin{eqnarray}
  \label{eq:lim-n1,q-lin}
  v_q^{\rm\,lin}(N_1,N_2) = V & \quad \leadsto \quad &
    \hat{n}_1(n_2) \equiv \hat{n}_{1,q}^{\rm\,lin} (n_2)  \\
  \label{eq:lim-n1-nl}
  \mbox{or} \quad
  v^{\rm\,nl}(N_1,N_2) = V & \quad \leadsto \quad &
     \hat{n}_1(n_2) \equiv \hat{n}_1^{\rm\,nl} (n_2) ~,
\end{eqnarray}
which represent the domains of the two pressure formulae in
the $n_2$--$n_1$-plane.
The explicit fomulae are discussed in App.\ \ref{CE_appd}\,.

\begin{figure}[ht!]
  \begin{center}
    \epsfig{file=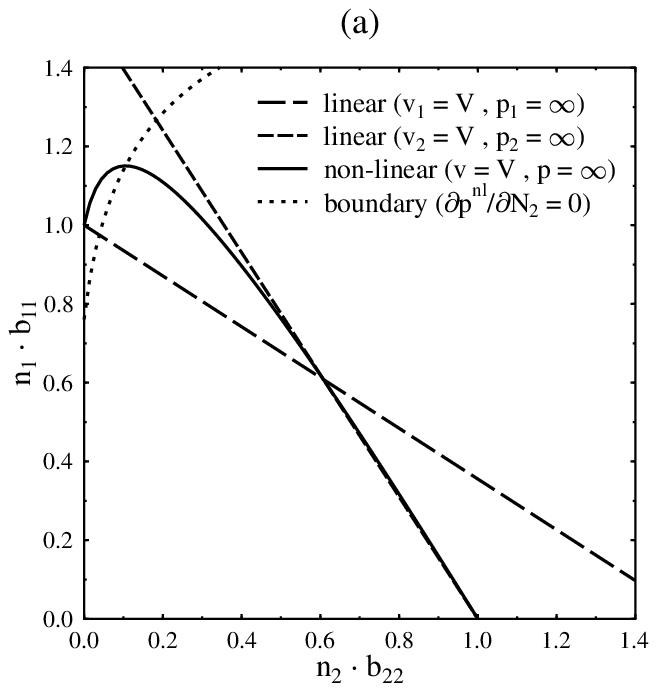, width=8.0cm}
    \epsfig{file=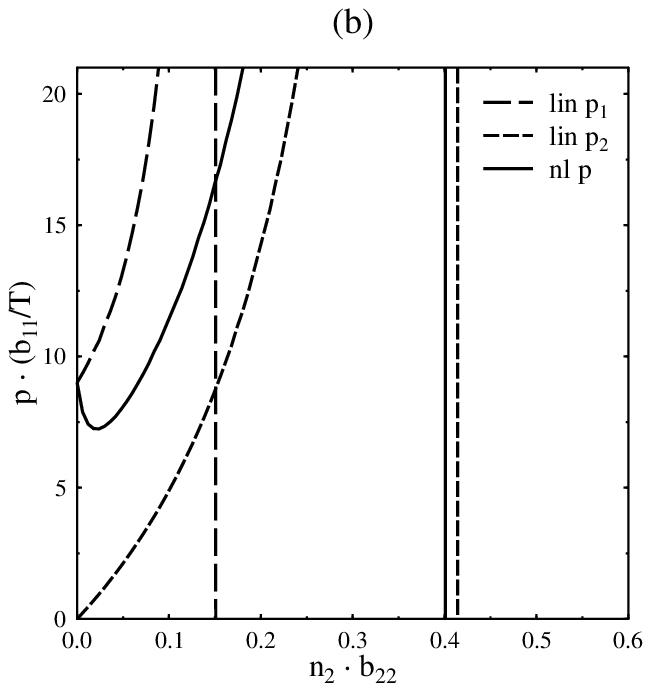, width=8.0cm}
  \end{center}
  \caption{
    \label{figs:1}
    (a) Domains of the linear and non-linear approximation
    for $R_2/R_1=0.4$\,:
    limiting densities $\hat{n}_1$ (isobars for $p_q(n_1,n_2)=\infty$)
    and the lower boundary $\hat{n}_1^{\rm\,nl,\,bd}$ to the region
    of negative partial derivatives of the non-linear pressure.
    The dashed lines correspond to the two poles of the linear pressure,
    and the solid line corresponds to the pole of the non-linear pressure.
    For given $n_2$ the possible densities $n_1^{\rm\,lin}$
    are below both dashed lines, whereas the possible densities
    $n_1^{\rm\,nl}$ are below the solid line.
    Negative derivatives $\pd p^{\rm\,nl}/\pd N_2 < 0$ occur only
    above the dotted line.
   \newline 
    (b) Pressure profiles in dimensionless units for $R_2/R_1 = 0.4$
    as in (a) at fixed $n_1 b_{11}= 0.9$\,.
    The dashed lines show the partial pressures of the linear approximation
    $p_1^{\rm\,lin}$ and $p_2^{\rm\,lin}$\,, while the solid line shows the
    total pressure of the non-linear approximation $p^{\rm\,nl}$
    with initial decrease due to negative $\pd p^{\rm\,nl} / \pd N_2$\,.
    }
\end{figure}

In Fig.~\ref{figs:1}\,(a) an example of these limiting densities is shown
for $R_2/R_1=0.4$\,.
It is clearly seen that the non-linear domain (below the solid line) is
larger than the linear domain (below both dashed lines), which is
generally the case for $R_2 \neq R_1$\,.
Especially for $R_2 \ll R_1$ the non-linear domain is distinctly larger
for high densities of the large component, $n_1 b_{11}>n_2 b_{22}$\,,
whereas both domains are similar for high densities of the small
component, $n_2 b_{22}>n_1 b_{11}$\,.

The linear approximation is constructed in traditional { VdW\/} spirit;
the densities $n_q^{\rm\,lin}$ achieved in this approximation are
below the maximum density of the corresponding {\em one-component\/}
{ VdW\/} gas $\max(n_q^{\rm\,oc})=1/b_{qq}$\,, which is defined by
the pole of $p_q^{\rm\,oc}\equiv p_{\rm\,VdW}(T,V,N_q;b_{qq})$ from
Eq.~(\ref{eq:p_VdW})\,.

In the non-linear approximation, however, the possible densities of
the larger particles $n_1^{\rm\,nl}$ can exceed  $1/b_{11}$ due to the
occurence of negative partial derivatives of the pressure,
$\pd p^{\rm\,nl}/\pd N_2<0$\,.
In this context it is necessary to state that this behavior does not
lead to a thermodynamical instability of the non-linear approximation
as proven in App.\ \ref{nl_appd}\,.
The linear approximation shows no such behavior, it is always
$\pd p^{\rm\,lin}/\pd N_1>0$ and $\pd p^{\rm\,lin}/\pd N_2>0$\,.

The condition $\pd p^{\rm\,nl}/\pd N_2=0$ defines the boundary
$\hat{n}_1^{\rm\,nl,\,bd}(n_2)$ of the region of negative partial
derivatives of the non-linear pressure.
In Fig.~\ref{figs:1}\,(a) this boundary is shown by the dotted line
for $R_2/R_1=0.4$\,;
the values of $\pd p^{\rm\,nl}/\pd N_2$ are negative above this line.

Densities larger than $n_1^{\rm\,nl}=1/b_{11}$ can only occur, if $R_2$
is smaller than a critical radius,
\begin{equation}
  \label{eq:R2crit_CE}
  R_2 < R_{2,\rm\,crit}(R_1) = (\sqrt[3]{4} - 1)\,R_1 \approx R_1/1.7 ~.
\end{equation}
Then, the boundary $\hat{n}_1^{\rm\,nl,\,bd}(n_2)$ starts inside the
non-linear domain, see App.\ \ref{CE_appd} for details.

The reason for this behavior is the ratio of the amounts of small
and large particles.
There are much more small than large particles in the system for
densities $n_1, n_2$ along the boundary $\hat{n}_1^{\rm\,nl,\,bd}(n_2)$
at high densities $n_1$\,: here, the fewer large particles are surrounded
by many small particles.
Therefore, the excluded volume interaction of the large particles
in the non-linear pressure (\ref{eq:p-nl}) is governed not by the
simple term $b_{11}$ but by the mixed term $b_{12}$\,, which is distinctly
smaller than $b_{11}$ for $R_2 \ll R_1$\,.
The maximum density achieved in the non-linear approximation
$\max(\hat{n}_1^{\rm\,nl})=4/b_{11}$ is obtained for $R_2 \to 0$ and 
$N_2 \gg N_1$\,, i.\,e.~these formulae go far {\em beyond\/} the
traditional { VdW\/} results in the corresponding situation.

An example of pressure profiles for $p_1^{\rm\,lin}$\,, $p_2^{\rm\,lin}$ and
$p^{\rm\,nl}$ for $n_1 b_{11} = 0.9$ is shown in Fig.~\ref{figs:1}\,(b)\,,
where it is $R_2/R_1=0.4$ as in Fig.~\ref{figs:1}\,(a)\,.
The non-linear pressure (solid line) firstly decreases as the densities
$n_1, n_2$ correspond to the region of negative partial derivatives,
see Fig.~\ref{figs:1}\,(a)\,.
The partial pressures of the linear approximation are shown by dashed lines.
The non-linear domain is seen to be larger since it is one of the linear
partial pressures which diverges first for increasing $n_2$\,.

%

I conclude that the linear and non-linear approximation show a drastically
different behavior for high values of the large component's density $n_1$\,.
In the linear approximation (\ref{eq:p-lin}) the possible density values
are below $1/b_{11}$ and $1/b_{22}$\,, respectively, and the derivatives
$\pd p^{\rm\,lin}/ \pd N_q$ are always positive.
Whereas in the non-linear approximation (\ref{eq:p-nl}) higher densities
$n_1 > 1/b_{11}$ are possible due to the occurence of negative derivatives
$\pd p^{\rm\,nl}/ \pd N_2 < 0$\,.
This may be considered as pathological -- or used as an advantage to
describe special situations, e.\,g.~densities
$1/b_{11}<n_1<\hat{n}_1^{\rm\,nl}$ for $R_2 \ll R_1$
(see App.~\ref{CE_appd}).

However, the use of any { VdW\/} approximation is in principle problematic
for densities near $1/b_{qq}$\,.
For low densities the non-linear and linear approximation are
practically equivalent, and the non-linear approximation is preferable
since the formulae are essentially simpler.


\section{Grand Canonical Treatment of Binary Mixtures}

Let me  now turn to the GCE.
The grand canonical partition function is built using the CPF\,,
\begin{eqnarray}
  \label{eq:GCPF}
  {{\cal Z}(T,V,\mu_1,\mu_2)}  
  & =& {\textstyle \sum\limits_{N_1=0}^{\infty}
            \,\sum\limits_{N_2=0}^{\infty} }
            \exp\l[{\textstyle \frac{\mu_1 N_1 + \mu_2 N_2}{T} }\r]
            Z(T,V,N_1,N_2) ~,
\end{eqnarray}
whereas the chemical potentials $\mu_1$ and $\mu_2$ correspond to the
components 1 and 2\,, respectively.
For the { VdW\/} CPF (\ref{eq:twoc_CPF_nl-D}) or (\ref{eq:twoc_CPF_lin})
there are limiting particle numbers $\hat{N}_1(N_2)$ or $\hat{N}_2(N_1)$\,,
where each CPF becomes zero.
For this reason the above sum contains only a finite number of terms.
Then it can be shown that in the thermodynamical limit
(i.\,e.~in the limit $V\to\infty$ for $N_q/V={\rm const.}$)
the grand canonical pressure
$p(T,\mu_1,\mu_2)\equiv T\,\ln[{\cal Z}(T,V,\mu_1,\mu_2)]/V$
depends only on the {\em maximum term\/} of the double sum
(\ref{eq:GCPF})\,, where $N_1=\N_1$ and $N_2=\N_2$\,.
One obtains
\begin{eqnarray}
  \label{eq:p_GCE}
{p(T,\mu_1,\mu_2)} 
  & =& \lim_{V \to \infty} \frac{T}{V} \,
      \ln\biggl[\exp\l[{\textstyle \frac{\mu_1 \N_1 + \mu_2 \N_2}{T} }\r]
      Z(T,V,\N_1,\N_2) \biggr]~, 
\end{eqnarray}
wheras $\N_1$ and $\N_2$ are the {\em average\/} particle numbers.

\subsection{The Two { VdW} Approximations.}
For the non-linear { VdW\/} approximation (\ref{eq:twoc_CPF_nl-D})
the last expression takes the form
\begin{eqnarray}
  \label{eq:pnl_GCE}
 {p^{\rm\,nl}(T,\mu_1,\mu_2)} 
  & =&
    \lim_{V \to \infty} \frac{T}{V} \,
    \ln \l[{\textstyle \frac{A_1^{\,\N_1}}{\N_1!}
           \,\frac{A_2^{\,\N_2}}{\N_2!} }
            \l({\textstyle  V - \N_1 b_{11} - \N_2 b_{22}
          + \frac{\N_1\,\N_2}{\N_1+\N_2}\,D }\r)^{\!\N_1+\N_2} \r]~,
\end{eqnarray}
where $A_{q} = A(T,\mu_q;~m_q, g_q) \equiv \exp[\mu_q/T]\,\phi_q$\,.
The evaluation of both maximum conditions for the grand canonical pressure
\begin{eqnarray}
  0 &\stackrel{\textstyle !}{=}&
      \frac{\pd}{\pd \N_q}
      \l\{ \ln\l[{\textstyle  \frac{A_1^{\,\N_1}}{\N_1!}
                  \,\frac{A_2^{\,\N_2}}{\N_2!} } 
      \l({\textstyle  V - \N_1 b_{11} - \N_2 b_{22}
          + \frac{\N_1 \, \N_2}{\N_1+\N_2}\,D }\r)^{\!\N_1+\N_2} \r] \,
      \r\} ~,  \nonumber
\end{eqnarray}
yields a system of two coupled transcendental equations,
\begin{eqnarray}
  \label{eq:xi1}
{\xi_1^{\rm\,nl} (T, \mu_1 , \mu_2)}  
  & =&
    A_1 ~ \exp\l[{\textstyle  -(\xi_1^{\rm\,nl} + \xi_2^{\rm\,nl})\, b_{11}
                    + \frac{{\xi_2^{\rm\,nl}}^{\,2}}
                           {\xi_1^{\rm\,nl} + \xi_2^{\rm\,nl}}\,D }\r] \,,  \\
  \label{eq:xi2}
{\xi_2^{\rm\,nl} (T, \mu_1 , \mu_2)} 
  & =&
    A_2 ~ \exp \l[{\textstyle  -(\xi_1^{\rm\,nl} + \xi_2^{\rm\,nl})\, b_{22}
                   + \frac{{\xi_1^{\rm\,nl}}^{\,2}}
                          {\xi_1^{\rm\,nl} + \xi_2^{\rm\,nl}}\,D }\r] \,,
\end{eqnarray}
where $\xi_1^{\rm\,nl}$ and $\xi_2^{\rm\,nl}$ are defined as
\begin{eqnarray}
  \label{eq:xi1_Def}
  \xi_1^{\rm\,nl}
    &\equiv& \frac{\N_1}{V - \N_1 b_{11} - \N_2 b_{22}
                         + \frac{\N_1 \, \N_2}{\N_1 + \N_2} \, D} \,~,  \\
  \label{eq:xi2_Def}
  \xi_2^{\rm\,nl}
    &\equiv& \frac{\N_2}{V - \N_1 b_{11} - \N_2 b_{22}
                         + \frac{\N_1 \, \N_2}{\N_1 + \N_2} \, D} \,~.
\end{eqnarray}
In the thermodynamical limit the average particle numbers $\N_1$ and $\N_2$
are proportional to $V$ as $\N_q = n_q^{\rm\,nl}\,V$\,.
Then the volume $V$ disappears in the definitions of $\xi_1^{\rm\,nl}$
and $\xi_2^{\rm\,nl}$ given by Eqs.~(\ref{eq:xi1_Def}, \ref{eq:xi2_Def})\,,
and they can be solved for either the density $n_1^{\rm\,nl}$ or
$n_2^{\rm\,nl}$\,,
\begin{eqnarray}
  \label{eq:n1}
  n_1^{\rm\,nl} (T, \mu_1 , \mu_2)
    &=& \frac{\xi_1^{\rm\,nl}}
             {1 + \xi_1^{\rm\,nl} b_{11} + \xi_2^{\rm\,nl} b_{22}
              - \frac{\xi_1^{\rm\,nl} \, \xi_2^{\rm\,nl}}
                     {\xi_1^{\rm\,nl} + \xi_2^{\rm\,nl}}\, D} \,~, \\
  \label{eq:n2}
  n_2^{\rm\,nl} (T, \mu_1 , \mu_2)
    &=& \frac{\xi_2^{\rm\,nl}}
             {1 + \xi_1^{\rm\,nl} b_{11} + \xi_2^{\rm\,nl} b_{22}
              - \frac{\xi_1^{\rm\,nl} \, \xi_2^{\rm\,nl}}
                     {\xi_1^{\rm\,nl} + \xi_2^{\rm\,nl}}\, D} \,~.
\end{eqnarray}
The $\xi_q^{\rm\,nl} = \xi_q^{\rm\,nl}(T,\mu_1,\mu_2)$ are the
solutions of the coupled Eqs.~(\ref{eq:xi1}) and (\ref{eq:xi2})\,,
respectively.

Hence, the pressure (\ref{eq:pnl_GCE}) can be rewritten in terms of
$\xi_1^{\rm\,nl}$ (\ref{eq:xi1}) and $\xi_2^{\rm\,nl}$ (\ref{eq:xi2})\,,
\begin{equation}
  \label{eq:pxi}
  p^{\rm\,nl} (T,\mu_1 ,\mu_2) = T\,\l(\xi_1^{\rm\,nl} + \xi_2^{\rm\,nl} \r) ~,
\end{equation}
supposed that Eqs.~(\ref{eq:n1}, \ref{eq:n2}) are taken into account.
If the definitions (\ref{eq:xi1_Def}) and (\ref{eq:xi2_Def}) are used
for $\xi_1^{\rm\,nl}$ and $\xi_2^{\rm\,nl}$\,,
the pressure formula (\ref{eq:pxi}) coincides with
the canonical expression (\ref{eq:p-nl}) for $N_1=\N_1$ and $N_2=\N_2$\,.

Since the formulation is thermodynamically self-consistent the
identity $n_q \equiv \pd p(T,\mu_1,\mu_2)/ \pd \mu_q$
leads to Eqs.~(\ref{eq:n1}, \ref{eq:n2}) as well.

The grand canonical formulae of the linear approximation
\cite{Gor:99} are obtained exactly as presented for the
non-linear case in Eqs.~(\ref{eq:pnl_GCE}--\ref{eq:pxi})\,.
In the linear case the system becomes
\begin{eqnarray}
  \label{eq:xi1-lin}
  \xi_1^{\rm\,lin} (T, \mu_1 , \mu_2)
    &=& A_1 ~ \exp \l[ -\xi_1^{\rm\,lin}\, b_{11}
                       - \xi_2^{\rm\,lin}\, \tilde{b}_{12} \r] ~,  \\
  \label{eq:xi2-lin}
  \xi_2^{\rm\,lin} (T, \mu_1 , \mu_2)
    &=& A_2 ~ \exp \l[ -\xi_2^{\rm\,lin}\, b_{22}
                       - \xi_1^{\rm\,lin}\,  \tilde{b}_{21} \r] ~,
\end{eqnarray}
\begin{eqnarray}
  \label{eq:xi1-lin_Def}
{\rm  where~~}  \xi_1^{\rm\,lin}
    &\equiv& \frac{\N_1}{V -\N_1 b_{11} -\N_2 \tilde{b}_{21}} \,~,  \\
  \label{eq:xi2-lin_Def}
  \xi_2^{\rm\,lin}
    &\equiv& \frac{\N_2}{V -\N_2 b_{22} -\N_1 \tilde{b}_{12}} \,~.
\end{eqnarray}
The particle densities are found by solving
Eqs.~(\ref{eq:xi1-lin_Def}, \ref{eq:xi2-lin_Def}) for either
$n_1^{\rm\,lin}$ or $n_2^{\rm\,lin}$\,,
\begin{eqnarray}
  \label{eq:n1-lin}
{n_1^{\rm\,lin} (T, \mu_1, \mu_2)} 
  & =&
    \frac{\xi_1^{\rm\,lin} (1 + \xi_2^{\rm\,lin} \,[b_{22} - \tilde{b}_{21}]) }
         {1 + \xi_1^{\rm\,lin} b_{11} + \xi_2^{\rm\,lin} b_{22}
            + \xi_1^{\rm\,lin} \xi_2^{\rm\,lin}
              \,[b_{11} b_{22} - \tilde{b}_{12} \tilde{b}_{21}]} ~,  \\
  \label{eq:n2-lin}
{n_2^{\rm\,lin} (T, \mu_1, \mu_2)}
  & =&
    \frac{\xi_2^{\rm\,lin} (1 + \xi_1^{\rm\,lin} \,[b_{11} - \tilde{b}_{12}]) }
         {1 + \xi_1^{\rm\,lin} b_{11} + \xi_2^{\rm\,lin} b_{22}
            + \xi_1^{\rm\,lin} \xi_2^{\rm\,lin}
              \,[b_{11} b_{22} - \tilde{b}_{12} \tilde{b}_{21}]} ~.
\end{eqnarray}

For the linear case the pressure (\ref{eq:p_GCE}) can be rewritten
in terms of $\xi_1^{\rm\,lin}$ (\ref{eq:xi1-lin}) and
$\xi_2^{\rm\,lin}$ (\ref{eq:xi2-lin})\,,
\begin{eqnarray}
  \label{eq:pxi-lin}
  p^{\rm\,lin} (T, \mu_1 , \mu_2)
    = T \,\l(\xi_1^{\rm\,lin} + \xi_2^{\rm\,lin} \r) ~,
\end{eqnarray}
if Eqs.~(\ref{eq:n1-lin}, \ref{eq:n2-lin}) are taken into account,
like in the non-linear case.



Let me  brief\/ly return to the usual { VdW\/} case,
the one-component case.
The corresponding transcendental equation is obtained from either
Eqs.~(\ref{eq:xi1}, \ref{eq:xi2}) or (\ref{eq:xi1-lin}, \ref{eq:xi2-lin})
by setting $R_1=R_2\equiv R$ and $A_1=A_2\equiv A$\,,
\begin{equation}
    \xi^{\rm\,oc}(T,\mu) = A \,\exp\l[ -\xi^{\rm\,oc}\,b \r] ~,
\end{equation}
whereas $b \equiv b_{11}=b_{22}$\,.
The {\em transcendental factor\/} $\exp[-\xi^{\rm\,oc}\,b]$ has the form of a
suppression term, and the solution $\xi^{\rm\,oc}\equiv p^{\rm\,oc}/T$
of this transcendental equation evidently decreases with increasing $b$
for constant $T$ and $\mu$\,.
Then in turn, the corresponding particle density
$n^{\rm\,oc}=\xi^{\rm\,oc}/(1+\xi^{\rm\,oc}\,b)$
is suppressed in comparison with the ideal gas due to the
lower $\xi^{\rm\,oc}$ {\em and\/} the additional denominator.
Thus, a  transcendental factor corresponds to a suppression
of particle density.

Now it can be seen from Eqs.~(\ref{eq:xi1}, \ref{eq:xi2}) and
(\ref{eq:xi1-lin}, \ref{eq:xi2-lin}) that the transcendental factors of
both {\em two-component\/} approximations contain as well this usual
{\em one-component-} or { VdW}-{\em like\/} suppressive part
$\exp[-(p/T)\,b_{qq}]$\,.
But since it is $D \ge 0$ and $\tilde{b}_{pq} < b_{pp}$\,, respectively,
there is furthermore an attractive part in each corresponding transcendental
factor.

In the non-linear approximation the attractive part can even dominate the
suppressive part for the smaller component, e.\,g.~in Eq.~(\ref{eq:xi2})
for $R_2<R_1$\,.
Then the larger component can reach densities $n_1^{\rm\,nl}$ higher
than $1/b_{11}$\,, analogous to the CE\,.

High densities in the canonical treatment correspond to large
values of the chemical potentials in the grand canonical treatment.
In the limit
\begin{equation}
  \label{eq:mu1lim}
  \mu_1/T \to \infty ~~ (T, \mu_2 = \mbox{const.})
    \quad \mbox{or} \quad
  \xi_1^{\rm\,nl} \to \infty
\end{equation}
the solution of Eq.~(\ref{eq:xi2})\,, $\xi_2^{\rm\,nl}$\,, can be enhanced
for increasing $\xi_1^{\rm\,nl}$ instead of being suppressed, if $R_2$ is
sufficiently small.
This may be called the {\em non-linear enhancement\/}.
The behavior of the non-linear approximation in the limit (\ref{eq:mu1lim})
depends only on the ratio of the two radii $R_1/R_2$ and is characterised by
the coefficient
\begin{equation}
  \label{eq:a2-Def}
  a_2 \equiv \sqrt{D/b_{22}} - 1 ~.
\end{equation}
A negative $a_2$ relates to a suppressive transcendental factor in
Eq.~(\ref{eq:xi2})\,.
For equal radii $R_2=R_1$ it is $a_2=-1$\,, and the suppression is evidently
not reduced but { VdW}-like.
For $-1<a_2<0$ this suppression is reduced, the most strongly for
$a_2\approx 0$\,.

In the case $a_2=0$ the suppression for $\xi_2^{\rm\,nl}$
(\ref{eq:xi2}) disappears in the limit (\ref{eq:mu1lim})\,,
on has $\xi_2^{\rm\,nl}\to A_2={\rm const.}$
This case provides the critical radius
$R_{2,\rm\,crit}$ (\ref{eq:R2crit_CE})\,.

For $a_2>0$ or $R_2<R_{2,\rm\,crit}$ the non-linear enhancement
of $\xi_2^{\rm\,nl}$ occurs for increasing $\xi_1^{\rm\,nl}$\,;
it is the stronger the larger $a_2$ is.
Then $n_1^{\rm\,nl}$ (\ref{eq:n1}) can exceed
$\max(n_1^{\rm\,oc})=1/b_{11}$\,,
whereas $n_2^{\rm\,nl}$ (\ref{eq:n2}) does not vanish.
The density $\max(\hat{n}_1^{\rm\,nl})=4/b_{11}$ is achieved for
$a_2\to\infty$ or $R_2\to 0$\,.

The suppression in the transcendental factor of $\xi_1^{\rm\,nl}$
(\ref{eq:xi1}) is generally reduced for $R_2<R_1$\,, the more strongly
the smaller $R_2$ is, but there is no enhancement possible in the limit
(\ref{eq:mu1lim})\,.




\subsection{Relativistic Excluded Volumes.}
In this section I will investigate the influence of re\-lativistic
effects on the excluded volumes of fast moving particles by accounting
for their ellipsoidal shape due to { Lorentz\/} contraction.
In Ref.~\cite{Bugaev:RVDW2} a quite simple, ultra-relativistic approach has
been made to estimate these effects:
instead of ellipsoids two cylinders with the corresponding radii have been
used to calculate approximately the excluded volume term $b_{pq}$ for the
two-component mixture.
The resulting relativistic excluded volumes depend on the temperature
and contain the radii and the {\em masses\/} as parameters.
The simple, non-mixed term reads \cite{Bugaev:RVDW1}
\begin{eqnarray}
  \label{eq:bqqT}
  b_{qq} (T) &=& \alpha_{qq} \l(\frac{37 \pi}{9} \,\frac{\sigma_q}{\phi_q}
                   + \frac{\pi^2}{2} \r) R_q^{\,3} ~,
\end{eqnarray}
where $\sigma_q \equiv \sigma (T; m_q, g_q)$ denotes
the {\em ideal gas\/} scalar density,
\begin{equation}
  \sigma (T; m, g)
    = \frac{g}{2 \pi^2} \int\limits_0^{\infty} \dd k \, k^2 \,
      \frac{m}{E(k)} \,\exp\l[{\textstyle -\frac{E(k)}{T} }\r] ~.
\end{equation}

The expression for the mixed case can be derived similarly
from \cite{Bugaev:RVDW2}\,,
\begin{eqnarray}
  \label{eq:b12T}
  b_{12}(T)~ =~ \alpha_{12} 
   \l\{{\textstyle
          \l(\frac{\sigma_1}{\phi_1}\,f_1
             + \frac{\pi^2}{4} \frac{R_2}{R_1}\r) R_1^{\,3} } 
             {\textstyle
          + \l(\frac{\sigma_2}{\phi_2}\,f_2
                 + \frac{\pi^2}{4} \frac{R_1}{R_2}\r) R_2^{\,3} }\r\} ~,
\end{eqnarray}
whereas the abbreviations $f_1$ and $f_2$ are dimensionless functions
of both radii,
\begin{eqnarray*}
  f_1 = {\textstyle \frac{\pi}{3} \!\l(2 + \frac{3R_2}{R_1}
                             + \frac{7R_2^{\,2}}{6R_1^{\,2}}\r) } ~,
  &\quad&
  f_2 = {\textstyle \frac{\pi}{3} \!\l(2 + \frac{3R_1}{R_2}
                             + \frac{7R_1^{\,2}}{6R_2^{\,2}}\r) }
 ~.
\end{eqnarray*}
The normalization factors
\begin{eqnarray}
  \alpha_{11} = \alpha_{22}
    &=& {\textstyle \frac{16}{\textstyle \frac{37}{3} + \frac{3\pi}{2}} } ~,  \\
  \alpha_{12}
    &=& {\textstyle \frac{{\textstyle \frac{2\pi}{3} } \l(R_1+R_2\r)^3}
                         {\l(f_1 + {\textstyle \frac{\pi^2}{4}
                                     \frac{R_2}{R_1} }\r) R_1^{\,3}
                          + \l(f_2 + {\textstyle \frac{\pi^2}{4}
                                       \frac{R_1}{R_2} }\r) R_2^{\,3}} }
\end{eqnarray}
are introduced to normalise the ultra-relativistic approximations
(\ref{eq:bqqT}, \ref{eq:b12T}) for $T=0$ to the corresponding
non-relativistic results.
For the hadron gas, however, these { Boltzmann\/} statistical formulae
will only be used at high temperatures, where effects of quantum statistics
are negligible.

Note that it is {\em not appropriate} to consider temperature dependent
hard core radii $R_p(T)$ or $R_q(T)$ since the $b_{pq}(T)$ terms give
the { Lorentz}-contracted excluded {\em volumes\/} and are involved
functions of $T, m_p, m_q, R_p$ and $R_q$\,.
However, for a given value of $b_{pq}(T)$ the necessary hard core radii
$R_p$ and $R_q$ will obviously depend on the temperature.

It is evident that the formulae (\ref{eq:bqqT}, \ref{eq:b12T}) suffice
already for the multi-component case, because even a multi-component
{ VdW\/} formulation contains only $b_{pq}$ terms.

For both approximations the expressions for the pressure (\ref{eq:pxi})
or (\ref{eq:pxi-lin}) and corresponding particle densities
(\ref{eq:n1}, \ref{eq:n2}) or (\ref{eq:n1-lin}, \ref{eq:n2-lin})
remain unchanged.
However, due to the temperature dependence of the relativistic excluded
volumes the entropy density is modified
\begin{eqnarray} \label{eq:s}
  s(T,\mu_1,\mu_2)      
    &\equiv& \frac{\pd}{\pd T} \, p(T,\mu_1,\mu_2) 
    \equiv s_{\rm nrel}~
             +~ s_{\rm rel} (\pd_T b_{11}, \pd_T b_{22}, \pd_T b_{12}) ~.
\end{eqnarray}
The additional term $s_{\rm rel}$ depends on temperature derivatives
of the relativistic excluded volumes,
$\pd_T b_{pq}\equiv\pd b_{pq}/\pd T$\,,
which represent their thermal compressibility.

Furthermore, the term $s_{\rm rel}$ generates additional terms
for the energy density, according to $e\equiv T s-p+\mu_1 n_1+\mu_2 n_2$\,.
In the non-linear approximation one obtains
\begin{eqnarray}
  \label{eq:e}
{e^{\rm\,nl}(T,\mu_1,\mu_2)} 
 =
    n_1^{\rm\,nl} \, \frac{\epsilon_1}{\phi_1}
    + n_2^{\rm\,nl} \, \frac{\epsilon_2}{\phi_2}
    - (n_1^{\rm\,nl} + n_2^{\rm\,nl}) \ T^2   \l({\textstyle \xi_1^{\rm\,nl} \,\pd_T b_{11}
                + \xi_2^{\rm\,nl} \,\pd_T b_{22}
                - \frac{\xi_1^{\rm\,nl}\,\xi_2^{\rm\,nl}}
                       {\xi_1^{\rm\,nl} + \xi_2^{\rm\,nl}}\,\pd_T D }\r) ~,
\end{eqnarray}
and the linear approximation yields
\begin{eqnarray}
  \label{eq:e-lin}
  e^{\rm\,lin}(T,\mu_1,\mu_2)
   ~ =~ n_1^{\rm\,lin} \, \frac{\epsilon_1}{\phi_1}
        ~+~ n_2^{\rm\,lin} \, \frac{\epsilon_2}{\phi_2}  
      &  - & n_1^{\rm\,lin} \, T^2 \, \l( \xi_1^{\rm\,lin} \, \pd_T b_{11}
             ~+~ \xi_2^{\rm\,lin} \, \pd_T \tilde{b}_{12} \r)   \nonumber \\
             &-& n_2^{\rm\,lin} \, T^2 \, \l( \xi_2^{\rm\,lin} \,\pd_T b_{22}
            ~ +~  \xi_1^{\rm\,lin} \,\pd_T \tilde{b}_{21} \r) ~,  
\end{eqnarray}
whereas $\epsilon_q\equiv\epsilon(T;m_q,g_q)$ denotes the {\em ideal gas\/}
energy density
\begin{equation}
  \epsilon (T; m, g) = \frac{g}{2 \pi^2} \int\limits_0^{\infty} \dd k\,k^2\,
    E(k)\,\exp \l[{\textstyle -\frac{E(k)}{T} }\r] ~.
\end{equation}
The additional terms in the entropy density (\ref{eq:s}) and in the
energy density (\ref{eq:e}) or (\ref{eq:e-lin}) which contain temperature
derivatives do evidently not occur in the case of the usual non-relativistic,
i.\,e.~constant excluded volumes.

Let me  now study the hadronic equation of state generated
by each of the two-component { VdW\/} approximations and 
their modifications due to relativistic excluded volumes.
When used to describe {\em hadrons\/}, the hard core radii $R_q$
should be considered as {\em parameters\/} rather than particle radii.
 The first component  is  identified as nucleons ($m_1 \equiv m_{\rm n}=939$ MeV\,,
$\mu_1 \equiv \mu_{\rm n} = \mu_{\rm B}$ and $g_1 \equiv g_{\rm n}=4$
for symmetric nuclear matter) and the second as pions
($m_2 \equiv \overline{m}_\pi=138$ MeV\,, $\mu_2 \equiv \mu_\pi = 0$ and
$g_2 \equiv g_\pi=3$)\,.
Quantum statistical effects other than the degeneracy factors $g_q$ are
neglected. 
To reproduce experimental data, however, it would be necessary to consider
all hadrons and hadronic resonances as well as the contributions from
hadronic decays into daughter hadrons.

\begin{figure}[ht!]
  \begin{center}
    \epsfig{file=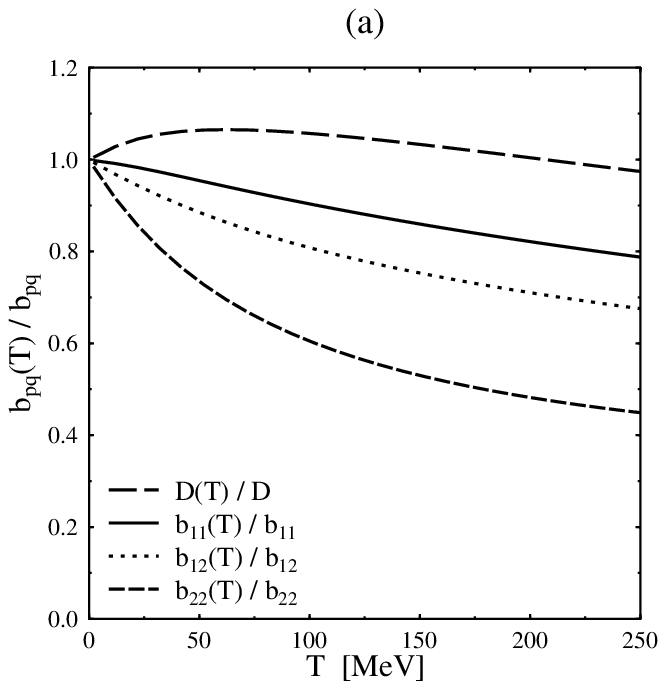, width=8.0cm}
    \epsfig{file=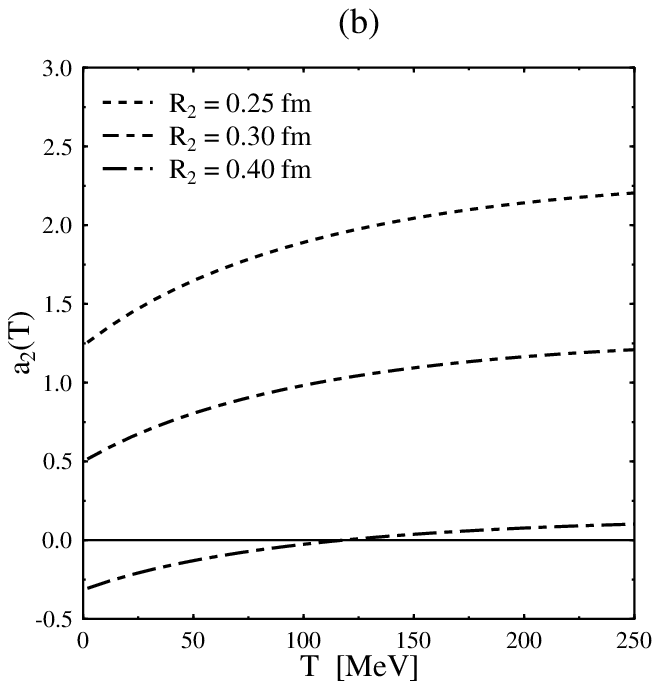, width=8.0cm}
  \end{center}
  \caption{
    \label{figs:2}
    Temperature dependence of the relativistic excluded volume terms
    for $m_1=m_{\rm n}, m_2=\overline{m}_\pi, R_1=0.6$ fm\,.  
\newline
    (a) Relative values for $R_2=0.3$ fm\,:
        $b_{11}(T)/b_{11}$\,, $b_{12}(T)/b_{12}$\,,
        $b_{22}(T)/b_{22}$ and $D(T)/D$ (solid line, dotted line,
        short and long dashes, respectively).
        The relativistic excluded volume of light species ($b_{22}(T)/b_{22}$)
        is affected more strongly by temperature. 
\newline
    (b) The characteristic coefficient of the non-linear approximation
        $a_2(T)=(\sqrt{D(T)/b_{22}(T)}-1)$ for various
        $R_2=0.25$\,, 0.3 and 0.4 fm\,.
        The non-linear enhancement ($a_2(T)>0$) becomes stronger due to
        the decrease of the relativistic excluded volumes with increasing
        temperature.
    }
\end{figure}

For some examples the temperature dependence of the relativistic excluded
volumes is shown in Fig.~\ref{figs:2}\,(a)\,, given in units of the
corresponding non-relativistic terms, $b_{pq}=b_{pq}(0)$\,.
The solid line and the short dashes show the basic excluded volumes
$b_{11}(T)$ and $b_{22}(T)$\,, respectively.
In these relative units the decreases of $b_{11}(T)$ and $b_{22}(T)$
depend only on the corresponding masses.
It is apparent that the pion excluded volume $b_{22}(T)$ is affected much
stronger than the excluded volume of the nucleons, $b_{11}(T)$\,.
The dotted line shows the mixed volume term $b_{12}(T)$\,,
and the long dashes show the compound volume term
$D(T) \equiv b_{11}(T)+b_{22}(T)-2b_{12}(T)$\,.
These two terms depend obviously on both masses and both radii.


The curves for the generalised excluded volume terms of the linear
approximation $\tilde{b}_{12}(T)$ and $\tilde{b}_{21}(T)$ behave
similarly to $b_{12}(T)$\,.

Introduction of the relativistic excluded volumes $b_{pq}(T)$\,, however,
has two effects.
First, the maximum densities become larger since it is generally
$1/b_{qq}(T)>1/b_{qq}$ as seen in Fig.~\ref{figs:2}\,(a)\,.
Furthermore, the balance between the lighter and the heavier species is
changed because the lighter species is affected more than the heavier
at the same temperature:
For the above parameters it is $b_{22}(T)/b_{22}\le b_{11}(T)/b_{11}$\,.
%

In the non-linear approximation this balance is characterised by the
coefficient $a_2$ defined by Eq.~(\ref{eq:a2-Def})\,.
In Fig.~\ref{figs:2}\,(b) the temperature dependence of
$a_2(T)\equiv(\sqrt{D(T)/b_{22}(T)}-1)$ is shown for three different
values of $R_2$\,.
The {\em relativistic\/} coefficient $a_2(T)$ increases with $T$\,,
i.\,e.~the non-linear enhancement becomes stronger for higher temperatures.
For some values of $R_2$\,, e.\,g.~$R_2=0.4$ fm\,, a primary suppression
$a_2(0)\equiv a_2<0$\,, turns into an enhancement $a_2(T)>0$\,, when the
temperature is sufficiently high.
For temperature dependent excluded volumes $R_{2,\rm\,crit}$ looses its
meaning; here, only $a_2(T)>0$ is the valid condition for the
occurrence of the non-linear enhancement or densities
$n_1^{\rm\,nl}>1/b_{11}(T)$\,.

The linear coefficient, $\tilde{a}_2(T)=-2b_{12}(T)/(b_{11}(T)+b_{22}(T))$\,,
is not strongly affected by temperature for the above choice of hadronic
parameters:
It increases slightly with $T$ but remains negative.
Hence, changes in the balance between the lighter and the heavier species
play a minor role for the linear approximation.

\begin{figure}[ht!]
  \begin{center}
    \epsfig{file=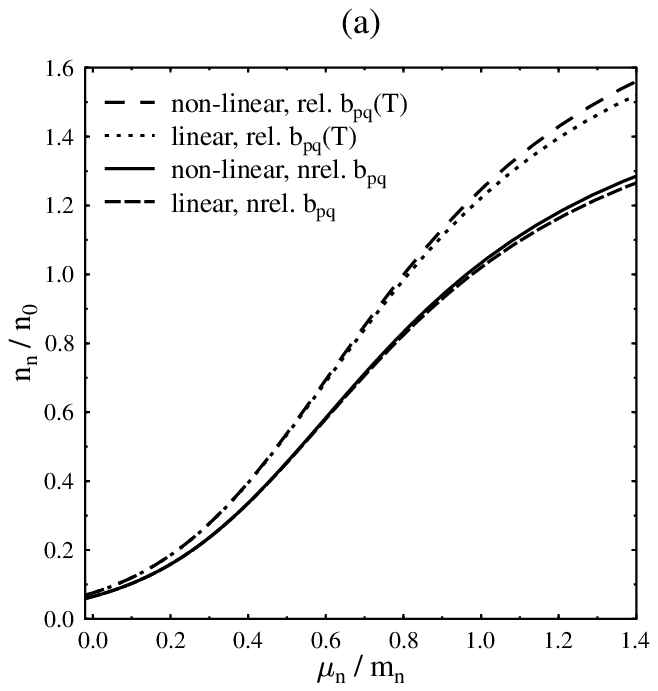, width=8.0cm}
    \epsfig{file=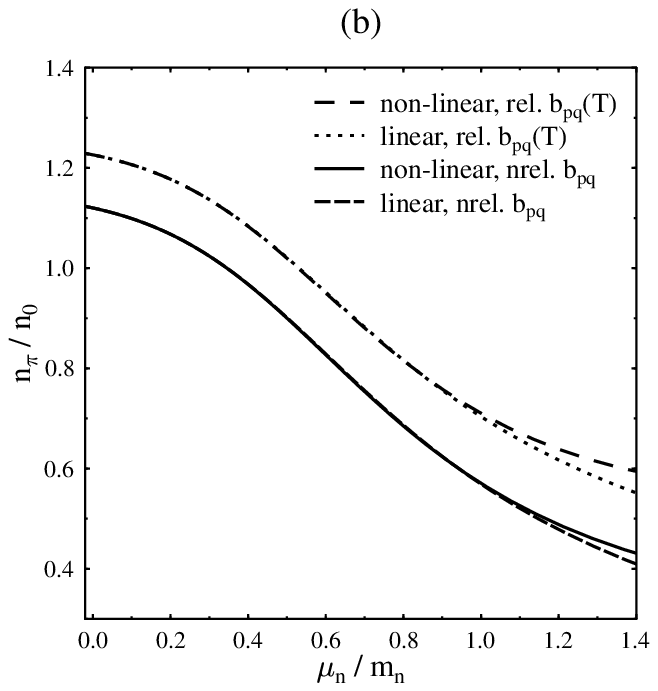, width=8.0cm}
  \end{center}
  \caption{
    \label{figs:3}
    Comparison of the model predictions for the nucleon (a) and
    pion (b) density $n_{\rm n}$ and $n_\pi$\,, respectively,
    vs.~$\mu_{\rm n}/m_{\rm n}$
    ($R_1=0.6$ fm\,, $R_2=0.3$ fm and $T=185$ MeV\,,
    densities in units of $n_0=0.16$ fm$^{-3}$).
    In both figures the two upper lines correspond to relativistic
    excluded volumes $b_{pq}(T)$ and the two lower lines to non-relativistic
    excluded volumes $b_{pq}$\,.
    The results of the linear and non-linear approximation coincide --
    only for extremely large $\mu_{\rm n}/m_{\rm n}$ the non-linear results
    lie slightly higher than the corresponding linear results.
    The deviations due to relativistic excluded volumes are significant.
    }
\end{figure}

Particle densities for nucleons and pions in units of
$n_0=0.16$ fm$^{-3}$ vs.~$\mu_1/m_1 \equiv \mu_{\rm n}/m_{\rm n}$ are
shown in Figs.~\ref{figs:3}\,(a) and (b) for $T = 185$ MeV\,.
The linear and non-linear results are shown for constant excluded volumes
with short dashes and solid lines, respectively, and further for
relativistic excluded volumes with dotted lines and long dashes,
respectively.
At this high temperature the relativistic results are significantly higher
than the non-relativistic result.
A difference between the linear and the non-linear approximation due to
the non-linear enhancement becomes noticeable only for high
$\mu_{\rm n}/m_{\rm n} > 0.8$\,.
Thus, for $R_{\rm n}=R_1=0.6$ fm from above, the linear and non-linear
approximation are practically equivalent for nucleon densities below
$n_{\rm n} \approx 0.8\,n_0$\,, i.\,e.~for densities below about $n_1 \approx 1/(2\,b_{11})$\,.
On the other hand, due to the strong decrease of $b_{22}(T)$ with
increasing temperature, the influence of the relativistic excluded volumes
is essential for temperatures of the order of $T \approx m_\pi$\,.

The presence of the additional terms containing temperature
derivatives in the energy density (\ref{eq:e}) or (\ref{eq:e-lin})
makes it impossible to convert a { VdW\/} gas with relativistic
excluded volumes into a gas of free streaming particles.
Therefore, it is problematic to use these formulae for the post-freeze-out
stage.
For the latter the quantities of the free streaming particles without any
interaction should be used, see discussion in \cite{Bugaev:96,Mag:99,Andl:99}
and references therein.
However, these equations of state may be used to describe the stage between
chemical and thermal freeze-out, i.\,e.~a pre-freeze-out stage in terms of
Refs.~\cite{Bugaev:96,Mag:99,Andl:99}\,.
This is examplified in the next section.

%


\subsection{Hard-core Radii  from Particle Yield Ratios.}
As a simple application of the equations of state presented above,
let me re-evaluate the thermal model fit parameters for particle yield
ratios of Ref.~\cite{Yen:97}\,, namely the hard core radii of pions $R_\pi$
and other hadrons $R_{\rm o}$\,.
A two-component { VdW\/} excluded volume model has been used there
to explain the pion abundance in A+A-collisions by a smaller hard core
radius for the pions than for the other hadrons.
The ratios has been fitted to BNL AGS (Au+Au at 11 A\,GeV) and CERN SPS
(Pb+Pb at 160 A\,GeV) data \cite{PrQM:96} within a thermal model, including
all resonances up to 2 GeV and using quantum statistics.

The applied model, however, corresponds to the incorrect separated model
as pointed out in previous section.
For convenience I give these formulae 
in { Boltzmann\/} approximation.
Within the previously defined  notation the two coupled transcendental
equations read
\begin{eqnarray}
  \label{eq:xi1-sp}
  \xi_1^{\rm\,sp} (T, \mu_1, \mu_2)
    &=& A_1 \,\exp \l[ -(\xi_1^{\rm\,sp} + \xi_2^{\rm\,sp})\, b_{11}\r] ~,  \\
  \label{eq:xi2-sp}
  \xi_2^{\rm\,sp} (T, \mu_1, \mu_2)
    &=& A_2 \,\exp \l[ -(\xi_1^{\rm\,sp} + \xi_2^{\rm\,sp})\, b_{22} \r] ~,
\end{eqnarray}
wheras $p^{\rm\,sp}(T,\mu_1,\mu_2)=T\,(\xi_1^{\rm\,sp}+\xi_2^{\rm\,sp})$\,.
In this context $A_1$ represents a sum over the contributions of all hadron
species but pions, while $A_2$ corresponds to the pions only.

The expressions for the particle densities are obtained from
$n_q^{\rm\,sp} \equiv \pd p^{\rm\,sp}/ \pd \mu_q$\,,
\begin{eqnarray}
  n_1^{\rm\,sp} (T, \mu_1, \mu_2)
    &=& \frac{\xi_1^{\rm\,sp}}
             {1 + \xi_1^{\rm\,sp} b_{11} + \xi_2^{\rm\,sp} b_{22}}\, ~,  \\
  \label{eq:n2-sp}
  n_2^{\rm\,sp} (T, \mu_1, \mu_2)
    &=& \frac{\xi_2^{\rm\,sp}}
             {1 + \xi_1^{\rm\,sp} b_{11} + \xi_2^{\rm\,sp} b_{22}}\, ~.
\end{eqnarray}
Solving these equations for $\xi_1^{\rm\,sp}$ and $\xi_2^{\rm\,sp}$
one recovers the canonical pressure formula of the separated model
(\ref{eq:p-sp_CE}) as announced earlier.

Due to the separation of both components in this model there is no excluded
volume term $b_{12}$ for the interaction between different components at all.
This is an essential difference to both the linear and the non-linear
approximation.
Note that the separated model is {\em not\/} a two-component
{ VdW\/} {\em approximation\/} because it cannot be obtained
by approximating the CPF (\ref{eq:twoc_CPF})\,.

The transcendental factors of the formulae (\ref{eq:xi1-sp}, \ref{eq:xi2-sp})
exhibit a constant { VdW}-like suppression $\exp[-(p/T)\,b_{qq}]$\,.
There is a reduction of this suppression in the linear and in the
non-linear approximation, as shown above.
The { VdW}-like suppression is reduced, if $b_{12}$ appears in the
corresponding formulae since $b_{12}$ is smaller than $b_{11}$ for $R_2<R_1$\,.
It is evident that the deviation of the linear and non-linear approximation
from the separated model is the larger the more $R_1$ and $R_2$ differ from
each other.

In the first step of the fit procedure of Ref.~\cite{Yen:97}
only the hadron ratios excluding pions have been taken to find the
freeze-out parameters.
For AGS  $T \approx 140$ MeV\,, $\mu_{\rm B} \approx 590$ MeV and for SPS
$T \approx 185$ MeV\,, $\mu_{\rm B} \approx 270$ MeV have been obtained.
In the second step, a parameter introduced as the
{\em pion effective chemical potential\/} $\mu_\pi^{\,*}$ has been fitted to
the pion-to-hadron ratios.
Using { Boltzmann\/} statistics it can be shown that the pion enhancement
is thoroughly regulated by the value of $\mu_\pi^{\,*}$ \cite{Yen:97}\,;
one has obtained $\mu_\pi^{\,*} \approx 100$ MeV for AGS and
$\mu_\pi^{\,*} \approx 180$ MeV for SPS data, respectively.

The pion effective chemical potential depends explicitly on the
excluded volumes but also on the pressure.
The pressure itself is a transcendental function depending solely on the
excluded volumes since $T$ and $\mu_{\rm B}$ are already fixed by step one.
In Ref.~\cite{Yen:97} the formula 
$\mu_\pi^{\,*} \equiv (v_{\rm o} - v_\pi)\,p(v_{\rm o}, v_\pi)$
has been obtained for the separated model, where $v_\pi \equiv b_{22}$
and $v_{\rm o} \equiv b_{11}$ are the excluded volumes corresponding
to the hard core radii of pions $R_\pi \equiv R_2$ and other hadrons
$R_{\rm o} \equiv R_1$\,, respectively.
Thus, the $\mu_\pi^{\,*}$ values for AGS and SPS data define two curves
in the $R_\pi$--$R_{\rm o}$-plane.
The main conclusion of Ref.~\cite{Yen:97} is that the intersection point
of these two curves ($R_\pi=0.62$ fm\,, $R_{\rm o}=0.8$ fm) gives the correct
pair of hard core radii for pions and for the other hadrons, i.\,e.~AGS and
SPS data are fitted simultaneously within the 
applied model.

In Ref.~\cite{Hgas:2} these values of $R_{\rm o}$ and $R_\pi$ have been
criticised for being unreasonably large.
There, a complete fit of solely SPS data has been performed within
a separated model.
The best fit has been obtained for equal hard core radii,
$R_\pi=R_{\rm o}=0.3$ fm\,, motivated by nucleon scattering data.
Good agreement has been found as well for a {\em baryon\/} hard core radius,
$R_{\rm Bar}=0.3$ fm\,, and a common hard core radius for {\em all mesons\/},
$R_{\rm Mes}=0.25$ fm\,, choosen in accord with the above ratio
of radii, $R_{\rm o}/R_\pi=0.8/0.62$\,.
Larger hard core radii, especially those of Ref.~\cite{Yen:97}\,, are
quoted as giving distinctly worse agreement.

\begin{figure}[ht!]
  \begin{center}
    \epsfig{file=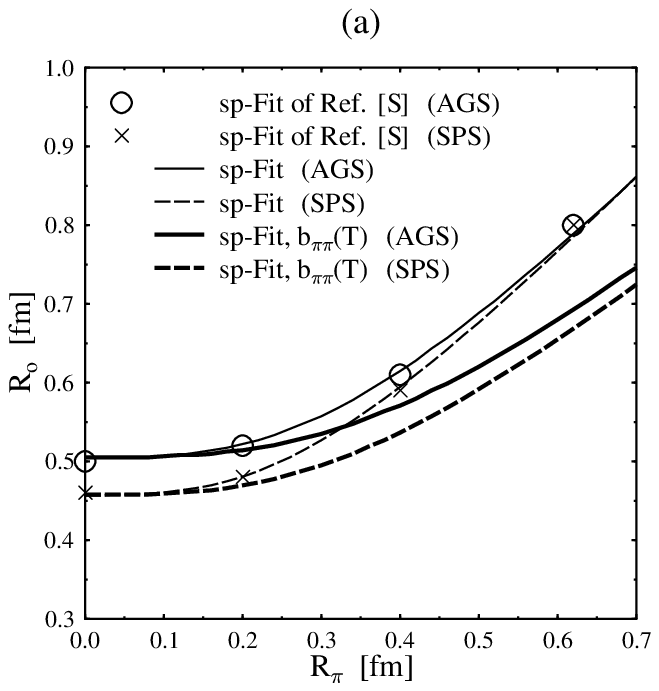, width=7.5cm}
    \epsfig{file=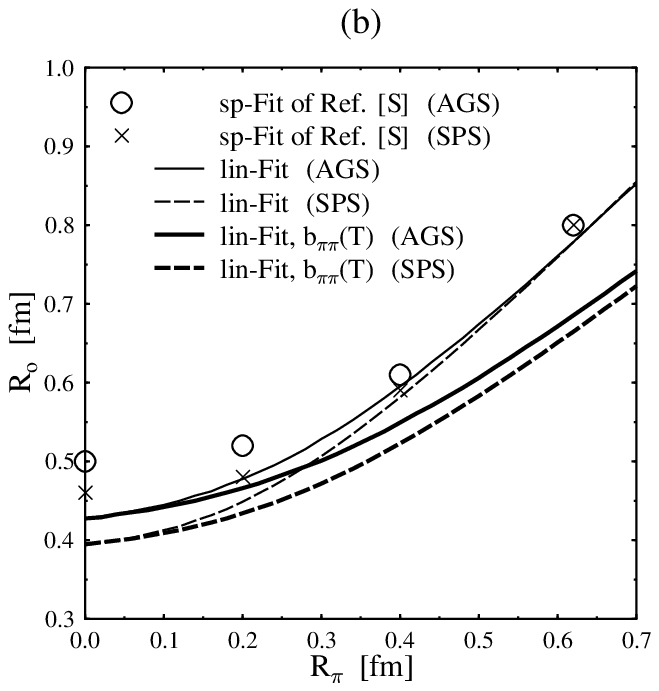, width=7.5cm} 
    \end{center}
  \caption{
    \label{figs:4}
    Fits of particle yield ratios for AGS and SPS
    data \protect\cite{PrQM:96} with the separated model
    and the linear approximation.
    The thin lines show the fits for the separated model (a)
    and the linear approximation (b)\,;
    the non-linear approximation gives identical results for
    the latter case.  
\newline
    The thick lines in (a) and (b) show the corresponding curves for
    relativistic excluded volumes $v_\pi=b_{22}(T)$\,:
    there is no intersection for either of the three models.
    In both figures the results of the fit from Ref.~\protect\cite{Yen:97} 
    for AGS and SPS data are indicated by circles and crosses, respectively, and are referred as Ref. [S].
    }
\end{figure}

Assuming the validity of { Boltzmann\/} statistics,
I have re-calculated the $R_{\rm o}(R_\pi)$-curves for the above
$\mu_\pi^{\,*}$ values;
firstly in the separated model (\ref{eq:xi1-sp}--\ref{eq:n2-sp})\,,
i.\,e.~as presented in \cite{Yen:97}\,.
The resulting curves, shown as thin lines in Fig.~\ref{figs:4}\,(a)\,,
naturally match the results of the underlying fit of Ref.~\cite{Yen:97}\,,
which are indicated by markers.

Then I have considered the linear and the non-linear approximation.
Due to the occurence of $b_{12}$ terms in these two cases, both functional
forms of $\mu_\pi^{\,*}$ are essentially different from the separated case.
Consequently, the shapes of the $R_{\rm o}(R_\pi)$-curves are different
as well.
Thus,  one finds  distinct deviations from the separated model, especially for
$R_\pi \to 0$\,, and the values for the intersection point are slightly
lower; see thin lines in Fig.~\ref{figs:4}\,(b) for the linear exrapolation.
The  non-linear approximation gives identical results for this purpose because
the hadron densities are too small for a noticeable non-linear enhancement.


The crucial point is now to turn on the relativistic temperature
dependence of the excluded volumes.
To keep the analysis simple I treat only pions this way since they
give the strongest effect.

Although the other hadrons are assumed to have equal hard core radii,
their relativistic excluded volumes would be different for $T > 0$\,,
according to their different masses.
To check the influence of relativistic excluded volumes for {\em all\/}
particles, I have used {\em one average\/} mass of 1~GeV for
{\em all other\/} hadrons.
The corresponding change in the $R_{\rm o}$-values are below 5\%\,.

The results of the fit for relativistic excluded volumes for pions
are shown in Figs.~\ref{figs:4}\,(a) and (b) as thick lines.
Though this approach is more realistic, there is {\em no\/} intersection
point for {\em any\/} of the three models even for very large radii
$R_{\rm o}, R_\pi \gg 0.5$ fm\,.
For the approximated case of a single averaged hadron mass there is no
intersection either.
Because of the different freeze-out temperatures for AGS and SPS
the $v_\pi(T) \equiv b_{22}(T)$ values are changed differently
in both cases, and so are the scales for the corresponding $R_\pi$\,.

Due to the errors in experimental data one ought to obtain a corridor instead
of a curve for each set of data.
Consequently, the particle yield ratios can be reproduced well by
e.\,g.~$R_{\rm o} \approx 0.4$ fm\,, $R_\pi \approx 0.2$ fm or larger values
for any of the models with relativistic excluded volumes.
Therefore, I conclude that the fit procedure proposed in Ref.~\cite{Yen:97}
is {\em not\/} suitable to find a {\em unique\/} pair of hard core radii for
pions and other hadrons, as long as a best fit is searched for just two sets
of data of particle yield ratios.
The use of a relativistic excluded volume for pions along with a correct
approximation reduce the value of the necessary nucleon hard core radius
essentially towards more realistic values.


\subsection{Intermediate Conclusions.}
In this section  
several EOS  for the two-component VdW
excluded volume model are derived and investigated in details.
 Two essentially different formulations have been discussed, the linear and
the non-linear approximation.

The non-linear approximation is the simplest possibility.
Here, the large component can reach higher densities $n_1$ than the
usual limiting { VdW\/} density $1/b_{11}$\,, if the other component
has a suffiently small hard core radius, $R_2 < R_{2,\rm\,crit}$\,.
%
In the linear approximation the densities cannot exceed the usual
limiting { VdW\/} densities $1/b_{11}$ and $1/b_{22}$\,, but
generalised excluded volume terms have to be introduced.

For both approximations the suppression factors of the grand canonical
formulae contain a { VdW}-like term, proportional to
$\exp[-(p/T)\,b_{qq}]$\,, which however is reduced non-trivially.
In the linear case there is a slight reduction, wheras in the non-linear case
this reduction can turn the suppression even into an enhancement for the
smaller component, which leads to exceeding of $1/b_{11}$ for the
density of the larger component $n_1$\,.

The commonly used formulae of the separated model are shown to be not
suitable for the two-component case, because they correspond to a system
where both components are separated from each other and cannot mix.
In this model the grand canonical suppression factor is just { VdW}-like
and has no reduction of the suppression.

Furthermore relativistic, i.\,e.~{ Lorentz}-contracted, excluded volumes
have been introduced.
Naturally, the relativistic excluded volume per particle decreases with
rising temperature.
This effect is the stronger the lighter the particle species is.
The suppression of particle densities in { VdW\/} models is lower for
a component of smaller excluded volume in comparison with a component
of larger excluded volume.
Therefore, the temperature dependence of the relativistic excluded volumes
causes a reduction of the particle densities suppression.

The full equations of state have been presented, for both the linear
and non-linear approximation, with constant and with relativistic
excluded volumes.
For the entropy density and energy density there are additional terms
containing temperature derivatives of the relativistic excluded volume
terms due to their 'thermal compressibility'.
In comparison with the non-relativistic case, the expressions for the
pressure and the particle densities remain unchanged, but the possible
range of values is obviously wider, since it is generally
$1/b_{11}(T) \ge 1/b_{11}$ and $1/b_{22}(T) \ge 1/b_{22}$\,.

As an application of the derived formulations a fit of particle yield
ratios for SPS and AGS has been re-evaluated.
In Ref.~\cite{Yen:97} this fit had been done in a separated model
by adjusting the hard core radii for the pions $R_\pi$ and for the other
hadrons $R_{\rm o}$\,.
The results of the new fit are essentially different from the separated
model but coincide for both the linear and non-linear approximation.
The picture changes drastically, however, if relativistic excluded volumes
are adopted for pions.
The basic idea of the fit -- one pair of hard core radii suffices to fit
AGS and SPS data simultaneously -- does not lead to a result anymore.
This is the case for the separated model and for both approximations,
linear and non-linear.
Experimental uncertainties lead to a {\em region\/} of possible values
in the $R_{\rm o}$--$R_\pi$-plane;
one could describe the data for $R_{\rm o} \ge 0.4$ fm and
$R_\pi \ge 0.2$ fm\,.

I  conclude that there are two causes of an {\em enhancement\/} of particle
densities, e.\,g.~the thermal pion abundance, in { VdW\/} descriptions:
First, the density suppression is generally lower for the smaller component
in two-component models.
Second, there is a further reduction of the density suppression
due to the relativistic excluded volumes.
The latter are essentially smaller for light hadron species than for
heavy species, especially for temperatures $T \gg 50$ MeV\,.

When applied to the hadron gas, the linear and non-linear results almost
coincide for nucleon densities up to $n_0 \approx 0.16$ fm$^{-3}$ (for
$R_{\rm o} \le 0.6$ fm) since the non-linear enhancement does not appear
there, but the deviation from the incorrect separated model is distinct.
However, the formulae of the non-linear approximation are essentially
simpler than these of the linear approximation.

The influence of relativistic effects on the excluded volumes becomes
indispensable for temperatures typical for heavy ion collisions.
However, the EOS presented above become  acausal  above the cross-over
where the usual hadrons can coexist with quarks and gluons   \cite{Karsch:03}.
Therefore, in order to extend the hadron gas description  above the cross-over
 it is necessary to return back to the VdW extrapolation suggested in \cite{Bugaev:RVDW1}  and find the reason why 
the relativistic excluded volumes do not vanish at high pressures.





%
%
%
\def\vdw{{\it VdW}\,}
\def\vdwful{{\it  Van-der-Waals}\,}
\def\cs{{\it CS}\,}
\def\csful{{\it contracted spheres}\,}
\def\req#1{(\ref{#1})}
\def\vp{v_{\rm o}}

\def\bra{\langle}
\def\ket{\rangle}

%
%
\let\a=\alpha \let\b=\beta \let\g=\gamma \let\d=\delta
\let\e=\varepsilon \let\z=\zeta \let\h=\eta \let\th=\theta
\let\dh=\vartheta \let\k=\kappa \let\l=\lambda \let\m=\mu
\let\n=\nu \let\x=\xi \let\p=\pi \let\r=\rho \let\s=\sigma
\let\t=\tau \let\o=\omega \let\c=\chi \let\ps=\psi
\let\ph=\varphi \let\Ph=\phi \let\PH=\Phi \let\Ps=\Psi
\let\O=\Omega \let\S=\Sigma \let\P=\Pi \let\Th=\Theta
\let\L=\Lambda \let\G=\Gamma \let\D=\Delta

%
%
%
\def\noi{{\noindent}}

\def\nn{\nonumber \\}
\def\bc{\begin{center}}
\def\ec{\end{center}}
\def\1#1{{\bf #1}}
\def\Rsm{$R_{\odot}$}
\def\Rs{R_{\odot}}
\def\lp{\left(}
\def\rp{\right)}
\def\pe{\frac{p}{E}}
\def\ep{\frac{E}{p}}
\def\pp{\sqrt{E^2 - m^2}}

\section{Relativization of the VdW EOS }

As was shown in the preceding section the relativistic  VdW equation obtained in the traditional way 
leads to the reduction of the second virial coefficient (analog of the
excluded volume)
compared to  nonrelativistic case.
However, in the high pressure limit the second virial coefficient
remains finite. This fact immediately leads to the problem
with causality in relativistic mechanics - the
speed of sound exceeds the  speed of light \cite{Raju:92}.

At the moment there are  few  guesses on how to formulate the 
statistical mechanics of  this state, but it is possible that the relevant 
quasiparticle degrees of freedom may   include the dressed constituent 
quarks or/and their  hadron-like  bound states. 
In this case one can think about the possibility to describe  such states 
in terms of relativistic particles with the  hard core repulsion on short distances. 
In fact, according  to Shuryak \cite{Shur:05} the recent study of the strongly coupled 
colored classical plasma does include the short range repulsion of the  inverse distance square type. 
Also it is quite possible that the huge values of the partonic  cross-sections which are 
necessary to reproduce the  values of elliptic flow observed at RHIC energies 
\cite{GyulassyMolnar}
do evidence that during the course of heavy ion collision  the partons are  acquiring   some  finite transversal size.

In addition,
 the VdW EOS 
which obeys the causality condition in the limit of high density 
and simultaneously reproduces the correct low density behavior  has a significant   theoretical 
value  for the  relativity because, due to some unclear reasons,  such an EOS  
was not formulated during the century that passed after the special relativity  birth.
This section is devoted to
the investigation of the necessary  assumptions  to formulate   such an equation of state.

Similarly to the nonrelativistic VdW case \cite{Raju:92} 
this leads to the problem with causality at very high pressures.
Of course, in this formulation the superluminar speed of sound
should  appear at  very high  temperatures which   are unreachable in hadronic phase. 
Thus the simple  ``relativization''  of the  virial expansion is much more realistic 
than the nonrelativistic description used in Refs. \cite{Hgas,Hgas:2}, but it   does not
solve the problem completely.

The reason why the simplest generalization \req{vdwgc}  fails is rather trivial.
Eq. \req{vdwgc} does not take into account the fact that at high densities
the particles disturb the motion of their neighbors.
The latter leads to the more compact configurations than predicted by
Eqs. \mbox{(\ref{vdwgc} - \ref{etmu}),} i.e.,
the motion of neighboring  particles becomes
correlated due to a simple geometrical reason.
In other words, since the $N$-particle distribution
is a monotonically decreasing function of the
excluded volume, the most probable state
should correspond to the configurations of smallest
excluded volume of all neighboring particles.
This subject is, of course, far beyond the present paper. 
Although I will touch this subject slightly while discussing  the limit $\mu/T \gg 1$ below,
my  primary task here will be  to give  a 
relativistic generalization of the VdW EOS,
which at low pressures behaves in accordance with the relativistic 
virial expansion presented above, and  at the same time  is  free
of the causality paradox at high pressures.  

In the  treatment below  I  will  completely neglect the angular rotations of the Lorentz contracted spheres
because their correct analysis can be done only  within the frame work of  quantum scattering theory which is not used here.
However, it is clear that the rotational effects can be safely neglected at low densities because there are 
not so many collisions in the system. At the  same time the rotations of the Lorentz contracted spheres
at very high pressures, which are of my  main interest, can be neglected too, because at  so high densities the particles should be 
so close to each other, that they  must  prevent the rotations of   neighboring particles. 
Thus,  for these two limits one can safely ignore the  rotational effects and proceed further on like 
for  the usual VdW EOS.

Eq. \req{vdwgc} is only one of many possible VdW extrapolations to high density.
As in nonrelativistic case
one can write many expressions which will give the first two terms
of the full virial expansion exactly, and
the difference will appear in the third virial coefficient.
In relativistic case there is an additional ambiguity:
it is possible to perform the momentum integration, first, and 
make the VdW extrapolation next, or vice versa. The result will,
evidently, depend on the order of operation.

As an example let me give a brief ``derivation'' of Eq. \req{vdwgc}, and its 
counterpart in the grand canonical ensemble. The two first terms of the standard  cluster expansion 
read as
\cite{May:77,Bugaev:RVDW1}  
\begin{equation} 
p =  T \, \rho_t (T) \,\,e^{\textstyle \frac{\m }{T} }   
\lp  1 - a_2 \, \rho_t (T) \, e^{\textstyle \frac{\m }{T} }  \rp 
\,\,. 
\end{equation} 
\label{presi}
Now I approximate the last term on the  right hand side as  
$\rho_t (T) \, e^{\textstyle \frac{\m }{T} } \approx  \frac{p}{T}$. Then I
extrapolate it to high pressures by  moving this term into the  exponential  function as
\begin{equation} \label{presii}
p \approx  T \, \rho_t (T) \,\,e^{\textstyle \frac{\m }{T} }
\lp  1 - a_2 \,  \frac{p}{T}  \rp 
\approx T \, \rho_t (T) \, \exp\lp \frac{\m - a_2\,p }{T}  \rp 
\,\,.
\end{equation} 

\noi
The resulting expression coincides with Eq. \req{vdwgc}, but the above manipulations make it simple
and transparent. 
Now I will repeat all  the above steps while keeping
both momentum integrations fixed 
\begin{eqnarray} 
p & \approx & \frac{T\, g^2\, e^{\textstyle \frac{\m }{T}  }}{ \rho_t (T)}
\int
\frac{d{\1 k_1}}{(2\pi)^3}
\frac{d{\1 k_2}}{(2\pi)^3}
\,\,e^{\textstyle - \frac{E(k_1) + E(k_2) }{T} }
\lp  1 - \frac{ v (\1 k_1, \1 k_2)\, p }{T}   \rp
\nonumber \\
\label{presiii}
& \approx & 
\frac{T\, g^2 }{ \rho_t (T)}
\int
\frac{d{\1 k_1}}{(2\pi)^3}
\frac{d{\1 k_2}}{(2\pi)^3}
\,e^{\textstyle \frac{\m - v (\1 k_1,\1 k_2)\, p\, -\,E(k_1)\, -\,E(k_2) }{T} }
\,\,.
\end{eqnarray} 

The last expression contains the relativistic excluded volume \req{etmu} explicitly
and, as can be shown, is free of the causality paradox. 
This is so because at high pressures the main contribution to the momentum 
integrals corresponds to the  smallest values of the excluded volume \req{etmu}.
It is clear  that such  values are reached when the both spheres are ultrarelativistic and 
their velocities are collinear.

With the help of the following notations 
\begin{eqnarray}
\bra {\cal O} \ket  & \equiv  & \frac{ g }{ \rho_t (T) } \int \frac{d{\1 k}}{(2\pi)^3} \,\, {\cal O}  \,\,e^{\textstyle - \frac{E(k) }{T} } \,,
\\
\bra\bra {\cal O} \ket \ket & \equiv  & \frac{ g^2 }{ \r^2_t (T) } \int \frac{d{\1 k_1}}{(2\pi)^3} \frac{d{\1 k_2}}{(2\pi)^3}
\,\, {\cal O}  \,\,e^{\textstyle - \frac{ v (\1 k_1,\1 k_2)\, p\, +\,E(k_1)\, +\,E(k_2) }{T} }
\label{presiiii}
\end{eqnarray} 
for the averages one can define all other thermodynamic functions as 
\begin{eqnarray} \label{iiizeroA}
n (T, \m) & = & 
 \frac{ \partial p (T, \m) }{ \partial \m }   =   \frac{ p}{T \lp 1 + e^{\textstyle \frac{\m}{T}}  \rho_t(T) 
  \bra\bra  v(\1 k_1, \1 k_2)  \ket \ket \rp } \,, \\
s (T, \m) & = & 
 \frac{ \partial p (T, \m) }{ \partial T }   =  \frac{ p}{T} +  \frac{ 1}{T} 
 \frac{ \lp 2\,e^{\textstyle \frac{\m}{T}}  \rho_t(T)  \bra\bra E \ket \ket  - [\mu +  \bra E \ket ]\, p\, T^{-1}  \rp}{
 1 + e^{\textstyle \frac{\m}{T}}  \rho_t(T)  \bra\bra v(\1 k_1, \1 k_2) \ket \ket } \,, \\
\varepsilon (T, \m) & = & 
T\,s (T, \m) + \m\,n (T, \m) - p (T, \m)  =  \frac{  2\,e^{\textstyle \frac{\m}{T}}  \rho_t(T) 
 \bra\bra E \ket \ket  - [ \mu +  \bra E \ket ] \, p\, T^{-1}  }{
 1 + e^{\textstyle \frac{\m}{T}}  \rho_t(T)  \bra\bra  v(\1 k_1, \1 k_2)  \ket \ket } \,.
\end{eqnarray}
Here  $n (T, \m)$ is the particle density, while  $s (T, \m)$ and $\varepsilon (T, \m) $ denote the entropy 
and energy density, respectively.

\begin{figure}[ht]

\centerline{
\epsfig{figure=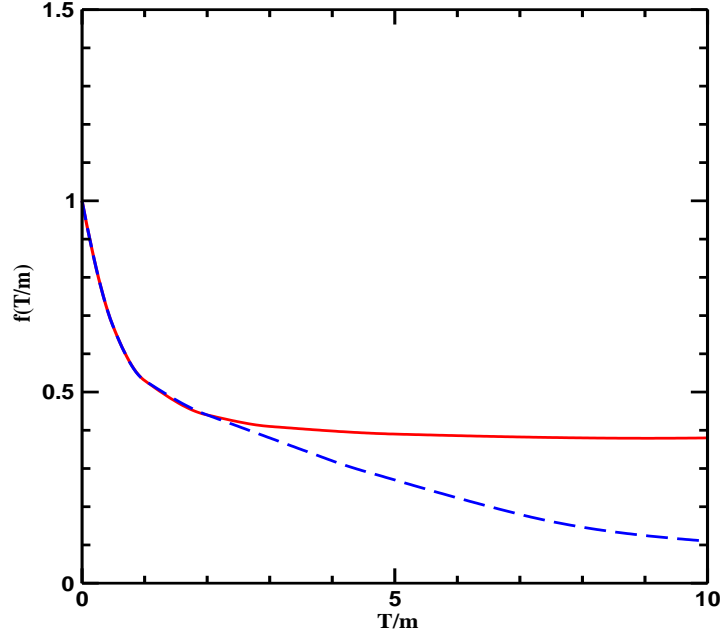,height=8.4cm,width=9.45cm}
}

\vspace*{0.05cm}

\caption{\label{Vcomp1}  
Comparison of the exact value of the second virial coefficient $a_2 (T)/a_2(0)$ (solid curve)
 with the averaged    value of the relativistic excluded volume 
$\alpha \bra\bra  v^{Urel}(\1 k_1, \1 k_2)  \ket \ket / (2 a_2(0))$ (dashed curve) given by Eq. \req{iiione}
for $\mu = 0$. 
The  normalization coefficient 
$\alpha \approx 1/1.065$  \req{vcorr} is introduced  to reproduce the low density results.
}

\end{figure}

In the low pressure limit $ 4 \, p \, v_{\rm o}  T^{-1} \ll 1$ the corresponding exponent  in 
\req{presiiii} can be expanded and the mean value of the relativistic excluded volume can be 
related to the second virial coefficient  $a_2 (T)$ as follows
\begin{equation} \label{iiizeroB}
\bra\bra v(\1 k_1, \1 k_2) \ket \ket  \approx   a_2 (T) ~ - ~ 
\frac{p}{T} \bra\bra v^2(\1 k_1, \1 k_2) \ket \ket  \, ,
\end{equation} 
which shows that at low pressures the average value of the relativistic excluded volume
should match the second virial coefficient  $a_2 (T)$, but should be smaller than $a_2 (T)$ at 
higher pressures and this behavior is clearly seen in Fig.~\ref{Vcomp1}. 



\vspace*{0.0cm}

A comparison of the particle densities \req{ntmu} and \req{iiizeroA} shows that despite the different 
formulae  for pressure the particle densities of these models have a very similar expression, but  in \req{iiizeroA} 
the second virial coefficient
is replaced by  the averaged value of the relativistic excluded volume $  \bra\bra  v(\1 k_1, \1 k_2)  \ket \ket  $.
Such a  complicated dependence of the particle density  \req{iiizeroA} on  $T$ and $\mu$ requires a nontrivial  analysis for  the limit of  high pressures.

To analyze  the high pressure limit $p \rightarrow \infty$ analytically I need an analytic  expression for  
the excluded volume. For this purpose I will use the ultrarelativistic expression derived in the Appendix \ref{RelEV_appd}: 
\begin{equation} \label{iiione}
v(\1 k_1, \1 k_{2})  \approx \frac{v^{Urel}(\1 k_1, \1 k_{2})}{2} \equiv   \frac{  v_{\rm o} }{2 }  
\lp  \frac{m}{E(\1 k_1)} + \frac{m}{E(\1 k_2)} \rp
\lp  1 + \cos^2 \lp \frac{\Theta_v}{2} \rp \rp^2  
 +  \frac{3 \, v_{\rm o} }{2}  \sin \lp \Theta_v \rp   \,\,.
\end{equation} 
As usual the total excluded volume $v^{Urel}(\1 k_1, \1 k_{2})$ is taken per particle. 
Eq. \req{iiione} is valid for $0 \le \Theta_v \le \frac{\pi}{2} $, to use it for 
$ \frac{\pi}{2} \le \Theta_v \le \pi $ one has to make a replacement 
$ \Theta_v \longrightarrow \pi -  \Theta_v$ in  \req{iiione}. 
Here  the coordinate system  is chosen in such a way that the angle $\Theta_v$ between the 3-vectors of 
particles' momenta $\1 k_1$ and  $\1 k_{2}$ coincides with the usual spherical angle $\Theta$ of  spherical
coordinates (see the  Appendix \ref{RelEV_appd}). 
To be specific,  the OZ-axies of the momentum space  coordinates of  the second particle  is chosen to coincide with the 3-vector of the momentum $\1 k_1$ of the first particle.



\begin{figure}

\mbox{
\hspace*{0.0cm}\psfig{figure=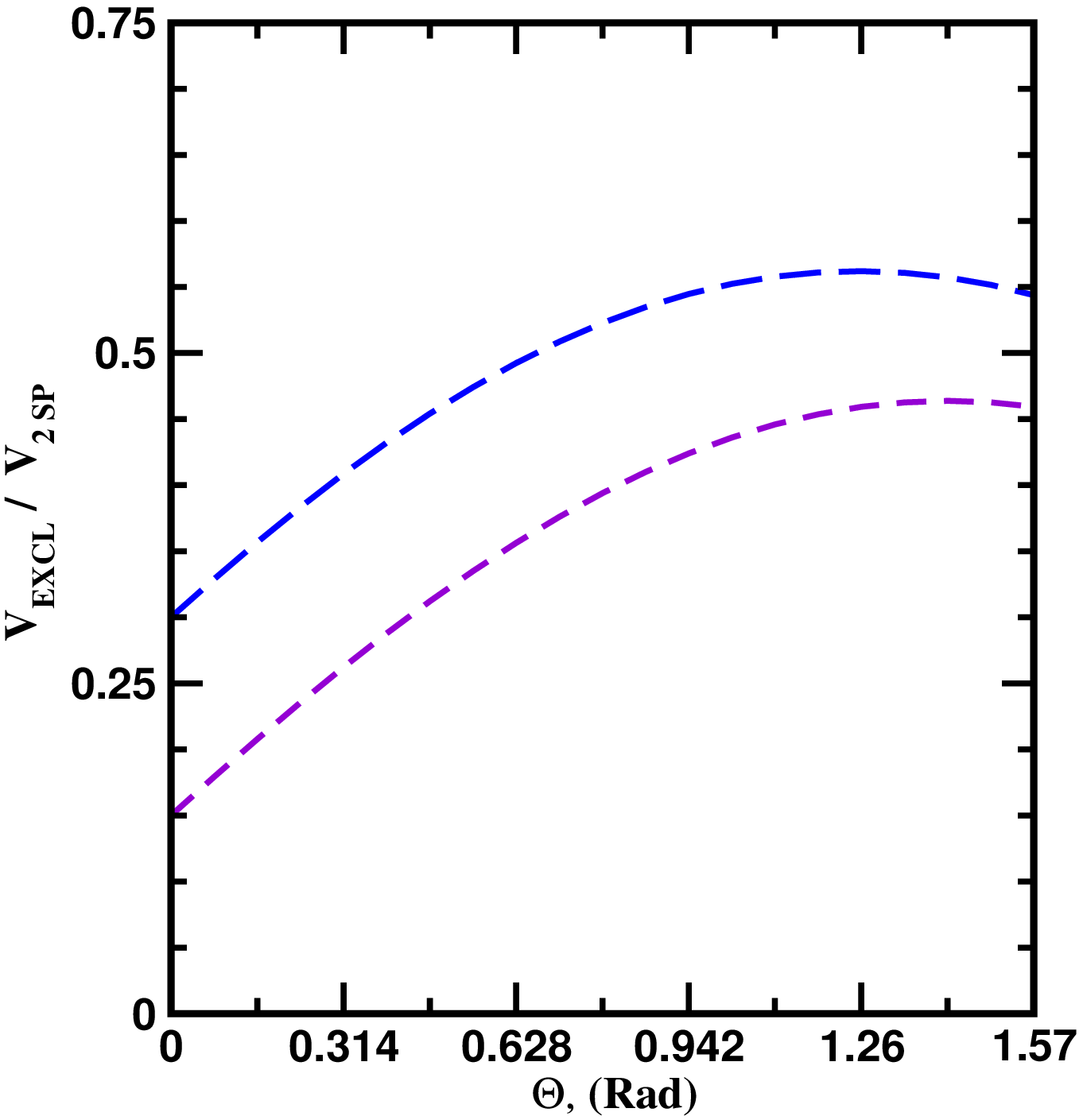,height=7.0cm,width=8.cm} 
\hspace*{-0.5cm} \psfig{figure=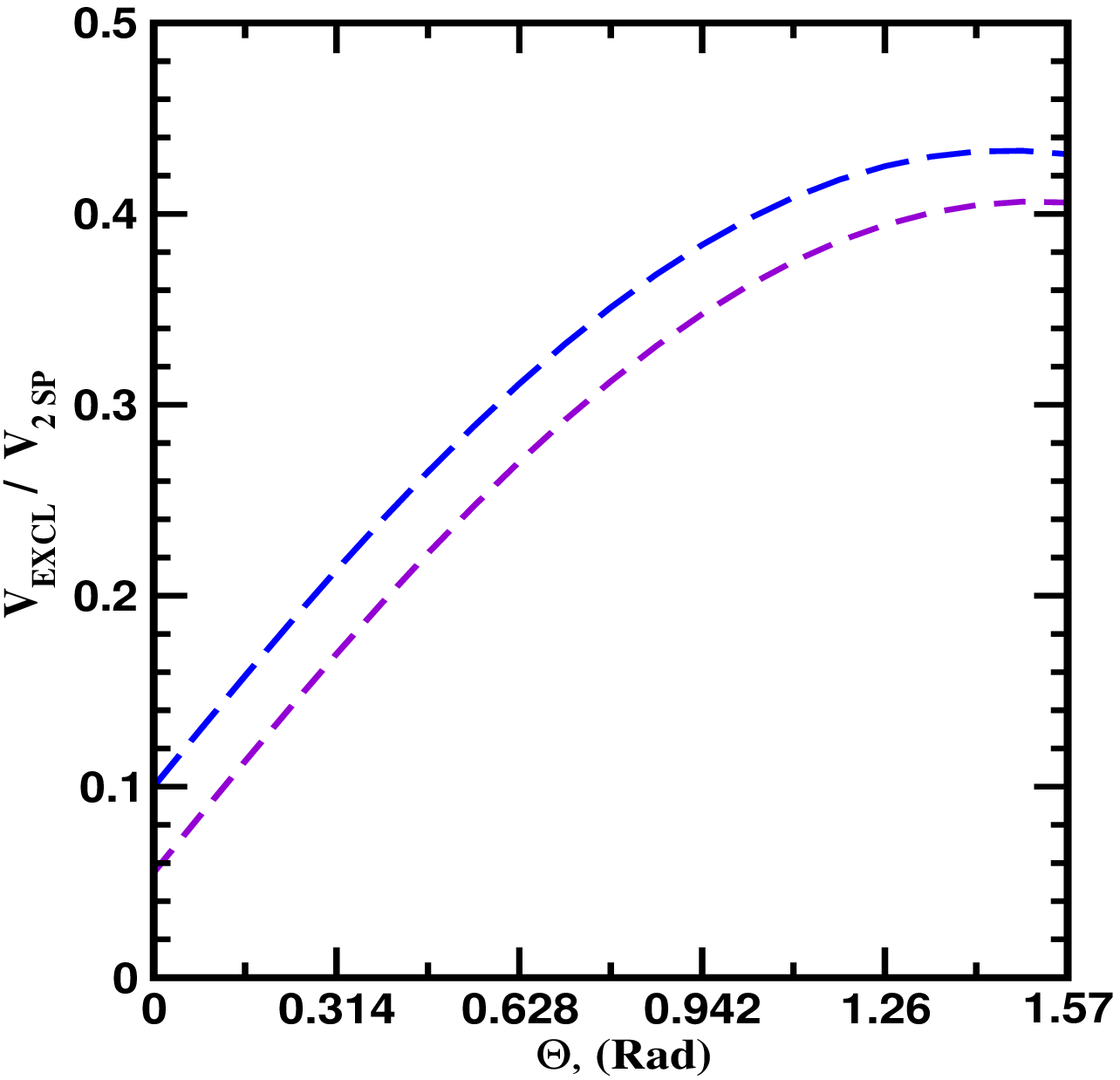,height=7.0cm,width=8.cm}
}

\vspace*{0.03cm}

\caption{\label{FigCyl}
Comparison of the relativistic  excluded volumes 
for highly contracted spheres. 
In the left panel the long dashed curve corresponds  to 
$\frac{E(k_1)}{m} = 2$ and  $\frac{E(k_2)}{m} = 10$ whereas 
the short dashed curve is found for  
$\frac{E(k_1)}{m} = 5$ and  $\frac{E(k_2)}{m} = 10$.
The corresponding values in the right panel 
are $\frac{E(k_1)}{m} = 10$,  $\frac{E(k_2)}{m} = 10$ (long dashed curve) and  
$\frac{E(k_1)}{m} = 10$,   $\frac{E(k_2)}{m} = 100$ (short dashed curve).
It shows that the  excluded volume for  $\Theta_v$ close to $\frac{\pi}{2}$ is 
finite always, while for the collinear velocities the excluded volume   approaches zero,  
if  both spheres are ultrarelativistic. 
}
\end{figure}

The Lorentz frame is chosen to be the rest frame of the whole system because 
otherwise the expression for pressure becomes cumbersome.  Here $v_{\rm o}$ stands for the eigen volume of particles which, for simplicity, are assumed to
have the same hard core radius and the same mass. 

Despite the fact that this 
equation  was obtained for  ultrarelativistic limit, it is to a within few per cent accurate in the whole range of parameters
(see Fig.~\ref{Vcomp1} and Appendix \ref{RelEV_appd}  for the details),
and, in addition,  it is  sufficiently simple to allow  one  the analytical manipulations.

\subsection{Proof of Causality in the  High Pressure Limit.}
As it is seen from the  expression for the relativistic excluded volume \req{iiione} 
for very high pressures only the  smallest values of the relativistic excluded volume will 
give a non-vanishing contribution to  the angular integrals of thermodynamic functions.
This means that only  $\Theta_v$-values  around $0$ and around $\pi$ will contribute  into the  thermodynamic functions (see Fig.~\ref{FigCyl}). 
Using the variable $x = \sin^2 \lp \Theta_v/2 \rp$, one can rewrite the
$\1 k_{2}$ angular integration as follows
\begin{eqnarray} \label{iiitwo}
\hspace*{-0.5cm}&&I_{\Th} ( k_1)  =  
\int
\hspace*{-0.1cm}
\frac{d{\1 k_2}}{(2\pi)^3} 
e^{\textstyle - \frac{  v ( \1 k_1, \1 k_2) p  }{T} } 
=
4 \int
\hspace*{-0.1cm}
\frac{d\,k_2  k^2_2 }{(2\pi)^2} 
\int_0^{0.5} \hspace*{-0.1cm} d\,x
\,\,e^{ -   \lp A C \lp 1 - \frac{x}{2} \rp^2  + B \sqrt{x (1 - x )} \rp   }
\,\,,\, \quad\quad \\
\label{iiithree}
\hspace*{-0.5cm}&&{\rm with}\hspace*{0.5cm}    A = 2 v_{\rm o}  \frac{p}{T} \,;
\hspace*{0.3cm} B =  \frac{3}{2} A\,;
\hspace*{0.3cm} C = 
\lp \frac{m}{E( k_1)} +  \frac{m}{E( k_2)} \rp\,\,,
\end{eqnarray}
where I have  accounted for the fact  that  the integration over the polar angle gives a factor $2\pi$
and that one should double the integral value in order to integrate  over a half of the $\Theta_v$ range.

Since $C \le 2$ in \req{iiithree} is a decreasing function of the momenta, then in the limit $A \gg 1$ one can 
account only for  the $\sqrt{x}$ dependence in the exponential in \req{iiitwo} because it is the leading one. 
Then integrating by parts one obtains 
\begin{equation} \label{iiifour}
\hspace*{-0.5cm}I_{\Th} ( k_1) \approx 
4 \int
\hspace*{-0.1cm}
\frac{d\,k_2  k^2_2 }{(2\pi)^2}
\,\,e^{ -  A C}
\int_0^{0.5} \hspace*{-0.1cm} d\,x
\,\,e^{ -  B \sqrt{x }    }
\approx
8 \int
\hspace*{-0.1cm}
\frac{d\,k_2  k^2_2 }{(2\pi)^2}
\,\,e^{ -  A C} \frac{1}{B^2}
\,\,.\, 
\end{equation}

Applying the result above to the pressure \req{presiii}, in the limit under consideration  one finds 
that the momentum integrals are decoupled and  one gets the following equation for pressure
\begin{equation} \label{iiifive}
\hspace*{-0.5cm} p (T, \m)  \approx  \frac{16\, T^3 e^{\frac{\mu}{T}}  }{9 \,v_{\rm o}^2 \, p^2 \, \rho_t(T) }
\left[  g  \int
\hspace*{-0.1cm}
\frac{d\,k  k^2}{ (2 \pi)^2 }
\,\,e^{\textstyle - \frac{E(k)}{T} - \frac{2 \, v_{\rm o} \, m  }{ T E(k)}\, p  }
\right]^2   \,.
\end{equation}

Our next step is to perform the gaussian integration in  Eq.  \req{iiifive}.
Analyzing the function 
\begin{equation} \label{iiisix}
F \equiv 2 \ln k - \frac{E (k)}{T} - A  \frac{m}{E (k)} \,
\end{equation} 
for  $A \gg 1$, one can safely use the ultrarelativistic approximation for particle momenta  
$k \approx E(k) \rightarrow \infty $. Then it is easy to see  that 
the function F in  \req{iiisix} has an extremum at 
\begin{equation} \label{iiiseven}
\frac{\partial F}{\partial E} =   \frac{2}{E} - \frac{1}{T} + A  \frac{m}{E^2} = 0 \quad  \Rightarrow \quad E = E^*  \approx  \frac{A\, m}{\sqrt{1 + \frac{A\, m}{T} }  - 1 }  \equiv  T \lp  \sqrt{1 + \frac{A\, m}{T} } + 1 \rp \,,
\end{equation}
which turns out to be a maximum, since the second derivative of F   \req{iiisix} is negative
\begin{equation} \label{iiieight}
\biggl. \frac{\partial^2 F}{\partial E^2}\biggr|_{E = E^*} \approx  -  \frac{2}{(E^*)^2} - 2\,A  
\frac{m}{(E^*)^3}  < 0
 \,.
\end{equation}

There are two independent ways to increase  pressure: one can increase 
the value of chemical potential while keeping
temperature fixed and vice versa.  I will consider the high chemical potential limit $\mu/T \gg 1$ for finite 
$T$ first,  since this case is rather unusual.  In this limit the above expressions can be simplified further on 
\begin{equation} \label{iiieightB}
E^* \approx \sqrt{2\, m\, v_{\rm o} \, p }\,, \quad \Rightarrow \quad  
\biggl. \frac{\partial^2 F}{\partial E^2}\biggr|_{E = E^*} \approx -  \frac{2}{T \, \sqrt{2\, m\, v_{\rm o} \, p } } \,.
\end{equation}
Here in the last step I explicitly substituted   the  expression for $A$. 
Performing the gaussian integration for momenta in \req{iiifive}, one arrives at
\begin{equation} \label{iiinine}
\int
\hspace*{-0.1cm}
\frac{d\,k  k^2}{ (2 \pi)^2 }
\,\,e^{\textstyle - \frac{E(k)}{T} - \frac{2 \, v_{\rm o} \, m  }{ T E(k)}\, p  } \approx
\frac{(E^*)^2}{ (2 \pi)^2 }\, \sqrt{\pi \, T \, E^*}\, e^{\textstyle - \frac{2 \, E^*}{T} }\,,
\end{equation}
which leads to the following equation for the most probable energy of particle $E^*$
\begin{equation} \label{iiiten}
E^* \approx  D\, T^4 \,  \,e^{\textstyle  \frac{\mu - 4 E^*}{T} }\,, \quad 
D \equiv \frac{ 8\, g^2 m^3 v_{\rm o}}{9\, \pi^3 \rho_t(T)}\,.
\end{equation}
As one can see, Eq. \req{iiiten} defines pressure of the system. Close inspection shows that the high pressure limit 
can be achieved, if the exponential in \req{iiiten}  diverges much slower than $\mu/T$. The latter defines the 
EOS in the leading order as 
\begin{equation} \label{iiieleven}
E^* \approx  \frac{\mu}{4}\,, \quad \Rightarrow \quad  
p \approx \frac{\mu^2}{32\, m \, v_{\rm o}}  \,.
\end{equation}
The left  hand side equation above demonstrates that in the $\mu/T \gg 1$ limit the natural energy scale is given by  chemical potential.  This is a new and important feature  of the relativistic VdW EOS compared to the previous findings.  

The right hand side Eq. \req{iiieleven}
allows one to find all other thermodynamic functions in this limit  from thermodynamic  identities:
\begin{eqnarray} \label{iiitwelve}
s  \approx  0 \,, \quad
n  \approx  \frac{2 \, p}{\mu} \,, \quad
\varepsilon  \equiv  T s + \mu n - p \approx p \,.
\end{eqnarray} 
Thus, it was shown  that for $ \mu/T \gg 1$  and finite $T$  the speed of sound $c_s$ in the leading order does 
not exceed the speed of light since
\begin{equation} \label{iiithirteen}
c_s^2 =  \biggl. \frac{\partial p}{\partial \varepsilon}\biggr|_{s/n} =  \frac{d\, p}{d\, \varepsilon} = 1\,.
\end{equation}
From Eq. \req{iiiten} it can be shown  that the last result holds  in all orders. 

It is interesting  that the left hand side  Eq.  \req{iiieightB} has a simple kinetic interpretation.  
Indeed, recalling that the pressure is the change of momentum during the collission time one can write  \req{iiiseven} as follows
(with $E^* = k^*$)
\begin{equation} \label{iiifourteen}
p =  \frac{(k^*)^2}{ 2\, m\,  v_{\rm o}} =  \frac{2\, k^*}{\pi R_{\rm o}^2}\cdot \frac{3\,v^* \gamma^*}{8\, R_{\rm o}} \cdot  \frac{1}{2}\,.
\end{equation}
In the last result the change of momentum during the collision with the wall  is $2\, k^*$, which takes the time 
$\frac{8\, R_{\rm o}}{3\,v^* \gamma^*}$. The latter is  twice of the Lorentz contracted height ($4/ 3 R_{\rm o}$)  of  the cylinder of the base $\pi R_{\rm o}^2$ which is  passed  with the speed $ v^* $.  
Here the particle velocity $v^*$ and the corresponding gamma-factor 
$\gamma^*$ are defined as $ v^* \gamma^* = k^* / m$. The rightmost factor $1/2$ in \req{iiifourteen}
accounts  for the fact that only a half of particles moving perpendicular to the wall has the momentum $- k^*$.   Thus, Eq. \req{iiifourteen} shows that in the limit under consideration the pressure is generated by the particle momenta which are perpendicular
to the wall.
This, of course, does not mean that all particles in the system have the momenta which are perpendicular to a
single wall. 
No, this means  that  in those places near the wall where the particles' momenta are not perpendicular (but are parallel) to it, 
 the  change of   momentum  $2 k^*$ is transferred  to the wall  by  the particles  located in  the inner regions of the system  whose 
momenta are perpendicular to the wall. 
Also it is easy to deduce that such a situation is possible, if 
the system is divided into the rectangular cells or boxes inside which the particles are moving along the height of the box and  their  momenta  are collinear, but they are  perpendicular 
to the
particles' momenta in all surrounding  cells. 
Of course,  inside of such a box each Lorentz contracted  sphere would generate an excluded volume which is  equal to a volume of a cylinder of height  $\frac{2\, R_{\rm o}}{\gamma^*} $ and base  $\pi R_{\rm o}^2$. This cylinder differs from the one involved 
in Eq. \req{iiifourteen}, but one notes that exactly the hight $\frac{4 R_{\rm o}}{3\, \gamma^*}$ is used in the derivation of the ultrarelativistic  
limit for  the relativistic excluded volume \req{vunnor} (see the Appendix  \ref{RelEV_appd}  for details).
Thus,  it is remarkable that the  low density  EOS extrapolated to very high values of the chemical potential, at which it is not supposed to be valid,  gives  a reasonable estimate 
for the pressure at high densities.

Another interesting conclusion that follows  from this limit is   that  for the relativistic VdW  systems existing in the nonrectangular 
volumes the relativistic analog of the dense packing  maybe  unstable.

The analysis of the limit $T/\mu \gg 1$ and finite $\mu$ also starts from Eqs. \req{iiifive}--\req{iiiseven}.
The function F from \req{iiisix} again has the maximum at $E^* \equiv E (k^*) = k^*$  defined by the 
right hand side Eq. \req{iiiseven}.
Now  the second derivative of function F becomes 
\begin{equation} \label{iiisixteen}
\biggl. \frac{\partial^2 F}{\partial E^2}\biggr|_{E = E^*} \approx  -  \frac{2}{(E^*)^2} - 2\,A  \frac{m}{(E^*)^3} = 
-  \frac{2 \, \sqrt{ 1 + \frac{A\, m}{T}   } }{ (E^*)^2}  \,.
\end{equation}
This result  allows one to perform the gaussian integration for momenta in \req{iiifive} for this limit  as
%
\vspace*{-0.2cm}
\begin{equation} \label{iiiseventeen}
\int
\hspace*{-0.1cm}
\frac{d\,k  k^2}{ (2 \pi)^2 }
\,\,e^{\textstyle - \frac{E(k)}{T} - \frac{3 \, v_{\rm o} \, m  }{ T E(k)}\, p  } \approx
\frac{(E^*)^3 \, e^{\textstyle -  2 \, \lp 1 + \frac{A\, m}{T}   \rp^{\frac{1}{2}}  } }{ (2 \pi)^2\, 
\lp 1 + \frac{A\, m}{T}   \rp^{\frac{1}{4}} }\,  I_\xi \lp 1 + \frac{A\, m}{T}   \rp \,, 
\end{equation}
where the auxiliary integral $I_\xi$ is defined as follows
\vspace*{-0.2cm}
\begin{equation} \label{iiieighteen}
 I_\xi (x) \equiv  \int\limits_{-\sqrt{x}}^{+ \infty} d \xi \,\,e^{- \xi^2} \,.
\end{equation}
The  above results can be used to find  the thermal density $\rho_t(T)$ in the limit $T\rightarrow \infty$ by the substitution $A = 0$. Using \req{iiiseventeen}, one can rewrite the equation for pressure \req{iiifive} as the 
equation for the unknown variable  $z \equiv A\, m / T$
\vspace*{-0.2cm}
\begin{equation} \label{iiinineteen}
z^3 \approx \,  e^{\textstyle  \frac{\mu}{T}  } \phi (z)\,, \quad 
\phi (z) \equiv \frac{3\, g\,v_{\rm o} \, m^3\, I_\xi^2 (1+z) \lp 1 + (1 + z )^{\frac{1}{2}} \rp^3  }{
 \lp 2\,\pi \,e^{\sqrt{1+z} }  \rp^2  I_\xi (1)  (1 + z )^{\frac{1}{2}} } \,.
\end{equation}

Before  continuing this  analysis  further on it is necessary to make the two comments concerning 
Eq. \req{iiinineteen}. 
First, rewriting  the right hand side Eq. \req{iiinineteen} in terms of pressure, one can see that
the value of  chemical potential is  formally  reduced exactly in three times. In other words, it looks like that in the limit of high temperature and finite $\mu$ the pressure of the relativistic VdW gas is created by the particles
with the charge being equal to the one third of their original charge. 
Second,  due to the nonmonotonic  dependence of  $\phi(z)$ in the right hand side Eq. \req{iiinineteen} it is
possible that the left hand side Eq. \req{iiinineteen} can have several solutions for some values of parameters.   
Leaving aside the discussion of this possibility,  I will further  consider  only 
such a solution of \req{iiinineteen}  which corresponds to the 
largest value of  the pressure \req{iiifive}. 

Since the function $\phi(z)$ does not have any explicit dependency on $T$ or  $\mu$, one can establish  a very convenient  relation 
\begin{equation} \label{iiitwenty}
 \frac{ \partial z }{ \partial T }   =  -  \frac{ \mu }{  T }  \,  \frac{ \partial z }{ \partial \mu }  
\end{equation}
between the partial derivatives  of  $z$ given by the left hand side Eq. \req{iiinineteen}. Using it,
one can calculate all the  thermodynamic functions  from the pressure  $ p = \alpha\, T^2 z $ (with $ \alpha \equiv (3\, m \, v_{\rm o})^{-1}$)
as: 
%
\vspace*{-0.2cm}
\begin{eqnarray} \label{iiitwone}
n  & \approx &  \alpha\, T^2 \, \frac{ \partial z }{ \partial \mu }\,,    \\
 \label{iiitwtwo}
s  & \approx &   \alpha\, \left[  2\, T \, z + T^2  \frac{ \partial z }{ \partial T } \right]  = \frac{2\, p - \mu n }{T}  \,,  \\
\varepsilon &  \equiv &   T s + \mu n - p \approx p \,.
 \label{iiitwthree}
\end{eqnarray} 
\vspace*{-0.2cm}
The last result  leads to  the causality condition  \req{iiithirteen} for the limit $T/\mu \gg 1$ and finite $\mu$ as well.
In fact, the above result can be extended to any $\mu > - \infty$ and any value of $T$ satisfying  the inequality
\begin{equation} \label{iiitwfour}
E^*  \approx   T \lp  \sqrt{1 + z } + 1 \rp \gg m \,,
\end{equation}
which is sufficient to derive Eq. \req{iiinineteen}. To show this it is sufficient to see that for $z = 0 $ one has 
the inequality $z^3 < e^{\textstyle  \frac{\mu}{T}  } \phi (z)$, which changes to the opposite inequality 
$z^3 > e^{\textstyle  \frac{\mu}{T}  } \phi (z)$ for $z = \infty$. Consequently, for any value  of $\mu $ and $T$
satisfying  \req{iiitwfour} the left hand side Eq. \req{iiinineteen} has at least one solution $z^* > 0$ for which 
one can establish Eqs. \req{iiitwenty}--\req{iiitwthree} and prove the validity of  the  causality condition \req{iiithirteen}.

The  model  \req{presiii} along with the analysis of high pressure limit  can be straightforwardly generalized to include several particle species.  
For the pressure $p (T, \{ \mu_i\} )$ of the mixture of  $N$-species  with  masses  $m_i$ 
$(i = \{1, 2,.., N\})$, degeneracy $g_i$,  hard core radius $R_i$  and chemical potentials $\mu_i$ is defined as a solution of the following equation
\begin{equation} \label{iiizeroC}
p (T, \{ \mu_i\} ) =
\int
\frac{d^3{\1 k_1}}{(2\pi)^3}
\frac{d^3{\1 k_2}}{(2\pi)^3}  
\sum_{i, j = 1}^N  \frac{T\, g_i\, g_j }{ \rho_{tot} (T, \{ \mu_l\}  )} 
\,e^{\textstyle \frac{\mu_i + \mu_j - v_{ij} (\1 k_1,\1 k_2)\, p\, -\,E_i(k_1)\, -\,E_j(k_2) }{T} } \,,
\end{equation} 
where the relativistic excluded volume per particle of species $i$ 
(with the  momentum $\1 k_1$) and $j$ (with the  momentum $\1 k_2$) is denoted as 
$v_{ij} (\1 k_1,\1 k_2)$,  $E_i(k_1) \equiv \sqrt{k_1^2 + m_i^2}$ and 
$E_j(k_2) \equiv \sqrt{k_2^2 + m_j^2}$ are the corresponding energies, and the total 
thermal density is given by the expression 
\begin{equation} \label{iiioneC}
 \rho_{tot} (T, \{ \mu_i\}  ) =
\int
\frac{d^3{\1 k}}{(2\pi)^3} 
\sum_{i = 1}^N ~  g_i  
\,e^{\textstyle \frac{\mu_i  -\,E_i(k) }{T} } \,. 
\end{equation} 
The excluded volume  $v_{ij} (\1 k_1,\1 k_2)$  can be accurately  approximated 
by  $\alpha\, v^{Urel}_{12}(R_i, R_j) / 2$ defined by Eqs. (\ref{vunnor}) and 
(\ref{vcorr}).  

The multicomponent generalization (\ref{iiizeroC}) is obtained in the same sequence 
of steps like the one component expression (\ref{presiii}). The only difference is  in
the definition of the total thermal density (\ref{iiioneC}) which now includes  the 
chemical potentials. 
Note also that the expression (\ref{iiizeroC})  by construction  recovers the  virial expansion up to the  second order 
at low
particle densities,  but 
it cannot be reduced to any of two extrapolations which are   suggested in  \cite{VdW:1889} and 
\cite{Gor:99} for  the multicomponent 
mixtures  and    carefully analyzed  in Ref. \cite{Zeeb:02}. Thus, the expression 
(\ref{iiizeroC}) removes the non-uniqueness of the VDW extrapolations to high densities,
if one requires  a  causal behavior in this limit.

\subsection{A Few Remarks on Possible Observables.}
In the preceding section   I considered  a relativistic analog of the VdW EOS  which  reproduces the 
virial expansion for the gas of  the Lorentz contracted rigid spheres at low particle  densities and, 
as was shown above,  
is  causal  at high densities. 
As one can see from the expression for particle density \req{iiizeroA} the one-particle momentum distribution function has a more complicated energy dependence than the usual Boltzmann 
distribution function, which would be interesting to check up experimentally. 
This, of course, is a very difficult task since the particle spectra measured in high energy nuclear 
collisions involve a strong collective flow which can easily hide or smear the additional 
energy  dependence.  
However, it is possible that  such a  complicated energy dependence of the momentum spectra and 
excluded volumes of  lightest hardons, i.e. pions and kaons, can be  verified  for highly accurate 
measurements, if the collective flow is correctly taken into account. 
The latter is  a tremendously difficult task  because it  is related to  the freeze-out 
problem in  relativistic hydrodynamics \cite{Bugaev:96} or hydro-cascade approach 
\cite{Bugaev:02HC}. 

Perhaps, 
it would be more realistically   to incorporate the developed approach into the effective models of 
nuclear/hadronic matter  
\cite{Dirk:91,Na:61,Gu:X,chirm}, and check the obtained EOS on the huge amount of data 
collected by the nuclear physics of intermediate energies. Since the suggested 
relativization  of  the VdW EOS makes it softer at high densities and, hence,  one can hope to  improve 
the description of the  nuclear/hadronic matter  properties (compressibility constant, elliptic flow, effective nucleon masses etc) at low temperatures and high baryonic 
densities \cite{IncompressibilityNew:2}. 

Also it is possible that the momentum spectra of this type  can help to extend 
the hydrodynamic description  
into the  region of large transversal momenta of hadrons ($p_T > 1.5 - 2 $ GeV)
which are  usually  thought to be too large to 
follow the  hydrodynamic regime  \cite{Heinz:05}. 

Another possibility to verify the suggested model is to study the angular  correlations 
of the hard core  particles emitted from the neighboring regions and the enhancement of
the particle yield  of those hadrons   which appear due to  the coalescence of the constituents with the strong short range repulsion.  
As was proven  in the  preceding  sections (see also Fig.~\ref{FigCyl}) the present model  predicts 
that the probability to find the neighboring  particles  with the collinear   velocities
is higher than to find the neighboring  particles with the non-collinear   velocities. 
Due to this reason the coalescence of particles  with the collinear velocities 
should be enhanced.  This effect   gets  strong, if   pressure  is high  and if particles are
relativistic in the local rest frame.
Therefore, it would be interesting to study the coalescence of any relativistic  constituents 
with hard core repulsion 
(quarks or hadrons)
at high pressures  in a spirit of the recombination model of Ref. \cite{Mull:03} and extend  it  results
for light hadrons to  lower values of transversal momenta.  
Perhaps,   the inclusion of  such an effect into consideration  may  essentially 
improve not only our understanding  of   
the quark  coalescence process,
but also the formation of deuterons and other nuclear fragments in relativistic nuclear collisions.  
This subject is, however, outside the scope of the present work. 

As a typical VdW  EOS the present model  should be   valid for  the low particle densities,
but 
the above  analysis of the limit   $\mu / T \gg 1$ for fixed $T$  led to  a surprisingly clear  kinetic expression 
for  the system's pressure \req{iiifourteen}.   Therefore, it is possible  that  this low density result
may provide 
us with  the correct hint to study  the relativistic analog of the dense
packing problem.  
Thus,  it would be interesting to verify, whether the above approach remains valid  for relativistic quantum treatment  because  there  are  several unsolved problems  
for the  systems of  relativistic bosons and/or  fermions which, on the one hand, are related to the problems
discussed here and which, on the other hand, 
maybe potentially  important for relativistic nuclear collisions and   for  nuclear astrophysics.

\section{Conclusions} 

In this  chapter I  presented the generalization of the virial and cluster expansions for the momentum 
dependent interparticle potentials.  This is necessary generalization of the corresponding nonrelativistic 
expansions. Such a generalization allowed me to formulate a new class of the hadronic matter EOS 
for the Lorentz contracted rigid spheres. 
The first representative from this class, Eq.~(\ref{vdwgc}),    is obtained as the usual VdW extrapolation to high densities. It  accounts for  the Lorentz  contraction of  the relativistic hard core repulsion at low densities, but breaks causality at high densities.
Since the Lorentz contraction stronger affects the lighter particles,  the hadronic EOS of this type require the  statistical treatment with  the  multicomponent  hard core repulsion between the  different particle species. 

A detailed analysis of  canonical and grand canonical descriptions of the hadronic mixtures with  the  multicomponent  hard core repulsion shows that the  two  VdW extrapolations, considered in this chapter,  give practically the  identical  results for the particle ratios observed at AGS and highest SPS  energies.  Also such an analysis  showed 
that the major conclusion of  Ref. \cite{Yen:97} on the values of hadronic  hard core radii is wrong, whereas 
the relativistic treatment of pion hard core repulsion presented here, is consistent with the results 
reported in Ref. \cite{Hgas:2}.

Further on I considered another VdW extrapolation for  the Lorentz  contracted rigid spheres, 
Eq.~(\ref{presiii}), and showed that in the limit of  high pressure  it obeys a causality.  The suggested  EOS, of course,  is not  a rigorous  result  based on the exact summation of the relativistic  virial  expansion, but,  similarly  to its  nonrelativistic counterpart, it  gives a  qualitatively  correct estimate for  high pressures. 
A new element of this EOS is that its  average excluded volume  coincides with the relativistic  second  
virial coefficient at small pressures only, whereas  at high pressures the average excluded volume  is 
smaller than  the relativistic  second  virial coefficient. As I showed above this fact leads to an interesting 
kinetic interpretation of the systems pressure (\ref{iiifourteen})  at $\mu/T \gg 1$ and finite  $T$, which 
corresponds to the relativistic analog  of dense packing.  Thus, the suggested model  EOS  
for  the Lorentz  contracted rigid spheres  predicts the existence of  relativistic polycrystalline structure 
in  hadronic matter  at  $\mu/T \gg 1$ and finite  $T$, if the deconfinement PT to QGP occurs at sufficiently high pressure.  Perhaps,  this can be the case for the effective nucleons in the Walecka-like models 
(see chapter 1) with the relativistic hard core repulsion which can be applied to various astrophysical objects.

Since the suggested EOS obeys a  causality in high pressure limit, it  can be used for the liquid phase of nuclear matter   within the GSMM concept introduced in the chapter 2. Also this EOS  is a good candidate to describe the EOS of  lightest hadrons (pions, kaons e.t.c.), which, according to lattice quantum chromodynamics (see a discussion in the next chapter),  can coexist with heavy QGP bags at high temperatures  above the cross-over transition.   

At high pressures the suggested EOS for  the Lorentz  contracted rigid spheres  favors the collinear velocities 
for neighboring particles because they occupy less volume and, hence, the surrounding media is  ``disturbed" less, whereas the configurations with the  perpendicular velocities of  neighboring particles are highly suppressed. Perhaps, such an  angular asymmetry  can  improve our understanding of statistical  aspects of a coalescence process of any relativistic constituents (nucleons or quarks) in a dense media. 

Finally, the relativistic  analog of the VdW EOS~(\ref{presiii}) resolves the problem of non-uniqueness  of 
the VdW extrapolation to the systems with the multiple hard core repulsion between the constituents, if 
one requires the causal behavior of the EOS at high pressures. 



\chapter{Exactly Solvable Phenomenological EOS of  the Deconfinement PT}


A natural step to generalize  the hadron gas  model analyzed in the previous chapter could be its extension to infinitely heavy hadronic states. Thus, one could include into statistical description 
all known hadronic  states and  the hypothetical hadronic states which may be  arbitrarily heavy. 
Then, using the properties of the cluster models like the FDM and SMM, one could analyze the obtained model 
for the PT existence. But historically this happened differently: the infinitely heavy hadronic states were introduced
into statistical mechanics by Rolf Hagedorn \cite{hagedorn-65, hagedorn-68}, first,  and then it was realized 
that the limiting temperature, the Hagedron temperature,   generated by the statistical bootstrap model (SBM) 
\cite{Frautschi:71} for the systems of all hadrons consisting of other hadrons, evidences for a new physics
beyond the Hagedron temperature.  Based on the Hagedorn results, Cabibbo and  Parisi concluded that 
the limiting temperature is the phase transition temperature  \cite{Parisi:75} to the state  of  partonic degrees of freedom, to quarks and gluons. This was the beginning of  relativistic nuclear physics.

Thus, from the very beginning the SBM \cite{hagedorn-65, hagedorn-68, Frautschi:71}  gave the first evidence
that an exponentially growing hadronic  mass spectrum
$g_H (m) =  \exp[ m/ T_H] ~ (m_{\rm o}/ m)^{a}$
for $m \rightarrow \infty$ (the constants $ m_{\rm o} $ and $a$ will be defined later) could lead to  new thermodynamics above the Hagedorn temperature
$T_H$. 
Originally, the divergence of thermodynamic functions at temperatures $T$  above $T_H$
was interpreted as the existence of a limiting temperature for hadrons.
In other words,
it is impossible
to build the hadronic thermostat above $T_H$.
A few years later an  exponential form of the  asymptotic mass spectrum  was found in
the MIT bag model \cite{BagModel} and the associated limiting temperature was correctly 
interpreted as the phase transition temperature  to the partonic degrees of freedom \cite{Parisi:75}.
These results initiated  extensive studies of hadronic thermodynamics within the framework of
the GBM  \cite{Kapusta:81,Kapusta:82}.
The SBM with a non-zero  proper volume \cite{Vol:1}
of hadronic bags was solved analytically \cite{Goren:81}
by the Laplace transform  to the isobaric ensemble (see chapters 1 and 2)  and the existence of phase transition
from hadronic to partonic matter, nowadays  known as  QGP,
was shown.
Since then  the  more sophisticated formulations of the SBM  \cite{SBM:new, Blaschke:04} were suggested.

The major achievement of the SBM  is  that it naturally explains why the
temperature of secondary hadrons created in  collisions of elementary particles at high energies  
cannot exceed $T_H$.
However,  this  result  is based  on {\it two related  assumptions.}
First, the grand canonical formulation for the SBM is appropriate, and
second, the resonances of infinite mass should contribute
to thermodynamic functions.  

However,  the story with the SBM is not over yet because 
very recently, using the microcanonical  formulation,  my collaborators and I  showed  
\cite{HThermostat:1, Thermostat:3}  that 
in the absence of any restrictions  on the mass,   resonances with the  Hagedorn 
mass spectrum behave as  perfect thermostats  and perfect chemical reservoirs,
i.e. they impart the Hagedorn temperature $T_H$  to  particles which are in thermal
contact and force them to be in chemical equilibrium. 
These findings  led  to  the following  significant  conclusions  \cite{HThermostat:1}:  
(i) canonical and/or grand canonical formulations of the statistical mechanics 
of any system coupled to a Hagedorn thermostat  are not equivalent to the 
microcanonical one;  (ii)  in the presence of  the Hagedorn thermostat 
{\it it is improper to include 
any temperature other than $T_H$}  into the   canonical and/or grand canonical  
description; (iii) the Hagedorn thermostat  generates a volume independent 
concentration of the particles  in chemical equilibrium  with it \cite{HThermostat:1}.

The first of these results  was obtained by  R. D. Carlitz \cite{Carlitz:72}, who 
analyzed the nonequivalence conditions for  the  canonical and microcanonical formulations
of the SBM. However,  Carlitz's  somewhat complicated and detailed  mathematical analysis
of the problem prevented  the observation of  the  consequences (ii) and (iii)
regarding  the themostatic properties of  Hagedorn systems. 
Therefore, 
the entire  framework of the SBM
and GBM, which is also based on the two major  assumptions discussed above,
should  be revisited. 
In other words,   it is necessary to return to the foundations of the statistical mechanics of hadrons
and study the role of the Hagedorn mass spectrum 
for  finite masses  of hadronic resonances above the cut-off value $m_{\rm o}$, 
below which the hadron mass spectrum  is discrete.  
Such an analysis for an arbitrary  value of  $a$  in  $g_H (m)$
was  done   in \cite{HThermostat:2}.

The results on thermostatic properties of  heavy hadronic resonances maybe important for the elementary 
particle collisions, in which the short range repulsion, perhaps, can be ignored, but  to use it to model A+A 
collisions one must get rid of the artificial singularities in the GCE. For this purpose it is necessary to use some additional physical input. In the chapter 2  I considered  the GBM solution in finite volume.  This model 
generalizes the SBM and accounts for the hard core repulsion between hadronic bags. In the GCE  such a treatment (at least in the simplest case) generates an exponential decreasing mass spectrum  which removes 
an artificial singularities of the SBM. This exactly solvable model \cite{Goren:81}, however, is just a toy model 
which cannot be used to describe simultaneously  the 1$^{st}$ deconfinement PT  and a cross-over. 
A recent try \cite{Goren:05}   to revitalize the GBM  does not look reasonable because the  main assumption
of Ref.  \cite{Goren:05}  about the existence of the line along which the order of PT gradually increases 
contradicts the whole concept of 
critical phenomena  which is based on a assumption that  any PT occurs due to the break down of certain symmetry in the system under consideration.  However, as I will show below the GBM lacks the surface  free energy of large hadronic bags.
If the latter is included into statistical treatment, then such a model 
\cite{Bugaev:07, Bugaev:07new}, the quark-gluon bags with surface tension (QGBST) model, is able to   simultaneously  describe the 1$^{st}$ deconfinement PT  and cross-over. 

Another possibility to remove the artificial singularities of the SBM in the GCE is to use the properties of the Mott transition and  introduce the special mass dependent  width of heavy resonances 
\cite{Blaschke:03, Blaschke:04, Blaschke:05}.  Such a model allows one to describe the data on thermodynamic functions obtained by the lattice quantum chromodynamics simulations. 
Also it explains the reason why very heavy hadronic resonances are not observed 
experimentally. 

The chapter is based on the following works \cite{ Bugaev:07b, HThermostat:1, HThermostat:2, Moretto:05, Blaschke:03, Blaschke:04, Blaschke:05, Bugaev:07, Bugaev:07new, Bugaev:08new}.

\section{Microcanonical Analysis of the Hagedorn Model}

The microcanonical analysis of the Hagedorn model 
is important 
for  understanding  the differences and similarities
between  A+A and elementary particle 
collisions at high energies.
There are two temperatures measured in A+A collisions that are very close to
the transition temperature $T_{Tr}$ from hadron gas to QGP  calculated 
from  the lattice quantum 
chromodynamics  \cite{Lattice}  at vanishing baryonic density. 
The first  is the 
chemical freeze-out 
temperature
at vanishing baryonic density 
 $T_{Chem}  \approx 175 \pm 10$ MeV of the most  abundant hadrons (pions, kaons, nucleons 
{\it etc})  extracted from particle multiplicities  
at highest  SPS \cite{Hgas, Hgas:2} and all RHIC \cite{RHIC:Chem, RHIC:Chem2} energies.
Within the error bars $T_{Tr} \approx T_{Chem}$ 
is  also  very close to 
the kinetic freeze-out temperature  $T_{Kin} $ (i.e. hadronization temperature)
found from the transverse mass spectra 
of   heavy,  weakly interacting hadrons
such as
$\Omega$ hyperons, $J/\psi$ and $\psi^\prime$ mesons at the highest SPS  energy 
\cite{Bugaev:01d, Bugaev:02, Bugaev:02a}, 
and
$\Omega$-hyperons 
\cite{Bugaev:02b, Bugaev:03, QM02, RHIC:Hadronization2,RHIC:Hadronization3} 
and $\phi$ meson \cite{Bugaev:02b, Bugaev:03, RHIC:Hadronization3,RHIC:Hadronization4}    
at   $\sqrt{s} = 130$ A$\cdot$GeV and $\sqrt{s} = 200$ A$\cdot$GeV
energies of  RHIC.
The existence of the deconfinement  transition naturally explains the same value
for all these temperatures.

Can 
the same logic  be  applied to the collisions  of elementary particles, 
where  the  formation of a deconfined quark gluon matter 
 is rather problematic? 
The fact that 
the  hadronization  temperature \cite{Becattini:1} 
and inverse slopes of the transverse mass spectra  of  various 
hadrons \cite{Gazdzicki:04} found in  elementary particle  collisions at high energies 
are similar to those ones for A+A collisions
is tantalizing.

Hagedorn noted that the hadronic mass spectrum (level density) has the asymptotic ($m \rightarrow \infty$) form
\begin{equation}
	\rho_{\cal H}(m) \approx \exp \left({m}/{T_{\cal H}}\right) ,
\label{hagedorn}
\end{equation}	
where $m$ is the mass of the hadron in question and $T_{\cal H}$ is the parameter (temperature) controlling the exponential rise of the mass spectrum \cite{hagedorn-65,hagedorn-68}. The question of the mass range over which (\ref{hagedorn}) is valid is still under discussion \cite{Blaschke:04}.

The M.I.T. bag model \cite{BagModel} of partonic matter produces the same behavior via a constant pressure $B$ of the containing ``bag'' \cite{Kapusta:81,Kapusta:82}. In the absence of conserved charges the bag pressure $B$ forces a constant temperature $T_B$ and energy density $\epsilon$ from which it follows that the bag entropy is
\begin{equation}
	S = {\epsilon V}/{T_B} = {m}/{T_B}
\label{bag-entropy}
\end{equation}
where $V$ and $m$ are the volume and mass of the bag respectively. This leads to a bag mass spectrum $\exp \left( S \right)$ identical to Eq.~(\ref{hagedorn}) \cite{Kapusta:81,Kapusta:82}. 
This property implies the lack of any surface energy associated with the bag.

A variety of experiments with high energy ($\sqrt{s} \ge 30-50 $ GeV) elementary particle collisions on very different systems indicate a constant temperature characterizing both chemical and  thermal equilibrium  at vanishing baryonic densities  \cite{alexopoulos-02,Gazdzicki:04,Becattini:1}. It is interesting to explore the connection of these empirical temperatures with the Hagedorn temperature $T_{\cal H}$ on one hand and the bag temperature $T_{B}$ on the other.

I will show that the temperature of any such ${\cal H}$ system is not affected by the extrinsic injection of energy into the system but it is encoded and strictly enforced by the fixed temperature of the mass spectrum.

The insertion of an exponential spectrum such as Eq.~(\ref{hagedorn}) in the partition function
\begin{equation}
	{\cal Z} \left( T \right) = \int\limits^\infty_{E_{min}} \rho_{\cal H} \left( E \right) e^{-\frac{E}{T}} dE
\label{part-fcn}
\end{equation}
led to the incorrect conclusion that the entire range of temperatures $0 \le T < T_{\cal H}$ would be accessible and that $T_{\cal H}$ is the limiting temperature of the system. 

In order to see the origin of this erroneous conclusion, let me  consider the following illuminating exercise.
Consider a system $A$ composed of ice and water at standard pressure. For such a system the temperature (kelvin) is $T_A=273$ K. Because of coexistence, one can feed or extract heat to/from the system without changing $T_A$. The system $A$ is a thermostat.

If a quantity $Q$ of heat is added to the sytem, the change in entropy is
\begin{equation}
	\Delta S = {Q}/{T_A}.
\end{equation}
The level density of $A$ is then 
\begin{equation}
	\rho(Q) = S_0e^{Q/T_A}\approx Ke^{E/T_A}.
\end{equation}

The level density, or spectrum, is exponential in $E$ and depends only on the intrinsic ``parameter" $T_A$. Let us calculate the partition function of $A$:
\begin{equation}
\label{4}
	Z(T) = \int e^{E/T_A} e^{-E/T} dE = \int e^{-\left(\frac{1}{T}-\frac{1}{T_A}\right) E} dE 
	= \frac{T_A T}{T_A-T}
\end{equation}
This seems to indicate that $A$ can assume {\em any} temperature $0\le T<T_A$. But, by hypothesis, the only temperature possible for $A$ is $T_A$. What is the trouble?

Let me  consider two systems $A, B$ with level densities $\rho _A$ and $\rho _B$. Let the systems be thermally coupled to each other with total energy $E$.  Now I can  calculate the distribution in energies between the two systems,
\begin{equation}
\label{5}
	\rho _T(x) = \rho_A(E-x)\rho _B(x)
\end{equation}
Let $A$ be a ``thermostat", i.e. $\rho_A=e^{\epsilon /T_A}$. Then 
\begin{equation}
\label{6}
	\rho_T(x) = \exp\left(\frac{E-x}{T_A}\right) \rho_B(x) = e^{E/T_A}e^{-x/T_A}\rho_B(x).
\end{equation}
Let me  integrate over $x$ for macroscopic systems and macroscopic value of $E$
\begin{equation}
\label{7}
	\int\rho_T(x) dx = e^{E/T_A}\int e^{-x/T_A}\rho_B(x) dx = e^{E/T_A} Z_B(T_A).
\end{equation}
This is the origin of the partition function $Z_B(T_A)$ and the meaning of ``implicit" thermostat. By changing ``thermostat"  one  can change $T_A$ and the temperature of $B$.

Thus, every time we  construct a partition function, we imply the gedanken experiment of connecting the system to a thermostat, and that this experiment is actually possible for the system we are studying. Does this always work?

To see this, let us look for the most probable value of the distribution $\rho_T(x)$, which defines the equilibrium partition, by taking the log and differentiating:
\begin{equation}
\label{eq1}
	\ln \rho_T(x) = \ln \rho_A(E-x) + \ln \rho_B(x)
\end{equation}

\begin{equation}
\label{eq2}
	{\partial\ln\rho_T(x)}/{\partial x} = -\left.{\partial\ln\rho_A}/{\partial x}\right| +
	\left.{\partial\ln\rho_B}/{\partial x}\right| = 0 
\end{equation}
or
\begin{equation}
\label{eq3}
	{1}/{T_A} = {1}/{T_B}.
\end{equation}
For this to be possible, it is necessary that $\rho_A$ and $\rho_B$ admit the {\em same} logarithmic derivative somewhere in the allowed range of energy $x$.

Usually, and always for concave functions, $S(x) = \ln\rho(x)$ and $T=(\partial S/\partial x)^{-1}$ is such that  $0\le T\le \infty $.  Thus, for such systems it is possible to match derivatives for whatever value of $E$. Thermal equilibrium is achievable over a broad range of temperatures.

However, if $S_A(E) = \ln\rho_A(E)$ is linear in $E$, then $T_A=(\partial S/\partial E)^{-1}$ is a constant, independent of $E$.\ \ In this case, it is up to $B$ to look for the value of $x$ at which its logarithmic derivative matches $1/T_A$. The system $A$ is a ``thermostat" at $T=T_A$ and $B$ can only try to assume the value $T=T_B=T_A$, if it can do so.

Now suppose that also $S_B(E) = \ln\rho_B(E)$ is linear in $E$ with an inverse slope $T_B$. This means that only if $T_A=T_B$ is equilibrium possible, and the partition function of $B$, $Z_B$ is meaningfully defined only for $T=T_B$ and not for $0\le T\le T_B$.  One  cannot force a temperature $T\ne T_B$ on a thermostat. It can only have its own intrinsic temperature $T_B$.

Placing systems $A$ and $B$ into contact will lead to a continuous heat flow from one system to the other. Thermal equilibrium is not achievable.

Summarizing: it is permissible to calculate a system's partition function only,  if its $S(E)$ admits as inverse derivatives values such as one wants to impose through the Laplace transform. Failing that, the resulting partition function does not satisfy any thermodynamic criterion.

Carlitz  noticed \cite{Carlitz:72}  that Eq.~(\ref{hagedorn}) leads to a nonequivalence between the (grand)canonical  and microcanonical descriptions.\ \ However,  the striking consequences  of this fact reported in above  were not noticed and appreciated.

I  show here that the exponential form of the mass spectrum in Eq.~(\ref{hagedorn}) forces the \underline{unique} temperature $T_{\cal H}$ on both the chemical and thermal
equilibria associated with it. Below 
 the consequences of this hitherto unappreciated fact are discussed.

To begin, I  show that a system $\cal H$ possessing a Hagedorn-like spectrum, characterized by an entropy of the form (\ref{bag-entropy}), not only has a unique microcanonical temperature $T_{\cal H}$
\begin{equation}
	T_{\cal H} = \left( {d S}/{d E}\right)^{-1} = T_B ,
\label{temperature}
\end{equation}
but also imparts this same temperature to any other system to which $\cal H$ is coupled. In the language of thermodynamics: $\cal H$ is a perfect thermostat with the constant temperature $T_{\cal H}$.

Incidentally, it is worth noting that a perfect thermostat is indifferent to the transfer of any portion of its energy to any parcel within itself, no matter how small. In other words, it is at the limit of phase stability and the internal fluctuations of its energy density are maximal.  Therefore it does not matter whether this thermostat is one large bag or it is fragmented in an arbitrary number of smaller bags or, equivalently, it is a system of hadrons with a spectrum given by Eq.~(\ref{hagedorn}). This has no consequences on the properties of ${\cal H}$ as one can  see below.

\subsection{Harmonic Oscillator Coupled to $\cal H$.}
In order to demonstrate the thermostatic behavior of a Hagedorn system, let me  begin by coupling $\cal H$ to a one dimensional harmonic oscillator and use a microcanonical treatment. The unnormalized probability $P(\varepsilon)$ for finding an excitation energy $\varepsilon$ in the harmonic oscillator out of the system's total energy $E$ is
\begin{equation}
	P(\varepsilon)  \sim  \rho_{\cal H}(E-\varepsilon)  \rho_{\rm osc} (\varepsilon)
	= \exp \left( \frac{E-\varepsilon}{T_{\cal H}} \right) = \rho_{\cal H}(E) \exp \left(
	-\frac{\varepsilon}{T_{\cal H}} \right).
\label{sho}
\end{equation}
Recall that for a one dimensional harmonic oscillator $\rho_{\rm osc}$ is a constant. The energy spectrum of the oscillator is canonical up to the upper limit ${\varepsilon}_{max} = E$ with an inverse slope (temperature) of $T_{\cal H}$ independent of $E$. The mean value of the energy of the oscillator is
\begin{equation}
	\overline{\varepsilon} = T_{\cal H} \left[ 1 - \frac{E / T_{\cal H}}{\exp\left( E / T_{\cal H} \right)-1} 
	\right] .
\label{ave-ho-e}
\end{equation}
Thus in the limit that $E \rightarrow \infty$: $\overline{\varepsilon} \rightarrow T_{\cal H}$, i.e. no temperature other that $T_{\cal H}$ is admitted. { In the standard language of statistical mechanics
this example means that a one dimensional harmonic oscillator  can be used as an ideal thermometer. }

\subsection{An Ideal Vapor Coupled to $\cal H$.}
For a physically more relevant example, let me  consider a vapor of $N\gg1$ non-interacting particles of mass $m$ coupled to $\cal H$. The microcanonical level density of the vapor with kinetic energy $\varepsilon$ is
\begin{equation}
	\rho_{\rm vapor}(\varepsilon) = \frac{V^N}{N!\left( \frac{3}{2}N \right)!} 
	\left( \frac{m \varepsilon}{2\pi} \right)^{\frac{3}{2}N} ,
\label{vapor-part}
\end{equation}
where $V$ is the volume. The microcanonical partition of the total system is 
\begin{equation}
	\rho_{\rm total}(E,\varepsilon)  =  \rho_{\cal H}(E-\varepsilon)\rho_{\rm vapor}(\varepsilon) 
	=  \frac{V^N}{N! \left( \frac{3}{2}N \right)!} \left( \frac{m \varepsilon}{2\pi} \right)^{\frac{3}{2}N}
	e^{\frac{E-mN-\varepsilon}{T_{\cal H}}} .
\label{full-part}
\end{equation}
The distribution of the vapor is exactly canonical up to $\varepsilon_{max}=E$, if the particles are independently present, or $\varepsilon_{max}=E-mN$, if the particles are generated by $\cal H$. In either case, the temperature of the vapor is always $T_{\cal H}$.

At fixed $N$ the maximum of $\rho_{\rm total}(E,\varepsilon)$ with respect to $\varepsilon$ gives the most probable kinetic energy per particle as
\begin{equation}
	\frac{\partial \ln \rho_{\rm total}(E,\varepsilon)}{\partial \varepsilon} = 
	\frac{3N}{2\varepsilon} - \frac{1}{T_{\cal H}} = 0 \Rightarrow 
	\frac{\varepsilon}{N} = \frac{3}{2}T_{\cal H} ,
\label{max-01}
\end{equation}
provided that $E \ge mN + \frac{3}{2}N T_{\cal H}$. For $mN < E < mN + \frac{3}{2}N T_{\cal H}$, the most probable kinetic energy per particle value is $\frac{\varepsilon}{N} = \frac{E}{N} - m < \frac{3}{2} T_{\cal H}$; for $E \le mN$, $\frac{\varepsilon}{N}=0$. $T_{\cal H}$ is the sole temperature characterizing the distribution up to the microcanonical cut off, which may be above or below the maximum of the distribution, since the form of $\rho_{\rm total}(E,\varepsilon)$ is $E$-independent.

The maximum of $\rho_{\rm total}(E,\varepsilon)$ with respect to $N$ at fixed $V$ is given by
\begin{equation}
	\frac{\partial \ln \rho_{\rm total}(E,\varepsilon)}{\partial N} = -\frac{m}{T_{\cal H}} + \ln\left[\frac{V}{N} \left( \frac{mT_{\cal H}}{2 \pi}\right)^{\frac{3}{2}}\right]= 0,
\label{max-02}
\end{equation}
where Eq.~(\ref{max-01}) was used for $\varepsilon$. Thus the most probable particle density of the vapor is 
{\it independent of} $V$:
\begin{equation}
	\frac{N}{V} = \left( \frac{m T_{\cal H}}{2 \pi} \right)^{\frac{3}{2}} e^{-\frac{m}{T_{\cal H}}} \equiv n_{\cal H}  .
\label{numberpp}
\end{equation}
Equation~(\ref{numberpp}) demonstrates that not only is $\cal H$ a perfect thermostat but also a perfect particle reservoir. Particles of different mass $m$ will be in chemical equilibrium with each other. At equilibrium, particles emitted from $\cal H$ form a saturated vapor at coexistence with $\cal H$ at temperature $T_{\cal H}$. This describes a first order phase transition (hadronic to partonic). Coexistence occurs at a single temperature fixed by the bag pressure. 
{ Different arguments lead to a similar conclusion concerning the existence of a 
phase transition \cite{Carlitz:72}. }

These key results explain the common value of: the hadronization temperatures obtained within the statistical hadronization model  at vanishing baryonic densities \cite{Becattini:1}; 
the inverse slopes of the transverse mass spectra of hadrons observed in high energy elementary particle collisions  
with the transverse momenta $p_T \le 1$ GeV \cite{alexopoulos-02,Gazdzicki:04}; and the transition temperature from lattice QCD calculations for low baryonic density \cite{Lattice}. For further discussion see \cite{HThermostat:2}.

Let me  consider the case in which the vapor particle mass $m$ is free. The system's level density $\rho_{\rm total}(E,\varepsilon)$ is still given by Eq.~(\ref{full-part}). Using Eqs.~(\ref{max-01}) and~(\ref{numberpp}), one finds the most probable value of the system's level density as $\rho_{\rm total}^*(E,\varepsilon) \approx \exp\left[ S^* \right]$, where the entropy is $S^* = E/T_{\cal H} + N $. Differentiating $\rho_{\rm total}^*(E,\varepsilon)$ with respect to $m$ and applying Eq.~(\ref{numberpp}) gives

\vspace*{-0.1cm}

\begin{equation}
	\frac{\partial \ln \rho_{\rm total}^*(E,\varepsilon)}{\partial m} = N  \left[ \frac{3}{2 m}- 
	\frac{1}{T_{\cal H} } \right]  ~ =~ 0 \Rightarrow  m = \frac{3}{2}T_{\cal H} ,
\label{masspp}
\end{equation}

\vspace*{-0.1cm}
\noindent
i.e. the last equality provides the maximum of level density for $N \neq 0$. Since all the intrinsic statistical weights in $\rho_{\rm total}^*(E,\varepsilon)$ are factored into a single $\cal H$, the system breaks into fragments with $m=\frac{3}{2}T_{\cal H}$ except for one whose mass is determined by mass/energy conservation.

Substituting the most probable value of $\varepsilon$ and $m$ into the most probable value of $N$ gives the vapor concentration

\vspace*{-0.5cm}

\begin{equation}
	\frac{N}{V} = \left( \frac{3}{4 \pi e} \right)^{\frac{3}{2}} T_{\cal H}^{3} .
\label{concpp}
\end{equation}

\vspace*{-0.1cm}
\noindent
The vapor density of nonrelativistic particles acquires the typical ultrarelativistic limit form.

If the mass given by Eq.~(\ref{masspp}) does not exist, then the level density's most probable value $\rho_{\rm total}^*(E,\varepsilon)$ corresponds to the mass $m^*$ nearest to $\frac{3}{2}T_{\cal H}$ and $N (m^*)$ given by Eq. (\ref{numberpp}). The value of $m^*$ that maximizes the level density $\rho_{\rm total}^*(E,\varepsilon)$ is the pion mass.


\subsection{$\cal H$ as a Radiant Bag.}
Do the emitted particles need to remain in the proximity of $\cal H$ to insure equilibrium? Let me  assume that $\cal H$ is a bag thick enough to absorb any given particle of the vapor striking it. Then, detailed balance requires that on average $\cal H$ radiates back the same particle. Under these conditions particles can be considered to be effectively emitted from the surface of $\cal H$. {\it Thus the relevant fluxes do not depend in any way upon the inner structure of $\cal H$, nor on the presence of the outer vapor.}

The results in equations (\ref{max-01}) and (\ref{numberpp}) show that the saturated vapor concentration depends only on $m$ and $T_{\cal H}$ as long as $\cal H$ is present. A decrease in $V$ does not increase the vapor concentration, but induces a condensation of the corresponding amount of energy out of the vapor  into $\cal H$. An increase in $V$ keeps the vapor concentration constant via evaporation of the corresponding amount of energy out of $\cal H$  into the vapor. This is reminiscent of liquid-vapor equilibrium at fixed temperature, except that here coexistence occurs at a single temperature $T_{\cal H}$, rather than over a range of temperatures as in ordinary fluids.

The bag wall is Janus faced: one side faces the partonic world, and, aside from conserved charges, radiates a partonic black body radiation responsible for balancing the bag pressure; the other side faces the hadronic world and radiates a hadronic black body radiation, mostly pions. Both sides of the bag wall are at temperature $T_{\cal H}$. It is tempting to attribute most, if not all, of the hadronic and partonic properties to the wall itself, possibly even the capability to enforce conservation laws globally (quantum number conductivity). Despite the fact that this wall is an insurmountable horizon, hadronic measurements such as bag size and total radiance can yield some properties of the partonic world, e.g. the number of degrees of freedom \cite{alexopoulos-02}.

One can estimate an upper limit for the emission time using the outward energy flux of particles radiated from the bag. At equilibrium the in-going and out-going fluxes must be the same, thus the outward flux of particles 
in the nonrelativistic approximation using Eq.~(\ref{numberpp}) is
\vspace*{-0.2cm}
\noindent

\begin{equation}
	\varphi_{n_{\cal H}} \simeq \frac{n_{\cal H}}{4} \left( \frac{m}{m+2T_{\cal H}}\right) \sqrt{8\frac{T_{\cal H}}{\pi m}} . 
\label{part-flux}
\end{equation}
\vspace*{-0.1cm}
\noindent
Using the techniques of \cite{Bugaev:96,Bugaev:99}, one finds the energy flux $\varphi_{E_{\cal H}}$ and momentum flux 
$p_{\rm rad}$ as
\begin{equation}
	\varphi_{E_{\cal H}} \simeq \left( m +2 T_{\cal H}\right) \varphi_{n_{\cal H}}  , \quad 
	p_{\rm rad} = n_{\cal H} T_{\cal H} / 2.
\label{enrgy-flux}
\end{equation}
The pressure $p_{\rm rad}$ exerted on the bag by its radiation can be compared to the intrinsic bag pressure $B$: for pions $p_{\rm rad} \sim 0.02 B$. The time $\tau$ for the bag to dissolve into its radiation is
\vspace*{-0.2cm}
\noindent
\begin{equation}
	\tau \simeq \frac{3 \pi \exp \left( \frac{m}{T_{\cal H}} \right) E_0}{g_m \left( m^2 T_{\cal H}^{2} 
	\right ) R_{0}^{2}} ,
\label{bag-time}
\end{equation}

\vspace*{-0.2cm}
\noindent
$g_m$ is the particle degeneracy and $R_0$ and $E_0$ are the radius and total energy of the initial bag.

The fluxes written in Eqs.~(\ref{part-flux}) and (\ref{enrgy-flux}) (particle or energy per unit surface area) 
are integrated over an assumed spherical bag to give the result in Eq.~(\ref{bag-time}). However, because 
of the lack of surface tension, the bag's maximum entropy corresponds to either an elongated (cylinder) or 
a flattened shape (disc). Thus, Eq.~(\ref{bag-time}) should be interpreted as an upper limit.

The decoupling between the vapor concentration and $m$ and $T_{\cal H}$ occurs when $\cal H$ has completely evaporated (i.e. $E-Nm-\frac{3}{2}NT_{\cal H}=0$) at a volume of 
\vspace*{-0.25cm}
\noindent
\begin{equation}
	V_{d} \simeq \frac{1}{n_{\cal H}} \frac{E}{m+\frac{3}{2}T_{\cal H}} .
\label{v-decoup}
\end{equation}

\vspace*{-0.25cm}
\noindent
The disappearance of $\cal H$ allows the vapor concentration to decrease as ${N}/{V} = {n_{\cal H}V_{d}}/{V}$. 
%

%

For $V > V_{d}$ due to energy and particle number conservation the temperature is fixed at $T_{\cal H}$. This assumes the Hagedorn spectrum extends to $m=0$. However, there may be a lower cut off at $m_0$ which modifies the results as follows. For energies $E - mN - \varepsilon \gg m_0$ and $V<V_d$ the above results hold. For $V \gg V_d$, the situation is different: $\cal H$ evaporates until its mass is $m_0$. If the entire mass of $\cal H$ is transformed into vapor particles as the volume increases further, then the excess particles increases the concentration and decreases the temperature. As the volume increases further, the concentration varies as

\vspace*{-0.0cm}
\noindent
\begin{equation}
	\frac{N}{V} = \frac{n_{\cal H} V_{d}+\frac{m_0}{m}}{V} ,
\label{conc-dec2}
\end{equation}

\vspace*{-0.1cm}
\noindent
\mbox{while the temperature remains constant at}

\vspace*{-0.2cm}
\noindent
\begin{equation}
	T = \frac{n_{\cal H} V_d}{n_{\cal H} V_{d}+\frac{m_0}{m}} T_{\cal H} .
\label{conc-temp}
\end{equation}

\vspace*{-0.15cm}
\noindent
The threshold $m_0$ is absolute regardless of bag multiplicity. Many bags in equilibrium have a global Hagedorn threshold $m_0$ so particle-particle collisions are identical to heavy ion reactions.

\subsection{Fragmentation of $\cal H$.}
What is the stability of $\cal H$ against fragmentation? Is the situation of a vapor of $\cal H$-like particles different than that of a single $\cal H$-like particle? If the translational degrees of freedom are neglected, $\cal H$ is indifferent to fragmentation into an arbitrary number of particles of arbitrary mass (within the constraints of mass/energy conservation). This follows from the exponential mass spectrum of Eq.~(\ref{hagedorn}) and the consequent lack of surface energy.


If $\cal H$ fragments totally into a number $N$ of equal sized fragments but one all with translational degrees of freedom  and with an  exponential mass spectrum (\ref{hagedorn}), 
 then
\begin{equation}
\rho_T  = 
 \frac{  e^\frac{ E - N\,m - \varepsilon }{  T_{\cal H}  }    \, e^\frac{N\,m }{ T_{\cal H} } \, 
 V^N }{N! \left( \frac{3}{2}N\right)!} 
 \left[ \frac{m \varepsilon}{2 \pi} \right]^{\frac{3}{2}N}
=~
 \frac{ {e^{\frac{E}{ T_{\cal H} } } } ~ V^N}{N! } \left[ \frac{m T_{\cal H}}{2 \pi} \right]^{\frac{3}{2}N} .
\label{equation}
\end{equation}
In the last step I substituted the kinetic energy's most probable value (\ref{max-01}) and used the Stirling formula for $\left( \frac{3}{2}N\right)!$. Equation~(\ref{equation}) shows that all Hagedorn factors collapse to a single one with the $m$-independent argument $E$. Maximization of (\ref{equation}) with respect to $m$ leads to
\begin{equation}
	{\partial \ln \rho_T}/{\partial m} = {3N}/{2m} = 0 ,
\label{max-eq}
\end{equation}
which is consistent with $N=0$ and $m=E$: a single $\cal H$ particle with all available mass. 

This again illustrates the indifference of $\cal H$ toward fragmentation. Of course Eq.~(\ref{max-01}) gives directly the mass distribution of the Hagedorn fragments under the two conditions discussed above. These results justify the assumption of the canonical formulation of the statistical hadronization model that smaller clusters appear from a single large cluster \cite{Becattini:Can}.

{
It can be shown similarly  that for  $N$  identical  $\cal H$ resonances of mass $m \gg T_{\cal H}$ the most probable value of  kinetic energy of the system is $ \varepsilon \approx \frac{3}{2} N T_{\cal H}$. 
Furthermore, the most probable value of  $N$  can be found by the equation (\ref{numberpp}), in which 
one has to insert  the exponential  degeneracy  of  $\cal H$ and approximate the $\cal H$ mass  as 
$m \approx \frac{E}{N}$. 
 
}


As was shown above  $\cal H$ system is a perfect thermostat at fixed temperature $T_{\cal H}$ and a perfect particle reservoir. Particles in equilibrium with or emitted by $\cal H$ are in thermal and chemical equilibrium with themselves and with $\cal H$. They constitute a saturated vapor. This defines a first order PT in a finite system  and a phase coexistence completely controlled by the bag pressure. The hadronic side of $\cal H$ radiates particles in preexisting thermal and chemical equilibrium just as a black body radiates photons in thermal and chemical equilibrium. An $\cal H$ system is nearly indifferent to fragmentation into smaller $\cal H$ systems. This invariance with respect to fragmentation makes this model  relevant to elementary particle and heavy ion collisions. A lower cut off in the mass spectrum does not alter the above results.


\section{Hagedorn Thermostat Model}

In this section  I  would like to  show that
the results of  A+A and elementary particle collisions can be understood and explained  on the same footing  by 
the Hagedorn thermostat concept. 
For this purpose 
let me 
study  the properties of microcanonical  equilibrium  of  Boltzmann particles which are in contact
with  Hagedorn thermostats  and  elucidate the effect  of the mass cut-off
$m_{\rm o}$ of the Hagedorn spectrum
on the  temperature of  a system at  given energy. 
The microcanonical formulation of the Hagedorn thermostat model (HTM) is given in section \ref{sect:1}
Section \ref{sect:2} is devoted to the analysis of the most probable state of 
a  single heavy Hagedorn thermostat  of mass  $m \ge m_{\rm o}$ and  $N_B$  Boltzmann particles. 
Section \ref{sect:3}  contains my conclusions and possible experimental consequences.


\subsection{A Single Hagedorn Thermostat Case.}
\label{sect:1}
Let me  consider the microcanonical ensemble of $N_B$ Boltzmann point-like particles 
of mass $m_B$ and degeneracy $g_B$, 
and $N_H$  hadronic  point-like resonances of mass $m_H $ with a 
mass spectrum 
\begin{equation}
g_H (m_H) = \exp[ m_H/ T_H ] (m_{\rm o}/ m_H )^a ~~{\rm  for}~~  m_H \ge m_{\rm o}\,,
\label{HspectrumFull}
\end{equation}
which obeys the inequalities
$m_{\rm o} \gg T_H$ and $m_{\rm o} > m_B$.
A recent analysis \cite{Bron:04}  suggests  that the  Hagedorn mass spectrum  can be established  for $m_{\rm o}  < 2$ GeV.

In the SBM \cite{Frautschi:71} and the  MIT bag model  \cite{Kapusta:81}
 it was found that for $m_H \rightarrow \infty$  the parameter $a \le 3$.
For finite resonance masses  the  value of $a$  is  unknown, so   it will be   considered as a fixed parameter.

The microcanonical partition 
of the system,  with  volume $V$,  total energy $U$ and zero total momentum,  can be written as follows 
\begin{eqnarray}   \label{mone}
\hspace*{-0.9cm}\Omega  ( U, N_H, N_B) 
&\hspace*{-0.1cm}= \hspace*{-0.1cm}& 
\frac{V^{N_H} }{ N_H !}  \left[ \prod_{k = 1}^{N_H}  g_H(m_H)\hspace*{-0.1cm} \int \hspace*{-0.1cm} 
\frac{ d^3 Q_k}{(2 \pi)^3 } \right] 
%
\frac{V^{N_B} }{ N_B !}  \left[ \prod_{l = 1}^{N_B} g_B \hspace*{-0.1cm}\int  \hspace*{-0.1cm}
\frac{ d^3 p_l}{(2 \pi)^3 } \right] 
\delta\biggl( U - \sum\limits_{i = 1 }^{N_H} \epsilon^H_i - \sum\limits_{j = 1 }^{N_B} \epsilon^B_j \biggl) ,
\end{eqnarray}
\vspace*{-0.0cm}

\noindent
where the quantity $ \epsilon^H_i~=~\varepsilon(m_H, Q_i )$ $\left( \epsilon^B_j = \varepsilon(m_B, p_j) \right.$ with the notation
$\left. \varepsilon(M, P) \equiv  \sqrt{M^2 + P^2}  \right)$  denotes  the energy  of
the Hagedorn (Boltzmann) particle with the 3-momentum ${\vec Q}_i$ (${\vec p}_j$).
In order to simplify the  presentation of  my  idea,  Eq. (\ref{mone}) accounts  for   energy conservation only 
and neglects  momentum conservation.

The microcanonical partition (\ref{mone}) can be evaluated by the Laplace transform 
in total energy $U$ \cite{Pathria}.
Then the momentum integrals in (\ref{mone}) are factorized and can be performed
analytically. The inverse Laplace transform in the conjugate variable $\lambda$ can be 
 done analytically for
 the nonrelativistic and ultrarelativistic approximations  of 
the one-particle momentum distribution  function ($K_2(z) $ is the modified Bessel function)
\begin{eqnarray}\label{mfour}
\hspace*{-0.cm} 
 \int\limits_0^\infty \hspace*{-0.1cm}  
\frac{d^3 p ~  {\textstyle e^{-\lambda \varepsilon(M , p ) } }   }{ ( 2 \pi)^3 }  & = & 
\frac{M^2\,  K_2 (M\, \lambda)  }{2\, \pi^2\,\lambda} \,
 \approx  
 \left\{
\begin{tabular}{ll}
\vspace{0.1cm} \hspace*{-0.1cm}$ \left[ \frac{ 2 M }{\lambda} \right]^{\frac{3}{2} }\hspace*{-0.2cm} I_{\frac{1}{2} }  e^{- M \lambda } \,,$
&  \hspace*{-0.1cm}$  M  Re (\lambda)  \gg 1$\,, \\
\hspace*{-0.1cm}$ \frac{ 2 }{  \lambda^3 } ~ I_{ 2 }\, e^{- M \lambda }  \,, $  &\hspace*{-0.1cm}$  M  Re (\lambda)  \ll 1$\,,
\end{tabular}
\right.
\end{eqnarray}
where the auxiliary integral can be expressed in terms of the gamma function as follows 
\vspace*{-0.1cm}
\begin{equation}\label{mfive}
I_b ~\equiv ~ \int\limits_0^\infty \hspace*{-0.0cm}  
\frac{d \xi }{ (2 \pi)^2 } ~ \xi^b
~{ e^{-\xi } }~  = ~\frac{\Gamma(b+1)}{ (2\,\pi)^2 }  \,.
\end{equation}
\vspace*{-0.30cm}

Since the formal steps of further evaluation  
are similar for  both cases, I discuss 
in detail the nonrelativistic limit only,  and later 
present  the results for the other case.  
The  nonrelativistic approximation ($  M  Re (\lambda)  \gg 1$) for  Eq. ~(\ref{mone}) is as follows
\begin{eqnarray} \label{meight}
&&
\hspace*{-0.3cm} 
\Omega_{nr} 
= 
\frac{ \left[  V g_H(m_H)  \left[ 2 m_H  \right]^{ \frac{3}{2} }    I_{ \frac{1}{2} }  
\right]}{N_H!}^{N_H}  %
 \frac{ \left[  V g_B \left[ 2 m_B \right]^{ \frac{3}{2} }  I_{ \frac{1}{2} } \right] }{N_B!}^{N_B} 
\hspace*{-0.2cm} 
{\textstyle
\frac{ E_{kin}^{\frac{3}{2} (N_H + N_B) - 1}  }{  \left( \frac{3}{2} (N_H + N_B) - 1 \right)! }
}
\,,
\end{eqnarray}
where $E_{kin} = U - m_H N_H - m_B N_B $ is the kinetic energy of the system.

As shown below, the most realistic case 
corresponds to the nonrelativistic treatment of the Hagedorn resonances because 
the resulting temperature is  much smaller than 
their masses. Therefore, it is sufficient to consider the ultrarelativistic
limit for the Boltzmann particles only. In this case 
 ($  M  Re (\lambda)  \ll 1$)
 the equation (\ref{mone}) can be approximated as 
\begin{eqnarray}\label{mten}
&&\hspace*{-0.6cm} \Omega_{ur} 
= 
\frac{ \left[  V g_H(m_H)  \left[ 2 m_H  \right]^{ \frac{3}{2} }   I_{ \frac{1}{2} }  
\right]}{N_H!}^{N_H}   
 \frac{ \left[  V g_B  ~2  I_{ 2 } \right] }{N_B!}^{N_B} 
{\textstyle
\frac{ E_{kin}^{\frac{3}{2} (N_H + 2 N_B) - 1}  }{  \left( \frac{3}{2} (N_H + 2 N_B) - 1 \right)! }
}
\,,
\end{eqnarray}
where the  kinetic energy does not include the rest energy  of the Boltzmann particles, i.e.
$E_{kin} = U - m_H N_H $.

Within the adopted  assumptions 
 the above results are general
and can be used for any number of particles,  provided $N_H + N_B \ge 2$.
It is instructive to consider first  the  simplest case $N_H = 1$.  
This oversimplified model, in which a Hagedorn thermostat is always present,  allows
one  to study the problem rigorously.
For  $N_H = 1$ and $N_B~\gg~1$ I  treat  the mass of Hagedorn
thermostat $m_H$  as a free parameter and determine the  value which maximizes the entropy of the system.  The 
solution $ m_H^*~>~0$  of 
%
\begin{eqnarray}\label{meleven}
\hspace*{-0.5cm}&& \frac{ \delta \ln \Omega_{nr} (N_H = 1) }{\delta~ m_H } ~ =   { \frac{1}{T_H}~ +~ \left( \frac{3}{2}~ - ~ a \right) \frac{1}{m_H^*} ~ - ~  
\frac{3 (N_B + 1) }{2~ E_{kin} } ~ = ~ 0 } 
\end{eqnarray}
provides the maximum of the system's entropy, if  for $m_H = m_H^*$ the second derivative is negative
\begin{eqnarray}\label{mtwelve}
\hspace*{-0.5cm}&& \frac{ \delta^2 \ln \Omega_{nr} (N_H = 1) }{\delta~ m_H^2 } ~ =   { - ~ \left( \frac{3}{2}~ - ~ a \right) \frac{1}{m_H^{*\,2} } ~ - ~  \frac{3 (N_B + 1) }{2~ E_{kin}^2 } ~ < ~ 0 \,. }
\end{eqnarray}
If  the inequality  (\ref{mtwelve}) is satisfied, then the extremum condition (\ref{meleven})
defines the  temperature of the system of $(N_B + 1)$ nonrelativistic particles
\begin{equation}\label{mthirteen}
\hspace*{-0.25cm} T^* (m_H^*)  \equiv  
\frac{ 2 ~ E_{kin} }{  3 (N_B + 1) } = \frac{T_H}{ 1 ~ + ~ \left(  \frac{3}{2}~ - ~ a \right) \frac{T_H }{m_H^{*} }    }  \,.
\end{equation}
Thus, as $m_H^* \rightarrow \infty$ it follows that $T^*(m_H^*) \rightarrow T_H$, while for 
finite $m_H^* \gg T_H$  and $ a >  \frac{3}{2} $  ($ a <  \frac{3}{2} $) the temperature 
of the system is 
slightly larger  (smaller) than the Hagedorn temperature, i.e. $ T^* > T_H$ ($ T^* < T_H$).
Formally, the temperature of the system in equation (\ref{mthirteen}) may differ  essentially 
from $T_H$ for  a  light thermostat, i.e. for $m_H^* \le T_H$.
However, it is assumed that 
the Hagedorn mass spectrum exists
above the cut-off mass $m_{\rm o} \gg T_H$, thus $m^* \gg T_H$.


\subsection{The Role of the Mass Cut-off.}
\label{sect:2}
Now I would like to  study the effect of the mass cut-off of the Hagedorn spectrum on the
inequality (\ref{mtwelve}) in more detail. 
For $ a \le  \frac{3}{2} $ the condition (\ref{mtwelve}) is  satisfied. For $ a >   \frac{3}{2} $ the inequality (\ref{mtwelve})
is equivalent to 
\begin{equation}\label{mfourteen}
\hspace*{-0.25cm}
 \frac{ m_H^{*\,2}  }{   \left( a -  \frac{3}{2} \right)  ~ T^*(m_H^*)    }  ~ > ~ 
\frac{ 3 }{ 2 } ~  (N_B + 1) ~ T^*(m_H^*)  
\,,
\end{equation}
which  means that a  Hagedorn thermostat should be massive  compared to the kinetic energy of the system.  

A more careful analysis shows that 
for a negative value of the determinant $ D_{nr} $  $( \tilde{N} \equiv N_B - \frac{2}{3} a )$  
\begin{eqnarray}\label{mfifteen}
\hspace*{-0.67cm}
&D_{nr}  \equiv  {\textstyle  \left( U - m_B N_B - \frac{3}{2}~ T_H ~ \tilde{N} \right)^2 - ~} 
 {\textstyle 4 \left( a - \frac{3}{2} \right)T_H \left( U - m_B N_B \right) ~ < ~ 0\,, }
\end{eqnarray}
equation (\ref{meleven}) has two complex solutions, while  for $D_{nr}~=~ 0$ there exists a single 
real solution of (\ref{meleven}).
Solving (\ref{mfifteen}) for $(U - m_B N_B)$, shows that  for $\tilde{N} > \frac{2}{3} a - 1$,
 i.e. for ${N_B} > \frac{4}{3}  a - 1$ the inequality (\ref{mfifteen})  does not hold and $D_{nr} > 0$.
Therefore, in what follows I  will assume that ${N_B} > \frac{4}{3} a - 1$ and  only analyze  the case $D_{nr} > 0$. 
For this case  equation (\ref{meleven}) has two real solutions 
\begin{equation}\label{msixteen}
m_H^\pm = {\textstyle \frac{1}{2} \left[ U - m_B N_B - \frac{3}{2}~ T_H ~ \tilde{N}~ \pm ~ \sqrt{ D_{nr} } \right]\,.} 
\end{equation}
For $ a \le  \frac{3}{2} $ only $m_H^+$ solution is positive and
corresponds to a maximum of  the microcanonical partition $\Omega_{nr}$. 

For $ a >  \frac{3}{2} $ both solutions of (\ref{meleven}) are positive, but only  $m_H^+$  is a maximum.
From the two limiting cases:
\begin{eqnarray}\label{mseventeen}
\hspace*{-0.5cm}
\frac{ \delta \ln \Omega_{nr} (N_H = 1) }{\delta~ m_H } & \approx &  
{\textstyle \left( \frac{3}{2} -  a \right) \frac{1}{m_H}  \quad {\rm for} \quad m_H \approx 0\,,} 
\\
\label{meighteen}
\hspace*{-0.5cm}
%
\frac{ \delta \ln  \Omega_{nr} (N_H = 1)  }{\delta~ m_H }  & \approx &
{\textstyle 
\frac{3 (N_B + 1) }{2~ E_{kin} } \quad {\rm for} \quad E_{kin} \approx 0\,,}
\end{eqnarray}

\newpage
\clearpage

\vspace*{-4.5cm}
%
\begin{figure}[ht]
\centerline{\hspace*{4.5cm}\epsfig{figure=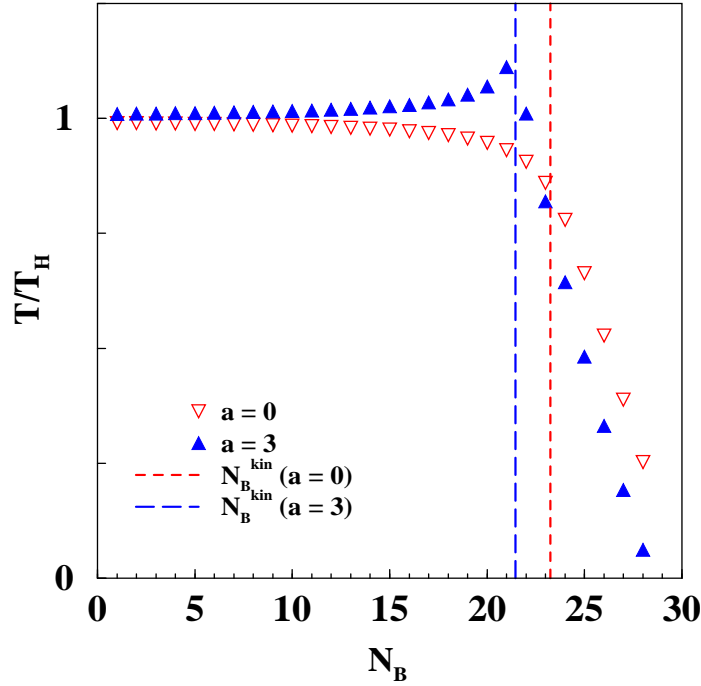,width=14.7cm,height=14.cm}
}
\caption{ \label{fig:41} 
A typical behavior  of the system's temperature  as the function of the  number 
of Boltzmann particles $N_B$ for $ a = 3$ and $a = 0$ for the same value of the total energy 
$U = 30 \,m_B$.
Due to the thermostatic properties of a Hagedorn resonance
the system's temperature  is nearly constant 
up to the kinematically allowed value $N_B^{kin}$ given by (\ref{mtwenty}). 
}     
\end{figure}

\noindent
and the fact that $ m_H^\pm $ 
obey the   inequalities
\begin{equation}\label{mnineteen}
0 ~ < ~  m_H^- ~  \le ~ m_H^+ ~ < ~ U - m_B N_B \,, 
\end{equation}
it is clear that $ m_H^* = m_H^-$ is a local minimum  of the microcanonical partition $\Omega_{nr}$,
while  $ m_H^* = m_H^+$ is  a local maximum of the partition $\Omega_{nr}$.

Using Eq. (\ref{msixteen}) for $m_H^+$, it is clear that 
for any value of  $a$ 
the constraint $m_H^+ \ge m_{\rm o} $
is equivalent to the  inequality 
\begin{equation}\label{mtwenty}
N_B ~ \le ~ N_B^{kin} \equiv  { \frac{ U ~ - ~ [ \frac{m_{\rm o} }{T_H}~ - ~a ] ~ T^*(m_{\rm o})  }{ m_B \, + \, \frac{3}{2} ~ T^*(m_{\rm o})  }  } \,.
\end{equation}
Thus, at fixed energy $U$   for all  $N_B \le N_B^{kin}$ 
 at  $m_H^*  = m_H^+$ there is a local maximum of the microcanonical partition  $\Omega_{nr}$ with
the temperature $ T = T^*( m_H^+) $. For $N_B > N_B^{kin}$   the maximum of the partition $\Omega_{nr}$ cannot be reached due to the cut-off constraint
and, consequently,   the most probable state corresponds to $m_H = m_{\rm o}$
with 
$T  \le T^*(m_{\rm o} )$ from Eq. (\ref{mthirteen}). 
In other words, for $N_B > N_B^{kin}$  the amount of energy $U$ is insufficient for  the mass of the Hagedorn thermostat to be above the cut-off  $m_{\rm o} $ 
and   simultaneously  maintain  the  temperature of the Boltzmann particles according to  
Eq. (\ref{mthirteen}).  
By assumption there is a single Hagedorn thermostat in the system, therefore,  
as $N_B$ grows the temperature of the system decreases
from $T^*(m_{\rm o} )$ value.
Thus, the equality (\ref{mtwenty}) defines the kinematical limit 
for reaching the maximum of the microcanonical partition.

To prove that the maximum of the microcanonical 
partition at $m_H = m_H^+$  is  
global  it is sufficient to show that 
the  constraint $m_H^+ \ge m_{\rm o}$
is not consistent with the condition $m_H^- > m_{\rm o}$. 
For $a \le \frac{3}{2}$ the maximum is  global  because for $ 0 <  m_H < m_H^+$ 
($m_H >  m_H^+$ ) the partition $\Omega_{nr} (N_H =1, m_H ) $ monotonically increases (decreases)
with $m_H$. 
For $a > \frac{3}{2}$  it is clear that the maximum at $m_H = m_H^+$ is local, 
if  the state with mass  $m_H =  m_{\rm o}$ is more probable, i.e. 
$\Omega_{nr} (N_H =1, m_{\rm o}) >  \Omega_{nr} (N_H =1, m_H^+ ) $.  Due to (\ref{mnineteen})  this can occur, if $m_H^- > m_{\rm o}$.  Substituting Eq.  (\ref{msixteen})  into the last inequality,
shows  that this inequality
reduces  to the condition $N_B > N_B^{kin}$.
This contradicts  the  constraint $m_H^+ \ge m_{\rm o}$ in the form of  Eq. 
(\ref{mtwenty}). 
Thus, the maximum of the
microcanonical partition is  global.

To complete this consideration of the  nonrelativistic case 
let me  express the partition (\ref{meight}) in terms
of the temperature (\ref{mthirteen}). 
 Applying the Stirling approximation  to  the  factorial $(\frac{3}{2}(N_B + 1) -1 )!$
for  $N_B^{kin} > N_B \gg 1$ and reversing the
integral representations (\ref{mfour})  and (\ref{mfive}) for $\lambda = 1/ T^*(m_H^+)$, one finds 
\begin{eqnarray}\label{mtwone}
\hspace*{-0.5cm}\Omega_{nr} (N_H=1)  &=&
%
%
\frac{  V \, g_H(m_H^+)  }{  T^*(m_H^+) } ~  \hspace*{-0.0cm}  \int  \hspace*{-0.0cm}  
\frac{d^3 Q}{ (2 \pi)^3 } 
~{\textstyle e^{ - \frac{  \sqrt{m^{+\,2}_H + Q^2}   }{  T^*(m_H^+) }  } }  
%
\frac{ e^{\frac{U}{ T^*(m_H^+) } }  }{N_B!}  \left[ V \, g_B  \hspace*{-0.0cm}  \int  \hspace*{-0.0cm}  
\frac{d^3 p}{ (2 \pi)^3 } 
~{\textstyle e^{ - \frac{  \sqrt{m^2_B + p^2}   }{  T^*(m_H^+) }  } }  \right]^{N_B}  .~
\end{eqnarray}
\vspace*{-0.2cm}

\noindent
This is just 
the  grand canonical partition of $(N_B + 1)$ Boltzmann
particles with temperature $ T^*(m_H^+) $. 
If $N_B > N_B^{kin} \gg 1$, then $ T^*(m_H^+)$
in (\ref{mtwone}) should be 
replaced  by 
$T_{\rm o} (N_B) \equiv  \frac{2(U - m_B N_B - m_{\rm o} ) }{3 (N_B + 1) } $. 

Fig.~\ref{fig:41} shows that 
for  $a >  \frac{3}{2}  $
the system's temperature $T = T^*(m_H^+)$  as a function of $N_B$ remains almost  constant    for 
 $N_B < N_B^{kin}$, reaches a maximum  at  $N_B^{kin} $  and rapidly decreases like 
 $T  = T_{\rm o}  (N_B) $  for  $N_B > N_B^{kin} $.
For $a <  \frac{3}{2} $ the temperature  has a plateau $T =  T^*(m_H^+)$ for $N_B <  N_B^{kin}$,
and rapidly    decreases  for 
$N_B > N_B^{kin} $ according to  $T_{\rm o}  (N_B) $. 

The same results  are  valid for the ultrarelativistic treatment of Boltzmann 
particles.  Comparing the nonrelativistic and ultrarelativistics expressions for
the microcanonical partition, i.e. equations (\ref{meight}) and (\ref{mten}), respectively,
one  finds that   the derivation of the  ultrarelativistic limit   requires only 
the substitution $N_B \rightarrow 2 N_B$ and $m_B / T_H \rightarrow 0$
in  equations (\ref{meleven} -- \ref{mtwone}). 
Note that  this  substitution does not alter  the expression for the temperature
of the system, i.e. the right hand side of (\ref{mthirteen}).

Finally,  it can be  shown  that for a heavy Hagedorn thermostat  ($m_H^+  \gg  m_{\rm o}$) these results remain 
valid for  a single Hagedorn thermostat split  into $N_H$ pieces of the same mass.
 
Substituting $m_H \rightarrow m_H N_H$ in the nonrelativistic expressions 
(\ref{meight}) and minimizing  it with respect to $m_H$,  
one finds that 
the temperature of the system in the form of  equation (\ref{mthirteen})  is $ T^*(m_H^* N_H) $, where the mass of $N_H$ Hagedorn thermostats $m_H^*$ is related to the solution $m_H^+ $ of equation (\ref{msixteen})   as  
$m_H^* = m_H^+ / N_H$. Since the original single thermostat of mass $m_H^+ $ was assumed to be heavy,  it follows $T^*(m_H^* N_H) = T^*(m_H^+) \rightarrow T_H$. 
A more careful study (see also   \cite{HThermostat:1})  using
an exact expression for the microcanonical partition of $N_H$ Hagedorn thermostats of the same mass
$m_H$ gives the same result, if  $m_H  \gg  m_{\rm o} $. 
A generalization of these statements to  the case of $N_H$ heavy 
Hagedorn thermostats of different masses  also leads to the same result.  
Thus,  splitting  a single
heavy Hagedorn thermostat into an arbitrary number of heavy resonances (heavier than $ m_{\rm o} $) does not
change the temperature of the system.


\subsection{Concluding Remarks on the  HTM.}
\label{sect:3}
In the present section I  generalized the  SBM  results \cite{Frautschi:71}  to  systems  of finite energy by showing  explicitly
that  even  a single resonance with  the Hagedorn mass spectrum degeneracy,
i.e. {\it a Hagedorn thermostat,}  keeps an almost constant   
temperature  close to $T_H$  for  any  number of Boltzmann particles $3 < N_B  \le N_B^{kin} $. 
For the  high energy limit  $U \gg m_{\rm o}$ this means that 
a single  Hagedorn resonance defines 
the temperature of the system to be only slightly  different from $T_H$ until 
 the energy of the Hagedorn thermostat is almost negligible compared to $U$.
In contrast to the grand canonical formulation of the original SBM  \cite{Frautschi:71},
in the presence of a Hagedorn thermostat 
the  temperature $T_H$ 
can be reached  at  any energy density.

The thermostatic  nature of a Hagedorn system  
obviously
 explains the ubiquity of both the inverse slopes of  measured transverse mass spectra 
\cite{Gazdzicki:04} and hadronization 
temperature found in  numerical simulations 
of hadrons
created in elementary particle collisions at high  energies \cite{Becattini:Can,Becattini:1,Becattini:2}.
By a direct evaluation of the microcanonical partition
I  showed that  in the presence of a single Hagedorn thermostat  the energy spectra of particles become 
exponential 
 with no  additional assumptions, e.g. {\it  phase space dominance} \cite{PSDominance} or
{\it string tension fluctuations} \cite{Strings, Strings:2}.
Also the limiting temperature found in the URQMD calculations made in a  finite box 
\cite{URQMD:Box} can be explained by the effect of the Hagedorn thermostat. 
One expects that, if the string parametrization of the URQMD in a  box \cite{URQMD:Box} was done microcanonically instead of grand canonically,  then  the same behavior  would be found.

The Hagedorn thermostat model
generalizes the statistical hadronization model which successfully
describes the particle multiplicities in nucleus-nucleus and elementary collisions \cite{Becattini:Can,Becattini:1,Becattini:2}.
The statistical hadronization model  accounts for   the decay of heavy resonances 
(clusters in terms of Refs. \cite{Becattini:Can,Becattini:1,Becattini:2}) only  and does not consider the additional
particles, e.g. light hadrons, free quarks and gluons, or other heavy resonances.
As I  showed,
the splitting of a single
heavy Hagedorn resonance into several  does not
change the temperature of the system. 
This finding  justifies the main assumption of 
the canonical formulation of the statistical hadronization model \cite{Becattini:Can} that smaller clusters   
may be reduced 
to a single large cluster. 
Also the present model explains the results of a new statistical model of resonance decay
\cite{Pal:05} that the only temperature available to the system is the Hagedorn 
temperature $T_H$.

Thus,  
recalling the  MIT Bag model interpretation of 
the Hagedorn mass spectrum  \cite{Kapusta:81,Kapusta:82}, I  conclude that 
 quark gluon matter confined in  heavy resonances (hadronic bags)  
controls the  temperature of surrounding particles close to $T_H$, and, therefore,
this temperature can  be considered as 
a coexistence  temperature for  confined color matter and hadrons. 
Moreover, as I showed, the emergence  of a coexistence temperature does
not require the actual deconfinement of the color degrees of freedom,
which, in terms of the GBM \cite{Goren:81}, is equivalent to 
the formation of the infinitely large
and infinitely heavy hadronic bag.  

A similar conclusion on the phase transition existence was given in Ref. \cite{Carlitz:72}.
However, 
Carlitz's  argument for phase transition is based on the condensation properties of heavy resonances, 
assuming that  
the Hagedorn mass spectrum is truncated from the above.
On contrary, the approach that I
formulate  here refers  exclusively  to the thermostatic properties of the Hagedorn
systems.

Within the framework of  the Hagedorn thermostat model it was  
found that,  even for a single Hagedorn thermostat and 
 $a >  \frac{3}{2} $,
the system's temperature $T = T^*(m_H^+)$  as a function of $N_B$ remains almost  constant    for 
 $N_B < N_B^{kin}$, reaches a maximum  at  $N_B^{kin} $  and rapidly decreases  
 for  $N_B > N_B^{kin} $ (see Fig.~\ref{fig:41}). 
For $a <  \frac{3}{2} $ the temperature  has a plateau $T =  T^*(m_H^+)$ for $N_B <  N_B^{kin}$,
and rapidly    decreases  for 
$N_B > N_B^{kin} $.
If such a characteristic behavior 
of the hadronization temperature or the hadronic inverse slopes 
can be measured as a function of event multiplicity, it may be possible to experimentally
determine  the value of  $a$. 
For  quantitative predictions
it is necessary to include more hadronic species into the model, but this will not change 
qualitatively  these   result.

If  one  applies  the HTM to elementary particle collisions at high energies, and thus
at  vanishing baryonic densities,
then, as  shown above, 
the temperature of  created particles will  be  defined by the
most probable mass of the Hagedorn thermostat. 
If the most probable resonance mass 
grows  with  the energy of collision,  then the hadronization  temperature should decrease (increase)
to $T_H$ for $a >  \frac{3}{2}  $ $(a <  \frac{3}{2}  )$. 
Such a decrease is  observed   in 
reactions of elementary particles  at  high energies, see Table 1 in Ref.
\cite{Becattini:2}.

In order to apply 
these results in a more realistic  fashion to the quark gluon plasma formation in relativistic 
nucleus-nucleus collisions (where
the excluded volume effects  are known to be important  
\cite{Vol:1,Goren:81, Hgas, Hgas:2,Hgas:22, Zeeb:02}  
for all hadrons) 
 the eigen volumes of all particles  should be incorporated into the model.
For pions this should be done in relativistic fashion \cite{Bugaev:RVDW1, Bugaev:RVDW2}, as was discussed in the chapter 3.
Also, as will be shown in the next section,  the effect of finite  width of Hagedorn resonances may be important \cite{Blaschke:04}  and should  be studied.



%

\section{Mott-Hagedorn Model for QGP }

{The lattice quantum chromodynamics  simulations  not only provide} the strongest
theoretical support of the QGP existence,
but they also give  detailed information on the properties of
strongly interacting matter over a wide range of temperatures.
A recent analysis \cite{Tawfik03} of the lattice energy density  showed
that a hadron gas model with lattice value of hadron masses  can perfectly explain the steep rise in the number
of degrees of freedom at $T \approx T_c$.
On the other hand, lattice quantum chromodynamics  has also revealed that hadronic correlations
persist for $T>T_c$ \cite{Petreczky, Hatsuda}.
The question arises whether it is more appropriate to describe 
the hot quantum chromodynamics  matter
in terms of hadronic correlations rather than in terms of quarks and gluons.
Therefore, in the present  section, I  would like to discuss  a generalization \cite{Blaschke:03, Blaschke:04, Blaschke:05} of
the Hagedorn  model, i.e.  the SBM, 
which allows for  the extension of a hadronic description above $T_c$.

The SBM \cite{hagedorn-65}
is based on the hypothesis that hadrons are made of hadrons,
with constituent  and compound hadrons being treated on the same footing.
This implies an exponentially growing  form of the hadronic mass spectrum
(\ref{HspectrumFull})
for $m \rightarrow \infty$. The  the Hagedorn temperature  $T_H$ was
interpreted as a limiting temperature reached at infinite energy density.
As was discussed at the beginning of this chapter, 
the extensive investigation of the SBM has led to a formulation of
both the important physical ideas and the mathematical methods for
modern statistical mechanics of strongly interacting matter \cite{SBM:new}.

However, up to now the formulation of the SBM had some severe  problems.
The first one is the absence of a width for the heavy resonances.
From the Particle Data Group \cite{PDG}  it is  known  that heavy
resonances with masses $m \ge 3.5$ GeV may have widths  comparable to
their masses. Taking the widths into account will effectively reduce
the statistical weight of the resonance.
The second problem arises while discussing the results of the hadron
gas  model \cite{Hgas, Hgas:2, RHIC:Chem}.
The hadron gas  model   accounts for all strong decays of resonances
according their partial width given in \cite{PDG}, and, hence, it
describes remarkably well
the light hadron  multiplicities measured in nucleus-nucleus collisions
at CERN SPS \cite{Hgas, Hgas:2} and BNL RHIC \cite{RHIC:Chem} energies.
This model is nothing else than the SBM of light hadrons which accounts for the
proper volume of hadrons with masses below $2.5$ GeV,
but neglects the  contribution of the exponentially
growing mass spectrum.

Thus, one immediately faces a severe  problem:
``Why do the  heavy resonances with masses above $2.5$ GeV predicted by the
SBM not  contribute  in the particle spectra measured in heavy-ion collisions
at SPS and RHIC energies?''
Note that the absence of heavy resonance contributions in the particle ratios
cannot be due to the statistical suppression of the Hagedorn mass spectrum
because the latter should not be strong in the quark-hadron  phase transition
region, where those ratios are believed to be formed \cite{Hgas, Hgas:2,RHIC:Chem}.

In the present section  I  discuss the possibility  that the introduction of a finite
width of the resonances can solve the above problems of the SBM.
In   the subsection  \ref{sect:431}  I formulate a simple statistical model that incorporates  
besides of the Hagedorn mass spectrum also medium dependent resonance widths
due to the hadronic Mott effect, and analyze its mathematical structure.
In the subsection \ref{sect:432} I discuss a model fit to recent lattice data of quantum chromodynamics  
thermodynamics \cite{Tawfik03} and some possible consequences for heavy-ion 
physics, 
{whereas the subsection \ref{sect:433} is devoted to the discussion of the J$/\psi$ anomalous suppression
within the model developed here. }  
%

 
\subsection{Resonance Width Model: Mott Transition.}
\label{sect:431}
According to quantum chromodynamics, hadrons are not elementary, point-like objects but rather 
color singlet bound states of quarks and gluons with a finite spatial extension
of their wave function. While at low densities a hadron gas description can
be sufficient, at high densities and temperatures, when hadronic wave functions
overlap, nonvanishing quark exchange matrix elements between hadrons 
occur in order to fulfill the Pauli principle. This leads to a Mott-Anderson 
type delocalization transition with frequent rearrangement processes of color 
strings (string-flip \cite{stringflip, stringflip:2}) so that hadronic resonances become 
off-shell with a finite, medium-dependent width. 
Such a Mott transition has been thoroughly discussed for light hadron systems
in \cite{huefner} and has been named {\it soft deconfinement}.
The Mott transition for heavy mesons may serve as the physical mechanism behind
the anomalous J/$\psi$ suppression phenomenon \cite{burau}.

For this purpose   the width $\Gamma$ of a resonance in the statistical model
with the Hagedorn mass spectrum is introduced through  the spectral function 
\begin{equation}
A(s,m)=N_s \frac{\Gamma~m}{(s-m^2)^2+\Gamma^2~m^2}~,
\end{equation} 
a Breit-Wigner distribution of virtual masses with a maximum at  $\sqrt{s}=m$
and the normalization factor
\begin{equation}
\label{two4}
%
%
N_s = \left[ \int_{m_B^2}^\infty {ds}
\frac{ \Gamma m }{ ( { s} - m^2)^2 + \Gamma^2 m^2  } \right]^{-1} =
\frac{1}{ \frac{\pi}{2} + 
\arctan \left( \frac{m^2 - m^2_B }{ \Gamma m} \right) }\,.
\end{equation}
The energy density of this model with zero resonance proper volume
for given temperature $T$ and baryonic chemical potential $\mu$
can be cast in the form
\begin{eqnarray}
\hspace*{-0.7cm}\varepsilon(T,\mu) &=& 
\sum_{i = \pi, \, g,...} g_i ~\varepsilon_M (T,\mu_i;m_i)
+ 
\sum_{A = M,B}  \int_{m_A}^\infty dm \int_{m_B^2}^\infty {ds}
~\, g_H(m)~A(s,m)~\varepsilon_A (T,\mu_A;\sqrt{s}),
\label{one4}
\end{eqnarray}
where the energy density per degree of freedom with a mass $m$ is
\begin{equation}
\label{three4}
%
%
\varepsilon_A (T,\mu_A;m)  = 
\int  \frac{d^3 k}{ (2 \pi)^3 }  
\frac{\sqrt{k^2+m^2}}{\exp \left(\frac{\sqrt{k^2 +m^2} - \mu_A}{T}
\right)  + \delta_A } \, ,
\end{equation}
with the degeneracy  $g_A$ and the baryonic chemical potential $\mu_A$ 
of hadron $A$. For mesons, $\delta_{M} = -1 ~$,  $\mu_{M} = 0$ and 
for baryons $~ \delta_{B} =  1$~and $\mu_{B}=\mu$, respectively.
According to Eq. (\ref{one4})  the energy density of hadrons consists of
the contribution of light hadrons for $m_i < m_A$ ~ and the contribution 
of the Hagedorn mass spectrum $\, g_H(m)$  for $m \ge  m_A$.

A new element of Eq. (\ref{one4}) in comparison to the SBM
is the presence of the $ \sqrt{s} $-dependent spectral function.
The analysis shows that, depending on the behavior of the resonance width 
$\Gamma$ in the limit $m \rightarrow \infty$,
there are the following possibilities:
\begin{itemize}
\item For vanishing resonance  width, $\Gamma = 0 $, Eq. (\ref{one4})  
evidently reproduces the usual SBM.
\item For  final values of the resonance width, $\Gamma =$ const, 
Eq. (\ref{one4}) diverges for all temperatures $T$ because, in contrast to the 
SBM, the statistical factor  in Eq. (\ref{one4}) behaves as 
$\left\{\exp \left[ (m_B - \mu_A) / T \right] + \delta_A \right\}^{-1}$
so that it cannot suppress the exponential divergence of the Hagedorn mass 
spectrum $\, g_H (m)$.
\item For a resonance width growing with mass like the Hagedorn spectrum
$\Gamma \sim  C_{\Gamma} \exp \left[ \frac{ m}{T_H} \right] $  or faster,
Eq. (\ref{one4}) converges again. 
\end{itemize}
Indeed, in the latter case  the Breit-Wigner spectral function behaves as 
\begin{equation}
\label{four4}
%
N_s \frac{ \Gamma m}{ (s - m^2)^2 + \Gamma^2 m^2 }~ 
\biggl|_{m \rightarrow \infty} \rightarrow  
\frac{2}{\pi ~ \Gamma}  \sim  \exp \left( - \frac{m }{ T_H } \right)
\end{equation}
and cancels the exponential divergence of the Hagedorn mass spectrum.
Hence, the energy density remains finite.
Note that both  the analytical properties of model (\ref{one4}) and
the right hand side of Eq. (\ref{four4}) remain the same, if a Gaussian shape 
of the spectral function is chosen instead of the Breit-Wigner one.

It can be shown that the behavior of the width at finite resonance masses is 
not essential for the convergence  of the energy density (\ref{one4}). 
In other words, for a convergent energy density (\ref{one4})
above $T_H$ it is sufficient to have
a very small probability density (\ref{four4}) (or smaller)
for a resonance of mass $m$ to be found in the state with the virtual mass 
$\sqrt{s}$.
Since there is no principal difference between the high and low mass 
resonances, one  can use the same functional dependence of the width 
$\Gamma$ for all masses.
Thus, for the following model  ansatz
\begin{equation}
\label{five4}
%
%
\Gamma (T) =
\left\{ \begin{array}{ll}
 0\,,  & {\rm for} \hspace*{0.3cm} T \le T_H  \,,  \\
 & \\
C_{\Gamma}~ \left( \frac{ m}{T_H} \right)^{N_m}
\left( \frac{ T}{T_H} \right)^{N_T} \exp \left( \frac{ m}{T_H} \right)\,,
& {\rm for } \hspace*{0.3cm} T >  T_H  \,,
\end{array} \right.
\end{equation}
the energy density (\ref{one4}) is finite for all temperatures and 
the divergence of the SBM is removed.
At $T = T_H$, depending on choice of parameters,  it may have 
either a discontinuity or its partial $T$ derivative may be discontinuous.
As discussed above, for $T \le T_H$ such a model corresponds to the usual SBM, 
but for high temperatures
$T > T_H$ it remains finite for a wide choice of powers $N_m$.

The idea behind the exponentially mass dependent width of resonances (\ref{four4}) is as follows:
to have a possibility for resonances of different masses to transform into each other one 
has at least  to allow the transformations of resonances of a  given  mass $m$ 
into the resonances  being located within  some finite  interval of masses $\Delta_m$ about $m$
and assume a finite (non-vanishing) partial width $\Gamma_p$ for each mass state within this interval;  
since in the immediate vicinity of a given mass  $m$ the number of all resonance states is given 
by the Hagedorn spectrum, i.e. is roughly  about $ \exp \left( \frac{ m}{T_H} \right) $, then the total width of 
the resonances  of  mass  is  approximately $\Gamma_p \exp \left( \frac{ m}{T_H} \right) \Delta_m$. 

{
As was shown in the preceding sections the CGE  should be  applied to the SBM with some care.  Thus, although the ansatz  (\ref{five4}) removes the artificial singularity above $T_H$, it, strictly speaking, cannot be used below  $T_H$. Therefore, a more suitable ansatz for the width reads as:
}
\begin{equation}
\label{fiveB}
%
%
\Gamma (T) =
\left\{ \begin{array}{ll}
C_{\Gamma}^\prime~ \left( \frac{ m}{T_H} \right)^{N_m}
\left( \frac{ T}{T_H} \right)^{N_T} \exp \left( \frac{ m}{T_H} \right)\,, 
 & {\rm for} \hspace*{0.3cm} T \le T_H  \,,  \\
 & \\
C_{\Gamma}~ \left( \frac{ m}{T_H} \right)^{N_m}
\left( \frac{ T}{T_H} \right)^{N_T} \exp \left( \frac{ m}{T_H} \right)\,,
& {\rm for } \hspace*{0.3cm} T >  T_H  \,,
\end{array} \right.
\end{equation}
with $C_{\Gamma}^\prime \gg C_{\Gamma}$.  The latter inequality ensures that the continuous spectrum  generated by  the ansatz  (\ref{fiveB}) does not contribute into energy density  (\ref{one4})
for $T \le T_H$. Also  the  ansatz  (\ref{fiveB}) numerically  
matches the original parameterization  of 
Ref. \cite{Tawfik03} for $T < T_H$, where the hadron gas model can be described by  the lightest hardon species (lighter than 1 GeV), but  with their lattice values of  masses. 
Moreover, a numerical  analysis of (\ref{five4}) and (\ref{fiveB}) parameterizations 
shows that, although the ansatz (\ref{fiveB}) is  physically correct, their results for the energy densities are the indistinguishable. 
Therefore, in what follows I will use the data obtained with  (\ref{five4}) because they require less efforts.

{
Note that for heavy resonances having the widths (\ref{five4}) or  (\ref{fiveB})
 the resulting mass distribution
will be a power law which is seen both in hadron-hadron reactions \cite{GazdGor}
and nucleus-nucleus reactions \cite{McLerran}
at high  energies.
}

Also one should remember that the representation of energy density (\ref{one4}) by the spectral function 
is chosen for the sake of simplicity. In fact, such a representation can be assumed for  pressure or entropy density. The latter is preferable, if one needs to build thermodynamically consistent description of all available lattice data, but this is not  the main point of the present discussion. Therefore, in what follows I will always choose  the most convenient thermodynamic function just for simplicity.  


%
\subsection{Applications to Lattice Quantum Chromodynamics  and Heavy-Ion Collisions.}
\label{sect:432}
\begin{figure}[t]
\centerline{\includegraphics[width=10cm,angle=-90]{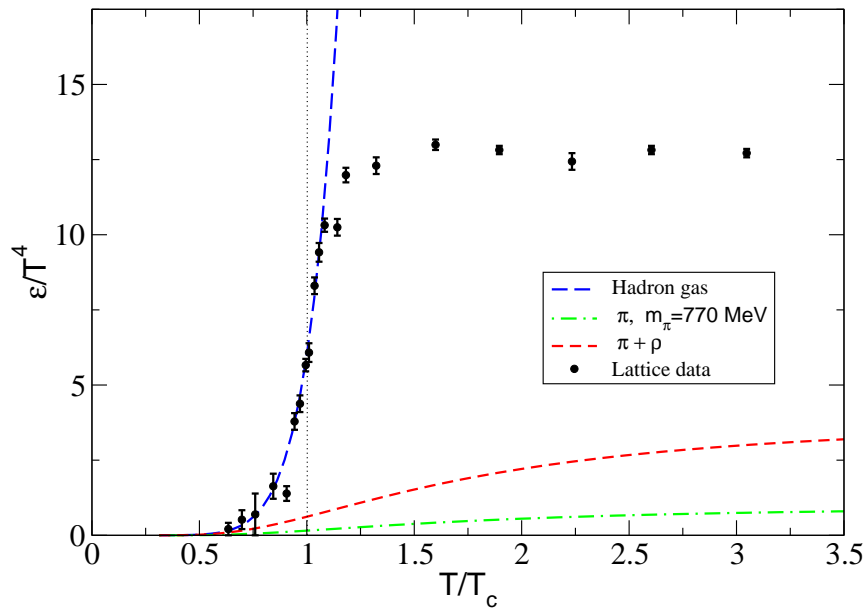}}
\caption{\label{fig420}
Fit of the lattice quantum chromodynamics  data \cite{Tawfik03} with the hadron gas   model of Ref. \cite{Tawfik03}. 
}
\end{figure}

\begin{figure}[t]
\centerline{\includegraphics[width=10cm,angle=-90]{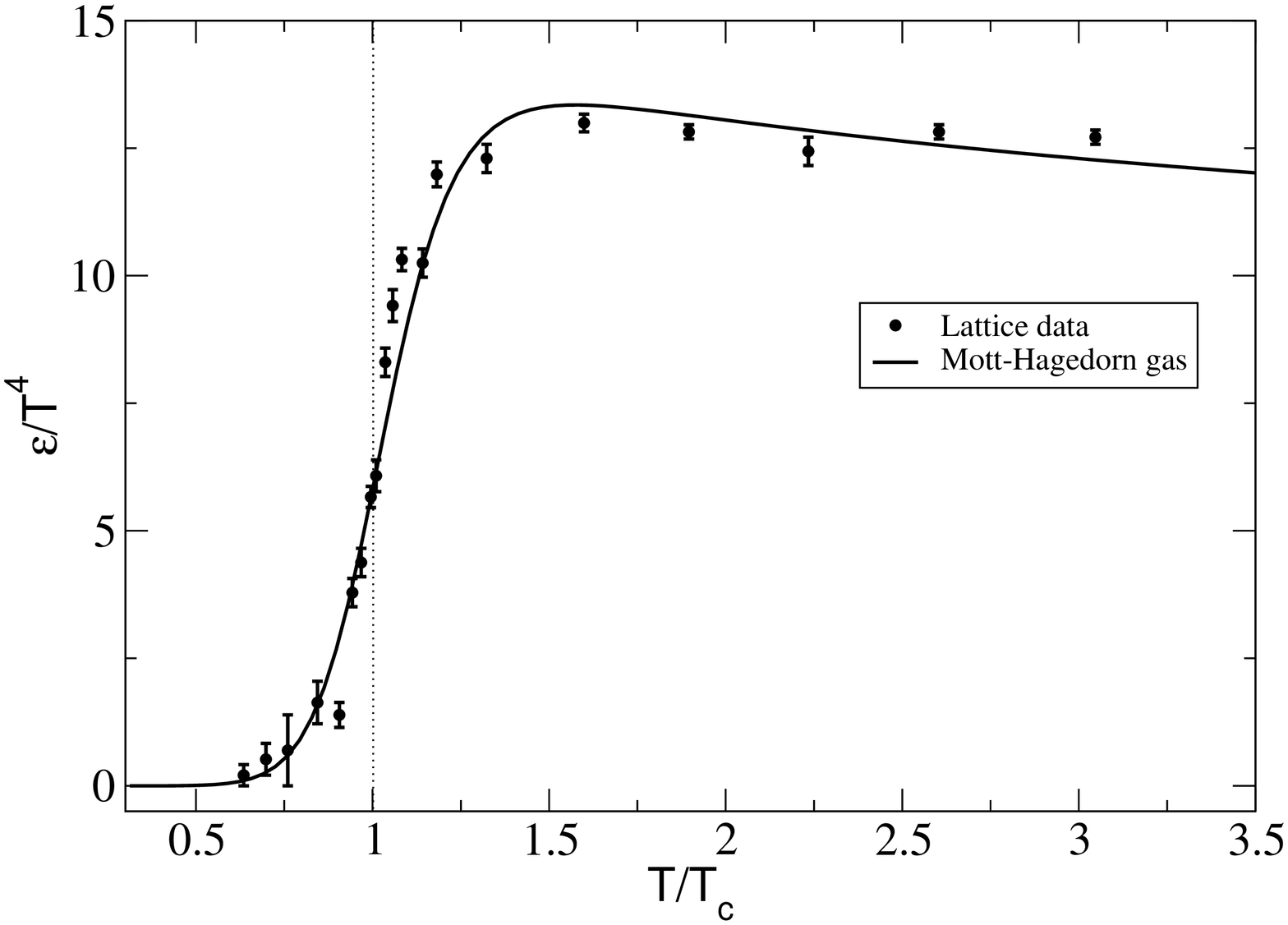}}
\caption{\label{fig421}
Fit of the lattice quantum chromodynamics  data \cite{Tawfik03} with the Mott-Hagedorn resonance 
gas model (\ref{five4}). For details see text.
}
\end{figure}

{As one can see from Fig. \ref{fig420} the hadron gas  model  \cite{Tawfik03} correctly 
reproduces the
lattice quantum chromodynamics  results below the critical temperature $T_c$ and just in a
vicinity above $T_c$, but not for large temperatures.  
On the other hand Fig. \ref{fig421} shows a comparison of the same lattice quantum chromodynamics  data  
\cite{Tawfik03}
with the Mott-Hagedorn gas (\ref{five4}) } where the parameters of the spectral 
function are  $N_T=2.325$, $N_m = 2.5$ and $T_H=165$ MeV and $m_A=m_B=1$ GeV. 
The successful description of the lattice energy density \cite{Tawfik03} 
indicates that above $T_c$ the strongly interacting matter may be well 
described in terms of strongly correlated hadronic degrees of freedom.
This result is based on the concept of soft deconfinement and provides an 
alternative to the conventional explanation of the deconfinement transition
as the emergence of quasifree quarks and gluons.

%

Another interesting  feature of the model (\ref{five4}) is 
that it allows to explain naturally the absence of
heavy resonance contributions to the particle yields measured at
highest SPS and all RHIC energies, where QGP conditions are expected 
\cite{Hgas, Hgas:2,RHIC:Chem}.
In order to find out whether a given resonance has a chance to survive till 
the freeze-out it is necessary to compare its lifetime with the typical 
timescale in the system.
There are two typical timescales usually discussed in nucleus-nucleus 
collisions, the equilibration time $\tau_{eq}$ and the formation time $\tau_f$.
The equilibration time tells when the matter created in collision process
reaches a thermal equilibrium which allows to use the hydrodynamic and 
thermodynamic descriptions.
For Au + Au collisions at RHIC energies it was estimated to be about 
$\tau_{eq} \approx 0.5$ fm \cite{Taueq}.  
On the other hand in transport calculations  the formation time is used:
the time for constituent quarks to form a hadron.  
The formation time depends on the momentum and energy of the created hadron,
but is  of the same order $\tau_f \approx 1 - 2$ fm \cite{Tauf} as the 
equilibration time.

Since within  this  model the QGP is equivalent to a resonance gas with medium
dependent widths, all hadronic resonances with life time $\Gamma^{-1} (m)$ 
shorter  than  $\max\{\tau_f, \tau_{eq}\} $
will have no chance to be formed in the system. 
Therefore, the upper limit of the the integrals over the resonance mass $m$ 
and over the virtual mass $\sqrt{s}$ in  Eq. (\ref{one4})
should be reduced to a resonance mass defined by 
\begin{equation}
\label{Asix}
%
%
\Gamma (m)^{-1} =  \max\{\tau_f, \tau_{eq}\} \,.
\end{equation}
\noindent
This reduction may essentially weaken the energy density gap at the transition
temperature  or even make it vanish.
Thus, the explicit time dependence should be introduced into
the resonance width model (\ref{one4}) while applying it to nuclear collisions, 
and this finite time (size) effect, as I  discussed, may change essentially
the thermodynamics of the hadron resonances formed in the nucleus-nucleus 
collisions.

\subsection{Anomalous J/$\psi$ Suppression.}
\label{sect:433}
The phenomenon of anomalous J/$\psi$ suppression as observed by the NA50
collaboration at CERN-SPS with ultrarelativistic Pb-Pb collisions at 158 GeV/A
did not yet find a satisfactory explanation.
It has been demonstrated that the extrapolation from results with pA collisions
and light ion beams (O, S) fails to describe the E$_T$ dependence of the
J/$\psi$ survival probability (production cross section) above E$_T\sim 40$ GeV. 
Successful fits to the data are obtained with models assuming a critical
phenomenon where the conditions for the onset of the new phase are fulfilled
at the above E$_T$ value. Among these models are in particular:
\begin{itemize}
\item percolation (Satz) \cite{Satz, Satz:2}
\item J/$\psi$ Mott transition (Blaizot/Ollitrault) \cite{Blaizot}
\item D-meson Mott transition (Blaschke/Burau)  \cite{burau}
\end{itemize}
Here I  would like to generalize the kinetic theory approach to J/$\psi$
suppression as it was formulated in Ref. \cite{burau} and study
charmonium dissociation in a Mott-Hagedorn resonance gas.
Besides the impact by off-shell $\pi$ and $\, \rho$ mesons which has been studied
in \cite{burau}, the resonace gas (\ref{one4})-(\ref{five4}) consists of more massive resonances so that no
reaction threshold occurs for the breakup into open-charm hadrons
(D-mesons, $\Sigma_c$, $\Lambda_c$, excited states).

The key quantities for the solution of a kinetic equation of the charmonium
distribution function are the rate coefficients which can be estimated within
the binary collision approximation.
This approximation shall be applicable since the first collision is expected
to destroy the charmonium and thus determines the lifetime as long as the
charm density is too low for the reverse process of charmonium gain
(fusion of $D\bar D$, $\Lambda_c ~ \bar D$, ...).

One can  estimate the charmonium lifetime to be
\begin{equation}
\tau^{-1}_\psi(T)=\sigma_{\psi R \to X} v n_R(T)
\end{equation}
%
\begin{figure}[ht]
\centerline{\includegraphics[width=10cm,angle=-90]{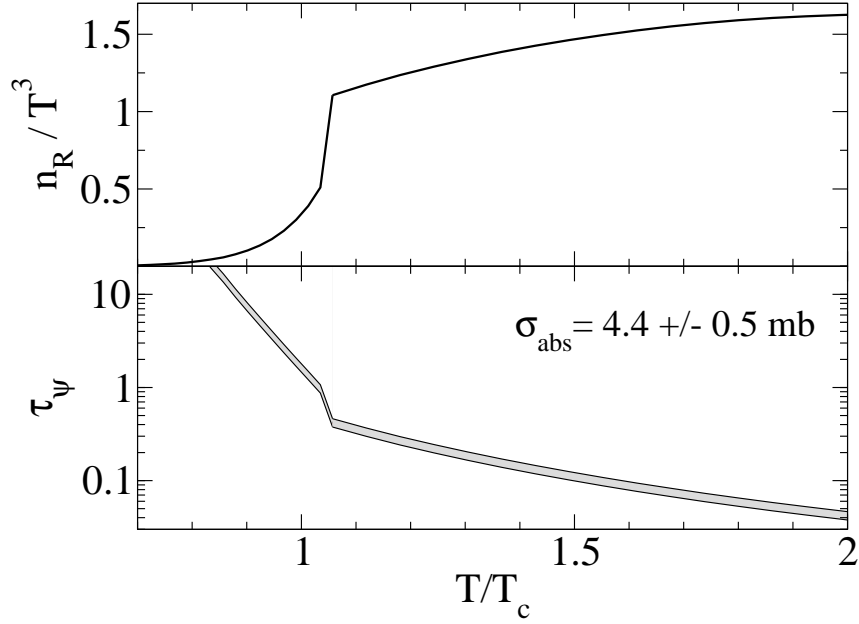}}
\caption{\label{dens}
Density of resonances  (upper panel)
and mean J/$\psi$ lifetime (lower panel)
in a Mott-Hagedorn resonance gas.
}
\end{figure}
%
under the assumption that the charmonium breakup cross section is rather
universal and may be estimated by the value for $\psi$ absorption measured in
pA collision experiments of charm hadroproduction.
In pA and S-U collisions an absorption cross section 
$\sigma_{\rm abs}$ of $4.4 \pm 0.5$ mb has been extracted from the data by using
a Glauber model analysis \cite{na50}.
The velocity of a typical hadron resonance is of the order of 0.5  and the
number density of resonances at a given temperature is given by
\begin{eqnarray}
\hspace*{-0.5cm} n_R(T)&=&\sum_{i=\pi, \, g, ..} g_i~ n_M(T,\mu;m_i)
+ \sum_{A=M,B}\int_{m_A}^\infty dm \int_{m_B^2}^\infty {ds} ~ \, g_H(m)~
A(s,m)~ n_A(T,\mu_A;\sqrt{s})~,
\end{eqnarray}
%
\begin{figure}[ht]
\centerline{\psfig{figure=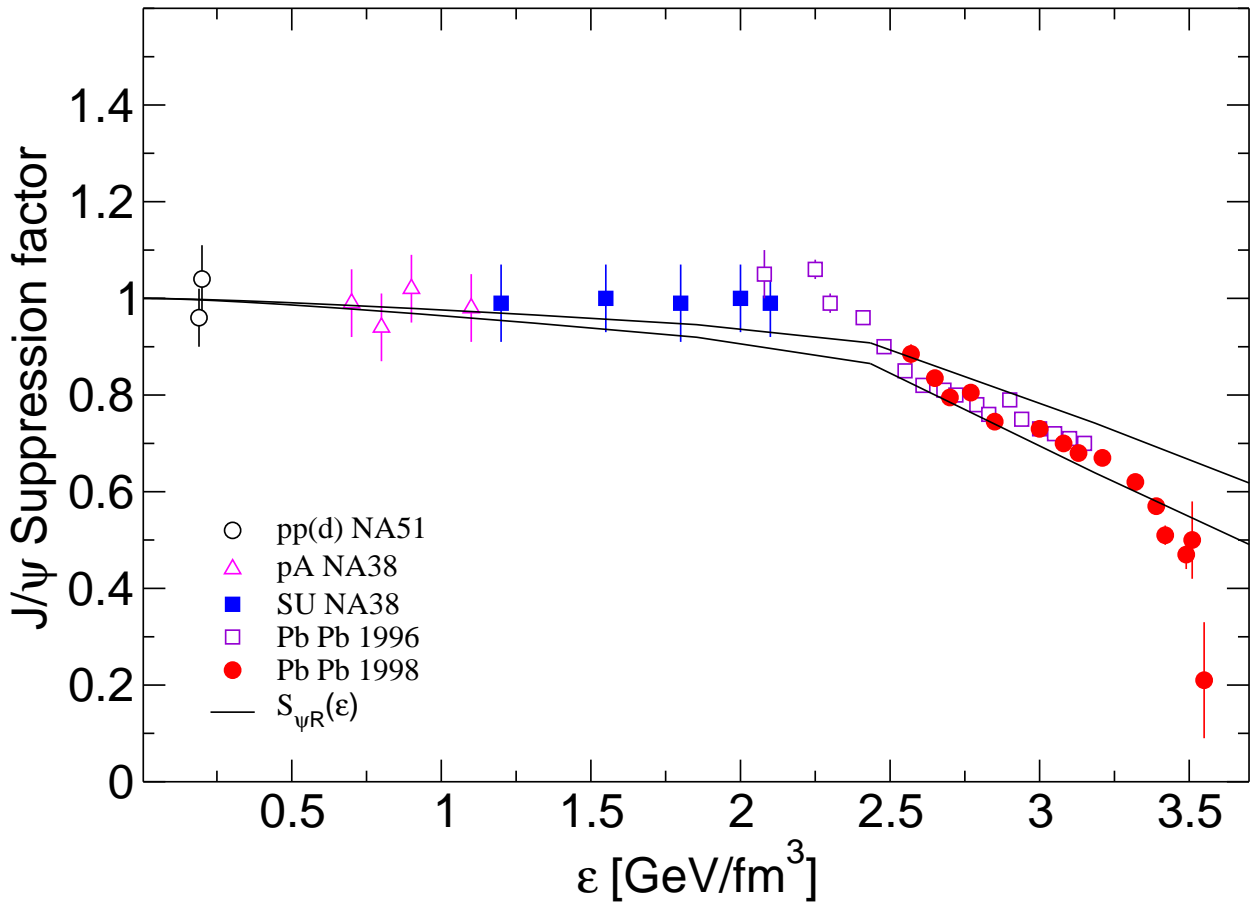,width=14.7cm,angle=-90}
}
\caption{\label{suppr}
J/$\psi$ survival probability in a Mott-Hagedorn resonance gas as a
function of the energy density. Data are from the NA50 collaboration
\cite{na50}.}
\end{figure}
%
where the number density per degree of freedom is as follows
\begin{equation}
n_A(T,\mu_A;m)=\int\frac{d^3k}{(2\pi)^3}\frac{1}{{\rm
e}^{(\sqrt{k^2+m^2}-\mu_A)/T}+\delta_A}~,
\end{equation}
see Fig. \ref{dens} (upper panel).
The result for the charmonium lifetime is given in Fig. \ref{dens}
and exhibits a sharp drop at the critical temperature for the Mott
transition in the resonance gas.
This behavior is then reflected in  a threshold-like behavior of the
charmonium survival probability 
\begin {equation}
S(T_0)=\exp\left(-\int_{\tau_0}^\infty d\tau ~\tau_\psi^{-1}(T(\tau))\right)~,
\end{equation}
{
estimated within the one-dimensional hydrodynamic expansion (Bjorken scenario)
with the following dependence $T^3(\tau)\tau=T^3(\tau_0)\tau_0$
of the temperature of the system on the
proper time $\tau$. 
}

The result is shown in Fig. \ref{suppr}. As soon as the initial temperature
$T_0=T(\tau_0)$ is above the critical one, the survival probability drops
according to the increase in the effective density of degrees of freedom
measured by $n_R (T)$, see Fig. \ref{dens}.
Therefore,  the present model suggests that  the step-like drop in the J/$\psi$ production
cross section in  Pb-Pb collisions at SPS energies can be explained by the
dramatic increase in the number of hadronic resonances at the 
Hagedorn temperature in the resonance gas.

Thus, the statistical bootstrap model allows one to interpret the QGP as the hadron
resonance gas dominated by the state of infinite mass (and infinite volume).
As was shown, it is necessary to include the resonance width into the SBM
in order to avoid the contradiction with  experimental data on hadron
spectroscopy.
One can see  that the simple model (\ref{one4})-(\ref{five4})
may not only eliminate the
divergence of the thermodynamic functions above $T_H$, but it is also able to
successfully describe the lattice quantum chromodynamics  data \cite{Tawfik03}
for energy density.
Such a model also
explains  the absence of heavy resonance contributions in the
fit of the experimentally measured particle ratios at SPS and RHIC energies.

However, such a modification of the SBM requires an essential change in our
view on QGP: it is conceivable that hadrons of very large masses which should
be associated with a QGP cannot be formed
in nucleus-nucleus collisions because of their very short lifetime.

{
As also  was shown  the
dramatic increase in the number of hadronic resonances at the 
Hagedorn temperature of this model
can explain the step-like drop in the J/$\psi$ production
cross section in  Pb-Pb collisions at SPS energies.
}

It is also necessary to mention  that the presented model should be applied to
experimental data with care: it can be successfully applied
to describe either the quantities associated with the chemical freeze-out,
i.e. particle ratios or spectra of $\Omega$ hyperons, $\phi$, $J/\psi$ and
$\psi^\prime$ mesons that are freezing out
at hadronization \cite{Bugaev:01d, Bugaev:02, Bugaev:02a, Bugaev:02b,  BD:00, TLS:01}.
But as discussed in Refs.
\cite{Bugaev:96, Bugaev:02HC, Bugaev:04HC} the model  presented  here   should not
be used for the post freeze-out momentum spectra of other hadrons
produced in the nucleus-nucleus collisions.
Perhaps only such weakly interacting hadrons like $\Omega$, $\phi$, J/$\psi$
and $\psi^\prime$ will allow us to test the model presented here.







\section{Quark Gluon Bags with Surface Tension}

This section is devoted to the reformulation of the GBM \cite{Goren:81,Vol:1,Kapusta:81},
which was analyzed in  the chapter 2 for  finite volumes. Its present generalization, 
the QGBST model \cite{Bugaev:07,Bugaev:07new, Bugaev:08new}, 
is more realistic than the GBM and  maybe useful  for the experimental  location of the 
(tri)critical endpoint of  quantum chromodynamics which is hoped to be done at  BNL RHIC, CERN SPS, GSI FAIR 
and, perhaps, at JINR NICA.

The GBM \cite{Goren:81,Vol:1,Kapusta:81}
itself contains  two
well-known models of deconfined and confined phases:
the bag model of   QGP \cite{BagModel, bag-qgp} and
the hadron gas model \cite{Hgas, Hgas:2}.  
Since, on the one hand,  the MIT Bag Model  \cite{BagModel} is able to simultaneously describe the hadron mass spectrum, i.e.
the hadron masses and their proper volumes, and the properties of the deconfined phase \cite{bag-qgp} and, on the other hand,  the  hadron gas model is rather successful in describing the experimental 
particle yields  \cite{Hgas, Hgas:2, RHIC:Chem, RHIC:Chem2},
there were hopes \cite{Goren:05}  that an exact analytical solution of the GBM found in 
\cite{Goren:81} could  be helpful in understanding the properties of strongly interacting matter. However, this solution does not allow one to introduce the critical end point 
of  the strongly interacting matter phase diagram.  Also,  a complicated construction 
of the line, along which the phase transition  order  gradually increases, suggested 
in  \cite{Goren:05}, does look too artificial. Therefore, the present GBM formulation 
lacks an important physical input and 
is  interesting only as a toy example which can be solved analytically.

On the other hand,   the models, which can 
correctly reproduce the  expectation \cite{Misha,fodorkatz,karsch} 
that the end point of the 1$^{st}$ 
order phase transition  (PT)  line to QGP  should 
be the 2$^{nd}$ order critical point, are indeed  necessary for heavy ion phenomenology. 
In addition, such phenomenological models can provide us with the information 
about the phase structure and equation of state of strongly interacting matter which is located   between the critical endpoint and the region of the 
color superconductivity because such an information is unavailable otherwise. 
Therefore, the present section  is devoted to the extension of the GMB. 
I  think that  the GMB  can be drastically improved   by the inclusion of such a vitally important element as  the surface tension of the quark-gluon bags.

The dynamical surface tension of the quark-gluon bags was estimated long ago \cite{Jaffe:1, Jaffe:2},
but it was never used in statistical description of the equation of state.  Moreover, the estimate of the bag surface tension made in  \cite{Jaffe:2} is negligible for $u$ and $d$ quarks of and, hence, can be safely neglected in  the present  treatment.
 The situation with the  surface tension of the quark-gluon bags is somewhat 
unclear:  the early estimates within the MIT Bag Model \cite{Svet:1, Svet:2} indicate that 
small hadronic bubbles can exist in the hot QGP, whereas the analysis based on the effective potential  of  the 1$^{st}$  order PT in early Universe  \cite{Ignat:1} does not support the results of Refs.  \cite{Svet:1, Svet:2}. 
Thus, it turns out that the surface energy may play an important role for  the properties of 
hadronic bubbles  \cite{Svet:1, Svet:2, Ignat:1}  and  QGP bags \cite{Moretto:05},  
but
the surface tension of large bags was not included into a consistent  statistical description of QGP.
Therefore,   the present paper  is devoted  to the investigation and  analysis of the critical
properties of the model of quark-gluon bags with surface tension (QGBST model hereafter). 

In statistical mechanics there are several exactly solvable cluster  models with 
the 1$^{st}$ order PT which describe the critical point properties  very well.
These models are built on the assumptions that 
the difference of  the bulk  part (or the volume dependent part) of  free energy  
of two phases disappears at  phase equilibrium and that, in addition, 
the difference of the surface part (or the surface tension) of  free energy  vanishes 
at the critical point. 
The most famous of them is  the Fisher droplet model (FDM) \cite{Fisher:67,Thermostat:3}
which has been successfully used to analyze the condensation  of a gaseous phase 
(droplets of all sizes)   into a liquid (see chapter 2). 
The  FDM has been applied to many different systems,
including nuclear multifragmentation  \cite{Moretto:97}, nucleation of real fluids \cite{Dillmann},
the compressibility factor of real fluids \cite{Kiang}, clusters of the Ising model \cite{mader-03}
and percolation  clusters \cite{Percolation}.

As was discussed in the chapter 2,  the analytical results \cite{Bugaev:00,Reuter:01}  for
a simplified SMM version \cite{Gupta:98, Gupta:99}   give a very strong evidence that the SMM, and, thus, 
the nuclear matter,  has a tricritical endpoint rather than a critical  endpoint.

Such a   success of  the SMM initiated  
the studies of the surface partitions of large clusters  
within the Hills and Dales Model
 \cite{Bugaev:04b,BugaevElliott} and led to a discovery of the origin  of
the temperature independent surface entropy similar to the FDM.  
As a consequence, the surface tension coefficient of large 
clusters consisting of the discrete constituents should linearly depend 
on the temperature of the system \cite{Bugaev:04b} and  must vanish at the critical endpoint.
However, the present formulation of the Hills and Dales Model
\cite{Bugaev:04b,BugaevElliott}, which successfully 
estimates the upper and lower bounds of the surface deformations of the discrete 
physical clusters, does not look  suitable for  quark-gluon bags.
Therefore, in this section I will assume a certain  dependence of the surface 
tension coefficient on temperature and baryonic chemical potential,  and  concentrate  on the impact  of  surface tension of the quark-gluon bags on  the properties of 
the deconfinement  phase diagram and 
the quantum chromodynamics critical endpoint. 
A discussion of the origin of the surface tension
is a subject of a future  research. 


Here I  will show that the existence of a cross-over at  low values of
the  baryonic chemical potential along with  the 1$^{st}$ order deconfinement PT
at high baryonic chemical potentials  leads to the existence of an additional PT 
of the 2$^{nd}$ or higher order  along the curve where the surface tension coefficient  vanishes. Thus, it turns out that the QGBST model predicts 
the existence of the tricritical rather than critical endpoint.

In the subsection \ref{sect:4.4.1} I   formulate the basic ingredients of 
the QGBST model  and  analyze  all possible singularities of its  isobaric partition for vanishing
baryonic densities. This analysis is generalized to non-zero  baryonic densities in the subsection \ref{sect:4.4.2}.
The subsection \ref{sect:4.4.3}  is devoted to the analysis of the surface tension induced  PT which exists above the deconfinement PT. 
The conclusions and research perspectives are summarized in the subsection \ref{sect:4.4.4}

\subsection{The Role of Surface Tension.}
\label{sect:4.4.1}
Again, like in the chapter 2, I consider the isobaric partition (the notations are the same as in chaprer II):
\begin{align}\label{Zs}
& \hat{Z}(s,T) \equiv \int\limits_0^{\infty}dV\exp(-sV)~Z(V,T) =\frac{1}{ [ s - F(s, T) ] }  
\end{align}
where the function $F(s, T)$  is defined as follows
\begin{align}
F(s,T)&\equiv F_H(s,T)+F_Q(s,T) = \sum_{j=1}^n g_j e^{-v_js} \phi(T,m_j) 
+ ~ u(T)
 ~  \int\limits_{V_0}^{\infty}dv~\frac{ \exp\left[-v\left(s -s_Q(T)\right)\right] }{v^{\tau}}
~.
 \label{FQs}
\end{align}

At the moment the particular choice of function $F_Q(s,T)$ in (\ref{FQs}) is not important. 
The key point of my treatment  is that it should have the form of Eq.  (\ref{FQs}) which 
has a singularity  at  $s=s_Q$ 
because for $s<s_Q$ the integral over the bag   volume $v$  diverges at its upper limit. 
Note that 
the exponential in (\ref{FQs}) is nothing else, but a difference of the bulk 
free energy of a bag of volume $v$, i.e. $ -T s v$,  which is  under external pressure 
$T s$,   and  the bulk  free energy of 
the same bag filled with QGP, i.e.  $ -T s_Q v$. 
At phase equilibrium this difference of the bulk free energies  vanishes. 
Despite  all  positive features,  Eq. (\ref{FQs}) lacks the surface part of  free energy of bags, which will be called  a surface energy hereafter. 
In addition to the difference of the bulk free energies
the realistic statistical models which demonstrated  their validity, 
the FDM \cite{Fisher:67} and SMM \cite{Bondorf:95}, 
have the contribution of the surface  energy which plays an important role in  defining the  phase diagram structure \cite{Bugaev:00,Bugaev:07b}. 
Therefore, I modify Eq. (\ref{FQs}) by introducing the surface   energy of the bags in a general fashion \cite{Reuter:01}:
\begin{align}\label{FQsnew}
&F_Q (s, T) = u(T)
   \int\limits_{V_0}^{\infty}dv~\frac{ \exp\left[\left(s_Q(T)-s\right)v - \sigma(T)\, v^{\varkappa}\right] }{v^{\tau}}\,,
 \end{align}
where the ratio of the  temperature dependent surface tension coefficient  to $T$
(the reduced surface tension coefficient hereafter) 
which has the form $\sigma(T) = 
\frac{\sigma_o}{T} \cdot
\left[ \frac{ T_{cep}   - T }{T_{cep}} \right]^{2k + 1} $  ($k =0, 1, 2,...$).  
Here $\sigma_o > 0$ can be a smooth function of the temperature, but for simplicity I  fix it to be a constant.  
For $k = 0$ the two terms in the surface (free) energy of a $v$-volume bag  have 
a simple interpretation \cite{Fisher:67}: thus, the surface energy of such a bag is
$\sigma_0 v^{\varkappa}$, whereas the free energy, which comes from  the surface entropy $\sigma_o T_{cep}^{-1} v^{\varkappa} $,  is  
$- T \sigma_o T_{cep}^{-1} v^{\varkappa}$.  
Note that   the surface entropy of a  $v$-volume  bag
counts its degeneracy factor or the number of ways to make 
such a bag with all possible surfaces. This   interpretation 
can be extended to 
$k >  0$  on the basis of  the Hills and Dales Model 
\cite{Bugaev:04b,BugaevElliott}.

In choosing such a simple surface energy parameterization I 
follow the original Fisher idea \cite{Fisher:67}  which allows one to account for 
the surface energy by considering 
some mean bag of volume $v$ and surface $v^{\varkappa}$. The 
consideration of 
the general mass-volume-surface bag spectrum I  leave  for the future investigation. 
The power  $\varkappa < 1$ which describes the bag's effective  surface is a constant which,  in principle, can differ from the typical FDM and SMM value 
$\frac{2}{3}$.
This is so because  near  the deconfinement PT region  QGP  has  low density and, hence, 
like in the low density  nuclear matter \cite{Ravenhall},  
the non-sperical bags (spaghetti-like or lasagna-like \cite{Ravenhall})  can be  favorable
(see also \cite{Svet:1, Svet:2} for  the bubbles of complicated shapes).
A similar idea of  ``polymerization" of gluonic quasiparticles was introduced recently 
\cite{Shuryak:05a}.

The second  essential  difference with the FDM and SMM surface tension 
parameterization is that I  do not require the  vanishing of $\sigma(T)$ above the CEP. 
As will be shown later,  this is the most important assumption which, in contrast to the GBM,  allows one to naturally describe the cross-over  from hadron gas to QGP.  
Note that  negative value of the reduced surface tension coefficient $\sigma(T)$ above the CEP 
does not mean anything wrong. As was discussed above, 
the  surface  tension coefficient consists of energy and entropy parts which  have   opposite signs \cite{Fisher:67,Bugaev:04b,BugaevElliott}. 
Therefore, $\sigma(T) < 0 $ does not mean that the surface energy changes the sign, but it
rather  means that the surface entropy, i.e. the logarithm of the degeneracy of bags of a fixed volume, simply  exceeds their  surface energy.  In other words, 
the number of  non-spherical bags of a fixed volume becomes so large that the Boltzmann exponent, which accounts for the energy "costs" of these bags,  cannot
suppress them anymore.

Finally, the third essential difference with the FDM and SMM is that I assume 
that the surface tension in the QGBST model  vanishes at some line in $\mu_B-T$ plane, 
i.e. $T_{cep} = T_{cep} (\mu_B)$. However,  below I will 
consider $T_{cep} = Const $ for simplicity and   will  discuss the necessary modifications of the model  with $T_{cep} = T_{cep} (\mu_B)$.

The surface energy should, in principle, be introduced into a  discrete part of 
the mass-volume spectrum $F_H$, but a successful  fitting of the particle yield ratios \cite{Hgas, Hgas:2} with 
the experimentally determined hadronic spectrum $F_H$ does not indicate 
such a necessity. 

{
In principle, besides  the bulk and surface parts of  free energy,  the  spectrum (\ref{FQsnew}) could include the curvature part as well, which may be important for small hadronic bubbles \cite{Svet:1, Svet:2} or for cosmological PT \cite{Ignat:1}.  
It is necessary to  stress, however,
that the critical properties of the present model are defined by the infinite bag, therefore
the inclusion in (\ref{FQsnew}) of a curvature term  of any  sign could  affect 
the thermodynamic quantities of this  model at $s = s_Q(T)$ and $\sigma (T) = 0$,
which is possible  at (tri)critical  endpoint only (see below). 
If, the curvature term was really  important for the cluster models like the present one, then  it should have been seen also at  (tri)critical  points of  the FDM, SMM and many systems  described by the FDM \cite{Moretto:97, Dillmann, Kiang, mader-03, Percolation}, but this is not the case. 
Indeed,  recently  the  Complement method   \cite{Complement} was applied to 
the analysis of the  largest, but still mesoscopic  drop of a radius $R_{dr}$  representing the liquid in equilibrium with its vapor.  The method allows one to find out  the concentrations 
of  the vapor clusters in finite system in a whole range of  temperatures and determine 
the free energy  difference of  two phases with high precision. The latter enables me 
not only to  extract  the critical temperature, surface tension coefficient and even the  value of Fisher index $\tau$ of the  infinite system,  but also such a delicate effects as 
the Gibbs-Thomson correction  \cite{Gibbs-Thomson}  to the free energy of a liquid 
drop.  Note that  the Gibbs-Thomson correction  behaves as   $R_{dr}^{-1}$, but the  Complement method   \cite{Complement}  allows one to  find it, whereas the curvature part of free energy, which is proportional to 
$R_{dr}$,  is not seen  both for a drop and for smaller clusters. 
Such a result   is  directly related  to  the QGP bags because 
quantum chromodynamics  is expected to be in the same universality class \cite{Misha} as 
the 3-dimensional Ising model whose clusters were analyzed  in \cite{Complement}. 
Therefore, admitting that for  finite QGP bags  the curvature effects may be essential,  
I leave them out  because the critical behavior of the present model is defined  by the properties of  the  infinite bag.

}

According to the general theorem \cite{Goren:81} the analysis of PT  existence of the GCP  is now reduced to the analysis of the rightmost singularity of 
the isobaric partition (\ref{Zs}).  Depending on the sign of the reduced surface tension coefficient, there are three possibilities.  Since at this point the present model  essentially  differs from
the FDM and GBM, it is necessary to study all possibilities in details. 

\begin{figure}[ht]
\centerline{\includegraphics[width=7.3cm,height=7.3cm]{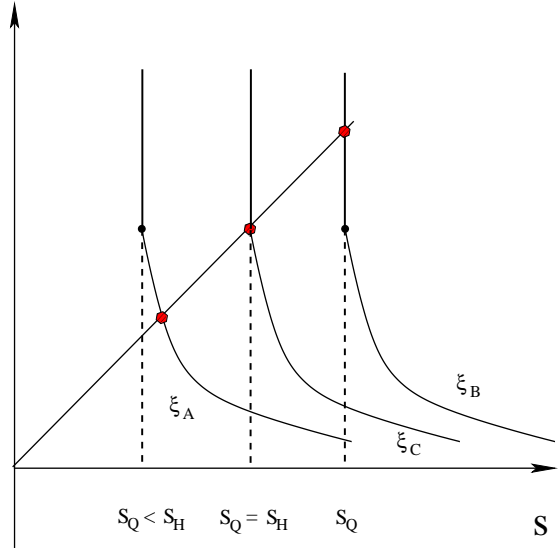}
}
\vspace*{-0.3cm}
\caption{
Graphical solution of Eq. (\ref{s*vdw}) which corresponds to a PT.
The solution of Eq. (\ref{s*vdw}) is shown by a filled hexagon.
The function $F(s, \xi)$ is shown by a solid curve for a few 
values of  the parameter $\xi$.  The function  $F(s, \xi)$ diverges 
for $s < s_Q( \xi)$ (shown by dashed lines), but is finite at $s = s_Q( \xi)$ (shown by black circle).  
At low values of  the parameter $\xi = \xi_A$, which can be either $T$ or $\mu_B$, 
the simple pole $s_H$ is the rightmost singularity and it corresponds to hadronic phase. 
For  $\xi = \xi_B \gg \xi_A$ the  rightmost singularity is an essential singularity $s = s_Q( \xi_B)$, 
which describes QGP. 
At intermediate  value $\xi = \xi_C$ both singularities coincide $s_H( \xi_C) = s_Q( \xi_C)$ and 
this condition is a Gibbs criterion. 
}
  \label{fig441}
\end{figure}

\noindent
({\bf I}) The first possibility corresponds to $\sigma(T) > 0$. Its treatment is  very similar 
to the GBM choice (\ref{FQs})  with $\tau > 2$ \cite{Goren:81}. In this case at low 
temperatures the QGP  pressure $T s_Q(T)$ is negative and, therefore, the   rightmost singularity is a simple pole of the isobaric partition  
$s^* = s_H (T) = F(s_H(T), T) > s_Q(T)$, which is mainly defined by a discrete part of  the mass-volume spectrum $F_H(s,T)$. 
The last inequality provides the convergence of the volume integral in (\ref{FQsnew})
(see Fig. ~\ref{fig441}). On the other hand at very high $T$ the QGP pressure dominates and, hence, the rightmost singularity is the essential  singularity  of the isobaric partition  $ s^* = s_Q(T)$.  
The phase transition occurs, when the singularities coincide:
\begin{align}\label{PTI}
&  s_H (T_c) \equiv  \frac{p_H (T_c)}{T_c} =  s_Q (T_c) \equiv  \frac{p_Q (T_c)}{T_c}\,,
 \end{align}
which is nothing else, but the Gibbs criterion. 
The graphical solution of Eq. (\ref{s*vdw})  for  all  these possibilities is shown in Fig.~\ref{fig441}.  
Like in the GBM \cite{Goren:81,Goren:05}, the  necessary condition for the PT existence is  the finiteness of $F_Q(s_Q(T), T) $ at $s = s_Q(T)$.
It can be shown that the sufficient conditions are   the following 
inequalities: $F_Q(s_Q(T), T) > s_Q(T) $ for low temperatures and 
  $F(s_Q(T), T) < s_Q(T) $  for  $T \rightarrow \infty$.  These 
conditions provide that at low $T$ the rightmost singularity  of the isobaric partition is a simple pole,
whereas for hight $T$ the essential singularity 
 $s_Q(T) $ becomes  its  rightmost one   (see Fig.~\ref{fig441} and a detailed 
analysis of case $\mu_B \neq 0$).

%
%

\vspace*{-0.0cm}

The PT order can be found from the $T$-derivatives of  $s_H (T)$. 
Thus,  differentiating  (\ref{s*vdw}) one finds 
\vspace*{-0.2cm}
\begin{align}\label{sHprime1}
s_H^{\prime}~=~
\frac{G~+~u\,{\cal K}_{\tau-1}(\Delta, - \sigma) \cdot  s_Q^{\prime}}{1~+~u\,{\cal K}_{\tau-1}(\Delta, - \sigma)} \,,
\end{align}

\vspace*{-0.2cm}

\noindent
where the functions $G $ and ${\cal K}_{\tau -a} (\Delta, - \sigma) $ are defined as
\begin{align}\label{G1}
&\hspace*{-0.25cm}G \equiv F_H^{\prime}+ \frac{u^{\prime}}{u} F_Q 
+  {\textstyle\frac{ (T_{cep} - 2k T) \sigma(T)}{(T_{cep} - T)\, T } }\,u \,
{\cal K}_{\tau- \varkappa}  (\Delta, - \sigma)\,, \\
\label{KQ}
&\hspace*{-0.25cm}{\cal K}_{\tau -a} (\Delta, - \sigma) \equiv  \hspace*{-0.0cm}
\int\limits_{V_o}^{\infty}\hspace*{-0.05cm}dv~\frac{\exp\left[-\Delta v - \sigma(T) 
v^{\varkappa} 
\right] }{v^{\tau-a}} \,,
\end{align}
where $\Delta \equiv s_H - s_Q$. 

Now it is easy to see that the transition is of the 1$^{st}$ order,
i.e. $s_Q^{\prime}(T_c)>s_H^{\prime}(T_c)$, provided  $ \sigma(T) > 0$ for any $\tau$.
The 2$^{nd}$ or higher order phase transition takes place
provided $s_Q^{\prime}(T_c)=s_H^{\prime}(T_c)$ at $T=T_c$.
The latter condition is satisfied  when $ {\cal K}_{\tau-1}$ diverges to infinity
at $T\rightarrow (T_c-0)$, i.e. for $T$ approaching $T_c$ from below.
Like for the GBM choice (\ref{FQs}), 
such a situation can exist for   $ \sigma(T_c) = 0$ and $\frac{3}{2} < \tau \le  2 $. 
Studying the higher $T$-derivatives  of $s_H(T)$ at $T_c$, one can show 
that  for  $ \sigma(T) \equiv  0$  and  for $(n+1)/n \le \tau < n/(n-1)$ ($n=3,4,5,...$) there is a $n^{th}$ order phase  transition
\begin{align}\label{nth}
& s_H(T_c)~ =~ s_Q(T_c)~,~~
s_H^{\prime}(T_c)~ =~ s_Q^{\prime}(T_c)~,~...~ \nonumber \\
& s_H^{(n-1)}(T_c)~=~ s_Q^{(n-1)}(T_c)~,~~
s_H^{(n)}(T_c)~\ne~ s_Q^{(n)}(T_c)~,
\end{align}
 with $ s_H^{(n)}(T_c)=\infty$
for $(n+1)/n < \tau < n/(n-1)$
and  with a finite value of $s_H^{(n)}(T_c)$ for $\tau =(n+1)/n$.\\

%
%
\begin{figure}[ht]
\centerline{\includegraphics[width=7.3cm,height=7.3cm]{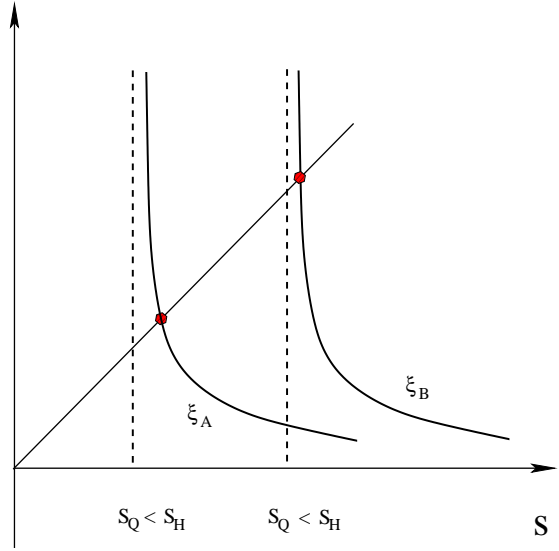}
}
\vspace*{-0.3cm}
\caption{
Graphical solution of Eq. (\ref{s*vdw}) which corresponds to a cross-over.
The notations are the same as in Fig.~\ref{fig441}. 
Now the function  $F(s, \xi)$ diverges 
at $s = s_Q( \xi)$ (shown by dashed lines). 
In this case the simple pole $s_H$ is the rightmost singularity for any value of $\xi $. 
}
\label{fig442}
\end{figure}

\vspace*{-0.0cm}

\noindent
({\bf II}) The second possibility, $\sigma(T) \equiv 0$, described in the preceding paragraph,  does not give anything new 
compared to the GBM \cite{Goren:81,Goren:05}. 
If the  PT exists, then
the graphical picture of  singularities is basically similar to Fig. 1. The only difference
is that, depending on the PT order,   the derivatives of  $F(s,T) $ function with respect to $s$  should  diverge at $s = s_Q(T_c)$.\\

\noindent
({\bf III})  A principally new possibility exists for $T > T_{cep}$, where $\sigma(T) < 0$.
In this case there exists a  cross-over, if for  $T \le T_{cep}$ the  rightmost 
singularity is $s_H(T)$, which corresponds to the leftmost curve in Fig.~\ref{fig441}. 
  Under the latter, its existence can be shown as follows. 
Let me  solve the equation for singularities (\ref{s*vdw}) graphically (see Fig.~\ref{fig442}). 
For  $\sigma(T) < 0$ the function $F_Q(s,T)$ diverges at $s = s_Q(T)$. 
On the other hand, the partial derivatives $\frac{\partial F_H(s,T)}{\partial s} < 0$ and 
$\frac{\partial F_Q(s,T)}{\partial s} < 0$ are always negative. Therefore, the  function 
$F(s,T) \equiv  F_H(s,T) +  F_Q(s,T)$ is a monotonically decreasing function of 
$s$, which vanishes at $s \rightarrow \infty$. Since the left hand side of Eq.  (\ref{s*vdw})  is a  monotonically increasing function of $s$, then there can exist a single intersection 
$s^*$ of $s$ and $F(s,T)$ functions. Moreover, for finite  $s_Q(T)$ values this 
intersection can occur  on 
the right hand side of the point $s = s_Q(T)$, i.e.  $s^* > s_Q(T)$ (see Fig.~\ref{fig442}). 
Thus, in this case the essential singularity $s = s_Q(T)$ can become the rightmost one
for infinite temperature only.  In other words, the pressure of the pure QGP can be reached 
at infinite $T$, whereas for finite $T$ the hadronic mass spectrum gives a non-zero  contribution into all thermodynamic functions. 
Note that such a behavior is typical for the lattice quantum chromodynamics data at zero baryonic chemical 
potential \cite{Karsch:03}.

It is clear that in terms of the present model  a cross-over  existence means
a  fast transition of energy or entropy density in a narrow $T$ region  from a dominance of  the discrete mass-volume spectrum of light hadrons  to
a dominance of  the 
continuous spectrum of heavy QGP bags.  This is exactly the case for  $\sigma(T) < 0$ because  in the right  vicinity of the point $s = s_Q(T)$ the function $F(s,T)$ decreases very  fast and then it gradually decreases as function of $s$-variable. Since,  $F_Q(s,T) $  changes  fast from $F(s,T) \sim F_Q (s,T) \sim s_Q(T)$ to 
$F(s,T) \sim F_H (s,T) \sim s_H(T)$,  their $s$-derivatives should change fast as well. Now, recalling that the change from 
$F(s,T) \sim F_Q (s,T)$ behavior to $F(s,T) \sim F_H (s,T)$ in $s$-variable 
corresponds to the cooling of the system (see Fig.~\ref{fig442}), I conclude that
that there exists a narrow region of temperatures, where the $T$ derivative of    system pressure, i.e. the entropy density, 
drops down from $\frac{\partial p}{\partial T}  \sim  s_Q(T) + T \frac{d s_Q(T)}{d T} $ to  
$\frac{\partial p}{\partial T}  \sim  s_H(T) + T \frac{d s_H(T)}{d T} $
very fast  compared  to other regions of $T$, if system cools.
If, however, in the vicinity of $T= T_{cep} -0$ the rightmost singularity is $s_Q(T)$, 
then for $T > T_{cep}$  the situation is different  and the cross-over does not  exist. A detailed analysis of
this situation is given in the subsection \ref{sect:4.4.3}

Note also that all these nice properties would vanish, if  the reduced surface tension coefficient is  zero or positive above $T_{cep}$. This is  one of the crucial points of the present model which puts forward certain doubts about the vanishing of 
the reduced  surface tension coefficient  in the FDM 
\cite{Fisher:67} and SMM \cite{Bondorf:95}. These doubts are also supported by the first principle results obtained by the Hills and Dales Model  \cite{Bugaev:04b,BugaevElliott}, because the surface entropy simply  counts the degeneracy of a cluster of a fixed volume and  it  does not physically affect  the surface energy of this cluster.

\subsection{Generalization to Non-Zero Baryonic Densities.}
\label{sect:4.4.2}
The possibilities  ({\bf I})-({\bf III}) discussed in the preceding section 
remain unchanged for non-zero baryonic numbers. The latter should be 
included into consideration  to make this  model more realistic. To keep 
the presentation simple, I  do not account for  strangeness. 
The inclusion of the baryonic charge of the quark-gluon bags 
does not change the two types of singularities of the isobaric partition (\ref{Zs})
and the corresponding equation for them (\ref{s*vdw}), but it 
leads to 
the following modifications of the $F_H$ and $F_Q$ functions:

\vspace*{-0.35cm}
\begin{align}\label{FHTmu}
F_H&(s,T,\mu_B)= \sum_{j=1}^n g_j     e^{\frac{b_j \mu_B}{T} -v_js} \phi(T,m_j)\,,
\\
F_Q &(s,T,\mu_B) = {\textstyle u(T, {\mu_B})} 
   \int\limits_{V_0}^{\infty}dv~ \frac{ \exp\left[\left(s_Q(T,\mu_B)-s\right)v - \sigma(T) 
v^{\varkappa}\right] }{v^{\tau}}\,. 
\label{FQTmu}   
\end{align}
Here the baryonic chemical potential is denoted as $\mu_B$,  the baryonic charge of 
the $j$-th hadron in the discrete part of the spectrum is $b_j$. The continuous part 
of the spectrum, $F_Q$ can be obtained from  some spectrum $\rho(m,v, b)$
in the spirit of Ref. \cite{Goren:82,CGreiner:06}, but this will lead the discussion  away from the main  subject. 

The QGP pressure $p_Q = T s_Q(T,\mu_B)$ can be also chosen in several ways. 
Here I  use the bag model pressure 
\begin{align}\label{sQB}
&\hspace*{-0.2cm}p_Q = \frac{\pi^2}{90}T^4 \left[
 \frac{95}{2} +
\frac{10}{\pi^2} \left(\frac{\mu_B}{T}\right)^2 + \frac{5}{9\pi^4}
\left(\frac{\mu_B}{T}\right)^4 \right]
- B \,, 
\end{align}
but the more complicated model pressures, even with the
PT of other kind like the transition between the color superconducting QGP 
and the usual QGP, can be, in principle,  used. 

\vspace*{-0.0cm}

%
%
\begin{figure}[ht]
\centerline{\includegraphics[width=9.4cm,height=7.3cm]{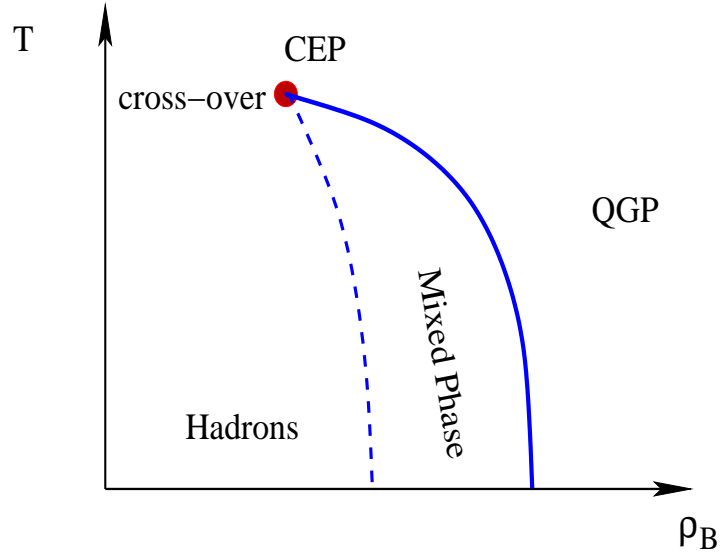}
}
\vspace*{-0.3cm}
\caption{
A schematic picture of the deconfinement phase transition diagram 
in the plane of baryonic density $\rho_B$  and  $T$ for 
the 2$^{nd}$ order PT at the critical endpoint (CEP), i.e. for $\frac{3}{2} < \tau \le 2$. 
For the 3$^{rd}$ (or higher) order PT  the  boundary of the mixed 
and hadronic phases (dashed curve) should have the same slope as 
the boundary of the mixed phase and QGP (solid curve) at the CEP. 
}
  \label{fig443}
\end{figure}

\vspace*{0.0cm}

The sufficient conditions for a PT  existence are  
\begin{align}\label{SufCondI}
&\hspace*{-0.25cm}
{\textstyle F((s_Q(T,\mu_B\hspace*{-0.05cm}=\hspace*{-0.05cm}0)\hspace*{-0.05cm}+\hspace*{-0.05cm}0),  T,\mu_B=0)  >  s_Q(T,\mu_B=0), } \\
&\hspace*{-0.25cm}
F ((s_Q(T,\mu_B )\hspace*{-0.05cm}+\hspace*{-0.05cm}0),  T,\mu_B)  < s_Q(T,\mu_B)\,,  \forall \mu_B > \mu_A.
\label{SufCondII}
\end{align}
The  condition (\ref{SufCondI})  provides that the simple pole singularity 
$s^* = s_H(T,\mu_B=0)$ is the rightmost 
one at vanishing $\mu_B=0$ and given $T$, whereas  the condition (\ref{SufCondII}) 
ensures that $s^* = s_Q(T,\mu_B)$ is the rightmost singularity of the isobaric partition for 
all values of the baryonic chemical potential above some positive  constant $\mu_A$. 
This can be seen in Fig.~\ref{fig441} for  $\mu_B$ being a variable.  
Since  $F (s, T,\mu_B)$, where it exists,  is a continuous function of its  parameters,
one concludes that, if the conditions (\ref{SufCondI}) and (\ref{SufCondII}), are fulfilled,
then at some chemical potential $\mu_B^c (T)$ the both singularities should  
be equal. Thus, one arrives at the Gibbs criterion (\ref{PTI}), but for two variables
\begin{align}\label{PTII}
&  s_H (T, \mu_B^c(T))  =  s_Q (T, \mu_B^c(T)) \,.
 \end{align}

\noindent 
It is easy to see that the  inequalities  (\ref{SufCondI}) and (\ref{SufCondII}) are  the  sufficient conditions  of a PT existence
for  more complicated functional dependencies of $F_H (s,  T,\mu_B)$ and 
$F_Q (s,  T,\mu_B)$ than the ones used here.

\vspace*{0.4cm}

%
%
\begin{figure}[ht]
\centerline{\includegraphics[width=9.4cm,height=7.3cm]{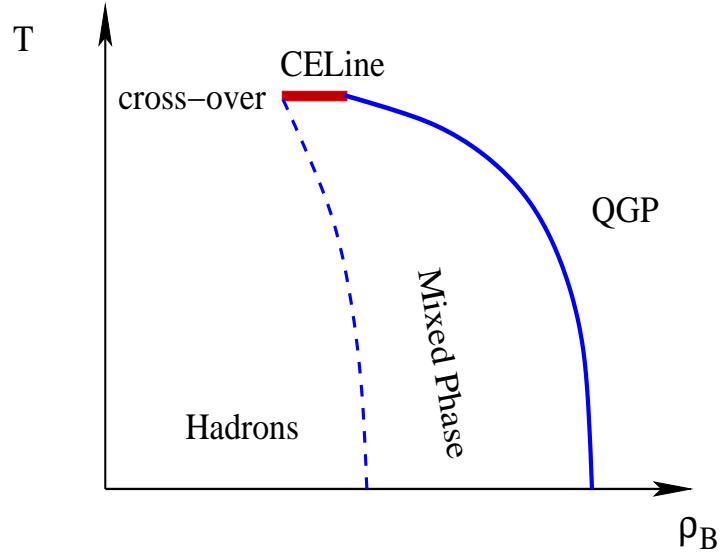}
}
\vspace*{-0.3cm}
\caption{
A schematic picture of the deconfinement phase transition diagram 
in the plane of baryonic density $\rho_B$  and  $T$ for $ \tau >  2$.
The critical endpoint in the $\mu_B-T$ plane generates 
the critical end line (CELine) in the $\rho_B-T$ plane shown by the thick horizontal line. 
This occurs because of the discontinuity of the partial derivatives of $s_H$ and $s_Q$ 
functions with respect to $\mu_B$ and $T$. 
}
  \label{fig444}
\end{figure}

For the  choice (\ref{FHTmu}), (\ref{FQTmu}) and (\ref{sQB})  of 
$F_H (s,  T,\mu_B)$ and $F_Q (s,  T,\mu_B)$ functions the  PT exists at 
$T < T_{cep}$,  because the sufficient conditions (\ref{SufCondI}) and (\ref{SufCondII}) 
can be  easily fulfilled   by a proper choice of the bag constant $B$ and the function 
${\textstyle u(T, \mu_B)} > 0$ for the interval $T \le T_{up}$ with the constant $T_{up} > T_{cep}$. 
Clearly, this is the  1$^{st}$ order PT, since the surface
tension is finite and it provides the convergence of the  integrals (\ref{G1}) and  (\ref{KQ})
in the expression (\ref{sHprime1}), where  the  usual $T$-derivatives should be now 
understood as the partial ones for $\mu_B = const$.

Assuming that the conditions (\ref{SufCondI}) and (\ref{SufCondII}) 
 are fulfilled by the correct choice of the  model parameters $B$ and  
 ${\textstyle u(T, \mu_B)} > 0$,   one  can see now that at $T = T_{cep}$ there exists 
a PT as well, but its order is defined by the value of $\tau$. As was discussed in the 
preceding section for  $\frac{3}{2} < \tau \le 2$ there  exists the 2$^{nd}$ order PT. 
For  $1 < \tau \le \frac{3}{2}$  there  exist  the  PT of higher order, defined 
by the conditions formulated in Eq.  (\ref{nth}). 
This is a new possibility, which, to  my  best knowledge,  does not contradict  to any 
general physical principle (see Fig.~\ref{fig443}). 

The case $ \tau > 2$ can be ruled out because  there must  exist
the first order PT for  $T \ge T_{cep}$, whereas for  $T < T_{cep}$ there exists 
the cross-over. Thus,  the critical endpoint in $T-\mu_B$ plane  will correspond to 
the critical  interval  in the   temperature-baryonic density plane.  
Since such a structure 
of the  phase diagram in the variables temperature-density  has, to my  knowledge,
never  been observed, I  conclude that  the case  $ \tau > 2$ is unrealistic (see Fig.~\ref{fig444}). 
Note that a similar phase diagram exists in the FDM with the only  difference  that 
the boundary of the mixed and liquid phases (the latter in the QGBST model corresponds 
to QGP) is moved to infinite particle density.

\subsection{Surface Tension Induced Phase Transition.}
\label{sect:4.4.3}
Using the results for the  case ({\bf III}) of the preceding subsection, I conclude that 
above $T_{cep}$ there is a cross-over, i.e. the QGP and hadrons coexist together 
up to the infinite values of $T$ and/or $\mu_B$. Now, however, it is necessary to  answer
the question: How can the two different sets of singularities  that exist on two sides of 
the line $T =  T_{cep}$ provide the continuity of the solution of Eq. (\ref{s*vdw})?

It is easy to  answer  this question for $\mu_B < \mu_B^c(T_{cep})$  because 
in this case all partial $T$ derivatives of  $s_H(T, \mu_B) $, which is the rightmost singularity, 
exist and are finite at any point of the line $T =  T_{cep}$.  This can be seen from the
fact that  for the considered  region of parameters  $s_H(T, \mu_B) $ is the rightmost singularity and, consequently, $s_H(T, \mu_B) > s_Q (T, \mu_B)$. The latter inequality 
provides the existence and finiteness of the volume integral in $F_Q(s,T,\mu_B)$.
In combination with  the power $T$ dependence of  the reduced surface tension 
coefficient $\sigma(T)$
the same inequality provides   the existence and finiteness of 
all its partial  $T$ derivatives of $F_Q(s,T,\mu_B)$ regardless  to the sign of 
$\sigma(T)$.  Thus, using the Taylor expansion in powers of $(T - T_{cep})$ 
at any point of the interval
$T =  T_{cep}$ and $\mu_B < \mu_B^c(T_{cep})$, one can calculate 
$s_H(T , \mu_B)$ for the values of  $T > T_{cep}$ which are  inside the convergency  radius of  the Taylor expansion.  

The other situation is for $\mu_B \ge \mu_B^c(T_{cep})$ and $T > T_{cep}$,
namely
in this case above 
the deconfinement PT there must exist a weaker PT 
induced by the disappearance of the reduced  surface tension coefficient. 
To demonstrate this I  have solved Eq. (\ref{s*vdw}) in the limit, when $T$ approaches 
the curve $T= T_{cep}$ from above, i.e. for  $T \rightarrow  T_{cep}+0$, and study the 
behavior of $T$ derivatives of the  solution of Eq. (\ref{s*vdw}) $s^*$ for fixed values of 
$\mu_B$.
For this purpose  I  have to evaluate the  integrals ${\cal K}_\tau (\Delta,\gamma^2)$
introduced in Eq. (\ref{KQ}). 
Here  the notations 
$\Delta \equiv s^* - s_Q(T, \mu_B)$ and $\gamma^2 \equiv - \sigma (T) > 0$ are  introduced for convenience. 

To avoid the unpleasant behavior for $\tau \le 2$ it is convenient to transform  (\ref{KQ}) 
further on by integrating by parts: 
\begin{align}\label{KQ2}
{\cal K}_\tau (\Delta,\gamma^2) \, \equiv & ~ g_\tau(V_0)  - \frac{\Delta}{(\tau-1)} {\cal K}_{\tau -1} (\Delta,\gamma^2) + 
\frac{\varkappa \, \gamma^2}{(\tau-1)} {\cal K}_{\tau-\varkappa} (\Delta,\gamma^2) \,,
\end{align}
where the regular function $g_\tau(V_0) $ is defined as
\begin{align}\label{gtau}
&  g_\tau(V_0) \equiv  \frac{1}{(\tau-1)\, V_0^{\tau-1}} \exp\left[ -\Delta V_0 + 
\gamma^2 V_0^{\varkappa}\right] \,. 
 \end{align}
For $\tau - a > 1$ one can change the variable of integration 
$v \rightarrow z / \Delta$ and  rewrite $ {\cal K}_{\tau-a} (\Delta,\gamma^2)$ as 
\begin{align}
\hspace*{-0.25cm}
 {\cal K}_{\tau- a} (\Delta,\gamma^2)  = \Delta^{\tau- a-1} \hspace*{-0.15cm} 
 \int\limits_{V_0 \Delta}^\infty  \hspace*{-0.15cm}
dz ~\frac{\exp\left[- z  + \frac{\gamma^2}{\Delta^\varkappa} z^{\varkappa} 
\right] }{z^{\tau- a}}  ~ \equiv ~
%
\Delta^{\tau-a-1} \, {\cal K}_{\tau-a} \left(1, \gamma^2\Delta^{-\varkappa} \right) \,.&
\label{KQ3}
 \end{align}
This result shows that  in the limit $\gamma \rightarrow 0$, when the rightmost 
singularity must  approach $s_Q(T,\mu_B)$ from above, i.e. $\Delta \rightarrow 0^+$, the function (\ref{KQ3}) behaves as  $ {\cal K}_{\tau- a} (\Delta,\gamma^2) \sim \Delta^{\tau-a-1} + O(\Delta^{\tau-a})$. This is so because for $\gamma \rightarrow 0$ 
the ratio $\gamma^2\Delta^{-\varkappa}$ cannot go to infinity, otherwise  the function 
${\cal K}_{\tau-1} \left(1, \gamma^2\Delta^{-\varkappa} \right) $,  which enters into the right hand side of  (\ref{KQ2}),   would diverge exponentially 
and this   makes impossible an existence of the solution of Eq. (\ref{s*vdw}) for 
$T = T_{cep}$. The analysis shows that for $\gamma \rightarrow 0$  there exist  two possibilities: either  $\nu \equiv \gamma^2\Delta^{-\varkappa} \rightarrow Const$ or 
$\nu \equiv \gamma^2\Delta^{-\varkappa} \rightarrow 0$.
The most straightforward  way to analyze these possibilities for  $\gamma \rightarrow 0$ is to assume the following behavior 
\begin{align}
&  \Delta  =  A\, \gamma^\alpha +  O(\gamma^{\alpha+1}) \,, 
\label{Dasgamma} \\
& \frac{ \partial \Delta}{\partial T}   =   \frac{ \partial \gamma}{\partial T} 
 \left[A\,\alpha\,  \gamma^{\alpha-1} +  O(\gamma^{\alpha})\right] 
 \sim \frac{(2\,k +1) A \,\alpha\,  \gamma^{\alpha}}{2\, (T - T_{cep}) },
\label{DTDasgamma} 
\end{align}
and find out the $\alpha$ value by equating (\ref{DTDasgamma}) with  the $T $ derivative (\ref{sHprime1}). 

Indeed, using  (\ref{sHprime1}), (\ref{G1}) and (\ref{KQ}), one can write
\begin{align}\label{Dsprime}
& \hspace*{-0.25cm}
\frac{ \partial \Delta}{\partial T}   = 
 \frac{ G_2  + 
 u\, {\cal K}_{\tau-\varkappa}(\Delta,\gamma^2) \,2 \,\gamma \gamma^\prime}{ 1 + u\, {\cal K}_{\tau-1}(\Delta,\gamma^2)} \approx 
 \frac{ \Delta^{2-\tau} G_2} {  u\, {\cal K}_{\tau-1}(1, \nu )}  
 + \nonumber \\
&  \frac{  
2\, \gamma \gamma^\prime
 \Delta^{1-\varkappa} \left[ \nu\, \varkappa \, {\cal K}_{\tau-2\varkappa}(1, \nu) - 
 {\cal K}_{\tau-1-\varkappa}(1, \nu) \right]}{ (\tau-1-\varkappa)\, {\cal K}_{\tau-1}  (1, \nu )  }\,,
 \end{align}
where the prime denotes the partial $T$ derivative. Note that  the function
$G_2 \equiv F^\prime+u^\prime {\cal K}_\tau(\Delta,\gamma^2)-s_Q^\prime $ can vanish for a few  values of $\mu_B$ only.
In the last step of deriving (\ref{Dsprime}) I used the  identities (\ref{KQ2}) and (\ref{KQ3})
and dropped the non-singular terms. As was discussed above, in the limit 
$\gamma \rightarrow 0$  the function $\nu$ either remains a constant or vanishes, then the term 
$ \nu\, \varkappa \, {\cal K}_{\tau-2\varkappa}(1, \nu) $ in (\ref{Dsprime})
is either of the same order 
as the  constant $ {\cal K}_{\tau-1-\varkappa}(1, \nu)$ or vanishes. 
Thus, to reveal  the behavior of  (\ref{Dsprime}) for $\gamma \rightarrow 0$  it is sufficient to find a leading term out of $\Delta^{2-\tau}$ and $\gamma \gamma^\prime  \Delta^{1-\varkappa} $ and compare it with the assumption (\ref{Dasgamma}).

The analysis shows that for  $\Delta^{2-\tau} \le  \gamma \gamma^\prime  \Delta^{1-\varkappa}$ the last term in the right hand side of (\ref{Dsprime}) is the leading one. 
Consequently, equating the powers of $\gamma$ of the leading terms in 
(\ref{DTDasgamma}) and (\ref{Dsprime}), one finds
\begin{align}\label{alpha1}
&  
%
\gamma^{\alpha -2} \sim \Delta^{1-\varkappa} \Rightarrow ~ \alpha \varkappa = 2 ~~{\rm for}~~ \tau \le 1 + \frac{\varkappa}{2 k + 1} \,, 
 \end{align}
where the last inequality follows from the fact that the term  
$\gamma \gamma^\prime  \Delta^{1-\varkappa}$  in (\ref{Dsprime}) is the dominant one.

Similarly, for  $\Delta^{2-\tau} \ge  \gamma \gamma^\prime  \Delta^{1-\varkappa}$ one 
obtains $\gamma^{\alpha -1} \gamma^\prime \sim \Delta^{2 -\tau}$ and, consequently,
\begin{align}\label{alpha2}
& 
\alpha  = \frac{2}{(\tau-1)(2k+1)} ~~{\rm for}~~ \tau \ge 1 + \frac{\varkappa}{2 k + 1} \,.
 \end{align}

Summarizing the above  results for $\gamma \rightarrow 0$ as
\begin{eqnarray}\label{DsdTtot}
\frac{ \partial \Delta}{\partial T}   \hspace*{-0.0cm}  \sim 
{\textstyle \frac{T_{cep}\gamma^{\alpha }}{T-T_{cep}} } =
\left\{
\begin{tabular}{ll}
\vspace{0.1cm}  ${\textstyle \left[ \frac{T-T_{cep}}{T_{cep}}\right]}^{\frac{2k+1}{\varkappa}-1} $\,, &  \hspace*{-0.2cm} $ \tau \le 1 + \frac{\varkappa}{2 k + 1}$\,, \\
& \\
${\textstyle \left[\frac{T-T_{cep}}{T_{cep}} \right]}^{\frac{2-\tau}{\tau-1}  } $\,, &  \hspace*{-0.2cm} $\tau \ge 1 + \frac{\varkappa}{2 k + 1} $\,,
\end{tabular}
\right.  \hspace*{-0.3cm}
\end{eqnarray}
one can also write the expression for the second derivative of $\Delta$  as 
\vspace*{-0.3cm}
\begin{eqnarray}\label{D2sdT2tot}
\frac{ \partial^2 \Delta}{\partial T^2}   \hspace*{-0.0cm}  \sim 
\left\{
\begin{tabular}{ll}
\vspace{0.1cm}  ${\textstyle \left[ \frac{T-T_{cep}}{T_{cep}} \right]}^{\frac{2k+1}{\varkappa}-2} $\,, &  \hspace*{-0.2cm} $ \tau \le 1 + \frac{\varkappa}{2 k + 1}$\,, \\
& \\
${\textstyle \left[ \frac{T-T_{cep}}{T_{cep}} \right]}^{\frac{3-2\tau}{\tau-1} } $\,, &  \hspace*{-0.2cm} $\tau \ge 1 + \frac{\varkappa}{2 k + 1} $\,.
\end{tabular}
\right.  \hspace*{-0.3cm}
\end{eqnarray}
The last result shows  that, depending on $\varkappa$ and $k$ values, 
the second derivatives of $s*$ and $s_Q(T,\mu_B)$ can differ from each other  for 
$ \frac{3}{2} < \tau < 2$ or can be equal for $ 1 < \tau \le \frac{3}{2}$.
In other words, one finds  that at  the line $T = T_{cep}$ there exists  the 2$^{nd}$ order PT for $ \frac{3}{2} < \tau < 2$ and the higher order PT   for $ 1 < \tau \le  \frac{3}{2}$,
which separates the pure QGP phase from the region  of  a cross-over, i.e.  the mixed states of hadronic and QGP bags. Since it exists at the line of a zero surface tension, 
this PT  will be called the {\it surface induced PT.} 
For instance,  from (\ref{D2sdT2tot}) it follows that for $k = 0$ and  $\varkappa > \frac{1}{2}$ there is  the 2$^{nd}$ order PT, whereas  for   $k = 0$ and  $\varkappa = \frac{1}{2}$ or for $k > 0$ and  
$\varkappa < 1 $ there is  the 3$^{d}$ order PT, and so on.

\begin{figure}[ht]
\centerline{\includegraphics[width=9.5cm,height=9.5cm]{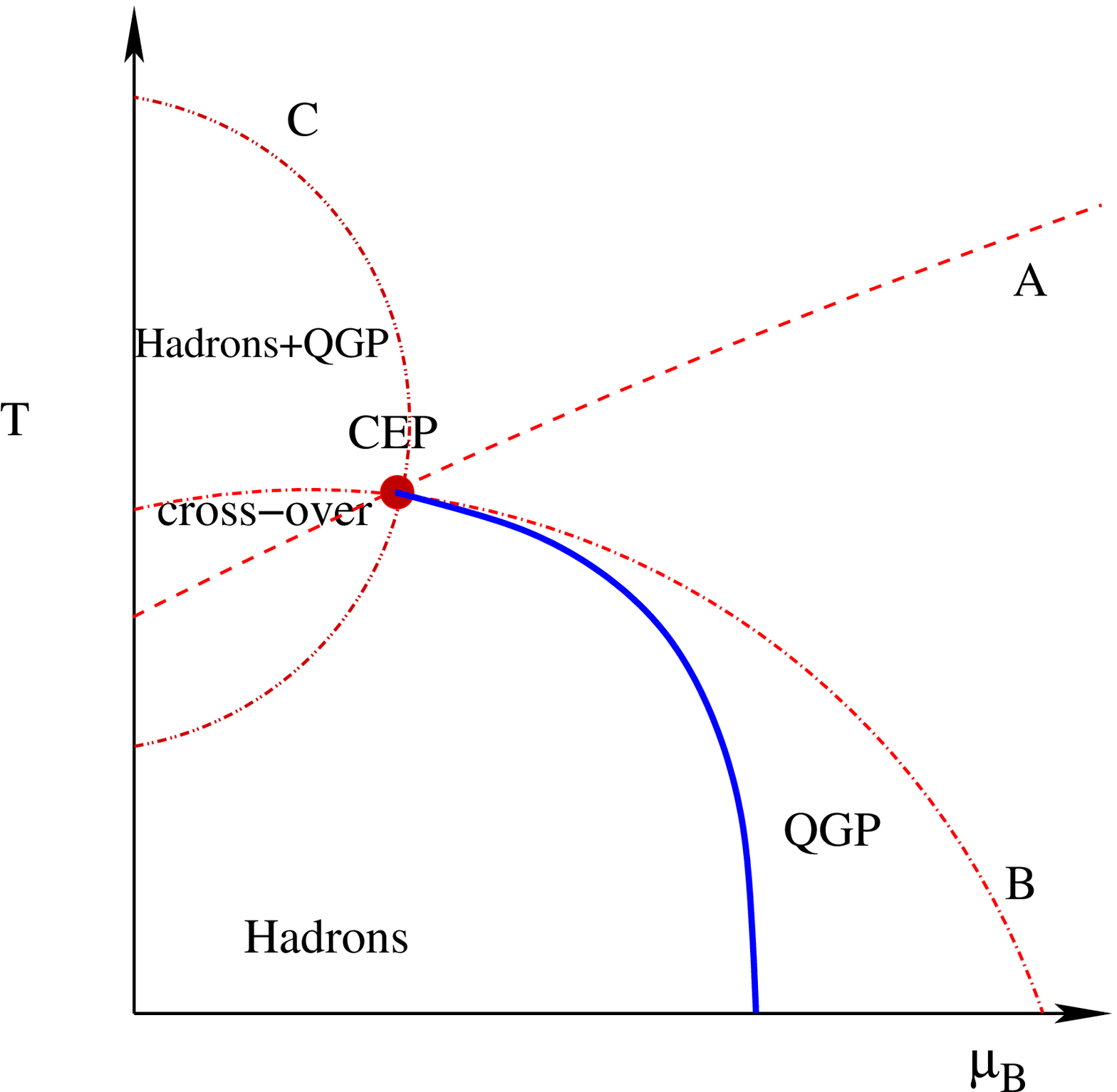}
}
\vspace*{-0.3cm}
\caption{
A schematic picture of the deconfinement phase transition diagram (full curve)
in the plane of  baryonic chemical potential $\mu_B$  and  $T$ for 
the 2$^{nd}$ order PT at the tricritical endpoint (CEP).
The model predicts an existence of the  surface induced PT of the 2$^{nd}$ or
higher order (depending on the model parameters). This PT starts at the CEP and goes 
to higher values of $T$ and/or $\mu_B$. Here it  is shown by the dashed curve CEP-A, if the phase diagram is endless, or
by the dashed-dot curve CEP-B, if the phase diagram ends at $T= 0$, or by 
the dashed-double-dot curve CEP-C,  if the phase diagram ends at $\mu_B= 0$.
Below (above) each of A or B curves the reduced surface tension coefficient 
is positive (negative).  For  the curve C the surface tension coefficient is positive 
outside of it. 
}
  \label{fig445}
\end{figure}

Since the analysis performed in  the present section did not include any $\mu_B$ derivatives 
of $\Delta$, it remains valid for   the $\mu_B$ dependence of the 
reduced surface tension coefficient, i.e. for $T_{cep} (\mu_B)$.
Only it is  necessary to make a few comments on a possible location of  
the {\it surface tension null line}  $T_{cep} (\mu_B)$.  
In principle, such a null line can be located anywhere, if its location does not contradict 
to the sufficient conditions (\ref{SufCondI}) and (\ref{SufCondII}) of   the  1$^{st}$ deconfinement PT existence.  Thus, the surface tension null line must cross the 
deconfinement line in the $\mu_B-T$ plane at a single point which is  the tricritical endpoint
$(\mu_B^{cep}; T_{cep} (\mu_B^{cep})) $, whereas for  $\mu_B > \mu_B^{cep} $ the 
null line should have higher temperature for the same $\mu_B$   than the deconfinement one, i.e. 
$T_{cep} (\mu_B) > T_c  (\mu_B) $ (see Fig~\ref{fig445}).  Clearly, there exist  two distinct cases
for the surface tension null line: either it is endless, or it ends at 
zero temperature. But recalling  that at low temperatures and high values of the baryonic chemical potential there may exist the Color-Flavor-Locked   phase 
\cite{Krishna:98},
it is possible that the null line may also cross the boundary of the  Color-Flavor-Locked   phase and, perhaps, it may create another special point at this intersection. 
From the present lattice quantum chromodynamics data the case C in Fig.~\ref{fig445} is the least possible. 

One may wonder why this surface induced PT was not observed so far. 
The main reason is that the lattice quantum chromodynamics calculations at non-zero $\mu_B$ are very 
difficult, and  because of this   the identification  of the precise  location of the critical 
endpoint  is  highly nontrivial task \cite{fodorkatz,karsch,Misha}. Therefore, 
the identification of the 2$^{nd}$ or higher order PT which might be located in 
the vicinity of  the deconfinement PT could be a real challenge. 
In addition, for all  $\mu_B > \mu_B^{cep} $ the surface induced PT  may lie so  close to the deconfinement PT  line that 
it would be extremely difficult to observe it at the present lattices.

\vspace*{-0.0cm}

%
%

\vspace*{-0.0cm}

To understand the meaning of the  surface induced PT it is instructive to
 quantify the difference between phases by looking into the mean size of the bag:
\begin{align}\label{BagSize}
& 
\langle v \rangle \equiv - \frac{\partial \ln F(s, T, \mu_B)}{\partial~ ~s} \biggl|_{s = s^*-0} \,.
 \end{align}
As was shown 
in hadronic phase  $\Delta > 0$ and, hence, it consists of the bags of finite mean volumes, whereas, 
by construction, the QGP phase is a single infinite bag. For the cross-over states 
$\Delta > 0$ and, therefore, they are the bags of finite mean volumes, which 
gradually increase, if  the rightmost singularity approaches $s_Q(T,\mu_B)$, i.e.
at very large values $T$ and/or $\mu_B$. Such a classification is useful 
to distinguish  quantum chromodynamics phases of present model: it shows that hadronic and cross-over 
states are separated from the QGP phase by the 1$^{st}$ order deconfinement PT and 
by the 2$^{nd}$ or higher order PT, respectively.


\section{Conclusions and Perspectives}
\label{sect:4.4.4}

In this chapter I discussed three related models - the SBM (or rather its microcanonical version, the HTM), the Mott-Hagedorn resonance gas model and the QGBST model.   The latter two  originate 
from  the Hagedorn model. All these models are exactly solved. Whereas the Mott-Hagedorn gas model  was just constructed in a way that  the expressions for 
thermodynamic functions are analytical, the QGBST  required  an involved analysis of its PTs. 
The most unusual surprise was given by the microcanonical  analysis of the Hagedorn model. 
Surprisingly, after so many years of studies it was possible to rigorously prove 
\cite{HThermostat:1, HThermostat:2, Thermostat:3} by the finite volume treatment  that this model has, indeed, has a PT. Thus, the old  guesses  based on a formal analogy \cite{Carlitz:72} and  on a deep understanding of  physics \cite{Parisi:75}  turned out to be right.  Of course,  at first glance this result
seems  simply  curious, but  at the second glance one realizes that  the underlying  physics of this model  is highly  non-trivial and this is the reason of  why it took so many years to clarify the thermostatic properties of heavy Hagedorn resonances, although the nonequivalence of microcanonical and (grand)canonical ensembles for the exponential mass spectrum of resonances is known for a while \cite{Carlitz:72}. 

Also it is possible that further development of  statistical mechanics of multihadron production in 
elementary particle collisions on the basis of  statistical hadronization model  \cite{becattini:97, Becattini:1, Becattini:Can, Becattini:2}  or comparable models  \cite{Greiner:04}  can  lead to further refinement of the HTM  after  which  the model can be used for  quantitative analysis of the data. 

One generalization of the SBM, the Mott-Hagedorn gas model,  which employes  the properties of  the Mott  transition to generate 
the large  width  of heavy resonances  \cite{Blaschke:03, Blaschke:04, Blaschke:05} was presented above. This model  removes the singularity of the SBM in the GCE by a similar mass dependence  of a width of heavy resonances. Such a mathematical trick might look somewhat artificial, but this model 
naturally explains the reason why the heavy resonances which are expected to be produced in A+A and 
elementary particle collisions are not observed experimentally.  Also, using a few parameters, this model 
reproduces the energy density found by the lattice quantum chromodynamic simulations  \cite{Tawfik03}
and is able to describe the anomalous suppression  of $J/\psi$ mesons which is observed experimentally at SPS energies \cite{na50}.  Therefore, I conclude that  the medium dependent  resonance width 
should be incorporated in a more elaborate statistical description of the cross-over states and 
the role of the Mott effect should elucidated   by both  the lattice quantum chromodynamic simulations and experimental studies.

Then  I discussed  an analytically solvable  statistical model, 
the QGBST model \cite{Bugaev:07,Bugaev:07new, Bugaev:08new},   which simultaneously  describes 
the 1$^{st}$ and 2$^{nd}$ order PTs with a cross-over. The suggested  approach is general 
and can be used 
for  more complicated parameterizations of  the hadronic mass-volume spectrum, if in the vicinity of
the deconfinement PT region the discrete and continuous parts  of this spectrum  can be expressed in the form of Eqs. (\ref{FHTmu}) and  (\ref{FQTmu}), respectively. Also the actual parameterization of 
the QGP pressure $p = T s_Q(T,\mu_B)$ was not used so far, which means that these  result can be extended to more complicated functions, that can contain other phase transformations (chiral PT,
or the PT to color superconducting phase)   provided that the sufficient  conditions (\ref{SufCondI}) and (\ref{SufCondII})  for the deconfinement PT existence  are satisfied. 

In this model the desired properties of the deconfinement phase diagram are achieved by accounting for the temperature dependent surface tension of the quark-gluon bags. 
As I showed, it is crucial for the cross-over existence that at  $T= T_{cep}$ 
the reduced surface tension 
coefficient vanishes and remains negative for temperatures above  $T_{cep}$.
Then the deconfinement $\mu_B-T$  phase diagram has the 1$^{st}$ PT  at  
$\mu_B > \mu^c_B( T_{cep})$ for 
$ \frac{3}{2} < \tau \le  2$ , which degenerates into the 2$^{nd}$ order PT 
(or higher order PT for  $ \frac{3}{2} \ge \tau >1 $)  at 
$\mu_B = \mu^c_B( T_{cep})$, and a cross-over for $0 \le \mu_B < \mu^c_B( T_{cep})$.
These two ingredients  drastically change the critical properties of the GBM 
\cite{Goren:81} and resolve the long standing problem of  a unified description of  the 
1$^{st}$ and 2$^{nd}$ order PTs and  a cross-over,
which, despite all claims,  was not resolved in Ref. \cite{Goren:05}. 
In addition, I  found that at the null line of the surface tension there must exist 
the surface induced  PT of the 2$^{nd}$ or higher order, which separates 
the pure QGP from the mixed states of hadrons and QGP bags, that coexist above
the cross-over region (see Fig.~\ref{fig445}).  Thus, the QGBST model predicts that
for $\tau \le 2$
 the quantum chromodynamics critical endpoint is the tricritical endpoint. It would be interesting  to verify this prediction
with the help of the lattice quantum chromodynamics analysis. 
For this it will be necessary to study the behavior of the bulk and surface contributions to 
the free energy of the  QGP bags  and/or  the string connecting 
the static quark-antiquark pair.

In contrast to  popular mean-field models the PT mechanism in  the QGBST model is clear: it happens due to  the competition of the rightmost singularities of the isobaric partition function. Since the  GCP function of the QGBST model does not depend on any  (baryonic or entropy or energy) density, but depends exclusively  on $T, \mu_B$ and $V$, its phase diagram  does not 
contain any back bending and/or spinodal instabilities  \cite{Randrup:04} which are typical for the mean-field (= classical) models.
The found exact analytical solution does not require a complicated and  artificial 
procedure of conjugating  the two parts of the equation of state  in the vicinity of the critical endpoint
like it is done by hands in Refs. \cite{Nonaka:05,Antoniou:05} because  all this is automatically included in the statistical description. 

Also in the QGBST model the pressure of the deconfined phase is generated by the infinite  bag,
whereas the discrete part of the mass-volume spectrum plays  an auxiliary role even above the cross-over region. Therefore,  there is no reason to believe that any quantitative changes of the properties of low lying hadronic states  generated by the surrounding  media (like the mass shift of the $\omega$ and $\rho$ mesons \cite{Shuryak:05}) would be the robust  signals of the deconfinement PT. 
On the other hand,  the QGP bags created in the experiments  have finite mass and volume and, hence,  the strong discontinuities  which are typical for the 1$^{st}$ order  PT should be smeared out 
which would  make them hardly distinguishable from the cross-over. Thus, to seriously discuss 
 the signals of the 1$^{st}$  order deconfinement PT and/or  the tricritical endpoint,  one needs to 
solve the  finite volume version of the QGBST model like  it was done for the CSMM \cite{Bugaev:04a} and the GBM \cite{Bugaev:07b} in the chapter 2.  This, however, is not sufficient because, in order to make any reliable prediction for experiments, the finite volume EOS  must be used  in   hydrodynamic equations which, unfortunately, are not suited for such a  purpose. 
As was discussed at the end of the chapter 2 the short lived metastable modes with decay/formation 
time $\tau_n$ being shorter than the expansion time  cannot  be described by hydrodynamics. 
Also at the moment it is  unclear how to build up the corresponding kinetic description. 
Thus, the studies of the strongly interacting matter EOS in finite systems  face us with 
a necessity to return to the foundations of heavy ion phenomenology and to modify them according to the requirements of the experiments.  

The QGBST model indicates  how difficult it will be to locate the tricritical point of QGP. 
In principal it can be found, if the mass spectrum of heavy resonances  becomes  a power like. 
This occurs only, if the system is in a mixed phase and if the surface tension coefficient of heavy bags 
vanishes. Unfortunately, there are two complications: the first one is that for finite systems 
the 1$^{st}$ order PT  behaves like a cross-over, and, the second complication is that  the heavy 
QGP bags cannot be detected directly, but through the resonance decays only. Therefore,  it is possible 
that, although the mass distribution of  heavy  QGP bags  is reconstructed,  due to the  little difference 
between the 1$^{st}$ order PT and a cross-over in finite systems
the power law  will exist, perhaps,  not  only at the tricritical endpoint, but at some line of  
thermodynamic parameters.

Although the QGBST model has a great advantage compared to other models because, in principle,  
it can be formulated on the basis of the experimental data on the degeneracies, masses and eigen volumes  of hadronic resonances in the spirit of Ref. \cite{Goren:82}, 
it requires further improvements to make it  suitable for the quantitative estimates. 
Thus, above the surface tension null line 
the hadrons can coexist with QGP  at high temperatures. Consequently, the nonrelativistic 
consideration of hard core repulsion in the present model should be modified to
its  relativistic treatment  for light hadrons like it was discussed in the chapter 3. 
Also,  the realistic EOS requires the inclusion of the temperature and mass  dependent 
width of heavy resonances into a  continuous part of the mass-volume spectrum 
which  may essentially modify our understanding 
of the cross-over mechanism  in a spirit of the Mott-Hagedorn resonance gas model considered in this 
chapter. 

Finally, a precise  temperature dependence of the  surface tension  coefficient 
along with the role of the curvature part of free energy 
of the bags should be investigated and their  relation to the interquark string tension should be studied in detail. 
For this  purpose I would like   to modify  the Hills and Dales Model (see chapter 2) 
in order to include the surface deformations with the base of  arbitrary size  whereas its present formulation is suited for  discrete clusters and, hence, for discrete bases of surface deformations.



\renewcommand{\l}{\left} 
\renewcommand{\r}{\right}

\renewcommand{\N}{{\cal N}}

\renewcommand{\dd}{{\rm d}}
\renewcommand{\pd}{\partial}

\def\vdw{{\it VdW}\,}
\def\vdwful{{\it  Van-der-Waals}\,}
\def\cs{{\it CS}\,}
\def\csful{{\it contracted spheres}\,}
\def\req#1{(\ref{#1})}
\def\vp{v_{\rm o}}

\def\bra{\langle}
\def\ket{\rangle}

%
%
\let\a=\alpha \let\b=\beta \let\g=\gamma \let\d=\delta
\let\e=\varepsilon \let\z=\zeta \let\h=\eta \let\th=\theta
\let\dh=\vartheta \let\k=\kappa \let\l=\lambda \let\m=\mu
\let\n=\nu \let\x=\xi \let\p=\pi \let\r=\rho \let\s=\sigma
\let\t=\tau \let\o=\omega \let\c=\chi \let\ps=\psi
\let\ph=\varphi \let\Ph=\phi \let\PH=\Phi \let\Ps=\Psi
\let\O=\Omega \let\S=\Sigma \let\P=\Pi \let\Th=\Theta
\let\L=\Lambda \let\G=\Gamma \let\D=\Delta

%
%
%
\def\noi{{\noindent}}

\def\nn{\nonumber \\}
\def\bc{\begin{center}}
\def\ec{\end{center}}
\def\1#1{{\bf #1}}
\def\Rsm{$R_{\odot}$}
\def\Rs{R_{\odot}}
\def\lp{\left(}
\def\rp{\right)}
\def\pe{\frac{p}{E}}
\def\ep{\frac{E}{p}}
\def\pp{\sqrt{E^2 - m^2}}

\chapter{Freeze-out  Problem of Relativistic Hydrodynamics and Hydrokinetics}

\def\CF{{CF}\,\,}   
\def\CFful{{Cooper-Frye}\,\,}  
\def\CO{{\it CO}\,\,}   
\def\COful{{ cut-off}\,\,}   
\def\SF{\Sigma_{fr}}
\def\SI{\S_{in}}

\def\xfr{x^{1*}}

\def\gfp{the gas of free particles\,}

\def\beqs{\begin{eqnarray}}
\def\eeqs{\end{eqnarray} \vspace*{-0.5cm}}
\def\nn{\nonumber \\}

%
\def\vfrfg{v_{f\1 G}}
\def\vgrfg{v_{s\1 G}}
\def\vgrff{v_{s\1 F}}
\def\vvcm{{\bf v}_{f\, CM}}
\def\vicm{v^i_{f\, CM}}
\def\vxcm{v^1_{f\, CM}}
\def\vycm{v^2_{f\, CM}}
\def\vzcm{v^3_{f\, CM}}

\def\egtil{\widetilde\varepsilon_g}
\def\pgtil{\widetilde p_g}
\def\ngtil{\widetilde n_{c.g}}

\def\ptt{\tilde{p_t}}
\def\tin{T_{in}}

\def\nmuupa{n^\m (f_a)}
\def\nmudowna{n_\m (f_a)}
\def\nmuupl{n^\m (f_l)}
\def\nmudownl{n_\m (f_l)}
\def\nmuupr{n^\m (f_r)}
\def\nmudownr{n_\m (f_r)}

\def\vm{\vspace*{-0.4cm}}

From its birth relativistic hydrodynamics \cite{LANDAU:53}  became the  most powerful theoretical tool to study
the dynamics of PTs in A+A collisions at 
high energies. 
During last 20 years  it was successfully used 
to model  the deconfinement  PT
between  QGP  and hadronic matter \cite{StockerGreiner, Mishustin89,Rischke1,QM:04}.
So far,  only within  the hydro approach it was possible to predict the three major 
signals of the deconfinement transition seen at SPS energies, i.e. the Kink \cite{Kink},
the Strangeness Horn \cite{Horn} and the Step \cite{Step}. 
Nevertheless, the hydro  modeling of relativistic heavy ion collisions is difficult 
and even not  straightforward. 

Relativistic hydrodynamics is a set of
the partial differential equations which describe  the local 
energy-momentum and charge conservation \cite{llhydro}
\begin{eqnarray}
\label{one51}
\partial_\mu T^{\mu\nu}_{f} ( x,t) & = & 0\,\,,  \quad \quad T^{\mu\nu}_{f} ( x,t) =   \lp \epsilon_f + p_f \rp 
u_f^\mu u_f^\nu - p_f g^{\mu\nu}\,\,, \\
\label{two51}
\partial_\mu N^{\mu}_{f} ( x,t) & = & 0\,\,,  \quad \quad N^{\nu}_{f} ( x,t)  =  n_{f} u_f^\nu \,\,.
\end{eqnarray}
Here the components of the energy-momentum tensor $T^{\mu\nu}_{f}$ of the perfect fluid
and its (baryonic) charge 4-current $ N^{\mu}_{f} $ are given in terms of energy density $\epsilon_f$,
pressure $p_f$, charge density $n_f$ and 4-velocity of the fluid $u_f^\nu $.
This is a simple indication  that hydrodynamic description directly probes
the equation of state of the matter under investigation.

As usual to complete the system (\ref{one51}) and (\ref{two51}) it is necessary to provide

{\bf (A)}  {\it the initial conditions } on some hypersurface and 

{\bf (B)} {\it equation of state} (EOS).

The tremendous complexity  of   {\bf (A)} and {\bf (B)}  transformed each of them into a specialized direction of  research of  relativistic  heavy ion community.
However, there are several specific features of relativistic hydrodynamics which have to be mentioned.
In contrast to  nonrelativistic hydrodynamics which is an exact science,
the relativistic one  applied to collisions of hadrons or/and heavy nuclei faces  a few problems from the very beginning.  
Since collisions occur in vacuum there are no specific boundary conditions.
Moreover, 
since the system created during the collision process  is small and short lived there were always 
questions whether hydro description is good and accurate  and whether the system  thermalizes 
sufficiently fast in order that hydro description can be used \cite{Rosenthydro}. 

It is clear that the last two questions cannot be answered within the  framework  of hydrodynamics.
One has to study these problems in a wider frame, and there was some progress achieved
on this way \cite{QM:04}.  However, to be applicable and realistic, relativistic hydrodynamics requires
the knowledge of 

{\bf (C)} {\it boundary conditions} which must be consistent with conservation laws (\ref{one51}) and 
(\ref{two51}). \hfill\\
The latter  is known as the {\it freeze-out problem} (FO), and 
it has   two basic aspects \cite{LANDAU:53}: 

{\bf (C1)} hydrodynamic equations should be terminated at some freeze-out hypersurface (FO HS)
$\SF( x,t)$;

{\bf (C2)} at the FO HS $\SF( x,t)$ all interacting  particles should be converted into the free-streaming
particles which go to detector.

The complications come from the fact that the FO HS cannot be found a priory without 
solving hydrodynamic equations (\ref{one51}) and (\ref{two51}).  This is a consequence  of relativistic causality on the time-like (t.l.)   parts of the FO HS  defined by  positive 4-interval  $d s^2 = d t^2 - d x^2  > 0 $ in Landau-Lifshitz convention  \cite{llhydro}. 
Therefore, the {\it freeze-out criterion} is usually formulated
as an additional equation (constraint) $F(x^*, t) = 0$ with the solution  $\{ x^* \} \,\, \Leftrightarrow \,\, 
x^{1*} = x^1(x^2,x^3, t)\,,$ 
 which has to be 
inserted   into conservation laws and solved simultaneously with them. In what follows I will mostly use the 1+1 dimensional formulation for simplicity and convenience, but the main results will be formulated in 3+1 dimensional space-time. 

There were many unsuccessful  tries  \cite{Mi58, Mi61, CF74, Si:89, Si:89b, namiki}  
to solve this problem by imposing 
the form of  the FO HS a priory, but all of them led to severe difficulties - 
either to negative number of particles or break up of  conservation laws.
The major difficulty  is that the hydro equations should be terminated in such a way, 
that their solution remains unmodified by this very fact. This is known as the recoil problem. 
Clearly  this problem 
cannot be postponed to later times, as it is usually done in hydro simulations 
\cite{Mishustin89, Rischke1, namiki, Bernard:95, Heinz:05},  
because at the boundary with  vacuum the particles
start to evaporate from the very beginning of hydro expansion, and this fact should be 
accounted by equations as well. 
Thus,  one has to account for the emission of particles from the FO HS and modify 
the hydro equations accordingly. 

However  this is only a part of the trouble. Another  one
comes from the problem of calculating the  particle spectra on a t.l. 
hypersurface. For a space-like (s.l.)  FO HS (i.e. with negative 4-interval $ s^2 < 0 $)
the correct answer for the spectra of particles is given by the
formula of Cooper and Frye \cite{CF74}. However, one cannot use this formula
for t.l.  hypersurfaces, since it leads to negative
numbers of particles. This is due to the fact that it was obtained
only for the s.l. case, where the decay of one element of gas
does not affect the decay of adjacent elements.

The freeze-out  problem was completely  solved in \cite{Bugaev:96}  and developed further in 
Refs. \cite{Bugaev:99, Bugaev:99a, Bugaev:99b,Bugaev:09}.  
Although Sinykov's  effort   \cite{Si:89, Si:89b}  did not resolve this problem, was  a very descent  try to solve it  by purely hydrodynamic approach and, hence, it was very useful to me.
The  solution of the FO problem was  found after a realization of 
a  fact that at  the t.l.  parts of FO HS
there is a {\it fundamental difference between the particles of fluid and the particles
emitted  from its surface:} the  EOS of the fluid can be anything, but it implies a zero
value for the mean free path, whereas the emitted particles cannot  interact at all because
they have an infinite mean free path. 
Therefore, it was necessary to extend the  conservation laws (\ref{one51}) and 
(\ref{two51}) from a fluid alone to a  system 
consisting of a  fluid and  the particles of gas emitted  (gas of free particle) from the FO HS. 
As I will prove below  on the t.l. parts of the FO HS 
these solution inevitably  includes a new type of shock wave, 
which I called the FO shock \cite{Bugaev:96}.
As I showed,  such a  FO shock not only resolves the recoil problem, but also it  provides 
an absence of the causal paradoxes at the t.l. parts of FO HS.

After the solution of the freeze-out problem  was found in Ref. \cite{Bugaev:96}, 
there were two basic directions to 
develop it further on the basis of the transport simulations. 
Thus, there were two hydro groups which wanted to improve the solution \cite{Bugaev:96}. 
The Bergen group wanted to justify and improve my  approach using
kinetic ``equations'' \cite{FO2, FO3, FO4, FO5, LASZLO}. 
All these tries seems to be  unsuccessful because none of those ``equations'' was ever derived.
The Bergen group  used an intuition to postulate their equations and 
the hand waving arguments to justify them. 
These tries, of course,  cannot be even closely  compared with the rigorous analysis 
which was worked out in 
Ref. \cite{Bugaev:96, Bugaev:99, Bugaev:99a,Bugaev:99b,Bugaev:09}.
The San Paolo group \cite{FO1, Grassi:04} 
tried to establish the continuous emission of particles in space and time. 
But again the suggested  equations \cite{FO1, Grassi:04}  are used without any physical justification. 
Moreover, until the  work  of  Sinyukov and collaborators \cite{SIN:02},   those equations for the escape probability had elementary mathematical mistakes.

Furthermore, 
from the very beginning it was also clear that the relaxation type equations considered
as an improvement  of the formalism developed in my works 
\cite{Bugaev:96,Bugaev:99, Bugaev:99a,Bugaev:99b,Bugaev:09}   cannot be applied to the  actual
hydro simulations  because they do not conserve energy, momentum and baryonic charge. 
In this sense, the approach of the Bergen group was  a  step back because even 
a  naive use of the Cooper-Frye formula respects  all conservation laws. 

The San Paolo group went ahead with the  idea of the volume emission (see the 
full list of references in \cite{Grassi:04}).
However, I think  this group  failed 
to understand three major points:

{\bf (I)} if the mean free path of the fluid particles is small compared to the characteristic size 
of the system (usually it is transversal size), then the volume emission of the particles is reduced
to the surface emission and the previous analysis becomes a perfect approximation; 

{\bf (II)} if, on contrary,
the mean free path of the fluid particles is compared to the size of the system, then there is
no reason to believe  in thermal equilibrium at all and one has to abandon the whole hydro
treatment pursuing by the San Paolo group;

{\bf (III)}  a real improvement of the hydrodynamic solution 
\cite{Bugaev:96,Bugaev:99, Bugaev:99a,Bugaev:99b,Bugaev:09}  of the freeze-out problem requires a far more fundamental approach which should be based on  a new
formalism.

A more solid attempt  to study the validity of the solution \cite{Bugaev:96} was presented
in Ref. \cite{MOLNAR:00}. However, the conclusions on the applicability of the {\it freeze-out shock}
model were too general. In contrast, the results of comparable simulations \cite{BRAVINA:99} 
indicate that the freeze-out shock approximation  is valid within 15 \% at the t.l.  parts of the FO HS 
because about 85 \% of all particles are born within a narrow region of the width about 0.5 - 1.2 fm,
whereas the main problem with the narrow freeze-out hypersurface exists at the s.l. parts of the FO HS. 
The latter is a typical region where the  usual hydrodynamics and the Cooper-Frye formula \cite{CF74} were  traditionally believed to be valid.  
These conclusions of Ref.  \cite{BRAVINA:99} were the first evidence that the hadronic rescattering and 
decay of heavy hadronic resonances should be treated with the transport approach.

Again there appeared two major directions of research: a simplified  theoretical approach for hadronic rescattering
was suggested in \cite{SIN:02}, whereas Bass and Dumitru suggested a principally new numerical approach to this problem \cite{BD:00} - 
the {\it ``hydro-cascade''} model (or BD model) which was 
further developed in \cite{TLS:01} (TLS model).
This  approach is a combination of hydro simulations with kinetics (hydrokinetisc)  and it   assumes that  the nucleus-nucleus collisions
proceed in three stages: hydrodynamic
expansion (``hydro'') of the QGP, phase transition from the QGP to
the hadron gas  and the stage of hadronic
rescattering and resonance decays (``cascade''). The switch from hydro to
cascade modeling takes place
at the boundary between the mixed  and hadronic phases.
The spectrum of hadrons
leaving this hypersurface of the  transition between  QGP and hadron gas is taken as an  input for the
cascade.

This  approach  incorporates  the best features of  both the hydrodynamic and cascade
descriptions.  It  allows for, on one hand, the calculation of  the phase transition
between  the quark gluon plasma  and  hadron gas using hydrodynamics
and, on the other hand,  the freeze-out of hadron spectra
using  the cascade description.
This  approach
allows  one to  overcome  the  usual
difficulty of  transport models  in modeling    phase transition phenomenon.
For this reason, this approach has been rather successful in explaining
a variety of collective phenomena that has been observed
at the CERN SPS   and
Brookhaven RHIC  energies.
However, both the BD and TLS models face some
fundamental  difficulties
which cannot be ignored {(see a detailed discussion in \cite{Bugaev:02HC, Bugaev:04HC}).}
Thus, within the BD approach
the initial distribution for the cascade is found using  the \CFful formula \cite{CF74},
which takes into account particles with all possible velocities,
whereas in the TLS model the initial cascade distribution is given by the \COful formula
\cite{Bugaev:96,Bugaev:99a},
which accounts for only those particles that can leave the phase boundary.
As shown in Ref. \cite{Bugaev:02HC}  the \CFful formula leads to  causal and
mathematical problems in the present version of the  BD model because the
 phase boundary between QGP and hadron gas  inevitably has t.l.  parts.
On the other hand, the TLS model  does not conserve
energy, momentum and number of charges and this, as proved in \cite{Bugaev:02HC, Bugaev:04HC},
is due to the fact that the equations of motion used in \cite{TLS:01}
are incomplete and, hence, should be modified.

These difficulties are likely in part responsible for the fact
that the existing  hydro-cascade  models, like  the more simplified ones,
fail to  explain  the {\it HBT puzzle} \cite{QM:04}, i.e.
the fact that
the experimental HBT radii at RHIC   are  very similar to those
found at SPS, even though  the center of mass energy is larger  by an order of
magnitude. Therefore, it turns out that  the  hydro-cascade approach
successfully {\it parameterizes}  the one-particle momentum spectra and their
moments, but does not {\it  describe} the space-time picture of the
nuclear collision  as probed by  two-particle interferometry. 

After I realized that the main  difficulty of the hydro-cascade  approach looks  similar to
the traditional problem of
freeze-out  in relativistic hydrodynamics \cite{Bugaev:96, Bugaev:99, Bugaev:99a,Bugaev:99b,Bugaev:09},
I started to work out the correct  set of hydro-cascade equations.
It turned out that in both cases the  domains (subsystems) have t.l.  boundaries
through which the exchange of particles  occurs  and this fact should be taken into account.
In relativistic hydrodynamics this problem was solved by the
constraints which appear  on the FO HS 
hypersurface and provide  the global energy-momentum and
charge conservation \cite{Bugaev:96, Bugaev:99, Bugaev:99a,Bugaev:99b,Bugaev:09} between 
the perfect fluid  domain and the domain of the gas of free particles.
Similarly, it was necessary to generalize  the usual
Boltzmann equation in order to  account for  the exchange of particles
through  the t.l.   boundary between the domains. 
As derived in  \cite{Bugaev:02HC}
the generalized Boltzmann equations  necessarily  contain  the $\delta$-like  source terms.
By integrating the Boltzmann equation  of the  fluid domain over particle momenta, I derived 
the desired set of hydro-cascade equations \cite{Bugaev:02HC} and 
analyzed them in \cite{Bugaev:04HC}.  The  approach developed in \cite{Bugaev:02HC, Bugaev:04HC},
is a new tool to model the PT dynamics  in finite systems. Already the present formulation allows one, in principal, to consider the  phase evolution of many domains of one phase being 
surrounded by the other phase. For instance,  it can be used  to model a  hadronization of  any number of  the QGP  droplets or bags which cannot be described the usual hydro equations.

In this chapter I present the FO  model  which is based on the conservation laws of
energy and momentum between the perfect  fluid and the gas of free particles
and give a derivation of the hydrokinetic equations from the first principles  and 
explain their analysis in details. 

This chapter is based on the following works  
\cite{Bugaev:96, Bugaev:99, Bugaev:99a,Bugaev:99b,Bugaev:09,Bugaev:02HC, Bugaev:04HC}.

\section{Freeze-out  Problem in Relativistic Hydrodynamics}

Before  reformulating the hydrodynamic equations I will consider the problem of negative numbers
of particles which was a nightmare of  hydro simulations since the \CFful  formula's invention \cite{CF74}. This is necessary in order to understand  the source of problem which puzzled the researches 
for almost four decades. 

\subsection{Decay of  Perfect Fluid into the Gas of Free Particles.}
In order to obtain the particle spectra for the gas, I will use the
method suggested by Gorenstein and Sinyukov \cite{SFO2}.
Suppose there is a boundary between fluid and gas.
Let me  consider the decay of a small element $\Delta x$ of the gas of
free particles in its rest frame. The gas is supposed to be located
in the left hemisphere and to have the
freeze-out temperature $T = T^*$.
(I suppose that the derivative to the freeze-out
hypersurface $v_\sigma$ in the $t-x^1$ plane is positive). Note that
this frame is the rest frame of the gas {\em before\/} decay. Hereafter
I will  call it the reference frame of the gas (RFG).
Suppose the particles in the gas have an equilibrium distribution function
$\phi\left(\frac{p_0}{T^*}\right)$.

\begin{figure}[ht]

\centerline{\psfig{figure=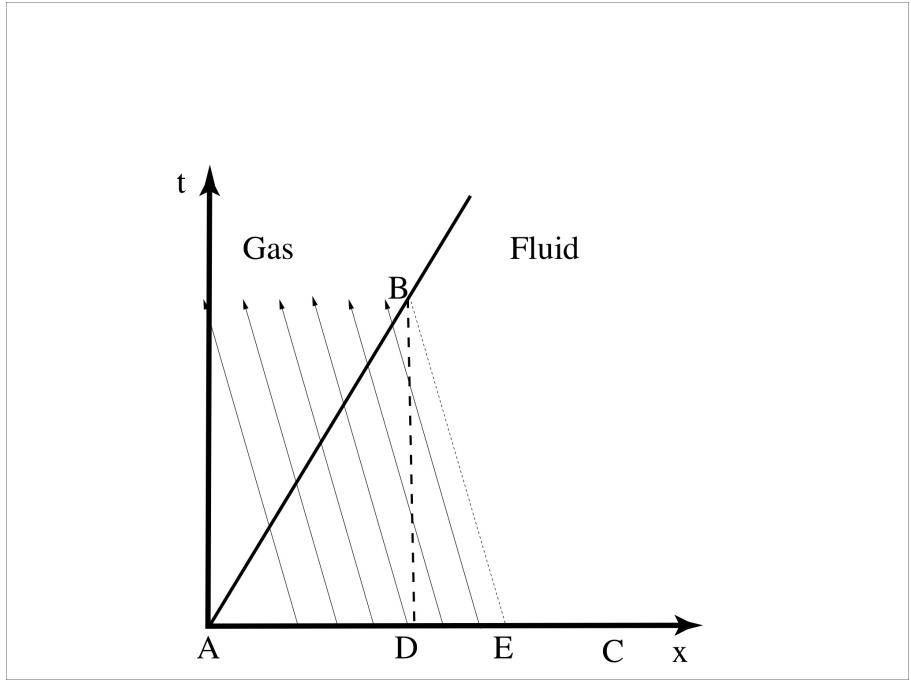,height=9cm,width=12cm}}

\caption{ \label{Decay1}
 Decay of the element  $\Delta~x = AD $. 
Particles with negative momenta are 
leaving  elements $AD$  and  $ DE = - p/p_0 ~ \Delta~t$. 
The trajectories   of free streaming particles are  indicated by the  lines 
with small  arrows. 
}
\end{figure}

First I consider the contribution from particles with negative
momenta that leave the element $\Delta	x$ (see Fig.~\ref{Decay1})

\begin{equation}
\frac{d N_1}{d^2 p_{\perp} \Delta S_{\perp}} =
\phi\left(\frac{p_0}{T^*}\right) \Delta x\, dp\, \Theta(-p)\,\,\, ,
\end{equation}

\noindent
where $p_{\perp}$ is the transverse momentum of the particle, and
$\Delta S_{\perp}$ the transverse size of the element.
The second contribution is given by particles with negative momenta
from the element $-\frac{p}{p_0} \Delta t$
\begin{equation}
\frac{d N_2}{d^2 p_{\perp} \Delta S_{\perp}} =
- \phi\left(\frac{p_0}{T^*}\right) \frac{p}{p_0}\,
\Delta t \, dp \, \Theta(-p)\,\, .
\end{equation}

\begin{figure}[ht]

\centerline{\psfig{figure=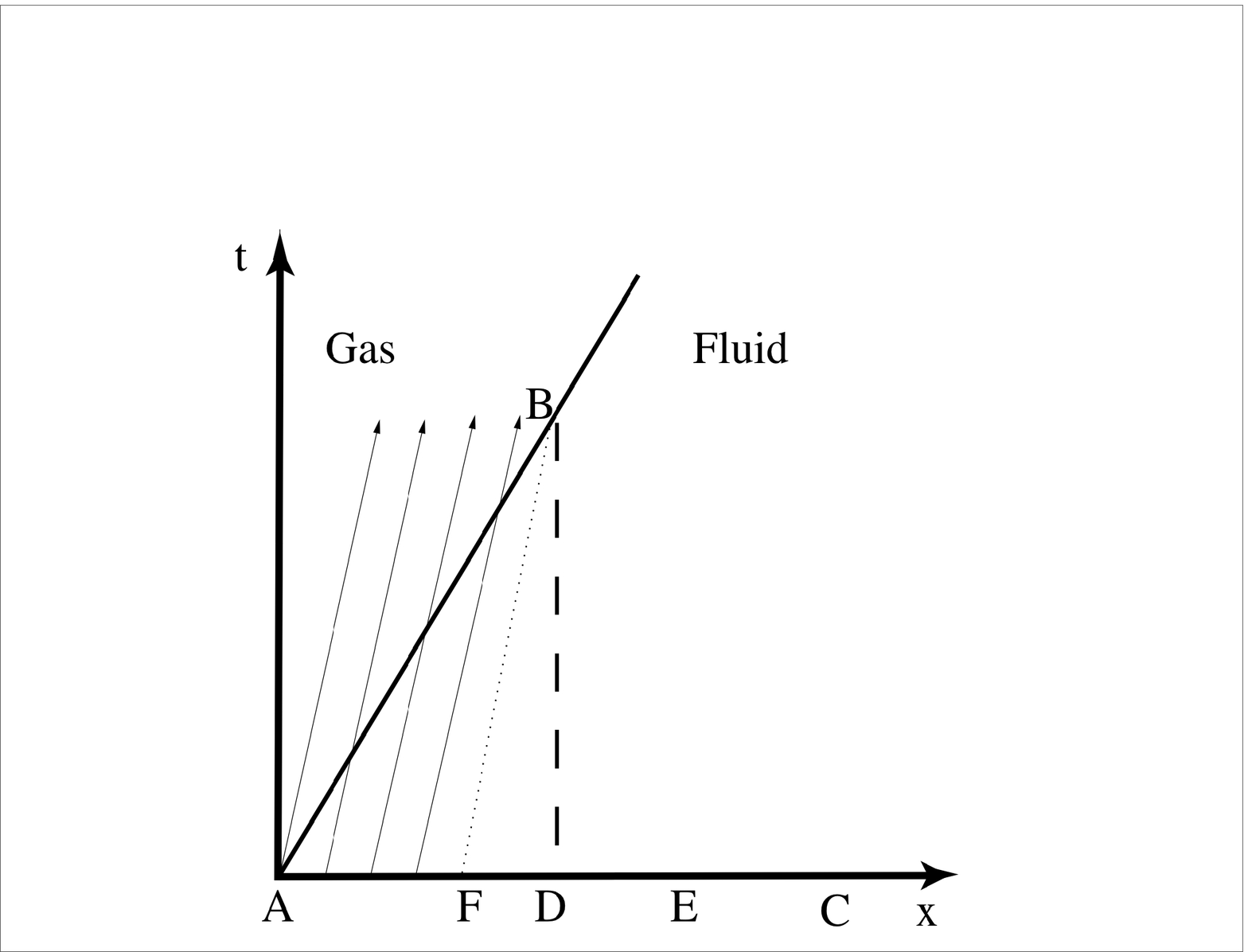,height=9cm,width=12cm}}

\caption{ \label{Decay2}
 Decay of the element  $\Delta~x = AD $. 
Particles with positive momenta are 
leaving  elements $AF = \Delta~x  - p/p_0 ~ \Delta~t$. 
Note that particles emitted from the element $FD$ are returning to fluid's interior. 
The trajectories   of free streaming particles are  indicated by the  lines 
with small  arrows. 
}
\end{figure}

\noindent
Finally, the third contribution comes from particles with positive momenta
from the element $\Delta x -\frac{p}{p_0} \Delta t$. However, those
particles will cross the freeze-out hypersurface only if their velocity is
smaller than the derivative to the hypersurface $v_\sigma$ in the $t-x$ plane.
Thus, the third term reads as follows (see Fig.~\ref{Decay2}):
\begin{equation}
\frac{d N_3}{d^2 p_{\perp} \Delta S_{\perp}} =
\phi\left(\frac{p_0}{T^*}\right) \left[\Delta x - \frac{p}{p_0}
\Delta t \right] \, dp \, \Theta(p) \, \Theta\left(v_\sigma - \frac{p}{p_0}
\right)\,\, .
\end{equation}

\noindent
After some simple algebra one obtains the formula for the spectrum of the
gas of free particles
\begin{equation}
\frac{d N_{tot}}{d^2 p_{\perp} d p \Delta S_{\perp}} =
\phi\left(\frac{p_0}{T^*}\right) \left[\Delta x - \frac{p}{p_0}
\Delta t \right] \Theta\left(v_\sigma - \frac{p}{p_0}\right)\,\, .
\end{equation}

\noindent
As will be shown below,
the modification of the spectrum due to the $\Theta$-function
will lead to an energy-momentum tensor that differs from the equilibrium case.

Our last step is to write the formula for the spectrum in a fully relativistic
form. For that one has  to change the energy of the particles appearing in the
distribution function in the RFG
to the product of the four-vectors of momentum and hydrodynamic
velocity, $p_\mu u^\mu$, and
change the integration over the hypersurface of freeze-out to the product
of the four-vectors of momentum and normal vector
to the freeze-out hypersurface, $p_\mu d\sigma^\mu$. Finally, one has
\begin{equation}  \label{CutDistrib}
p_0 \frac{d N_{tot}}{d^3 p } =
\phi\left(\frac{p_\mu u^\mu}{T^*}\right) p_\nu d \sigma^\nu\,
\Theta\left(p_\rho d \sigma^\rho \right)\,\, ,
\end{equation}

\noindent
where the vector $d \sigma_\mu = (v_\sigma,- 1) dt\,  \Delta S_{\perp} $ is the normal vector to
the freeze-out hypersurface in the left hemisphere.
It is, however, easy to check that
the above formula is valid for the right hemisphere as well.

Integrating (\ref{CutDistrib}) over the whole FO HS, one gets
\begin{equation}  \label{CutDistribI}
p_0 \frac{d N_{tot}}{d^3 p } = \int\limits_{\Sigma_{fr}}
\phi\left(\frac{p_\mu u^\mu}{T^*}\right) p_\nu d \sigma^\nu\,
\Theta\left(p_\rho d \sigma^\rho \right)\,\, ,
\end{equation}

The meaning of this nice result is that this  is the formula of Cooper and
Frye \cite{CF74}, but without negative particle numbers! 
Thus, Eq. (\ref{CutDistrib}) 
accounts for outgoing particles only (see Figs.~\ref{Scheme1} and~\ref{Scheme2} for more details), whereas other particles are returning to fluid 
and, hence, should not be taken into account!  For this reason
I named the distribution function (\ref{CutDistrib}) as   {\it the cut-off distribution function.}

It is easy to see that for a s.l.  hypersurface, where $v_\sigma > 1$,
the above expression gives the result obtained by Cooper and Frye \cite{CF74}.

The energy-momentum tensor of free
particles in the RFG reads as:
\begin{equation}
T^{\mu\nu}_g(v_\sigma) = \int \frac{d^3 p}{p_0}
p^\mu p^\nu \Theta\left(v_\sigma - \frac{p}{p_0}\right)
\phi\left(\frac{p_0}{T}\right)\,\, .
\end{equation}

\noindent
It is easy to calculate this tensor for the case
of noninteracting massless particles; it
has the form:
\clearpage
\begin{figure}[ht]
\centerline{\hspace*{1.5cm}\psfig{figure=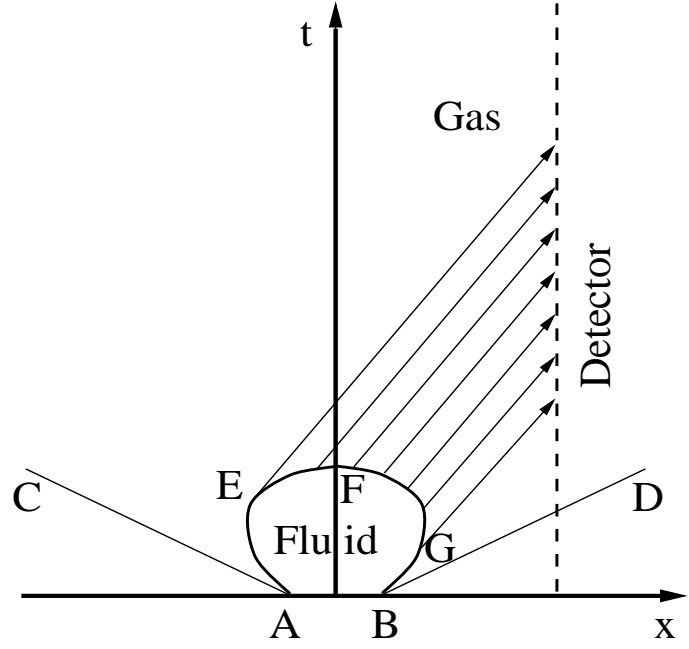,height=12cm,width=12cm}}

 \vspace*{-3.0cm}
 
\caption{ \label{Scheme1}
Decay of the fluid into the gas of free particles on the convex freeze-out hypersurface 
$AEFGB$.
Trajectories of the particles are indicated by the lines with arrows.
Dashed line represents the detector's  world line.
Lines $AC$ and $BD$ denote the light cones. 
For the given particle velocity $v = \frac{p^x}{p^0}$ 
the integration limits in coordinate space, points $E$ and $G$, are found
from the condition $p_\rho d \sigma^\rho = 0$.
The geometrical meaning of points $E$ and $G$ is evident from the construction: 
they are the tangent points of the particle velocity to the freeze-out hypersurface
in $t-x$ plane.
In contrast to the \COful\, formula for invariant spectra which ensures such limits automatically,
the \CFful one takes into account the particles emitted
from all points of the FO HS  $AEFGB$. 
}
\end{figure}

\begin{eqnarray}
T^{00}_g(v_\sigma) = \epsilon\left(T^*\right) \frac{1 + v_\sigma}{2}\,\, , \\
T^{01}_g(v_\sigma) = \epsilon\left(T^*\right)  \frac{v_\sigma^2 - 1}{4}\,\, ,
\\
T^{11}_g(v_\sigma) = \epsilon\left(T^*\right)  \frac{v_\sigma^3 + 1}{6}\,\, ,
\end{eqnarray}

\noindent
where $\epsilon$ is the usual energy density. The above result is valid for
the left hemisphere. The corresponding formulae
for the right hemisphere can be obtained in the same way.
The~general 
~\clearpage
\begin{figure}[ht]
\centerline{\hspace*{1.5cm}\psfig{figure=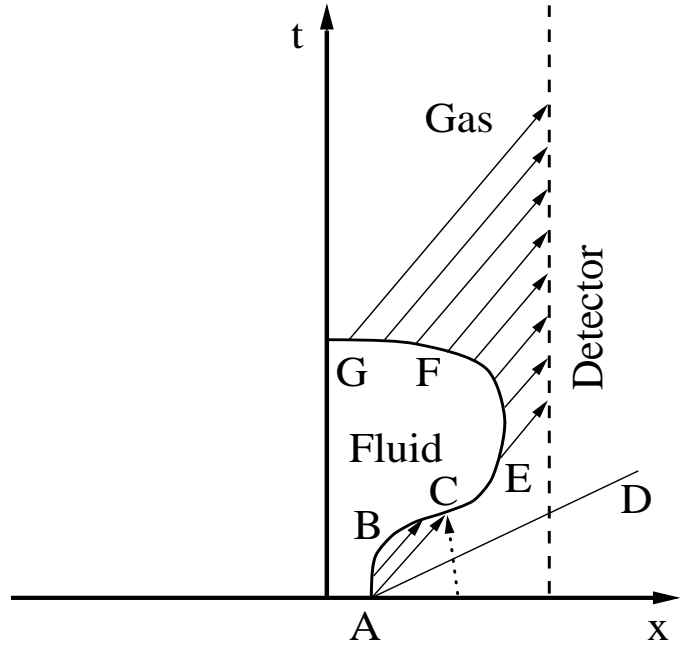,height=12cm,width=12cm}}

\vspace*{-3.0cm}

\caption{\label{Scheme2} 
Decay of the fluid into the gas of free particles on the concave freeze-out hypersurface
$ABCEFG$.
Notations correspond to the previous figure.
In this case, however, one has to take into account the particles feeding back into the fluid
on the part $ABC$ of the freeze-out hypersurface.
These particles were emitted earlier and do not appear from the rescattering.
The latter is forbidden by the assumption that the particle spectra are "frozen" once
they belong to the gas of free particles.
Hence, the particle trajectories, like the one shown by the dotted line, are not allowed
in the freeze-out picture because those particles appear from nothing.
}
\end{figure}

\noindent
formulae for  $T^{\mu\nu}_g(v_\sigma)$ can be found in \cite{Bugaev:99}. 
Now one can see that for the case
$v_\sigma = 1$ the above expressions give the usual formula for
the ideal gas. For  $v_\sigma = - 1$ no particle can cross such a HS and, hence, all   components of the gas tensor vanish.

In Sinyukov's works  \cite{Si:89, Si:89b} another model
for freeze-out was suggested.
The author of \cite{Si:89, Si:89b} considered the decay from a box and did not account for the additional
contributions from the intrinsic volume of the gas, namely from the element
$-\frac{p}{p_0} \Delta t$ (see Fig.~\ref{Decay1}). 
Due to that, the energy-momentum tensor obtained
in these papers  is not symmetric which  means that the orbital momentum of the
considered system is not locally  conserved!

On the other hand,  in  \cite{Si:89, Si:89b}  considered an {\em ad hoc\/} model of 
deflagration from
a very hot pionic matter into the gas of free particles.
However,  such an  oversimplified picture does not correspond 
to the real situation in A+A collisions,
since the solution of the hydrodynamic equations \cite{Mishustin89, Rischke1, namiki, Bernard:95}
does not exhibit the pure  shock-like transition in the expansion of hot and dense
pionic matter. Moreover,  in contrast to the assumption of Refs.  \cite{Si:89, Si:89b}. the hydro solutions  must  necessarily have the s.l. parts of the FO HS. 

Now I would like to clarify an important question: "What is the difference
between a perfect fluid and the gas of free particles at the FO?"
It seems that the main difference is that they have different values of
the mean free path. Due to that, there are collisions in the fluid which
lead to thermodynamical equilibrium. In contrast, there are about no collisions
in the gas of free particles (and I will neglect them completely), because the
mean free path
is very large. 
Of course, fluid and gas have somewhat different values of temperature,
but due to the fact that the mean free-path strongly depends on the temperature,
this difference should not be too large.
It was found \cite{SFO4} that for pions the mean free-path depends on the
fifth power of the inverse temperature: $ \lambda \approx \frac{const}{T^5} $.
Thus, the difference between the temperatures of the fluid and the gas
should be small, but
due to the strong dependence of the collision cross-section on the
temperature, their mean free-paths should be very different.
Such a  conclusion was also demonstrated by  \cite{BRAVINA:99} with the transport 
simulations which show that  at the t.l.  parts of the FO HS  
about 85 \% of all particles are born (or scattered last time)  within a narrow region 
of the width about 0.5 - 1.2 fm. Also these simulations showed that  the main problem 
with the validity  of  the narrow freeze-out hypersurface existence comes from   the s.l. parts of the FO HS  due to decay of hadronic resonances.  Thus, keeping these facts in mind, I will 
assume that the FO HS has a zero width. As one sees these is reasonable assumption 
for the t.l. parts of the FO HS, whereas the decay of resonances can be accounted 
after the  FO, since  this is necessary to work out 
{\it a purely 
hydrodynamic solution of the freeze-out problem.}
As will be seen  this concept is 
based on the conservation laws at  the discontinuity
between the perfect  fluid and the gas of free particles.

However, before completing this subsection it is necessary to generalize  the obtained results
(\ref{CutDistrib}) and (\ref{CutDistribI}) in order to account for feed back of the particles emitted earlier
from the concave  parts of the FO HS (see Fig.~\ref{Scheme2} for details). 
This can be done by accounting for the incoming particles with 
$\phi_{g.fback}\left(\1 p \right) \Theta\left(-p_\rho d \sigma^\rho \right)$. In order to distinguish the corresponding distributions 
of the gas of free particles, hereafter they are labeled by additional subscript:
\begin{eqnarray} \label{five51}
&&  \phi_{g} \lp \frac{ p_\r u_g^\r }{T_g} , d \sigma_\mu \rp \bigg|_{\xfr} ~= ~
   \Bigl( \phi_{g.emit}\left(\1 p \right)
\Th \left( p^\m d \s_\m \right)  +
\phi_{g.fback}\left(\1 p \right)\Th \left (- p^\m d \s_\m \right) \Bigr) \bigg|_{\xfr}\,,
\end{eqnarray}
where I  have dropped the space-time dependence in the distribution functions above 
for the sake of convenience. An important assumption of this approach is that the post FO 
distribution function at the moment of emission, i.e. $\phi_{g.emit}\left(\1 p \right)$, 
is an equilibrium one.  This assumption implies  that the outgoing
free particles  are emitted from the region in which the collision rate is sufficient high to  keep 
thermodynamic equilibrium.  This might look strange, but it seems  to be very reasonable because the
hydrodynamic approach  is valid in the same range as  equilibrium thermodynamics 
\cite{LANDAU:53, llhydro}.

\subsection{Conservation Laws at the Freeze-out  Hypersurface.}
The FO  problem was solved in \cite{Bugaev:96}  and developed further in 
Refs. \cite{Bugaev:99, Bugaev:99a,Bugaev:99b,Bugaev:09}. 
It was found after a realization of a  fact that at  the t.l.  parts of FO HS
there is a fundamental difference between the particles of fluid and the particles
emitted  from its surface: the  EOS of the fluid can be anything, but it implies a zero
value for the mean free path, whereas the emitted particles cannot  interact at all because
they have an infinite mean free path. 
Mixing  these two different objects  the  researches were confused for decades. 
However, once they are distinguished it clear that
it is necessary to extend the  conservation laws (\ref{one51}) and 
(\ref{two51}) from a fluid alone to a  system 
consisting of a  fluid and  the particles of gas emitted  (gas of free particle) from the FO HS. 
The resulting  energy-momentum tensor  and baryonic current (for a single particle species)  
of the  system  can be, respectively,  cast as 
\begin{eqnarray}
\label{three51}
T^{\mu\nu}_{tot} ( x,t) & = & \Theta_f^*~T^{\mu\nu}_{f} ( x,t) +  ~\Theta_g^*~T^{\mu\nu}_{g} ( x,t)\,, \\
\label{four51}
N^{\mu}_{tot} ( x,t) & = & \Theta_f^*~N^{\mu}_{f} ( x,t)~ +  ~\Theta_g^*~N^{\mu}_{g} ( x,t)\,, 
\end{eqnarray}
where at the FO HS the energy-momentum tensor of the gas $T^{\mu\nu}_{g} $
and its baryonic current $N^\mu_g$  are  given in terms of {\it the cut-off}  distribution function of particles  that have the 4-momentum $p^\mu$, i.e. (\ref{five51}), 
\begin{eqnarray}
\label{six51}
T^{\mu\nu}_{g}\left( x^*, t \right)  & = & \int \frac{d^3 { p}}{p_0} \, p^\mu p^\nu \, \phi_{g}\left( x^*, t, p, d \sigma_\mu \right)    
%
\,\,, \\
\label{seven51}
N^{\mu}_{g}\left( x^*, t \right)  & = & \int \frac{d^3 { p}}{p_0} \, p^\mu ~ \, \phi_{c.g}\left( x^*, t, p, d \sigma_\mu \right)    
%
 \,\,.
\end{eqnarray}
Here  $\phi_{g}\left( x^*, t, p \right)$  and    $\phi_{c.g}\left( x^*, t, p \right)    $ denote the equilibrium distribution function of number of particles and charges, respectively.
$d \sigma_\mu$ are the components of the external normal 4-vector to the FO HS  $\SF( x^*,t)$ 
\cite{ Bugaev:96,Bugaev:99a,Bugaev:99b,Bugaev:09}. 

The important feature of equations (\ref{five51})-(\ref{three51}) is the  presence of several  
$\Theta$-functions. 
The $\Theta_g^* = \Theta(F(x,t)) $ function of the gas
and $\Theta_f^* = 1 - \Theta_g^* $ function of the fluid   can be  explicitly expressed in
terms of the freeze-out criterion and can automatically 
ensure that the energy-momentum tensor of the 
gas (liquid) is not vanishing  only in the domain where the gas (liquid) exists.

The  equations of motion of  the full system  are just the conservation laws:
\begin{equation}\label{eight51}
\partial_\mu~ T^{\mu\nu}_{tot} (x, t)  = 0\,, \quad  \partial_\mu~ N^{\mu}_{tot} (x, t) = 0 
\end{equation}
Let me study {\it the boundary conditions} first. I will do it for the energy-momentum tensor only, since 
for the charge it can be done similarly. 
Integrating  equations (\ref{eight51}) over the 4-volume  around the vicinity of
the FO HS, namely 
$ x^1 \in [\xfr - \delta^2\,; \xfr  + \delta^2 ] $ and the corresponding 
finite limits for the other coordinates, 
and applying the Gauss theorem as it was done in previous subsection, one obtains  
the energy-momentum conservation  for  the corresponding part  of the FO HS $\SF$
\begin{equation}  \label{emconservationint}
\int\limits_{\SF - \d^2} d \sigma_\mu T^{\mu\nu}_{tot} = 
\int\limits_{\SF + \d^2} d \sigma_\mu T^{\mu\nu}_{tot}  \,\, ,
\end{equation}
where in both sides of the equality $d \sigma_\mu $ is the external
normal with respect to the fluid.

Then in the limit $\delta \rightarrow 0 $ 
one easily gets     the energy-momentum 
(the charge) conservation  at  the  FO HS from  Eqs. (\ref{six51}) 
\beqs \label{emboundaryi}
d \sigma_\mu  T^{\mu\nu}_f \bigg|_{\xfr}& = &
d \sigma_\mu  T^{\mu\nu}_g\, \bigg|_{\xfr} \,\, , \\
d \sigma_\mu  N^{\m}_{c.f} \bigg|_{\xfr}& = &
d \sigma_\mu  N^{\m}_{c.g} \bigg|_{\xfr} \,\, ,
\label{cboundaryi}
\eeqs

\vm
\noindent
or, writing it explicitly,
\beqs \label{emboundaryii}
&& d \s_\m \int \frac{d^3 \1 p}{p^0} \,  p^\m
 p^\n  \, \phi_{f} \lp \frac{ p_\r u_f^\r}{T} \rp \bigg|_{\xfr}
=  \nn 
&&
 d \s_\m 
\int \frac{d^3 {\1 p}}{p_0} \, p^\m p^\n \, \Bigl( \phi_{g.emit}\left(\1 p \right)
\Th \left( p^\m d \s_\m \right)  +
\phi_{g.fback}\left(\1 p \right)\Th \left (- p^\m d \s_\m \right) \Bigr) \bigg|_{\xfr}\,, \\
&& d \s_\m \int \frac{d^3 \1 p}{p^0} \,\,  p^\m
 \,\, \phi_{c.f} \lp \frac{ p_\r u_f^\r}{T} \rp \bigg|_{\xfr}
= \nn 
\label{cboundaryii}
&& d \s_\m
\int \frac{d^3 {\1 p}}{p_0} \,\, p^\m \, \Bigl( \phi_{c.g.emit}\left(\1 p \right)
\Th \left( p^\m d \s_\m \right)  +
\phi_{c.g.fback}\left(\1 p \right)\Th \left (- p^\m d \s_\m \right) \Bigr) \bigg|_{\xfr}\,,
\eeqs

\noindent
with the evident notations for fluid distribution functions.

From the expressions (\ref{emboundaryii}) and  (\ref{cboundaryii}) it is clear that
there are two distinct cases, namely {\bf (i)} when the distribution
function of \gfp\,  coincides with that one of the fluid (or the EOS 
is the same for both) and {\bf (ii)} when they are entirely different.
In the latter case there is no criterion which equation of state is
preferable.  This should be defined by the physics of the considered
task.  I believe that the full solution of the problem can be given
within the kinetic approach only.  However, since we adopted the
hydrodynamical approach which implies also the thermodynamical
equilibrium,  in what follows  it is
assumed that at least the
emission part is described by the equilibrium distribution function.

{\it Statement 1:}
If the fluid and the gas have same EOS, then on the s.l. parts of the FO HS
their distribution functions are identical, on the convex parts of the t.l. 
FO HS there exists a {\it freeze-out shock} transition, and on the
concave ones there is a {\it parametric freeze-out shock} transition (see below). 

Let me demonstrate this.  Indeed, a simple analysis shows that for the
case {\bf (i)} there are the following possibilities:

{\bf (i.a)} if $\Th \lp p^\m d \s_\m \rp = 1$ for $\forall\,\,\, \1
p$, i.e., on {\it the s.l.} parts of the FO HS or on the light
cone, there is no feedback and then there is only a trivial solution
given by the \CFful  formula $\phi_{f} \lp \frac{ p_\r u_f^\r}{T} \rp
\bigg|_{\xfr} = \phi_{g.emit}\left(\1 p \right) \bigg|_{\xfr}$ with
the fluid temperature being equal to the FO one.  Note that the
discontinuity through such a hypersurface (or the t.l. shocks, as
the authors of Ref.  \cite{timeshock} call it) is impossible in this
case as it was shown in  \cite{marik1}.

{\bf (i.b)} for {\it the convex t.l.} parts of the FO HS the
feedback term vanishes as was discussed before (see  Fig.~\ref{Scheme1}) and
hence there is a {\it freeze-out shock} between the fluid and \gfp
(actually, the same fluid, but with the FO\, temperature and \COful
distribution function; for details see later).  It is reasonable to
call it this way because the pressure and the energy density of \gfp\,
are not the usual ones, but have extra dependence on the parameters of
the FO HS.  The hydrodynamic parameters of the fluid should be found from
the conservation laws on the discontinuity 
\beqs 
d \s_\m \int
\frac{d^3 {\1 p}}{p_0} \,\, p^\m p^\n \,\, \phi_{f} \lp \frac{ p_\r
u_f^\r}{T} \rp \bigg|_{\xfr} & = & d \s_\m \int \frac{d^3 {\1 p}}{p_0}
\, p^\m p^\n \, \phi_{f} \lp \frac{ p_\r u^{\prime\r}}{T^*} \rp \Th
\left( p^\m d \s_\m \rp \bigg|_{\xfr}\,, \\
 d \s_\m \int\, \frac{d^3 {\1 p}}{p_0}   \,\,  p^\m
  \,\, \phi_{c.f} \lp \frac{ p_\r u_f^\r}{T} \rp \bigg|_{\xfr}
& = & 
 d \s_\m
\int \frac{d^3\, {\1 p}}{p_0} \,\, p^\m \, \phi_{c.f} \lp \frac{ p_\r u^{\prime\r}}{T^*} \rp
\Th \left( p^\m d \s_\m \rp 
\bigg|_{\xfr}\,. 
\eeqs

{\bf (i.c)} for {\it the concave t.l.} parts of the FO HS the
feedback term does not vanish (see also Fig.~\ref{Scheme2}) and hence it is a new
type of shock between the fluid and \gfp \, which can be called {\it a
parametric freeze-out shock} with the contribution of the feedback
particles being a parameter in the general meaning.  In this case the
equations on the discontinuity, i.e.  Eqs. (\ref{emboundaryii})  and the
similar one for the conserved charge, should be studied.  However, it
requires the knowledge of the expressions for the feedback particles
emitted from arbitrary hypersurface which is outside the scope of this
dissertation.
In addition, the assumption of the instant thermalization 
of the feedback particles has to be investigated,
but this is far beyond the hydrodynamical approach.

{\it Statement 2:} If the fluid and the gas have entirely different
EOS, then on the s.l. parts of the FO HS the 
t.l. shocks \cite{timeshock} can exist, whereas on the convex parts of the
t.l. FO HS there is a {\it freeze-out shock} transition, and on
the concave ones there is a {\it parametric freeze-out shock}.

The proof is similar to the previous consideration.
If the fluid and the gas EOS \, are different, [case {\bf (ii)}],
one has the following cases:

{\bf (ii.a)} if $\Th \lp p^\m d \s_\m \rp = 1$ for $\forall\,\,\, \1 p$,
i.e., on {\it the s.l.} parts of the  FO HS or on the light cone,
again the feedback  contribution is zero, but  
there is no trivial solution in contrast to the case {\bf (i.a)},
and the t.l. shocks \cite{timeshock} are possible 
as it was discussed in Ref. \cite{marik1}.
They are defined by the following conservation laws
\beqs
 d \s_\m \int \frac{d^3 {\1 p}}{p_0}  \,\,  p^\m
 p^\n  \,\, \phi_{f} \lp \frac{ p_\r u_f^\r}{T} \rp \bigg|_{\xfr}
& = &
 d \s_\m
\int \frac{d^3 {\1 p}}{p_0} \, p^\m p^\n \,  \phi_{g.emit} \lp \frac{ p_\r u^{\prime\r}}{T^*} \rp
\bigg|_{\xfr}\,, \\
 d \s_\m \int\, \frac{d^3 {\1 p}}{p_0}   \,\,  p^\m
  \,\, \phi_{c.f} \lp \frac{ p_\r u_f^\r}{T} \rp \bigg|_{\xfr}
& = &
 d \s_\m
\int \frac{d^3\, {\1 p}}{p_0} \,\, p^\m \, \phi_{c.g.emit} \lp \frac{ p_\r u^{\prime\r}}{T^*} \rp
\bigg|_{\xfr}\,,
\eeqs

\noindent
where  the equilibrium distribution functions for the gas 
$\phi_{g.emit} \lp \frac{ p_\r u^{\prime\r}}{T^*}\rp$ and 
$\phi_{c.g.emit} \lp \frac{ p_\r u^{\prime\r}}{T^*}\rp$
are used.

More recent results on this subject can be found in Refs.
\cite{laslo4, SFO11}.  However, this kind of solutions can probably
exist only under very special conditions, namely for the supercooled
quark-gluon plasma, since 
it is difficult to imagine any reason for the fluid near the FO state (when all
interactions between particles are very weak) to convert suddenly
without any cause into \gfp \, with entirely different EOS.

{\bf (ii.b)} for {\it  the convex t.l.} parts of the FO HS the feedback term  again 
vanishes
and therefore  there is a shock transition between
the fluid and \gfp  but now fluid and  gas being completely different states! 
Therefore, the gas should be described by the 
\COful distribution function.
In this case the hydrodynamic  parameters of the fluid 
are defined by the conservation laws on the discontinuity
\beqs
\hspace*{-0.7cm} d \s_\m \int \frac{d^3 {\1 p}}{p_0}  \,\,  p^\m
 p^\n  \,\, \phi_{f} \lp \frac{ p_\r u_f^\r}{T} \rp \bigg|_{\xfr}
& = &
 d \s_\m
\int \frac{d^3 {\1 p}}{p_0} \, p^\m p^\n \,  \phi_{g.emit} \lp \frac{ p_\r u^{\prime\r}}{T^*} \rp
\Th \left( p^\m d \s_\m  \rp
\bigg|_{\xfr}\,, \\
\hspace*{-0.7cm}
 d \s_\m \int\, \frac{d^3 {\1 p}}{p_0}   \,\,  p^\m
  \,\, \phi_{c.f} \lp \frac{ p_\r u_f^\r}{T} \rp \bigg|_{\xfr}
& = &
 d \s_\m
\int \frac{d^3\, {\1 p}}{p_0} \,\, p^\m \, \phi_{c.g.emit} \lp \frac{ p_\r u^{\prime\r}}{T^*} \rp
\Th \left( p^\m d \s_\m \rp
\bigg|_{\xfr}\,.
\eeqs

{\bf (ii.c)} for {\it the concave t.l.} parts of the FO HS the
feedback term does not vanish and therefore the {\it parametric
freeze-out shock} introduced above should exist.  This kind of the
shock solution should satisfy the conservation laws in the most
general form of Eqs. (\ref{emboundaryii}) and (\ref{cboundaryii}).
However, the detailed consideration of such a  solution would lead us 
too far off the main topic of the present discussion, although a
similar discontinuity will be discussed in the section on hydrokinetics. 

Thus,  the full analysis of all possible boundary
conditions is performed.  The conservation laws discussed above have to
be solved together with the equations of motion  of  the fluid in order to find out
the FO HS.  Next subsection is devoted to the derivation of the
equations of motion of the fluid alone and their consistency with the
boundary conditions.

There are several papers \cite{Si:89, SFO10, SFO11} on a shock-like treatment of the
freeze-out problem. However, in my  opinion these approaches are {\em ad hoc},
since the existence of such a shock is postulated and not obtained
as a result of  equations of motion.
For instance,  the conservation laws on the boundary between the fluid and \gfp \,
discussed in \cite{FO2} (see Eqs. (6), (7) therein)
were just postulated and are not related to any hydrodynamical evolution of
the fluid at all. Moreover, those equations, in the integral
form as presented in Ref. \cite{FO2}, cannot be used 
to solve them together with the hydrodynamical equations
of the fluid.

To summarize, I have presented above the full and complete analysis
of the possible boundary conditions, following the original idea of
the paper \cite{Bugaev:96}.  The derived formalism allows one  not only to
find out the new class of the shock transitions, i.e. {\it the
parametric freeze-out shock}, but also to formulate the hydrodynamical
approach in the way consistent with the emission of the particles from
an arbitrary FO HS.

\subsection{Consistency Theorem.}
Let me study the consistency of the equations of motion  for the whole system with the
boundary conditions derived in the previous subsection.  At the moment
 the explicit form of the energy-momentum tensor  and  
4-current for \gfp is not important, but  one  should remember that these
quantities have to satisfy the conservation laws in the differential
form, since the gas particles  move freely along the straight lines:
\vm
\beqs \label{emconservationgasi}
\partial_\mu T^{\mu\nu}_{g}& = &0\,\, , \\
\partial_\mu N^{\m}_{c.g}& = &0\,\, 
\label{cconservationgasi}
\eeqs

\vm
\noindent
in the domain where \gfp exists.
Exploiting this fact, one can rewrite the original conservation laws
  (\ref{eight51}) as the equations of motion  of the fluid alone 
\beqs \label{emfluidi}
\hspace*{-0.5cm} \partial_\mu \Bigl\{ 
\Theta^*_f~
T^{\mu\nu}_{f} \Bigr\}
& = & - \sum_{a} \d \lp F( x^*_a, t) \rp \nmudowna
 \times \nn  
&& 
\int \frac{d^3 {\1 p}}{p_0} \, p^\m p^\n \, \Bigl( \phi_{g.emit}\left(\1 p \right)
\Th \left( p^\m \nmudowna \right)  +
\phi_{g.fback}\left(\1 p \right)\Th \left (- p^\m \nmudowna \right) \Bigr) \bigg|_{\xfr_a}\,,
\\
\hspace*{-0.5cm} \partial_\mu \Bigl\{
\Theta^*_f~N^{\mu}_{f} \Bigr\}
& = & - \sum_{a} \d \lp F(x^*_a, t) \rp \nmudowna \times \nn
\label{cfluidi}
&&  
\int \frac{d^3 {\1 p}}{p_0} \,\, p^\m \, \Bigl( \phi_{c.g.emit}\left(\1 p \right)
\Th \left( p^\m \nmudowna \right)  +
\phi_{c.g.fback}\left(\1 p \right)\Th \left (- p^\m \nmudowna \right) \Bigr) \bigg|_{\xfr_a}\,,
\eeqs

\vm
\noindent
where  $\nmudowna$ is the external normal 4-vector with respect to the fluid
which is  defined as 
\begin{equation}
n_\mu (f_a) ~ \equiv  ~\partial_\mu~ { F} ( x^*_a, t ) 
\end{equation}
for  all solutions of the FO criterion equation that labeled by  the subscript a.

Eqs.  (\ref{emfluidi}) and (\ref{cfluidi})  look like the hydrodynamical equations of motion with the source terms,
the first of them describes {\it the loss} of the energy-momentum  flux through the FO HS 
due to emission of the "frozen" particles
and the second one is responsible for {\it the gain} by the reentering particles 
at the concave parts of the t.l. FO HS.
It is evident because  the $\Th$-function in the  first term takes into account 
only positive values of the product \mbox{$p^\m \nmudowna > 0$,} while the 
second one is nonvanishing only for its  negative values, i.e., \mbox{$p^\m \nmudowna < 0$}.

However, it is easy to show that,  in fact,   there are no source terms because they
disappear due to the boundary conditions studied in the previous subsection!
Indeed, taking derivatives of the remaining $\Th$-functions in the left hand
side of  Eqs. (\ref{emfluidi})  and (\ref{cfluidi}), one obtains the source-like terms with
the fluid energy-momentum tensor   and with the conserved 4-current, respectively, 
which exactly cancel the source terms in the r.h.s of (\ref{emfluidi}) and (\ref{cfluidi}). 
Then the equations of motion of the fluid acquire a familiar form:
\begin{eqnarray}
\label{nine51}
 \Theta_f^*~\partial_\mu T^{\mu\nu}_{f} ( x,t) & = & 0\,\,,   \hspace*{1.81cm} \quad 
  \Theta_f^*~\partial_\mu N^{\mu}_{f} ( x,t) ~ = ~ 0\,\,,     
\end{eqnarray}

These equations complete the proof of the following {\it Theorem 1:}
If \gfp\, 
with the emission part defined by \COful distribution and
with known feedback part 
is described by the equations of motion  (\ref{emconservationgasi}) and (\ref{cconservationgasi}),
then the  equations of motion  
of the whole system consisting
of the perfect fluid and \gfp\, split up
into the system of the fluid's equations of motion 
(\ref{nine51}) 
and of the
boundary conditions in the general form of  Eqs. (\ref{emboundaryii}) and (\ref{cboundaryii})
on the FO HS.

There is  a fundamental difference between the equations of motion (\ref{one51})  of traditional hydrodynamics  and the corresponding equations (\ref{nine51}) of hydrodynamics with particle emission although they look very similar:
if the FO HS is found, then, in contrast to usual hydrodynamics,  the equations (\ref{nine51}) automatically
vanish in the domain where the fluid is absent. 
Evidently,
a solution of the traditional hydrodynamic equations under the
given additional conditions  is also a solution of  Eqs. (\ref{nine51}).  Moreover, the solution of the latter has to be understood
this way. Thus, the effect of the particle emission is implicitly taken
into account  by the  equations of motion  of the fluid alone, and is expressed
explicitly in the boundary conditions between the fluid and the gas.

{\it Theorem 1} leads also to the fact that inclusion of the 
source terms which are proportional to the $\d$-function on 
the FO HS  into the  equations of motion  of the fluid  will always require their vanishing,   
unless the derivatives of the fluid's energy-momentum tensor   or the gas one 
contain similar singularities.
This is a very important consequence of the {\it Theorem 1} which 
is easy to understand recalling the fact that singularities like 
$\d$-function and its derivatives (if they appear) are of different
order and should be considered independently.
Therefore, if these singularities enter the same equality, then
singularities of the same order will generate independent equations.

Now we are ready  to consider the consistency problem of
the energy-momentum  and the current conservation of the fluid given by  Eqs.   
(\ref{five51})
with the emission of the "frozen" particles of the gas described by 
the \COful distribution function.

{\it Theorem 2:}
Energy, momentum  and charge of the initial fluid together with  
the corresponding contribution of the reentering particles from 
the concave parts of the FO HS are equal to the corresponding
quantities of the emitted from this hypersurface
free particles 
with the \COful momentum distribution function, i.e., those quantities are conserved.

{\it Proof.} First  I integrate  
the derived equations  (\ref{nine51}) of the fluid evolution over the 4-volume 
and use the Gauss theorem to
write the  4-volume integral over the integral of the corresponding closed HS. 
 Then by splitting the closed HS into the initial and FO parts  
one obtains an integral form of the energy-momentum conservation 
for the fluid alone
\begin{equation} \label{Pnu}
P^\n_{f.in} \equiv
- \int\limits_{\SI} d \s_\m~ T^{\mu\nu}_{f} ( x,t) ~ = ~
\int\limits_{\SF} d \s_\m~ T^{\mu\nu}_{f} ( x^*, t) \,. 
\end{equation} 
Due to the boundary conditions (\ref{emboundaryii}) on the FO HS   one can rewrite 
the r.h.s of  Eq. (\ref{Pnu}) in terms  of the gas distribution function  and  obtain the following
equality
\beqs
&-& \int\limits_{\SI} d \s_\m T^{\mu\nu}_{f} ( x,t) 
- \int\limits_{\SF} d \s_\m\int \frac{d^3 {\1 p}}{p_0} \, p^\m p^\n \, 
\phi_{g.fback}\left(\1 p \right)\Th \left (- p^\m d \s_\m \right)  \bigg|_{\xfr} =
\nn
&&\int\limits_{\SF} d \s_\m
\int \frac{d^3 {\1 p}}{p_0} \, p^\m p^\n \, \phi_{g.emit}\left(\1 p \right)
\Th \left( p^\m d \s_\m \right) \bigg|_{\xfr} \,, 
\eeqs

\vm
\noindent
which actually states the  energy-momentum conservation in the integral form and  also shows
that the sum of the energy-momentum of the fluid and the particles reentering the fluid
during its evolution are exactly transformed into the energy-momentum  of the
emitted particles with the correct distribution function!

A proof for the conserved current can be obtained in a similar way.

Recently there appeared at attempt to resolve the FO problem by intoducing  the 
sources into the hydrodynamic equations \cite{Ivanov:06}. 
As it is seen from the text of Ref. \cite{Ivanov:06} the source terms 
modify the solution of the hydrodynamic equations essentially and such a modification should have been long ago  found in the experiments on nonrelativistic hydrodynamics. In addition the problem of 
negative numbers of particles is not resolved in  \cite{Ivanov:06}  and in this approach there exist a typical causal paradox  that the post FO state in the inner part of a fluid 
is reached before it is reached in its outer part.  Consequently, the gas of free particles ``waits'', when the outer parts of the fluid ``tells'' it to  freely go 
to the detector.  Therefore, I conclude that the  attempt of Ref. \cite{Ivanov:06} is
unsuccessful.

\section{Equations for  the Freeze-out Hypersurface}

The conservation laws   (\ref{emboundaryii}) and  (\ref{cboundaryii}) 
are the partial differential equation for the  FO HS. Here I will  discuss 
the general  scheme of how to solve these equations together with the hydro equations. 
In what follows I will
neglect the contribution of the feedback particles in order to
simplify presentation, and, hence,   the
freeze-out  and the  parametric freeze-out shocks  will not be  distinguished.

From academic and numeric  points of view it is worth to closely  inspect  
these equations. 
As a good example to apply the derived scheme, I   consider 
an important problem of relativistic hydrodynamics --  the freeze-out
of a  simple wave.

\subsection{Freeze-out Calculus: General Scheme.}
Let me obtain now the full system of all necessary equations.  First,
I  suppose that the FO HS exists and then I will  derive its equations.
For this it is convenient to evaluate the boundary
conditions in the rest frame of the fluid (hereafter RFF) before the FO shock, 
where the energy-momentum tensor  of the fluid is diagonal. 
I
choose the local coordinate system with one of the axis,
let it be the $X$-axis, being parallel to the normal 3-vector. The latter
then is reduced to 
\begin{equation}
\nmudowna = \lp \,- \partial_0 ~\xfr_a ; \,\,\,
1 ; \,\,\, 0 ; \,\,\, 0 \rp \,
\left\{ \begin{array}{rl}
 -1\,,  & \hspace*{0.3cm} a \in {\rm left ~hemisphere} \,,  \\
 & \\
 +1\,,  & \hspace*{0.3cm} a \in {\rm right~hemisphere}  \,.
\end{array} \right. \label{newnormali}
\end{equation}
%

Next, I  mention that the velocity of \gfp\, cannot have nonzero projection
on any tangential direction to the normal 3-vector in this frame.
It can be shown directly by manipulation with formulae, but it is evident from 
the simple reason that an oblique shock (see corresponding chapter in Ref. \cite{llhydro})
should have continuous tangent velocities on the both sides of shock.
Then it follows for the RFF that the gas velocity can be parallel or 
antiparallel to the normal 3-vector only.

This statement is valid for the RFG as well, and the expression for
the normal 4-vector in this frame is evidently similar to 
Eq. (\ref{newnormali}).  In order to distinguish them, hereafter I will
write the corresponding subscript.

Boundary conditions 
(\ref{emboundaryii})  and (\ref{cboundaryii})  have a simplest representation in the RFG 
since the moments of the \COful  distribution  function do not look too much 
complicated in this frame:
\beqs
T^{X0}_f(\vfrfg) - T^{00}_f (\vfrfg) \vgrfg& = &
T^{X0}_g(\vgrfg) - T^{00}_g (\vgrfg) \vgrfg \,\, , 
\label{emboundaryiii} \\
T^{XX}_f(\vfrfg) - T^{X0}_f (\vfrfg) \vgrfg& = &
T^{XX}_g(\vgrfg) - T^{X0}_g (\vgrfg) \vgrfg \,\, , \\
N^{X}_{c.f}(\vfrfg) - N^{0}_{c.f} (\vfrfg) \vgrfg& = &
N^{X}_{c.g}(\vgrfg) - N^{0}_{c.g} (\vgrfg) \vgrfg \,\, , \label{cboundaryiii}
\eeqs

\vm

\noindent
where the energy-momentum tensor  and 4-current of the fluid have standard form of 
Eqs. (\ref{one51}) and (\ref{two51}), respectively, with the 3-velocity $\vfrfg$.  In
the above formula the velocity $\vgrfg = \partial_0 \xfr$ is the time
derivative of the FO hypersurface in the RFG, or velocity of the
shock in this frame.  Note, however, that in contrast to the usual shocks the above equations look like
the conservation laws in the arbitrary Lorentz frame (not the rest frame of the gas!)
where the nondiagonal components of the energy-momentum tensor are nonzero.

Introducing the following notations for the "effective" energy density, pressure and 
charge density of \gfp   
\beqs
\egtil (\vgrfg) & = &
T^{00}_g(\vgrfg) -  T^{X0}_g (\vgrfg) \vgrfg^{-1} \,\, , 
\label{epsnew} \\
\pgtil (\vgrfg) & = &
T^{XX}_g(\vgrfg) - T^{X0}_g (\vgrfg) \vgrfg \,\, , 
\label{pnew} \\
\ngtil (\vgrfg) & = &
N^{0}_{c.g}(\vgrfg) - N^{X}_{c.g} (\vgrfg) \vgrfg^{-1} \,\, , \label{nnew}
\eeqs

\vm

\noindent
one can transform the right-hand side of  Eqs. (\ref{emboundaryiii}) -- (\ref{cboundaryiii})
to the familiar expressions of the relativistic shocks \cite{llhydro} written in the rest 
frame of the matter behind the discontinuity.
In contrast to the usual shock, however, 
Eqs. (\ref{emboundaryiii}) -- (\ref{cboundaryiii})
do not form the closed system together with the EOS, but they  are dynamical
equations for the  FO HS.
Then velocities of fluid and shock in the RFG can be expressed 
by the standard relations  
\beqs
\vfrfg^2 & = &
\frac{ \lp \varepsilon_f - \egtil (\vgrfg) \rp \lp p_f - \pgtil (\vgrfg) \rp }
{ \lp \varepsilon_f + \pgtil (\vgrfg) \rp \lp p_f + \egtil (\vgrfg) \rp } \,\, , 
\label{velvelfrfg} \\
\vgrfg^2 & = &
\frac{ \lp p_f - \pgtil (\vgrfg)  \rp \lp \varepsilon_f + \pgtil (\vgrfg) \rp }
{\lp \varepsilon_f - \egtil (\vgrfg) \rp \lp p_f + \egtil (\vgrfg) \rp } \,\, .   \label{velvelgrfg}
\eeqs

\vm

\noindent
Now it is clearly seen that last relation is a transcendental equation for the $\vgrfg$ --
velocity of the FO HS in the RFG. It cannot be solved analytically  for an arbitrary EOS. 
In addition it is necessary to transform it to the fluid rest frame in order  
to complete it with the solution of the hydrodynamical equations for the fluid
\begin{equation} \label{velgrff}
\vgrff = \frac{\vgrfg - \vfrfg}{1 - \vgrfg \,\,\vfrfg}\,\,.
\end{equation}
Fortunately, there exist a simple expression for the square of this velocity,
namely 
\begin{equation}  \label{velvelgrff}
\vgrff^2 =  
\frac{ \lp p_f - \pgtil (\vgrfg)  \rp \lp p_f + \egtil (\vgrfg) \rp }
{\lp \varepsilon_f - \egtil (\vgrfg) \rp \lp \varepsilon_f + \pgtil (\vgrfg) \rp } \,\, , 
\end{equation}
which can be easily understood if one recalls that in the theory 
of relativistic shocks the above relation has a meaning of the shock velocity 
in the  rest frame of the initial fluid.

Equation for the charge density becomes 
\begin{equation} \label{nnfluid}
n_{c.f}^2 = \ngtil^2 (\vgrfg)
\frac{ \lp\,\,p\,_f \,\,\, +  \,\,\,\varepsilon\,_f \,\,\, \rp }
{\lp p_f + \egtil (\vgrfg)  \rp } \cdot
\frac{ \lp \,\,\, \pgtil\, (\,\vgrfg\,\,) \,\,\, + \,\,\,\varepsilon\,_f  \,\,\, \rp }
{\lp \pgtil (\vgrfg) + \egtil (\vgrfg) \rp } \,\,.
\end{equation}
Evidently, it can be cast in the form of usual Taub adiabate \cite{Taub}. 
Together with equations for the shock velocity in RFG and RFF,
Eqs. (\ref{velvelgrfg}) and (\ref{velvelgrff}) respectively, it forms a  
complete system of boundary conditions.

Let me  discuss the boundary
conditions and how to solve these equations together with the hydrodynamic
equations for the fluid.  In what follows I will assume that the
solution of hydrodynamical equations (\ref{five51})  for
the fluid is known in the center of mass frame (hereafter \, CM) for the
whole available space-time volume, and hydrodynamical quantities, for
instance, $T_f, \m_f \,{\rm and}\,\,\, u^\n_{f\, CM} = \lp
1;\vvcm\rp/\sqrt{1 - \vvcm^2} $, are given in each space-time point
$X^\n_{f\, CM}\,\,{\rm with}\,\, \n\, \in \{0;1;2;3\}$\,.  Having this
solution, one can map it into the RFF by the Lorentz transformation
\beqs 
X^i_{\1 F} & = & \frac{1}{\sqrt{1 - \vvcm^2}} \lp X^i_{\, CM} -
X^0_{\, CM} \vicm \rp \,\, , \label{xnewrff} \\
X^0_{\1 F} & = &
\frac{1}{\sqrt{1 - \vvcm^2}} \lp X^0_{\, CM} - X^i_{\, CM} \vicm \rp
\,\,, \label{tnewrff}
\eeqs

\vm
\noindent
with $i\, \in \{1;2;3\}$\,. Then all hydrodynamical quantities are defined in the RFF.

After this transformation into the RFF, the obtained coordinate system
does not necessarily coincide with the original system  which  was used
for the derivation of the shock-like expressions (\ref{velvelgrfg}),
(\ref{velvelgrff}) and (\ref{nnfluid})  at  the boundary between fluid and gas.
Let assume  that in the  RFF the 3-vector of the shock velocity $\vgrff$
is described by the standard set of spherical coordinates with the
angles $\phi$ and $\th$.  Exploiting this fact,
one can derive a differential equation for the FO HS  in the RFF 
 \cite{Bugaev:99, Bugaev:99b}.
Projecting the 3 vector of FO shock velocity  onto the spacial coordinates (\ref{xnewrff}) in the RFF, rewriting them in terms of the standard spherical  coordinates and 
getting rid of spherical  angles, one obtains the following  partial differential equation 
for  $X^{*1} \lp X^2; X^3; X^0\rp $  \cite{Bugaev:99}:
\begin{equation} \label{eqi}
\frac{\partial X^{*1}}{\partial X^0}\bigg|_{s\1 F}^2  
\left[ 
1 + \frac{\partial X^{*1}}{\partial X^3}\bigg|_{s\1 F}^2 \lp 1 +
\frac{\partial X^{*1}}{\partial X^2} \bigg|_{s\1 F}^{-2}  \rp 
\right]
=
\vgrff^2 \,\, \frac{\partial X^{*1}}{\partial X^3}\bigg|_{s\1 F}^2  
\,\,,
\end{equation}
which has to be solved  in the  RFF along  with the  boundary conditions   (\ref{velvelgrfg}),
 (\ref{velvelgrff}) and  (\ref{nnfluid})  and
with the solution of the hydrodynamic equations for the fluid.

To show how to find the solution of this system, let me rewrite it in a more convenient form 
with the help of both EOS:
\beqs
\hspace*{-1.3cm}
F_{nf}\hspace*{-0.1cm}\lp T_f, \m_f, \m_g, \vgrfg \rp
~\equiv~
\frac{n_{c.f}^2 }{ \ngtil^2 (\vgrfg)}
& \hspace*{-0.2cm}- \hspace*{-0.2cm}&
\frac{ \lp\,\,p\,_f \,\,\, +  \,\,\,\varepsilon\,_f \,\,\, \rp }
{\lp p_f + \egtil (\vgrfg)  \rp } \cdot
\frac{ \lp \,\,\, \pgtil\, (\,\vgrfg\,\,) \,\,\, + \,\,\,\varepsilon\,_f  \,\,\, \rp }
{\lp \pgtil (\vgrfg) + \egtil (\vgrfg) \rp } ~=~ 0\,\,,  \label{eqii} \\
F_{\vgrff} \lp T_f, \m_f,  \m_g, \vgrfg, \vgrff \rp
~\equiv~
\vgrff^2 & \hspace*{-0.2cm}- \hspace*{-0.2cm}&
\frac{ \lp p_f - \pgtil (\vgrfg)  \rp \lp p_f + \egtil (\vgrfg) \rp }
{\lp \varepsilon_f - \egtil (\vgrfg) \rp \lp \varepsilon_f + \pgtil (\vgrfg) \rp } 
~= ~0\,\,,  \label{eqiii} \\
F_{\vgrfg} \lp T_f, \m_f,  \m_g, \vgrfg \rp
~\equiv~
\vgrfg^2 & \hspace*{-0.2cm}- \hspace*{-0.2cm}&
\frac{ \lp p_f - \pgtil (\vgrfg)  \rp \lp \varepsilon_f + \pgtil (\vgrfg) \rp }
{\lp \varepsilon_f - \egtil (\vgrfg) \rp \lp p_f + \egtil (\vgrfg) \rp }
~= ~0\,\,, \label{eqiv}
\eeqs

\vm 

\noindent
where  only the most important arguments are shown.

The preceding  analysis can be summarized  in the following 
{\it Theorem 3:} If the solution of the transcendental Eqs. (\ref{eqii}) -- (\ref{eqiv}) exists, 
then the FO HS at  the t.l. boundary between the  fluid and the gas of free particles  is given by the differential equation (\ref{eqi}).

To prove this theorem let me, first, resolve the transcendental equation  (\ref{eqiv}) 
for the solution of  hydrodynamic equations given in the RFF. 
Suppose it exists and is denoted as 
\begin{equation}
\label{soliv}
\vgrfg = \vgrfg \lp T_f \lp X^{*1} \rp , \m_f \lp X^{*1}\rp , \m_g   \rp\,\,.
\end{equation}
Substituting it into the equation of  charge conservation (\ref{eqii}), one finds the relation between the chemical potential
of the  gas of free particles $\mu_g$, the chemical potential of fluid $\mu_f$, and fluid's  temperature $T_f$. 
Assume it can be expressed as follows
\begin{equation}
\label{solii}
\m_g  = \m_g \lp T_f \lp X^{*1} \rp, \m_f\lp X^{*1}\rp  \rp\,\,.
\end{equation}
Inserting  this solution into Eq. (\ref{soliv})  and then substituting the  both obtained equations  
into the expression \mbox{(\ref{eqiii}),} 
one defines the shock velocity $\vgrff \lp T_f \lp X^{*1} \rp, \m_f\lp X^{*1}\rp\rp$ in terms of
the hydrodynamical solution mapped into the RFF. 
The last result completes the differential equation (\ref{eqi})
for the t.l. parts of the FO HS  in the RFF.
The s.l. ones should be obtained in accordance with the traditional 
\CFful\, prescription. 
On the light cones both parts should match and this is a requirement to 
choose the correct root of the transcendental equations for the  
t.l. FO HS.

In the case of zero charge the above consideration simplifies  
because both chemical potentials and 
equation of  charge conservation should be left out.

Thus, the principal way how to find the t.l. FO HS  is described.
Now let me  consider the FO  problem in $1+1$ dimensions.


\subsection{Freeze-out in $1+1$ Dimensional Hydrodynamics.}
According to {\it Theorems 1 {\rm and} 3} one has  to conjugate the
{\it freeze-out shock} with the hydrodynamical  solution for the fluid. 
In 1+1 dimensions the original system of equations (\ref{eqi}) -- (\ref{eqiv}) for the FO HS  
is greatly simplified.
Setting formally $d X^{*2}, d X^{*3} \rightarrow 0$ in Eq. (\ref{eqi}), one 
finds then in the RFF
\begin{equation} \label{swi}
\frac{d X^{*} }{d X^0}\bigg|_{s\1 F}  = 
\vgrff ( T_f \lp X^{*} \rp ), 
\end{equation}

\noi
or, rewriting it in the  CM frame, one gets
\begin{equation} \label{swia}
\frac{d X^{*} }{d X^0}\bigg|_{s\, CM}  = 
\frac{ v_{f\, CM} + \vgrff } { 1 + v_{f\, CM}  \vgrff }    
\end{equation}

\noi
by the relativistic addition of the fluid velocity  in the CM frame.

Dividing Eq. (\ref{eqiv}) by Eq. (\ref{eqiii}), one obtains  
the following important expression after getting rid of the squares 
\begin{equation} \label{swii}
\vgrfg^\pm = \pm 
 \vgrff \frac{  \varepsilon_f + \pgtil (\vgrfg) }{  p_f + \egtil (\vgrfg) } \,\,.
\end{equation}

Substituting  Eq. (\ref{swii}) into Eq. (\ref{eqiv}), one obtains the second equation for velocities 
\begin{equation} \label{swiii}
\vgrff \,\, \vgrfg^\pm   = \pm 
\frac{  p_f - \pgtil (\vgrfg)    }
{ \varepsilon_f - \egtil (\vgrfg)  } \,\,,
\end{equation}

\noindent
where I denote the possible sign values by the corresponding superscript.

For the sake of simplicity I  will consider the fluid and gas without charge.
Then Eq. (\ref{eqii}) becomes an identity.
In what follows I  impose that  
the EOS of the fluid is  $p_f = c_s^2 \s_f T_f^{\frac{1 + c_s^2}{c_s^2}} = c_s^2
\varepsilon_f $,
and gas of free particles has the EOS
of the ideal gas of massless particles $p_g = \frac{\s_g}{3} T_g^4 = \varepsilon_g / 3$.

Such an example is meaningful because intuitively it is  clear that
once the interaction
is not important for the gas and its particles are nearly free, then the  most 
natural choice
for its EOS is the ideal gas one. 
In  order to simplify presentation
I consider the massless gas.
However, this simple choice will demonstrate the conceptually most important physical features of
the FO model.

The same might be valid for the fluid as well, but I would like to
study a more general case of the fluid EOS.  Then, the case when both
EOS are the same is included by the proper choice of the speed of sound
$c_s = \sqrt{\frac{d p}{d \varepsilon}}$ of the fluid.

The effective energy density and the
pressure of \gfp\, can be cast as:
\beqs
\egtil & = & - \varepsilon_g (T^*) \frac{\lp 1 - \vgrfg \rp^2}{4 \vgrfg}\,\,,
\label{swiv} \\
\pgtil & = & \,\,\,\, \varepsilon_g (T^*) \frac{\lp 1 - \vgrfg \rp^2 \lp 2 + \vgrfg \rp}{12 }\,\,,
\label{swv}
\eeqs

\vm
\noindent
where I give the results for the right hemisphere.
In what follows, for definiteness, 
I will consider only a single boundary between the fluid and \gfp,
assuming that at the beginning of the FO process the fluid occupies the left hemisphere,
whereas   the gas  starts to fill  the  right hemisphere.

In this subsection I will use the simplest FO  criterion  $T_g (x, t) = T^* = const$.
Of course,  one could
consider  different freeze-out criteria. For  instance, one could fix the FO criterion  by the constant  value of  energy density in the Landau-Lifshitz frame (see Appendix B of Ref. \cite{Bugaev:99}), or by the constant value of effective energy density  (\ref{swiv}). My choice is aimed to 
simplify  the  presentation, while keeping its essence.

Introducing the new variable
\begin{equation}
R \equiv  \frac{ \left[ \s_f \lp T_f \rp \right]^{\frac{1 + c_s^2}{c_s^2}} }{\s_g\lp T^* \rp^4}  > 0\,\,,
\end{equation}

one can rewrite
\begin{equation} \label{riv}
\vgrfg^\pm = \mp \,\, 2 \vgrff \,\, \frac{1 +  c_s^2}{\vgrff^2 + c_s^2} - 3
\,\,.
\end{equation}

\noindent
For the left hemisphere one has to change only the overall sign in 
the left-hand side of the above equation.

Then I will analyze the solution $\vgrfg^{+}$ only, because other case can 
be obtained by the substitution $\vgrff \rightarrow - \vgrff$.
From the inequality  $|\vgrfg^+| < 1$  the  available range of  
the velocity $\vgrff $  follows
\begin{equation} \label{rv}
- 1 < \vgrff < - c_s^2\,\,.
\end{equation}

\begin{figure}[ht]
\centerline{\hspace*{3.0cm}\epsfig{figure=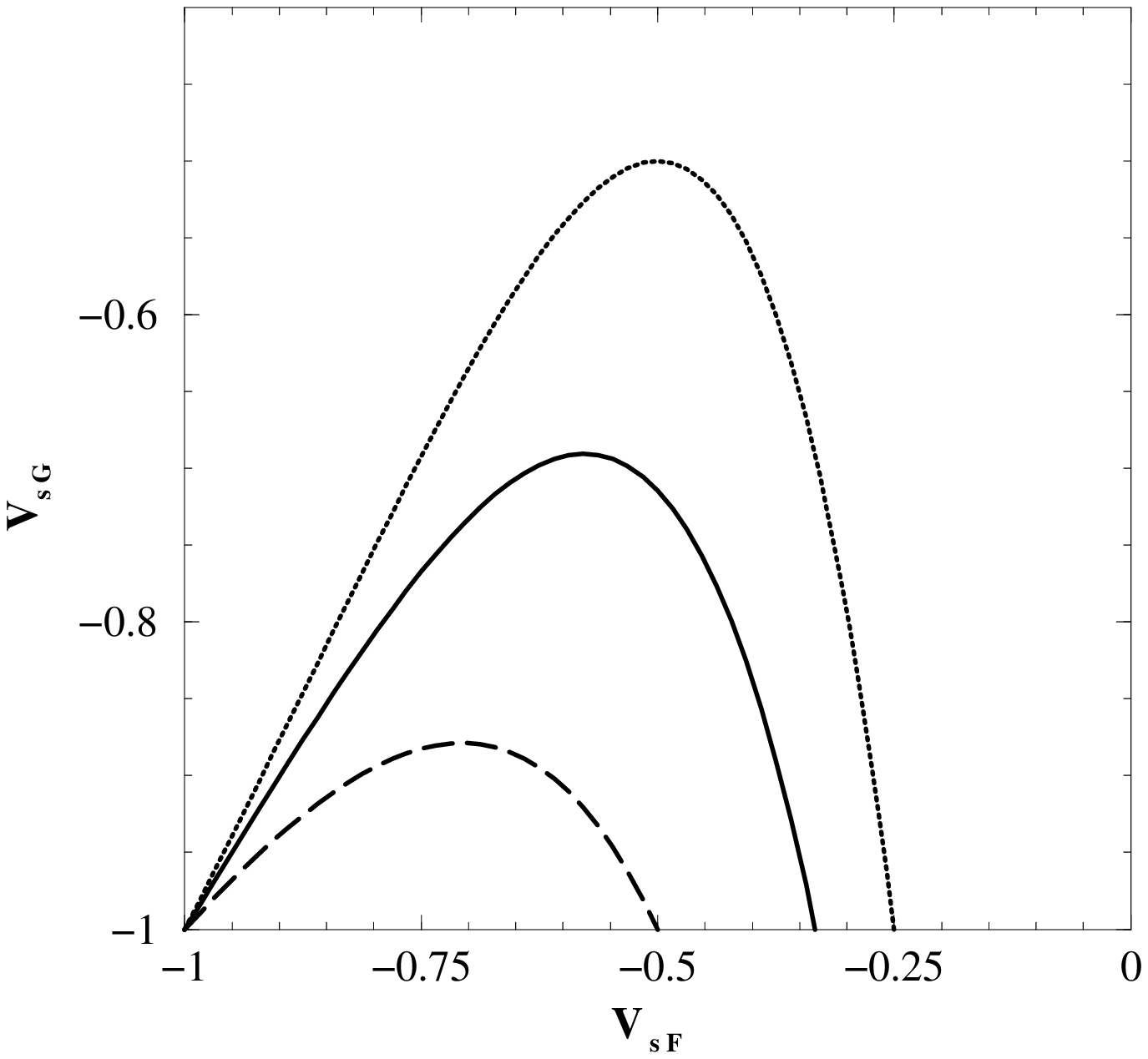,height=10cm,width=12cm} 
\hspace*{-3.4cm}  
\epsfig{figure=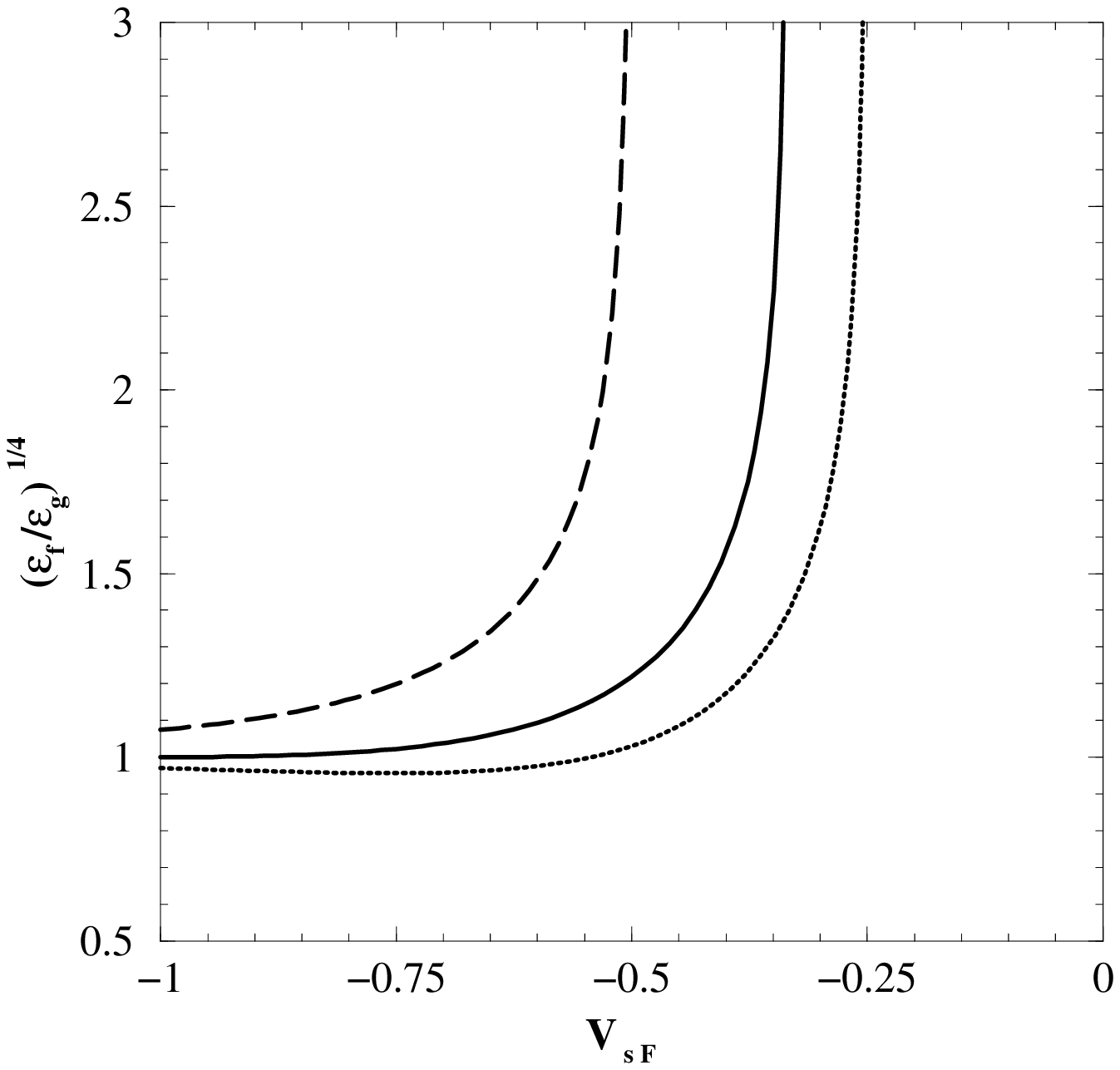,height=10cm,width=12cm}
}

\caption{ \label{Fvsigg}
{\bf Left panel.} Dependence of the FO  shock velocity $\vgrfg$ in the RFG upon the 
the shock velocity $\vgrff$ in the RFF for the three different EOS of fluid.
The values of the fluid's  speed of sound are: 
$c_s = \frac{1}{\sqrt{3}}$ (solid line), 
$c_s = \frac{1}{\sqrt{2}}$ (dashed line)
and
$c_s = \frac{1}{\sqrt{4}}$ (dotted line).
\newline
{\bf Right panel.} Ratio of the energy densities on the both 
sides of the FO shock as the function of  
the shock velocity $\vgrff$ in the RFF for the same EOS of fluid as in 
the left panel.
}
\end{figure}

\noi
The maximal value of  $\vgrfg^+ $ corresponds to $\vgrff = - c_s$
 (see  the left panel of Fig.~\ref{Fvsigg}).

Inequalities (\ref{rv}) show the limiting values of the shock velocity
in the RFF which are derived by the conservation laws and 
relativistic causality condition.
However, the entropy growth condition will give the narrower interval 
for the allowed values of the velocity $\vgrff$ (see below).

After some algebra from Eq. (\ref{rv})  one can  get  the following expression 
for the unknown $R$  \cite{Bugaev:99, Bugaev:99a, Bugaev:99b}:
\begin{equation} \label{rvi}
R^+ =  
\frac{\lp 2 c_s^2  + 2 \vgrff^2 + \vgrff \lp 1 + c_s^2 \rp \rp^2 \lp \vgrff - 1  \rp} 
{ 3 \,\lp \vgrff + c_s^2 \rp \, \lp \vgrff^2 + c_s^2 \rp^2}
\,\,.
\end{equation}

\noindent
Finally, the result for the unknown fluid temperature on the boundary with the gas reads as
\begin{equation} \label{rvii}
T_f = \left[
\frac{ \s_g \lp T^*\rp ^4 
\lp 2 c_s^2  + 2 \vgrff^2 + \vgrff \lp 1 + c_s^2 \rp \rp^2 \lp \vgrff - 1  \rp
}
{3\, \s_f\,\,
\lp \vgrff + c_s^2 \rp \, \lp \vgrff^2 + c_s^2 \rp^2
}
\right]^{\frac{c_s^2}{1 + c_s^2}}
\,\,.
\end{equation}

The formal solution of the freeze-out problem
in 1+1 dimensions
follows from the last equation: 
solving it for $\vgrff (T_f (X^*))$ 
and integrating Eq. (\ref{swia}) with the known hydrodynamical solution for the fluid,
one finds the desired answer.

From Eq. (\ref{rvii}) it is seen that FO criteria $T_g = T^* = const$ and $T_f = T^* = const$ are 
not equivalent in general. 
The only exception is when the FO HS is a straight line in $(X^0,X)$ plane. 
This occurs in the  FO  process of the simple wave 
(see later).
The criterion $T_f = T^*$ looks technically simpler because 
the time derivative of the FO HS
$\vgrff (T_F (X^*))$ 
is defined by the hydrodynamical solution for the fluid. 
Then one has to find the gas temperature.
However, it might be that
under a ''bad choice'' of the FO criterion,  the gas temperature may drop  too low 
(in the limit $\vgrff (T_F (X^*)) \rightarrow - c_s^2$ it follows
$T_g \rightarrow 0$) for the  applicability of both thermodynamics and  
hydrodynamics.

This fact is a reflection of the relativistic causality which tells us
that the pre FO fluid cannot ``know'' that in the next moment is has to freeze-out into the gas of noninteracting particles. The information about the necessity
to freeze-out the particle distributions is ``stored'' in the local density and
cross-sections of the post FO state (gas), but there is no physical agent 
to  ``inform'' the pre FO fluid about this because the agent should move   backwards in time. 

Dependence of the fluid temperature  on $\vgrff$\, for $c_s^2 = \frac{1}{3}$ is presented
in Fig.~\ref{Fvsigg}.  From the right panel of  Fig.~\ref{Fvsigg} it is seen that
the fluid temperature is always larger than the gas one.  This fact is
born in a more general statement, namely the energy density of the
fluid $\varepsilon_f(T_f)$ always exceeds the gas one $\varepsilon_g(T^*)$ 
 one for $c_s^2 \ge \frac{1}{3}$ \cite{Bugaev:99, Bugaev:99b}.

\subsection{Entropy Production in the Freeze-out Shock.}
Next I would like to study the problem of the thermodynamical
stability, or in other words
the entropy production in the FO shock. 
This question  is of a special interest for me  because it 
clearly demonstrates  that the FO shock is a new kind of the discontinuity in which 
entropy increases in the rarefaction transition for thermodynamically 
normal media \cite{normalmed}.

The fluid entropy flux through the FO HS in an  arbitrary Lorentz frame is given by 
\beqs
s_f^{\m}   & = & s_f u^\m_f\,\,, \\
{\cal S}_f & = & \int\limits_{\SF} d \s_\m s_f^{\m} \,\,.
\eeqs

\noi
With the help of the previous section the entropy flux in RFG  can be written 
in the general form
\begin{equation}
s_f^{\m} n_\m  = s_f 
\lp \frac{ \lp p_f - \pgtil (\vgrfg) \rp \lp \egtil (\vgrfg) + \pgtil (\vgrfg) \rp }
{ \lp \varepsilon_f - \egtil (\vgrfg) \rp \lp p_f + \varepsilon_f \rp } \rp^{ \frac{1}{2}} \,\,, 
\end{equation}

\noi
where the normal vector $n_\m$ acquires the  form
$n_\m = (- \vgrfg ; 1) $  in  1 + 1 dimensions.
 
The corresponding expression for \gfp has to be found from the \COful distribution 
function.
In the RFG it is evident that the entropy of outgoing particles is accounted
by the change of the momentum integration volume $d^3 p \rightarrow \Th(p_\rho d \sigma^\rho ) d^3 p$
in the standard expressions for both classical and quantum statistics. 
For the Boltzmann  distribution function $\phi$ \cite{groot}, for example, 
it yields
\beqs
s_g^{\m} & = & \int \frac{ d^3 \1 p}{p^0} p^\m\,\, \phi\,\, [ \,\,1 - \ln \phi \,\,]\,\, \Th(p_\rho d \sigma^\rho ) \,\,, \\
{\cal S}_g & = & \int\limits_{\SF} d \s_\m\,\, s_g^{\m} \,\,.
\eeqs

From the expressions for the entropy flux of the massless gas
\beqs
s^\n_{g} (\vgrfg, T) & = & s_g ( T^* ) 
\lp \, \frac{1 - \vgrfg}{2} ; \, \frac{1 - \vgrfg^2}{4} \rp \, \,,\\
{\cal S}_g & = & \int\limits_{\SF} d X^0 \,\,n_\m \,\, s_g^{\m} \,\,,
\eeqs

\vm
\noi
where $s_g ( T^* )$ is the entropy density of the gas at the freeze-out temperature,
one can find  the ratio of the entropy flux  on the both sides of the FO shock:
\begin{equation}
P_s = \lp \frac{ s^\m_{g} n_\m }{ s^\m_f n_\m } \rp^4\,\,,
\end{equation}

\noi
which becomes
\begin{equation}
P_s^+ \bigg|_{ c_s^2 = \frac{1}{3}}  =  
 \frac{ \s_g \,\, (\,\, 3 \,\,\vgrff^2\,\, +\,\, 1\,\,)^2\,\, (\,\, 3\,\, \vgrff\,\, +\,\, 1\,\,) } 
{16 \s_f \vgrff^4 ( \vgrff - 1) }\,\,
\end{equation}

\noi
for the solution $\vgrff^+$ with $c_s^2 = \frac{1}{3}$.

\begin{figure}[ht]
\centerline{\hspace*{3.0cm}\epsfig{figure=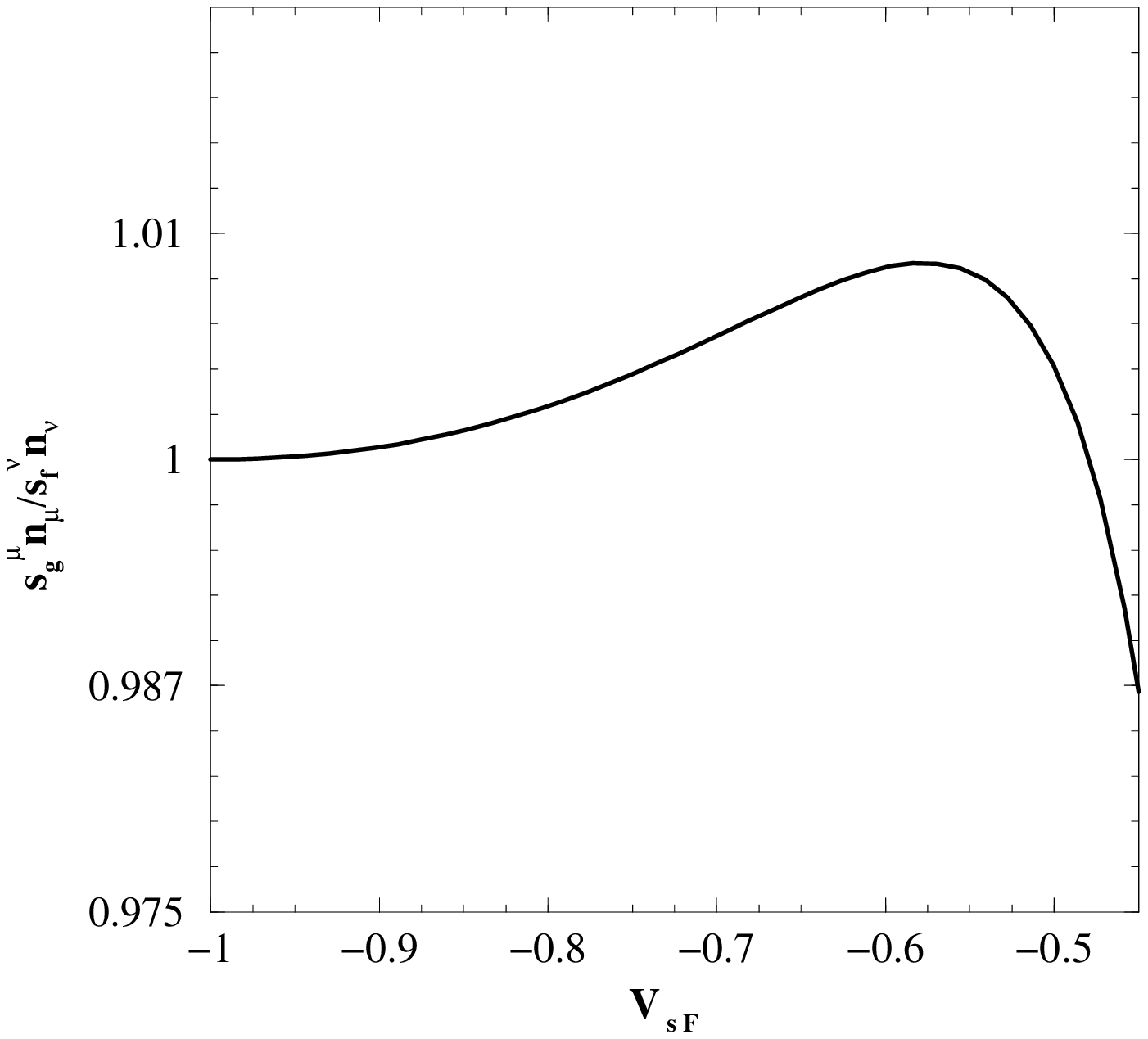,height=10cm,width=12cm} 
\hspace*{-3.4cm}  
\epsfig{figure=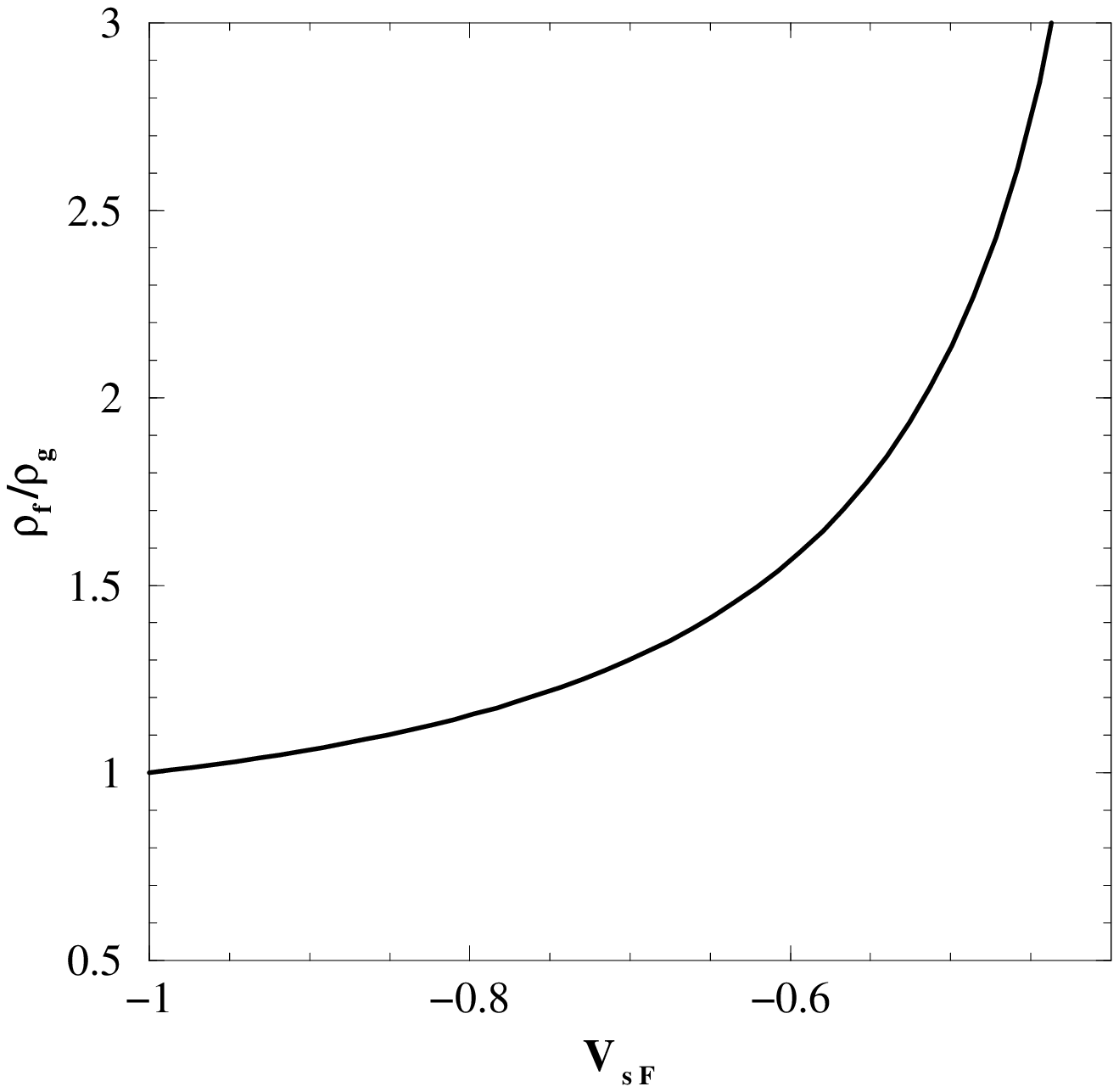,height=10cm,width=12cm}
}

\caption{ \label{Frentr103}
Ratio of the entropies (left panel) and particle densities (right panel) on the both
sides of the FO shock  as a function of
the shock velocity $\vgrff$ in the RFF  
for the fluid velocity of sound $c_s = \frac{1}{\sqrt{3}}$. 
In the left panel the states to the right of the maximum are mechanically unstable (see text).
}
\end{figure}

The left panel of Fig.~\ref{Frentr103}  shows the limiting values of  $\vgrff$
obtained by the  thermodynamical stability criterion
\begin{equation} \label{psiii}
P_s^+ \bigg|_{ c_s^2 = \frac{1}{3}} \ge 1 \,\,\,\, \Rightarrow \,\,\,\,  
- 1 < \vgrff \le - 0.479 \,\,\,\,{\rm for} \,\,\,\, \s_g = \s_f. 
\end{equation}

The maximal value of the entropy production in the FO shock for the same EOS of the fluid is 
\begin{equation}
max \lp P_s^+ \bigg|_{ c_s^2 = \frac{1}{3}} \rp^{ \frac{1}{4}} \approx 
1.01088 \lp \frac{\s_g}{ \s_f} \rp^{\frac{1}{4}}\,\,,
\end{equation}

\noi
i.e., is about one percent for the same number of the internal degrees of freedom 
of  fluid and gas.

Another problem of stability  is the mechanical stability (see, for example, Refs. \cite{rozhdyan, bugaevetal2, bugaevetal1, bugaevetal1b} 
and references therein)
of the {\it FO shock}.
It is of  crucial importance for the FO process
because it is related to the recoil problem of the FO process on the
fluid expansion. 
Usually one might argue that the FO of the small fluid element would affect 
the hydrodynamical solution in the whole future cone of this element.

However, it is known \cite{rozhdyan, bugaevetal2, bugaevetal1, bugaevetal1b}  that in
rest frame of  the thermodynamically normal matter 
the perturbations of  small amplitudes propagate with
the speed of sound  and the rarefaction shocks 
propagate with the subsonic velocity.  
Since the FO process corresponds  to a decompression (the gas  is more dilute than the fluid),
it follows that  in a normal media near the FO state
the information about the emission of particles from the FO HS can be transmitted 
interior fluid  by a simple wave or
a rarefaction shock, whereas the compression shock is ruled out.
Therefore, 
one concludes that  a  FO shock with  the shock velocity in the following range 
\begin{equation} \label{psiv}
- 1 \,\, <\,\, \vgrff^+ \,\,\le - \,\,c_s 
\end{equation}
is also  mechanically stable with respect to the small perturbations of the fluid state 
\cite{Bugaev:99a} (c.f. the left panel of  Fig.~\ref{Frentr103} ).
Thus, we  proved the {\it Statement 3}: For the EOS under consideration
the perturbations of the fluid state at the t.l. FO HS are slower  than
the supersonic { FO shock}
and, hence, they do not affect the hydrodynamical evolution of the fluid inside the FO HS.

Now one can see that, in contrast to the usual approach,  
the recoil problem is resolved  for the supersonic velocities
of the FO shock in the RFF.
This is one of the main goals of the suggested FO scheme.
However, an investigation of the mechanical stability of the FO shock  in full 
requires a special  consideration.

It is now important  to verify the validity of the adopted approximations 
 comparing of the mean free path $\l$ in the fluid and in \gfp.
Since $\l$ contains a cross-section, I would like  to study the ratio of the 
particle density on both sides of the FO shock.
Such an estimation gives the lower limit of applicability  because of the inequality
\begin{equation}
R_\r \bigg|_{ c_s^2 = \frac{1}{3}} = \frac{\rho_f}{\rho_g} \le \frac{\l_g}{\l_f}. 
\end{equation}

\noi
Substituting the densities found in the corresponding rest frames (for \gfp\, 
there used the results obtained in the Eckart frame \cite{groot}),
one gets  
\begin{equation}
R^+_\r \bigg|_{ c_s^2 = \frac{1}{3}} = 
\lp \frac{\s_f ( \vgrff - 1)^3 (3 \vgrff^2 + 1)^2}
{16\,\,\s_g \,\,(\,\,3\,\,\vgrff\,\, +\,\, 1\,\,)^3\,\, \vgrff^2} \rp^{\frac{1}{4}}
\end{equation}

\noi
for $ c_s^2 = \frac{1}{3}$.
It is easy to prove that for the same EOS with the same number of degrees 
of freedom the particle density in the fluid is larger than in \gfp, i.e., 
the following inequalities hold (see also the right panel of Fig.~\ref{Frentr103})
\begin{equation}
\frac{\l_g}{\l_f}\,\,\,\ge\,\, R^+_\r \bigg|_{ c_s^2 = \frac{1}{3}}\,\,\ge\,\,1\,\,.
\end{equation}

\noi
It would be a problem of internal consistency otherwise,
because the fluid with the larger mean free path should freeze-out
before \gfp\, and, hence,  the whole consideration would become 
questionable.

Now it is clear that a very large mean free path in \gfp\,
might lead to the reduction of the collision rate and, hence, 
the usage of the \COful\, equilibrium distribution function 
would not be justified.
However, such a region of the velocity $\vgrff \rightarrow - c_s^2$ is 
not allowed by the thermodynamical stability condition (c. f. Eq. (\ref{psiii}) ). 
Thus, it is shown that the considered freeze-out scheme in 1 + 1 dimensions does not 
have internal contradictions.

\subsection{Freeze-out of the Simple Wave.}
Let me  consider the FO of the semi-infinite homogeneous normal
matter without charge, occupying the left hemisphere in 1+1
dimensions.  Then the hydrodynamical solution for the fluid is known
-- it is a simple wave \cite{llhydro}. This simple isentropic flow
describes the propagation of the perturbations with small amplitudes
in the thermodynamically normal media \cite{normalmed}.  The
characteristics of the simple wave and, therefore, its isotherms are
just straight lines in the space-time variables originating at the initial position of the
boundary with the vacuum.  It is important to remind also that
characteristics are the t.l.  HS  and, therefore, are of
a special interest.  A short and clear description of the simple wave
can be found in the Appendix A of  Ref. \cite{bugaevetal2}.

Adopting the EOS of the previous  subsections one  can apply their  results
straightforwardly  to describe the FO of the simple wave.
Making the consideration  in the RFF, I  conclude that the shock velocity
in this frame, namely $\vgrff$, has to be equal to the velocity of the simple
wave there, i.e., to the velocity of sound $c_s$.
Since both waves move to the left-hand side in RFF, one can write
\begin{equation} \label{ci}
\vgrff = - c_s  \,\,.
\end{equation}

\noi
Consequently, the FO of the simple wave corresponds to a particular 
choice of the FO shock velocity
which one has to substitute in all necessary formulae of  
the preceding  subsection.

Thus, substitution of  Eq. (\ref{ci}) into Eq.  (\ref{riv}) yields 
\begin{equation} \label{cii}
\vgrfg^\pm = \pm \frac{1 +  c_s^2}{c_s} - 3
\,\,.
\end{equation}

\noi 
The latter equation
leads to the restrictions on the fluid EOS.
Indeed, requiring the validity of the inequality $|\vgrfg^+| < 1$,
one gets the lower limit of the velocity of sound in the fluid, i.e.,
\begin{equation} \label{ciii}
c_s^- \equiv 2 - \sqrt{3} \,\,  <  \,\, c_s \,\, < \,\, 1 < \,\,  c_s^+  \equiv 2 + \sqrt{3}
\,\,.
\end{equation}

\noi
The lower boundary of the velocity of sound corresponds to the extremely soft EOS
with  $ c_s^- \approx 0.26795$ or with the power of the temperature
in the expression for the pressure being $\frac{1 + c_s^2}{c_s^2} \approx  15 $.

The ratio of the energy densities becomes
\begin{equation} \label{civ}
R^+ \bigg|_{ S.W.} =
\frac{\lp 1 - 4 c_s + c_s^2 \rp^2 \lp 1 + c_s \rp}{12\,c_s^3\, \lp 1 - c_s \rp}
\,\,,
\end{equation}

\noi
and the temperature of the fluid in the simple wave reduces to
the expression
\begin{equation} \label{cv}
T_f \bigg|_{ S.W.} = \left[
\frac{ \s_g \lp T^*\rp ^4 \lp 1 - 4 c_s + c_s^2 \rp^2 \lp 1 + c_s \rp}
{12\, \s_f\,\,c_s^3\, \lp 1 - c_s \rp}
\right]^{\frac{c_s^2}{1 + c_s^2}}
\,\,.
\end{equation}

In order to illustrate the scale of quantities let me study the case
$c_s = \frac{1}{\sqrt{3}} \approx 0.57735 $.  Then the shock velocity in
the RFG is $\vgrfg^+ = \frac{4}{\sqrt{3}} - 3 \approx - 0.6906$,
and the ratio $R^+$ is  $R^+\bigg|_{ S.W.} =
\frac{8}{3\,\,\sqrt{3}} \approx 1.5396 $.  The freeze-out temperature
of the fluid in the simple wave is
\begin{equation} \label{rviii}
T_f \bigg|_{ S.W.} = \left[ \frac{ 8 \,\,\s_g \,\, } {3 \s_f\,\, \sqrt{3}}
\right]^{\frac{1}{4}} T^* \approx 
1.1139
\left[ \frac{ \s_g  } { \s_f } \right]^{\frac{1}{4}} T^*
\,\,.
\end{equation}

If the fluid and \gfp\, possess the same
number of the degrees of freedom,
then the fluid temperature exceeds the freeze-out one only by 11 percent.
Thus, no dramatical difference for the temperatures should be expected
when both EOS are same.

The entropy growth in the simple wave corresponds to the maximal possible value 
\begin{equation}
\lp P_s^+ \rp^{ \frac{1}{4}} \bigg|_{ S.W.} = max \lp P_s^+ \bigg|_{ c_s^2 = \frac{1}{3}} \rp^{ \frac{1}{4}} \approx
1.01088 \lp \frac{\s_g}{ \s_f} \rp^{\frac{1}{4}}\,\,.
\end{equation}

\noi
The ratio of the mean free path on the both sides of FO shock satisfies  the inequality 
\begin{equation}
\frac{\l_g}{\l_f}\bigg|_{ S.W.}\,\,\,\ge\,\, R^+_\r \bigg|_{ S.W.}\,\,\approx\,\,1.655\,
\lp \frac{\s_f}{ \s_g} \rp^{\frac{1}{4}}\,\,.
\end{equation}

\noi
Thus, the FO scheme of the simple wave is thermodynamically stable 
and  leads to the increase of the mean free path in \gfp\,
compared to the fluid.


The described  FO scheme of  the simple wave should  be used
in  relativistic nuclear collisions with more realistic EOS. 
However, for illustrative purpose I will analyze the FO of 
the expanding  pion rich matter with Landau initial conditions.
ALready this example will demostrate us
the reduction of the emission volume in comparison with the standard
\CFful\, FO picture.  This is so because the same FO temperature 
of the emitted particles  in the picture with FO shock
and without it corresponds to the different energy densities of the
pre FO fluid in the simple wave. In the FO shock the pre FO fluid temperature is 
higher than the FO  temperature,
whereas in the traditional approach they are equal. It follows now
that for the same initial condition the fluid velocity prior  the
FO is smaller for the larger energy density, i.e.,
in the picture with  the FO shock.

Due to this fact one should expect some reduction of the emisssion volume
for  the same expansion time for the suggested FO of the simple wave in comparison with the traditional \CFful estimations.
This is a clear indication of the "influence effect" of the particle emission
on the evolution of the fluid in the simple wave.

Let me now find
the spectra of massless particles (bosons to be specific) emitted by the
simple wave within the \COful\, FO scheme and compare it with
corresponding result obtained by the traditional \CFful\, one.
Adopting the standard  cylindric geometry
(X-axis is longitudinal) in momentum space, one gets the following
expression for the invariant spectra in the  CM  frame (after the angular integration)
\begin{equation} \label{myspectrumi}
\frac{d N^{\COful} }{p_t d p_t d y_{\, CM} d S_{\perp} d t} =
\frac{ E_t \cosh \lp y_{\, CM} \rp \lp \tanh \lp y_{\, CM} \rp  - v_\s \rp
\Theta \lp \tanh \lp y_{\, CM} \rp  - v_\s \rp
}
{ (2 \pi)^2 \lp e^{\frac{  E_t \cosh \lp y_{\, CM} -  \h_G \rp - \m_c }{T^*} }- 1 \rp } \,\,,
\end{equation}

\noi
where $E_t = \sqrt{p_t^2 + m^2}$ is the transverse energy of the particle,
$p_t$ is its transverse  momentum,
$y_{\, CM}$ is the  CM  particle rapidity, $\h_G$ is the rapidity of the RFG in the  CM  frame,
and $v_\s$ is a velocity of the {\it FO shock} in the CM frame.
For simplicity I consider only one internal degree of freedom.

For the massless particles without charge ($\m_c = 0$) it is convenient to introduce the dimensionless
variable $\ptt = \frac{p_t}{T^*}$. Then Eq. (\ref{myspectrumi}) becomes
\begin{equation} \label{myspectrumii}
T^{*-3}
\frac{d N^{\COful} }{  \ptt d \ptt d y_{\, CM} d S_{\perp} d t} =
\frac{ \ptt \cosh \lp y_{\, CM} \rp \lp \tanh \lp y_{\, CM} \rp  - v_\s \rp
\Theta \lp \tanh \lp y_{\, CM} \rp  - v_\s \rp
}
{ (2 \pi)^2 \lp e^{ \ptt \cosh \lp y_{\, CM} -  \h_G \rp  } - 1 \rp } \,\,.
\end{equation}

Let me assume that initial fluid has the temperature $\tin$.
Then the velocity $v_f$ and the temperature $T_f$ of  the fluid in the simple
wave
are related through the expression
\begin{equation} \label{myspectrumiii}
v_f \lp T_f \rp  = \tanh\lp c_s^{-1} \ln \lp \frac{\tin}{T_f} \rp \rp \,\,.
\end{equation}

The {\it FO shock} velocity in the  CM  frame is given by the formula
\begin{equation} \label{myspectrumiv}
v_\s = \frac{ v_f \lp T_f \rp  - c_s }{1 - c_s v_f \lp T_f \rp} \,\,,
\end{equation}

\noi
where the fluid temperature is taken on the FO boundary with the gas  and it is defined by
Eq. (\ref{rvii}).

For  $\vgrff = - c_s$   one finds  the  RFG velocity  in the  CM  frame 
\cite{Bugaev:99, Bugaev:99b}:
\begin{equation} \label{myspectrumv}
\tanh \lp \h_G \rp = \frac{ v_f \lp T_f \rp-\vfrfg^{+}(-c_s )}{1 - \vfrfg^{+}(-c_s ) v_f \lp T_f \rp}
\,\,.
\end{equation}

\begin{figure}[ht]
\centerline{\hspace*{2.8cm}\epsfig{figure=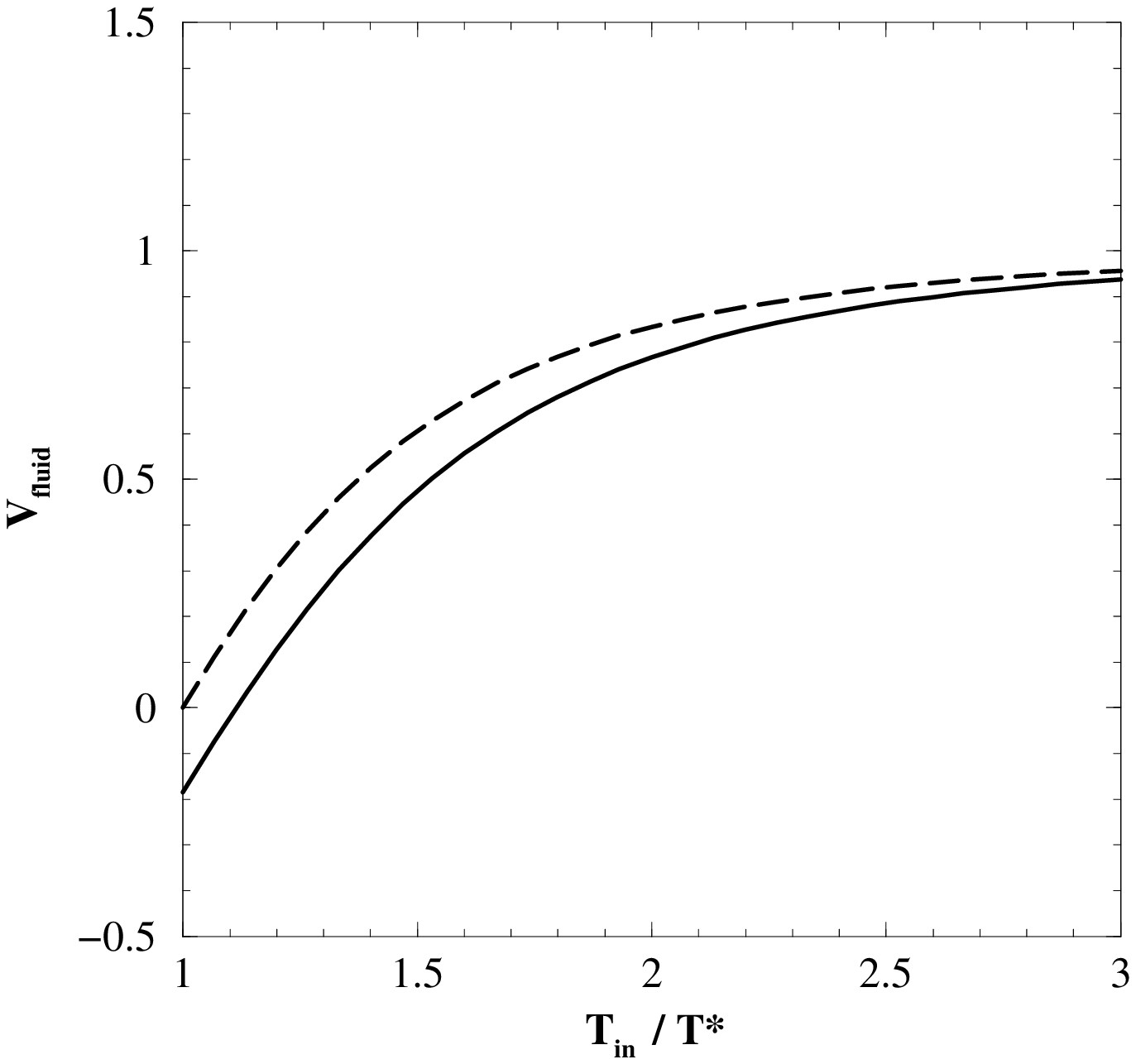,height=10cm,width=12cm} 
\hspace*{-3.4cm}  
\epsfig{figure=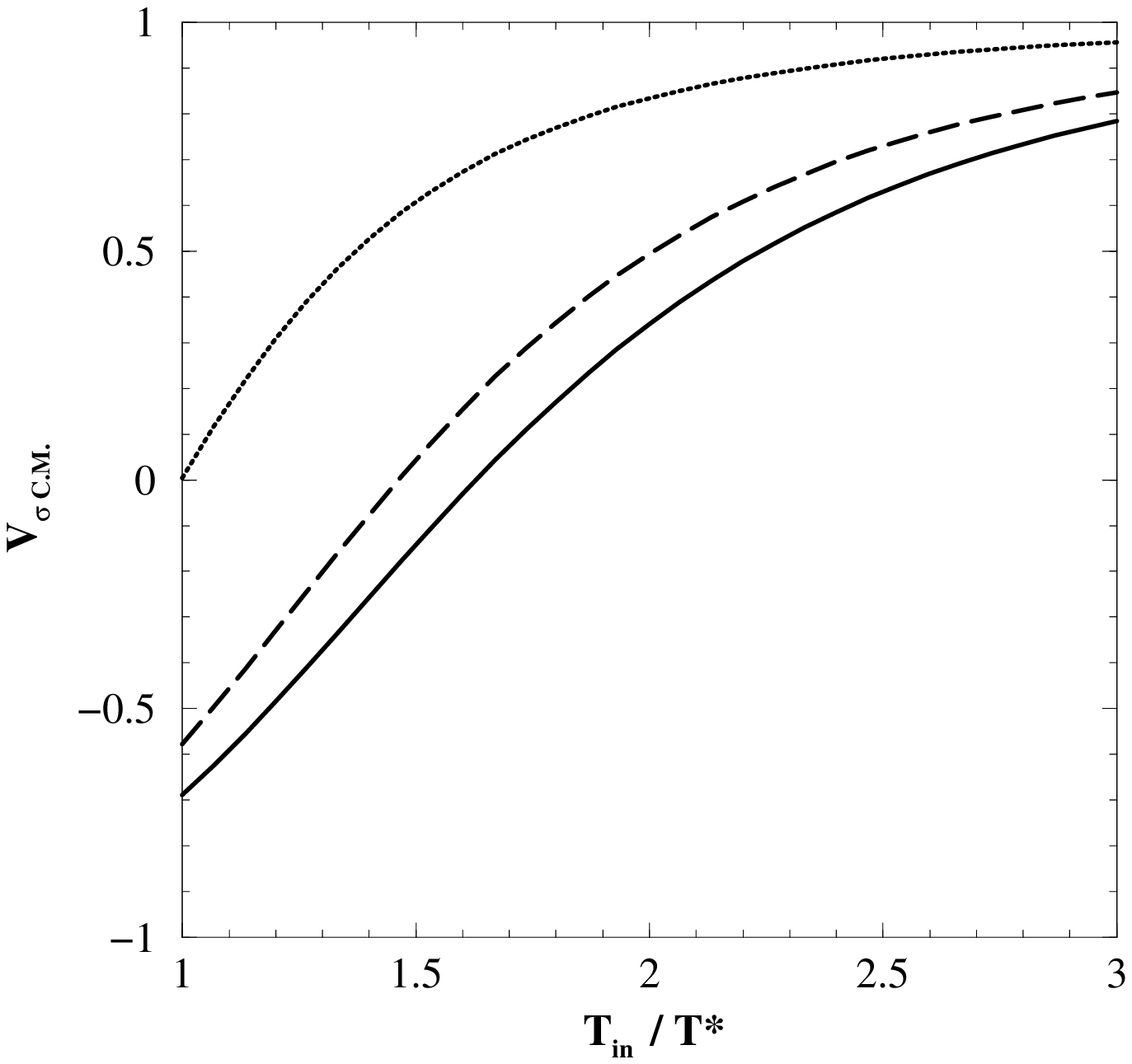,height=10cm,width=12cm}
}

\caption{ \label{Velsw}
{\bf Left panel.} 
Fluid velocity $v_f$ in the simple wave at the FO HS as a function of the initial temperature of the fluid $T_{in}$.
The dashed line corresponds to the \CFful\, FO scheme,
and the solid one corresponds to the \COful\, FO scheme.
\newline
{\bf Right panel.} 
Velocity of the FO boundary $v_{\s}$
in the CM frame
as a function of the initial temperature of the fluid $T_{in}$.
The dashed line corresponds to the \CFful\, FO scheme,
and the solid one corresponds to the \COful\, FO scheme.
The dotted line represents the velocity of the RFG
in the CM frame.
}
\end{figure}

\noi
Fig.~\ref{Velsw} shows $v_f$ and $v_\sigma$  velocities for the fluid EOS $c_s^2 = \frac{1}{3}$ as a function
of the initial temperature $\tin$.
For such an EOS the FO temperature of the fluid is given by Eq. (\ref{rviii}).
Assuming that fluid and  gas have the same number of degrees of freedom, one
obtains
\beqs
v_f \bigg|_{c_s^2 = \frac{1}{3}}^{\COful}
& \approx & \tanh\lp \sqrt{3}\,\, \ln \lp \frac{\tin}{1.1139 T^*} \rp \rp
\,\,, 
\label{myspectrumvib} \\
\vfrfg^{+} \bigg|_{c_s^2 = \frac{1}{3}}^{\COful} & \approx & -.18835 \,\,. \label{myspectrumvii}
\eeqs

The traditional FO scheme  based on the \CFful formula has several important differences.
First of all, spectrum does not contain  the  cut-off $\Theta$-function
\begin{equation}  \label{cfspectrumi}
T^{*-3}
\frac{d N^{\CFful} }{  \ptt d \ptt d y_{\, CM} d S_{\perp} d t} =
\frac{ \ptt \cosh \lp y_{\, CM} \rp \lp \tanh \lp y_{\, CM} \rp  - v_\s \rp
}
{ (2 \pi)^2 \lp e^{ \ptt \cosh \lp y_{\, CM} -  \h_G \rp  } - 1 \rp } \,\,.
\end{equation}

\noi
Second, the FO fluid temperature coincides with the gas one, i.e., $T_f = T^*$.
Thus, the FO in the traditional scheme happens at smaller energy density and at
larger velocity of the fluid in comparison with the one  explained above.
And, third, there is no difference between the rest frame of fluid and the frame of its decay.
Therefore, the fluid velocity and the relative velocity of the emitted particles read as 
\beqs
v_f \bigg|_{c_s^2 = \frac{1}{3}}^{\CFful}
& = & \tanh\lp \sqrt{3}\,\, \ln \lp \frac{\tin}{T^*} \rp \rp
\,\,, 
\label{myspectrumvi} \\
\vfrfg^{+} \bigg|_{c_s^2 = \frac{1}{3}}^{\CFful} & = & 0 \,\,. \label{cfspectrumii}
\eeqs

Now it is easy to see that for the same initial  temperature of the fluid the
hydrodynamic motion of the fluid in the \CFful  FO scheme is more developed 
(see Fig.~\ref{Velsw} for details), i.e.,
\begin{equation} \label{myspectrumviii}
v_\s \bigg|_{c_s^2 = \frac{1}{3}}^{\CFful} \lp \tin \rp
 >  v_\s \bigg|_{c_s^2 = \frac{1}{3}}^{\COful} \lp \tin \rp
\,\,, 
\end{equation}


\noi
whereas the  CM rapidities are nearly equal
\begin{equation} \label{myspectrumviv}
\h_G \bigg|_{c_s^2 = \frac{1}{3}}^{\CFful} \lp \tin \rp
 \leq  \h_G \bigg|_{c_s^2 = \frac{1}{3}}^{\COful} \lp \tin \rp \,\,.
\end{equation}

\vspace*{-0.5cm}

\begin{figure}[ht]
\centerline{\hspace*{2.8cm}\epsfig{figure=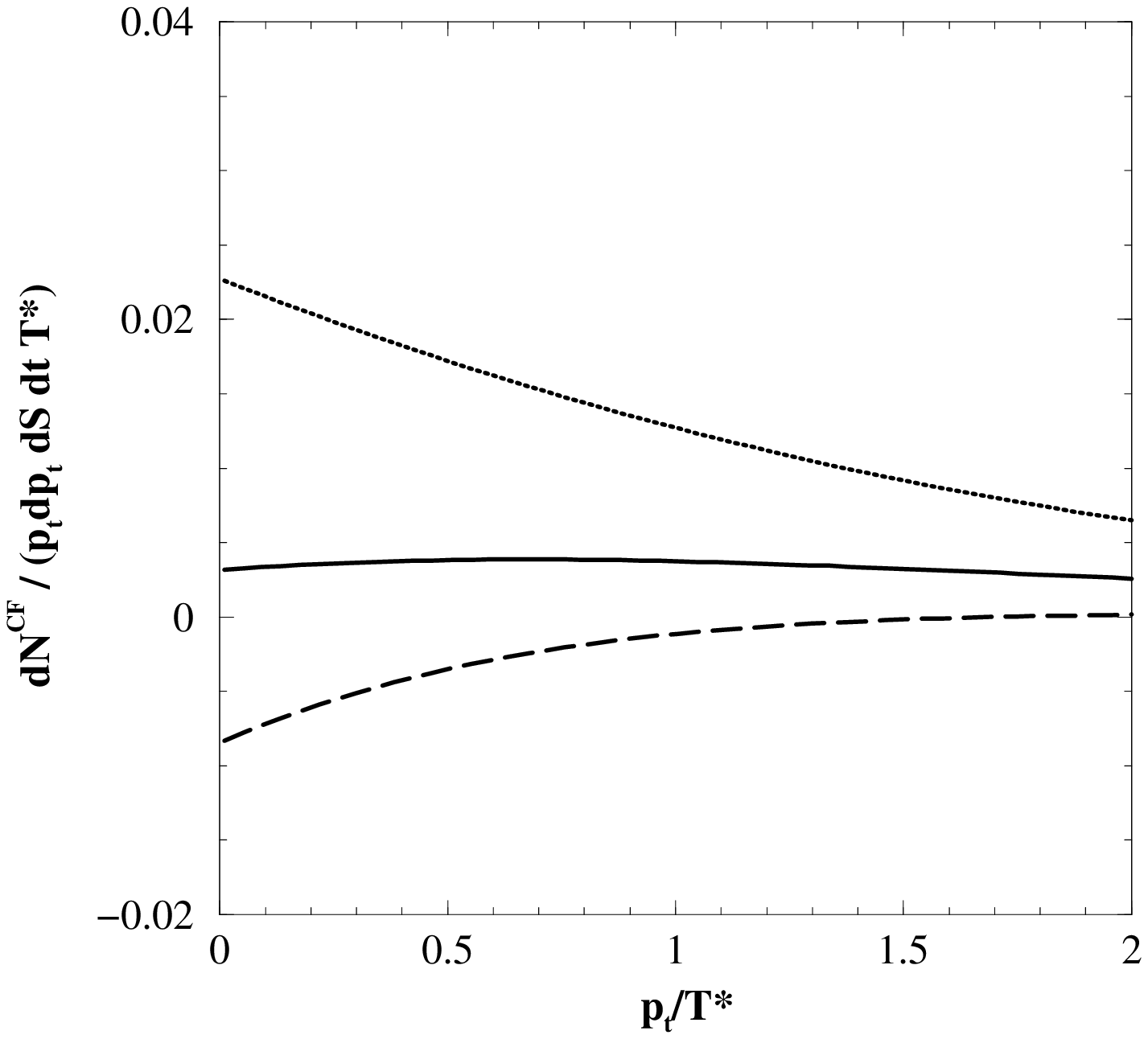,height=10cm,width=12cm} 
\hspace*{-3.4cm}\epsfig{figure=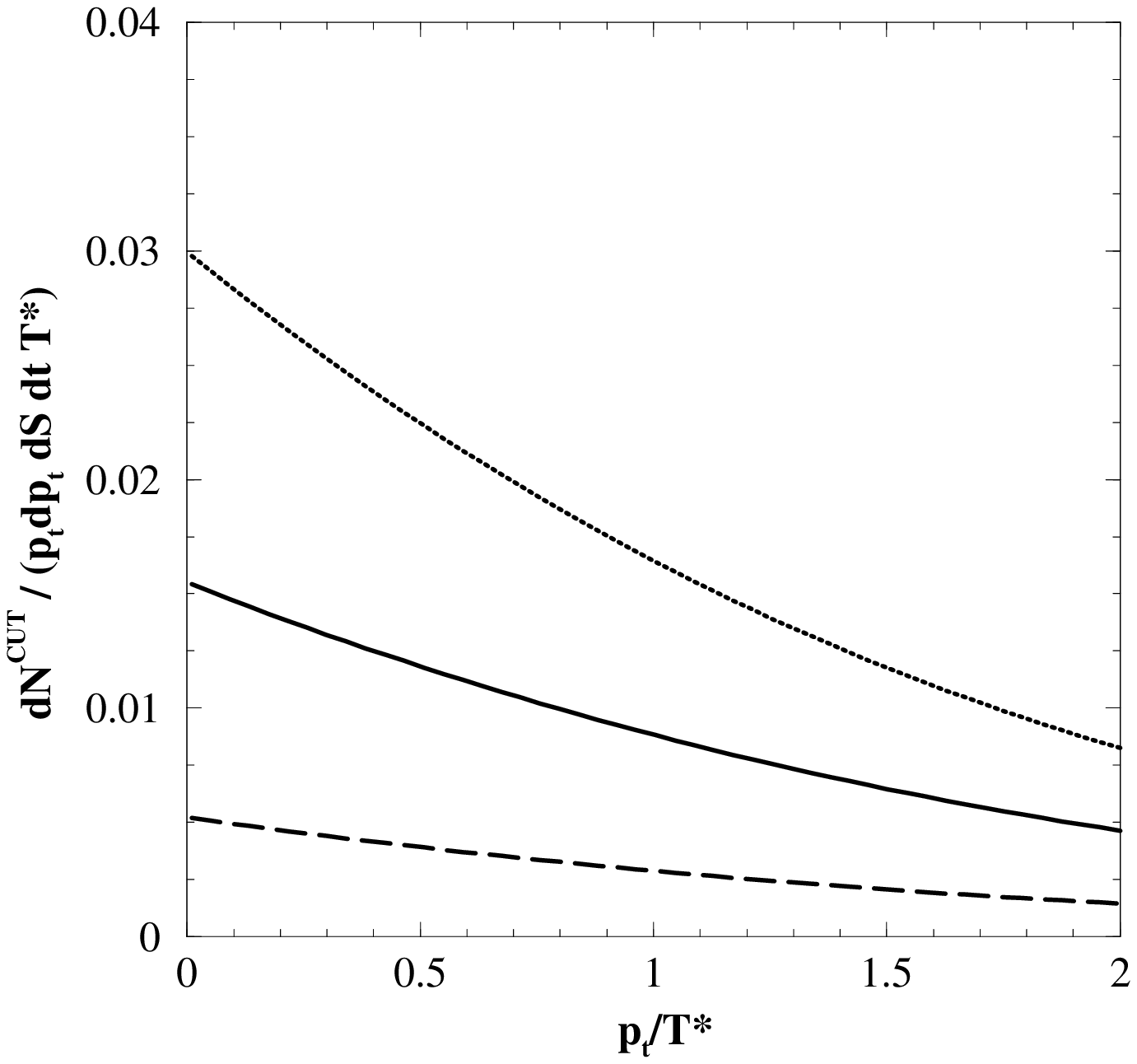,height=10cm,width=12cm}
}

\caption{ \label{PTspec1}
{\bf Left panel.} 
Momentum distribution function  of the \CFful\, FO scheme
integrated over the CM rapidity $y_{CM} \in [-1; \,\,1\,\,]$.
The spectra are found for three values of
the initial temperature of the fluid:
$T_{in} = 1.1 T^*$ (dotted line),
$T_{in} = 1.5 T^*$ (solid line),
and
$T_{in} = 1.9 T^*$ (dashed line).
\newline
{\bf Right panel.} 
Same as in the left panel, but for the \COful\,s FO scheme.
}
\end{figure}

\noi
The first inequality above leads to the important consequences: 
for the \COful  FO scheme
(i) the  energy emission  per unit time from the FO HS is larger,  
and (ii)
the C.M. rapidity interval with the positive values of the 
particle spectra is more broad.
Inequality (\ref{myspectrumviv}) has a negligible effect 
on the difference of the \COful and the  \CFful  results. 
Therefore, one immediately gets the following inequality
\begin{equation} \label{myspectrumviiv}
\frac{d N^{\COful} }{  \ptt d \ptt d y_{\, CM} d S_{\perp} d t} 
\,\, \geq \,\,
\frac{d N^{\CFful} }{  \ptt d \ptt d y_{\, CM} d S_{\perp} d t}\,\,,
\end{equation}

\noi
which, evidently, holds for the spectra integrated  over rapidity or transversal momentum. 

\begin{figure}[ht]
\centerline{\hspace*{2.8cm}\epsfig{figure=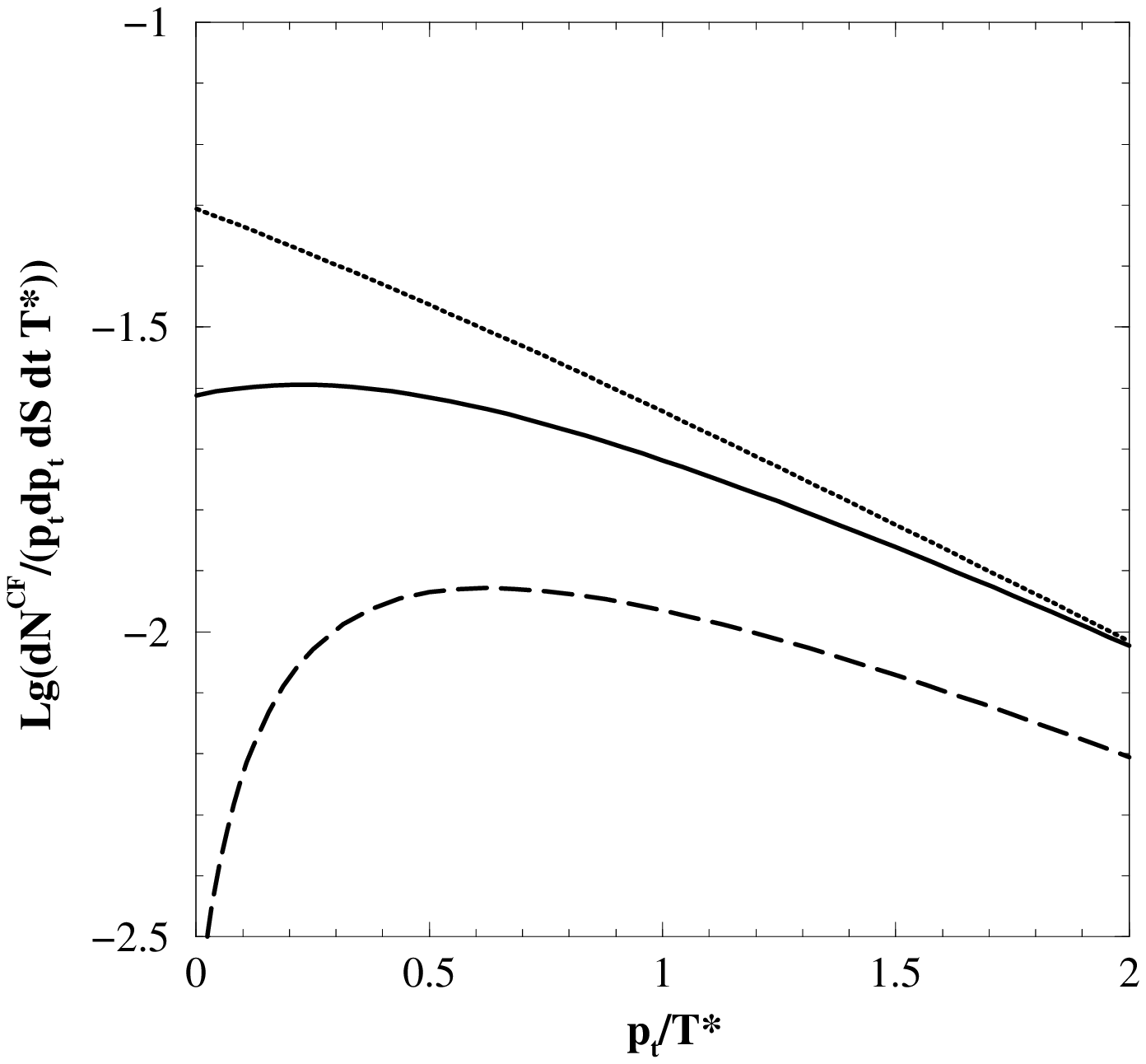,height=10cm,width=12cm} 
\hspace*{-3.4cm}\epsfig{figure=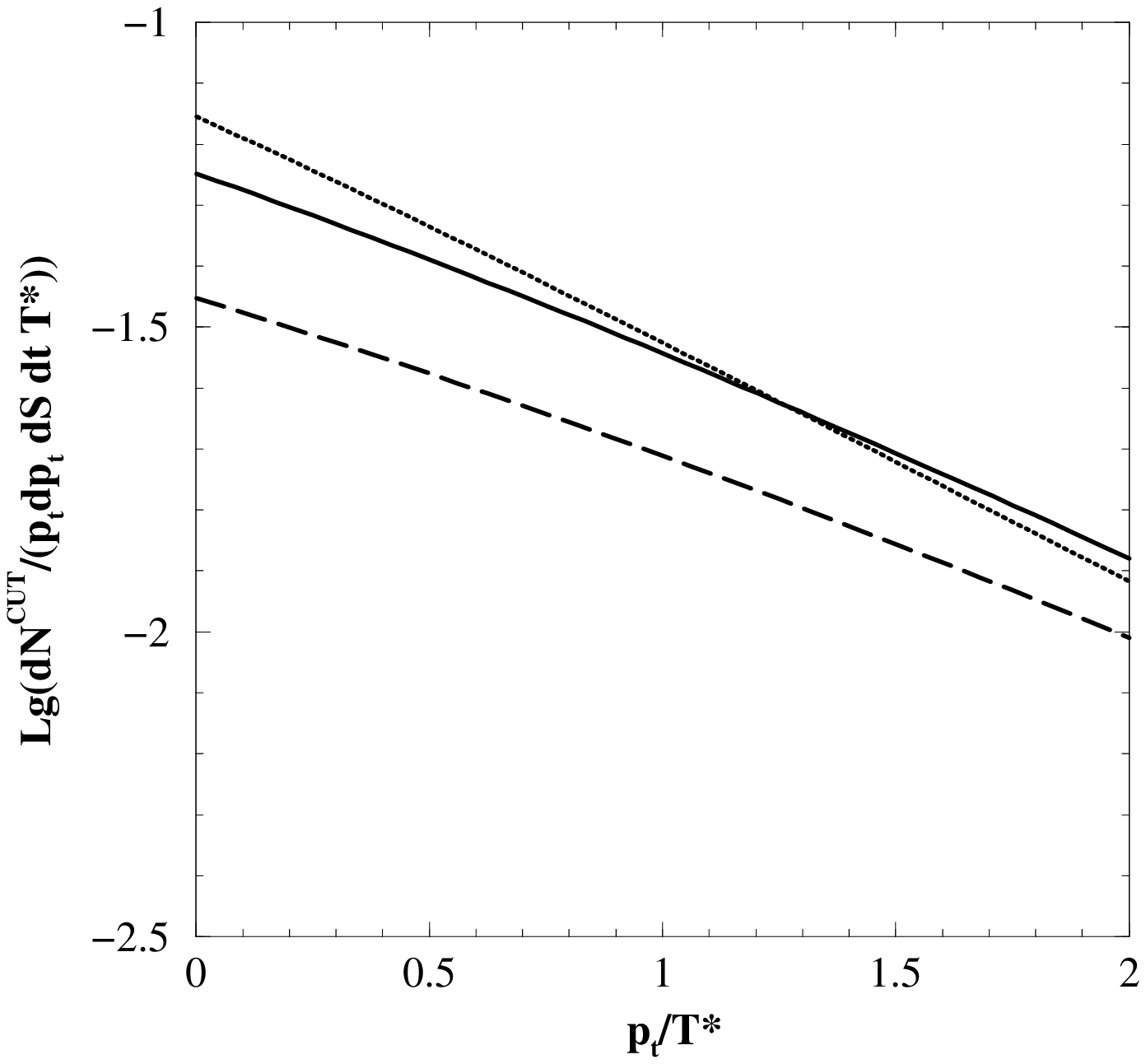,height=10cm,width=12cm}
}

\caption{ \label{PTspec2}
Comparison of the \CFful\, (left panel) and \COful\, (right panel) FO schemes.
The notations are the same as in Fig.~\ref{PTspec1}, but the spectra were
integrated over the CM rapidity $y_{CM} \in [-2; \,\,2\,\,]$.
}
\end{figure}

\begin{figure}[ht]

\centerline{\epsfig{figure=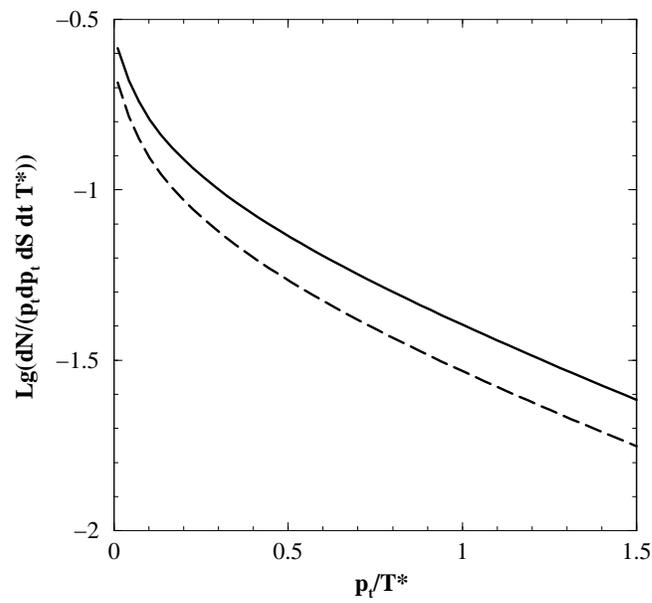,height=10cm,width=12cm}}
\caption{ \label{PTspec3} 
Comparison of the  momentum distribution of
the \CFful\, (dashed line) and the \COful\, (solid line) FO schemes
integrated over the CM rapidity $y_{CM} \in [-1; \,\,6\,\,]$.
Initial temperature of the fluid is
$T_{in} = 1.5 T^*$.
}
\end{figure}

\begin{figure}[ht]
\centerline{\hspace*{2.8cm}\epsfig{figure=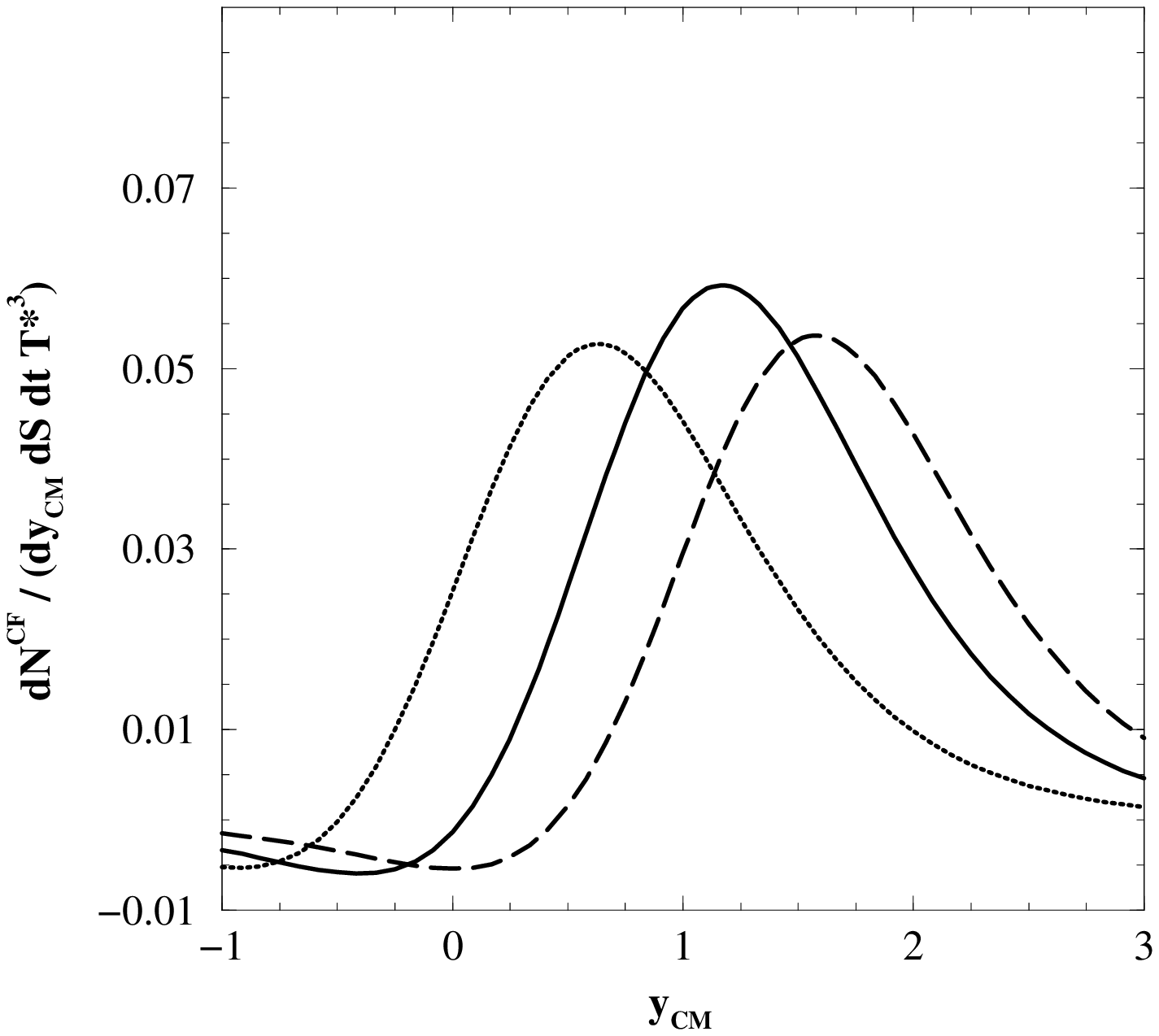,height=10cm,width=12cm} 
\hspace*{-3.4cm}\epsfig{figure=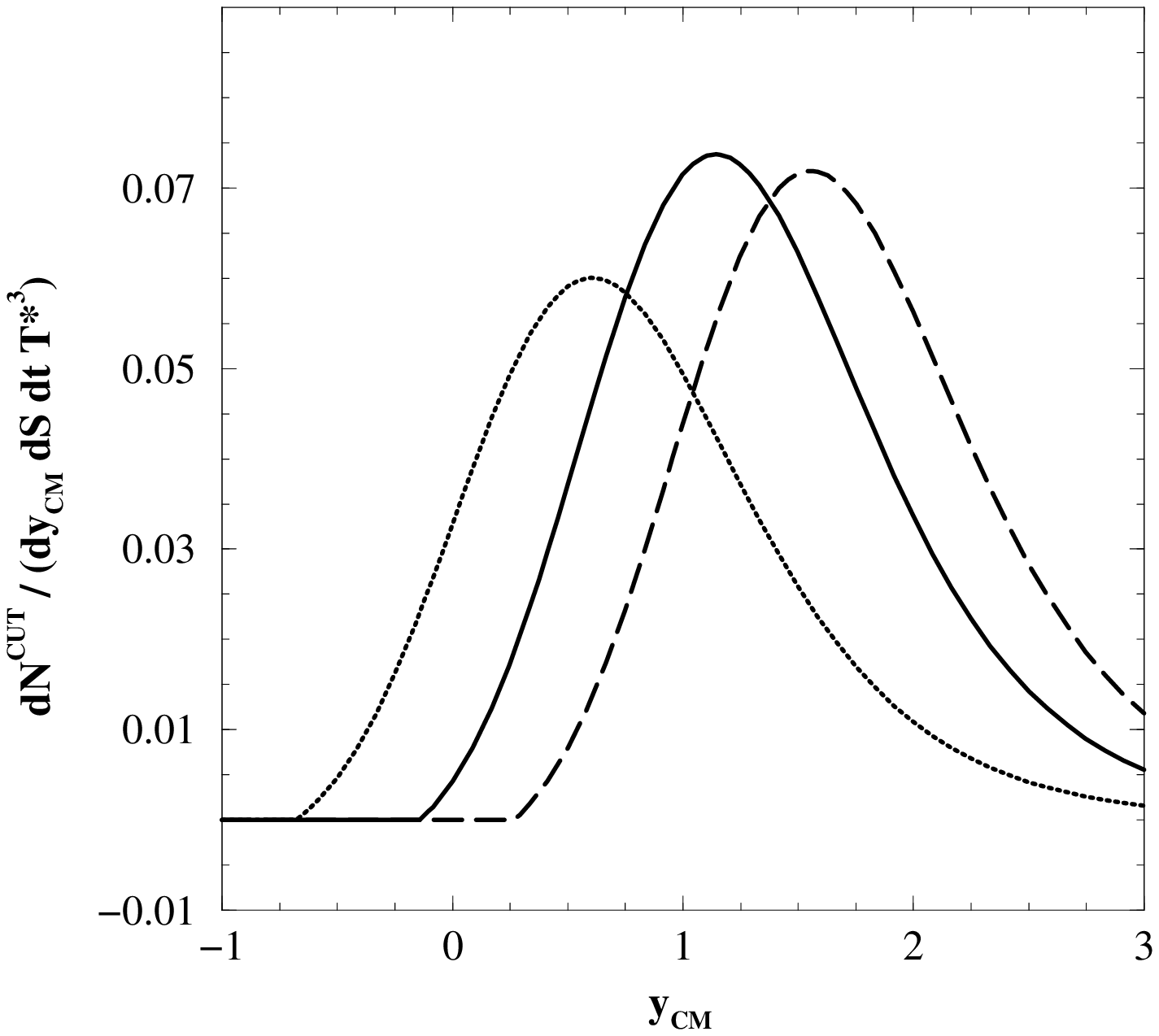,height=10cm,width=12cm}
}

\caption{ \label{Lspec1} 
Comparison of the \CFful\, (left panel) and \COful\, (right panel) FO schemes.
The momentum distribution of
both FO schemes was 
integrated over the  CM   transverse momentum  $p_{t} \in [0.01 ; \,\,6\,\,] T^*$.
Initial temperatures of the fluid are:
$T_{in} = 1.1 T^*$ (dotted line),
$T_{in} = 1.5 T^*$ (solid line),
and
$T_{in} = 1.9 T^*$ (dashed line).
The left panel clearly demonstrates the presence of negative particle
numbers in the  \CFful\, FO scheme, which  is not the case for the 
\COful\, FO scheme. Aslo one can see, that the hight of distribution function at maximum  are different for these FO schemes.
}
\end{figure}

Due to these reasons  the integrated spectra in both schemes are very different.
Figs.~\ref{PTspec1}--\ref{Lspec1}
show some typical  examples of the spectra integrated
over the  rapidity and  the transverse momenta for both schemes.
Evidently, the spectra integrated in the small rapidity window about $y_{\, CM} \approx 0$ are extremely
different - for the high initial temperatures  the \CFful\, formula gives negative
particle numbers everywhere (see Figs.~\ref{PTspec1} and \ref{PTspec2}).
This feature is hidden, if integration is carried for positive
values of the rapidity (see Fig.~\ref{PTspec3}). 
Nevertheless, 
the quantitative difference remains.
The spectra integrated over the transverse momenta have, basically, similar features.
However, these simple examples of the spectra  show that some previous 
results based on
the \CFful\, formula may be considerably revised while the correct FO scheme
is used.


\newpage

%


\clearpage

As discussed above  to describe the FO in 
finite systems requires  one needs to know
the hydrodynamic evolution of the system. 
Nevertheless, as it was  argued earlier 
the fluid volume for the \COful FO scheme 
is reduced on the magnitude of about 

$$\int \lp S_{\perp}^{\CF}(t_{em})\, v_\s^{\CF} - S_{\perp}^{\CO}(t_{em})\, v_\s^{\CO} 
\rp d t_{em} $$

\noindent
in comparison with the  \CFful\, one ($S_{\perp}(t_{em})$ is the area of the FO 3-surface
at the emission time $t_{em} $).
The maximal difference of the velocities is about $ v_\s^{\CF} - v_\s^{\CO} \approx 0.18 $ 
and it corresponds to the initial temperature 
$T_{in} \approx 1.5 T^*$.
Taking a typical emission time of 10~fm
one can roughly estimate
the reduction of emission transverse radius of fireball on about 2 fm 
compared to the \CFful FO scheme. 
Such a reduction of the emitting source size  is 
in a better agreement with the results of the HBT analysis of the
relativistic nuclear collisions.


In the considered example I gave a complete  solution of the FO problem for the 
simple wave.
Due to the fact that characteristics and isotherms of the fluid in the simple wave are just 
straight lines in any Lorentz frame, 
one can show that the ratio of the partial derivatives of the temperatures 
on both sides of the shock is the same, i.e., 
$\frac{\partial_x T_f}{\partial_t T_f} = \frac{\partial_x T^*}{\partial_t T^*}$
in any frame.
This is the simplest example against the statement of Ref. \cite{FO2} 
that such a condition is unphysical.
As was shown in this section, it is
unnecessarily  strong  for the derivation  of the equation for the FO HS as 
it was supposed in the paper \cite{Bugaev:96},
but it appears in particular applications like the freeze-out of the simple wave.

The momentum and rapidity spectra of the two compared schemes are different,
but the negative numbers of particles in the \CFful FO scheme are usually hidden
by the positive contributions which are coming from the s.l. parts of the FO 
HS.  
Therefore, one should look into some integrated characteristics of the 
hydrodynamic systems, like emission volume, which cannot be correctly described
by the \CFful FO scheme.

%


%
%
%
%
\def\mys#1{\Sigma_{ } { }_{\bf #1}}  
\def\ro{\rho_{\rm o}}  
\def\reff{$R_{eff}$}   
\def\lp{\left(}   
\def\rp{\right)}    
\def\bc{\begin{center}}  
\def\ec{\end{center}}  
\def\vp{\vspace*{0.1cm}}  
\def\vm{\vspace*{-0.2cm}}  
\def\CF{{\it CF}\,\,}   
\def\CFful{{Cooper-Frye}\,\,}  
\def\CO{{\it CO}\,\,}   
\def\COful{{\it cut-off}\,\,}   
%
%
%
%

\section{Relativistic Kinetic Equations for Finite Domains}

In the preceding sections I have considered the FO problem
in  relativistic hydrodynamics. 
Now I would like to analyze 
a  new approach 
to resolve this problem  
which was invented  by   Bass and Dumitru (BD model)  \cite{BD:00}
and further developed  by 
Teaney, Lauret and Shuryak (TLS model)  \cite{TLS:01}. 
These  hydrocascade   models assume that  the nucleus-nucleus collisions   
proceed in three stages: hydrodynamic   
expansion (hydro) of the quark gluon plasma (QGP), phase transition from the QGP to  
the hadron gas  and the stage of hadronic  
rescattering and resonance decays (cascade). The switch from hydro to  
cascade modeling takes place   
at the boundary between the mixed  and hadronic phases.   
The spectrum of hadrons  
leaving this HS of the QGP-hadron gas transition is taken as input for the  
cascade.  
  
This  approach  incorporates  the best features of  both the hydrodynamical and cascade 
descriptions.  It  allows for, on one hand, the calculation of  the phase transition 
between  the quark gluon plasma  and  hadron gas using hydrodynamics  
and, on the other hand,  the FO of hadron spectra   
using  the cascade description. 
This  approach 
allows  one to  overcome  the  usual 
difficulty of  transport models  in modeling    phase transition phenomenon. 
For this reason, this approach has been rather successful in explaining  
a variety of collective phenomena that has been observed
at the CERN  SPS  and  
Brookhaven RHIC  energies.  
However, both the BD and TLS models face some   
fundamental  difficulties    
which cannot be ignored {(see a detailed discussion in \cite{Bugaev:02HC}).}  
Thus, within the BD approach   
the initial distribution for the cascade is found using  the \CFful formula \cite{CF74},  
which takes into account particles with all possible velocities,   
whereas in the TLS model the initial cascade distribution is given by the \COful formula  
\cite{Bugaev:96,Bugaev:99},  
which accounts for only those particles that can leave the phase boundary.   
As shown in Ref. \cite{Bugaev:02HC}  the \CFful formula leads to  causal and  
mathematical problems in the present version of the  BD model because the  
QGP-hadron gas phase boundary inevitably has time-like parts.  
On the other hand, the TLS model  does not conserve  
energy, momentum and number of charges and this, as will be demonstrated later,  
is due to the fact that the equations of motion used in \cite{TLS:01}  
are incomplete and, hence, should be modified. 
 
These difficulties are likely in part responsible for the fact 
that the existing  hydrocascade  models, like  the more simplified ones,  
fail to  explain  the {\it HBT puzzle} \cite{QM:04}, i.e. 
the fact that  
the experimental HBT radii at RHIC   are  very similar to those
found at SPS, even though  the center of mass energy is larger  by an order of 
magnitude. Therefore, it turns out that  the  hydrocascade approach 
successfully {\it parameterizes}  the one-particle momentum spectra and their 
moments, but does not {\it  describe} the space-time picture of the  
nuclear collision  as probed by  two-particle interferometry.  
 
The main  difficulty of the hydrocascade  approach looks  similar to  
the traditional problem of 
FO  in relativistic hydrodynamics \cite{Bugaev:96,Bugaev:99}. 
In both cases the  domains (subsystems) have time-like boundaries 
through which the exchange of particles  occurs  and this fact should be taken into account.  
In relativistic hydrodynamics this problem was solved by the  
constraints which appear  on the FO  
HS and provide  the global energy-momentum and  
charge conservation \cite{Bugaev:96,Bugaev:99,Bugaev:99b}. 
{ A generalization of the usual} 
Boltzmann equation which accounts  
for  the exchange of particles 
on the time-like  boundary between domains in the relativistic kinetic theory 
was given recently  in Ref.  \cite{Bugaev:02HC}. 
It was  shown  that
the kinetic equations  
describing the exchange of particles   
on the time-like  boundary between subsystems   
should necessarily contain  the $\delta$-like  source terms. 
{ From these kinetic equations} 
the correct system of   hydrocascade  equations to  model the relativistic 
nuclear collision process was derived 
without specifying the properties of 
the separating HS.  
{ However, both an  explicit switch off} criterion 
from the hydro equation to the cascade one  
and the boundary conditions between them   
were not considered in \cite{Bugaev:02HC}. 
The  present section is devoted  
to the analysis of the boundary  
conditions for the system of   hydrocascade  equations. 
This is 
necessary to formulate the numerical algorithm 
for solving the hydrocascade  equations.

This  section is organized as follows. 
A brief derivation of the set of    kinetic equations 
is given and  source terms are obtained first. 
Then  the analog of the collision integrals 
is  discussed and a fully covariant formulation of the system of coupled 
kinetic equations is found.   The relation between the system obtained and 
the relativistic Boltzmann equation is also considered. 
The correct equations of motion for the  hydrocascade  approach 
and their boundary conditions 
are analyzed next. 
It is  also shown that the existence of  strong discontinuities across the space-like boundary,
the time-like shocks,  is in contradiction with  the basic assumptions of  a transport approach.
The solutions of   boundary conditions between  the hydro and cascade domains for a single degree of freedom
and for many degrees of freedom are discussed in details.

\subsection{Drift Term for Semi-Infinite Domain.} 
\begin{figure}
\centerline{
\includegraphics[width=8.0cm,height=7.5cm]{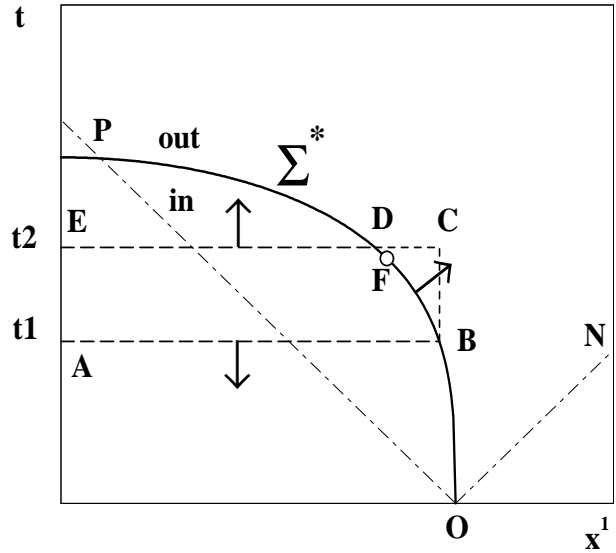}
}
\caption{
Schematic two dimensional picture
of the boundary HS $\Sigma^*$ (solid curve).
Arrows show the external normal
vectors. The light cone $NOP$ is shown by the dash-dotted line.
The point $F$ divides $\Sigma^*$ into the time-like  ($OF$) and space-like ($FP$) parts.
}
\label{fig531}
\end{figure}

Let me consider  two semi-infinite domains, ``in'' and ``out'',  
separated by the HS $\Sigma^*$  
which, for the purpose of presenting the idea, I assume to be  
given in (3+1) dimensions   
by a single valued function  
$t = t^*(\vec{x}) = x_0^*(\vec{x})$.  
The latter  
is assumed to be a unique solution of the equation 
${\cal F}^{*}(t, \vec{x}) = 0$ ({\it a switch off criterion}) 
which has  a positive time derivative 
$\partial_0 {\cal F}^{*}(t^*, \vec{x}) > 0$ on the HS $\Sigma^*$. 
Hereafter all quantities defined at $\Sigma^*$ will be marked with an asterisk.
The distribution function $\phi_{in}(x,p)$  for $t \le t^*(\vec{x})$    
is assumed to belong  
to the ``in'' domain,  
whereas $\phi_{out}(x,p)$ denotes the distribution function of the ``out'' domain   
for $t \ge t^*(\vec{x})$(see Fig.~\ref{fig531}).  
In this treatment it is assumed that the initial conditions for   
$\phi_{in}(x,p)$ are given, whereas   
on $\Sigma^*$ the  function $\phi_{out}(x,p)$ is allowed  
to differ from $\phi_{in}(x,p)$ and   
this will modify the kinetic equations for both functions.  
For simplicity I  consider a classical gas of point-like Boltzmann particles.  


Similar to Ref. \cite{groot} I derive the kinetic equations  
for $\phi_{in}(x,p)$ and $\phi_{out}(x,p)$ from the requirement of    
particle number conservation.   
Therefore, the particles leaving one domain and crossing the  HS $\Sigma^*$   
should be subtracted from the corresponding distribution function and  
added to the other.  
Now  consider the closed HS  of the ``in'' domain, $\Delta x^3$   
(shown as the contour $ABDE$ in Fig.~\ref{fig531}),    
which consists of  
two semi-planes $\sigma_{t1}$ and $\sigma_{t2}$ of constant time $t1$ and $t2$, respectively,  
that are connected from $t1$ to $t2 > t1$ by  
the arc $BD$  of the boundary $\Sigma^*(t1,t2)$    
in Fig.~\ref{fig531}.  
The original number of particles on the HS $\sigma_{t1}$ is given by   
the standard expression \cite{groot}  

\vspace*{-0.6cm}

\begin{equation}\label{N53one}  
N_1 = - \int\limits_{\sigma_{t1}} d \Sigma_\mu \frac{d^3 p}{p^0}~ p^\mu~ \phi_{in}(x,p)\,,   
\end{equation}  
\noindent
where $d \Sigma_\mu$ is  the external normal vector to $\sigma_{t1}$ and, hence,   
the product $p^\mu d \Sigma_\mu \le 0$ is non-positive.   
It is clear that these particles can   
cross either HS $\sigma_{t2}$ or $\Sigma^*(t1,t2)$.  
The corresponding numbers of particles are as follows  

\vspace*{-0.3cm}

\vm  
\begin{eqnarray}\label{N53two}  
N_2  \hspace*{0.0cm}=   \hspace*{-0.1cm}\int\limits_{\sigma_{t2}} \hspace*{-0.1cm}&& d \Sigma_\mu \frac{d^3 p}{p^0}~ p^\mu~ \phi_{in}(x,p)\,,\\   
\label{N53three}  
N_{loss}^*  = \hspace*{-0.2cm}\int\limits_{\Sigma^*(t1,t2)}\hspace*{-0.4cm}&&  d\Sigma_\mu   
\frac{d^3 p}{p^0}~ p^\mu~ \phi_{in}(x,p) ~\Theta(p^\nu d\Sigma_\nu) \,.  
\end{eqnarray}  

\vspace*{-0.cm}
  
\noindent  
The $\Theta$-function in the {\it loss} term (\ref{N53three}) is very important  
because it accounts for the particles leaving the ``in'' domain  (see also  
discussion in \cite{Bugaev:96,Bugaev:99b}).   
For the space-like parts of the HS  $\Sigma^*(t1,t2)$ which are   
defined by negative  sign $ds^2 < 0$ of   
the squared line  element, $ds^2 = dt^*(\vec{x})^2 - d\vec{x}^2 $,   
the product $p^\nu d\Sigma_\nu > 0$ is always positive and, therefore,   
particles with all possible  momenta can leave the ``in'' domain through  
the $\Sigma^*(t1,t2)$.  
For the time-like parts of $\Sigma^*(t1,t2)$ (with sign $ds^2 > 0$)   
the product $p^\nu d\Sigma_\nu $ can have either sign, and the $\Theta$-function  
{\it cuts off} those particles which return to the ``in'' domain.  
  
Similar one has to consider the particles coming to the  ``in'' domain   
from outside. This is possible through the time-like parts of the HS $\Sigma^*(t1,t2)$,   
if the  particle momentum satisfies  the inequality $ - p^\nu d\Sigma_\nu > 0$.   
In terms of the external normal $d \Sigma_\mu$ with respect to the ``in'' domain   
(this normal vector is   
shown as an arrow on the arc $BD$  in Fig.~\ref{fig531} and   
will be used hereafter for all integrals over the  HS $\Sigma^*(t1,t2)$)   
the number of gained particles    

\vspace*{-0.2cm}
  
\vm   
\begin{equation}\label{N53four}  
N_{gain}^*  = - \int\limits_{\Sigma^*(t1,t2)}\hspace*{-0.4cm}  d\Sigma_\mu  
\frac{d^3 p}{p^0}~ p^\mu~ \phi_{out}(x,p) ~\Theta(-p^\nu d\Sigma_\nu) \,  
\end{equation}  
\vm  
\vm  
   
\noindent  
is, evidently, non-negative.    
Since the total number of particles is conserved, i.e.   
$N_2 = N_1 - N_{loss}^* + N_{gain}^*$, one can use the Gauss theorem  
to rewrite the obtained integral over the closed HS $\Delta x^3$   
as an  integral over the  $4$-volume $\Delta x^4 $   
(area inside  the contour $ABDE$ in Fig.~\ref{fig531})   
surrounded  by $\Delta x^3$  
\vspace*{-0.2cm}
\begin{eqnarray}  
\int\limits_{\Delta x^4} \hspace*{-0.1cm}  d^4 x  
\frac{d^3 p}{p^0}~ p^\mu  ~{\partial}_\mu ~ \phi_{in}(x,p) = \int\limits_{\Sigma^*(t1,t2)}\hspace*{-0.4cm}  d\Sigma_\mu \frac{d^3 p}{p^0}~ p^\mu 
%
%
%
~[\phi_{in}(x,p) - \phi_{out}(x,p) ] \Theta(-p^\nu d\Sigma_\nu) \,.&&  
\label{N53five}
\end{eqnarray} 
\vm  
\noindent  
Note that in contrast to the usual case \cite{groot}, i.e. in the absence of 
a boundary $\Sigma^*$,     
the right-hand side (rhs) of Eq. (\ref{N53five})  does not vanish identically.  
  
The rhs of Eq. (\ref{N53five}) can be transformed further to  
a  $4$-volume integral in the following  
sequence of steps. First one has to express the integration element $d\Sigma_\mu$  
via the normal vector $n^*_\mu$ as follows $(dx^j > 0,$ for $ j =1,2,3)$   

\vspace*{-0.2cm}
  
\vm  
\begin{equation}\label{N53six}  
 d\Sigma_\mu = n^*_\mu dx^1 dx^2 dx^3;   
 \quad  n^*_\mu \equiv \delta_{\mu 0} - \frac{ \partial t^*(\vec{x}) }{\partial x^\mu} (1 - \delta_{\mu 0} )\,,   
\end{equation}  
\vm  
\vm  
   
\noindent  
where $\delta_{\mu \nu}$ denotes the Kronecker symbol.   
Then, using the identity  $\int\limits_{t1}^{t2} dt\, \delta (t - t3) = 1$    
for the Dirac $\delta$-function  
with  
$t1 \le t3 \le t2$, one can rewrite the rhs integral in (\ref{N53five}) as   

\vm  
\begin{equation}\label{N53seven}   
\int\limits_{\Sigma^*(t1,t2)}\hspace*{-0.4cm}  d\Sigma_\mu \cdots \equiv   
\int\limits_{V^4_\Sigma} d^4 x~\delta (t - t^*(\vec{x}) )~ n^*_\mu \cdots\,,   
\end{equation}  
\vm  
\vm  
   
\noindent  
where  short hand  notations  are introduced for 
the corresponding
 $4$-dimensional volume $V^4_\Sigma = (t2-t1) \int\limits_{\Sigma^*(t1,t2) }\,dx^1 dx^2 dx^3 $ which is  
shown as the rectangle $G B C D $ in Fig.~\ref{fig531} ( $|G B| = |C D|$).
Evidently, the Dirac $\delta$-function allows one  to extend integration in (\ref{N53seven}) to the  
unified $4$-volume  $V^4_U = \Delta x^4 \cup V^4_\Sigma$ of $\Delta x^4$ and $V^4_\Sigma$  
(the volume  $V^4_U $ is shown as the area $ABCE$ in Fig.~\ref{fig531}).  
Finally, with the help of notations  

\vspace*{-0.4cm}
  
\begin{equation}\label{N53eight}  
\Theta_{out} \equiv \Theta (t - t^*(\vec{x}) ); \quad \Theta_{in} \equiv 1- \Theta_{out}   
\end{equation}  
it is possible to extend the left hand side (lhs)  
integral in Eq. (\ref{N53five}) from $\Delta x^4$ to $ V^4_U$.  
Collecting all the above results, from Eq. (\ref{N53five}) one obtains  
\hspace*{-0.5cm}\begin{eqnarray}  
%
\int\limits_{ V^4_U} \hspace*{-0.1cm}  d^4 x  
\frac{d^3 p}{p^0}~ \Theta_{in}~ p^\mu  ~{\partial}_\mu ~ \phi_{in} =   
\int\limits_{V^4_U}\hspace*{-0.1cm}  d^4 x  
\frac{d^3 p}{p^0}~ p^\mu n^*_\mu 
%
%
~[\phi_{in} - \phi_{out} ]~ \Theta(-p^\nu n^*_\nu)   
~\delta (t - t^*(\vec{x}) ) \,.&&   \label{N53nine}  
\end{eqnarray} 
\noindent  
Since the  volumes $\Delta x^4$ and $V^4_U$ are arbitrary, one obtains   
the kinetic equation for the distribution function of the ``in'' domain   
\begin{eqnarray}  
\hspace*{-0.9cm}&&\Theta_{in}~ p^\mu  ~{\partial}_\mu ~ \phi_{in} (x,p) = C_{in} (x,p) +  
%
%
%
p^\mu n^*_\mu [\phi_{in}(x,p) - \phi_{out}(x,p) ] ~\Theta(-p^\nu n^*_\nu)  
~\delta (t - t^*(\vec{x}) ) \,.   \label{N53ten}  
\end{eqnarray} 
\noindent  
Note that the  general solution   of Eq. (\ref{N53nine})  
contains  
an arbitrary function $C_{in} (x,p)$  (the first term in the rhs of (\ref{N53ten}))  which identically vanishes while being 
integrated over the invariant momentum measure $ d^3 p /p_0 $.  
Such a property is  typical for a collision integral \cite{groot}, 
and I will discuss its derivation  in the next subsection.   
To shorten the notation,  the  domain of each distribution function will be denoted
as a subscripted Latin capital letter $A$ or $B$ ($A, B \in \{in, out\}$)
to avoid confusion with Greek 4-indices.

Similar one can obtain the equation for the distribution   
function of the ``out'' domain  
\begin{eqnarray}  
\hspace*{-0.9cm}&&\Theta_{out}~ p^\mu  ~{\partial}_\mu ~ \phi_{out} (x,p) = C_{out} (x,p) + 
%
%
p^\mu n^*_\mu [\phi_{in}(x,p) - \phi_{out}(x,p) ]~ \Theta(p^\nu n^*_\nu)  
~\delta (t - t^*(\vec{x}) ) \,,   \label{N53eleven}
\end{eqnarray}    
\noindent    
where the normal vector $n^*_\nu$ is  given by  (\ref{N53six}).  
Note the asymmetry between the rhs of Eqs. (\ref{N53ten})  
and (\ref{N53eleven}): for the space-like parts of HS $\Sigma^*$   
the  source term with $\Theta(-p^\nu n^*_\nu) $ vanishes identically because   $p^\nu n^*_\nu > 0$.  
This reflects the  causal properties of the equations above:    
propagation of particles faster than light is forbidden, and hence no particle  
can (re)enter the ``in'' domain.   

  
\subsection{Collision Term for Semi-Infinite Domain.}  
Since in the general case $\phi_{in}(x,p) \neq  \phi_{out}(x,p)$ on  $\Sigma^*$,  
the $\delta$-like terms in  the rhs of   
Eqs. (\ref{N53ten}) and (\ref{N53eleven}) cannot vanish simultaneously on this HS.  
Therefore, the functions $\Theta_{in}^*  \equiv \Theta_{in}|_{\Sigma^*} \neq 0$ and   
$ \Theta_{out}^*  \equiv \Theta_{out}|_{\Sigma^*} \neq 0$ do not vanish   
simultaneously on  $\Sigma^*$ as well.   
The $ \Theta(x)$ is not uniquely defined at $ x = 0$, and, therefore, 
there is some freedom to choose a convenient value at $ x = 0$.
Since there is no preference between ``in'' and ``out'' domains  
it is assumed that   
\begin{equation}\label{N53twelve}  
\Theta_{in}^* = \Theta_{out}^* = \Theta (0) = \frac{1}{2}\,,  
\end{equation}  
\noindent  
but the final results are independent of this choice.  
This result can be understood by considering the limit $a \rightarrow 0$ of
the following definition: 
$\Theta (x) \equiv \frac{1}{2} \lim_{a \rightarrow 0}  \left[ \tanh \left( {x}/{|a|} \right) + 1\right]$.

Now the collision terms for Eqs. (\ref{N53ten}) and (\ref{N53eleven}) can be readily obtained.    
Adopting the usual assumptions for   
the distribution functions \cite{BOGOL,groot,BALESCU}, one can   
repeat the standard derivation of the collision terms \cite{groot,BALESCU}  
and get the desired expressions.   
I will  not recapitulate this standard part, but only discuss  how to modify the derivation for   
this purpose.  
First of all, one has to start the derivation in the $\Delta x^4$ volume of  
the ``in'' domain and then  
extend it to the unified $4$-volume  
$V^4_U = \Delta x^4 \cup V^4_\Sigma$ similarly to the preceding section.  
Then the first part of the collision term for Eq. (\ref{N53ten}) reads ( $A, B \in \{in, out\}$)
\begin{eqnarray}\label{N53thirteen}  
\hspace*{-0.6cm}
C_{in}^{I} (x,p) & = & \Theta_{in}^2 \left( I^G [\phi_{in}, \phi_{in}] - I^L [\phi_{in}, \phi_{in}] \right)  
\,, \\  
\label{N53fourteen}  
\hspace*{-0.6cm}I^G [\phi_{A}, \phi_{B}] & \equiv & \frac{1}{2} \int D^9 P~    
\phi_{A}(p^{\prime} )~ \phi_{B}(p_1^{\prime})~ W_{p\,p_1^{} | p^{\prime}p_1^{\prime}}   
\,, \\   
\label{N53fifteen}   
\hspace*{-0.6cm} I^L [\phi_{A}, \phi_{B}] & \equiv & \frac{1}{2} \int D^9 P~  
\phi_{A}(p)~ \phi_{B}(p_1)~ W_{p\,p_1^{} | p^{\prime}p^{\prime}_1}\,,  
\end{eqnarray}  
  
\vm  
  
\noindent  
where the invariant measure of integration is denoted by   
$ D^9 P \equiv \frac{d^3 p_1}{p^0_1} \frac{d^3 p^{\prime} }{p^{\prime 0}}   
\frac{d^3 p^{\prime}_1 }{p^{\prime 0}_1} $ and $W_{p\,p_1^{} | p^{\prime}p^{\prime}_1}$   
is the transition rate in the elementary  reaction     
with energy-momentum conservation given in the form  
$p^\mu + p_1^\mu = p^{\prime \mu} + p^{\prime \mu}_1$.  
The rhs of  (\ref{N53thirteen}) 
contains
the standard gain and loss terms which are defined by Eqs. (\ref{N53fourteen}) and (\ref{N53fifteen}),
respectively, weighted by the ``probability'' of collision between particles from the ``in'' domain    
given by the square of the  $\Theta_{in}$-function.
The value $\Theta_{in}^2 \equiv 1$ is found inside of the `in'' domain, whereas 
$\Theta_{in}^2 = \Theta_{in}^{* 2} = 1/4$ at the boundary $\Sigma^*$ because
according to (\ref{N53twelve}), for  each value of the distribution function $\phi_{in}$ 
in the rhs of (\ref{N53thirteen}), only half of the boundary $\Sigma^*$ belongs to the ``in'' domain.
This can be better understood by considering, first, the above mentioned 
tangent representation for the $\Theta$-function, and then taking the limit 
$a \rightarrow 0$ next.
%

It is easy to understand that  
on $\Sigma^*$   
the second part of the collision term   
(according to Eq. (\ref{N53twelve})) is defined by the   
collisions between particles of ``in'' and ``out'' domains  
\begin{equation}\label{N53sixteen}  
\hspace*{-0.25cm}  
C_{in}^{II} (x,p)  =  \Theta_{in} \Theta_{out}  \left( I^G [\phi_{in}, \phi_{out}] - I^L [\phi_{in}, \phi_{out}] \right)  
.   
\end{equation}  
\noindent  
Again, the product $\Theta_{in} \Theta_{out} = 0$ everywhere, except at the HS $\Sigma^*$,
where it corresponds to the ``probability'' of collision at $\Sigma^*$ 
for the particles coming from both domains.   
This can be easily seen from the hyperbolic tangent representation of the  $\Theta$-function.

Combining (\ref{N53ten}), (\ref{N53thirteen}) and (\ref{N53sixteen}), one gets the kinetic  
equation for the  ``in'' domain  
\begin{eqnarray} 
\hspace*{-0.7cm}&&\Theta_{in}~ p^\mu  ~{\partial}_\mu  \phi_{in} (x,p) =  C_{in}^{I} (x,p) +  
C_{in}^{II} (x,p) + p^\mu n^*_\mu \times \nonumber \\ 
\label{N53seventeen} 
%
\hspace*{-0.7cm}&&[\phi_{in}(x,p) - \phi_{out}(x,p) ] ~\Theta(-p^\nu n^*_\nu) 
~\delta (t - t^*(\vec{x}) ) \,. 
\end{eqnarray} 
\noindent  
The kinetic equation for the ``out'' domain  
can be derived similarly and then it can be   
represented in the form  
\begin{eqnarray} 
\hspace*{-0.5cm}&&\Theta_{out}~ p^\mu  ~{\partial}_\mu  \phi_{out} (x,p) =  C_{out}^{I} (x,p) +  
C_{out}^{II} (x,p) + p^\mu n^*_\mu \times \nonumber \\ 
\label{N53eighteen} 
\hspace*{-0.5cm}&&[\phi_{in}(x,p) - \phi_{out}(x,p) ] ~\Theta(p^\nu n^*_\nu) 
~\delta (t - t^*(\vec{x}) ) \,, 
\end{eqnarray} 
where the evident notations for the collision terms 
$C_{out}^{I} \equiv \Theta_{out}^2 \left( I^G [\phi_{out}, \phi_{out}] - I^L [\phi_{out}, \phi_{out}] \right) $  
and  
$C_{out}^{II} \equiv \Theta_{in} \Theta_{out} \left( I^G [\phi_{out}, \phi_{in}] - I^L [\phi_{out}, \phi_{in}] \right) $ 
are used.  
  
The equations  (\ref{N53seventeen}) and (\ref{N53eighteen}) can be represented also 
in a covariant form with the help of the function ${\cal F}^{*}(t, \vec{x})$.  
Indeed, 
applying the definition of the derivative of the implicit function to  
$\partial_\mu t^*(\vec{x}) $, 
one can rewrite the external normal vector (\ref{N53six}) as  
$n^*_\mu \equiv \partial_\mu {\cal F}^{*}(t, \vec{x}) / \partial_0 {\cal F}^{*}(t, \vec{x})$. 
Now using the inequality $\partial_0 {\cal F}^{*}(t^{*}, \vec{x}) > 0$  and  
the following identities  
$\delta ( {\cal F}^{*}(t, \vec{x}) ) = \delta (t - t^*(\vec{x}) ) /   
\partial_0 {\cal F}^{*}(t^*, \vec{x})$,  
$\Theta_{A} \equiv \Theta ( S_A~{\cal F}^{*} (t, \vec{x})  )$  
one can write Eqs. (\ref{N53seventeen}) and (\ref{N53eighteen})   
in a fully  covariant form  
\begin{eqnarray} 
\hspace*{-0.5cm}&&\Theta_{A}~ p^\mu  ~{\partial}_\mu ~ \phi_{A} (x,p) =  C_{A}^{I} (x,p) +  
C_{A}^{II} (x,p) + p^\mu \partial_\mu {\cal F}^{*} \times 
\nonumber \\ 
\label{N53nineteen} 
\hspace*{-0.5cm}&&[\phi_{in}(x,p) - \phi_{out}(x,p) ] ~\Theta(S_A~p^\nu \partial_\nu {\cal F}^{*}) 
~\delta ( {\cal F}^{*}(t, \vec{x})  ) \,, 
\end{eqnarray} 
where the notations $A \in in$, $S_{in}=-1$ ($A \in out$,  $S_{out}=1$)  
are introduced for ``in'' (``out'') domain.

For the continuous distribution functions on $\Sigma^*$,  
i.e. $\phi_{out}|_{\Sigma^*} = \phi_{in}|_{\Sigma^*}$,     
the $\delta$-like source terms on the rhs of Eqs. (\ref{N53seventeen})    
and (\ref{N53eighteen})  
vanish and one  
recovers the Boltzmann equations.  
Moreover, with the help of the evident relations  
\begin{eqnarray}\label{N53twenty}  
\hspace*{-0.5cm}&&- {\partial}_\mu ~ \Theta_{in} = {\partial}_\mu ~ \Theta_{out} =  
~\delta ( {\cal F}^{*}(t, \vec{x})  ) ~\partial_\mu {\cal F}^{*} (t, \vec{x}) \,, \\ 
\label{N53twone}  
\hspace*{-0.5cm}&&C_{in}^{I} + C_{in}^{II} + C_{out}^{I} + C_{out}^{II} =    
I^G [\Phi, \Phi] - I^L [\Phi, \Phi]\,,   
\end{eqnarray}   

\noindent  
where  
$\Phi(x,p) \equiv \Theta_{in}~\phi_{in}(x,p) + \Theta_{out}~\phi_{out}(x,p) $,   
one can get the following result summing up  Eqs. (\ref{N53seventeen}) and (\ref{N53eighteen})   
\begin{equation}\label{N53twtwo}  
p^\mu  ~{\partial}_\mu ~ \Phi (x,p) = I^G [\Phi, \Phi] - I^L [\Phi, \Phi]\,.  
\end{equation}  

\noindent  
In other words, the usual Boltzmann equation follows from  
the system (\ref{N53nineteen})    
automatically  {\it  without any assumption} about the behavior   
of $\phi_{in}$ and $\phi_{out}$ on the boundary HS  
$\Sigma^*$.  
Also  Eq. (\ref{N53twtwo}) is valid    
not only under condition (\ref{N53twelve}),  
but for {\it any choice}  $0 < \Theta_{A}^{*} < 1$ obeying Eq. (\ref{N53eight}).  

In fact the system (\ref{N53nineteen}) generalizes   
the relativistic kinetic equation to the case of the strong   
temporal and spatial inhomogeneity, i.e.,  
for  $\phi_{in}(x,p) \neq \phi_{out}(x,p)$ on $\Sigma^*$.  
Of course, one has to be extremely careful while discussing  
the strong temporal inhomogeneity (or discontinuity on the space-like parts of $\Sigma^*$)   
such as the so called {\it time-like shocks} \cite{timeshock,TIMESHOCKb}  
because, as shown below, their existence  
contradicts  the usual assumptions \cite{BOGOL,groot,BALESCU}    
adopted for distribution functions.   
  
From the system (\ref{N53nineteen}) it is possible to derive the   
macroscopic equations of  motion for the  energy-momentum tensor
 by multiplying the corresponding equation with $p^\nu$   
and integrating it over the invariant measure. Thus,  Eq. (\ref{N53nineteen})   
generates the following expression  
($T^{\mu \nu}_{A} \equiv \int \frac{d^3 p }{p^ 0}~ p^\mu p^\nu \phi_{A}(x,p)$)  
\begin{eqnarray}  
\hspace*{-0.5cm}&&\Theta_{A}~ {\partial}_\mu ~ T^{\mu \nu}_{A}  =   \int \frac{d^3 p }{p^ 0}~  p^\nu    
C_{A}^{II} (x,p) + \int \frac{d^3 p }{p^ 0}~ p^\nu   
p^\mu \partial_\mu {\cal F}^{*} \times \nonumber \\  
\label{N53twthree}  
\hspace*{-0.5cm}&&[\phi_{in} (x,p) - \phi_{out} (x,p) ] ~\Theta(S_A~p^\rho \partial_\rho {\cal F}^{*})  
~\delta ( {\cal F}^{*}(t, \vec{x})  ) .  
\end{eqnarray}  
\noindent  
Similar to the usual Boltzmann equation  
the momentum integral of the collision term $C_{in}^{I}$ vanishes  
due to its symmetries \cite{groot},  
but it can be shown   
that the integral of the second collision term $C_{in}^{II}$ does     
not vanish because it involves two different 
distribution functions.   
 
The corresponding system of  equations  for the conserved current   
$N^\mu_A \equiv \int \frac{d^3 p }{p^ 0}~ p^\mu \phi_{A} (x,p)  $  
can be obtained by  direct integration of the system (\ref{N53nineteen})    
with the invariant measure  
\begin{eqnarray} 
\Theta_{A}~ {\partial}_\mu ~ N^{\mu}_{A} & = &      
\int \frac{d^3 p }{p^ 0}~   p^\mu \partial_\mu   {\cal F}^{*}  
[\phi_{in} (t, \vec{x}) - \phi_{out} (t, \vec{x}) ] \times  \nonumber \\ 
\label{N53twfour} 
&&\Theta(S_A~p^\rho \partial_\rho  {\cal F}^{*} ) 
~\delta ( {\cal F}^{*}(t, \vec{x})  ) . 
\end{eqnarray} 
\noindent 
The  above equation does not contain the contribution from antiparticles 
(just for simplicity), but  the latter  can be easily recovered.     
Note that  in contrast to (\ref{N53twthree}) the momentum integral 
of both collision terms vanish in  Eq. (\ref{N53twfour}) due to symmetries.   
    
   
 
\section{Derivation of Conservation Laws for Hydrokinetics }  
  
\begin{figure}
\centerline{
\includegraphics[width=8.0cm,height=7.5cm]{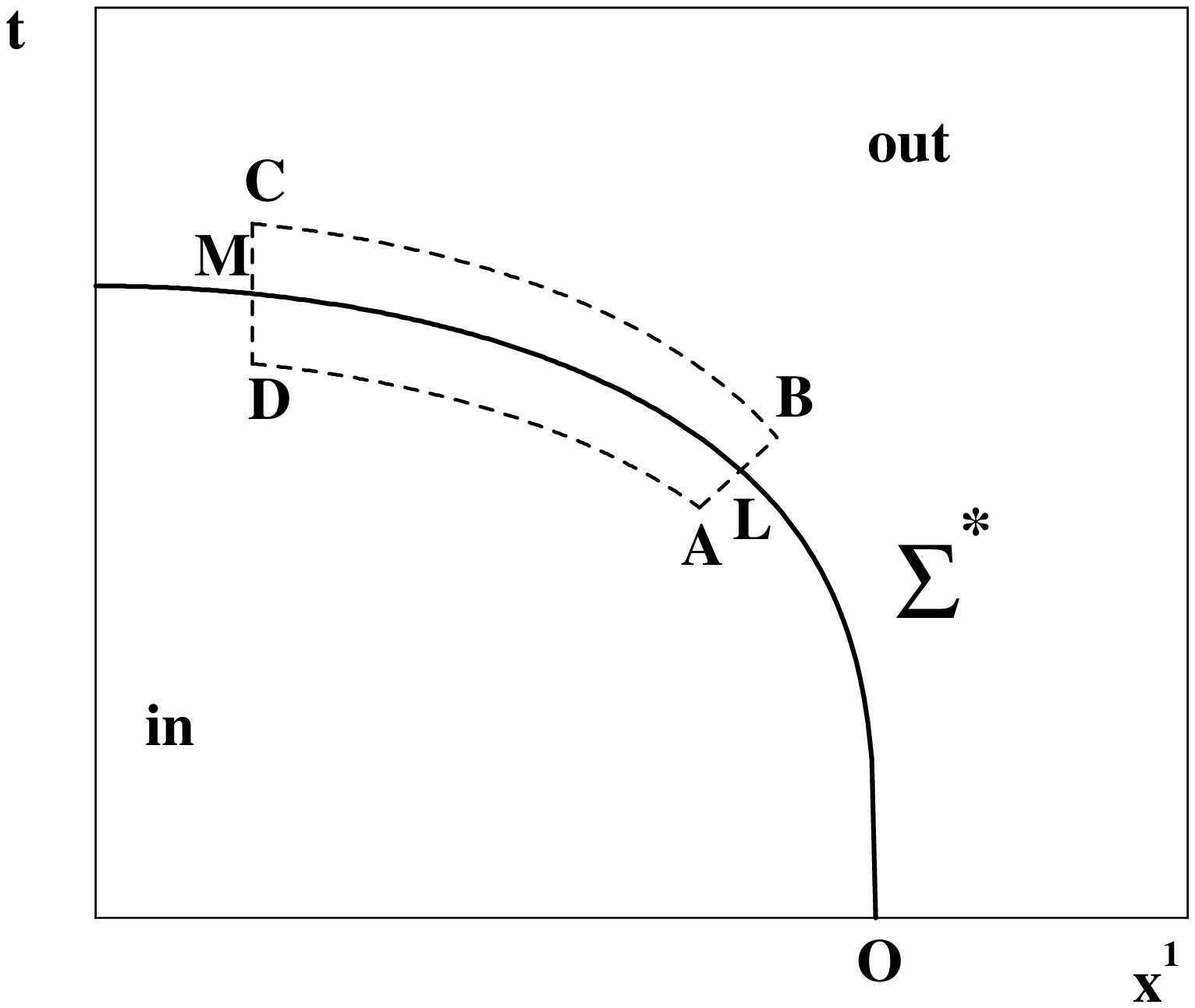}
}
\caption{
Schematic two dimensional picture
of the integration contour to derive the boundary conditions 
(\ref{N53twfive}) - (\ref{N53twseven}) between the ``in'' and
``out'' domains. In the limit of  a vanishing maximal distance
 $\Delta  \rightarrow 0$ between the HSs $AD$ and $BC$,  both of these
HSs are reduced to the part $\tilde\Sigma$ (an arc $LM$) of the boundary
 $\Sigma^*$ between domains.
}
\label{fig532}
\end{figure}

It is clear that  Eqs. (\ref{N53nineteen}), (\ref{N53twthree}) and (\ref{N53twfour})   
remain   
valid  both for finite domains and   
for a  multiple valued function $t = t^*(\vec{x})$ as well.  
To derive the whole system of these equations  
in the  latter case,  one has to divide the function  
$t^*(\vec{x})$ into the single valued parts, but this discussion is beyond the scope of this analysis.  
Using Eqs.  
(\ref{N53nineteen}), (\ref{N53twthree}) and (\ref{N53twfour})  
one can  analyze the boundary conditions  
on the HS $\Sigma^*$.  
The simplest way to get the boundary conditions 
is to integrate  Eqs. (\ref{N53twthree}) and (\ref{N53twfour}).  
Indeed, integrating (\ref{N53twthree})  
over the 4-volume $V^4_{ \tilde\Sigma}$ (shown as the area ABCD in Fig.~\ref{fig532}) containing part 
$\tilde\Sigma$ of the 
HS  $\Sigma^*$, 
one obtains the energy-momentum conservation.   
Before 
applying the Gauss theorem to the lhs of (\ref{N53twthree}),  
I  note that the corresponding $\Theta_A$-function reduces 
the 4-volume $V^4_{\tilde\Sigma}$ to its part which belongs to  the $ A$-domain. 
The latter is  
shown as area $ALMD$ ($BCML$) for $A \in in$ ($A \in out$) in Fig.~\ref{fig531}.  
Then 
in the limit of a vanishing maximal distance  $\Delta \rightarrow 0$ 
between the HSs $AD$ and $BC$ in Fig.~\ref{fig531}, 
the volume integral of the lhs of Eq. (\ref{N53twthree}) 
can be rewritten as the two integrals  $\int d\sigma_\mu T^{\mu\nu}_A$ : 
the first  integral is performed over the HS $\tilde\Sigma$ 
shown  as an arc 
$LM$ in Fig.~\ref{fig531},  and the second  integral  reduces to the same HS  but taken in the opposite direction, i.e. the $ML$ arc in Fig.~\ref{fig531}. 
Thus, the volume integral of the lhs of Eq. (\ref{N53twthree}) vanishes in this limit  
for tensors $T^{\mu\nu}_A$ being continuous functions of coordinates, and 
one obtains 
\begin{eqnarray} 
\hspace*{-0.5cm}&&0 \hspace*{-0.05cm} = \hspace*{-0.1cm} \int\limits_{ V^4_{\tilde\Sigma} } \hspace*{-0.1cm}  d^4 x~\Theta_A 
{\partial}_\mu  \lp  T^{\mu \nu}_A (x,p) \rp \hspace*{-0.01cm}  \equiv \hspace*{-0.1cm} 
\int\limits_{ V^4_{\tilde\Sigma} } \hspace*{-0.1cm}
  d^4 x~ \frac{d^3 p }{p^ 0}~  p^\nu   C_{A}^{II} (x,p) +   
\nonumber \\  
\hspace*{-0.5cm}&&  
\int\limits_{ V^4_{\tilde\Sigma} }  d^4 x~  \frac{d^3 p }{p^ 0}~ \delta ( {\cal F}^{*}(t, \vec{x})  )~  
p^\nu p^\mu \partial_\mu {\cal F}^{*}~ 
%
%
[\phi_{in} (x,p)   - \phi_{out}(x,p) ] ~ \Theta(S_A~p^\rho \partial_\rho  {\cal F}^{*})   \,.   \label{N53twfive} 
\end{eqnarray}  
\noindent 
Similarly  to the previous treatment, in the limit $\Delta \rightarrow 0$ the second integral on the  rhs  
of (\ref{N53twfive}) can be reexpressed as    
an integral over the closed HS.  
Since the latter  is arbitrary, then Eq. (\ref{N53twfive}) 
can be satisfied, if and only if the  energy-momentum conservation occurs  
for every point of the HS $\Sigma^*$ 
\begin{eqnarray}
\label{N53twsix}
&& \hspace*{-1.7cm} 
T^{\mu \nu}_{in \pm} ~
\partial_\mu {\cal F}^{*}(t^{*}, \vec{x})  =
T^{\mu \nu}_{out \pm} ~
\partial_\mu {\cal F}^{*}(t^{*}, \vec{x})  \,, 
%
~{\rm with}~~
%
 T^{\mu \nu}_{A \pm}  \equiv
\int \frac{d^3 p }{p^ 0}~ p^\mu p^\nu \phi_{A} (x,p)~
\Theta(\pm~p^\rho \partial_\rho {\cal F}^{*})
 \,. 
%
\end{eqnarray}
\noindent 
In deriving (\ref{N53twsix}) from (\ref{N53twfive})  
I used the fact that  
the 4-volume  integral of the second collision term  $C_{A}^{II}$   
vanishes for finite values of distribution functions  
because of the Kronecker symbols.  
The results for the conserved current follows similarly from Eq. (\ref{N53twfour}) 
after integrating it  
over the 4-volume $V^4_{ \tilde\Sigma}$ and taking the limit $\Delta \rightarrow 0$ 
\begin{eqnarray}
\label{N53twseven}
&& \hspace*{-1.7cm} 
 N^{\mu }_{in \pm}  ~
\partial_\mu {\cal F}^{*}(t^{*}, \vec{x})  =
 N^{\mu }_{out \pm}  ~
\partial_\mu {\cal F}^{*}(t^{*}, \vec{x}) \,,
%
%
%
~{\rm with}~~
%
 N^{\mu }_{A \pm}  \equiv
\int \frac{d^3 p }{p^ 0}~ p^\mu \phi_{A} (x,p)~ \Theta(\pm~p^\rho \partial_\rho {\cal F}^{*})
 \,. 
%
\end{eqnarray}
The fundamental  difference  between the conservation laws (\ref{N53twsix}),  (\ref{N53twseven})  
and the ones of the  usual hydrodynamics  is that the systems (\ref{N53twsix}) and   (\ref{N53twseven}) 
conserve the quantities of the outgoing from ($S_A = 1$)   and  incoming  
to ($S_A = -1$) ``in'' domain particles 
{\it separately}, whereas in the usual hydrodynamics only the  sum of these  contributions 
is conserved.    
 
\vspace*{0.3cm}

\vspace*{-0.2cm}

The trivial solution of Eqs. (\ref{N53twsix}) and (\ref{N53twseven}) corresponds to a continuous  
transition between ``in'' and ``out'' domains 
\begin{equation}\label{N53tweight}  
\phi_{out} (x, p) \Bigl|_{\Sigma^*} = \phi_{in} (x, p) \Bigr|_{\Sigma^*}\,.   
\end{equation}  
 
This choice  corresponds to the BD model \cite{BD:00}.   
The  BD model gives  a correct result for an oversimplified kinetics 
considered here. However, in the  case of the first order  phase  transition  
(or a strong cross-over) which was a prime  target of the hydrocascade  models     
 \cite{BD:00,TLS:01} the situation is different.  
In the latter case the speed of sound either vanishes (or becomes very small) \cite{Hung:94,Hung:98} 
and, hence, the rarefaction shock waves become  possible \cite{RAREF:1,RAREF:2,bugaevetal2}.    
The reason why the rarefaction shocks  may exist lies in the anomalous thermodynamic 
properties \cite{bugaevetal2} of the media near the phase transition region.    
In other words,  on the boundary between the  mixed  and  
hadronic phases the rarefaction shocks are mechanically 
stable  \cite{bugaevetal2}, whereas the compression shocks are mechanically unstable. 
This is also valid for the vicinity of  the generalized mixed phase of 
a strong cross-over.
 
One important  consequence of the shock  mechanical stability criterion
(see also the preceding subsections of this chapter)  
is that the stable  shocks necessarily are supersonic in the media where they propagate.  
The latter means that the continuous rarefaction flow  
in the region of phase transition  is mechanically unstable as well,  
since  a rarefaction shock, if  it appears, 
propagates inside the fluid faster than the sound wave and, hence,  
it should change the fluid's state.    
Due to this reason   the unstable hydrodynamic solutions simply do not appear \cite{llhydro,bugaevetal1}.

Applying these arguments to the BD model, one concludes:   
for the first order phase transition or strong cross-over the sound wave in 
the (generalized) mixed phase may be unstable and the strong discontinuities  
of the thermodynamic quantities  
are possible \cite{RAREF:1,RAREF:2,bugaevetal2}. The latter corresponds to   
the   non-trivial solution of the conservation 
laws (\ref{N53twsix}) and (\ref{N53twseven}), which allows    
a discontinuity of the   
distribution function on two sides of the HS $\Sigma^*$.  
Since there is twice the number of  conservation laws  compared to the  usual 
hydrodynamics,   
it is impossible, as shown below,   to build up  
the nontrivial solution of Eqs. (\ref{N53twsix}) and (\ref{N53twseven}),  
if the distribution functions on both sides of the HS $\Sigma^*$,  i.e. $\phi_{in}$ and $\phi_{out}$, are    taken to be 
the equilibrium ones.

Consider first the space-like parts of the HS $\Sigma^*$. 
Then Eqs. (\ref{N53twsix}) and (\ref{N53twseven}) for $S_A = -1$ vanish identically 
because of  the inequality $p^\mu \partial_\mu {\cal F}^{*}(t^{*}, \vec{x}) > 0$, 
whereas for $S_A = 1$ Eqs. (\ref{N53twsix}) and (\ref{N53twseven})  recover   
the usual hydrodynamical conservation laws at  the discontinuity.   
However,  
it can be shown that the existence of  strong discontinuities across the space-like HSs,  
the {\it time-like shocks} \cite{timeshock,TIMESHOCKb},  
is rather  problematic because it leads to a contradiction  of  the basic 
assumptions adopted for the distribution function,   
even though  the conservation 
laws (\ref{N53twsix}) and (\ref{N53twseven}) are formally fulfilled.

Indeed, according to the Bogolyubov's classification \cite{BOGOL},  
a one-particle treatment can be established for   
a   typical time $\Delta t$ 
which, on one hand,   should be much larger than  
the collision time $\tau_{Coll}$,  and, on the other hand,  
it should be much smaller than the relaxation time $\tau_{Relax}$  
\begin{equation}\label{N53twnine} 
\tau_{Coll} \ll \Delta t \ll  \tau_{Relax}\,.  
\end{equation} 
Similar to the usual Boltzmann equation (see also discussions in \cite{BOGOL,BALESCU}), 
in  deriving the collision  terms of  Eq. (\ref{N53nineteen}) one implicitly   
adopted the requirement that 
the distribution function does not change  substantially for times $\Delta t$ 
less than the relaxation time $\tau_{Relax}$. 
However, at  the discontinuities on the space-like parts of $\Sigma^*$, 
suggested in  \cite{timeshock,TIMESHOCKb},   
the distribution function changes suddenly, i.e. $\Delta t = 0$,   
and  
the left inequality (\ref{N53twnine}) cannot be fulfilled 
at the {\it time-like shock}. 
Therefore, according to the Bogolyubov's classification \cite{BOGOL},  such  a process,   
 which is 
shorter than the typical collision time,  belongs to a  prekinetic or chaotic stage
and, hence,  cannot be studied  at  the level of a one-particle distribution 
function. 
It would instead  require  the analysis of  a  hierarchy of  $N$-particle   
distribution functions, where $N$ is the number of particles in the system.     
Thus, the existence of time-like shocks contradicts  the 
adopted assumptions  for a one-particle distribution.
Their existence  should be  demonstrated  first 
within the higher order  distributions.  
This  statement  applies  to several  papers published by the
Bergen group during the last few years where   {\it time-like shocks }  were
attenuated  in time  using a  phenomenological quasi-kinetic approach \cite{LASZLO}.
For  the same reason, the use of equilibrium values  for temperature and chemical potential 
in an attenuated time-shock is rather problematic for time scales shorter than $ \tau_{Coll}$.
Note, however, that the discontinuities at  the time-like parts of $\Sigma^*$
(usual shocks)  have no such restrictions and, hence, in what follows I will analyze  
only   these discontinuities.

 
\subsection{Boundary Conditions at $\Sigma^*$ for a Single Degree of Freedom.} 
Now  it is necessary   to  find out  whether it is  possible 
to obtain 
the nontrivial solution of systems  (\ref{N53twsix}) and (\ref{N53twseven}) 
using the parts of  equilibrium distributions on  
the time-like segments  of the HS $\Sigma^*$.  
%
To simplify  the presentation,  first 
I consider  the same kind of particles in both domains.
It is convenient to transform  the coordinate system  
$(t^*(\vec x ); \vec x )$ into the special local frame
$(t^*_L (\vec x_L ); \vec x_L )$ which is
the rest-frame of  discontinuity   between the distributions $\phi_{in}$ and $\phi_{out}$.
This coordinate system will be indicated by the  subscript $L$.
The special local frame   is  defined as follows:   
the $x$-axis  should coincide with the local external normal  vector 
to the HS $\Sigma^*$,  $y$- and $z$-axes 
belong to  the tangent hyperplane of $\Sigma^*$.   
In this case the external  normal vector to the time-like parts of $\Sigma^*$ is
$n^*_\mu = (~0 ;~ \partial_1 {\cal F}^{*}_L; ~0; ~0)$, 
and one can readily check 
that the value of the derivative  $ \partial_1 {\cal F}^{*}_L$ plays
an important   role 
in the conservation laws (\ref{N53twsix}) and (\ref{N53twseven}) 
only through the   {\it cut-off} $\Theta$ function.
Then, as in the theory of  usual relativistic shocks \cite{llhydro,bugaevetal2,bugaevetal1}, 
it can be shown that equations for  the $y$- and $z$-components of system (\ref{N53twsix})     
degenerate into the identities because of the symmetries  of the  energy-momentum tensor. 
Therefore, the number of   independent  equations at the discontinuity is  7:  
a switch off criterion  
and six independent equations out of systems (\ref{N53twsix}) and (\ref{N53twseven}) 
($t$- and $x$-equations (\ref{N53twsix}) and one equation (\ref{N53twseven}) for 
two choices of $S_A = \{ -1; +1\}$).       
 
On the other hand the number of unknowns  is 6 only: temperature $T^*_{in}$ and  
baryonic chemical potential $\mu^*_{in}$ of the ``in'' domain,  
temperature $T^*_{out}$ and 
baryonic chemical potential $\mu^*_{out}$ of the ``out'' domain,   the collective velocity $v^*_{in}$
of the ``in'' domain  particles,
and the collective velocity $v^*_{out}$ of the particles of ``out'' domain,  
which  should be collinear   
to the normal vector $n^*_\mu$ in the rest-frame of the discontinuity.   
A formal counting of equations and unknown shows that it is impossible 
to satisfy the conservation laws (\ref{N53twsix}) and (\ref{N53twseven})  if  
the distribution functions on both sides are the  equilibrium ones.   
 
The last  result means that instead of  a traditional 
discontinuity one has to search for  a principally new boundary condition  
on the  HS $\Sigma^*$. The analysis  shows that there are two  such possibilities 
with the equilibrium  distribution function in the ``in'' domain and  
a special 
superposition 
of  
two  {\it cut-off} equilibrium  distributions for the ``out'' domain.    
The first possibility is to choose $\phi_{out}$  as follows: 
\begin{eqnarray}\label{N53thirty} 
\hspace*{-0.9cm}\phi_{out} \Bigl|_{\Sigma^*}  \hspace*{-0.2cm} & = &   
\phi_{in}~ (~T_{in}^*, ~\mu_{in}^*, ~v^*_{in} )~~ \Theta(~~ p^1  \partial_1 {\cal F}^*_L )  +    
%
%
%
\phi_{out} (T_{out}^*, \mu_{out}^*, v_{out}^*)~ \Theta(- p^1  \partial_1 {\cal F}^*_L )   
 \, ,   
\end{eqnarray} 
i.e. the distribution of outgoing particles from the ``in'' domain (the first term in the rhs of
Eq. (\ref{N53thirty})) is continuous on the HS $\Sigma^*$,
whereas  the distribution  of the  particles entering the  ``in'' domain   (the second term in  
the rhs of Eq. (\ref{N53thirty})) 
has a discontinuity on $\Sigma^*$ which 
conserves the energy, momentum and baryonic charge because of
the following boundary conditions
 ($ \nu = \{0; ~1\}$) 
\begin{eqnarray} 
\hspace*{-0.6cm}&&
T^{1 \nu}_{in -}~\, (T_{in}^*, \mu_{in}^*, v_{in}^*) ~ =~
\label{N53thone}
T^{1 \nu}_{out -}~(T_{out}^*, \mu_{out}^*, v_{out}^*)  \,, \\
\hspace*{-0.6cm}&&
N^{1 }_{in -} ~ (T_{in}^*, \mu_{in}^*, v_{in}^*) ~ =~
N^{1 }_{out -}~(T_{out}^*, \mu_{out}^*, v_{out}^*)
\label{N53thtwo} 
\, . 
\end{eqnarray} 
The above choice  of boundary conditions at  $\Sigma^*$  
reduces  systems (\ref{N53twsix}) and (\ref{N53twseven})  
for $S_A = 1$ to the identities, and, hence,   
from the systems (\ref{N53twsix}) and (\ref{N53twseven}) 
there remain only   three independent  equations   
(\ref{N53thone}), (\ref{N53thtwo}) for $S_A = - 1$.   
Along with a  switch off criterion,  these four equations can now be     
solved   for six  independent variables with 
the two variables  chosen to be free for a moment.  
Thus,   both  the outgoing and incoming parts of  
the distribution function (\ref{N53thirty}) 
can be chosen  as the equilibrium ones, but with different   
temperatures, chemical potentials and  non-zero relative velocity
 $v^*_{rel}  \equiv ( v^*_{out} - v^*_{in} ) / ( 1 -  v^*_{out}  v^*_{in} ) $
with respect to the distribution function $\phi_{in}$.  

Note a principal difference between this discontinuity  and all  the ones known  in
relativistic hydrodynamics:
the  ``out'' domain state consists, in general,  of two  different subsystems (fluxes) that have
individual hydrodynamic parameters. It is clear
that  it is impossible to reduce  three of those hydrodynamical parameters of one flux to 
those three of another flux 
because there are only two free variables out of six. 
Thus, together with the ``in'' domain flux  there are
in total  three fluxes involved  in this discontinuity.  Therefore, it is appropriate to
name it a {\it three flux discontinuity} in order to distinguish it from the ordinary shocks
that are defined by  maximum of two fluxes.

The outgoing component of the distribution (\ref{N53thirty}) 
coincides with the  choice of the boundary conditions 
suggested in  the  TLS model \cite{TLS:01}, whereas the 
equations (\ref{N53thone}) and (\ref{N53thtwo}) are missing in this model. 
For this  reason, the TLS model fails to conserve  energy, momentum and 
charge.  Note also that 
the lower values  of the  temperature $T_{out}^* \le T_{in}^* $ and 
baryonic  chemical potential $\mu_{out}^* \le \mu_{in}^* $,   
which are typical for the rarefaction process considered in \cite{TLS:01},  
should be compensated by an  extra flow from  the incoming particles to  
the ``in'' domain, i.e. $v_{rel}^*$ should be opposite to the external normal vector 
$n_\mu^*$ in the rest-frame of the  {\it three flux  discontinuity}.
Therefore, such a discontinuity is analogous 
to the compression shock wave in relativistic hydrodynamics,  
 and 
cannot appear in  the rarefaction process for    
any of the hadronic species considered in Ref. \cite{TLS:01}.

Similarly, one can find  
another non-trivial  solution of the systems (\ref{N53twsix}) and (\ref{N53twseven}) 
which   
corresponds to  opposite choice to Eq. (\ref{N53thirty})
\begin{eqnarray}\label{N53ththree} 
\hspace*{-0.9cm} 
\phi_{out} \Bigl|_{\Sigma^*} \hspace*{-0.2cm}  & = &   
\phi_{in}~ (~T_{in}^*, ~\mu_{in}^*, ~v^*_{in} )~~ \Theta(- p^1  \partial_1 {\cal F}^*_L )  +    
%
%
%
\phi_{out} (T_{out}^*, \mu_{out}^*, v_{out}^*)~ \Theta(~~ p^1  \partial_1 {\cal F}^*_L )   
 \, ,   
\end{eqnarray} 
i.e., the incoming to the ``in'' domain component of the 
distribution above (the first  term  in the rhs of Eq. (\ref{N53ththree})) 
is continuous on HS $\Sigma^*$, but 
the  component leaving  the ``in'' domain   has a  
discontinuity on $\Sigma^*$ which  obeys  
the following conservation laws ($ \nu = \{0; ~1\}$):   
\begin{eqnarray} 
%
\hspace*{-0.6cm}
&&
T^{1 \nu}_{in +}~\, (T_{in}^*, \mu_{in}^*, v_{in}^*) ~ =~
\label{N53thfour}
T^{1 \nu}_{out +}~(T_{out}^*, \mu_{out}^*, v_{out}^*)  \,, \\
%
%
\hspace*{-1.2cm}
&& 
N^{1 }_{in +} ~ (T_{in}^*, \mu_{in}^*, v_{in}^*) ~ =~
N^{1 }_{out +}~(T_{out}^*, \mu_{out}^*, v_{out}^*)
\label{N53thfive} 
\, . 
\end{eqnarray} 
It is  clear that both the outgoing and incoming components of the 
distribution (\ref{N53ththree}) can be chosen as the equilibrium distribution  
functions. A simple analysis of the system (\ref{N53thfour}), (\ref{N53thfive}) 
shows that for $T_{out}^* \le T_{in}^*$ and $\mu_{out}^* \le \mu_{in}^*$
the relative velocity $v_{rel}$  in the local  
frame should be collinear to the external normal vector $n_\mu^*$.
Such a discontinuity is analogous to the rarefaction 
shock wave in the relativistic hydrodynamics.  
 Thus,  in contrast to  the  TLS choice, Eq. (\ref{N53ththree}) should 
be used  as the initial conditions for  the ``out'' domain 
while studying  the rarefaction process of  
 matter  with anomalous thermodynamic properties. 
 
Now it is appropriate  to discuss   how  
the non-trivial solutions (\ref{N53thirty}) and (\ref{N53ththree})  
will modify the system of the hydrocascade equations  
(\ref{N53nineteen}), (\ref{N53twthree}) and (\ref{N53twfour}).   
In what follows I  assign  the hydrodynamic equations 
to the  ``in'' domain and the cascade  ones to the 
``out'' domain (the opposite case 
can also be considered).   Inserting  (\ref{N53thirty}), (\ref{N53thone}) and (\ref{N53thtwo})   
into the ``in'' Eqs. (\ref{N53twthree}), (\ref{N53twfour}),  
and into the ``out''  Eq. (\ref{N53nineteen}), one obtains the following  
system: 
\begin{eqnarray} 
\label{N53thsix} 
\hspace*{-0.6cm}&&\Theta_{in}~ {\partial}_\mu ~ T^{\mu \nu}_{in}  =   \int \frac{d^3 p }{p^ 0}~  p^\nu 
C_{in}^{II} (x,p) \, ,  \\   
\label{N53thseven} 
\hspace*{-0.6cm}&&\Theta_{in}~ {\partial}_\mu ~ N^{\mu}_{in} = 0 \, , \\ 
\label{N53theight} 
\hspace*{-0.6cm}&&\Theta_{out}~ p^\mu  ~{\partial}_\mu ~ \phi_{out} (x,p) =  C_{out}^{I} (x,p) + 
C_{out}^{II} (x,p) \, ,   
\end{eqnarray} 
i.e., due to the boundary conditions (\ref{N53thirty}) -- (\ref{N53thtwo})   
the $\delta$-like terms have disappeared from the original system of equations.  
It is clear also that the source  term in the rhs of Eq. (\ref{N53thsix}) 
does not play any role because it is finite on the HS $\Sigma^*$ 
and   vanishes everywhere outside  $\Sigma^*$.   
 
In order to obtain the system of hydrocascade equations  
(\ref{N53thsix}) -- (\ref{N53theight}) 
for the non-trivial solution defined by Eqs. (\ref{N53ththree}) -- (\ref{N53thfive}), 
the hydrodynamic description  
has to be extended  
to the  outer $\varepsilon$-vicinity  
($\varepsilon \rightarrow 0$)  of the 
HS $\Sigma^*$  
\begin{eqnarray} 
\label{N53thnine} 
&&\Theta_{out}~ {\partial}_\mu ~ T^{\mu \nu}_{out}  =   \int \frac{d^3 p }{p^ 0}~  p^\nu 
C_{out}^{II} (x,p) \, ,  \\ 
\label{N53forty} 
&& \Theta_{out}~ {\partial}_\mu ~ N^{\mu}_{out} = 0 \, ,  
\end{eqnarray} 
which  in practice  means that for Eqs. (\ref{N53ththree}) -- (\ref{N53thfive}) 
one has to  solve  the cascade equation 
(\ref{N53theight})  a bit inside of the ``out'' domain   
infinitesimally close to  $\Sigma^*$
in order to remove  the  $\delta$-like term in  (\ref{N53theight}) and 
move this term  to the discontinuity on the HS $\Sigma^*$.  
 
The remarkable feature of the system of hydrocascade equations  
(\ref{N53thsix}) -- (\ref{N53forty}) is that   
each equation automatically vanishes outside  the domain where it 
is specified. Also, by the construction, it is free of the principal 
difficulties of the BD and TLS models discussed above.   
The question how to conjugate the {\it three flux discontinuity} with the solution of the hydro 
equations  (\ref{N53thsix}), (\ref{N53thseven}), (\ref{N53thnine}) and (\ref{N53forty}) will be discussed
in the next section.


 
\subsection{Boundary Conditions at $\Sigma^*$ for Many Degrees of Freedom.} 
In order to apply the above results to 
the description of the QGP-hadron gas phase transition  occurring   in relativistic 
nuclear collisions  
it is necessary to take into account the fact that the real situation differs 
from the previous consideration in two respects. 
The first one is that  in the realistic case inside  the ``in'' domain there should  
exist the QGP, whereas it should not appear in the ``out'' domain.   
Of course, the discussion of the QGP kinetic theory is a much more complicated problem 
and  lies  beyond the scope  of this analysis.  
For my  purpose it is sufficient to generalize the equations of motion (\ref{N53thsix}) - (\ref{N53forty})  inside domains 
and the conservation laws (\ref{N53twsix}) and (\ref{N53twseven}) between these  domains to the realistic case.   
Such a generalization can be made  because in the case of the QGP-hadron gas phase transition
there will  also be an exchange
of particles between  the ``in'' and ``out''  domains which must  be accounted for by the $\delta$-like
source terms in the transport equations. The only important difference from the formalism
developed in the preceding   sections is  that QGP must   hadronize
while entering 
the ``out'' domain, whereas the  hadrons should  melt while entering the ``in'' domain. 
Note, however, that in relativistic hydrodynamics one has 
to assume that all reactions, i.e.  the QGP hadronization and melting of hadrons  in this case,
occur instantaneously.  Under this assumption  
one can justify the validity of 
the equations of motion (\ref{N53thsix}) - (\ref{N53forty})  and the conservation laws between
the QGP and the hadron gas at  the boundary $\Sigma^*$.
 
The second important fact to be taken into account is that some hadrons  
have the large scattering cross-sections with other particles and some hadrons  
have the small cross-sections,  because of this,   different hadrons   participate in the  
collective flow differently.  
A  recent effort \cite{Bugaev:01d,Bugaev:02} to classify the inverse slopes of the  
hadrons at SPS lab. energy  158 GeV$\cdot$A      
led to the  conclusion that   
the most abundant  hadrons, e.g.  pions, kaons,  (anti)nucleons, 
$\Lambda$ hyperons e.t.c., participate in the hadron  rescattering and resonance decay 
till the very late time of expansion,  
whereas $\Omega$ hyperons, $J/\psi$ and $\psi^\prime$ mesons    
practically do not interact with the  hadronic media and, hence,    
the FO of their transverse  momentum spectra  ({\it kinetic FO}) may occur just at 
hadronization temperature $T_H$. Therefore,  
the inverse slopes of the $\Omega$, $J/\psi$ and $\psi^\prime$ particles 
are a combination of the thermal motion and the transversal expansion of 
the media from which these particles are formed.   
  
These results for the $\Omega$ baryons and $\phi$ mesons were obtained  
within the BD and TLS models, whereas for the $J/\psi$ and $\psi^\prime$ mesons 
it was suggested for the first time in Refs.  \cite{Bugaev:01d,Bugaev:02}.  
Later   these results  
were  further refined in Ref. \cite{Bugaev:02a}  by the simultaneous fit with 
 only  
one free parameter (the maximal value of transversal velocity)    
of the measured  $\Omega$ \cite{Omega,Omega1}, $J/\psi$ and $\psi^\prime$ \cite{Jpsi}  
transverse momentum spectra  
in Pb+Pb collisions at 158 GeV$\cdot$A  
that are frozen-out at hadronization temperature $T_H$.  
The experimental  situation with the $\phi$ mesons at SPS   
is, unfortunately,  not clarified yet  because the  results of  
the NA49  \cite{phiA}  and NA50 \cite{phiB}  Collaborations  disagree.  
The  analysis of the  transverse momentum spectra of  
$\Omega$ hyperons  \cite{Bugaev:02b, Bugaev:03,RHIC:Hadronization3} 
and $\phi$ mesons  \cite{Bugaev:02b} 
reported   by the STAR Collaboration for  energies  
$\sqrt{s} = 130$ A$\cdot$GeV in Refs. \cite{RHIC:Hadronization2} and \cite{RHIC:Hadronization4}, respectively,  
  and for   
$\sqrt{s} = 200$ A$\cdot$GeV in Ref. \cite{RHIC:Hadronization3}    
shows that this picture remains valid for RHIC energies as well.  
 
It is easy to find  that for  particles like  
$\phi$, $\Omega$, $J/\psi$ and $\psi^\prime$,
which    weakly interact    with other hadrons, 
the distribution function $\phi_{out}$ 
should coincide with  $\phi_{in}$ 
\begin{equation}\label{N53foone} 
\hspace*{-0.5cm}
\phi_{out} \Bigl|_{\Sigma^*}    =   
\phi_{in}~ (~T_{in}^*, ~\mu_{in}^*, v^*_{in})~~ \Theta(~~ p^1  \partial_1 {\cal F}^*_L )  \,, 
\end{equation}  
where, in contrast to (\ref{N53thirty}), there  is no incoming component of 
the distribution because the non-interacting particles cannot rescatter  
and  change their velocity. 
Note also that a small modification of the   
incoming part of
$J/\psi$ momentum distribution  due to decay of heavier charmonia in the ``out'' domain
can be safely neglected.
Remarkably,  the cascade initial condition (\ref{N53foone}) exactly coincides  
with the one used in the TLS model. Therefore,  
the main TLS conclusions \cite{TLS:01} on the $\phi$ mesons and  $\Omega$ hyperons  
remain unchanged, whereas  for hadrons with large scattering cross-sections the TLS 
conclusions may change significantly.

Omitting the contributions of weakly interacting hadrons  from the components
of the energy-momentum tensor and baryonic  4-current,
one can generalize the boundary conditions
(\ref{N53twsix}) and (\ref{N53twseven}) on the  HS $\Sigma^*$ between  
the domains  and formulate the energy-momentum and charge conservation laws 
in terms of the parts  of the {\it cut-off}  distribution functions.  
For definiteness I will consider the first order phase transition between 
QGP and hadronic matter through out the rest of this section. The case of the second order  
phase transition can be analyzed similarly.  
In terms of the local coordinates $(t^*_L (\vec x_L ); \vec x_L )$, 
introduced previously, the conservation laws (\ref{N53twsix}) and (\ref{N53twseven}) 
can be generalized as follows  ($ \nu = \{0;~1\}$) 
\begin{eqnarray} 
\hspace*{-1.0cm}
\alpha_q 
\sum_{Q = q, \vec{q}, \ldots} 
T^{1 \nu}_{Q \pm} ~(~T_{in}^*, ~Z_Q\cdot\mu_{in}^*, ~ v^*_{in}) & + & 
%
(1 - \alpha_q)  \sum_{H = \pi, K, \ldots} 
T^{1 \nu}_{H \pm} (~T_{in}^*, ~Z_H\cdot
\mu_{in}^*, ~ v^*_{in})  = \nonumber \\
\label{N53fotwo} 
%
&& 
\sum_{H = \pi, K, \ldots} 
T^{1 \nu}_{H \pm} (T_{out}^\pm, Z_H\cdot\mu_{out}^\pm, v^\pm_{out})
\,, \\  
%
%
\hspace*{-1.0cm}
%
\alpha_q 
\sum_{Q = q, \vec{q}, \ldots} 
N^{1 }_{Q \pm} ~(~T_{in}^*, ~Z_Q\cdot\mu_{in}^*, ~ v^*_{in})  & + &
%
(1 - \alpha_q)  \sum_{H = \pi, K, \ldots} 
N^{1 }_{H \pm} (~T_{in}^*, ~Z_H\cdot\mu_{in}^*, ~ v^*_{in})  = \nonumber \\
\label{N53fothree} 
\hspace*{-1.0cm}
&& 
\sum_{H = \pi, K, \ldots} 
N^{1 }_{H \pm} (T_{out}^\pm, Z_H\cdot\mu_{out}^\pm, v_{out}^\pm)
\end{eqnarray} 
where $\alpha_q$ is the volume fraction of the QGP in a  mixed phase, and 
the $Q$-sums of the energy-momentum tensor and baryonic 4-current components,
denoted as 
%
\begin{eqnarray}
\hspace*{-0.6cm}&&  
\label{N53fofour}
 T^{\mu \nu}_{Q \pm}  \equiv
\int \frac{d^3 p }{p^ 0}~ p^\mu p^\nu ~ \phi_{Q} (x, p) ~  
\Theta(\pm~p^\rho \partial_\rho {\cal F}^{*}) \, , \\
\hspace*{-0.6cm}&& 
\label{N53fofive}
 N^{\mu }_{Q \pm}  \equiv
\int \frac{d^3 p }{p^ 0}~ p^\mu ~ Z_Q \phi_{Q} (x, p)~   
\Theta(\pm~p^\rho \partial_\rho {\cal F}^{*}) \, , 
\end{eqnarray} 
run over all corresponding  degrees of freedom of QGP.
The  $H$-sums  also run  
over all hadronic degrees of freedom. In Eqs. (\ref{N53fotwo}) and (\ref{N53fothree}) 
$Z_Q$ and $Z_H$ denote the baryonic  charge of the corresponding particle species.

Now from Eqs. (\ref{N53fofour}) and (\ref{N53fofive})  it is clearly seen that the correct 
hydrocascade approach  requires   more detailed  information 
about  the microscopic properties of QGP than  is  usually provided by  traditional  
equations of state.  To proceed further  I  assume that those 
components are known. The general approach to calculate  the angular and momentum
integrals in Eqs. (\ref{N53fofour}) and (\ref{N53fofive}) was developed in Ref. \cite{Bugaev:99}
and was  applied to  the massive Boltzmann gas description in \cite{LPC:TN,Bugaev:99}.

The important difference between  conservation laws 
(\ref{N53fotwo}), (\ref{N53fothree})  and (\ref{N53twsix}), (\ref{N53twseven}) is that 
in the ``out'' domain  
the temperature $T_{out}^-$, chemical potential $\mu_{out}^-$ and relative velocity $v_{out}^-$ 
of the incoming to $\Sigma^*$ hadrons should differ 
 from the corresponding quantities $T_{out}^+$, $\mu_{out}^+$ and $v_{out}^+$ of the outgoing 
from $\Sigma^*$ particles, 
and both sets should  differ from the quantities  $T^*_{in}$, $\mu_{in}^*$ and $v^*_{in}$ of the ``in'' domain.
In order to prove this statement,  it is necessary to 
compare  the number of equations and number of unknowns for the two distinct cases.  
Namely, (i) if the initial state  is in  the mixed QGP-hadron gas  phase, and (ii) if  the initial state 
belongs to the  QGP. 

In case (i) there are 10 equations and 10 unknowns:

$\bullet$ The  equations are as follows: 
6 conservation laws from Eqs. (\ref{N53fotwo}) and (\ref{N53fothree}); 
 value of the  initial energy;  value of the initial baryonic density; the relation between 
initial temperature $T^*_{in} $ and the baryonic chemical potential $\mu^*_{in}$ taken
at the phase boundary; and the switch off criterion.

$\bullet$ The unknowns are as follows: three temperatures $T^*_{in}$, $T_{out}^-$, $T_{out}^+$;
three chemical potentials $\mu_{in}^*$, $\mu_{out}^-$, $\mu_{out}^+$; three
velocities $v^*_{in}$, $v_{out}^-$, $v_{out}^+$ defined in the rest frame  of  a discontinuity;
and the QGP fraction volume $\alpha_q$.   

Thus, in this case,  one can find a desired solution of the system of ten transcendental equations,
which is the most general form of the three flux discontinuity introduced 
by Eqs. (\ref{N53thirty} ) - (\ref{N53thtwo}).

To  complete the solution of  hydro equations  (\ref{N53thsix}), (\ref{N53thseven}), (\ref{N53thnine})
and (\ref{N53forty}) one  must  find  the value of velocity  $v^*_{in}$ from the system of
ten transcendental equations discussed above.  This velocity then 
defines an ordinary  differential equation 
$d x^1_L / d t^*_L =  - v^*_{in}$  for the HS $\Sigma^*$ in the rest frame of matter of the ``in'' domain, which  must  be solved
simultaneously with the hydro equations.

If initial state belongs to the interior of the QGP phase, case (ii), 
then the usual  hydro solution will be valid till
the system reaches the boundary with the mixed phase, from which the non-trivial 
discontinuity described by Eqs. (\ref{N53fotwo}) and (\ref{N53fothree}) will  begin.
The differences  from  the previously considered case are now clear: 
in contrast to case (i),
the volume fraction of QGP  is fixed to unit $\alpha_q = 1$;
 the energy and baryonic charge densities are no longer independent, but
are completely  defined by the temperature and baryonic  chemical potential, which 
 are connected by the entropy conservation for the continuous hydro solution
in QGP.

Therefore, in case (ii) there are 9 equations and 9 unknowns, which are as follows:

$\bullet$ The  equations are: 
6 conservation laws from Eqs. (\ref{N53fotwo}) and (\ref{N53fothree}); 
temperature dependence of  the baryonic chemical potential $\mu^*_{in} = \mu^*_{in} (T^*_{in} )$   due to the
entropy conservation; the relation connecting  
 temperature $T^*_{in} $ and  baryonic chemical potential $\mu^*_{in}$, since they belong
to the phase boundary; and the switch off criterion.

$\bullet$ The unknowns, except for the fixed volume fraction $ \alpha_q = 1$,
 are the same as in case (i). 

Again the number of unknowns matches the number of
equations, and the procedure to solve  the system of hydro equations 
(\ref{N53thsix}), (\ref{N53thseven}), (\ref{N53thnine}) and (\ref{N53forty}) simultaneously with the
 boundary conditions (\ref{N53fotwo}) and (\ref{N53fothree}) is the  same as in case (i).


Now it is appropriate  to discuss  
the switch off criterion ${\cal F}^{*}(t, \vec{x}) = 0$  in more details.   
By the construction of  the hydrocascade  approach,   
the cascade treatment should be applied  when hydrodynamics   
starts to lose its applicability:
according to the original assumption
 the hydro equations (\ref{N53thsix}), (\ref{N53thseven}), (\ref{N53thnine}) and (\ref{N53forty}) work well
inside of the 4-volume surrounded by 
the HS $\Sigma^*$  and in the outer $\varepsilon$-vicinity 
($\varepsilon \rightarrow 0$) of  $\Sigma^*$
 [see also a discussion after Eq. (\ref{N53theight})],  whereas just   outside of 
this domain  the thermal equilibrium dismantles and  one 
cannot  use the {\it cut-off} equilibrium distributions  interior of 
the  ``out'' domain.   
Consequently,  a  switch off criterion should be formulated  
solely for some  quantity  defined in  the  outer $\varepsilon$-vicinity
of HS $\Sigma^*$, and it has to define the  bounds of applicability of 
thermal equilibration and/or hydrodynamic description.
Note that
in   the BD and TLS models this did not matter because  
both groups kept the cascade  initial conditions as close  as possible to the output  of 
hydro.  However, in the  case of the {\it three flux discontinuity}  on the time-like 
parts of  HS $\Sigma^*$ the proper use 
of the  switch off criterion   plays a decisive role in the  
construction  of the mathematically correct hydrocascade solution (see also a 
discussion of the FO criterion in Refs. \cite{Bugaev:96,Bugaev:99}).  
It is clear that, in contrast to the BD and TLS formulations,   
the switch off criterion
may generate a  sizable effect while applied to interior of  hadronic phase.
This is so,  because  even a small difference (just a few MeV)  between 
the temperature $T^*_{in}$, which belongs to the phase transition region, 
and temperatures $T_{out}^-$ and $T_{out}^+$ of the ``out'' domain may lead to a tremendous  flow
of outgoing  hadrons  because of the enormous latent heat of the QGP.

\vspace*{1.cm}

\section{Concluding  Remarks}  


In this chapter the solution of the FO problem
in relativistic hydrodynamics is given  within
a zero width approximation for  the  FO HS.
I analyzed the difference between the \CFful procedure
and its \COful generalization for the t.l. FO HS.
I showed  that
a reformulation of the traditional hydrodynamic  is
necessary in order
to include the particle emission from the t.l. parts of the FO HS.
The modified self-consistent hydrodynamics with the specific boundary
conditions is formulated and  different types
of shock-like FO are studied.
It is shown that the momentum  spectra
 of the particles emitted from the t.l. parts of the FO HS 
are described by the cut-off distribution function
and particle emission looks as a discontinuity
in the hydrodynamical motion (FO shock). This is the new 
kind of hydrodynamic discintinuities.

The correct boundary conditions enable me to derive the equations of
motion of the fluid alone, which do not contain any source term.  I
have also proved the energy-momentum  and charge conservation in the integral
form for the hydrodynamics with specific boundary conditions, if the
\COful  distribution function is used.

I analyzed the equations for the t.l. 
parts of the FO HS and showed how to solve them together with
the equations of motion of the fluid.  A complete
analysis of the FO problem for the t.l. s parts of the FO HS in
1+1 dimensions is presented.
The entropy growth and mechanical stability conditions
as well as the ``recoil problem''
for the shock-like FO are studied.
As an application of the general scheme, the
FO of the simple wave is considered, and analytical solution of
this problem for the massless gas of free particles is given.
The spectra of the emitted particles are calculated
and compared to those obtained by the \CFful  FO  procedure.

Then I considered a more complicated, but mathematically
similar problem: the correct boundary conditions for hydrokinetic 
approach. 
To reach this goal I have derived  a 
system of relativistic kinetic equations which describe the particle exchange 
between two domains separated by the HS of arbitrary properties.
I showed that the usual Boltzmann equation for the following sum of two distributions  
$\Phi(x,p) \equiv \Theta_{in}~\phi_{in}(x,p) + \Theta_{out}~\phi_{out}(x,p) $
automatically follows from the derived system, but not vice versa. 
Integrating the kinetic equations  I derived the system of the hydrocascade
equations for a single degree of freedom. Remarkably, 
the conservation laws on the boundary between two domains
conserve the incoming and outgoing components of the energy, momentum and
baryonic charge separately leading to  twice the number of  conservation laws
on the separating HS compared to the usual relativistic hydrodynamics.
Then  I showed that for a single degree of freedom 
these boundary conditions between domains
can  be satisfied  only by a special superposition of two {\it cut-off} equilibrium
distributions for the ``out'' domain. Since the obtained discontinuity
has three irreducible fluxes,  it is named a {\it three flux discontinuity},
in contrast to usual shocks defined by
two fluxes.   
It was also shown that the TLS-like choice of the boundary conditions, in contrast to
expectation of  \cite{TLS:01}, corresponds to an analog of the compression shock in traditional
hydrodynamics, and, therefore,  cannot be used to model the rarefaction process.

Then I showed that  existence of the {\it time-like shocks} \cite{timeshock,TIMESHOCKb},
formally rederived by this formalism,  contradicts  the usual assumptions adopted for the one-particle distributions
and, hence, the solution of this problem  requires the analysis of 
higher order distribution functions.   
Therefore, in the rest of this chapter  I concentrated on 
a detailed  analysis of the discontinuities at the
time-like HSs, i.e.  the space-like shocks in terms of Refs. \cite{timeshock,TIMESHOCKb}.  These results were then generalized to a more realistic
case:   when the mixed QGP and hadron gas  phase    exists  in the ``in'' domain and hadrons exist 
in the ``out'' domain. Such a generalization  also required the exclusion of the hadrons with  
the small scattering cross-section (like $\Omega$, $J/\psi$ and $\psi^\prime$ particles) from the boundary conditions between domains. 
As was showed in the preceding subsection, the presence of  the first order phase transition makes the resulting 
system of transcendental equations more complicated than in the case of a single degree of freedom.

It turns out  that 
a  minimal number of  variables in this discontinuity is either 9 or 10, depending on the
location of the initial state on the phase diagram.  Therefore, on the hadronic  side
the {\it three flux discontinuity}
 should have  two different flows with their own temperatures, chemical potentials and
collective velocities.
This    solution has a number of unique features in comparison with usual shocks:

\begin{itemize}
\item  This discontinuity may generate  a very strong, explosive like, flow of outgoing particles 
from the  ``in'' domain, first,  because  a huge latent heat of QGP is involved, and,
second, due to an extra  momentum associated with the {\it cut-off} distribution.
Indeed, considering the outgoing component of the distribution 
$\phi_{out} ~ \Theta( p^1  \partial_1 {\cal F}^*_L ) $  for massless pions  in the
frame where this function maximally resembles the non-cut Boltzmann distribution, 
i.e. in the rest-frame of the latter,
one finds  a nonvanishing  collective velocity $v_\pi = \frac{ (1 + v_\sigma) }{2}$.
Here  $v_\sigma \equiv  \frac{ d R_\perp }{d t} ~(|v_\sigma| \le 1$  for time-like parts of $\Sigma^*$) denotes 
 the transversal radius velocity in this frame.
\item The strong explosive flow of outgoing particles is localized at the time-like parts of the 
HS $\Sigma^*$, whereas  at the space-like parts of $\Sigma^*$
there will be a continuous flow. It is even  possible that for some choice of parameters
the space-like boundary may be absent.
\item The particle density  of outgoing pions will  strongly depend  on the 
speed of the transversal radius expansion. Thus,  for  massless pions
the particle density found according to the Eckart definition \cite{groot} is
$\rho_\pi = \frac{ \rho_\pi (T^+_{out})  }{4} \sqrt { (1 -  v_\sigma)^3 (3 + v_\sigma) }$, 
i.e. it is smaller for all $v_\sigma > -1 $ than the thermal particle density $\rho_\pi (T^+_{out}) $.
Therefore, the two particle correlations off  the low particle density regions should be
reduced. Since  the situation   $v_\sigma > > -1 $ is typical  for the beginning of 
the transversal expansion \cite{TLS:01},    the main contribution to the transversal pion correlations  
will come from the  later times of expansion. Thus, it is possible that the space-time region which 
defines 
the side and out  pion correlation radii will be essentially more localized both in space and time 
than in traditional hydrodynamic solutions.
\item Because there are two fluxes in the ``out'' domain, they will interact with each other.
The resulting distribution should be, of course, found by the cascade simulations, but
it is clear that the fastest  of them will decelerate and the cold one will reheat.
Besides the possibility  to accelerate or decelerate the outgoing transversal  flow more rapidly than in the BD and
TLS models,  the {\it three flux discontinuity} may naturally generate some {\it turbulence} patterns
in the ``out'' domain. 
\end{itemize}
Taking into account all these features alone with the fact that neither the BD nor TLS boundary
conditions have such a strong discontinuity, I conclude that the {\it three flux discontinuity}
opens a principally new possibility not only to resolve the HBT puzzle \cite{QM:04},
but also  to study some  new phenomena, like a turbulence pattern, associated with 
a new kind of shock, 
a {\it three flux discontinuity}, in relativistic hydrocascade approach.

Despite the reasonably good description of the one-particle
spectra of the most abundant hadrons, even such sophisticated model as the TLS one
badly overestimates both
of the transverse radii measured by pion interferometry  like other hydrodynamic models.
This is a strong indication that the hydro part of  all existing hydrocascade and
hydrodynamic  models requires an essential revision.
How this revision will affect the present BD and  TLS results is unclear at the moment,
but the solution of the HBT puzzle \cite{QM:04} should serve as a good
test for the correct picture of the space-time evolution during the post-hadronization stage.
The  additional tests for  the
correct  hydrocascade equations  should be 
the reproduction of  three  recently  established  signals of the deconfinement  phase transition,
i.e.
the pion Kink \cite{Horn,Kink} seen at lab. energy of $\sim 30$ GeV$\cdot$A, 
 the  $K^+/ \pi^+$ peak  at the same lab. energy  \cite{Horn}  
(the Strangeness Horn)
and the  plateau \cite{Step} in the inverse slope of the $K^\pm$  transverse momentum 
spectra 
at the whole range of the SPS energies (the Step in caloric curves)  
 measured by the  NA49 Collaboration  \cite{SPSKaonsA,SPSKaonsB}.
It is  also  necessary to check  other predictions of the Statistical Model of the Early Stage \cite{Horn}, 
namely the anomalies in  the  entropy to energy fluctuations  \cite{Fink} (the ``Shark Fin'') and in 
strangeness to energy fluctuations \cite{Well}  (the ``Tooth''), because  both the ``Shark Fin'' and  ``Tooth''
may be sensitive to the turbulence behavior due to  energy dissipation.

Note, however, that the completion of this task requires an additional research of the 
hydrocascade approach. First, it is necessary to develop further the microscopic models
of the QGP equation of state in order to find out the required by Eqs. (\ref{N53fotwo}) - (\ref{N53fofive})
components of the {\it cut-off}  energy-momentum tensor and baryonic 4-current.  
This can be done, for example, within 
the phenomenological extensions \cite{Goren:81,JL:94,Blaschke:03} of  
the  Hagedorn model. Second, a similar problem for hadrons should be solved as well, otherwise,  
as  I  discussed above,
the {\it switch off} criterion from the hydro to cascade cannot be formulated correctly 
within the hydrocascade approach.
And, finally,  for practical modeling it is necessary to formulate a mathematical algorithm to  solve simultaneously  the system of hydrocascade equations (\ref{N53thsix}) - (\ref{N53forty})  with the boundary conditions (\ref{N53fotwo}) and (\ref{N53fothree})
between the hydro and cascade domains. 
These problems, however,  are out of the scope of the present dissertation.


\def\jpsi{$J/\psi$\,}
\def\psip{$\psi^{\prime}$\,}

\def\freezeout{FO\,}


\chapter{Experimental Signals of the Deconfinement  Transition}

The  concepts of chemical (hadron multiplicities) and kinetic
(hadron momentum spectra) FOs were introduced to interpret data on
hadron production in relativistic A+A collisions.
The equilibrium hadron gas  model describes remarkably well
the light hadron
multiplicities measured in A+A collisions at the SPS \cite{Hgas, Hgas:2, Hgas:22}
and RHIC \cite{RHIC:Chem, RHIC:Chem2} energies, where the creation of the
QGP is expected.
Recently it was found \cite{GG:99} also that   charmonium
yield systematics in nuclear collisions at
the SPS \cite{na50} follow the pattern predicted by the hadron gas model.
The hadronization temperature parameter extracted from the fits to the
hadron multiplicities is
similar for both energies: $T_H=170\pm 10$~MeV.
It is close to an
estimate of the temperature $T_C$ for
the
QGP to hadron gas  transition at zero baryonic density.

Experimental results on inclusive hadron spectra and
correlations
 evidence for a hydrodynamic expansion of the matter
created in heavy ion collisions.
Strong transverse flow effects in Pb+Pb collisions at the SPS (average
collective transverse velocity is approximately $0.5\div 0.6$) are firmly
established from the combined analysis
\cite{Ka,PionFR,Wi:99} of the pion transverse mass spectra and
correlations.
The kinetic (`thermal') FO of pions and nucleons seems to
occur at a rather late
stage of an A+A reaction.
The thermal FO temperature parameter of pions
measured by the NA49 Collaboration \cite{na49} for central Pb+Pb collisions
at the SPS is
$T_f\cong 120$~MeV.

Further exploration of  idea of the statistical
$J/\psi$ production \cite{GG:99} led me and my collaborators to the formulation of the
hypothesis that the kinetic FO of $J/\psi$  and $\psi^{\prime}$
mesons takes place directly at hadronization \cite{Bugaev:01d,Bugaev:02}.
This means that  chemical and thermal
FOs occur simultaneously for those mesons and they, therefore,
carry information on the flow velocity of strongly interacting matter just
after
the  transition to the hadron gas  phase.
A possible influence of
the effect of rescattering
in the hadronic phase on the transverse momentum ($p_T$) spectra
was recently studied
within
a hydrocascade approach \cite{BD:00,TLS:01}.
As I discussed in the preceding chapter, 
the hydrocascade approach splits  the  
A+A collisions into three stages:
hydrodynamic QGP
expansion (``hydro''),  
transition from QGP to hadron gas  (``switching'') 
and the stage of hadronic rescattering and resonance decays
(``cascade'').
The switching from hydro to cascade takes place at $T=T_C$, where the
spectrum of hadrons leaving the surface of the QGP to hadron gas  transition is taken
as an input for the subsequent cascade calculations.
The results \cite{BD:00,TLS:01} suggest that
the $p_T$ spectrum of $\Omega$s is only weakly affected during
the cascade stage. The corresponding calculation for charmonia
are not yet performed within this model, but a similar result may be
expected due to their very high masses and low interaction cross sections.

In my next work \cite{Bugaev:02a} devoted to the analysis of the SPS data
it was demonstrated that the  measured
transverse mass ($m_T=\sqrt{p_T^2 + m^2}$)  spectra of
$\Omega^{\pm}$ hyperons \cite{Omega} and
$J/\psi$  and $\psi^{\prime}$ mesons \cite{Jpsi}
produced in Pb+Pb  at
158~A$\cdot$GeV  collisions
 can be reproduced within a hydrodynamical approach
using the same
FO parameters: hadronization temperature
$T\cong 170$~MeV and the mean transverse flow
velocity  $\overline{v}_T\cong
0.2$.
Within such an approach the value of the $\overline{v}_T$
parameter extracted in this way should be interpreted as
the mean flow velocity of the hadronizing QGP.

The next step was, of course, to verify such a hypothesis using the available  RHIC data. 
Employing the estimates of the hydrocascade model \cite{TLS:01} for the transverse velocty it was possible to predict \cite{Bugaev:02b} the  inverse slopes of the  early hadronizing particles which, within the error bars,  agreed with the results for the $\Omega^{\pm}$ hyperons reported by  the STAR Collaboration at the  conference ``Quark Matter 2002`` \cite{QM02} for the center of mass energy $\sqrt{s_{NN} } = 130$ GeV. 
Since up to now the charmonia data are not available at 
RHIC energies, to test the hypothesis of early hadronization it was possible 
to use the measured transverse spectra of multistrange
hadrons - $\phi$-meson and $\Omega$-hyperons.  Their fitting    
at fixed value of hadronization temperature $T = 170 \pm 5$~MeV, i.e. with a single parameter, allowed me
to determine the mean transverse collective velocity and predict the inverse slopes for 
the charmonia and charmed meson transverse spectra \cite{Bugaev:03}. Moreover,  it was possible to find out 
the mean emission volume at the QGP hadronization (see below) which agrees well  with 
the hydrocascade calculations estimate \cite{TLS:01}.

The analysis of the multistrange hadrons data once more indicated me a very special role
playing by the strange quark. The first solid evidence for such a role was demonstrated
by the Strangeness Horn  \cite{Horn}, i.e. by the peak of the ratio of  positive kaons to positive pions yields as the function of colliding energy. 
On the other hand, the existing hydrocascade  simulations \cite{TLS:01} demonstrated
that in the deconfinement region the inverse slopes of kaon transverse momentum spectra
are not affected by hadronic reactions after the QGP hadronization.
The reason for such a behavior was not clear and it initiated my   interest
in analyzing the experimental data. 

A qualitative analysis showed me that  the kaons  emitted from the deconfinement region
are affected  by the rescattering  and by the resonance decays. However,  it turned out that
both of these effects do modify the kaon distribution at soft momenta, but in opposite directions and, hence, they almost compensate each other. This is the qualitative  explanation  of the of the hydrocascade results on the kaon inverse slopes \cite{TLS:01}.
Then the existing data indicated a nonmonotonic  behavior of the  kaon inverse slopes
as the function of the colliding energy \cite{Step}.
The analysis of these data led me and my collaborators to the formulation of a
new signal of the deconfinement, known as the Step \cite{Step}. 

This chapter is based on the following works \cite{Bugaev:01d,Bugaev:02,Bugaev:02a,Bugaev:02b,Bugaev:03,Step}.

\section{Statistical Production of Particles.} 

\jpsi\, suppression in  nuclear collisions was suggested 
by T. Matsui and H. Satz to be a signal of the deconfinement PT.
Nowadays there are several models which mainly differ from the original 
idea of the \jpsi suppression in the mechanism of  charmonia dissociation
either in quark gluon  or in nuclear media. 
Such an approach I discussed in the subsection on the Mott-Hagedorn resonance gas
of the present dissertation.  
Recently a principally different picture of the statistical  
charm production was suggested \cite{GG:99}.
The next breakthrough in this field, the statistical coalescence model, suggests \cite{BMGK} 
that
charmonium is generated by the coalescence of earlier produced $c$ and $\bar{c}$
quarks. 
However, important questions, what happens after hadronization and what are 
the transversal mass spectra 
of \jpsi\, and \psip\, mesons,  were not considered in \cite{BMGK}. 
Therefore, I will mainly concentrate on the problem of the \jpsi and \psip\, FO
and will analyze their apparent temperature. Then I will answer
the question, ``What can we learn from experimental
data about hadronization of QGP?''

The idealized concept of chemical and thermal 
 FOs enables one to interpret data on hadron production
in relativistic A+A collisions.
The first experimental results on yields and transverse mass  spectra 
suggested the following scenario:
for the most abundant hadron species $(\pi, N, K, \Lambda)$ the chemical FO
which seems to coincide with the hadronization of the QGP, is followed by the 
thermal or kinetic FO
occurring at a rather late stage of the A+A reaction.
Thus,   
for the central Pb+Pb collisions at 158 A$\cdot$GeV the temperature of the chemical
FO was extracted from the fit of the multiplicity data to be $T_H = 175 \pm 10$ MeV
\cite{Hgas, Hgas:2}.
The  thermal FO parameters, i.e., temperature $T_F$ and averaged transversal velocity
$\langle v_T \rangle$,  for pions were determined
from the results of two pion correlations \cite{PionFR}
to be quite different

\vspace*{-0.3cm} 

\begin{equation}\label{pifr}
T_F = 120 \pm 12 \,\,{\rm MeV},\quad \quad \langle v_T \rangle = 0.5 \pm 0.12\,\,.
\end{equation}

On the other hand  from hadronic cascade simulations it is known that thermal FO\,   
of multistrange hadrons
($\phi, \Xi, \Omega$)
happens, probably, earlier than the kinetic FO of pions. 
Now the main question is, ``When does  the \jpsi FO occur? Late or early?'' 

The convolution of the transverse flow velocity of the 
matter element
with the thermal motion of hadrons in the rest frame of this element 
leads to a nearly  
exponential shape of final $m_T$ spectrum  
$\frac{1}{m_T} \cdot \frac{d~ N}{d~ m_T}
\approx C \cdot e^{\textstyle - m_T / T^* }$
with the apparent temperature (AT) $T^*$ which is defined
from the $p_T^2$ distribution at fixed longitudinal rapidity $y_p$ of the particle  as
\begin{equation}\label{tdef}
T^* = 
 - \left[ 2 m_T \frac {d} {d p_T^2} \ln \lp \frac {dN} {d y_p\, d p_T^2} \rp \right]^{-1}\,\,.
\end{equation}
In the limit  $m_T \gg T$ and $ \, p_T \rightarrow 0$  
the AT depends on the particle mass 
and $\langle v_T \rangle$
as follows 
\begin{equation}\label{tapp}
 T^*  = T_F + \alpha \cdot m \cdot \langle v_T \rangle^2 \,\,.
\end{equation}
For the spherical fireball Eq. (\ref{tapp}) was first 
found
in Ref.  \cite{Zim}.
For the cylindrical geometry, surprisingly,  
there are   two answers  which differ by the value of the coefficient $\alpha$: 
there is a {\it naive result} $\alpha = \frac{1}{2}$ \cite{heiz:99} and  the
{\it more elaborate} one $\alpha = \frac{2}{\pi}$ \cite{sin:99}. 
Therefore, before going further it is necessary to discuss the origin of
the difference between the results of Refs. \cite{heiz:99} and \cite{sin:99}.

\subsection{Particle Spectra and Apparent Temperature.}
The general  expression for the momentum distribution 
of the outgoing particle having the 4-vector of momentum
$p^\mu =$ $= (m_T\cosh y_p,\,\, p_T \cos(\phi_p),
\,\, p_T \sin(\phi_p), \,\, m_T \sinh y_p,)$ which is emitted from
the arbitrary FO HS  
$\Sigma$ by the 
fluid element having the hydrodynamical 4-velocity parameterized as 
$u^\mu = \gamma_T (\cosh y_L,\,\, v_T \cos(\phi_u),
\,\, v_T \sin(\phi_u), \sinh y_L,)$  is given by the {\it cut-off distribution function}  
\cite{Bugaev:96, Bugaev:99} (see also the preceding chapter)

\vspace*{-0.3cm} 

\begin{equation}\label{cutdf}     
E \frac{d^3 N}{d p^3} = \int_\Sigma p^\mu\,d \Sigma_\mu~
\varphi \lp {\textstyle \frac{p^\nu u_\nu}{T} } \rp~
{ \Theta\lp p^\rho\, d \Sigma_\rho \rp }
\,\,,
\end{equation}
where the standard  notations  are assumed for 4-vectors 
($y_p$, $y_T$ and $y_L$  are particle longitudinal, fluid transversal and 
fluid longitudinal rapidities, respectively, and $\phi$ denotes the  corresponding polar angles),  
$\gamma_T = 1/\sqrt{1-v_T^2}$ is the relativistic
 $\gamma$-factor, $d \Sigma_\mu$
denotes the 4-vector of external normal to the FO HS $\Sigma$
and $\varphi$ is the one-particle phase space distribution function.
The $\Theta$-function in (\ref{cutdf}) is of  crucial importance
because it makes sure that only outgoing particles are counted 
from both the t.l. and s.l. FO HS. 

For further analysis I will neglect the contributions coming from the t.l.
parts of the FO HS and  then comment on general case. 
Now the  product $ p^\rho\, d \Sigma_\rho > 0$
is  positive for any momentum and  the {\it cut-off distribution} (\ref{cutdf})
automatically reproduces the famous Cooper-Frye result \cite{CF74}. 
For the Boltzmann distribution $\varphi= \frac{g}{(2 \pi)^3} e^{\textstyle - \frac{p^\nu u_\nu}{T} }$
of particles 
emitted from the FO HS  $R^* = R^*(t, z)$
one obtains from (\ref{cutdf}) 

\vspace*{-0.5cm} 

\begin{eqnarray}\label{spectri}
\frac{d^2 N}{d y_p~d p_T^2} & = & \frac{g }{(2\pi)^2}\hspace*{-.1cm}\int_\Sigma~
dz~dt~ R^*~
e^{\textstyle - \frac{m_T\gamma_T \cosh (y_p-y_L)}{T}
  } m_T \cosh y_p~{ I_\phi} ({\textstyle \frac{p_T v_T \gamma_T}{T}}) \,\,,
\\
\label{iphi}
I_\phi ({\textstyle \frac{p_T v_T \gamma_T}{T} }) & = &
\int_{0}^{2\pi} d\phi_p~
\lp
{\textstyle \frac{\partial R^*}{\partial t} -
\tanh y_p \frac{\partial R^*}{\partial z} -
 \frac{p_T \cos (\phi_p) }{m_T \cosh y_p} } \rp
e^{\textstyle \frac{p_T  v_T \gamma_T \cos (\phi_p)}{T} }\,\,. 
\end{eqnarray}

\vspace*{-0.2cm} 

\noindent
Now it is clearly seen that 
the main contribution to the integral (\ref{iphi})
corresponds to the 
small angles $\phi_p$ between a 3-momentum of  particle and a 3-vector of the hydrodynamic velocity, 
whereas contributions coming from the large angles $\phi_p$ are suppressed exponentially. 

Neglecting the term with $\frac{p_T}{m_T}$  in Eq. (\ref{iphi}) for 
$p_T \ll m_T$ and expanding the Bessel function $I_0(x) \approx 1 + (0.5x)^2 $ 
for $  p_T v_T \gamma_T \ll T$, one finds  
from (\ref{tdef}) the following result for the AT 

\vspace*{-0.3cm} 

\begin{equation}\label{tnaive}
\frac{1}{T^*} \approx 
\left\langle\hspace*{-1.5mm}\left\langle 
\frac{ \gamma_T \cosh (y_p-y_L) - \frac{ m_T v_T^2 \gamma_T^2}{2\,T} }{T} 
\right\rangle\hspace*{-1.5mm}\right\rangle 
\approx 
\left\langle \frac{ \gamma_T  - \frac{ m_T v_T^2 \gamma_T^2}{2\,T} }{T} \right\rangle 
\end{equation} 
where the double averaging means a double integration over  
time and longitudinal  coordinate with the  weight function defined by
Eq. (\ref{spectri}). For  the longitudinal expansion     
depending on the rapidity $y_L$ only (e.g. Bjorken expansion)
the integration over $z$-coordinate   can be done  because  
for heavy particles the Boltzmann exponential behaves like a Kronecker $\delta$-function,
i.e., $e^{ - m_T \gamma_T \cosh (y_p-y_L) / T } \cong \delta(y_p-y_L) $.
This leads to a single averaging over the evolution time 
(last equality in Eq. (\ref{tnaive}) ) with a slightly
modified weight function.
The next approximation in (\ref{tnaive}) implies  
the  nonrelativistic transversal expansion and 
$m_T \rightarrow m$ 

\vspace*{-0.3cm}

\begin{equation}\label{tcond}
T^* \approx
T + \frac{m \left\langle v_T^2 \right\rangle }{2} 
 \approx
 T + \frac{m \left\langle v_T \right\rangle ^2 }{2}
\,\,, \quad {\rm for} \quad 
m v_T^2 \gamma_T \ll  T\,\,, 
\end{equation}
where the last equality is fulfilled  within 10-15 \% for the linear dependence of the velocity
on the transversal radius.
Note, however, that a popular expression
 (\ref{tcond}) was obtained under the
condition which is hardly fulfilled in practice for heavy particles and, hence, 
Eq. (\ref{tcond}) cannot be established from Eq. (\ref{tnaive}).
Moreover,  for particles of $m_T \ge 1 $  GeV,  freezing out  
under conditions (\ref{pifr}), Eq. (\ref{tnaive})  leads to  negative 
contributions to the AT and even to  negative values of the AT! 
The latter follows from the fact that contributions coming from the small 
transversal radii (and, hence, small $v_T$) in (\ref{spectri}) are suppressed.

Thus,  an accurate examination of the derivation of Ref. \cite{heiz:99} 
shows that {\bf negative AT} can be seen, if the 
FO  HS is a s.l. one.  
In the preceding chapter I dissussed some explicit examples  that
{\bf negative AT} exist, if the  Cooper-Frye formula is applied to
the t.l. HS, whereas the cut-off formula (\ref{cutdf})
generates positive AT  in that case.
Discussion of the AT for the arbitrary HS is
out of the scope of this chapter. I mention only that 
negative AT which are not seen experimentally  
may be an artifact of the Cooper-Frye formula.

In contrast to the rightmost equality in (\ref{tcond}), a more  sophisticated 
derivation of Ref. \cite{sin:99} accounts for the
finiteness of the system and is partly free of the problems discussed above. 
In \cite{sin:99}  the  Cooper-Frye formula is modified by
an additional factor $\exp\{  - a (\cosh y_T -1) \}$ 
(with $ a = 1/ ( \left\langle R^* \right\rangle  v_T^{\prime} (0) )^2 $)
which in case of nonrelativistic transversal expansion reduces to 
$\exp\{  - \frac{R^{*\,2}}{2 \left\langle R^* \right\rangle ^2} \}$. 
Here $\left\langle R^* \right\rangle$ is the mean transverse radius
and $v_T^{\prime} (0)$ is the transverse velocity derivative at the center of the fireball.
The large radii contributions in (\ref{tnaive}) are suppressed due to  
such a factor and, hence, they do not lead to negative AT values. 
Further evaluation  ends  in (\ref{tapp}) with $\alpha = \frac{2}{\pi}$ which 
appears from the additional factor and reflects the cylindrical symmetry. 
Therefore, the latter is used in the next subsection, whereas a more elaborate estimate
I will consider afterwards.

\subsection{Freeze-out of $J/\psi$ Meson.}
Recently it was found that 
the data on $J/\psi$ and $\psi^\prime$ yields in central Pb+Pb 
collisions at 158 A$\cdot$GeV are consistent with the  results of 
the statistical model for the typical value of
$T_H \cong 175$~MeV
extracted from light hadron systematics (see Refs. in \cite{GG:99, Bugaev:02b}).
The hypothesis of statistical $J/\psi$ production at hadronization can
be further tested using data on $m_T$ spectra.
New  NA50  data  \cite{Jpsi}
on transverse mass spectra of $J/\psi$ mesons
in central Pb+Pb collisions at 158 A$\cdot$GeV confirm this expectation: the
spectrum is nearly exponential with  
$T^*(J/\psi) =245\pm 5$~MeV.
The measured $T^*(J/\psi)$ value is significantly smaller than 
expected one from 
Eq. (\ref{tapp}) 
for 
the pion FO  parameters (\ref{pifr})  ($T^* \cong$ 610 MeV).

\vspace*{1.5cm}

\begin{figure}[ht]
\centerline{\hspace*{-3.6cm}\epsfig{figure=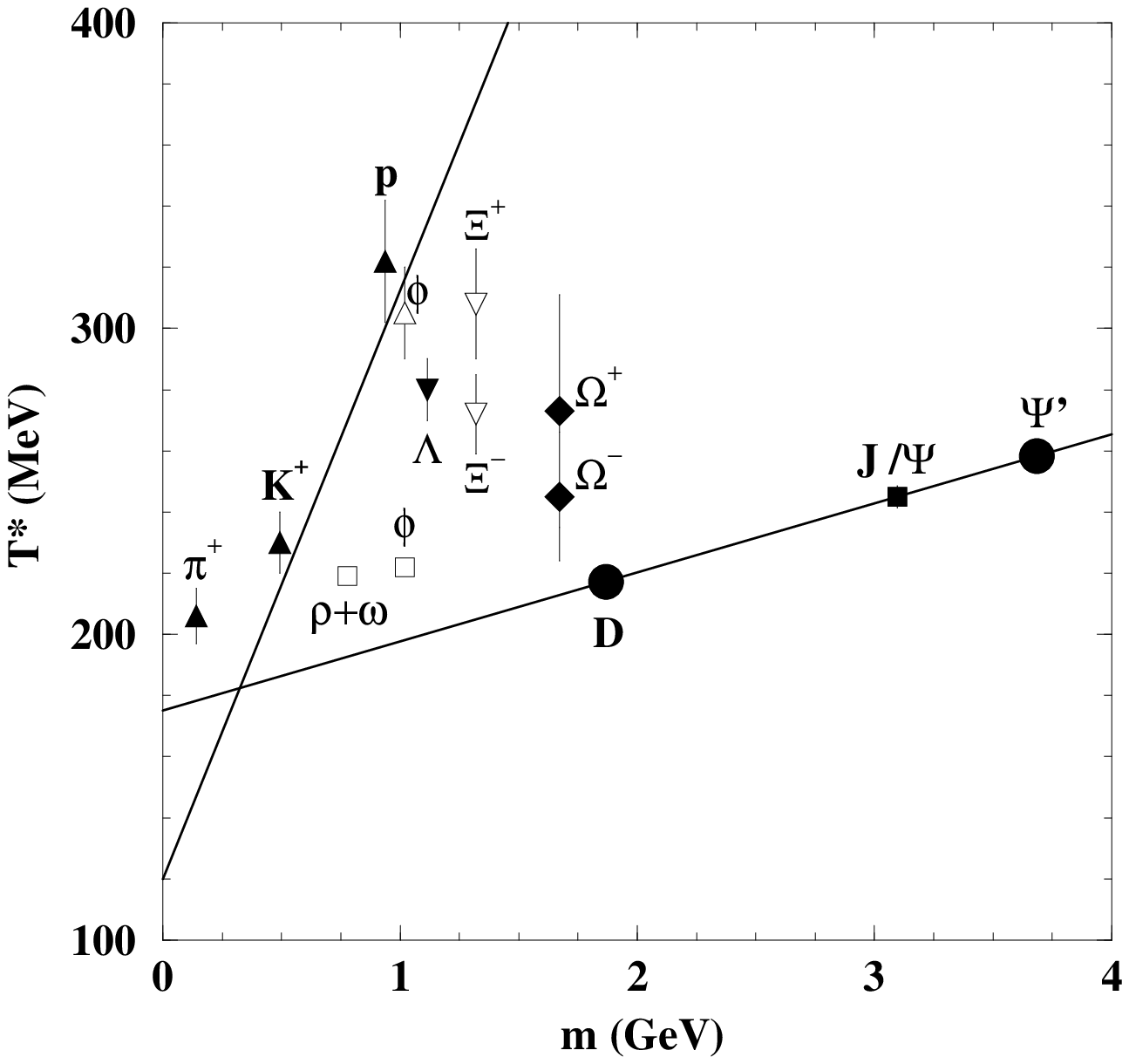,height=7.5cm,width=7.5cm}}

\vspace*{-1.0cm}

  \caption{\label{N61fig:one}
  The inverse slope parameter as a function of the particle mass for
  central Pb+Pb collisions at 158 A$\cdot$GeV. The results 
  are compiled from: 
  \cite{trUF}
  (filled $\bigtriangleup$ ), 
  \cite{phiB}
  (open $\Box$),
  \cite{phiA}
  (open $\bigtriangleup$),
  \cite{trDF}
  (filled $\bigtriangledown$), 
  \cite{trDO}
  (open $\bigtriangledown$), 
  \cite{diam} (filled $\Diamond$), 
  \cite{Jpsi}
  (filled $\Box$).
  The filled circles are predictions of the model for $\psi^\prime$ and D
  mesons. The upper solid line  is given by Eq.~(\ref{tapp}) with $\alpha = \frac{2}{\pi}$ 
  for the FO
  parameters (\ref{pifr}). 
  The lower solid line is given by Eq.~(\ref{slope1}) with  $T_H=175$~MeV and
  $\langle v_T^H \rangle = 0.19$
  which correspond to the QGP hadronization.
  }
  \end{figure}

The `low' value of $T^*$ for $J/\psi$ suggests its rather early thermal
FO.
The low interaction cross section of the $J/\psi$ meson with the
 most abundant hadrons \cite{cseci}
and its large mass lead to a very low probability of the $J/\psi$ rescattering on hadrons.
All this suggests  that
the thermal FO of $J/\psi$ coincides with the hadronization of QGP,
i.e., that the $J/\psi$ meson does not participate in the hadronic
rescattering after hadronization. It is, however, natural to expect that
there is a significant collective transverse flow of hadronizing QGP
developed at the early stage of partonic rescattering. Consequently the
AT of the $J/\psi$ meson as well as all other hadrons for
which chemical and thermal FOs coincide with hadronization can be
expressed as:

\vspace*{-0.3cm} 

\begin{equation}  \label{slope1}
T^*_H (m) ~=~T_H~+~\frac{2}{\pi}~ \cdot m~ \cdot \langle v^H_T\rangle ^2~,
\end{equation}
where $\langle v^H_T\rangle$ is the mean transverse flow velocity of the QGP
at the hadronization. Assuming $T_H=175$~MeV and using the measured value of
$T^*(J/\psi) =245$ MeV one finds from Eq.(\ref{slope1}): $\langle v^H_T\rangle
\cong 0.19$. As expected the obtained transverse flow velocity of QGP at
hadronization is significantly smaller than the transverse flow velocity of
pions ($\approx 0.5$). The linear $m$--dependence of $T^*_H$ (\ref{slope1})
is shown in Fig.~\ref{N61fig:one} by the lower solid line. Within the approach discussed
here, Eq.~(\ref{slope1}) can be used to obtain a next estimate of the lower
limit of the measured AT for all hadrons. In fact, the
values of the parameter $T^*$ for all light hadrons are higher than $T^*_H(m)$.   
The recent results \cite{Jpsi} on
the $m_T$ spectra of the $\psi^{\prime}$ meson
indicate that $T^*(\psi^{\prime}) \approx T^*_H(\psi^{\prime}) = 258 \pm 5$
MeV, which suggests that  the $\psi^{\prime}$ meson (like $J/\psi$) does
not participate in the hadronic rescattering either.

One may expect that the thermal FO may coincide with hadronization
also for $D$ and $\Upsilon$  mesons. Under this assumption one can calculate the value of the
apparent temperature for those mesons : $T^*(D) \cong T^*_H(D) \cong 217$ ~MeV
and $T^*(\Upsilon) \cong T^*_H(\Upsilon) \cong 392 $ ~MeV, respectively.

For the production of open and hidden charm particles
in A+A collisions at RHIC energies one can expect stronger transverse
collective flow effects than at the SPS. This will lead to a linear mass
dependence (\ref{slope1}) of the AT with approximately
the same value of $T_H\cong 175$~MeV, but with a larger value of $\langle
v_{T}^{H}\rangle $. A recent analysis \cite{Redl:01} 
of the RHIC data leads to the above
value $T_H\cong 175$~MeV of the chemical FO temperature. 

Since  the RHIC data  for $J/\psi$ mesons are not    
measured yet, 
one can  use the hydrodynamic calculations of Ref.~\cite{Dum:99} which predict
the value of $\langle v_{T}^{H}\rangle \cong 0.30$ at the QGP hadronization for the low 
collision energy at  RHIC. This will lead  to an increase of the AT of
charmed hadrons at RHIC in comparison to those values at SPS, e.g., $
T^*(J/\psi) \cong T_H^*(J/\psi)\cong 350$~MeV and $T^*(D) \cong T^*_H(D)
\cong 280$~MeV.
If bottomonium experiences the same FO conditions at RHIC like $J/\psi$ meson,
then its apparent temperature will be about 
$T^*(\Upsilon) \cong T^*_H(\Upsilon) \cong 720 $ ~MeV. 

To end this subsection, I conclude that the recent results on transverse mass spectra of $J/\psi$ and $%
\psi^{\prime}$ mesons in central Pb+Pb collisions at 158 A$\cdot$GeV are
considered. It is shown that these data support the hypothesis of the
statistical production of charmonia at hadronization and suggest a
simultaneous  hadronization and the thermal FO for $J/\psi$ and $%
\psi^{\prime}$ mesons. Based on this approach the collective transverse
velocity of hadronizing quark gluon plasma is estimated to be $\langle v^H_T
\rangle \approx 0.2 $. Predictions for transverse mass spectra of hidden and
open charm along with hidden bottom
mesons at SPS and RHIC are discussed.
Now I would like to perform a finer analysis of the data.


\subsection{$\Omega$, $J/\psi$ and $\psi'$ Early Hadronization at SPS Energies.}
The equilibrium hadron gas  model describes remarkably well the hadron
multiplicities measured in A+A collisions at top SPS \cite{Hgas, Hgas:2, Hgas:22}
and RHIC \cite{RHIC:Chem, RHIC:Chem2} energies, where the creation of QGP is expected. The
extracted hadronization temperature parameter is 
similar for both energies $T_H=170\pm
10$~MeV. This is close to an estimate of the temperature $T_C$ for
the
QGP to hadron gas  transition obtained in Lattice QCD simulations at zero baryonic
density (see e.g. \cite{Lattice}). One may therefore argue 
that the QGP created
in high energy heavy ion collisions  hadronizes
into an (approximately) locally equilibrated hadron gas  and the chemical
composition of this hadron gas  is weakly affected by rescattering during the
expansion of the hadronic system \cite{stock}.

As was discussed in the preceding subsection, I and my collaborators  
for the first time  
formulated the hypothesis that the
kinetic FO of $J/\psi$ and $\psi'$ mesons takes place directly at
hadronization and that those mesons therefore carry information on the
flow velocity of strongly interacting matter just after the transition to
the hadron gas. Based on the measured $J/\psi$ and $\psi'$ spectra in Pb+Pb
collisions at 158 A$\cdot$GeV \cite{Jpsi} and using the hypothesis of
the statistical production of charmonia at hadronization
\cite{GG:99,BMGK,Go:00}, it was possible to extract  a mean transverse collective flow
velocity of hadronizing matter: $\langle {v}_T^H \rangle \cong 0.2$.
Here I would like to verify these results more carefully and include 
the $\Omega^\pm$ hyperons into consideration. 

The effect of the rescattering in the hadronic phase was recently studied
within a hydrocascade approach \cite{BD:00,TLS:01}. 
The results of Refs.~\cite{BD:00,TLS:01}
suggest
that the transverse momentum 
($p_T$) spectrum of $\Omega$ may be weakly affected
 during the
cascade stage even for central Pb+Pb collisions at the top SPS energy.
This is because of the small hadronic cross section and large mass
of the $\Omega$
hyperon
\cite{Sorge}.
The corresponding calculations for charmonia are  not yet performed
within this model.

Thus,  we are faced with an intriguing problem: if the above considerations
for charmonia \cite{Bugaev:01d} and $\Omega$ ~\cite{BD:00,TLS:01} are correct, their
$p_T$ spectra should be simultaneously reproduced using the same
hydrodynamic parameters: the hadrinazation temperature  $T_H$ and the mean
transverse velocity $\overline{v}_T$ (the latter notation will be used hereafter to distinguish it from the notation of the  previous discussion).
In the present subsection I 
demonstrate that such a description is, indeed,  possible. The transverse
mass 
spectra around midrapidity in Pb+Pb collisions  at the SPS (158~A$\cdot$GeV) were
recently measured for $\Omega$
by WA97 Collaboration \cite{Omega, Omega1} and for $J/\psi$
and $\psi'$ by NA50 Collaboration \cite{Jpsi}. These spectra will be the
subject of the analysis below.

Assuming kinetic FO of matter
 at constant temperature $T$,
the transverse mass spectrum of $i$-th hadron species (with mass $m_{i}$)
in cylindrically symmetric and longitudinally boost invariant fluid
expansion within the blast wave model 
can be approximated as \cite{Heinz}:
\begin{equation}\label{hydro61}
\frac{dN_i}{m_T dm_T}~
\propto~
m_T~ \int_{0}^{R}r dr~ K_1\left({ \frac{m_T \cosh y_T}{T} }\right)~
I_0\left({ \frac{p_T\sinh y_T}{T}}\right)~,
\end{equation}
where $y_T=\tanh^{-1}v_T$ is the transverse fluid rapidity, $R$ is the
transverse system size, $K_1$ and $I_0$ are the modified Bessel functions.
The spectrum (\ref{hydro61}) is obtained under assumption that the
freeze-out occurs at constant longitudinal proper time $\tau =\sqrt{t^2-z^2}$,
where $t$ is the time and $z$ is the longitudinal coordinate. 
Thus, the     
FO time $t$ is independent of the transverse coordinate $r$
\cite{Heinz}. 
The analysis of the numerical calculations of Ref.~\cite{TLS:01} 
shows that the latter is approximately fulfilled. 
The quality of the approximation made gets better for considered here
heavy particles and small transverse flow velocities because a possible
deviation from Eq.~(\ref{hydro61})
is proportional to $p_T^2 v_T / (2 m_T T)$ 
and hence  it decreases with increasing particle mass at
constant $p_T$.

In order to calculate (\ref{hydro61}) the function
$y_T(r)$ has to be given. 
A linear flow profile, $y_T(r)=y_T^{max}\cdot r/R$, is often
assumed in phenomenological fits \cite{Heinz}.
The numerical calculations of Ref.~\cite{TLS:01} justify this assumption. For
heavy hadrons analysed in here 
the condition $m_i\gg T$ is always satisfied and, therefore, the asymptotic
form for large arguments of $K_1(x)\sim x^{-1/2}\exp(-x)$ can be used in
Eq.~(\ref{hydro61}). At SPS energies typical values of $v_T$ are small
($v_T^2 \ll 1$) and consequently $\cosh y_T\cong 1+ \frac{1}{2}v_T^2$~ and
$\sinh y_T\cong v_T$. 

The experimental $m_T$--spectra
are usually parametrised by a function:
\begin{equation}\label{T*}
\frac{dN_i}{m_T dm_T}~
\propto ~
\sqrt{m_T}~
\exp\left({ -~\frac{m_T}{T_i^*} }\right)~,
\end{equation}
where the inverse slope $T^*_i$ is extracted from the fit to the data.
Formally this corresponds  to neglecting the transverse flow
($v_T\equiv 0$ in Eq.~(\ref{hydro61})), but introducing instead an
``effective'' temperature.
However, when Eq.~(\ref{T*}) is considered as an
approximation of Eq.~(\ref{hydro61}), the inverse slopes $T_i^*$ 
should depend on both $m_i$ and $p_T$.
The limiting cases of $T_i^*$ behavior  at low and high $p_T$ can be easily
studied using the small and large argument asymptotic of the modified
Bessel function $I_0$ in 
Eq.~(\ref{hydro61}):
\begin{equation}\label{T1}
T^*_i(p_T\rightarrow 0)~=~\frac{T}{1~-~\frac{1}{2}~\overline{v}_T^2~
(m_i/T~-~1)}~\approx~ T~+ \frac{1}{2}~m_i~\overline{v}_T^2~,
\end{equation}
\begin{equation}\label{T2}
T^*_i(p_T\rightarrow \infty)~\equiv ~ T^*~=~\frac{T}{1~-~v_T^{max}~+~
\frac{1}{2}~(v_T^{max})^2}~,
\end{equation}
where the average velocity $\overline{v}_T$ in Eq.~(\ref{T1}) is defined as
$\overline{v}_T^2=\int_0^R rdr\, v_T^2(r)/\int_0^R rdr$. 
The maximum velocity $v_T^{max}$ in Eq.~(\ref{T2}) is related to
$\overline{v}_T$ as $\overline{v}_T^2=(v_T^{max})^2/2$
provided a linear flow profile is assumed, $v_T(r)= v_T^{max}\cdot r/R$.
Note that $T^*$ in Eq.~(\ref{T2}) is equivalent to the well known ``blue
shifted'' temperature, $T[(1+v_T^{max})/(1-v_T^{max})]^{1/2}$
(see e.g. \cite{TLS:01,Heinz}), calculated for $(v_T^{max})^2 \ll1$. The shape
of the ``high--$p_T$'' tail ($p_T \gg m_i$) of the $m_T$ distribution
(\ref{hydro61}) is ``universal'', i.e. $T^*$ given by Eq.~(\ref{T2}) is
independent of particle mass $m_i$. On the other hand, the inverse slopes
$T^*_i$ (\ref{T1}) at ``low--$p_T$'' are strongly dependent on $m_i$. Two
remarks are appropriate here. First, for heavy particles like $\Omega$ and
$J/\psi$ the term $\frac{1}{2}m_i\overline{v}_T^2/T$ 
in Eq.~(\ref{T1}) is not small compared
to unity and thus the second (approximate) equality in this equation is
violated. Second, a condition of the validity of Eq.~(\ref{T1}),
$p_T\overline{v}_T  \ll T$, is too restrictive for heavy hadrons, e.g. for
$T\cong 170$~MeV and $\overline{v}_T\cong 0.2$ discussed below it leads
to $m_T-m_i  \ll 0.3\, {\rm GeV}^2 /m_i$. This means that Eq.~(\ref{T1}) is
valid for the values of $m_T-m_i$ which are much smaller than
0.2~GeV for $\Omega$ and than 0.1~GeV for $J/\psi$.

Consequently: (a) none of the asymptotic regimes ({\ref{T1}) or
(\ref{T2})
can be clearly seen in the experimental $m_T$ spectra, i.e. neither
``low--$p_T$'' ($m_T-m_i  \ll 0.3\, {\rm GeV}^2/m_i$) nor ``high--$p_T$''
($m_T - m_i \gg m_i$) approximations are useful ones (at least for studying
the available $m_T$ spectra of $\Omega$ and charmonia); (b) fitting the
experimental $m_T$ spectrum of $i$--th hadron species by Eq.~(\ref{T*}) one
finds, in fact,  the ``average inverse slopes'' which depend not only on
particle mass $m_i$, but also on the $m_T-m_i$ interval covered in a given
experiment (see also Ref.~\cite{TLS:01}, where $T^*_i$ have been discussed
separately for $m_T-m_i< 0.6~$GeV and for $ 0.6~GeV < m_T-m_i< 1.6~$ GeV). 

For small values of $v_T$ relevant to our discussion
a  good approximation of Eq.~(\ref{hydro61}) at $m_T-m_i<m_i$ can be
obtained by substituting the $v_T$ distribution in Eq.~(\ref{hydro61})
by its average value $\overline{v}_T$ and by using large argument
$K_1$ asymptotic:
\begin{equation}\label{hydro1}
\frac{dN_i}{m_T dm_T}~
\propto ~
\sqrt{m_T}~
\exp\left({ -~\frac{m_T (1+\frac{1}{2}\overline{v}_T^2)}{T} }\right)
~I_0\left({ \frac{p_T\overline{v}_T}{T} }\right)~.
\end{equation}
It was  checked numerically that the values of the parameter
$\overline{v}_T$ extracted from the fits (see below) to Eqs.~(\ref{hydro61})
and  ~(\ref{hydro1}) assuming a linear velocity profile
differ by about 5\%.

Let me turn now to the test of the  hypothesis of the kinetic
FO of $J/\psi$, $\psi'$ and $\Omega$ occurring directly at hadronization
i.e. at $T = T_H = 170$ MeV.
The $m_T$--spectra of these hadrons are measured around midrapidity 
\cite{Jpsi,Omega} for Pb+Pb collisions at 158 A$\cdot$GeV.
The fit to these data performed using Eq.~(\ref{hydro1}) with
$T = T_H = 170$ MeV yields $\overline{v}_T=0.194 \pm 0.017 $ and
$\chi^2/dof = 1.3$.
The value of $\overline{v}_T$ varies by $\mp 0.016$ when $T_H$ is changed
within its uncertainty $\pm 10$ MeV.
Note that $T$ and $\overline{v}_T$ parameters are anti--correlated.
A surprisingly good agreement (see Fig.~\ref{N61Bfig1}) of this  model with the data on 
$m_T$--spectra serves as a strong support of the hypothesis of
statistical nature of $J/\psi$ and $\psi'$ production
\cite{GG:99} and their kinetic FO occurring directly at hadronization
\cite{Bugaev:01d}.

\begin{figure}[ht]
\centerline{\epsfig{file=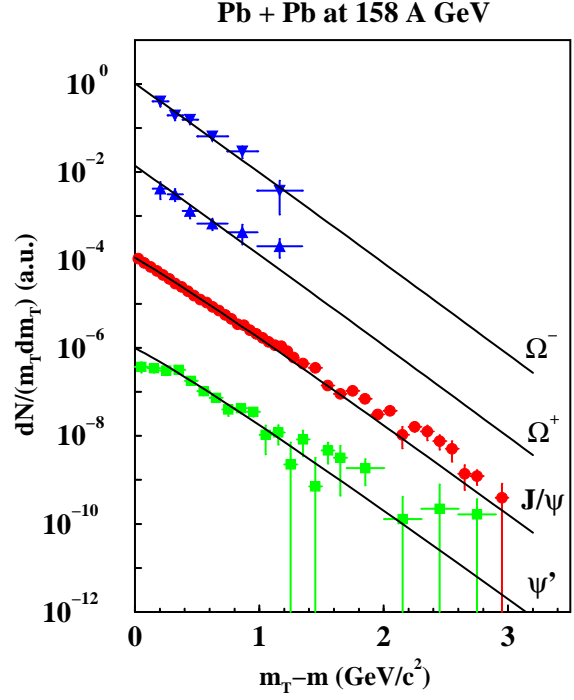,width=10cm}}

\caption{ \label{N61Bfig1}  The transverse mass spectra of
$\Omega^-$  (triangles down) 
and $\Omega^+$ (triangles up) \protect\cite{Omega} 
 as well as $J/\psi$ (circles)
and $\psi'$ (squares) \protect\cite{Jpsi} produced in Pb+Pb collisions at
158 A$\cdot$GeV. The solid lines indicate a prediction of the model
(\ref{hydro1}) assuming kinetic FO of heavy hadrons directly
after hadronization of expanding quark gluon plasma. The FO
parameters are: $T = 170$ MeV and $\overline{v}_T = 0.194$. } \label{fig1}
\end{figure}

\begin{figure}[ht]
\centerline{\epsfig{file=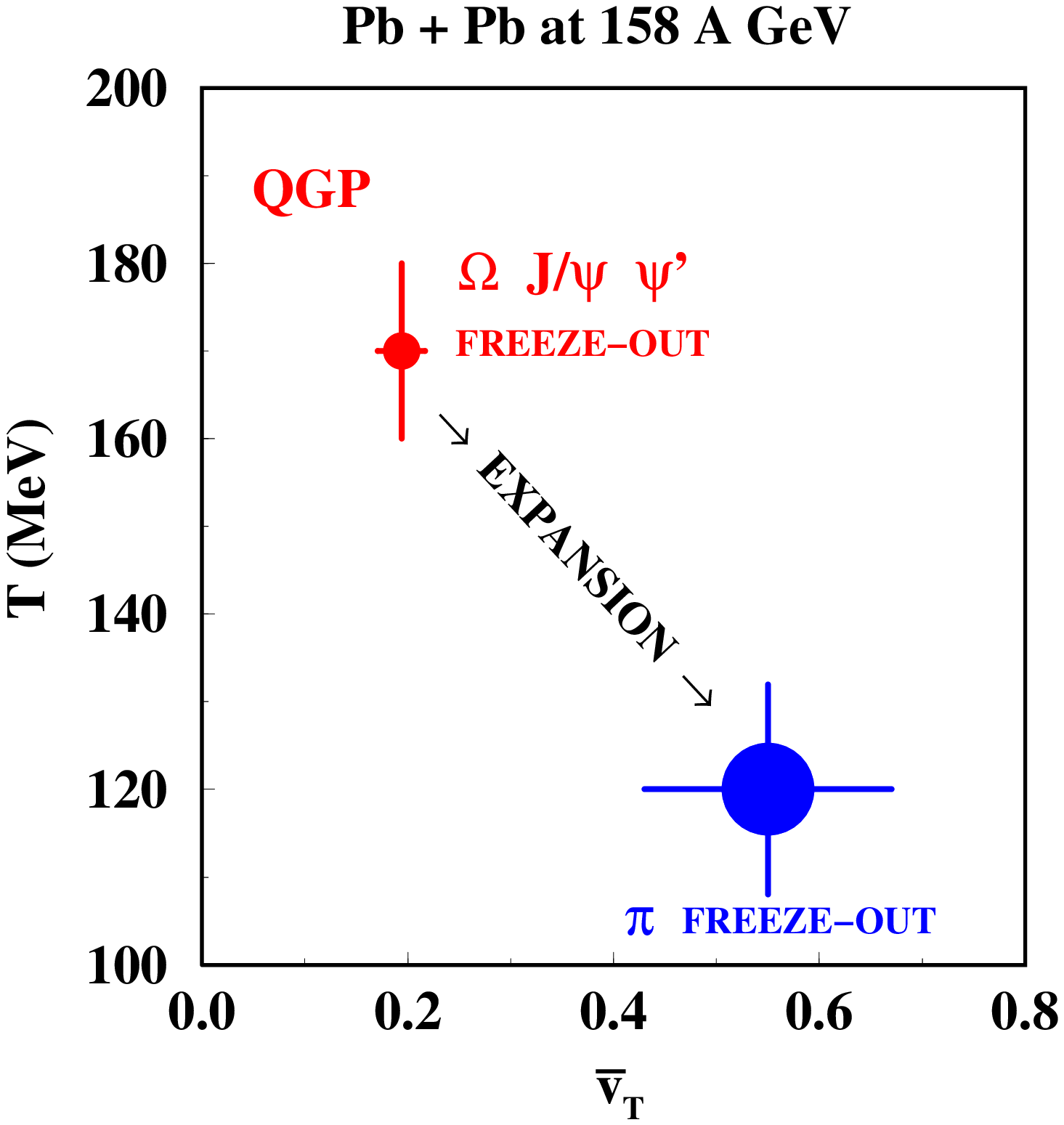,width=10cm}}

\caption{ \label{N61Bfig2} The expansion history of
strongly interacting matter created in Pb+Pb collisions at 158
A$\cdot$GeV. The points indicate the temperature $T$ 
and the mean transverse flow
velocity  $\overline{v}_T$ of matter at 
the time of $\Omega$, $J/\psi$ and $\psi'$
FO (upper point) 
and at the time of pion kinetic FO (lower
point). 
} 
\end{figure}

The dependence of the $J/\psi$ and $\psi'$ transverse mass spectrum on the
centrality (quantified by the neutral transverse energy $E_{T}$) of Pb+Pb
collisions at 158~A$\cdot$GeV was also measured by NA50 Collaboration
\cite{Jpsi}. An increase of $\langle p_{T} \rangle$ and $\langle p_{T}^2
\rangle$ from peripheral collisions to the most central collisions can be
explained, within the present  approach, by an increase of the model parameter
$\overline{v}_{T}$ with $E_{T}$. Note that the increase of mean flow
velocity and consequently an increase of $\langle p_{T} \rangle$ and
$\langle p_{T}^{2}\rangle$ with a centrality of the collision as well as
an increase of $\langle p_{T}\rangle$ and $\langle p_{T}^{2}
\rangle $ with particle mass $m_{i}$ are characteristic features of
hydrodynamics. In contrast to $J/\psi$ and $\psi^{\prime}$ mesons the
$m_{T}$--spectra of the Drell--Yan pairs (dileptons with invariant mass
$M>4.2$~GeV) do not show this type of hydrodynamical behavior. The
values of $\langle p_{T}\rangle$ and $\langle p_{T}^{2}\rangle$
\cite{Jpsi} for the Drell--Yan pairs are smaller than those for $J/\psi$
and $\psi^{\prime}$ and do not change significantly with $E_{T}$.

The kinetic FO parameters of pions were extracted
from the analysis of single-- and two--pion spectra measured for central
Pb+Pb collisions at 158 A$\cdot$GeV by the NA49 Collaboration
\cite{Ka,PionFR,Wi:99}. The results are: $T_{f} \cong 120$ MeV and
$\overline{v}_T \cong 0.55$, i.e. they are very different than those
obtained here from the analysis of heavy hadron spectra. In fact,  
it was checked that the
parameters obtained for pions lead to the $m_T$ spectra of heavy hadrons
which strongly disagree with the data. A decrease of temperature and an
increase of transverse flow velocity with time is a general property of
expanding systems. The different kinetic FO times of heavy
hadrons and pions allowed us to establish  the expansion history of the hadron
gas phase created in Pb+Pb collisions at 158 A$\cdot$GeV for the first time. The
FO points of heavy hadrons and pions extracted from the data are
plotted in Fig.~\ref{N61Bfig2} defining the path of the expanding hadron system in the
$T-\overline{v}_T$ plane.

Thus, the $m_T$--spectra of $\Omega$, $J/\psi$
and $\psi'$ produced in Pb+Pb collisions at 158 A$\cdot$GeV are analysed
within the hydrodynamical model of the QGP expansion and hadronization.
The spectra are in agreement with the hypothesis of kinetic FO of
these heavy hadrons occurring directly after the transition from the quark
gluon plasma to the hadron gas. A mean collective transverse flow of
hadronizing matter of $\overline{v}_T \cong 0.2$ is extracted from the
fit to the spectra using temperature $T_H \cong$170~MeV fixed by the
analysis of hadron multiplicities \cite{Hgas, Hgas:2, Hgas:22}. This result together with a
previously obtained parameters of pion FO ($T_{f} \cong 120$~MeV
and $\overline{v}_T\cong 0.55$) allowed  for the first time to establish the
history of the expanding hadron matter \cite{Bugaev:02a} created in nuclear collisions.

In the RHIC energy range the temperature parameter is approximately the
same $T=T_H\cong 170$~MeV \cite{RHIC:Chem, RHIC:Chem2}, whereas the transverse hydrodynamic
flow at $T=T_H$ is expected to be stronger. The model predictions for
the $m_T$--spectra of
$\Omega$ and charmonia  at the RHIC energies
will be presented in the next section.


%
%

\def\smh{\hspace*{0.3cm}}

\def\medh{\hspace*{0.6cm}} 

\def\bh{\hspace*{0.9cm}}

\section{ 
Hadron Spectra and
QGP Hadronization in Au+Au Collisions at RHIC
}

The recent measurements \cite{na49} of the energy dependence of pion and kaon
production in central collisions of heavy nuclei indicate 
\cite{Horn}
that the transient state of deconfined matter is created
at the early stage of these collisions for energies
higher than about 30 A$\cdot$GeV, i.e. at high SPS and
RHIC energies.
Analysis of the hadron multiplicities measured in 
these collisions 
within statistical models
at the SPS \cite{Hgas, Hgas:2, Hgas:22} and RHIC \cite{RHIC:Chem, RHIC:Chem2} energies
shows that
the chemical FO  takes place
near the boundary between the quark--gluon and hadron phases.
The values of the
temperature parameter extracted from the data are similar
for both energies, $T_H=170\pm 10$~MeV, and, as discussed above,  they are 
close to the value of the deconfinement transition temperature  
at zero baryonic density estimated 
in the lattice QCD (see e.g. Ref.~\cite{Lattice}).

The analysis of the experimental  data at the SPS 
\cite{Ka,Wi:99} 
and a numerical modeling of the hadron cascade stage in A+A collisions
at SPS and RHIC energies
\cite{BD:00,TLS:01} 
indicate that the kinetic (i.e., particle spectra) FO
of the most abundant hadrons
takes place at  temperatures significantly lower than $T_H$.
Nevertheless, as was argued in the preceding subsection, 
one may  expect that the kinetic FO
of some heavy and weakly interacting  hadrons 
(e.g. $\Omega$ hyperons and 
$\phi$, $J/\psi$, $\psi^{\prime}$ mesons) 
 occurs directly at the
QGP hadronization stage or close to it.
Thus, 
for these hadrons the chemical and kinetic FOs
coincide and are determined by the features of the QGP
hadronization.
For $\Omega$ hyperons and $\phi$ mesons this expectation is based on the
results of  hydrocascade approach \cite{BD:00,TLS:01}. For   
$J/\psi$ and  $\psi^{\prime}$ mesons such a result I obtained with my collaborators  
\cite{Bugaev:01d,Bugaev:02a,Bugaev:02}. 
It  is a straightforward consequence of the 
recently proposed
statistical mechanism of 
charmonia production at the QGP hadronization 
\cite{GG:99,BMGK,Go:00,rapp}.
As I showed above,
within this approach the transverse mass 
spectra of the $\Omega$,
$J/\psi$ and $\psi^{\prime}$ measured 
in Pb+Pb collisions at the SPS can be
described successfully. 
The transverse mass spectra of $\Omega$ hyperons \cite{QM02}
and $\phi$ mesons \cite{RHIC:Hadronization3,RHIC:Hadronization4} produced 
in central Au+Au collisions at $\sqrt{s_{NN}} = 130$~GeV 
were recently measured by the STAR Collaboration.
These data allow to test the present model in the new
energy regime.
They also give  a unique opportunity to extract
parameters of the QGP hadronization at RHIC energies 
and consequently predict
spectra of
$J/\psi$ and $\psi'$ mesons.

Within a hydrodynamical approach of the QGP hadronization
the transverse mass spectrum of $i$-th hadron in the central rapidity
region can be written as (see, e.g., Ref.~\cite{gy}): 
\begin{eqnarray}\label{hydro62}
&&\frac{dN_i}{m_T dm_T dy}~\biggl|_{y = 0}~
= 
\frac{d_i~ \lambda_i~ \gamma_i^{n_i} }{\pi}~\tau_H ~ R_H^2~ 
\int_{0}^{1} \xi~ d\xi~ K_1 \left(
\frac{m_T \cosh y_T}{T_H}  \right) ~  
I_0\left({ \frac{p_T\sinh y_T}{T_H}}\right)~,
\end{eqnarray}
%
where $y$ is the particle longitudinal rapidity and  $y_T(\xi)=\tanh^{-1}v_T$
is the fluid transverse rapidity.
$R_H$  and $\tau_H$ are, respectively,
the transverse system size and proper time at the hadronization (i.e., 
at the boundary between the mixed phase and hadron matter), $\xi=r/R_H$
is a relative transverse coordinate.
The particle degeneracy and fugacity are denoted as $d_i$ and $\lambda_i$,
respectively.
Parameter $\gamma_i$ in Eq.~(\ref{hydro62}) ($\gamma_S$ \cite{Raf} for
$i=\phi, \Omega$ and $\gamma_C$ \cite{BMGK,Go:00} for $i=J/\psi$,
$\psi^{\prime}$) describes a possible deviation of strange and charm hadrons
from complete chemical equilibrium ($n_i=2$ for
$\phi,J/\psi,\psi^{\prime}$ and $n_i=3$ for $\Omega$). 

The spectrum (\ref{hydro62}) is obtained under
the assumption that the hydrodynamic expansion is longitudinally boost
invariant and
that
the FO occurs at constant longitudinal proper time $\tau
=\sqrt{t^2-z^2}$   
(see Eq.~(\ref{hydro61}) for the notations), i.e. the
FO time $t$ is independent of the transverse coordinate $r$. 
The formula  (\ref{hydro62}) is a more careful expression of Eq. (\ref{hydro61}),
which I use here to extract $\tau_H ~ R_H^2$ value  from the absolute 
normalization of spectra. 

In order to complete Eq.~(\ref{hydro62}) the functional form of the 
transverse rapidity distribution of hadronizing matter
$y_T(\xi)$ has to be given. A linear flow profile,
$y_T(\xi)=y_T^{max}\cdot \xi$, 
used in this  model 
is justified by the numerical calculations of
Ref.~\cite{TLS:01} 
and by an excellent fit \cite{Bugaev:02a} 
of the SPS transverse momentum spectra  for $\Omega^{\pm}$,
$J/\psi$, and $\psi^{\prime}$ particles.


\begin{figure}
\centerline{\epsfig{file=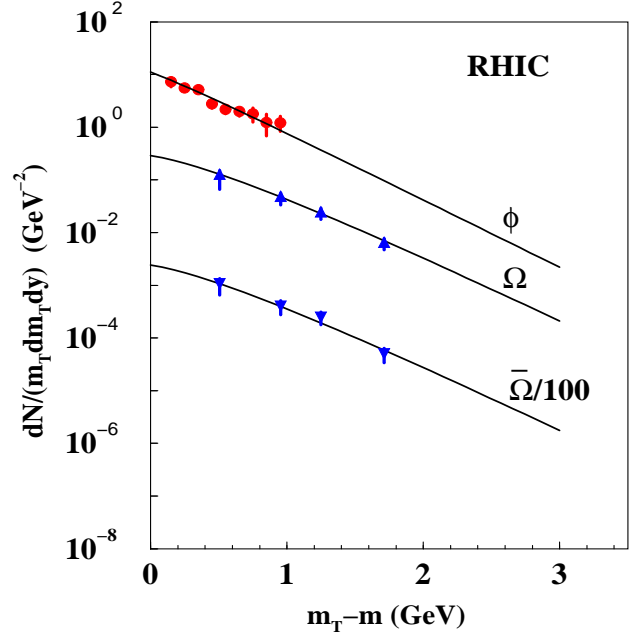,width=80mm,height=85mm}\hspace*{2.0cm}}
\caption{
The hadron transverse mass  spectra
in  Au+Au collisions at $\sqrt{s_{NN}} = 130$~GeV are shown.
The points indicate experimental data
for the $\Omega$
\protect\cite{QM02}
and
$\phi$
\protect\cite{RHIC:Hadronization4}
measured by STAR.
The model results 
are shown by full lines.
}
\label{N62fig:one}
\end{figure}


Thus, in the present  model,  the QGP hadronization is described  by the
following parameters:
temperature $T_H$, ``volume" $\tau_H R_H^2$,
maximum flow rapidity  $y_T^{max}$, fugacities
$\lambda_i$,  and saturation factors $\gamma_i$.
Note that the $\phi,J/\psi,\psi^{\prime}$ have no conserved charges
and $\lambda_i=1$ for these particles.
I use the fixed values of the parameters
$T_H = 170$~MeV, $\gamma_S = 1.0$, $\lambda_{\Omega^-} = 1/
\lambda_{\Omega^+} = 1.09$ 
(note that $\lambda_{\Omega^-}\equiv \exp[(\mu_B-3\mu_S)/T]$,
where $\mu_B$ and $\mu_S$ are, respectively, baryon and
strange chemical potentials). These
(average) values of the {\it chemical
freeze-out} parameters have been found in the 
hadron gas analysis \cite{RHIC:Chem, RHIC:Chem2} of the full set of 
the midrapidity particle number ratios measured in 
central Au+Au collisions at $\sqrt{s_{NN}} = 130$~GeV.
The fit to the  $m_T$-spectra of $\Omega^{\pm}$ hyperons \cite{QM02, RHIC:Hadronization2}
and $\phi$ mesons \cite{RHIC:Hadronization3,RHIC:Hadronization4}  measured in
central (14\% for $\Omega^{\pm}$ and
11\% for $\phi$) Au+Au collisions at $\sqrt{s_{NN}} = 130$~GeV 
is shown in Fig.~\ref{N62fig:one}.
The fit results are:   
$y_T^{max} = 0.74 \pm 0.09$, 
$\tau_H R_H^2 =275 \pm 70 $ fm$^3$    
and
$\chi^2/ndf \cong $ 0.46. 
In the calculation of errors of the two free parameters of the model 
the uncertainties of $T_H$ ($\pm 5$ MeV),
$\gamma_S$ ($\pm 0.05$) and $\lambda_{\Omega^-}$ ($\pm 0.06$)
were taken into account.


A simple exponential approximation 
of the spectra
is usually utilized to parameterize the experimental data: 
\vspace*{-0.2cm} 

\begin{equation}\label{T*1}
\frac{dN}{m_T dm_T dy}~\biggl|_{y = 0}~
= ~
C~
\exp\left( -~\frac{m_T}{T^*} \right)~.
\end{equation}


\begin{figure}
\centerline{\epsfig{figure=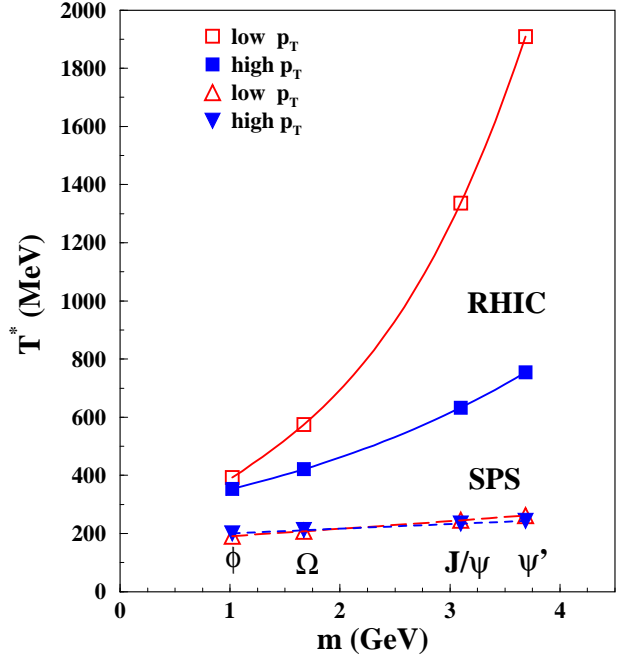,width=80mm,height=90mm}\hspace*{2.0cm}}
\caption{
The 
values of the inverse slope parameters $T^*$ for two different
$m_T$ domains -- ``low-$p_T$'' ($m_T-m < 0.6$~GeV)
and  ``high-$p_T$'' ($0.6~$GeV$ < m_T-m < 1.6$~GeV)  -- in
Au+Au collisions at $\sqrt{s_{NN}}=130$~GeV
are presented. They are found using  Eq.~(\ref{hydro62})  with $T_H=170$ MeV and
$y^{max}_T=0.74$.
For comparison, the values of $T^*$
extracted from fitting the data in Pb+Pb collisions at the
SPS (Eq.~(\ref{hydro62})  with $T_H=170$~MeV, $y_T^{max}=0.28$, see
Ref.~\protect\cite{Bugaev:02a})
are also shown.}
\label{N62fig:two}
\end{figure}

Note that in
 Refs. \cite{Bugaev:01d,Bugaev:02a,Bugaev:02}
an additional factor $m_T^{1/2}$ 
was present in the r.h.s. of Eq.~(\ref{T*1}).
It led  to smaller 
values of $T^*$ when fitting the same spectrum.
The $m_T$--spectrum (\ref{hydro62})  may, however, deviate significantly 
from a purely exponential one
and its shape depends on the
magnitude of the transverse flow and the mass of the particle.
The normalization factors $C$
and the inverse slope parameters $T^*$ in different intervals
of $m_T-m$ can be
found from the $\phi$,
$\Omega$, $J/\psi$ and $\psi^{\prime}$ spectra 
given by Eq.~(\ref{hydro62}) using the maximum likelihood method.
The average values of $T^*$ for the $m_T$ domains
of ``low-$p_T$'' ($m_T-m < 0.6$~GeV) and ``high-$p_T$'' 
($0.6~$GeV$ < m_T-m < 1.6$~GeV), 
discussed in Refs.~\cite{TLS:01,Bugaev:02},
are shown in Fig.~\ref{N62fig:two}.   
The values of $T^*$  determined in the preceding section by fitting the $\Omega^{\pm}$, $J/\psi$ and $\psi^{\prime}$
data in Pb+Pb collisions at
158 A$\cdot$GeV 
($T_H=170$~MeV, $y_T^{max}=0.28$)
are also shown 
for comparison.
The observed increase of $T^*$ with
increase of the hadron mass
is much stronger at RHIC than at SPS energies.
It is caused by  larger transverse flow velocity of
hadronizing QGP  at RHIC ($\overline{v}_T\cong 0.44$) than at
SPS ($\overline{v}_T\cong 0.19$).
The increase of $T^*$ is much more pronounced in ``low-$p_T$'' region
than in ``high-$p_T$'' one.
In this  model the $m_T$-spectra of charmonia
are extraordinary affected by the stronger
transverse flow at RHIC due to
enormous masses of these hadrons.
Thus, the data on $J/\psi$ and $\psi^{\prime}$ production
in Au+Au collisions, soon to be obtained at RHIC, 
should allow to test the hypothesis of their 
formation at the QGP hadronization.


\begin{figure}[ht]
\centerline{\epsfig{figure=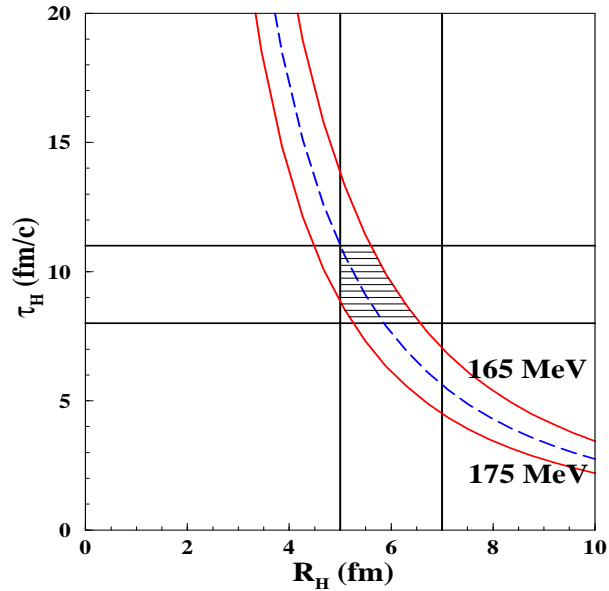,width=80mm,height=80mm}\hspace*{2.0cm}}
\caption{
The lines $\tau_H=A(T_H)\cdot R^{-2}_H$ of constant ``volume parameter''
$A(T_H)$
are shown: $T_H=170$~MeV corresponds to the dashed line, $T_H=165$~MeV and
$T_H=175$~MeV correspond to
the lower and upper solid lines, respectively.
The dashed area  is the intersection of the $R_H$--$\tau_H$
region between the $T_H=165$~MeV and $T_H=175$~MeV lines 
with the region of $R_H=5\div7$~fm and
$\tau_H=8\div11$~fm  estimated from Ref.~\protect\cite{TLS:01}.
 }
 \label{N62fig:three}
 \end{figure}


One should  note here that at present
there exists an uncertainty in the estimates of the $\gamma_C$ factor,
therefore,
the predictions concerning charmonia multiplicities
in Au+Au collisions at RHIC within
statistical approaches significantly vary
and their discussion goes beyond the scope of the present discussion.

The ``volume parameter'' $\tau_H R^2_H \equiv A(T_H)$
extracted from the fit to the $\Omega$ and $\phi$ 
spectra defines the line $\tau_H=A(T_H)\cdot R_H^{-2}$ in the 
$R_H$--$\tau_H$ plane.
The allowed region in 
the $R_H$--$\tau_H$ plane can be estimated by varying 
the temperature parameter within its limits, 
$T_H=165$~MeV and $T_H=175$~MeV.
The resulting lines are shown in Fig.~\ref{N62fig:three}. 
The transverse radius $R_H = 5\div 7 $~fm and
the proper time $\tau_H = 8\div 11$~fm at the QGP hadronization
can be estimated from 
the hydrodynamical  calculations of \cite{TLS:01} for central Au+Au collisions at
$\sqrt{s_{NN}}=130$~GeV 
(see Fig.~3 in Ref.~\cite{TLS:01}). 
These model boundaries and their intersection with the $R_H$--$\tau_H$
region found in my analysis are shown in
Fig.~\ref{N62fig:three}.

\begin{figure}
\centerline{\epsfig{figure=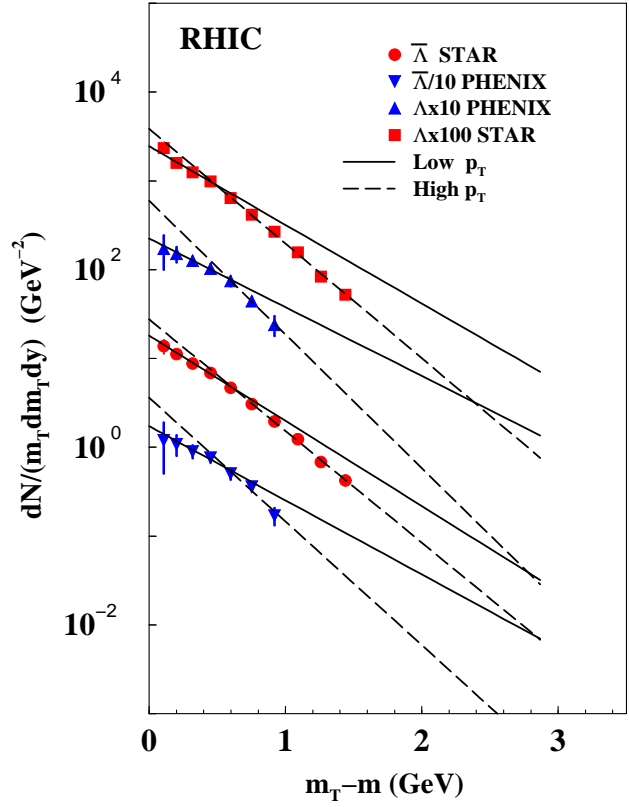,width=80mm,height=110mm}\hspace*{2.3cm}} 
\caption{
The points indicate the experimental $m_T$-spectra of
the $\Lambda$ and ${\bar \Lambda}$ in central Au+Au 
collisions  at $\sqrt{ s_{NN} }=130$~GeV measured by the STAR
\protect\cite{star2} and  PHENIX \protect\cite{phenix2}
Collaborations. The ``straight lines" are 
the exponential approximations of the spectra with 
Eq.~(\ref{T*1})
in the low-$p_T$ (solid lines) and high-$p_T$ (dashed lines) regions.}
\label{N62fig:four}
\end{figure}


Within the present  approach the $m_T$-spectra of
$\phi$, $\Omega$, $J/\psi,\psi^{\prime}$
are assumed to be frozen at the space-time hyper-surface
where the hadron phase starts.
This assumption is justified by
the small hadronic cross sections and large masses of these particles
(in addition, the $m_T$-spectra  of these hadrons are  almost not
affected by the resonance feeding).
However, the~ $m_T$-spectra\, of\, many other hadrons 
are expected
to be significantly modified by hadronic rescattering.
Contrary to this expectation it was recently postulated
\cite{BF1,BF2} that  the simultaneous chemical
and kinetic freeze-out in Au+Au collisions at RHIC
occurs for all hadrons
(a single freeze-out model).
Do experimental data allow us to distinguish between
these two approaches?

\begin{table}[h!] \label{VITable}
\caption{The values of inverse slope parameters $T^*$ for
(anti)protons and (anti)lambdas in Au+Au collisions at
$\sqrt{ s_{NN} }=130$~GeV are presented. The experimental
values are taken as the average ones over the STAR and PHENIX results
(a difference in the results for
particle and its
anti-particle is small).
}
\begin{center}
\begin{tabular}{|c|c|c|c|}\hline

 &  $\smh T_{low-p_T}^*$ \smh  & \smh  $T_{high-p_T}^*$ \smh & \smh Refs. \smh \\
 & (MeV) & (MeV) & \\
\hline
\hline
{\bf DATA}~~~$p,\bar{p}$  & $455\pm 105 $  & $290 \pm 40$ & \cite{star1,phenix1}\\
%
%
\hline
Hydro+RQMD  & 480 & 300 & \cite{TLS:01} \\
\hline
Single freeze-out  & 315  & 310 & \cite{BF1,BF2} \\
\hline
\hline
{\bf DATA}~~~{\footnotesize $\Lambda,\bar{\Lambda}$} & $505\pm 60 $  & $320\pm 30$ & \cite{star2,phenix2}\\
%
%
\hline
Hydro+RQMD & 440 & 310  & \cite{TLS:01} \\
\hline
Single freeze-out  & 360 & 330 & \cite{BF2}
\\
\hline
\hline
\end{tabular}
\\
%

\end{center}

\end{table}

The hydrocascade  approach \cite{TLS:01}
predicts for central  Au+Au
collisions at $\sqrt{ s_{NN} }=130$~GeV  that the hadron cascade
stage modifies the $m_T$-spectra of nucleons and
$\Lambda$ hyperons substantially. 
In particular a large increase of the inverse slope
parameter in the low-$p_T$ region is expected for these
hadrons as a result of hadronic rescattering and resonance
decay effects.
Thus, the  measurements of (anti)proton and (anti)lambda
$m_T$-spectra should allow to distinguish between
the single freeze-out model and models which
assume different kinetic freeze-out conditions for different hadrons.
I performed the $T^*$  analysis of the present RHIC data from STAR \cite{star1,star2} and
PHENIX \cite{phenix1,phenix2}.
The resulting  $T^*$ values are
summarized in Table~6.1.   together with the predictions
of the single FO model \cite{BF1,BF2}
and hydrocascade    model \cite{TLS:01}.
The $m_T$-spectra of the $\Lambda$ and ${\bar \Lambda}$
are also shown in Fig.~\ref{N62fig:four}.   
There are significant systematic differences between 
$T^*$ parameters obtained from the STAR and PHENIX data.
In view of this fact, the values quoted in the Table~6.1. are    
calculated as an arithmetic average of both results, whereas
the (systematic) error was estimated to be a half of the 
difference between them.

Despite 
the large uncertainties,
the data seem  to favor
the hydrocascade   model over
the single FO model.
Additional data in the low-$p_T$ region and their
theoretical analysis
would be helpful to clarify  presence of the hadron cascade stage and
its influence on $T^*_{low-p_T}$ of (anti)protons and (anti)lambdas.

The results on $m_T$--spectra of charmonia in central Au+Au
collisions at the RHIC energies
are expected to be available soon.
They should  allow to test 
a statistical approach to the charmonia production
at the QGP hadronization in high energy nuclear collisions.  
In particular, within this approach,
I predict a strong 
(a few times)
increase of the inverse slope
parameter $T^*$ of the charmonia $m_T$--spectra at RHIC
in comparison with that at SPS.
The higher is the energy the larger inverse slope is expected due to
increasing transverse flow of hadronizing QGP.
Thus, at $\sqrt{s_{NN}}=200$~GeV the increase
of $T^*$  should become even more
pronounced than at $\sqrt{s_{NN}}=130$~GeV. 
Due to strong sensitivity of
the charmonia spectra
to the hadronization temperature and transverse flow velocity,
their analysis should significantly improve the discussed  estimate of
these parameters.


\section{Transverse Caloric Curves of Kaons  
as   the Signal of  Deconfinement Transition 
 in  A+A Collisions 
}

The statistical model of the early stage of A+A
collisions suggests \cite{Horn} that the onset of
the deconfinement phase transition at the early stage of the collisions
may be signaled by the anomalous  energy dependence of several
hadronic observables.
In particular,
following earlier suggestions \cite{GaRo},
the behavior of strangeness and pion
yields in the transition region was studied in detail.
Recent
measurements \cite{na49} of pion and kaon
production in central Pb+Pb collisions at the CERN SPS, indeed, 
indicate that the transient state
of deconfined matter is created in these collisions for
energies larger than about 40 A$\cdot$GeV.
The present data show a  maximum of the
strangeness to pion ratio at this  energy. An exact position and the
detailed structure of this maximum is clarified
by the  recent results from
the 2002 Pb run at  30~A$\cdot$GeV \cite{DeconfOnset}.

In the present section I  discuss  another well known
observable, which may be sensitive to the onset of deconfinement,
the transverse momentum, $p_T$, spectra of produced hadrons.
It was suggested by Van Hove
\cite{van-hove} more than 20 years ago to identify the deconfinement phase
transition in high energy proton--antiproton interactions by
an anomalous
behavior (a plateau-like structure) of the average transverse momentum as a
function of hadron multiplicity.
Let me  briefly recall  Van Hove's arguments.
According
to the general concepts of the hydrodynamical approach the hadron
multiplicity reflects the entropy, whereas
the transverse hadron AT reflects
the combined effects of temperature and collective transverse expansion.
The entropy is assumed to be created at the early stage of the collision
and is approximately constant during the hydrodynamic expansion. The
multiplicity  is proportional
to the entropy, $S=s \cdot V$, where $s$ is the entropy density and
$V$ is the effective volume occupied by particles.
During the hydrodynamic expansion, $s$
decreases and $V$ increases with $s \cdot V$, being
approximately constant because, as I showed earlier, the FO shocks generate just 
about 1 \% of entropy.
 The large multiplicity  at high energies means a
large entropy density at the beginning of the expansion (and consequently
a larger volume at the end).
A large
value of $s$ at the early stage of the collisions
means  normally high  temperature $T_{0}$ at this stage.
This, in
turn, leads to an increase of transverse hadron AT, a
flattening of the transverse momentum spectra.
Therefore, with
increasing  collision energy
 one expects to observe an increase of both the
hadron multiplicity and average transverse momentum per hadron. 
In the original Van Hove
suggestion the correlation between average transverse momentum and hadron 
multiplicity was discussed for proton-antiproton collisions at fixed 
energy. Today it is possible  to study  A+A collisions at different 
energies.

However, the
presence of the deconfinement phase transition would change this
correlation. In the phase transition region the initial entropy density
(and hence the final hadron multiplicity) increases with collision energy, 
but temperature $T_{0} = T_C$ and pressure  $p_{0} = p_C$
remain constant. 
The equation of state presented in the form
$p(\varepsilon)/\varepsilon$ versus 
energy density $\varepsilon$ shows a minimum (the 
`softest
point' \cite{SZ79,Hung:94}) at the boundary of the 
({\it generalized} \cite{Hung:94}) mixed phase and the QGP. 
Consequently, 
the shape of 
the $p_T$ spectrum 
is approximately independent of the multiplicity or collision
energy.
The transverse 
expansion effect may even decrease when  crossing 
the transition region \cite{van-hove}. 
Thus one expects an anomaly in  
the energy dependence of transverse
hadron AT: the average transverse momentum increases with collision 
energy when the early stage matter is either in pure confined or in pure  
deconfined phases,
and it remains approximately constant when  the matter
is in the mixed phase.  


A simplified picture with $T=T_{C}=const$ inside the mixed phase is
changed, if the created early stage matter has
a non--zero baryonic density.
It was, however,  demonstrated \cite{Hung:98} that the main
qualitative features ($T\cong const$, $p\cong const$, 
and a minimum of the function
$p(\varepsilon)/\varepsilon$ $vs$ $\varepsilon$)
are present also in this case.
In the statistical model of the early stage  model \cite{Horn}, which  correctly predicted the energy dependence
of pion and strangeness yields, the modification of the equation of
state due to the deconfinement phase transition
is located between the lab. energy of 30 and  about 200 $A\cdot$GeV.
Thus, the anomaly in energy dependence of transverse hadron AT
may be expected in this energy range.
Can one  see this anomaly in the experimental data?

The experimental data  on transverse mass
spectra are usually parameterized by a
simple exponential dependence:
\begin{equation}\label{exp}
\frac{dN}{m_{T}~dm_{T}}~=~C\exp\left(-\frac{m_{T}}{T^{*}}\right)~,
\end{equation}
where the inverse slope parameter $T^{*}$ 
is sensitive to both the thermal and
collective motion in the transverse direction. In the parameterization
(\ref{exp}) the shape of the $m_T$ spectrum is fully determined by
a single parameter, the
inverse slope $T^*$.
In particular,
the average transverse mass, $\langle m_{T} \rangle$,
can be expressed as:
\begin{equation}\label{mt}
\langle m_{T}\rangle ~=~ T^{*}~+~m~+~\frac{(T^{*})^{2}}{m~+~T^{*}}~.
\end{equation}

The energy dependence of the inverse slope parameter fitted to
the $K^+$ and $K^-$ spectra for central Pb+Pb (Au+Au)
collisions is shown in Fig.~\ref{N63fig1}. 
The results obtained at AGS \cite{ags}, SPS \cite{na49} and
RHIC \cite{rhic} energies are compiled.
The striking features of the data can be summarized and interpreted as
follows.
\begin{itemize}
\item
The $T^{*}$ parameter increases strongly with collision energy up to the lowest
(40 A$\cdot$GeV) SPS energy point.
This is an energy region where the creation of confined matter at
the early stage of the collisions is expected.
Increasing collision energy leads to an increase of the
early stage  temperature and pressure.
Consequently,  the  transverse AT of produced hadrons,
measured by the $m_T$ spectra, increases with increasing energy.
\item
The $T^{*}$ parameter is approximately independent
of the collision
energy in the SPS energy range.
In this energy region the transition between confined and deconfined matter
is expected to be located.
The resulting modification of the  equation of state 
``suppresses'' the hydrodynamical transverse expansion and
leads to the observed plateau structure in 
the energy dependence of the $T^*$ parameter.
\item
At higher energies (RHIC data)  the $T^{*}$ again increases with collision
energy. The equation of state  at the early stage  becomes again stiff,
the  early stage temperature and pressure  increase with collision energy.
This results in increase of $T^{*}$ with  energy.
\end{itemize}

\vspace{0.2cm}
Surprisingly, such a plateau-like behavior is typical for 
the caloric curves of nuclear  multifragmentation data \cite{CalorCurve:Exp4}
and reflects the system's evolution within the mixed phase \cite{Bugaev:00, Bugaev:00b, Bugaev:01}.  Since the kaon AT behavior is similar in shape and in physics
to the usual caloric curves of nuclear multifragmentation, I suggest to call it the transverse caloric curve.  

The anomalous energy dependence of the $m_T$ spectra is a characteristic
feature of the kaon data.
Why is this the case?
How do the $m_T$ spectra of other hadrons look like?
The answer is rather surprising: among the measured hadron species the kaons
are the best and unique particles for observing the effect of the
modification of the equation of state due to the onset
of deconfinement.
The arguments are 
as follows.
\begin{itemize}
\item
The kaon $m_{T}$--spectra are only weakly affected by the hadron
re--scattering and resonance decays during the post--hydrodynamic hadron
cascade \cite{BD:00,TLS:01}. In fact, both effects should have modified 
the low $p_T$ range of kaon spectra, but into opposite directions. 
As a result, both effects almost compensate each other for kaons. 
\item
A simple one parameter exponential fit (\ref{exp}) is quite accurate up to $m_{T}-m
\cong 1$~GeV for $K^{+}$ and $K^{-}$ mesons in A+A collisions at all
energies. This means that the energy dependence of the average transverse
mass $\langle m_{T} \rangle$  and average transverse momentum
$\langle p_{T} \rangle$ for kaons is
qualitatively the same as that for the parameter $T^{*}$.
This simplifies the analysis of the experimental data.
\item
The high quality data on $m_T$ spectra of
$K^+$ and $K^-$ mesons in central Pb+Pb (Au+Au)
collisions
are available in the full range of
relevant energies.
\end{itemize}

\begin{figure}[ht]
\centerline{\epsfig{file=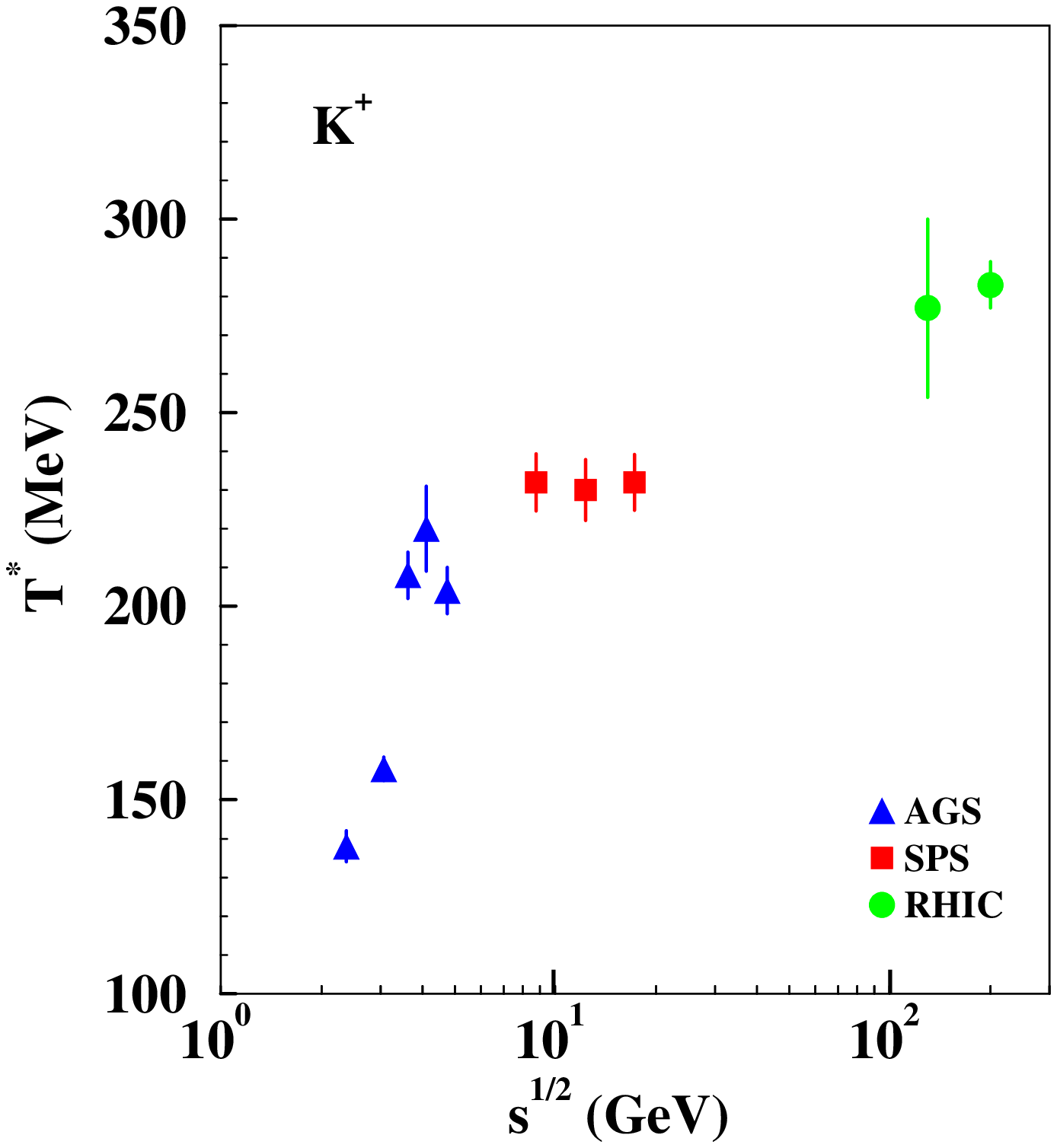,width=8cm}
\hspace*{0.7cm} \epsfig{file=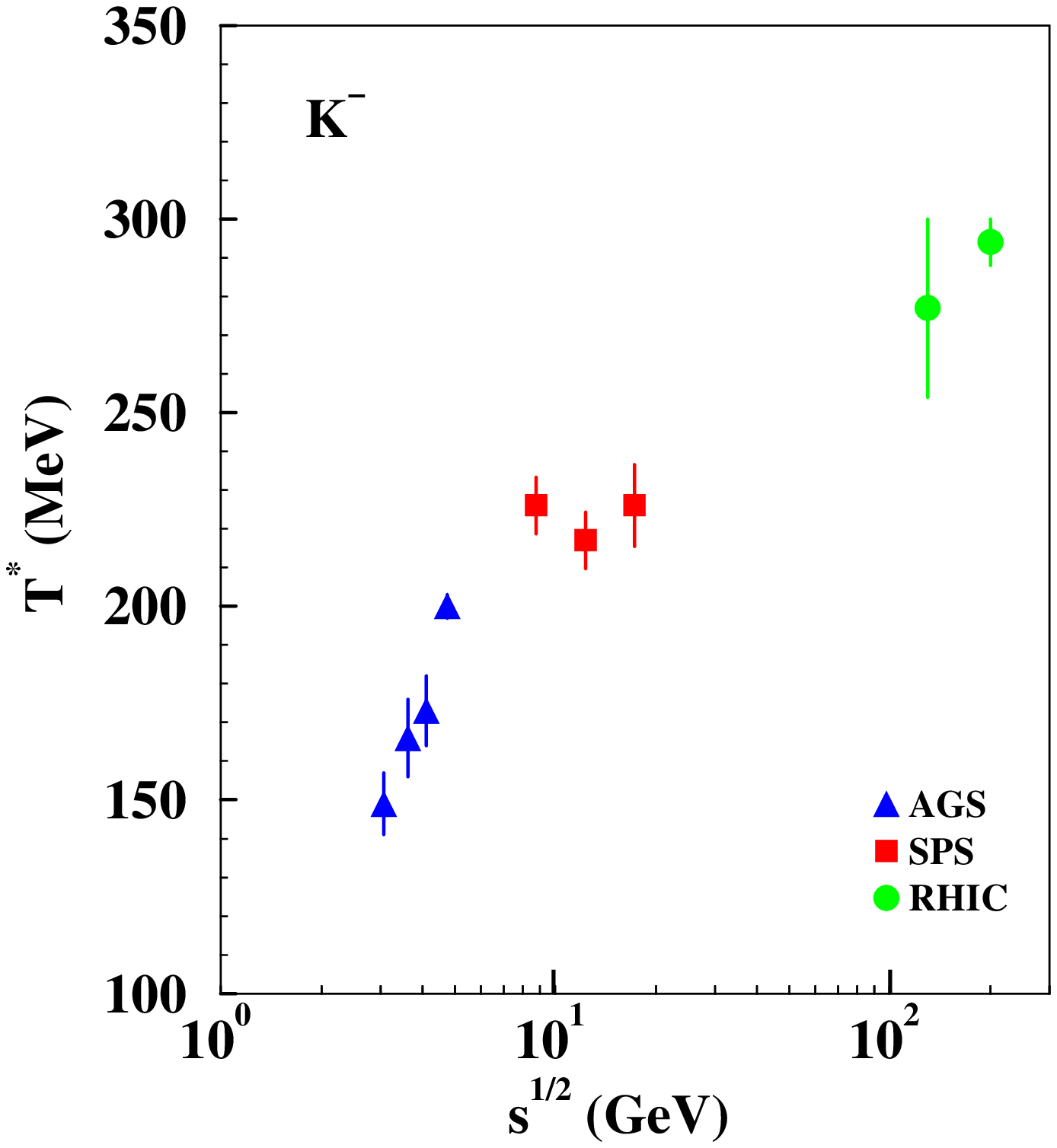,width=8cm}
}
\vspace{0.5cm}
\caption{
The energy dependence of the inverse slope parameter
$T^*$ for $K^+$ (left panel) and $K^-$ (right panel)
mesons produced at mid-rapidity in
central Pb+Pb (Au+Au) collisions at AGS 
\protect\cite{ags}
(triangles),
SPS 
\protect\cite{na49}
(squares) and RHIC 
\protect\cite{rhic}
(circles) energies.
Note that the abscissa  has a logarithmic scale because in the linear one 
it would be a huge plateau. 
}
\label{N63fig1}
\end{figure}

The hydrocascade approach \cite{BD:00,TLS:01} predicts a
strong modification of the $m_{T}$-spectra of protons and lambdas during
the hadron cascade stage in A+A collisions at both the SPS and RHIC. As
the hadron gas expands, the pions excite $\Delta$ and $\Sigma^*$
resonances and transform some part of their transverse energy in the
nucleon and hyperon sectors. Therefore, the hadron re--scattering and
resonance decays lead to significant increase (about 40$\%$ \cite{TLS:01}) of
the inverse slope parameters $T^*$ for (anti)nucleons and (anti)lambdas at the
expense of the pion transverse energy 
(also see discussion in Ref.~\cite{Bugaev:01d}). 
These changes of the slopes $T^{*}$
are not directly related  to the  equation of state
of the matter at the early stage.
It is rather difficult to separate these hadron-cascade effects and in
any case, this separation will be strongly model dependent.
Note also
that a simple exponential fit (\ref{exp}) neither works  for
$\pi$-mesons ($T^{*}_{low-p_{T}}>T^{*}_{high-p_{T}})$  \cite{star1, star2, phenix1}
nor for protons
and lambdas ($T^{*}_{low-p_{T}} < T^{*}_{high-p_{T}}$) \cite{phenix1, phenix2}. 
This means that the
average transverse masses, $\langle m_{T}\rangle$, and their energy
dependence are not connected to the behavior of the slope parameters in
the simple way described by Eq. (\ref{mt}): one should separately consider both
$T^{*}_{low-p_{T}}$ and $T^{*}_{high-p_{T}}$ slopes for these hadrons (see also 
Ref.~\cite{TLS:01} for details).

The transverse AT of $\Omega$ hyperons and $\phi$ mesons should, 
as in the case of kaons,
be sensitive to the matter equation of state
at the early stage of the collisions.
These particles seem to decouple just after the hadronization is over
and  they do not participate in the hadron
cascade stage \cite{BD:00, TLS:01,Bugaev:01d,Bugaev:02b}.
Unfortunately the spectra of  $\Omega$ hyperons are measured only at
top SPS \cite{Omega1} and RHIC energies \cite{QM02}.
More data exist for $\phi$ meson production \cite{phiA, phiB, phi2}.
However, 
the
large uncertainties in the experimental results
do not allow to draw a definite conclusion on the
possible anomaly in the energy dependence of $m_T$
spectra. 

Thus, 
one observes an anomalous energy
dependence of  transverse mass
spectra of $K^+$ and $K^-$ mesons
produced in central Pb+Pb (Au+Au) collisions.
The inverse slopes of the $m_T$--spectra increase with  energy
in the AGS and RHIC energy domains, whereas they remain constant
in the intermediate SPS energy range.
As was argued above,   this  anomaly
is  caused by a modification of the equation of state
in the transition region between confined and deconfined matter.
Now this observation is  considered  as a new signal, in addition to the
previously reported
anomalies in energy dependence of the pion and strangeness production,
of the onset of deconfinement  located at the low SPS energies.

\section{Concluding Remarks}

Here I discussed the possible experimental signatures of the deconfinement transition.
Note that none of the signals suggested in 80-th and 90-th was ever observed 
in a way as it was suggested. Threfore, the formulation of a realible signal 
of  the deconfinement transition is highly non-trivial task.
So far, there are only three signals of the deconfinement  which have been 
observed experimentally. 
They are: the Kink in the ratio of  pion yield  to number  of wounded nucleons  \cite{Kink}, the Strangeness Horn \cite{Horn} in $K^+$ to $\pi^+$ numbers ratio
and the Step \cite{Step} in kaon AT presented in this chapter. 
All these irregularities  are emerging at the very same lab.  energy
of about 30 GeV$\cdot$A.  

On the other hand the early hadronization of multistrange and charmed 
(open and hidden) hadrons is indirect, but, nevertheless, firm evidence for thermalization and collective flow of the QGP at the stage of its hadronization. 
A systematic study of A+A collisions with the nuclei of different sizes (including the small ones) based on the suggested hypothesis of early hadronization will, perhaps, find out
the threshold for the QGP collective flow. 




\chapter*{Conclusions}

\addcontentsline{toc}{chapter}{Conclusions}

In the present dissertation I discussed two main PTs that are studied in the strongly
interacting matter: the nuclear liquid-gas PT and the transition of liberating the color degrees of freedom from hadrons know as the deconfinement transition. 
Since the quantum chromodynamics, which is the fundamental theory of strong interaction,
has not reached yet the stage when it can be directly applied to the description 
of the experimental data  obtained in the relativistic heavy ion collisions, 
we are faced with the necessity to develop the phenomenological approaches to
model the EOS of strongly interacting matter in a very wide range of densities - 
from the one third of  normal nuclear density at which  the break up of
nuclear systems occurs in the reaction of nuclear multifragmentation  up to hundreds 
of normal nuclear densities which are expected to be created at LHC.   
In fact, the results of the experiments cannot be interpreted without the underling 
statistical descriptions. 

However, the construction of  the realistic EOS is not a straightforward  procedure
because it is necessary to take into account the specific features of strong interaction 
and combine them with some general  requirements, like thermodynamic self-consistency, 
causality e.t.c. One important additional requirement, which I adopted in my research, is 
to reduce as much as possible the mean-field  approximation for statistical models. 
So far, I suggested a single phenomenological mean-field EOS of nuclear matter from 
thermodynamically self-consisten class of Walecka-like models, but analytically I solved 
the SMM, the Mott-Hagedorn resonance gas  and QGBST models in thermodynamic limit, and the CSMM, the GSMM, the GBM, the HTM and the HDM for finite systems. Note that only two of these 
statistical models, the SMM and the GBM,  were formulated by other researchers, whereas all the rest 
was formulated  in my works with collaborators. 

This additional requirement to study and develop truly statistical, i.e. non mean-field,
models led to  several very  valuable results. 
The first of these results is that  the solution of a simplified SMM and calculation of
the critical exponents  of nuclear matter allowed me and my former student P. T. Reuter to predict  a narrow range  for the Fisher  index $\tau \approx 1.825 \pm 0.025$ which, in contrast to the  Fisher value  $\tau \approx  2.16 $,  is consistent with 
the  experimental results $\tau \approx 1.8 - 1.9$  obtained by  the  ISiS and EOS  Collaborations.   
Such a  value of $\tau$ shows that the nuclear matter  belongs to other 
universality class $(\tau < 2)$ than 
the Fisher droplet model $(\tau > 2)$  and that the nuclear liquid-gas PT has  the tricritical endpoint rather than the critical one.  

However, the analysis of the SMM critical indices showed me that the definition of the $\alpha$ 
index via the critical isochore is  not a well established definition.  For instance, 
the critical isochore in a simplified SMM belongs to the boundary of the mixed and liquid phases and, hence,  it explicitly  breaks down the implicit Fisher assumption which is necessary to  prove the validity of the scaling inequalities for critical exponents. 
Since there exist substances with rather  complicated phase diagrams, where the critical isochore (partially) lies out of the mixed phase, one  has  to  search for an alternative 
definition. Fisher's suggestion to use a specially defined index $\alpha_s$ ``saves'' the
scaling inequalities for known models, but it does not have such a simple and clear meaning
like $\alpha$ and has to be justified further. Perhaps, new models and new experimental tests
of the inequalities of critical indices will clarify this problem.   

The second result of high value obtained in this dissertation on a firm ground of statistical mechanics is 
an exact analytical solution of several statistical models for finite systems. 
This was possible because of the invention of a new and powerful mathematical method,
the Laplace-Fourier transform.  This method allows one to identically rewrite the GCE partition via the sum over the simple poles of the isobaric partition. 
The analysis of the behavior of
these singularities  for the case with a PT and without it led me  to the formulation of the 
finite volume analogs of phases and their phase diagram in the temperature -- real part of the effective chemical potential plane. 
These definitions based on the first principles of statistical mechanics allowed me to  clarify 
the pitfalls of the definition of phases suggested in works of T. Hill, Ph. Chomaz, F. Gulminelly and others for finite systems. 

Also such a representation shows that in the CSMM, the GSMM and the GBM  the low density phase, i.e the gas of nuclear fragments in the CSMM and the GSMM and hadron gas in the GBM,
are the only stable states which have a real value of free energy, whereas all other states
are metastable and they have complex values of free energy and effective chemical potential. 
The different values of the real part of the effective chemical potential generated by interaction indicate  that in finite system the states which belong to the same partition and have the same value of chemical potential are not in true chemical equilibrium with each other.  As discussed in the dissertation, the imaginary part of free energy of these $\lambda_n$ states  gives us an estimate  for the formation/decay time $\tau_n$  of  
the state $\lambda_n$ (n = 0, 1, 2, ...). 
At high densities  the  formation/decay time $\tau_n$ of the $\lambda_n$ state
of the system having volume $V$ 
is  $\tau_n (V) \approx \frac{V}{V_{\rm o} \pi \, n \, T}$. 
Here $V_{\rm o}$ denotes an eigen volume of the nucleon (minimal volume of the QGP bag)
for the CSMM or the GSMM (the GBM).   
The existence of the finite decay time puts some restriction on the application of 
hydrodynamic and even kinetic methods to describe such collective  $\lambda_n$ states  
in A+A collisions.  

Indeed, to apply the hydrodynamic description to the $\lambda_n$ state
defined in the local rest frame RF$^*$ it is necessary that its decay time $\tau_n$ being larger that the typical hydro time $t_H = [ \partial_\mu \cdot u^\mu ]^{-1} $, i.e.
$\tau_n \ge t_H$. In other words, the only $\lambda_n$ modes which live longer than 
the hydro expansion rate ($u^\mu $ is the 4-vector of hydrodynamic velocity) can 
be described  by relativistic hydrodynamics. Since the volume 
$V_h = \rho ^3 \, \frac{\partial\, x_1^*}{\partial\, \rho} \frac{\partial\, x_2^*}{\partial\, \rho} \frac{\partial\, x_3^*}{\partial\, \rho}$ here should be understood
as the homogeneity volume in the local RF$^*$ defined in terms of the spatial derivatives of the local particle (charge) density $\rho$,  one can see that for large $n$ 
and/or very high gradients $\frac{\partial\, \rho}{\partial\, x_i^*}$ the inequality $\tau_n (V_h)  \ge t_H$ cannot be fullfilled. 
The hope that such $\lambda_n$ states which cannot be described by hydro, but  can be described by relativistic Boltzmann equation, is also questionable because there may exist so large $n$ that 
its decay time $\tau_n = \rho ^3 \, \frac{\partial\, x_1^*}{\partial\, \rho} \frac{\partial\, x_2^*}{\partial\, \rho} \frac{\partial\, x_3^*}{\partial\, \rho} 
\frac{1}{V_{\rm o} \pi \, n \, T}$ will be shorter than the collision time $t_{Coll} \approx 
[ \sigma \langle v \rangle \rho ]^{-1} $ ($\sigma$ is the typical cross-section of
the elastic scattering and $\langle v \rangle $ is the mean thermal velocity of  particles
in the local RF$^*$) which, according to Bogolyubov classification, does not allow 
to apply the  one-particle kinetic treatment  and keep  thermodynamic equilibrium. 
Consequently, if the metastable $\lambda_{n>0}$ states with large $n$ are created in the relativistic A+A collisions, then they can be, in principle,  detected via the failure of the hydrodynamic description and/or via  
the absence of the local thermodynamic equilibrium.    
In addition, this discussion  indicates us the necessity to develop some  new microscopic methods 
to describe the metastable and short-lived collective states at  phase equilibrium.

Despite these  difficulties the invention of the Laplace-Fourier transform method 
allows one  to rigorously describe  the  PTs in the presence of the long-range interaction  
which  leads to the absence of thermodynamic limit.  As I discussed  in the chapter 2,
now such a problem can be reduced to the analysis of a PT in each finite subsystem
with nearly constant external field generated by the surrounding subsystems. 
Of course, the necessary condition for this method  is that the whole system under investigation can be divided into the set of subsystems with small gradients of the external
long-range  field.  
Such a condition can generate some restrictions on the application of this method, but
from the theoretical point of view it is  an important step towards the rigorous treatment  of PTs in  finite systems. 

At the moment it is unclear whether  with the help of the Laplace-Fourier transform method
one can  establish the experimental signals of the 2$^{nd}$ order PT 
in nuclear matter, predicted by a simplified SMM. 
But, perhaps, the refined theoretical  methods can help us in resolving this problem.

The Laplace-Fourier method allowed me to formulate and solve analytically the HDM and
find out the upper and lower bounds for the surface partition of large physical clusters consisting from the constituents of finite size. With the help of these exact results it 
was possible to derive and  improve the Fisher  parameterization for the surface tension
coefficient of large clusters. Also  these results clearly showed  me  that    the GBM lacks 
a very important ingredient -- the surface free energy of the QGP bags. 
An inclusion of the temperature and chemical potential dependent surface tension 
coefficient into the GBM led to a formulation of  a new and more realistic exactly solvable model, the QGBST model,  which  describes the 1$^{st}$ and 2$^{nd}$ order PTs with the cross-over   and demonstrates that for $\tau \le2$ the  quantum chromodynamics has the tricritical point.   This is the third principal results of the present dissertation. 

The QGBST model predicts an existence of an additional PT to  the deconfinement one  at 
the null line of the surface tension coefficient of the QGP bags 
(see chapter 4 for details). 
Perhaps,  its unique 
property, the power-like mass distribution of the QGP bags, 
at the tricritical point can be verified experimentally.  
Such a research requires the normalization of the QGBST model onto  existing 
thermodynamic functions obtained by the lattice quantum chromodynamic simulations and 
its extension to finite systems using the Laplace-Fourier transformation method.

An important feature of the statistical methods developed in the present dissertation 
is that they are similar for both of PTs discussed here. Thus, the suggested 
Laplace-Fourier transform method can be a good  starting point to create a common
theoretical language  for  the communities which are studying  the nuclear liquid-gas PT and
the deconfinement PT. 
Moreover, I believe that the results obtained in this dissertation on
the PTs in finite systems along with  the hydrokinetic equations derived here
are two key  elements to build up the microscopic kinetic theory of 1$^{st}$ PTs
in finite systems. As I discussed on several occations in the present work, such 
a truly  microscopic kinetic theory is vitally necessary for several fields. 
I think that it can be build up in a few years and can be verified experimentally. 
A similarity and difference  of the nuclear liquid-gas and deconfinement PTs 
can be used to successfully  work out  such a theory. Thus, the absence of a strong flow
in the   nuclear multifragmentation epxeriments is an attractive feature 
of these experiments  which can be used  for a detailed  justification 
of the  whole concept of  the microscopic kinetic theory of 1$^{st}$ PTs
in finite systems, which later on can be applied to  the more complicated 
searches for the QGP and deconfinement PT.

\newpage

The author is cordially thankful to the coauthors, friedns and colleagues
of the Department of High Energy Densities Physics at the 
Bogolyubov ITP of National Academy of Sciences   of Ukraine  
whose discussions and remarks were important and fruitfull to my research. 
The special  thanks are to the friends of mine  Blokhin A. L., Borisenko O. A. and Blaschke D. B., 
to my former teacher 
Gorenstein M. I. and to Zinovjev G. M.  for  their persistent  support. 
The most cordial thanks go to my dear wife Tanya  whose permanent care and 
efficient  help in preparing this manuscript were indispensible to me.


\chapter*{Appendices}

\addcontentsline{toc}{chapter}{Appendices}

\section*{Appendix A}

\addcontentsline{toc}{section}{Appendix A}

\addcontentsline{toc}{subsection}{The Canonical  Treatment of  the  Relativistic VdW EOS.}
\label{CE_appd}

In the following I would like to  study the differences between the linear and
the non-linear approximation:
the total excluded volumes $v_q=v_q(N_1,N_2)$ of the corresponding
partial pressures.
In the linear pressure formula (\ref{eq:p-lin}) each component has its own
total excluded volume given by
\begin{equation}
  \label{eq:v-lin1}
  v^{\rm\,lin}_1 \equiv N_1 b_{11} + N_2 \tilde{b}_{21} ~, \quad
  v^{\rm\,lin}_2 \equiv N_1 \tilde{b}_{12} + N_2 b_{22} ~,
\end{equation}
whereas in the non-linear pressure formula (\ref{eq:p-nl}) there is
a {\em common\/} total excluded volume for both components
\begin{equation}
  \label{eq:v-nl}
  v^{\rm\,nl}_1 = v^{\rm\,nl}_2 = v^{\rm\,nl}
    \equiv N_1 b_{11} + N_2 b_{22}
           - {\textstyle \frac{N_1\,N_2}{N_1+N_2} } D ~.
\end{equation}
It can be readily checked that it is either
$v^{\rm\,lin}_1 \le v^{\rm\,nl} \le v^{\rm\,lin}_2$ or 
$v^{\rm\,lin}_1 \ge v^{\rm\,nl} \ge v^{\rm\,lin}_2$\,,
i.\,e.~the pole of the non-linear pressure 
always lies between both poles of the linear pressure. 
Hence, there are values $N_1, N_2$\,, where the non-linear pressure
is still finite, but the linear pressure formula is yet invalid since
one of the partial pressures has already become infinite.
Consequently, the domain of the non-linear approximation is larger.

For given $V$ the two domains can be expressed by the limiting densities
(\ref{eq:lim-n1,q-lin}) and (\ref{eq:lim-n1-nl})\,, which are defined
by the poles $v_q(N_1,N_2)=V$ of the corresponding pressure.
In the linear approximation one obtains the expressions
\begin{eqnarray}
  \hat{n}_{1,1}^{\rm\,lin} (n_2)
    = \frac{1 - \tilde{b}_{21} n_2}{b_{11}}\, ~,
  &\quad&
    \hat{n}_{1,2}^{\rm\,lin} (n_2)
      = \frac{1 - b_{22} n_2}{\tilde{b}_{12}}\, ~.
\end{eqnarray}
For given $n_2$\,, therefore, the domain of $p^{\rm\,lin}$ (\ref{eq:p-lin}) is
\begin{equation}
\label{eq:avr_lin}
  0 \le n_1^{\rm\,lin} < \min \l\{ \hat{n}_{1,1}^{\rm\,lin} (n_2)\,,
                                       ~\hat{n}_{1,2}^{\rm\,lin}(n_2) \r\} ~.
\end{equation}

In the non-linear approximation there is solely one limiting density
\begin{eqnarray}
  \label{eq:n1xnl}
  \hat{n}_1^{\rm\,nl} (n_2)
    &=& {\textstyle
          \frac{1 - 2\,b_{12} n_2
                + \sqrt{(1-2b_{12} n_2)^2 + 4b_{11} n_2 \,(1-b_{22} n_2)}}
               {2\,b_{11}} }\, ~.
\end{eqnarray}
Consequently, for given $n_2$ the domain of $p^{\rm\,nl}$ (\ref{eq:p-nl}) is
\begin{equation}
\label{eq:avr_nl}
  0 \le n_1^{\rm\,nl} < \hat{n}_1^{\rm\,nl} (n_2) ~.
\end{equation}
In the non-linear approximation there is furthermore a region where
the pressure has negative partial derivatives with respect to the smaller
particles' number, $\pd p^{\rm\,nl}/\pd N_2<0$\,.
The condition $\pd p^{\rm\,nl}/\pd N_2=0$ defines the boundary
of this region
\begin{eqnarray}
  \label{eq:n1nl_bd}
  {\hat{n}_1^{\rm\,nl,\,bd} (n_2)} 
  & =&
    {\textstyle
      \frac{1 - 2\,(b_{12}-b_{22})\,n_2
            + \sqrt{(1 - 2\,(b_{12}-b_{22})\,n_2)^2 + 8\,(b_{11}-b_{12})\,n_2}}
           {4\,(b_{11}-b_{12})} }\, ~.
\end{eqnarray}
For given $n_2$ a negative derivative $\pd p^{\rm\,nl}/\pd N_2<0$
occurs only at a density $n_1>\hat{n}_1^{\rm\,nl,\,bd}$\,, while
the derivative $\pd p^{\rm\,nl}/\pd N_1$ is always positive
for $R_2\le R_1$ as readily checked.

In Fig.~\ref{figs:1}\,(a) the functions $\hat{n}_1(n_2)$
are presented in {\em dimensionless\/} variables $\hat{n}_1 b_{11}$
and $n_2 b_{22}$\,.
The properties of these dimensionless functions depend only on
the ratio of the two radii $R_2/R_1$\,.
The smaller this ratio is, the higher is the maximum value of
$\hat{n}_1^{\rm\,nl}$\,, while the region of negative derivatives
$\pd p^{\rm\,nl}/ \pd N_2$ becomes narrower.
The straight line $\hat{n}_{1,1}^{\rm\,lin} (n_2)$ starts
always at $1/b_{11}$\,, but its slope decreases for smaller $R_2/R_1$\,,
whereas $\hat{n}_{1,2}^{\rm\,lin}(n_2)$ ends at $1/b_{22}$ and its
slope increases.
The pressure of the {\em separated model\/} (\ref{eq:p-sp_CE}) would
yield one straight line from $n_1 b_{11}=1$ to $n_2 b_{22}=1$
in Fig.~\ref{figs:1}\,(a)\,, for any ratio of the radii $R_1$ and $R_2$\,.

For very small ratios $R_2/R_1$\,, i.\,e.~for $R_2\to 0$\,,
one finds from Eq.~(\ref{eq:v-nl}) that
$v^{\rm\,nl}\to N_1 b_{11}\,[1-\frac{3}{4} N_2/(N_1+N_2)]$\,.
This yields the maximum density $\max(\hat{n}_1^{\rm\,nl})=4/b_{11}$
for $N_2 \gg N_1$\,.
Thus $\hat{n}_1^{\rm\,nl}$ exceeds the maximum density of the linear
approximation or of the corresponding one-component {\sl VdW\/} gas,
$\max(\hat{n}_{1,1}^{\rm\,lin})=\max(n_1^{\rm\,oc})=1/b_{11}$\,, by a
factor of four in this case.

Note that the value $4/b_{11}$ appears in the linear approximation as well:
For $v_2^{\rm\,lin}\to V$ it is $\max(\hat{n}_{1,2}^{\rm\,lin})=4/b_{11}$
at $n_2=0$\,, but this density cannot be achieved because
$p_1^{\rm\,lin}(n_1,n_2)$ is infinite for $n_1\ge 1/b_{11}$\,.

Let us consider now the consequences of negative derivatives
$\pd p^{\rm\,nl}/\pd N_2<0$ in the non-linear approximation.
If a negative $\pd p^{\rm\,nl}/\pd N_2$ occurs for a density
$n_1^{\,\prime}={\rm const.}$ at $n_2=0$\,, the pressure
$p^{\rm\,nl}(n_1^{\,\prime},n_2)$ has a minimum at a certain density
$n_{2,\rm\,min}>0$\,, which is determined by the boundary
$\hat{n}_1^{\rm\,nl,\,bd}(n_2)$\,.
For increasing $n_1$ along the boundary, 
consequently, the non-linear pressure is always lower
than for increasing $n_1$ at fixed $n_2=0$\,.
Hence along the boundary higher densities can be achieved,
in particular $n_1>1/b_{11}$\,.

Therefore, exceeding of $n_1^{\rm\,nl}=1/b_{11}$ requires that
the boundary starts inside the the non-linear domain at $n_2=0$\,.
Thus the condition $\hat{n}_1^{\rm\,nl,\,bd}(0)<1/b_{11}$ provides
the critical radius $R_{2,\rm\,crit}$ (\ref{eq:R2crit_CE})\,,
\begin{equation}
  \label{eq:R2crit_appd}
  b_{11}<2 b_{12} \quad \leadsto \quad
  R_{2,\rm\,crit}(R_1) = (\sqrt[3]{4} - 1)\,R_1  ~.
\end{equation}
On the other hand the boundary starts at $n_1=8/(14\,b_{11})$ for $R_2\to 0$
at $n_2=0$\,, i.\,e.~for any density $b_{11} n_1\le 8/14\approx 1/2$
negative values of $\pd p^{\rm\,nl}/\pd N_2$ do not occur for any radii.

Although it is pathological that for {\em high\/} densities $n_1$
the non-linear pressure firstly decreases, if particles of the second
and smaller component are added to the system, there is a reasonable
explanation for the lowered pressure along the boundary (\ref{eq:n1nl_bd})
for small radii $R_2<R_{2,\rm\,crit}$\,.

Consider, for instance, $n_1 b_{11}=0.9$ in Fig.~\ref{figs:1}\,(a)\,.
Since it is $R_2/R_1=0.4$\,, the dimensionless density of the small
particles at the boundary $\hat{n}_1^{\rm\,nl,\,bd}(n_2)$ nearly
vanishes, $n_2 b_{22}\approx 0.05$\,, whereas the absolute amounts
of the small and large particles are about equal.
For the excluded volume interaction of the large particles in the
pressure formula (\ref{eq:p-nl})\,, therefore, the influence of the
mixed term $b_{12}$ becomes comparable to that of the distinctly larger
non-mixed term $b_{11}$\,.

For $R_2/R_1\ll 1$ one obtains $n_2\gg n_1$ near the boundary
at $n_1 b_{11}=0.9$\,, i.\,e.~the large particles are completely surrounded
by the smaller particles and interact mostly with these but hardly with
other large particles anymore.
In this situation, consequently, the excluded volume interaction of the
large particles is governed by the essentially smaller $b_{12}$\, and not by
$b_{11}\le 8\,b_{12}$\,.

One might interprete this behaviour as an effective attraction
between small and large particles, but it is rather a strong reduction
of the large particles' excluded volume suppression.

%

As {\sl VdW\/} approximations are low density approximations
they coincide for these densities, but they evidently become inadequate
near the limiting densities:
both discussed formulations do evidently not match the real
{\em gas of rigid spheres\/} there.

For high densities the linear approximation behaves natural,
i.\,e.~it is $\pd p^{\rm\,lin}/\pd N_q>0$ always.
However, one has to introduce the additional terms
$\tilde{b}_{12}$ and $\tilde{b}_{21}$\,.
For the choice (\ref{eq:bTi-s}) these terms provide a one-component-like
behaviour in the limits $R_2=R_1$ and $R_2=0$\,, but they have no concrete
physical meaning.

In the non-linear approximation there occur pathologic pressure derivatives
$\pd p^{\rm\,nl}/\pd N_2<0$ for $R_2\ll R_1$\,.
However, the non-linear formulae may be used for special purposes,
e.\,g.~for $n_1>1/b_{11}$ at intermediate $n_2 b_{22}$\,, where the linear
approximation is yet invalid.


\addcontentsline{toc}{subsection}{Stability of the Non-linear Approximation.}
\label{nl_appd}
The non-linear enhancement in the GCE or the occurence of
negative values for $\pd p^{\rm\,nl}/\pd N_q$ in the CE
suggest a further investigation concerning the thermodynamical
stability of the non-linear approximation.

One can readily check that in the CE it is $\pd p^{\rm\,nl}/\pd V<0$
generally, so there is no {\em mechanical\/} instability.
To investigate wether there is a {\em chemical\/} instability \cite{Reichl}
it is necessary to study partial derivatives with respect to the
particle numbers, $\pd/\pd N_q$\,, of the {\em chemical potentials\/}
\begin{eqnarray}
  \mu_p(T,V,N_1,N_2)
    &\equiv& - T\,{\textstyle \frac{\pd}{\pd N_p} }\,\ln[Z(T,V,N_1,N_2)] ~.
\end{eqnarray}
Partial derivatives of the {\em pressure\/} with respect to the
particle numbers $\pd p/\pd N_q$ have no relevance here.

For the examination of chemical stability it is appropriate
to switch from the {\em free energy\/} of the CE,
$F(T,V,N_1,N_2) \equiv - T \,\ln[Z(T,V,N_1,N_2)]$\,,
to the {\sl Gibbs\/} {\em free energy\/} or {\em free enthalpie\/}
\begin{eqnarray}
  G(T,p,N_1,N_2) &\equiv& F + p\,V = \mu_1 N_1 + \mu_2 N_2 ~,
\end{eqnarray}
where $\mu_q(T,p,N_1,N_2) \equiv \pd G / \pd N_q$\,.
This requires that $p(T,V,N_1,N_2)$ can be solved for $V(T,p,N_1,N_2)$\,,
which is the case for the non-linear approximation,
\begin{eqnarray*}
  V^{\rm\,nl}(T,p,N_1,N_2)
    &=& {\textstyle \frac{N_1+N_2}{p/T} } + N_1 b_{11}
        + N_2 b_{22} - {\textstyle \frac{N_1\,N_2}{N_1+N_2} }D\,.
\end{eqnarray*}
Further it is useful to introduce the molar free enthalpie
$g \equiv G/(N_1+N_2) = g(T,p,x_1)$ with the molar fractions
$x_1 \equiv N_1/(N_1+N_2)$ and $(1-x_1) = x_2 \equiv N_2/(N_1+N_2)$ of
component 1 and 2\,, respectively.
Then the chemical stability of a binary mixture \cite{Reichl}
corresponds to the condition
\begin{equation}
  \label{eq:ch-stab}
  {\textstyle \frac{\pd^2}{\pd x_1^{\,2}} } \,g(T,p,x_1)
    = {\textstyle \frac{\pd \mu_1(T,p,x_1)}{\pd x_1}
                  - \frac{\pd \mu_2(T,p,x_1)}{\pd x_1} }
    > 0 ~.
\end{equation}
For the non-linear approximation one obtains
\begin{eqnarray}
  g^{\rm\,nl}(T,p,x_1)  ~
    & = &
      x_1 \l\{ T\,\ln\l[{\textstyle \frac{x_1}{\phi_1} \frac{p}{T} }\r]
               + p \l( b_{11} - (1-x_1)^2\,D \r) \r\}  \nonumber \\
    &  + & (1-x_1) \l\{ T\,\ln\l[{\textstyle \frac{1-x_1}{\phi_2}
                                            \frac{p}{T} }\r]
                       + p \l( b_{22} - x_1^{\,2}\,D \r) \r\} ~,  
\end{eqnarray}
and thus condition (\ref{eq:ch-stab}) is satisfied:
\begin{eqnarray}
  {\textstyle \frac{\pd^2}{\pd x_1^{\,2}} } \,g^{\rm\,nl}(T,p,x_1)
    = {\textstyle \frac{T}{x_1} }
      + {\textstyle \frac{T}{1-x_1} } + p\,2D
    > 0 ~.
\end{eqnarray}
Therefore, the system described by the non-linear approximation is
thermodynamically stable -- despite the pathologic behaviour in special cases.
Due to the equivalence of the thermodynamical ensembles this is true for any
representation of the model.


\section*{Appendix B}

\addcontentsline{toc}{section}{Appendix B}



\addcontentsline{toc}{subsection}{Derivation of the Lorentz Contracted Excluded Volumes}
\label{RelEV_appd}

In order to study  the high pressure limit it is necessary to 
find out  the excluded volume of two ellipsoids which are obtained by the Lorentz 
contraction of the spheres.  
This is rather involved problem. Fortunately, the  analysis below 
requires to know only the 
ultrarelativistic limit when the mean  energy per particle  is very high  compared to  the mass of particle.
The problem can be simplified further  since 
it is sufficient to find an analytical expression for 
the relativistic excluded volume 
 the  collinear particle velocities because the  configurations with 
 the  noncollinear  velocities
have larger excluded volume and, hence, are suppressed.
Therefore, one can safely consider the excluded volume   
produced  by two contracted cylinders (disks)  having the same
eigen volumes as the ellipsoids.  
For this purpose the cylinder's height in the local rest frame is fixed to be   
$\frac{4}{3}$ of a  sphere radius.   

Let me introduce the  different radii $R_1$ and $R_2$ for  the cylinders, and 
consider a zero height for the second cylinder $h_2 = 0$  and  
non-zero height $h_1 $ for  the first one.
Suppose that  the center of the coordinate system coincides with the geometrical 
center of the first cylinder  and  the  $OZ$-axies is perpendicular to the cylinder's  base.
Then the angle $\Theta_v$ between the  partcile velocities is also  the angle between 
the  bases two cylinders.
To simplify the expression for pressure the Lorentz frame is chosen to be the rest frame of the whole system. 

In order to estimate the excluded volume one should  fix the particle velocities 
and translate the second cylinder around the first one keeping the angle $\Theta_v$ 
fixed. The desired excluded volume is obtained as the volume 
occupied by the center of the
second cylinder under these transformations.
Considering the projection on the $XOY$ plane 
(see the panel a) of   Fig.~\ref{CylDeriv}),   
one should transform the ellips of the  radii \mbox{$R_x = R_2 \cos \lp \Theta_v \rp$} 
and $R_y = R_2 $ around the circle of radius $R_1$. 
We approximate it by the circle of the averaged radius of the ellips  
$\bra R_{XOY} \ket = R_1 + (R_x + R_y)/2 = R_1 + R_2 ( 1 + \cos \lp \Theta_v \rp )/2$. 
Then the  first contribution to the excluded volume 
is the volume of the cylinder of the radius $\bra R_{XOY} \ket$ and 
the height  $h_1 = CC_1 $ of the cylinder $OABC$ in the panels .a) and b) 
of  Fig.~\ref{CylDeriv}), 
i.e.,   
\begin{equation}
v_{I} (h_1) = \pi \lp R_1 + \frac{R_2 ( 1 + \cos \lp \Theta_v \rp )}{2} \rp^2 h_1\,\,.
\end{equation}

Projecting the picture onto the $XOZ$ plane as it is shown in the pabel b) of Fig.~\ref{CylDeriv},
one finds that the translations of a  zero width disk   over the upper and lower 
bases of the first cylinder 
(the distance between the center of the disk and the base $CA$ is, 
evidently, $ CD_1 =  R_2 | \sin \lp \Theta_v \rp | $) 
generate the second conrtibution
to  the excluded volume
\begin{equation}
v_{II} (h_1) = \pi  R_1^2 \,2 \, R_2 | \sin \lp \Theta_v \rp | \,\,.
\end{equation}

The last contribution follows from the translation of the disk  from  
the cylinder's base to the cylinder's side as it is shown for the
$YOZ$ plane in  the panel c) of  
Fig.\ref{CylDeriv}.
The  area $BB_1F$ is the  part of
the ellips segment which magnitude  depends on the x-coordinate. 
However, one can approximate it as the  quarter of the disk area  
projected onto the  $YOZ$ plane and can get a simple answer
$\pi  R_2^2 | \sin \lp \Theta_v \rp | / 4$. 
Since there are four of such transformations and one has to consider them for  
all x-coordinates of the first cylinder (the length is $2\, R_1$), then 
one finds the third contribution as follows
\begin{equation}
v_{III} (h_1) = \pi  R_1^2 \,2 \, R_1 | \sin \lp \Theta_v \rp | \,\,.
\end{equation}


\begin{figure}
\centerline{\epsfig{figure=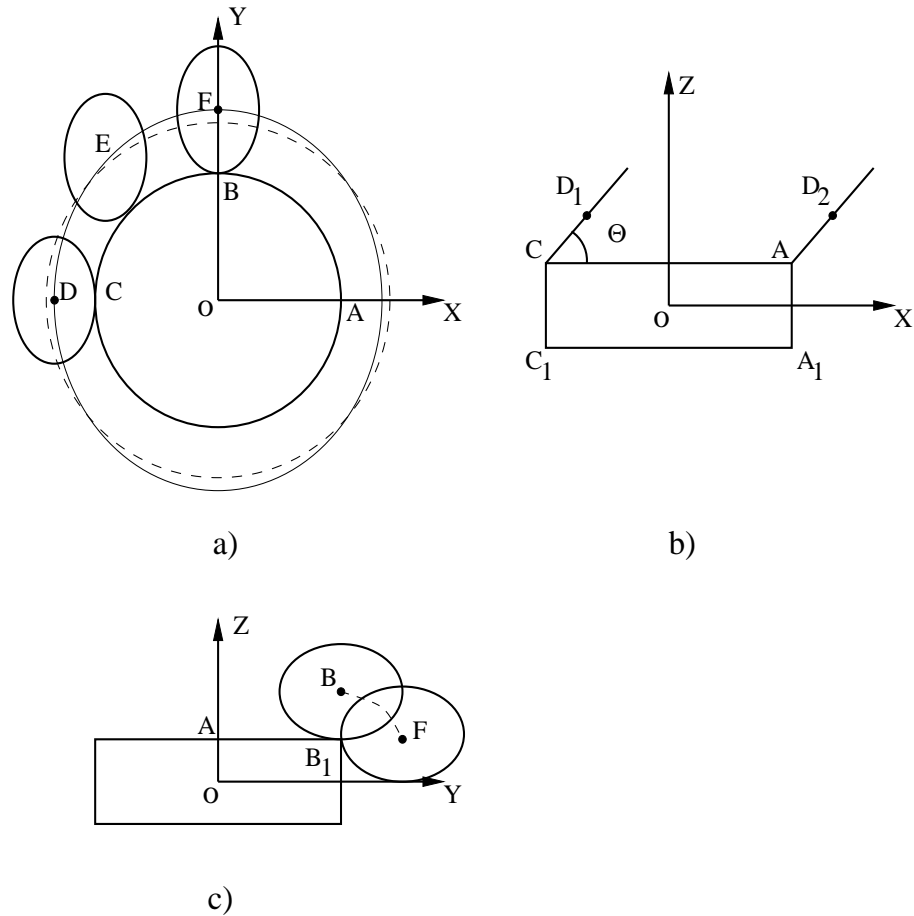,height=12cm,width=12cm}}

\vspace{-0.15cm}

\caption{\label{CylDeriv}
Relativistic  excluded volume derivation for relativistic 
cylinder $OABC$ and 
ultrarelativistic cylinder (disk)  $DC$ with radii $R_1$ and $R_2$, respectively.
$\Theta$ is the angle between their velocities.
Pictures a - c show the projections onto different planes.
The transformation of the cylinder $DC$  around the side of  
the cylinder $OABC$  is depicted in Fig. 3.a. 
The solid curve $DEF$ corresponds to the exact result, 
whereas  the dashed curve corresponds to the
average radius approximation
$\bra R_{XOY} \ket= OA + (DC + BF)/2 = R_1 + R_2 ( 1 + \cos \lp \Theta \rp)/2 $.
\newline
%
The transformation of the cylinder $DC = DC_1 = AD_2$  
along  the upper base of the cylinder $OABC = ACC_1A_1$ 
is  shown in \mbox{panel b.} 
Its contribution to the excluded volume
is a volume of the cylinder with the base  $AC = 2 R_1$ and the height  
$CD_1 \sin \lp \Theta \rp = R_2 \sin \lp \Theta \rp $. 
A similar contribution corresponds to the disk transformation along 
the lower base of the cylinder $A_1C_1$.
\newline
The third contribution to the relativistic excluded volume  
arises from the transformation of the  cylinder $DC = BB_1 = FB_1$  
from the upper base of the cylinder  $OABC = AB_1O$  to its side, and it
is schematically shown in Fig. 3.c.
The area $BB_1F \approx \pi / 4 R_2^2 \sin \lp \Theta \rp$  
is approximated as the one quarter of the area of the ellipse $BB_1$.
}
\end{figure}

Collecting all the contributions,  one obtains the excluded volume for the two cylinders of  
zero  and non-zero heights  
\begin{equation}\label{vcuns}
v_{2c} (h_1) = \pi \lp  R_1 + R_2  \cos^2 \lp \frac{\Theta_v}{2} \rp \rp^2 h_1 + 
2 \, \pi  R_1 R_2 (R_1 + R_2) | \sin \lp \Theta_v \rp | \,\,.
\end{equation}

\noi
The above equation, evidently,  gives an exact result for a zero angle and arbitrary height of 
the first cylinder.
Comparing it with the exact answer for  $\Theta_v = \frac{\pi}{2}$
\begin{equation}\label{vcyldisk}
v_{2c}^{E} \lp h_1, \Theta_v = \frac{\pi}{2} \rp = 
R_1 \lp \pi  R_1 + 4\, R_2   \rp  h_1 +
2 \, \pi  R_1 R_2 (R_1 + R_2)  \,\,,
\end{equation}

\noi
one finds that the dominant terms (the second terms in \req{vcuns} and \req{vcyldisk})   
again are  exact, whereas the corresponding corrections which are proportional to   $h_1$
are related to each other  as
$\frac{28,27}{28,56} \approx 0.9897$ (approximated to exact and for  $R_2 = R_1$).
Therefore, Eq. \req{vcuns} also gives a good approximation for the intermediate angles and 
small heights.

In order to get an expression for the non-zero heights of the second 
cylinder I note that the answer should be symmetric under the 
permutation of indexes 1 and 2. The lowest order correction 
in powers of the height  comes from the contribution $v_{I} (h_1)$. 
Adding the symmertic contribution $v_{I} (h_2)$ to  $v_{2c} (h_1) $ \req{vcuns}, 
one obtains the full answer 
\begin{eqnarray}
v_{2c} (h_1, h_2) & = & \pi \lp  R_1 + R_2  \cos^2 \lp \frac{\Theta_v}{2} \rp \rp^2 h_1 +
\pi \lp  R_2 + R_1  \cos^2 \lp \frac{\Theta_v}{2} \rp \rp^2 h_2  \nn
& + & 2 \, \pi  R_1 R_2 (R_1 + R_2)  \sin \lp \Theta_v \rp  \,\,.
\end{eqnarray}

\noi
The above expression  gives an exact result for a zero angle and arbitrary heights of cylinders.  
It also gives nearly exact answer for $\Theta_v = \frac{\pi}{2}$ in either limit
$h_1$ or $h_2 \rightarrow 0$.



\begin{figure}

\mbox{
\hspace*{0.0cm}\epsfig{figure=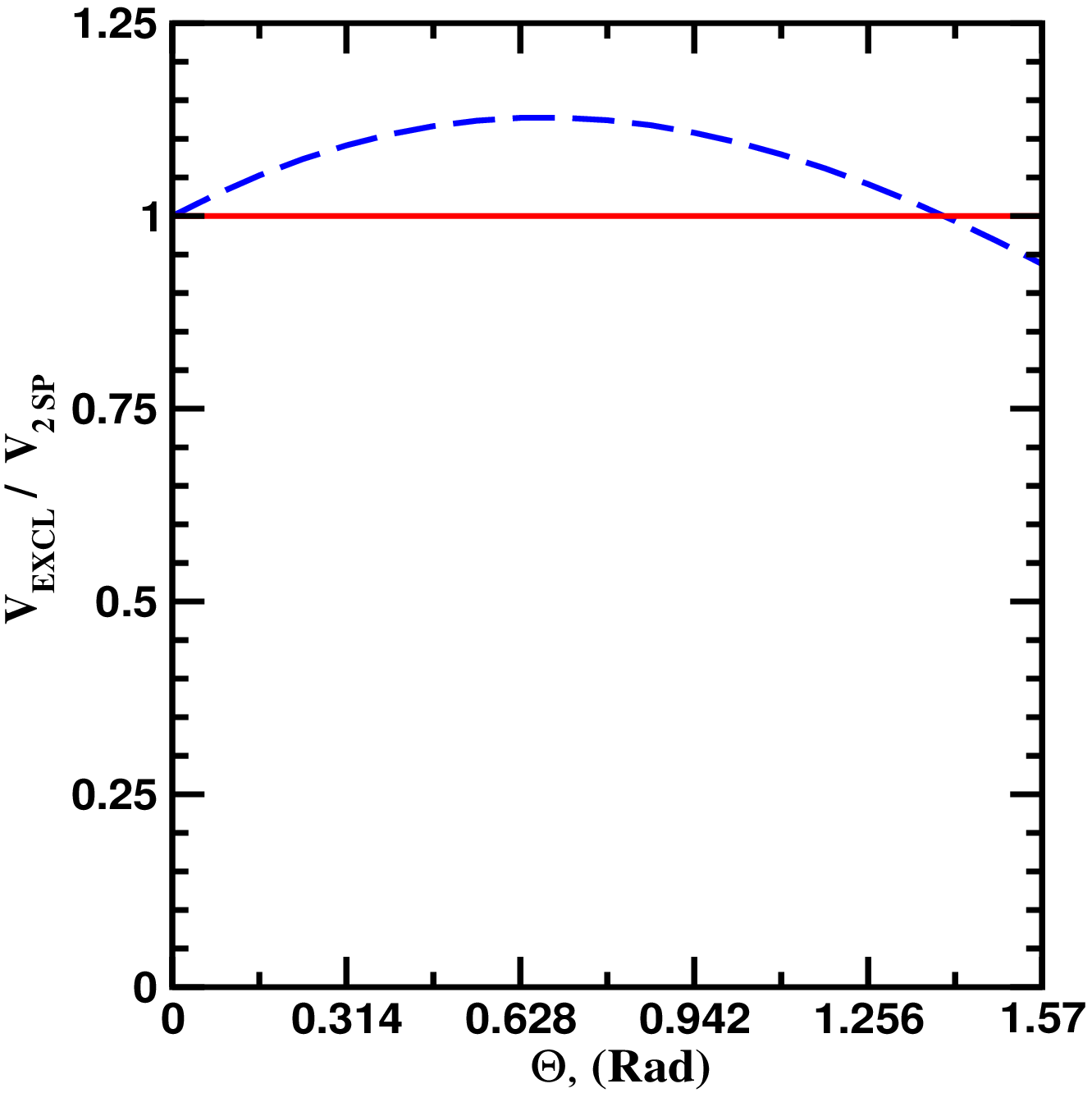,height=7.0cm,width=8.cm} 
\hspace*{-0.5cm} \epsfig{figure=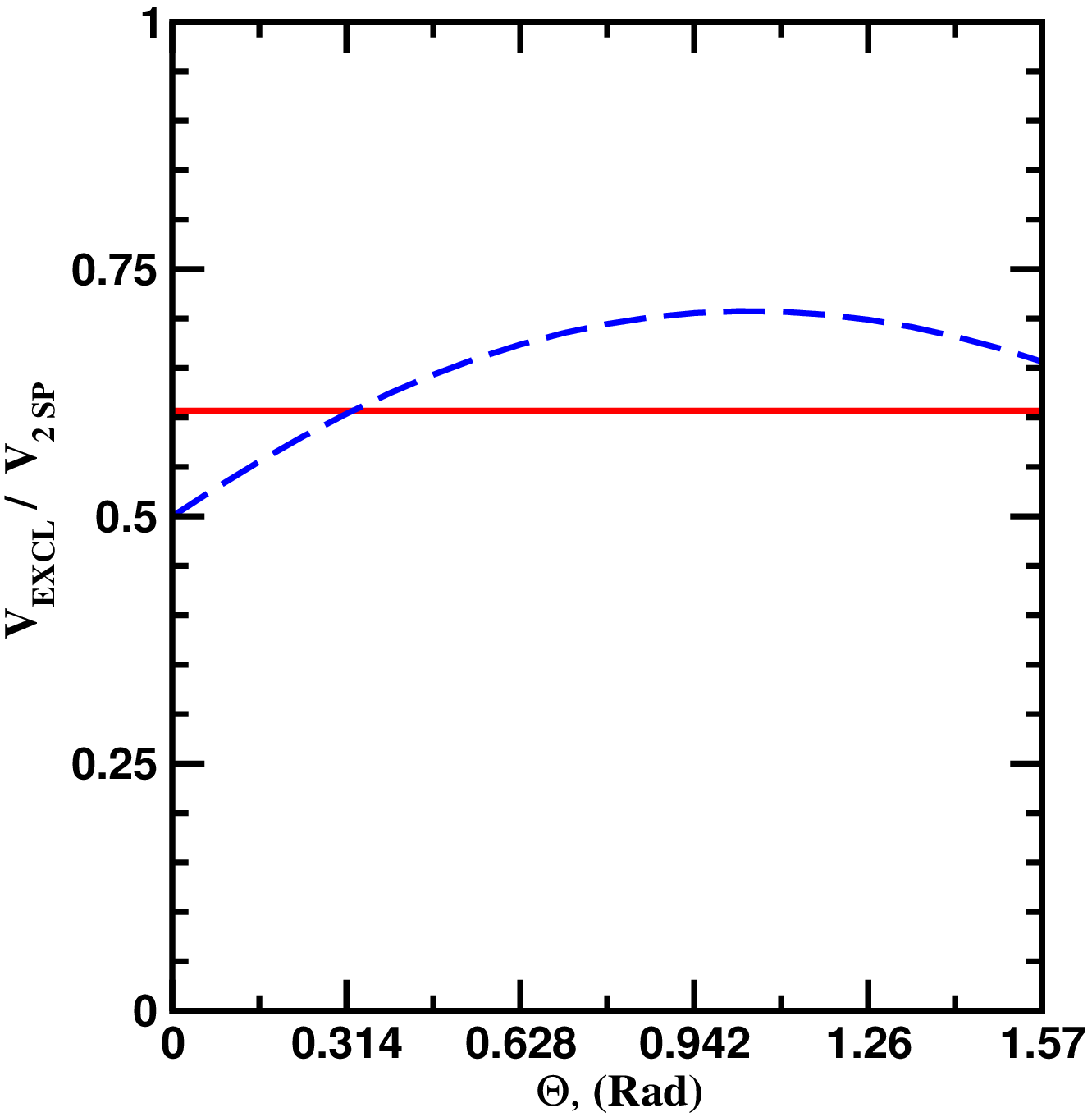,height=7.0cm,width=8.cm}
}

\vspace*{0.3cm}

\caption{\label{Comp2} 
Comparison of the relativistic  excluded volume obtained by the 
approximative ultrarelativistic formula 
with the exact results. 
The left panel shows the quality of the approximation $V_{EXCL} \equiv v^{Urel} (R, R) $ 
\req{vunnor} to 
describe the excluded volume 
of  two nonrelativistic spheres $V_{2 SP}$ of same radius $R$ as  
a function of the spherical angle $\Theta$, whereas  
the right panel depicts the approximation to the excluded volume 
of the nonrelativistic sphere and disk. 
In both panels the solid curve  corresponds to the exact result and the long dashed one 
corresponds 
to the ultrarelativistic approximation by two cylinders.   
The averaged ultrarelativistic  excluded volume 
in the left panel is 
 $\frac{ \bra V_{EXCL} \ket_{\Theta} }{ V_{2 SP} } \approx 1.065 $.
The corresponding averaged value for the  right  panel 
is $\frac{ \bra V_{EXCL} \ket_{\Theta} }{ V_{2 SP} } \approx 0.655 $, which
should be compared with the exact value 
$\frac{ \bra V_{EXCL} \ket_{\Theta} }{ V_{2 SP} } \approx 0.607 $.
}
\end{figure}

Choosing the heights to reproduce the proper volume of each of the Lorentz contracted 
spheres, one gets an approximation for the excluded volume of contracted spheres
in ultrarelativistic limit
\begin{eqnarray}
v^{Urel} (R_1, R_2) & =  &  \frac{4 }{3 } \pi  \frac{ R_1}{ \g_1}  \lp  R_1 + 
R_2  \cos^2 \lp \frac{\Theta_v}{2} \rp \rp^2  +
\frac{4}{3}  \pi \frac{R_2}{\g_2} 
 \lp  R_2 + R_1  \cos^2 \lp \frac{\Theta_v}{2} \rp \rp^2  \nn
\label{vunnor}
& + & 2 \, \pi  R_1 R_2 (R_1 + R_2)  \sin \lp \Theta_v \rp  \,\,.
\end{eqnarray}
The  corresponding $\g_q$-factors ($ \g_q \equiv E(\1 k_q)/ m_q$,  $q = \{1, 2\}$) are defined  in the local rest frame of the whole system for particles of mass $m_q$.
The last result  is valid for $0 \le \Theta_v \le \frac{\pi}{2} $, to use it for 
$ \frac{\pi}{2} \le \Theta_v \le \pi $ one has to replace  
$ \Theta_v \longrightarrow \pi -  \Theta_v$ in it. 

It is necessary  to mention that the above formula gives a surprisingly  
good approximation even in nonrelativistic limit for 
the excluded volume of two spheres.
For $R_2 = R_1 \equiv R$ one finds that the  maximal excluded volume corresponds 
to the  angle $\Theta_v = \frac{\pi}{4}$ and its value is \mbox{$\max \{ v^{Urel} (R, R) \}
\approx \frac{36}{3} \pi R^3$},  whereas  an  exact result for nonrelativistic
spheres is $ v_{2s} = \frac{32}{3} \pi R^3$, i.e., 
the ultrarelativistic formula \req{vunnor}  describes a nonrelativistic
situation with the maximal deviation of about $10 \%$ (see the left panel in
 Fig.~\ref{Comp2}).

Eq. \req{vunnor} also describes the excluded volume $v_{sd} = \frac{10 + 3 \pi}{3} R^3$ for a nonrelativistic sphere and ultrarelativistic ellipsoid  with the maximal deviation 
from the exact result
of about $15 \%$ (see the right panel in Fig.~\ref{Comp2}).

In order to improve its accuracy for the nonrelativistic
case, I introduce a factor $\a$ to normalize the integral
of the excluded volume \req{vunnor}
over the whole solid angle to the volume of two spheres
\begin{equation}\label{vcorr}
v^{Nrel} (R_1, R_2)
= \a \,\, v^{Urel} (R_1, R_2)\,; \hspace*{0.9cm}
\a = \frac{4  \p \lp  R_1 + R_2  \rp^3 }{\textstyle  3 \, \biggl. \int\limits_{\,0}^{\pi} 
d \Theta_v
\, \sin\lp \Theta_v \rp \,  v^{Urel} (R_1, R_2) \biggr|_{\g_1 = \g_2 = 1}
}
\,\,.
\end{equation}

For equal hard core radii  and equal masses of particles f the normalization factor reduces to the following value
$\a \approx \frac{1}{1.0654}$, i.e., it compensates the  most of the deviations
discussed above.
With such a correction the excluded volume \req{vcorr}  can be safely used for
the nonrelativistic domain because in this case 
the VdW excluded volume effect is itself  a correction to the ideal gas
and, therefore, the remaining deviation (less than 1.7 \%  for the whole available range of parameters)  from the exact result is
of  higher order.

It is useful to have the relativistic excluded volume expressed
in terms of 3-momenta 
\begin{eqnarray}
v^{Urel} (R_1, R_2) & = &   \frac{ v_{01}}{ \g_1}  \lp 1 +
R_2   
\frac{ |\1 k_1| |\1 k_2| + |\1 k_1 \cdot \1 k_2| }{2 \, R_1\,|\1 k_1| |\1 k_2|} 
\rp^2  
+
\frac{ v_{02}}{ \g_1}  \lp 1 +
R_1   
\frac{ |\1 k_1| |\1 k_2| + |\1 k_1 \cdot \1 k_2| }{2 \, R_2\,|\1 k_1| |\1 k_2|} 
\rp^2 
\nn
& + & 2 \, \pi  R_1 R_2 (R_1 + R_2) 
\frac{ |\1 k_1 \times \1 k_2| }{|\1 k_1| |\1 k_2|}
\,\,,
\end{eqnarray}

\noi
where $v_{0q}$ denote the corresponding eigen volumes 
$v_{0q} = \frac{4}{3} \pi R_i^3\,, \quad  q=\{ 1, 2\} $.  

For the practical calculations it is necessary to  
express the relativistic excluded volume in terms of the 
three 4-vectors - the two  4-momenta  of particles, $k_{q \, \mu}$,  and
the collective 4-velocity $ u^\mu = \frac{1}{\sqrt{1 - \1 v^2}} ( 1, \1 v)$.  
For this purpose one should reexpress the  gamma-factors and  at least one of    
 trigonometric  functions in \req{vunnor}  in a covariant form 
\begin{equation}\label{covar}
\g_q = \frac{\sqrt{m^2 + \1 k^2_q}}{m} = \frac{k_q^\m \, u_\m}{m}, \quad 
\cos\lp \Theta_v \rp = \frac{ k_1^\m \, u_\m \,\, k_2^\n \, u_\n - 
k_1^\m \, k_{2\,\m} }{
\sqrt{\lp (k_1^\m \, u_\m )^2 - m^2 \rp \lp (k_2^\m \, u_\m )^2 - m^2 \rp  } 
}\,\,.
\end{equation}
Using Eq. \req{covar},  one can express any trigonometric function of $ \Theta_v$ in a covariant form. 




















\def\np#1{{\it Nucl. Phys.} {\bf #1}}
\def\prl#1{{\it Phys. Rev. Lett.} {\bf #1}}
\def\jp#1{{\it J. of Phys.} {\bf #1}}
\def\zp#1{{\it Z. Phys.} {\bf #1}}
\def\pl#1{{\it Phys. Lett.} {\bf #1}}
\def\pr#1{{\it Phys. Rev.} {\bf #1}}
\def\hip#1{{\it Heavy Ion Physics} {\bf #1}}
\def\prep#1{{\it Phys. Rep.} {\bf #1}}
\def\preprint#1{{\it Preprint} {\bf #1}}


\renewcommand{\bibname}{The List of Used Sources}

\end{document}